\documentclass[preprint]{aastex}

\slugcomment{Submitted to ApJS}

\shorttitle{Planets orbiting M dwarfs}
\shortauthors{Tuomi et al.}

\begin{document}

\title{Frequency of planets orbiting M dwarfs in the Solar neighbourhood}

\author{M. Tuomi\altaffilmark{1,\star}, H. R. A. Jones\altaffilmark{1}, R. P. Butler\altaffilmark{2}, P. Arriagada\altaffilmark{2,3}, S. S. Vogt\altaffilmark{4}, J. Burt\altaffilmark{4}, G. Laughlin\altaffilmark{4}, B. Holden\altaffilmark{4}, S. A. Shectman\altaffilmark{5}, J. D. Crane\altaffilmark{5}, I. Thompson\altaffilmark{5}, S. Keiser\altaffilmark{2}, J. S. Jenkins\altaffilmark{6}, Z. Berdi\~nas\altaffilmark{6}, M. Diaz\altaffilmark{6}, M. Kiraga\altaffilmark{7}, J. R. Barnes\altaffilmark{8}}

\altaffiltext{1}{Centre for Astrophysics Research, School of Physics, Astronomy and Mathematics, University of Hertfordshire, College Lane, AL10 9AB, Hatfield, UK}
\altaffiltext{2}{Department of Terrestrial Magnetism, Carnegie Institution of Washington, 5241 Broad Branch Road NW, Washington D.C. USA 20015-1305}
\altaffiltext{3}{Department of Astronomy, Pontificia Universidad Cat\'olica de Chile, Casilla 306, Santiago 22, Chile}
\altaffiltext{4}{UCO/Lick Observatory, University of California, Santa Cruz, CA 95064, USA}
\altaffiltext{5}{The Observatories of the Carnegie Institution of Washington, 813 Santa Barbara Street, Pasadena, CA USA 91101}
\altaffiltext{6}{Departamento de Astronom\'ia, Universidad de Chile, Camino del Observatorio 1515, Las Condes, Santiago, Chile}
\altaffiltext{7}{Warsaw University Observatory, Aleje Ujazdowskie 4, 00-478 Warszawa, Poland}
\altaffiltext{8}{Department of Physical Sciences, The Open University, Walton Hall, Milton Keynes, MK7 6AA, UK}
\altaffiltext{$\star$}{Corresponding author E-mail: \texttt{mtuomi.astro@gmail.com}}

\begin{abstract}
The most abundant stars in the Galaxy, M dwarfs, are very commonly hosts to diverse systems of low-mass planets. Their abundancy implies that the general occurrence rate of planets is dominated by their occurrence rate around such M dwarfs. In this article, we combine the M dwarf surveys conducted with the HIRES/Keck, PFS/Magellan, HARPS/ESO, and UVES/VLT instruments supported with data from several other instruments. We analyse the radial velocities of an approximately volume- and brightness-limited sample of 426 nearby M dwarfs in order to search for Doppler signals of cadidate planets. In addition, we analyse spectroscopic activity indicators and ASAS photometry to rule out radial velocity signals corresponding to stellar activity as Doppler signals of planets. We calculate estimates for the occurrence rate of planets around the sample stars and study the properties of this occurrence rate as a function of stellar properties. Our analyses reveal a total of 118 candidate planets orbiting nearby M dwarfs. Based on our results accounting for selection effects and sample detection threshold, we estimate that M dwarfs have on average at least 2.39$^{+4.58}_{-1.36}$ planets per star orbiting them. Accounting for the different sensitivities of radial velocity surveys and \emph{Kepler} transit photometry implies that there are at least 3.0 planets per star orbiting M dwarfs. We also present evidence for a population of cool mini-Neptunes and Neptunes with indications that they are found an order of magnitude more frequently orbiting the least massive M dwarfs in our sample.
\end{abstract}




\newpage


\section{Introduction}

The smallest stars, called M dwarfs, are by far the most common stars in the Solar neighbourhood and in the whole Galaxy \citep{henry1994,chabrier2000,winters2015}. This means that the occurrence rate of planets around M dwarfs dominates the general occurrence rates of planets around main sequence stars.

Recent studies have reported high occurrence rates for low-mass planets orbiting M dwarfs \citep{bonfils2013,dressing2013,dressing2015,tuomi2014} of at least one planet per star. Based on radial velocity observations of an approximately volume- and brightness-limited sample, \citet{bonfils2013} reported that super-Earths are abundant around M dwarfs with occurrence rates of 0.36$^{+0.25}_{-0.10}$ and 0.52$^{+0.50}_{-0.16}$ planets per star in the period intervals of 1-10 and 10-100 days, respectively. Similarly, \citet{tuomi2014} estimated that there are 0.06$^{+0.11}_{-0.03}$ and 1.02$^{+1.48}_{-0.69}$ planets per star in the period intervals of 1-10 and 10-100 days, respectively, with minimum masses between 3-10 M$_{\oplus}$. However, although these values are consistent, they were both based on only 10 detected planetary signals and are thus highly uncertain.

Similar occurrence rates have been reported based on the \emph{Kepler} transit photometry. For instance, \citet{dressing2013} reported an occurrence rate of 0.90$^{+0.04}_{-0.03}$ for planets with radii in the range 0.5-4.0 R$_{\oplus}$ and orbital periods shorter than 50 days. In a more recent work, \citet{dressing2015} reported a cumulative occurrence rate of 2.5$\pm$0.2 planets per M dwarf with radii between 1-4 R$_{\oplus}$ and orbital periods below 200 days. This is consistent with the estimate of 1.9$\pm$0.5 based on combination of radial velocity and microlensing planet detections \citep{clanton2015}. Furthermore, half of the M dwarfs observed with \emph{Kepler} are surrounded by systems of high multiplicity with five or more planets on co-planar orbits \citep{ballard2016}. These results imply that virtually all M dwarfs are hosts to low-mass planets that can be categorised as Earths, super-Earths or mini-Neptunes and that there are likely, on average, more than two planets orbiting any individual M dwarf star. We will explain what we mean by when we call planets 'Earths', 'super-Earths', etc. in Section \ref{sec:what_is_a_planet}.

In contrast, giant planets do not appear to be very common orbiting M dwarfs \citep{butler2004,endl2006,johnson2007,cumming2008}. Based on a sample of 150 M dwarfs observed with \emph{Keck}, \citet{butler2004} discovered only one giant planet, enabling them to estimate the occurrence rate of planets with masses above 0.3 M$_{\rm Jup}$ with orbital periods below 1.5 years to be approximately 0.007 planets per star. This result was extended up to orbital distances of 2.5 AU by \citet{johnson2007}, who obtained an occurrence rate of 0.018$\pm$0.010 such planets per star. In a similar manner, \citet{endl2006} observed 90 M dwarfs with the \emph{Hobby-Eberly Telescope} but did not find any Jovian planets with semi-major axes below 1 AU yielding an upper limit for their occurrence rate of 0.0127. Moreover, \citet{cumming2008} estimated that 1\% of M dwarfs are hosts to giant planets with masses above 0.3 M$_{\rm Jup}$ and periods between 2-2000 days. Based on radial velocities from HARPS spectrograph, \citet{bonfils2013} obtained estimates of $\lesssim$ 0.01 and 0.02$^{+0.03}_{-0.01}$ planets per star for giant planets with 100 M$_{\oplus} < m_{p} \sin i < 1000$ M$_{\oplus}$ with orbital periods between 1-10 and 10-100 days, respectively. All these results appear to be consistent, as has been verified for giant planet occurrence rates from microlensing and radial velocity data \citep{clanton2015}, and also their combination with candidates from direct imaging \citep{clanton2016}, and imply that giant planets indeed are rarely companions to M dwarfs on orbits with periods up to 2000 days and beyond.

Planets appear to be more common around nearby G and K stars the smaller they are and their occurrence rate also increases as a function of orbital period in the range 0.5-50 days \citep{howard2012}. Moreover, \citet{howard2012} reported an effective cutoff period below which planets are extremely rare and estimated it to be somewhere between 2-7 days depending on the planetary mass. However, their results were limited to minimum planetary radii of 2R$_{\oplus}$ and thus only applicable for mini-Neptune type planets and the largest super-Earths and more massive objects. This gives rise to an occurrence rate of 0.19 planets per star for orbital periods of up to 50 days and radii in excess of 2R$_{\oplus}$ \citep{youdin2011}. Moreover, despite a reasonably high false positive rate of up to 18\% \citep{fressin2013}, the \emph{Kepler} transit photometry data has been used to estimate that the occurrence rate of small planets with radii in excess of 1R$_{\oplus}$ is likely as high as one per star for stars of the spectral types F, G, and K \citep{fressin2013}.

In this article, we extend the work of \citet{tuomi2014} and \citet{bonfils2013} by estimating the occurrence rates of planets around M dwarfs based on a larger sample. First, we include the volume-limited sample of 102 M dwarfs discussed in \citet{bonfils2013} and the UVES targets of \citet{zechmeister2009} analysed in \citet{tuomi2014}. We also extend the sample by including all M dwarfs that have been, according to our knowledge, observed by HARPS. This includes 327 HARPS targets out of which 225 have not been included in the HARPS-GTO programme \citep[see e.g.][]{bonfils2013}. Moreover, we have obtained data for 67 nearby M dwarfs with PFS and 159 with HIRES spectrographs extending the sample size to a total of 426. In addition to these four major data sets from four different instruments, we have obtained data from AAT, APF, CORALIE, ELODIE, HARPN, HET, LICK, and SOPHIE when such data was either available in the corresponding archive or published when discussing previously known planets orbiting the corresponding stars, or observed by us, which is especially the case with data from APF. The data, as well as the instruments and telescopes used to obtain them, including the acronyms in the current paragraph, are discussed in Section \ref{sec:data}.

If we detect a signal in the combined radial velocity data of a given target, we investigate whether stellar activity could be shown to be responsible for the radial velocity signal by analysing the \emph{All Sky Automated Survey} (ASAS) V and I-band (if available) photometry \citep[see][]{pojmanski2002}. Photometric signals at or close to the radial velocity ones are typically interpreted as an indication that the corresponding periodicities are in fact caused by magnetic activity, activity cycles, and/or stellar rotation coupled with active and/or inactive surface features rather than planets. Stellar activity can cause periodicities in radial velocities mimicking planetary signals and/or disabling their detection and accounting for all available information regarding stellar activity is therefore crucial \citep[e.g.][]{anglada2016,newton2016}.

Similarly, we study the variations in the spectral activity indices -- bisector span, full-width at the half-maximum and the CaII H\&K lines -- that reflect the activity of the star \citep[e.g.][]{santos2010} and could have counterparts in the radial velocity data \citep[e.g.][]{dumusque2011}.

Finally, we calculate the occurrence rate of planets for the sample of M dwarfs as a function of planetary minimum masses and orbital periods. Because the available Doppler data is sensitive for minimum masses as low as 1-2 M$_{\oplus}$ \citet[e.g.][]{anglada2016}, we can estimate the occurrence rate of low-mass planets in the stellar habitable zones (HZs). The existence of such planets is well-established for F, K and G stars \citep{kane2016} and there are also numerous examples of planet candidates labeled as ``habitable-zone super-Earths'' orbiting M dwarfs from Doppler spectroscopy surveys \citep{bonfils2013b,tuomi2013c,anglada2013,anglada2014,anglada2016,tuomi2014,wittenmyer2014,astudillo2015,wright2016} as well as from \emph{Kepler} transit photometry survey \citep{batalha2013,quintana2015,morton2016}.

As a consequence, it has been possible to estimate the occurrence rate of such planets orbiting M dwarfs based on radial velocity surveys. For instance, \citet{bonfils2013} reported a frequency of 0.41$^{+0.54}_{-0.13}$ planets with minimum masses between 1 M$_{\oplus}$ and 10 M$_{\oplus}$ in the stellar HZs whereas \citet{tuomi2014} estimated that there are 0.21$^{+0.03}_{0.05}$ HZ planets with minimum masses between 3 M$_{\oplus}$ and 10 M$_{\oplus}$. Comparable estimates based on transit photometry have been reported by \citet{dressing2015}. According to their results for planets in ``conservative HZ'', there are 0.16$^{+0.17}_{-0.07}$ Earth-size planets and 0.12$^{+0.10}_{-0.05}$ super-Earths per star. The corresponding numbers could be as high as 0.24$^{+0.18}_{-0.08}$ and 0.21$^{+0.11}_{-0.06}$ if the HZ is defined slightly more optimistically, respectively. However, these estimates depend on the selected planetary radius interval and the definition of the habitable zone. For ``simple'' HZ boundaries of \citet{petigura2013} there could be planets with radii between 1 R$_{\oplus}$ and 2 R$_{\oplus}$ for 83\% of M dwarfs \citep{dressing2015}. We aim at revising these estimates in the current work as well.

We summarise the properties of the data and the instruments used to obtain them in Section \ref{sec:data}, describe the statistical methods applied in our analyses in Section \ref{sec:methods} and discuss the properties of the sample stars in Section \ref{sec:sample}. We then discuss the planet population orbiting the target stars in Section \ref{sec:planets} and calculate the occurrence rate of low-mass planets for the whole sample in Section \ref{sec:occurrence} before discussing the results in Section \ref{sec:discussion}.

\section{Data}\label{sec:data}

We describe here the data obtained from the European Southern Observatory (ESO) archive for HARPS (Section \ref{sec:data_HARPS}), the data we observed with HIRES/Keck (Section \ref{sec:data_HIRES}) and PFS/Magellan (Section \ref{sec:data_PFS}), and the data that has been published previously in a number of papers reporting planets orbiting the stars in the current sample. We also discuss the ASAS photometry data, its general properties and quality, in Section \ref{sec:data_ASAS}. Selected statistics of the data sets are summarised in Table \ref{tab:data_velocities}.

Essentially, we include in this work all the Doppler spectroscopy data of M dwarfs that is available to us. This includes the data we have obtained with HIRES, PFS, and APF (and some data from HARPS and UCLES) but also all publicly available data (as far as we are aware) in the archives, most notably for HARPS, but also for SOPHIE and HARPN. Moreover, we also include all the data of nearby M dwarfs that (according to our knowledge) has been made public when e.g. announcing the discoveries of planets orbiting them.

As a total, our set of radial velocity data consists of 23473 individual observations of 426 different targets.

\subsection{HARPS radial velocity and activity data}\label{sec:data_HARPS}

\emph{High-Accuracy Radial velocity Planet Searcher} \citep[HARPS;][]{mayor2003} is one of the most precise instruments in the search for Doppler signals of planets orbiting nearby stars. We acknowledge the achievements of HARPS by referring to a number of interesting discoveries of planets orbiting nearby M dwarfs based on HARPS data \citep[e.g.][]{bonfils2005,bonfils2007,bonfils2013b,udry2007,forveille2009,forveille2011,mayor2009,anglada2012c,anglada2012,anglada2014,anglada2016,tuomi2013c,delfosse2013,locurto2013,tuomi2014}.

We obtained the HARPS data products from the European Southern Observatory (ESO) archive for all the targets that we considered M dwarfs\footnote{Based on their classifications as M dwarfs in SIMBAD Astronomical Database (\texttt{simbad.u-strasbg.fr}.)}. This resulted in data for 327 targets that have been observed with HARPS a total of 10311 times -- on average 32 observations per target. However, the median number of observations per target is only 11, which highlights the fact that half of the targets have only been observed a handful of times. We processed these data products by calculating the \emph{Template-Enhanced Radial velocity Re-analysis Application} velocities for HARPS (HARPS-TERRA) according to the techniques presented in \citet{anglada2012b}. The HARPS data sets have, on average, a baseline of 1770 days, which enables the detections of planets with orbital periods of up to 3-5 years. The resulting HARPS velocities were then accompanied with the corresponding line bisector span (BIS) and full-width at half-maxima (FWHM) measures that provide information on the contribution of the stellar activity on the corresponding line profiles and thus radial velocities \citep[e.g.][]{santos2010,boisse2011}.

Stellar activity cycles that are detectable in the activity indicators such as BIS, and FWHM, but also CaII H\&K lines, can introduce variations into radial velocity measurements of M dwarfs \citep{gomez2011,gomez2012}. We thus also obtained the resulting measures of CaII H\&K line emissions, the S-indices, for each observation and use the triplet (BIS, FWHM, S-index) as explanatory variables of the HARPS radial velocities to minimise the effects of activity and to avoid detecting signals that do not correspond to candidate planets but are caused by periodic activity-induced phenomena such as starspots coupled with stellar rotation and magnetic cycles.

With activity indicators available, it was then possible to exclude suspicious points corresponding to outliers in the activity indicators that might be affected by e.g. stellar flares or observational systematics \citep[see e.g.][]{anglada2016}. This typically resulted in the exclusion of 0-5 radial velocities for roughly 30\% of the targets, but did not make a significant difference on the obtained results in practice. However, this omission of outliers was essential in order to study the systematical relationships between the radial velocities and the activity indicators as discussed in e.g. \citet{gomez2011,gomez2012}.

A total of 103 of the HARPS targets were included in the HARPS \emph{Guaranteed Time Observations} (GTO) sample \citep{bonfils2013} used to study the statistics of the Doppler spectroscopy planets within an approximately volume-limited and brightness-limited sample. However, many more nearby M dwarfs have been observed with HARPS by the HARPS-ESO group, such as the well-known planet hosts GJ 676A \citep{forveille2011,anglada2012c} and GJ 163 \citep{bonfils2013b,tuomi2013c} but were not included in the HARPS-GTO search for nearby planets. Moreover, there are several stars that have been observed with HARPS -- data that is publicly available -- without any publicly available information regarding the selection of the targets. We thus simply downloaded the target lists of the proposals of the HARPS-ESO group aiming at detecting planets orbiting M dwarfs and downloaded the corresponding publicly available data from the archive.

\subsection{HIRES radial velocity and activity data}\label{sec:data_HIRES}

\emph{High Resolution Echelle Spectrometer} \citep[HIRES;][]{vogt1994} at the Keck I telescope in Hawaii, USA, has been used to conduct one of the longest-running Doppler spectroscopy surveys of nearby planets. HIRES observations have also been used to detect a large number of planets orbiting nearby stars \citep[again, references provided to discoveries of planets orbiting M dwarfs: e.g.][]{marcy2001,butler2004,rivera2005,butler2006,haghighipour2010,howard2010,howard2014,johnson2010,burt2014}. The HIRES spectrograph has been instrumental in conducting the \emph{Lick-Carnegie Exoplanet Survey} that is the longest running survey of nearby planets.

We include all the HIRES targets (159) in our analysis that were considered M dwarfs, or at most late K stars that are in practice difficult to differentiate from M dwarfs. On average, these targets have been observed 55 times with HIRES \citep[see also][]{butler2016}. We have therefore a total of 8687 HIRES observations of nearby M dwarfs in our combined data set. The HIRES data sets have an average baseline of 4660 days, which makes HIRES precision radial velocities an extremely important source of information for our purposes by providing data sets with long baselines enabling the detections of long-period planets. The HIRES data has been published by \citet{butler2016}. In the current work we also verify some of the signals of candidate planets reported by \citet{butler2016} by combining the HIRES observations with other independent data sets.

To enable us to distinguish between stellar long-period activity cycles and correspondingly long-period planetary orbits, we used the emission on CaII H\&K lines as tracers of stellar activity and calculated the S-indices for the data sets. These S-indices were used as explanatory variables of the variations in the radial velocities in an attempt to remove the activity-induced variations from the velocity data sets.

\subsection{PFS radial velocity data}\label{sec:data_PFS}

Mounted on the 6.5-m Magellan Clay telescope of the Las Campanas Observatory, Chile, the \emph{Planet Finder Spectrograph} (PFS) is similar to HARPS in precision even though, like HIRES, uses the iodine cell calibration technique to detect Doppler shifts \citep{crane2010}. Several candidate planets have been detected orbiting M dwarfs based on PFS observations over the recent years \citep[e.g.][]{arriagada2013,anglada2013,anglada2014,wittenmyer2014}.

We obtained an average of 25 PFS radial velocities per star and a total of 1759 of them for 69 stars in the solar neighbourhood that could be classified as M dwarfs. PFS is thus the third largest source of radial velocity information in the current work. However, the baselines of the PFS data sets are, on average, only 1240 days, which is the shortest out of the four main sources of data.

\subsection{UVES radial velocity data}\label{sec:UVES}

The \emph{Ultraviolet and Visual Echelle Spectrograph} \citep[UVES;][]{dekker2000} of VLT-UT2 has been used to observe the radial velocities of 41 nearby M dwarfs \citep{zechmeister2009}. This set consists of 1789 observations with an average baseline of 1300 days and was also analysed in \citet{tuomi2014}.

The code used to calculate the barycentric correction of the UVES velocities, as published in \citet{zechmeister2009}, appears to have had a mistake resulting in small but possibly significant biases in practice\footnote{Based on private communication with M. Zechmeister.}. However, as we cannot reprocess and publish the UVES velocities with a corrected code in this work, we use the velocities of \citet{zechmeister2009} with this caveat in mind. Therefore, if it appears that a signal in a combined data including UVES velocities shows evidence in favour of Keplerian periodicities, we require that the signals are supported by other instruments as well before accepting them as significant solutions. We note that members of our group are preparing a re-reduction of UVES velocities that should enable improving their precision and reliability by removing such biases as was done for the UVES velocities of GJ 551 \citep{anglada2016}.

The UVES velocities will still enable ruling out signals and providing phase-coverage as well as longer baselines than would be available without them. We thus estimate that they are useful for the purposes of the current work but do not rely on them when detecting signals in the radial velocity data.

\subsection{Other radial velocity data sets}


Some relatively old data from the \emph{Lick Observatory Hamilton Echelle Spectrometer} \citep{vogt1987} was also included in our analyses. Primarily this was done to take advantage of the resulting long baselines of the combined data sets even though, as one of the earliest instruments in Doppler spectroscopy planet searches, the data is not as precise as that from the other more modern spectrographs.

We included the Lick data of GJ 699 (Barnard's star; Section \ref{sec:GJ699}) in our analyses because it provided a baseline of 7000 days. Although not particularly precise and already analysed in \citet{choi2013}, we included the data set to better constrain the long-period acceleration of the GJ 699.


The \emph{Automated Planet Finder} \citep[APF;][]{radovan2014,vogt2014} of the \emph{Lick Observatory} in California, USA, was used to observe seven of the targes in our sample, including the planet host GJ 687 \citep{burt2014}. The other five APF targets included in our analyses are GJ 273 (Section \ref{sec:GJ273}), GJ 411 (Section \ref{sec:GJ411}), GJ 686 (Section \ref{sec:GJ686}), GJ 846 (Section \ref{sec:GJ846}), GJ 752A (Section \ref{sec:GJ752A}), and GJ 4070. There is a total of 272 APF radial velocities for these stars, on average 37 per star.

We note that the APF data provides a useful complementary source of information for the six targets by providing access to data with precision higher than that of HIRES for northern targets that typically cannot be observed with HARPS and for which data from HARPN, if it exists, has not been made available.


The \emph{University College London Echelle Spectrograph} \citep[UCLES;][]{diego1990} mounted on the \emph{Anglo-Australian Telescope} in Australia has been used to conduct one of the longest running searches for planets, the \emph{Anglo-Australian Planet Search} (AAPS), whose first detections were reported in 2001 \citep{butler2001,tinney2001}. More recently, Doppler spectroscopy observations of nearby stars with UCLES have resulted in detections of giant planets on long-period orbits due to the long duration of the AAPS \citep{tinney2011,wittenmyer2012}.

We obtained UCLES data for four targets in the sample, GJ 1, GJ 729, GJ 832, and GJ 887. Out of these targets, GJ 832 has been reported to be a planet host \citep{bailey2009,wittenmyer2014}. The UCLES data is useful in combination with data from other instruments due to its long baselines ranging from 3220 to 5500 days for the four targets in our sample.


We obtained two data sets from the \emph{Hobby Eberly Telescope} (HET) that has been used to search for planets orbiting nearby M dwarfs \citep{endl2003,endl2006}. Due to the fact that this data was mostly unavailable for us, we only have HET data for GJ 176 and GJ 179 that have been reported to be planet hosts based on HET observations \citep{endl2008,howard2010}, although GJ 176 b later turned out to have rather different orbital parameters \citep[][see also Section \ref{sec:GJ176}]{butler2009,forveille2009,robertson2015}.

Although the two HET data sets are not very large, we include them in the analyses because the corresponding targets are planet hosts and we do not wish to neglect any sources of information available to us.


We included the CORALIE and ELODIE data of GJ 876, a famous planet hosting M dwarf \citep{marcy1998,marcy2001,delfosse1998,rivera2005,rivera2010}, in our analyses. This was done because GJ 876 planetary system shows signatures of orbital evolution \citep{rivera2010} and we thus wished to have as long of a baseline of data as possible.


We obtained archived SOPHIE (\emph{Spectrographe pour l'Observation des Ph\'enom\`{e}nes des Int\'erieurs stellaires et des Exoplan\`{e}tes}) data \citep{perruchot2008} for two targets due to the fact that they showed some evidence for candidate planets (GJ 411 and GJ 686; Sections \ref{sec:GJ411} and \ref{sec:GJ686}), respectively. However, the SOPHIE data sets of these two targets did not contribute much to the respective orbital solutions due to the fact that larger and more precise data sets were also available for these two stars. Yet, we included the SOPHIE velocities in the analyses for complementarity and because they still might enable ruling out some solutions, thereby improving the robustness of the results.


HARPS-North \citep[HARPN;][]{cosentino2012} mounted on the \emph{Telescopio Nazionale Galileo} at La Palma, Spain, is a copy of the HARPS spectrograph but located in the northern hemisphere enabling a follow-up of the \emph{Kepler} objects of interest of suitable bright stars in the \emph{Kepler} field \citep[e.g.][]{hebrard2013,hebrard2014,bonomo2014,dumusque2014}. We obtained HARPN data of 7 of our targets from the archive, most notable GJ 300 that we identified as a planet host (Section \ref{sec:GJ300}). Although there was, on average, only 9 observations available for the seven targets, HARPN data is very precise enabling us to constrain the respective solutions better than without HARPN data. We also have a set of 50 HARPN velocities for GJ 273 obtained over a period of 5 nights to study the short-term radial velocity variability.



\subsection{ASAS photometry data}\label{sec:data_ASAS}

The \emph{All Sky Automated Survey} \citep[ASAS;][]{pojmanski1997,pojmanski2002} data that is publicly available in the ASAS archive\footnote{\texttt{www.astrouw.edu.pl/asas}} proved to be an excellent source of photometric information for the stars in our sample. Although some M dwarfs in the current sample are known to have periodic variations in their ASAS photometry data suggestive of stellar rotation -- we call these periodicities photometric rotation periods to differentiate between the actual rotation of the star and the photometrically measured periodicity -- for the majority of these targets, photometric rotation periods are not known. In \citet{kiraga2007}, the photometric rotation periods of a handful of M dwarfs were published, including some stars in our sample. We have summarised the results of \citet{kiraga2007} with respect to the 16 targets in our sample in Table \ref{tab:kiraga_rotations}. In this table, we also list the estimated rotation periods of sample stars that we could find in the literature.

ASAS photometry consists of high-quality time-series for a majority of the stars in our sample with typical baselines exceeding 2000 days and number of observations in excess of $\sim$300 -- up to 1000 points for some targets. The 360 targets for which ASAS photometry was available were observed on average 440 times. Twenty three of these showed evidence for contamination due to a nearby field star or a stellar companion. Regarding the rest of the ASAS targets in our sample, we have summarised the properties of the respective ASAS data sets in Table \ref{tab:asas}.

ASAS photometry data is observed simultaneously with five apertures that range in size from 2 to 6 pixels \citep{pojmanski2002}. This enabled us to choose the most suitable aperture for each target: the smallest one (aperture MAG0) for faintest objects and the largest (MAG4) for the brightest ones. In practice, we selected the aperture that showed the least variations. This resulted in us selecting the aperture number $n$ approximately according to $n = 12.5-V$.

We note that when discussing the number of photometric measurements from ASAS for a given target, we use the number of grade A measurements that are considered to have the highest quality. However, this is the number before removing the 5-$\sigma$ outliers and is thus, typically, 10-20 greater than the number of points used in the analyses. Yet, we use the number of grade A measurements because the number of outliers varies between apertures and between targets and this number is thus the most robust way of comparing and discussing the different data sets.

\section{Statistical and computational techniques}\label{sec:methods}

To consistently extend the work presented in \citet{tuomi2014}, we broadly applied the same statistical and numerical approaches they did to obtain results that can be easily compared with the occurrence rate estimates and the very existence of planetary signals in given data sets reported in \citet{tuomi2014}. However, we also introduced some important improvements in the statistical analysis techniques and statistical modelling of the radial velocity data (Section \ref{sec:rv_model}).

Moreover, we paid attention to the possibility that signals in the radial velocities might be connected to stellar activity rather than planets \citep[compare e.g.][]{anglada2014,anglada2016b,robertson2015b,jenkins2013,jenkins2014,santos2014} and analyse the data with an improved statistical model that takes into account the possibility of the radial velocity variations being connected to the corresponding variations in measures of stellar surface activity and brightness. While independent confirmation of the existence of the signals should ideally be required to claim that they are genuine Doppler signatures of \emph{bona fide} planets, such confirmation is rarely possible because high-precision spectrographs that could compete with the stability of HARPS, such as the PFS, APF, and HARPN, have not been available for long enough to provide sufficient amounts of data for independent confirmation of the majority of the weak signals that are present in the HARPS data\footnote{We also note that APF and HARPN are in the northern hemisphere making the overlap of targets with HARPS rather limited.}.

To rule out the stellar activity as a source of the periodic signals caused by e.g. the co-rotation of active and/or inactive regions on the stellar surface, we generalise the statistical model of \citet{tuomi2014} by adding a component describing the linear correlations between the velocities and activity indicators. For the HARPS data, the three activity indices we use for this purpose are the line bisector span (BIS) and full-width at half-maximum (FWHM) that are measures of the average spectral line width and asymmetry, respectively, and the Ca II H\&K index (S-index) that measures the photospheric variations. Therefore, the reference model we compare to models with $k>0$ Keplerian signals has seven free parameters for HARPS data when $k=0$. These parameters are the reference velocity ($\gamma$), linear acceleration ($\dot{\gamma}$), standard deviation of the excess white noise for each instrument ($\sigma_{l}$), correlation between the deviations from the mean of $i$th and $i-1$th measurement ($\phi$), and the parameters quantifying the linear dependence of the velocity data on BIS, FWHM, and S-index ($c_{1}$, $c_{2}$, and $c_{3}$, respectively). However, in the absence of such correlations with activity data, i.e. when the parameters $c_{1}$, $c_{2}$, and $c_{3}$ are not statistically significantly different from zero and their inclusion in the model does not increase the posterior probability of the model significantly, we use a simpler version of the model with only the first four parameters also used in \citet{tuomi2014}. For other data sets, activity information is typically not available apart from HIRES data, for which we accounted for the linear dependence of velocities on the S-indices \citep[see also][]{butler2016}. The modelling and analyses of the radial velocities are described in detail in Section \ref{sec:rv_model}.

\subsection{What is a radial velocity planet?}\label{sec:what_is_a_planet}

We describe here what we mean by a planet candidate detected by using radial velocity data. If the given conditions are satisfied for a given signal in a data set, we call it a planet candidate. Therefore, we propose the following definition for the purpose of determining whether a planet has been detected with radial velocities or not.

{~}\\
\textit{\textbf{Definition.} Doppler planet candidate is a radial velocity signal satisfying the below conditions 1-3.}

\begin{itemize}
  \item[1.] \textbf{Signal detection.} The data ($m$) contains a signal in accordance with the signal detection criteria discussed in e.g. \citet{tuomi2012} and \citet{tuomi2014}. First, the model containing $k$ Keplerian signals, model $\mathcal{M}_{k}$, satisfies $P(\mathcal{M}_{k} | m) > \alpha P(\mathcal{M}_{k-1} | m)$ for some threshold level $\alpha$. Here $P$ denotes the model probability and the condition implies that the probability ratio exceeds the \emph{a priori} set threshold $\alpha$. We choose this threshod conservatively to be $\alpha = 10^{4}$, or on the logarithmic scale that we adopt when discussing the significances of the signals, $\ln \alpha \approx 9.2$. Second, we require that the period and amplitude parameters are well-constrained from above and below. This requirement is made to ensure that the signal indeed is periodic (period constrained from above and below) and to ensure that the amplitude is indeed statistically significantly different from zero (constrained from below). Without this condition it could not be demonstrated that a given signal is periodic and significant.\footnote{In practice, this criterion can be replaced by other suitable signal detection criteria such as exceeding some sufficient likelihood-ratio threshold \citep[e.g.][]{anglada2012c,butler2016}.}
  \item[2.] \textbf{No activity-related counterparts.} A given signal should not have counterparts in spectral activity indicators, i.e. $S$-index, BIS, FWHM, etc., and modelling the correlations\footnote{We restrict our modelling to linear correlations but accounting for non-linearities justified by a astrophysical factors might be desirable in some cases.} with the radial velocities and the activity indices should not affect the signal, i.e. its detection and/or properties. It might naturally be the case that a given signal in radial velocity data has strong counterparts in activity data. Thus, if accounting for the correlations between them alters the solution, it can be concluded that the signal is most likely related to activity-induced variability rather than a planet. This is also the case if there is a periodicity in the activity data at or very near a corresponding periodicity in the radial velocities. Yet, the lack of such relation to activity does not mean that a given signal is not caused by activity. It does, however, decrease the probability of such a possibility because some radial velocity signals are well-known to have counterparts in activity data.
  \item[3.] \textbf{No photometric correspondence.} To be of planetary origin, a given radial velocity signal should not have periodic counterparts in photometric data. Such counterparts would indicate that activity-cycles, magnetic phenomena corresponding to variations in brightness, and/or stellar rotation coupled with active and/or inactive regions co-rotating on the stellar surface contribute to periodic and/or quasiperiodic variations in the radial velocities that we interpret as periodic signals. In such a case, the signals are the most likely caused by the surface of the star rather than planets orbiting it. Again, such correspondence does not conclusively rule out the possibility that a given radial velocity signal is caused by a planet but it does decrease the probability of such a hypothesis considerably.
\end{itemize}

{~}\\
The third criterion is not directly related to Doppler spectroscopy data but corresponds to obtaining photometric data and subjecting this data to a search for periodic phenomena that might coincide with the periodicities in radial velocities of the target star. We thus only apply it to increase the confidence in the planet candidates in the sample and to obtain alternative interpretations for some of the signals. For instance, if photometric data are not available, we cannot determine whether the third criterion is satisfied. In such cases, we only rely on the first two criteria when determining whether a given signal should be called a planet candidate or not and leave the question of photometric variations open for future work.

In the current work we define ``photometric counterpart'' to be a significant photometric signal with a period of $P_{\rm phot} \pm \delta P_{\rm phot}$ (we estimate conservatively that $\delta P_{\rm phot} = 0.1P_{\rm phot}$) such that the period of the radial velocity signal $P$ satisfies $P \in [P_{\rm phot} - \delta P_{\rm phot}, P_{\rm phot} + \delta P_{\rm phot}]$. We choose such a loose definition for a counterpart because photometric rotation periods and their radial velocity counterparts are not necessarily found at the same periods due to phenomena such as differential rotation and dependence of the rotation-induced signals on wavelength. Although active stars can show variation in the photometric rotation signal of up to 20\% \citep{messina2003}, we choose 10\% as the limit because our sample consists of quiescent main-sequence M dwarfs selected for their high radial velocity stability.

We note that there are well-known examples that demonstrate how periodic variations in photometric data can reveal a radial velocity signal to be caused by e.g. the co-rotation of starspots on the stellar surface. For instance, based on the ASAS photometry, \citet{kiraga2007} identified such periodic variations in both GJ 205 and GJ 358 that \citet{bonfils2013} found to have counterparts in the corresponding HARPS radial velocities (see Sections \ref{sec:GJ205} and \ref{sec:GJ358}, respectively). We do not expect that these stars are exceptions in any way and thus use the available photometric data to determine whether the radial velocity signals we find have photometric counterparts.

It is important to note that all such counterparts, whether in photometry or spectral activity indicators, only imply correlation, not causality. A given radial velocity signal, even in the presence of a photometric periodicity at or near the same period, can still be due to a candidate planet orbiting the star. A famous such case is the Solar magnetic cycle that roughly coincides with the orbital period of Jupiter. A notable exception in terms of a correlation between photometric and radial velocity signals is AD Leo (GJ 388) examined in Section \ref{sec:GJ388} and in \citet{tuomi2018}.

We also note that ideally the radial velocity signals are present in data from at least two independent instruments to be able to rule out the instruments and their potential biases and instabilities as sources of the detected signals. In practice, however, this is not possible for the majority of the targets in the current sample because differences in the instrument precisions, telescope sizes, and observational campaigns are such that high enough cadence for a detection is rarely available from two or more instruments. Moreover, with targets from both northern and southern skies, it is not always possible to observe a given target with two high-precision instruments suitably many times. However, there are some well-known exceptions to this, such as GJ 176 (Section \ref{sec:GJ176}), GJ 205 (Section \ref{sec:GJ205}), GJ 433 (Section \ref{sec:GJ433}), GJ 581 \citep{mayor2009,vogt2010}, and GJ 667C \citep{anglada2012,anglada2013}.

One often assumed criterion for the detection of radial velocity signals, although rarely explicitly stated, is a requirement that it has to be unique in the period space such that there is no doubt about which one of the periods corresponds to the orbital period of the planet. The violation of such a requirement, such as the detection of two or more signals caused by a unique underlying periodicity via aliasing \citep[such as the daily aliases in the HARPS data of GJ 3543;][see Section \ref{sec:GJ3543}]{astudillo2015}, typically requires the investigation of the origin of all the corresponding periodicities in order to be able to claim that a there indeed is a significant signal present in the data. We follow this approach only very loosely.

While a unique signal present as a probability maximum in the period space (power or likelihood maximum in periodograms) is easy to interpret when the local maxima are considerably lower, this might not be the case in reality with data sets containing gaps on annual and daily time-scales and due to the fact that observing time cannot necessarily be acquired in an optimal manner because of competition between different observational projects. It might thus be the case that we have two or more signals, together with their possible aliases, that appear almost equally probable. In such a case, we choose the global maximum as our solution but acknowledge the presence of alternative solutions as well while attempting to interpret all the maxima according to the best of our knowledge. This might lead to us adopting the alias of a signal as our solution rather than the signal itself, as was the case with e.g. GJ 581 d \citep{udry2007,mayor2009}, but we consider such incidences to be unlikely and rare enough such that they might affect significantly the orbital parameters of a single candidate planet but not the overall statistics of planets orbiting nearby M dwarfs whose quantification is the objective of this work.

Finally, should there be evidence in favour of more than one candidate planet orbiting a given target star, it is typically considered necessary to verify that the signals correspond to planet candidates on dynamically stable orbits. We do not attempt to verify the stability of the multiplanet systems diuscussed in this work. The reason for this choice is that most of the known multiplanet systems around the sample stars have already been subjected to stability analyses \citep[e.g.][]{mayor2009,rivera2010,anglada2013,anglada2014,bonfils2013b,tuomi2013c}. Moreover, all the newly detected candidate planets discussed in the current work in systems with $k>1$ have sufficient orbital spacings and such low minimum masses that we consider the corresponding planetary systems to be essentially dynamically stable. Naturally, we encourage detailed dynamical analyses of all such systems but consider such a task to be beyond the scope of the current work.

\subsubsection{Classification of planets}\label{sec:planet_classification}

Given that a signal is detected and can be called a planet candidate because it satisfies the above criteria, we classify it as a planet according to the following rules.

First, we divide the planets into categories based on their semi-major axes in relation to the estimated liquid-water habitable zones (HZs). These HZs, as estimated according to the formulae of \citet{kopparapu2013} for ``moist'' and ``maximum greenhouse'' limits, approximate the minimum and maximum distances from the star between which water could exist, under suitable atmospheric conditions, in its liquid form on the planetary surfaces (given that there is a solid surface). Planets inside this interval are called ``warm'' candidates whereas those inside (outside) the inner (outer) edge are called ``hot'' (``cool'') candidates.

With respect to minimum masses, we classify the candidate planets according to Table \ref{tab:mass_class}. This classification is merely a subjective choice, but it helps comparing the different candidate planets in the sample with one another. However, it is not our intention to suggest that planets can be categorised with a simple classification scheme such as that in Table \ref{tab:mass_class}. This merely enables us to differentiate between planets with likely solid surfaces (``Earths'' and ``super-Earths'') and those with considerable gaseous envelopes. We also differentiate between planets resembling Neptune and the more massive Solar System bodies that we collectively classify as ``Giants'' because they are rather rare around M dwarfs even when accounting for the handful of well-known examples, such as GJ 876 \citep{delfosse1998,marcy2001,rivera2005,rivera2010}, GJ 676A \citep{forveille2011,anglada2012c}, and GJ 832 \citep{bailey2009,wittenmyer2014}. We choose 10 M$_{\oplus}$ as approximately the limiting mass between planets with solid envelopes and gaseous ones even though \citet{weiss2014} argue that such a transition might rather happen at roughly 5 M$_{\oplus}$ corresponding to radii of approximately 1.5 R$_{\oplus}$ \citep[see also][]{dressing2015b,gillon2017}. We note that there is recent evidence that even larger planets might have masses and radii consistent with rocky compositions \citep{espinoza2016} but the mass reported by \citet{espinoza2016} of 16.3$^{+6.0}_{-6.1}$ is roughly consistent with 10 M$_{\oplus}$.

\begin{table*}
\caption{The planet classification sequence with maximum \emph{a posteriori} estimated minimum mass $m_{p} \sin i \in [m_{\rm min}, m_{\rm max}] = \Delta m$, where the interval $\Delta m$ is the 99\% Bayesian credibility interval of the minimum mass. We denote Neptune mass, Saturn mass, and Jupiter mass with M$_{\rm Nep}$, M$_{\rm Sat}$, and M$_{\rm Jup}$, respectively.}\label{tab:mass_class}
\begin{minipage}{\textwidth}
\begin{center}
\begin{tabular}{lll}
\hline \hline
 Criteria & Classification \\
\hline
 $m_{\rm max} <$ 2M$_{\oplus}$, M$_{\oplus} \in \Delta m$ & Earth \\
 $m_{\rm max} <$ 10M$_{\oplus}$, not Earth & super-Earth \\ 
 $m_{\rm max} <$ M$_{\rm Nep}$, not super-Earth or Earth & mini-Neptune \\
 $m_{\rm max} <$ M$_{\rm Sat}$, M$_{\rm Nep} \in \Delta m$, not mini-Neptune or smaller & Neptune \\
 $m_{\rm max} <$ M$_{\rm Sat}$, not Neptune or smaller & super-Neptune \\
 $m_{\rm max} <$ 13M$_{\rm Jup}$, not super-Neptune or smaller & Giant \\
 \hline \hline
\end{tabular}
\end{center}
\end{minipage}
\end{table*}

\subsection{Analysis of radial velocities}\label{sec:rv_model}

We analysed the radial velocities by following the techniques presented in \citet{tuomi2014} and \citet{butler2016} with some rather minor modifications. These techniques are presented here in detail for complementarity.

\subsubsection{Statistical modelling and posterior samplings}

Our statistical model expresses the radial velocity measurement $m_{i,l}$ of the $l$th instrument made at time $t_{i,l}$ as
\begin{equation}\label{eq:statistical_model}
  m_{i,l} = \gamma_{l} + \dot{\gamma} t_{i,l} +f_{k}(t_{i,l}) + \sum_{j} c_{l,j} \xi_{i,l,j} + \phi_{l} \exp \bigg\{ \frac{t_{i-1,l}-t_{i,l}}{\tau_{l}} \bigg\} r_{i-1,l} + \epsilon_{i,l} .
\end{equation} 
In this expression, the measurement is modelled to consist of a reference velocity ($\gamma_{l}$); linear acceleration $\dot{\gamma}$, excluding perspective acceleration that was subtracted from all data sets; superposition of $k$ Keplerian signals described with the function $f_{k}$; linear dependence on the measures of activity of the stellar surface $\xi_{i,l,j}$ quantified by parametetrs $c_{l,j}$; the moving average (MA) component parameterised with $\phi_{l}$ and exponential smoothing in the time-scale of $\tau_{l}$; and Gaussian white noise $\epsilon_{i,l}$ with a zero mean and a variance of $\sigma_{i,l}^{2} + \sigma_{l}^{2}$ where $\sigma_{l}$ is called the radial velocity jitter of the $l$th instrument. The term $r_{i-1,l}$ represents the residual after subtracting various components from the measurement $m_{i-1,l}$. We note that parameter $\tau_{l}$ was fixed equal to 4 days because its exact value would not have much effect on the results as long as it accounts for correlations in the data in a time-scale of a few days \citep{tuomi2014}.

This model gives rise to a likelihood function for the measurements $m_{i,l}, i=1, ..., N_{l}$, such that
\begin{eqnarray}\label{eq:likelihood}
  && l(m_{1,l}, ..., m_{N_{l},l} | \theta) = l(m_{1,l} | \theta) \times l(m_{2,l} | m_{1,l}, \theta) \times \cdots \nonumber\\
  && \times l(m_{N_{l},l} | m_{N_{l}-1,l}, ..., m_{N_{l} - p,l}, \theta)
\end{eqnarray}
where $\theta$ represents the parameter vector of the model containing all the free parameters. This implies a recursive way of calculating the full likelihood function because the $i$th likelihood depends only on the previous measurement due to the MA(1) component. 

The model accounts for the Keplerian periodicities caused by planetary companions including those on long-period orbits only apparent in the radial velocities as a linear acceleration. Moreover, in an attempt to model the potential contributions of stellar activity, the linear dependence of velocities on available activity indicators (BIS, FWHM, and S-index) is taken into account. Although this dependence might be of non-linear nature, we do not move beyond a linear dependence that is the first-order approximation of any non-linear function and is what has been reported for (moderately) active stars \citep[e.g.][]{boisse2011,dumusque2011,dumusque2012,gomez2012,hebrard2014,robertson2014,robertson2015}. The model in Eqs. (\ref{eq:statistical_model}) and (\ref{eq:likelihood}) also accounts for potential correlated noise that is known to be present in radial velocity data \citep[e.g.][]{baluev2009,tuomi2013b,feng2016} and is needed to minimise the number of false positive and negative detections \citet{dumusque2016}.

We selected priors such that they were set equal to unity for all parameters except $\sigma_{\rm J}$ and orbital eccentricity $e$ quantifying excess white noise and orbital eccentricity, respectively. Because a majority of the target stars in the current sample are inactive ones with low levels of radial velocity variability, we selected a prior for the excess white noise parameter such that $\pi(\sigma_{\rm J}) = \pi(\sigma_{\rm J} | \mu_{\sigma}, \sigma_{\sigma}^{2}) = \mathcal{N}(\mu_{\sigma}, \sigma_{\sigma}^{2})$ where we set $\mu_{\sigma} = \sigma_{\sigma}^{2} = 2.0$ ms$^{-1}$. This implies that we expect to see on average 2.0 ms$^{-1}$ excess white noise in the data sets but consider high levels of excess noise exceeding 8.0 ms$^{-1}$ to correspond to 3-$\sigma$ events in practice \citep{tuomi2013c,tuomi2014}. The eccentricity prior was selected such that $\pi(e) = \pi(e | \sigma_{e}^{2}) = \mathcal{N}(0, \sigma_{e}^{2})$ where we set $\sigma_{e} = 0.1$ \citep{tuomi2013c,anglada2013}. Although this eccentricity prior implies that we assume high eccentricities to be unlikely \emph{a priori}, high eccentricities are still allowed as solutions if they are supported by data \citep[e.g.][]{butler2016,jenkins2016}.

\subsubsection{Search for signals}

Searches for signals were conducted by applying the delayed-rejection adaptive Metropolis (DRAM) algorithm of \citet{haario2006} that is a generalisation of the common Metropolis-Hastings sampling algorithm \citep{metropolis1953,hastings1970}. The idea is to add a Keplerian signal to the model and investigate the posterior probability density as a function of the period parameter of the signal. If this chain visits all the areas in the period space repeatedly it is then possible to identify the highest posterior maxima in the period space suggestive of the presence of periodic Keplerian signals in the data. This can be compared to the periodogram analyses \citep[e.g.][]{lomb1976,scargle1982,cumming2004,anglada2012c} that aim at identifying the periods that minimise the sum of squared residuals or maximise the likelihood function. However, because such periodograms require local optimization that quickly becomes computationally too expensive as the number of free parameters increases, we use posterior samplings that enamble probing the probability landscape with respect to all parameters simultaneously with reasonable efficiency.

The algorithm of our choice, DRAM, has been shown efficient and reliable in practice \citep[e.g.][]{butler2016}. This algorithm works by proposing a new parameter vector $\theta_{i}$ by drawing it from a multivariate Gaussian proposal density. Based on the drawn sample, the proposal density is then updated by updating the mean and by calculating the covariance matrix $C_{i+1}$ to be used to draw the next vector, $\theta_{i+1}$, according to the recursive formula
\begin{equation}\label{eq:update_covariance}
  C_{i+1} = \frac{i-1}{i}C_{i} + \frac{s}{i} \bigg[ i \bar{\theta}_{i-1}\bar{\theta}_{i-1}^{T} - (i+1) \bar{\theta}_{i}\bar{\theta}_{i}^{T} + \theta_{i}\theta_{i}^{T} + \epsilon I \bigg] ,
\end{equation}
where we use $\bar{(\cdot)}$ and $(\cdot)^{T}$ to denote the mean and the transpose of a vector, respectively, and $\epsilon$ is some small positive number that ensures the resulting chain retains the correct ergodic properties \citep{haario2001}. In this expression the parameter $s$ is chosen such that $s = 2.4^{2}K^{-1}$, where $K$ is the dimension of the parameter vector, for optimal mixing of the chain \citep{gelman1996}.

The proposed vectors are then accepted according to the acceptance probability $\alpha_{1}$ defined as
\begin{equation}\label{eq:acceptance}
  \alpha_{1}(\theta_{i}, \theta_{i+1}) = \min \Bigg\{1, \frac{\pi(\theta_{i+1} | m) q_{i+1,1}(\theta_{i+1}, \theta_{i})}{\pi(\theta_{i} | m) q_{i+1,1}(\theta_{i}, \theta_{i+1})} \Bigg\} ,
\end{equation}
where $q_{i+1}$ is the updated proposal density. This is equivalent to the common Metropolis-Hastings acceptance probability and the algorithm differs only by constantly updating the proposal to adapt to the information gathered during the previous $i$ draws from the posterior. However, if the proposed vector $\theta_{i+1}$ is rejected we do not set $\theta_{i+1} = \theta_{i}$ as is the case in the adaptive Metropolis algorithm \citep{haario2001}, but propose a new vector $\theta_{i+2}$ from a modified proposal density $q_{i+1,2}(\theta_{i}, \theta_{i+1}, \theta_{i+2})$ where it is notable that this new proposal depends also on the proposal $\theta_{i+1}$ that was just rejected. This newly proposed vector has the acceptance probability of
\begin{eqnarray}\label{eq:acceptance2}
  \alpha_{2}(\theta_{i}, \theta_{i+1}, \theta_{i+2}) = \min \Bigg\{1, \frac{\pi(\theta_{i+2} | m) q_{i+1,1}(\theta_{i+2}, \theta_{i+1}) q_{i+1,2}(\theta_{i+2}, \theta_{i+1}, \theta_{i})}{\pi(\theta_{i} | m) q_{i+1,1}(\theta_{i}, \theta_{i+1}) q_{i+1,2}(\theta_{i}, \theta_{i+1}, \theta_{i+2})} \nonumber\\
  \times \frac{[1 - \alpha_{1}(\theta_{i+2}, \theta_{i+1})]}{[1 - \alpha_{1}(\theta_{i}, \theta_{i+1})]} \Bigg\} .
\end{eqnarray}
Given that this proposed vector, too, is rejected, additional proposal densities can be used in sequence according to the general formulae of \citet{haario2006}.

We take advantage of the delayed rejection technique by decreasing the variance of the proposal density in the dimension of the period parameter of the signal. That is, when proposed vector $\theta_{i+1}$ has been rejected, we modify the proposal such that the variance of its component corresponding to the period parameter of the signal is decreased by a factor of 10 enabling us to probe the immediate surroundings of the current state in the period space. This is continued until the new proposal is rejected for the third time after which we cease to propose new values in the vicinity of $\theta_{i}$ and consider the proposal rejected. This technique enables us to sample the period space in greater detail around probability maxima with only slight additional computational cost arising from additional proposals and the corresponding evaluations of the likelihood function. As an example of our period search technique, we refer to \citet{butler2016} where the technique was show to detect reliably signals of all known exoplanets in the Keck data.

We also performed another step during the samplings to identify all the maxima reliably in the period space. If the acceptance rate decreased below $10^{-2}$ indicating that the current value was much more probable than any values proposed in its neighbourhood in the parameter space, we decreased the width of the proposal density in the dimension of the period parameter by a factor of 1000 and continued the sampling sampling with this modified proposal density that only proposed values from the vicinity of the current state in the period space. This enabled us to sample efficiently the various maxima in the period space giving a more reliable mapping of the landscape of the most significant periodicities in the data. Although this modification to the sampling algorithm has the consequence that the corresponding DRAM samplings are no longer ergodic and that parameter inferences are thus not possible given such samples because we have violated the assumptions of \citet{haario2006}, we only applied such modified proposal density in order to estimate the global and local maxima correctly.

Finally, if the samplings of the parameter space were inefficient such that acceptance rates were consistently below 0.1, we used tempered Markov chains in the periodicity senarches to obtain higher acceptance rates that enabled the chains to visit all areas in the period space consistently and repeatedly \citep[e.g.][]{butler2016,feng2016}. In practice, we followed the approach of \citet{butler2016} and sampled a posterior $\pi_{_{\beta}}(\theta | m) \propto \big[ l(m | \theta) \pi(\theta) \big]^{\beta}$ where parameter $\beta \in [1,0)$ was selected to be lower than unity. In practice, if acceptance rate was below 0.1, we decreased parameter $\beta$ by a factor of 1.1 as long as necessary to obtain acceptance rates in excess of 0.1. Although using such tempered samplings means we were not sampling the actual posterior but a version where the maxima in the likelihood function were scaled down to enable the chain to jump between all of them efficiently, such an apporach enables finding the positions of the most significant maxima because they are not changed due to the scaling.

\subsubsection{Parameter space}

We searched for signals in the period space between periods of $P_{\rm min}$ and $P_{\rm max}$ that were set to 1 day and $T_{\rm obs}$, where the latter is the ``effective baseline'' of the observations. We call it the effective baseline because we defined it as
\begin{equation}\label{eq:baseline}
  T_{\rm obs} = t_{N} - t_{1} - T_{\rm gap} ,
\end{equation} 
where
\begin{equation}\label{eq:gap}
  T_{\rm gap} = \max_{i=1, ..., N-1} \{t_{i+1} - t_{i}\}
\end{equation}
represents the longest gap in the data. However, if $T_{\rm gap} < \frac{1}{2}(t_{N} - t_{1})$ we set $T_{\rm gap} = 0$. This definition is justified as follows. If there is a gap in the data that is longer than half of the difference of the first and last observation (the full data baseline), the effective baseline is not equal to this difference because the gap does not allow constraining long periods that are comparable to the data baseline. Instead, subtracting the largest gap enables us to obtain an approximate upper limit below which signals can be constrained in the period space.

In practice, we observed several data sets for which a majority of the data was taken within a period of two weeks but that also had one or few additional points taken within a single night during a different observing season. In such a case, the effective baseline is thus only two weeks, which is what our definition attempts to quantify.

\subsubsection{Parameter estimation}

The parameter estimation problem is similarly solved by applying the adaptive Metropolis posterior sampling algorithm \citep{haario2001}. We started samplings with untempered chains and set the initial state of the period parameter equal to the period (or periods for $k>1$) observed by the DRAM periodicity searches. We then subjected the obtained posterior samples to the detection criteria in order to see if the global maximum corresponded to a signal detection and obtained estimates for the model parameters if a signal was indeed detected.

Throughout the analyses of radial velocity data sets we use maximum \emph{a posteriori} (MAP) estimates and 99\% credibility sets, or intervals in a single dimension, when discussing or tabulating the values of parameters. The MAP values are simple maxima of the respective posterior probability densities defined according to
\begin{equation}\label{eq:MAP}
  \hat{\theta}_{\rm MAP} = \arg \max_{\theta \in \Omega} \pi(\theta | m) ,
\end{equation}
that is, the value of $\theta$ in the set $\Omega \subset \mathbb{R}^{K}$ that maximises the posterior probability density function $\pi(\theta | m)$ and $\Omega$ denotes the whole parameter space. Although typically identical to the posterior mode and/or mean, it can in practice be very different from these in case of multimodal or highly skewed probability densities.

Similarly, we do not apply ``standard errors'' or 1-$\sigma$ uncertainty estimates under the assumption that the posterior probability density is Gaussian. The reason is that this not even valid as a good approximation as can be seen when thinking about e.g. typical posterior densities of the eccentricities of low-mass planets that are limited from below by the fact that they have to be positive and whose mean estimate is typically very different from the mode or the MAP ones. We use the $\delta$-significance Bayesian credibility intervals, i.e. sets $\mathcal{D}_{\delta}$ such that the credibility set of a posterior density $\pi(\theta)$ is
\begin{equation}\label{eq:BCS}
  \mathcal{D}_{\delta} [\pi(\theta)] = \Bigg\{ \theta \in B \subset \Omega \Bigg| \pi(\theta)|_{\theta \in \delta B} = c, \int_{\theta \in B} \pi(\theta) d \theta = \delta \Bigg\} ,
\end{equation}
where $c$ is a positive constant and $\delta B$ is used to denote the edge of the bounded set $B$. This set is bounded by the equiprobability surface corresponding to the edge. In practice, we apply $\delta = 0.99$.

The practicality of these choices is apparent when considering a sample of $n$ random vectors such that $\theta_{1}, ..., \theta_{n} \sim \pi(\theta)$ based on a posterior sampling. It is then easy to arrange them in order based on the corresponding values $\pi(\theta_{i})$ and use the highest value as the MAP estimate based on the sampling. We estimate the credibility intervals by obtaining estimates for the marginalised probability density functions of the parameters as distributions of values and select the intervals such that they contain 99\% of the obtained posterior sample. In practice, however, it should be remembered that this uncertainty estimation method only provides the minimum and maximum bounds of $\mathcal{D}_{\delta}$ in each dimension and only the full sample can be used to estimate the true shape of the set of solutions sampled with the adaptive Metropolis algorithm.

\subsection{Model selection and comparison}\label{sec:significance}

The significances of the signals in the radial velocity data can be assessed by a number of methods. In the current work we apply three such approaches. As in \citet[e.g.][]{tuomi2012,tuomi2014}, we estimate the Bayes factors corresponding to models containing the signals ($k$ signals) with respect to models that do not contain them ($k-1$ signals). The Bayes factor can be written for the corresponding models $\mathcal{M}_{i}$ as
\begin{equation}\label{eq:bayes_factor}
  B_{k,k-1} = \frac{P(m | \mathcal{M}_{k})}{P(m | \mathcal{M}_{k-1})}
\end{equation}
where the marginalised likelihoods $P(m | \mathcal{M}_{k})$ have to be approximated in order to obtain estimates for the Bayes factors. In essence, we follow \citet{tuomi2014} and estimate these Bayes factors by applying the simple approximation discussed by \citet{newton1994} that applies a mixture of samples from both posterior and prior densities. When applying this estimation, we select a conservative threshold of $\alpha = 10^{4}$ \citet{tuomi2014} and require that $B_{k,k-1} > \alpha$ for every signal in our analyses.

However, we also derive estimates for the Bayes factors based on the Bayesian information criterion \citep[BIC; e.g.][]{liddle2007} that have been reported to be optimally adjusted against false positives and negatives when using a detection threshold of $\alpha=150$ \citet{feng2016}. This threshold has also been applied commonly in practice \citep[e.g.][]{feroz2011,tuomi2012} because it corresponds to ``strong evidence'' on the Jeffrey's scale as interpreted by \citet{kass1995}. For comparison, we also discuss significances of the signals based on likelihood-ratio tests, namely, a computation of likelihood ratios and their interpretation based on the $\chi^{2}$ statistics.

Because most of the signals in the sample are consistent with circular solutions, i.e. that there is no evidence that the orbital eccentricities are statistically significantly different from zero, we tune the detection thresholds for such three-parameter models. This means, when calculating Bayes factors based on BIC, whe assume that model $\mathcal{M}_{k}$ has three parameters more than model $\mathcal{M}_{k-1}$. The same applies to the likelihood-ratio test where we assume there are three degrees of freedom for every signal.

It is noteworthy that when calculating BIC and maximum-likelihood values, we still use the non-uniform priors for orbital eccentricity and the jitter parameter $\sigma_{\rm J}$. However, these priors can simply be considered a form of Tikhonov regularisation \citep{tikhonov1977} used to transform ill-posed inverse problems into problems with more unique solutions.

\subsection{Analysis of ASAS photometry}

The \emph{All-Sky Automated Survey} (ASAS) photometric data \citep{pojmanski2002,kiraga2007} was downloaded from the project website\footnote{\texttt{www.astrouw.edu.pl/asas}} for all the targets in our sample for which it was available. We first processed it by selecting only the highest quality data (grade A) and removing all 5-$\sigma$ outliers in an iterative manner according to the post-removal mean and variance \citep[as in e.g.][]{hartman2013}. In this process, we continued until all such outliers were removed with respect to the mean and variance that were calculated after the removal. This typically resulted in us removing 2-5\% of the measurements from the data sets that had roughly 300-500 measurements, sometimes considerably more. This removal of outliers was performed for all five ASAS apertures but we typically base our results on the data of the narrowest aperture (MAG0). However, other apertures were used for the brightest targets that could saturate the narrowest apertures in practice. This resulted in photometric time-series that were suitable for the search for stellar photometric periodicities.

Because of the reasonably uniform statistical properties in the ``cleaned'' photometric data, i.e. data sets from which outliers had been removed, we then searched for periodic signals by calculating likelihood-ratio periodograms in the following manner. First, we calculated the maximum likelihood of a two-parameter reference model consisting of a first order polynomial. This maximum likelihood was then compared to that of a linear four-parameter model where the photometric measurement was modelled as
\begin{equation}\label{eq:asas_model}
  m_{i} = a_{0} + a_{1} t_{i} + a_{2} \sin (\omega t_{i}) + a_{3} \cos (\omega t_{i}) + \epsilon_{i} ,
\end{equation}
where $\epsilon_{i}$ are assumed to be Gaussian random variables with zero mean and variance of $\sigma_{i}^{2}$ that we assume to be known based on the estimated photon-noise uncertainties in the ASAS data; $\omega$ is the signal frequency; and $a_{i}, i=0, ..., 3$, are the free parameters of the linear model. This enabled us to solve the parameters $a_{i}$ for a set of frequencies $\omega \in [2\pi P^{-1}_{\rm max}, 2\pi P^{-1}_{\rm min}]$ to estimate the likelihood ratio as a function of $\omega$ that we use to test whether there are significant periodicities in the ASAS data. This is effectively a generalisation of the Lomb-Scargle \citep{lomb1976,scargle1982} periodogram that also allows unknown baselines and linear trends. The reason we included a linear trend in the model (parameter $a_{1}$) is that we are merely interested in finding periodicities that can be constrained by the photometric data -- not potentially periodic variations with periods greatly in excess of the data baseline and thus only visible as linear trends. We chose $P_{\rm min} = 1$ day and $P_{\rm max} = 2T_{\rm obs}$, where $T_{\rm obs}$ is the data baseline.

We then calculated likelihood ratios corresponding to false alarm probabilities based on two parameter likelihood-ratio test between the two nested models. We consider a likelihood ratio exceeding the corresponding 0.1\% false alarm probability (FAP) at a frequency $\omega$ to correspond to a significant periodic signal in the data whereas those exceeding 5\% of 1\% ratios are merely ``suggestive''.

As a second test, we relaxed the assumption that the measurement uncertainties were known according to the values reported in the ASAS archive. We assumed that the ASAS measurements of a given target were identically distributed and used the variance of the whole data set as a measure representative of the uncertainties of all the measurements. This typically resulted in uncertainties that were smaller than the original ASAS uncertainties likely due to the fact that we had already removed the non-Gaussian outliers of the data distribution.

We note that because the first order polynomial term was included in the model, all long-period signals might be affected as the signals are only allowed to explain the excess curvature in the data that does not exist for signals with periods $P \gg T_{\rm obs}$. This choice was made to concentrate on periodicities that could be constrained, most notably rotation-induced photometric periodicities, rather than long-period activity cycles.

The periodicities in the ASAS photometry were then classified depending on the period parameters ($P$) of the corresponding sinusoids. We call the most significant periodicities with 2 days $<P<$ 150 days photometric rotation periods of the corresponding stars. The upper limit was selected because photometric rotation periods of M dwarfs are not likely to be longer than this \citep{mcquillan2014,newton2015}. However, the lower limit might result in false negatives when the target is a fast rotator. However, photometric periods below two days are the most likely due to stellar pulsations \citep{reinhold2015} and such stars are generally active and thus selected heavily against in radial velocity surveys. There was also another, more pragmatic reason, to limit the period space to periods above two days. This is because with ground-based observations, the daily aliases of signals near one day might be indistinguishable in significance, or even stronger than, the signals themselves. This prevents us from detecting periods below two days but we do not expect this to bias our results much in practice.

We identified several targets with evidence for photometric rotation periods in the sample, such as GJ 176 (Section \ref{sec:GJ176}), GJ 205 (Section \ref{sec:GJ205}), GJ 208 (Section \ref{sec:GJ208}) and GJ 358 (Section \ref{sec:GJ358}) to refer to some of the clearest examples. However, if $P>150$, we do not interpret the corresponding signals rotation periods but simply call them photometric periodicities that could be connected to stellar activity and magnetic cycles. Again, we identified several such cycles, such as shown when discussing the ASAS data of GJ 358, GJ 388 (Section \ref{sec:GJ388}), GJ 479 (Section \ref{sec:GJ479}) and GJ 628 (Section \ref{sec:GJ628}) to name but a few such cases.

\subsection{Analysis of spectral activity indicators}

The information in the spectral activity indicators of HARPS and HIRES was accounted for in two stages to see whether i) the activity indicators traced the variations in the radial velocities indicating that their variability was connected to stellar activity rather than planets and ii) to see if there were periodic signals in the activity indices suggestive of magnetic activity cycles and/or stellar rotation or other periodicities that might induce radial velocity variations mimicking Keplerian signals of planets.

The first stage consists of accounting for the linear correlations between the radial velocities and the activity indices according to the statistical model in Eq. \ref{eq:statistical_model}. These linear relationships are quantified by the parameters $c_{i}$ and we say that the velocities and the activity indicators are connected if the corresponding parametes have 99\% credibility intervals that do not contain zero. Such a relationship implies that some of the radial velocity variability is connected to stellar activity. Moreover, if the 99\% credibility interval of parameter $c_{i}$ contains zero, there is no evidence in favour of the velocities and activity indicators being linearly connected and it is thus possible to simplify the statistical model by setting $c_{i} = 0$. In such a case, we say that there is no evidence that the variations in the radial velocities and activity indices are connected and that they can thus be assumed to be independent. Moreover, they are then also independent of any signals we detect in the radial velocity data. In all our analyses of radial velocities, we treated parameters $c_{i}$ as free parameters of the statistical model and thus obtained samples describing their posterior probability densities as well and only fixed them to zero if simplifying the model was necessary to study the nature of some of the signals by comparing models with and without such linear dependences of velocities on the activity data.

In the second stage, we subject the time-series of the activity indicators to the same likelihood periodograms as the photometric time-series in a search for periodic signals. If no significant signals are found, we conclude that the radial velocity signals do not have activity-induced counterparts. If there are significant signals, however, we interpret them as activity-related cycles, such as magnetic cycles or rotation coupled with stellar surface inhomogeneities and interpret the signals in the radial velocities as arising from these cycles if they occur at or near the same periods (see Section \ref{sec:what_is_a_planet}).

\subsection{Occurrence rates}

We calculate the estimates for the occurrence rate of low-mass planets around the sample M dwarfs as a function of minimum mass and orbital period by estimating the joint detection probability function of these parameters as in \citet{tuomi2014} and by comparing that to the sample of candidate planets detected given the data. We thus normalise the sample of detected planets by the corresponding probability of being able to detect such planets in the sample to account for the selection bias of radial velocity surveys towards high-mass planets on short periods as well as the differing sensitivities for planets for different target stars.

This is performed by first obtaining an estimate for the number of significant signals defined to be $k_{i}$ in the $i$th data set, where $k_{i} \in \{0, 1, 2, ... \}$, and by then modelling the data with a model containing $k_{i}+1$ signals. Since this additional signal does not satisfy the detection criteria by definition, its amplitude is consistent with zero\footnote{Apart from a handful of cases dominated by activity-induced variability, this was always the case in practice.} enabling the Markov chains to probe the whole parameter space $A \subset \Omega$ this signal is allowed to have based on the data set. Therefore, the complement $A^{c} = \Omega \setminus A \subset \Omega$ represents the subset of the parameter space that can be ruled out -- i.e. provides an estimate of what signals cannot exist in the data. This enables us to estimate the detection threshold for each data set in the following way.

We divide the parameter space in period and minimum mass dimensions into a grid such that it can be examined where the parameters of the $k_{i}+1$th signal visit in the Markov chain samplings. Because this $k_{i}+1$th signal cannot be constrained from above and below in the amplitude and period space, if the chain visits a grid point corresponding to an interval $([M, M+\delta M], [P, P+\delta P]) = (\Delta M, \Delta P)$, we set $(\Delta M, \Delta P) \in A$ and approximate that the chain could also visit all grid points with $M' < \min \Delta M$ even though in practice this is not necessarily always the case with finite chain lengths\footnote{In practice, if chains visited the interval $\Delta M$ for some period interval $\Delta P$, they also always visited all the intervals $\Delta M'$ for $M_{0} < M' < M$ where the lower limit was set to $M_{0} = $1M$_{\oplus}$ because we do not expect to obtain occurrence information for planets below 1M$_{\oplus}$ due to the fact that no such planets were detected.}. We then assume that this is the case for the whole width of $\Delta P$ and thus set all $(\Delta M', \Delta P) \in A$. We set the corresponding detection probability function $p_{i}(\Delta M, \Delta P)=1$ if $(\Delta M, \Delta P) \in A$ and equal to zero otherwise. We also set $p_{i}(\Delta M, \Delta P)=1$ for period intervals outside the period space $[P_{\rm min}, P_{\rm max}]$ implying that signals cannot be detected in the period space where we do not search for them.

Repeating this for all $N$ data sets, we can approximate the joint detection probability function for all data sets  as
\begin{equation}\label{eq:detection_probability}
  f_{\rm p}(\Delta M, \Delta P) = 1 - \frac{1}{N} \sum_{i=1}^{N} p_{i}(\Delta M, \Delta P) .
\end{equation} 
Using this function, it is then possible to calculate the occurrence rate of planets with a much coarser grid for practical reasons -- occurrence rate can only be calculated for an interval in the parameter space if at least one planet is detected within that interval. We thus use an 8$\times$8 grid to estimate occurrence rates and adopt the average value of $f_{\rm p}$ as an estimate for each such interval. We estimate the corresponding uncertainties by using the standard deviation in the detection probability function in a given interval. The occurrence rate $f_{\rm occ}$ is thus calculated based on the observed number of planets $f_{\rm obs}$ in the interval as
\begin{equation}\label{eq:occurence_rate_approx}
  f_{\rm occ}(\Delta M, \Delta P) = \frac{f_{\rm obs}(\Delta M, \Delta P)}{N f_{\rm p}(\Delta M, \Delta P)}
\end{equation} 
in units of ``planets per star''. 

For intervals for which there are no candidates, we assume that the number of candidates detected is $<1$ enabling us to obtain an inequality estimating an upper limit for the corresponding occurrence rate estimate.

\section{Properties of the stars in the sample}\label{sec:sample}

In this section we discuss the general properties of the 426 stars in the sample. We have listed some observed and estimated properties of the stars in Tables \ref{tab:targets} and \ref{tab:targets_physics}, respectively. Table \ref{tab:targets} lists the estimated spectral classes, parallaxes, and $V$ and $J$-magnitudes of the targets as obtained from the website of the \emph{Strasbourg Astronomical Data Centre}\footnote{\texttt{cds.u-strasbg.fr}}. The spectral classifications are thus primarily based on estimations of \citet{torres2006} or \citet{koen2010}; parallaxes are primarily from Hipparcos \citep{vanleeuwen2007}; and V-magnitudes primarily from \citet{hog2000}, \citet{casagrande2008}, or \citet{koen2010}. The median of the masses based on the empirical equations of \citet{delfosse2000} is 0.43 M$_{\odot}$. We use this value when dividing the sample into low- and high-mass sub-samples for which we estimate the occurrence rates of planets independently.

\subsection{Physical properties}

\begin{figure}
\center
\includegraphics[angle=270, width=0.49\textwidth,clip]{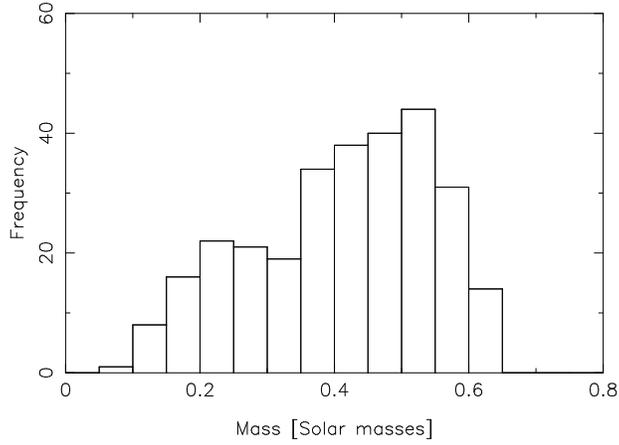}
\caption{Distribution of the masses of the targets in the sample estimated based on the empirical equations of \citet{delfosse2000} using $V$-band magnitudes.}\label{fig:mass_distribution}
\end{figure}

We estimated the stellar masses by using the empirical relations of \citet{delfosse2000} based on the stellar distances and their apparent $V$-band magnitudes. We assume a standard uncertainty of 10\% for the stellar mass estimates when calculating the minimum masses and semi-major axes of the planet candidates because they are consistent (and mostly the same) as those tabulated in \citet{bonfils2013}. The obtained values are in reasonably good agreement with more sophisticated estimates tabulated in e.g. \citet{mann2015}. Most of the stars in the sample have estimated masses ranging from 0.1 to 0.6 M$_{\odot}$, although a few targets are outside this range. This means that our results can be expected to apply within this range only. The mass distribution of the sample targets is shown in Fig. \ref{fig:mass_distribution}.

We estimated the effective temperatures of the sample stars by using the empirical equation of \citet{casagrande2008} based on the colour index $V-J$. The obtained temperatures were then used to estimate the luminosities based on \citet{boyajian2012} empirical equations. These estimates are listed in Table \ref{tab:targets_physics} together with the luminosities from \citet{gaidos2014}.

\begin{figure}
\center
\includegraphics[angle=270, width=0.49\textwidth,clip]{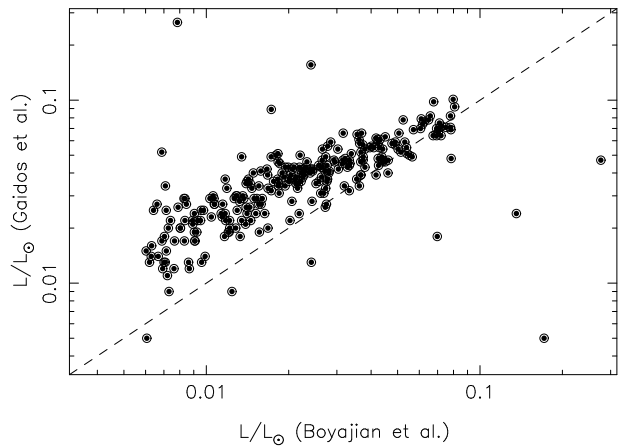}
\caption{Comparison of the estimated luminosities based on the empirical equations of \citet{boyajian2012} relying on estimated effective temperatures and the values listed in the catalog of \citet{gaidos2014}.}\label{fig:lum_comparison}
\end{figure}

When comparing the luminosity values of \citet{gaidos2014} with those estimated by using the equations of \citet{boyajian2012} based on the effective temperatures, we found the two to be in reasonably good agreement with a Pearson correlation coefficient of 0.66 (Fig. \ref{fig:lum_comparison}). However, it appears that, neglecting a few outliers, the estimates from \citet{gaidos2014} are systematically greater -- increasingly so for lower luminosities of around 0.01 L$_{\odot}$ and below. This means that because we use the luminosity estimate based on the empirical equation of \citet{boyajian2012} that ultimately relies on the $V-J$ colour index when approximating the distances of the stellar habitable zones, we might be underestimating their distances from the stellar surface.

\begin{figure}
\center
\includegraphics[angle=270, width=0.49\textwidth,clip]{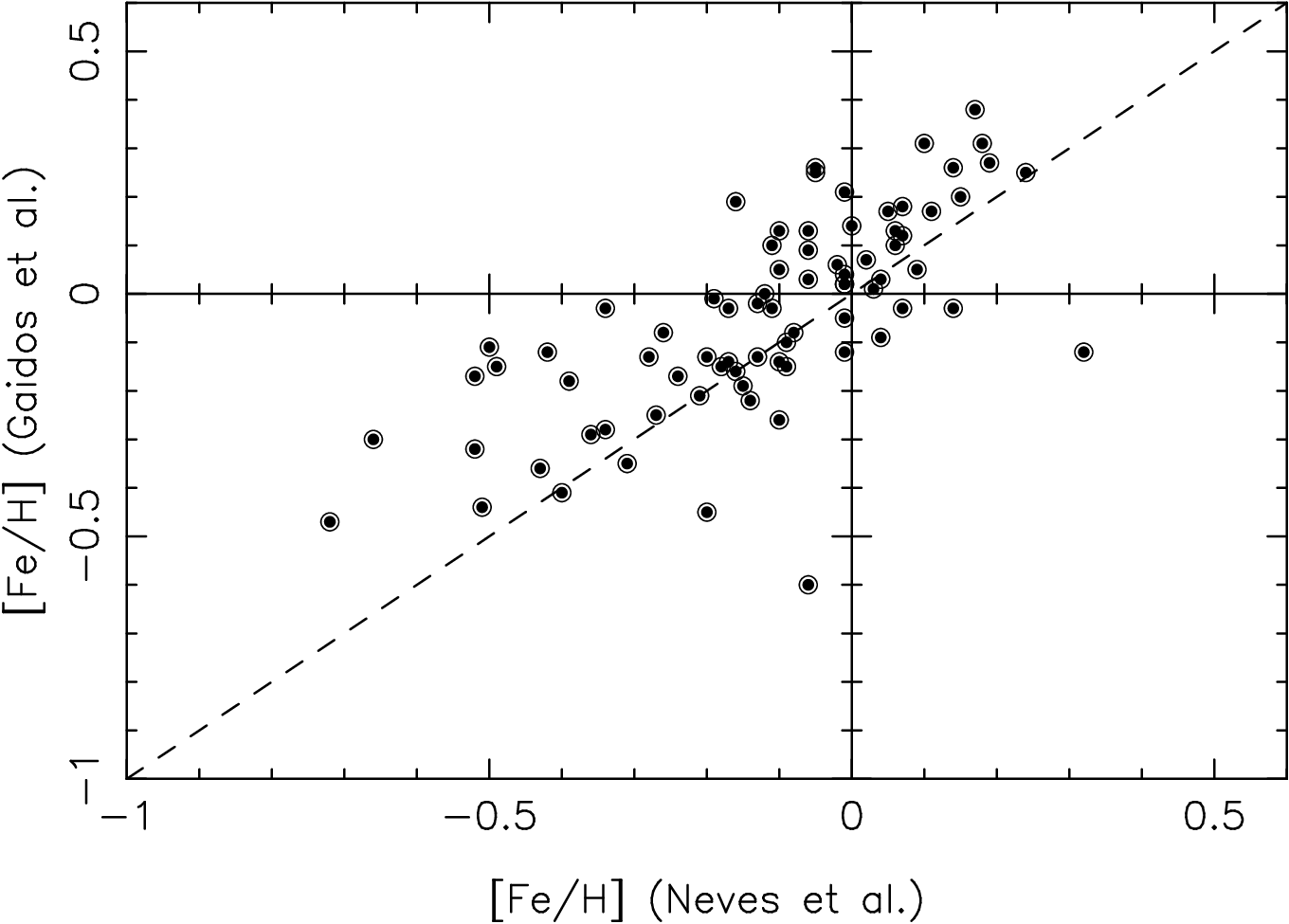}
\caption{Comparison of the metallicity estimates of \citet{gaidos2014} and \citet{neves2012,neves2013}.}\label{fig:met_comparison}
\end{figure}

\begin{figure}
\center
\includegraphics[angle=270, width=0.49\textwidth,clip]{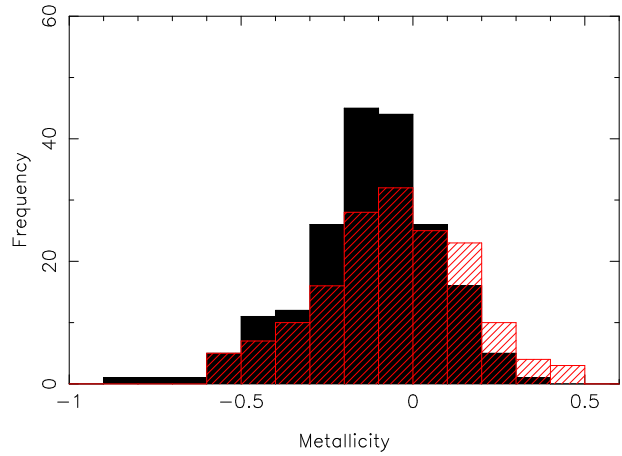}
\caption{The distribution of available metallicity estimates for the sample stars from \citet{neves2012,neves2013} (black histogram) and \citet{gaidos2014} (red histogram).}\label{fig:met_distribution}
\end{figure}

The stellar metallicities were also obtained from two sources -- from \citet{gaidos2014} and \citet{neves2012,neves2013}, respectively. For the stars in our sample, these two sources give median metallicities of -0.06 and -0.11 in a range from -0.9 to 0.5. Although the overlap of these two metallicity catalogs is only 77 targets, we could still compare the respective estimates and find that they are in a reasonably good agreement with a Pearson correlation coefficient of 0.70 (Figs. \ref{fig:met_comparison} and \ref{fig:met_distribution}).

We note that when studying the properties of the planet population around the sample stars as a function of stellar properties, we limit the analyses to subsamples for which the physical parameters have been, or can be, estimated. The stellar masses are an exception to this because they are needed when estimating the detection probability function of the sample as a function of planet mass and orbital period. The parallaxes of a handful of stars remain unknown and we thus estimate the masses of such targets by choosing the median mass of targets with the same spectral class in our sample, or if even the spectral class is unknown, use the median mass of the targets in the whole sample as estimated based on the equation of \citet{delfosse2000}. We accept that this will increase the uncertainty of the results but is likely to be within the assumed uncertainties in the stellar masses.

We summarise the physical properties of the target stars in Table \ref{tab:targets_physics}.

\subsection{Stellar habitable zones}

Estimating the semi-major axes corresponding to the inner and outer edges of the liquid-water habitable zones cannot be considered very precise -- especially when the HZs are calculated based on uncertain stellar properties, such as luminosity and effective temperature. Yet, such zones remain the best method available for estimating what orbital distances allow liquid water to exist on planetary surfaces under some reasonable assumptions regarding atmospheric conditions, surface gravity and planetary geochemical circumstances.

After pioneering work presented in e.g. \citet{kasting1993,selsis2007}, the most recent and likely the most reliable estimation methods for the stellar habitable zones of M dwarfs can be found in \citet{kopparapu2013}. The equations of \citet{kopparapu2013} have been applied in several recent works discussing the potential habitability of newly discovered low-mass planets \citep[e.g.][]{anglada2013,anglada2014,anglada2016,kopparapu2013b,tuomi2013a}. We thus adopted the estimates from \citet{kopparapu2013} representing the ``moist'' and ``maximum greenhouse'' limits of the HZ. These limits are also tabulated for the sample stars in Table \ref{tab:targets_physics}.

\subsection{Stellar rotation periods}

Knowledge regarding the stellar rotation periods is important when deciding whether periodic radial velocity variations are caused by a planet on a Keplerian orbit rather than starspots and active and/or inactive regions co-rotating on the stellar surface.

The rotation periods of dwarf stars can be mainly estimated based on two different kinds of data, namely, photometric and spectroscopic. Photometrically, rotation periods can be estimated by searching for periodic variations in a suitable period range \citep[e.g.][]{kiraga2007,hartman2013,walkowicz2013,mcquillan2014,mcquillan2014b}. Alternatively, it is possible to investigate the periodic variations in the spectral activity indices \citep[e.g.][]{robertson2014,robertson2015,astudillo2015,mascareno2015}. This latter method is especially useful because it is possible to estimate the rotation periods based on the same data that is used to search for planets without the need to obtain photometric observations as well. However, photometric rotation periods are generally considered more trustworthy and are typically reported when discussing newly discovered planets found with the radial velocity technique \citep[e.g.][]{haghighipour2010,vogt2010,astudillo2015,anglada2016}

\begin{figure}
\center
\includegraphics[angle=270, width=0.19\textwidth,clip]{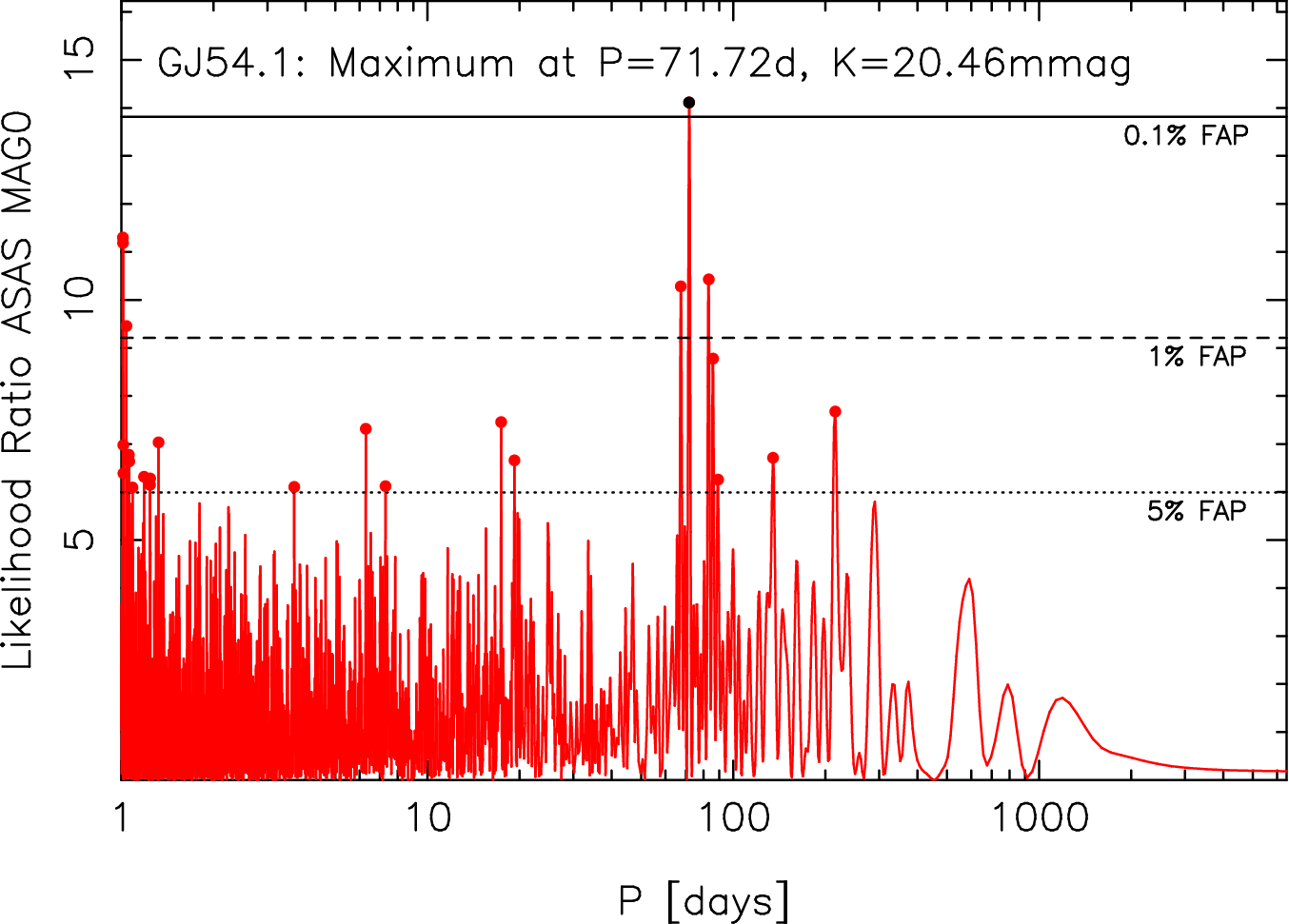}
\includegraphics[angle=270, width=0.19\textwidth,clip]{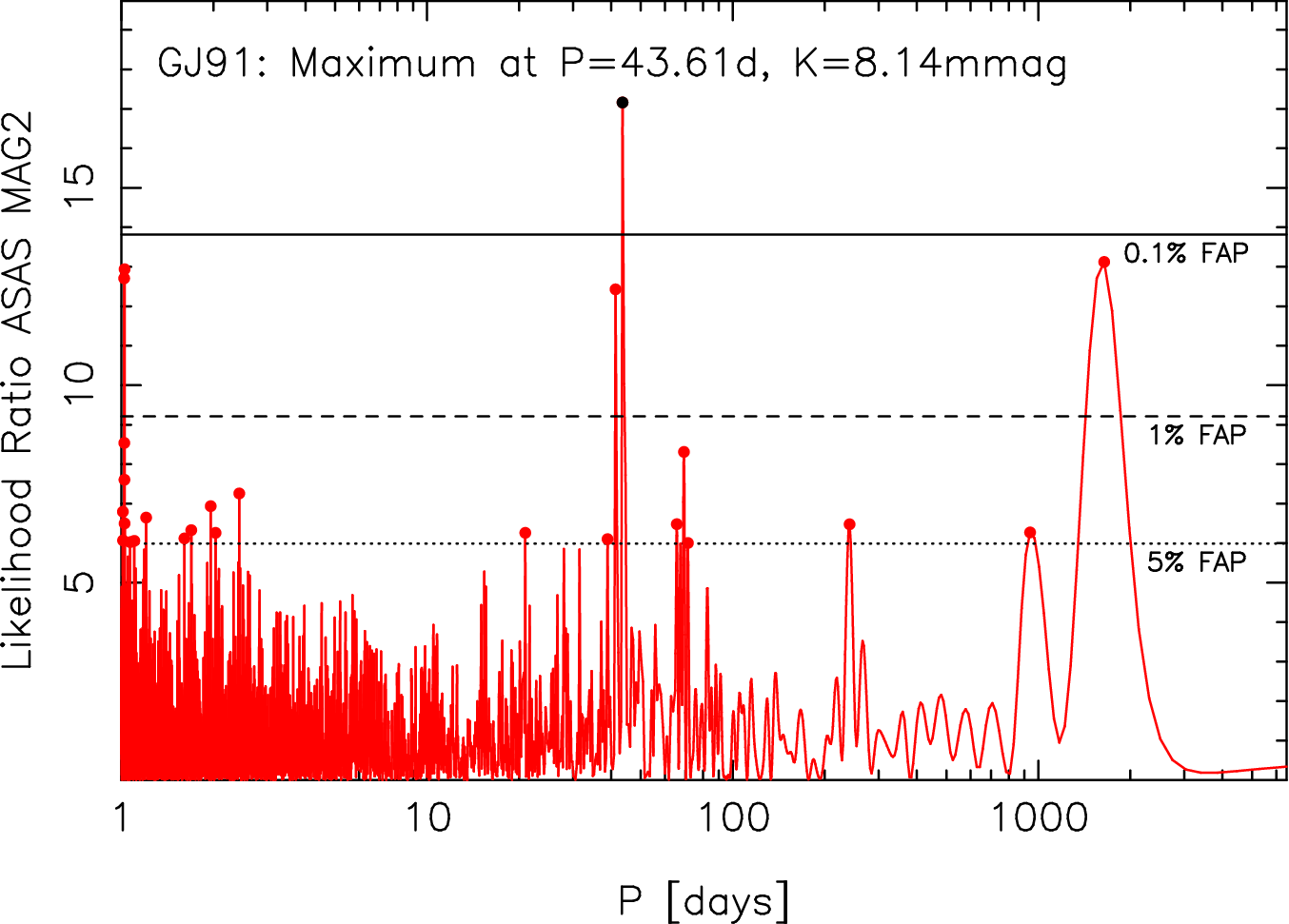}
\includegraphics[angle=270, width=0.19\textwidth,clip]{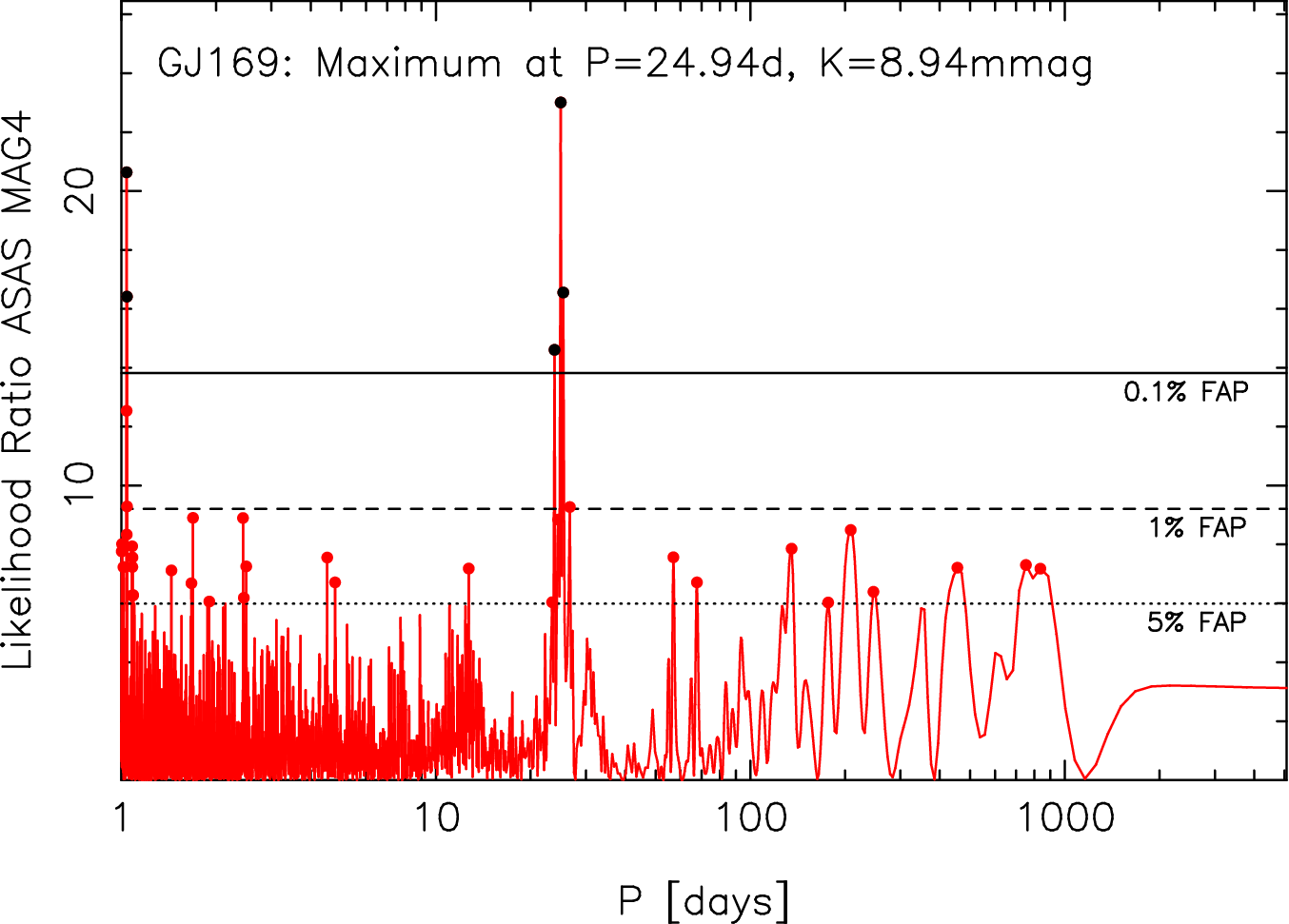}
\includegraphics[angle=270, width=0.19\textwidth,clip]{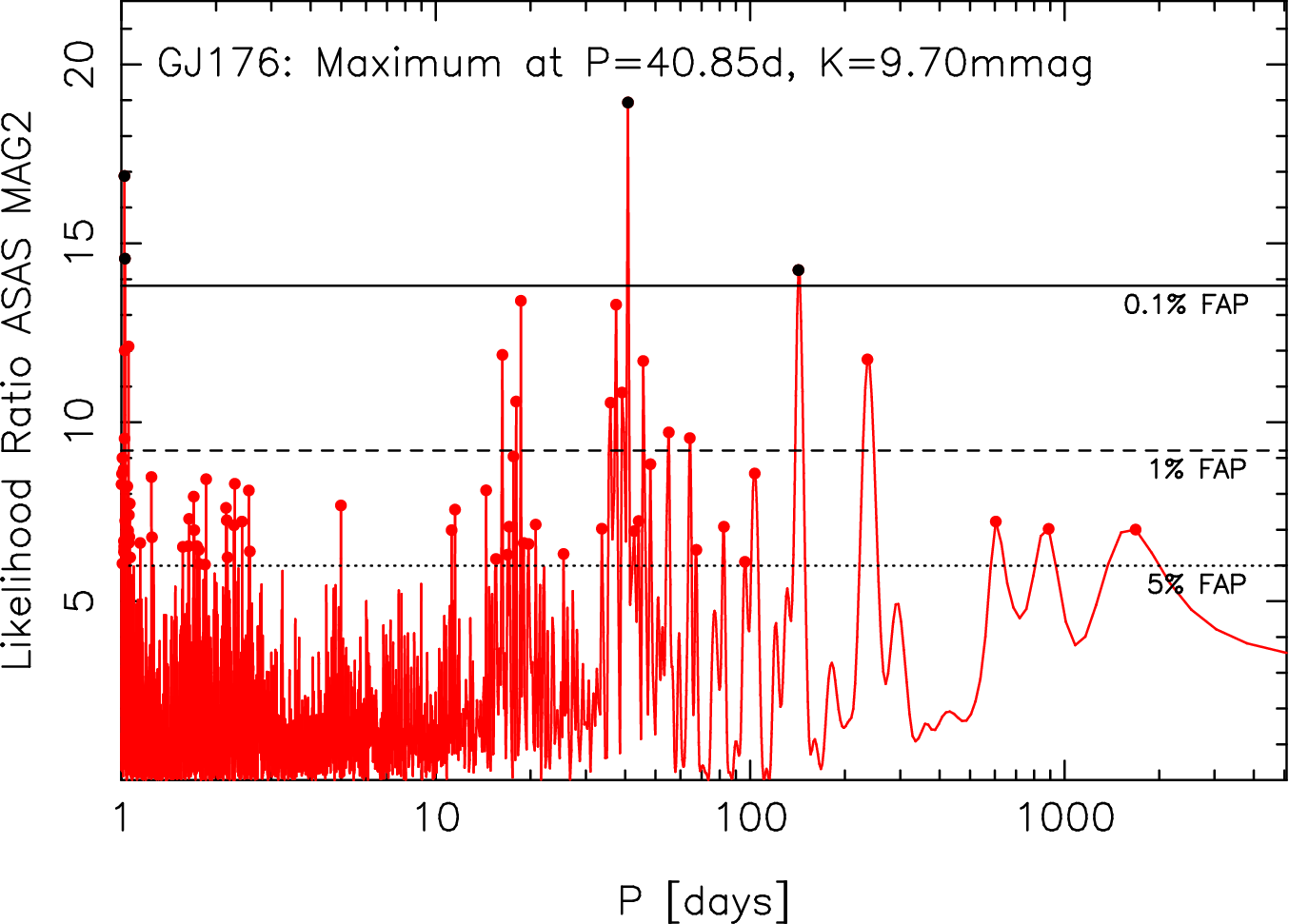}
\includegraphics[angle=270, width=0.19\textwidth,clip]{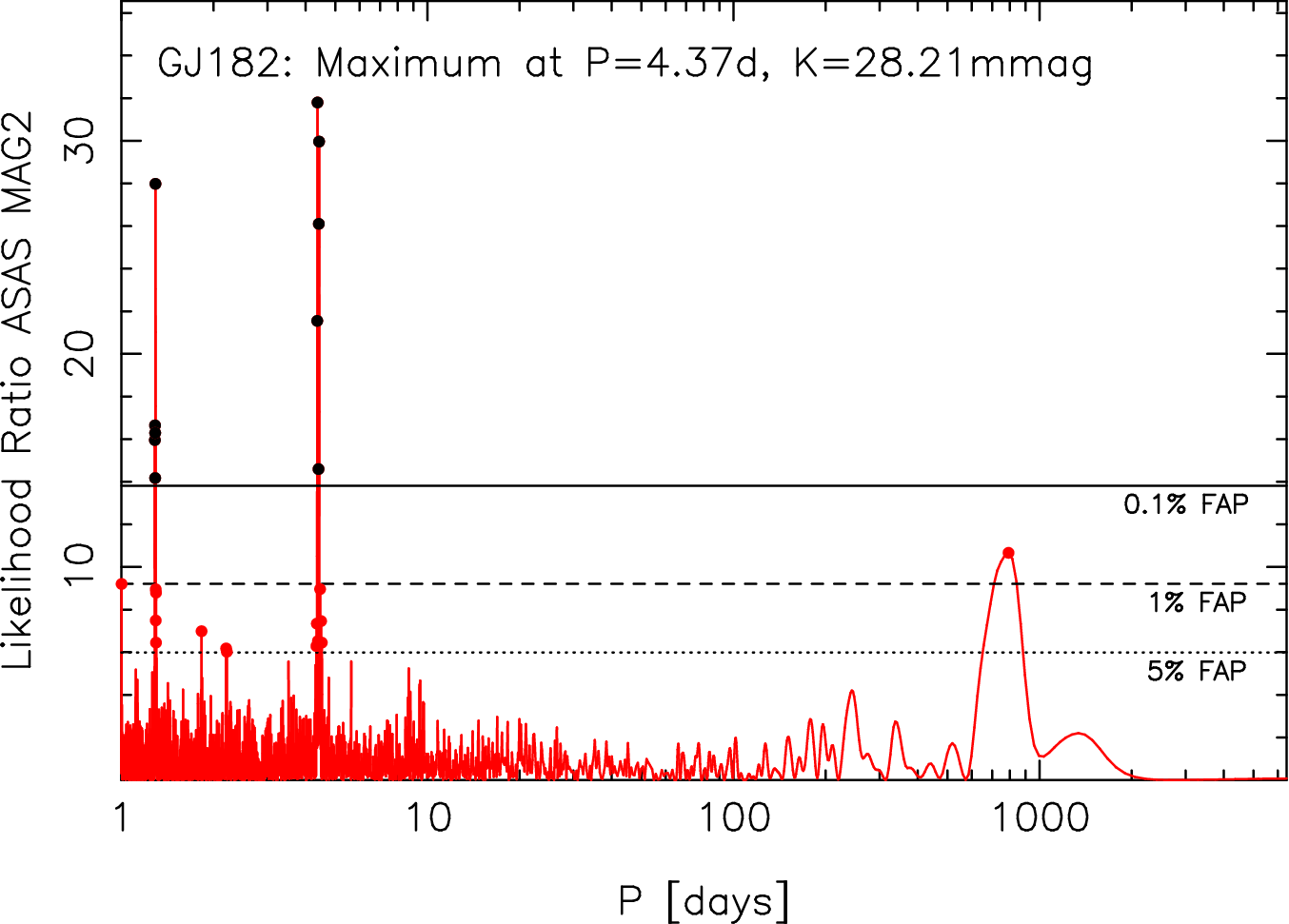}
\includegraphics[angle=270, width=0.19\textwidth,clip]{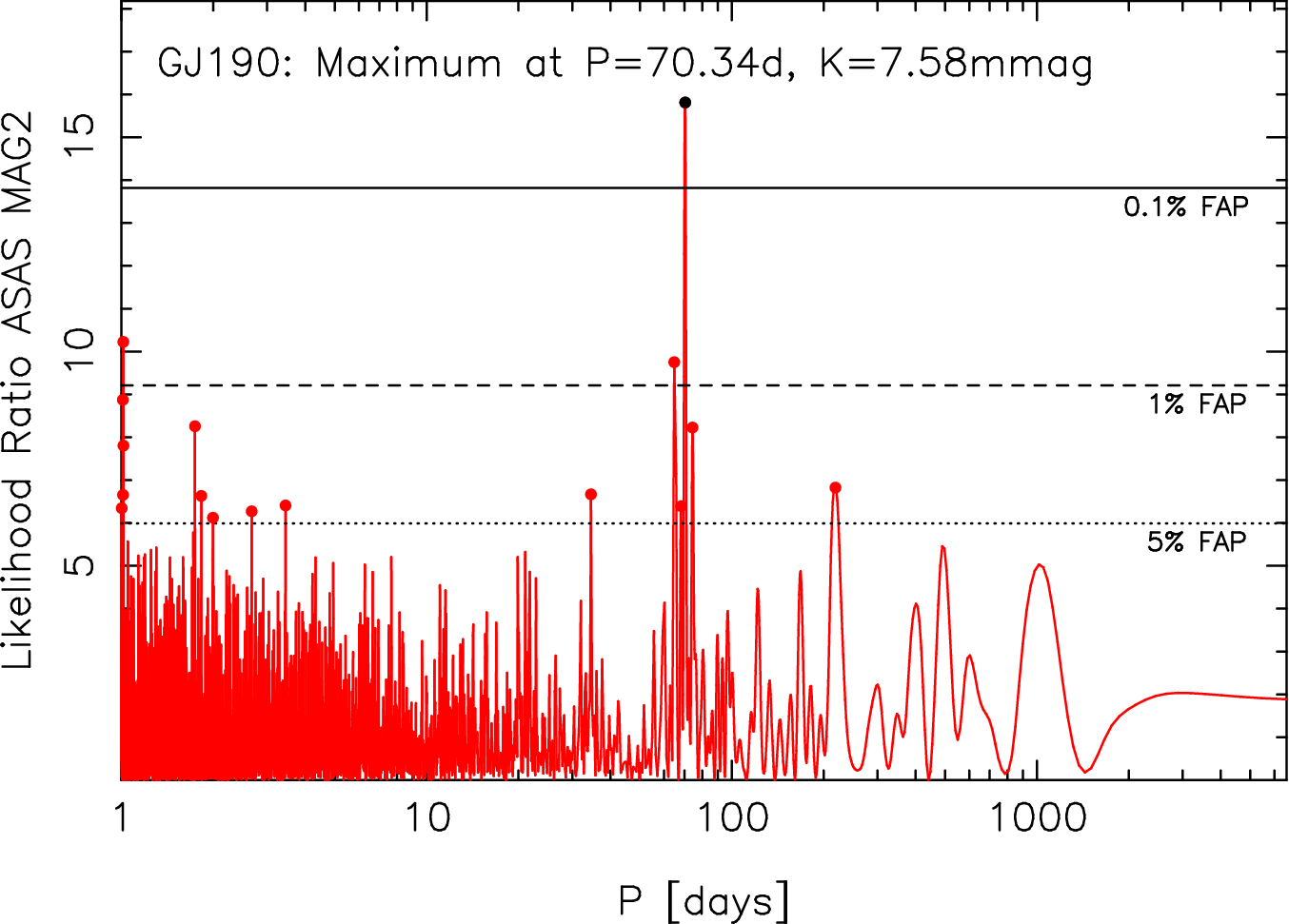}
\includegraphics[angle=270, width=0.19\textwidth,clip]{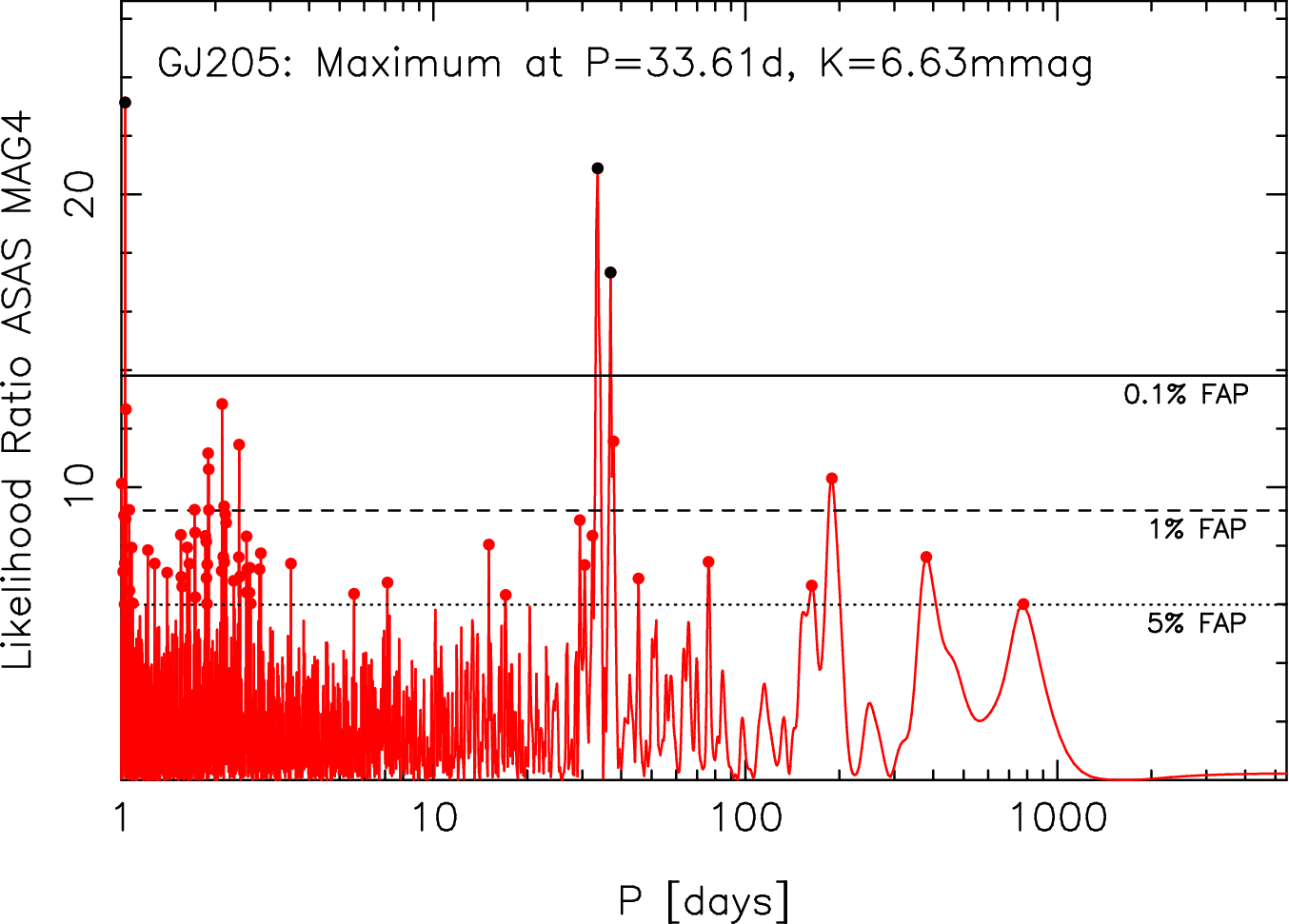}
\includegraphics[angle=270, width=0.19\textwidth,clip]{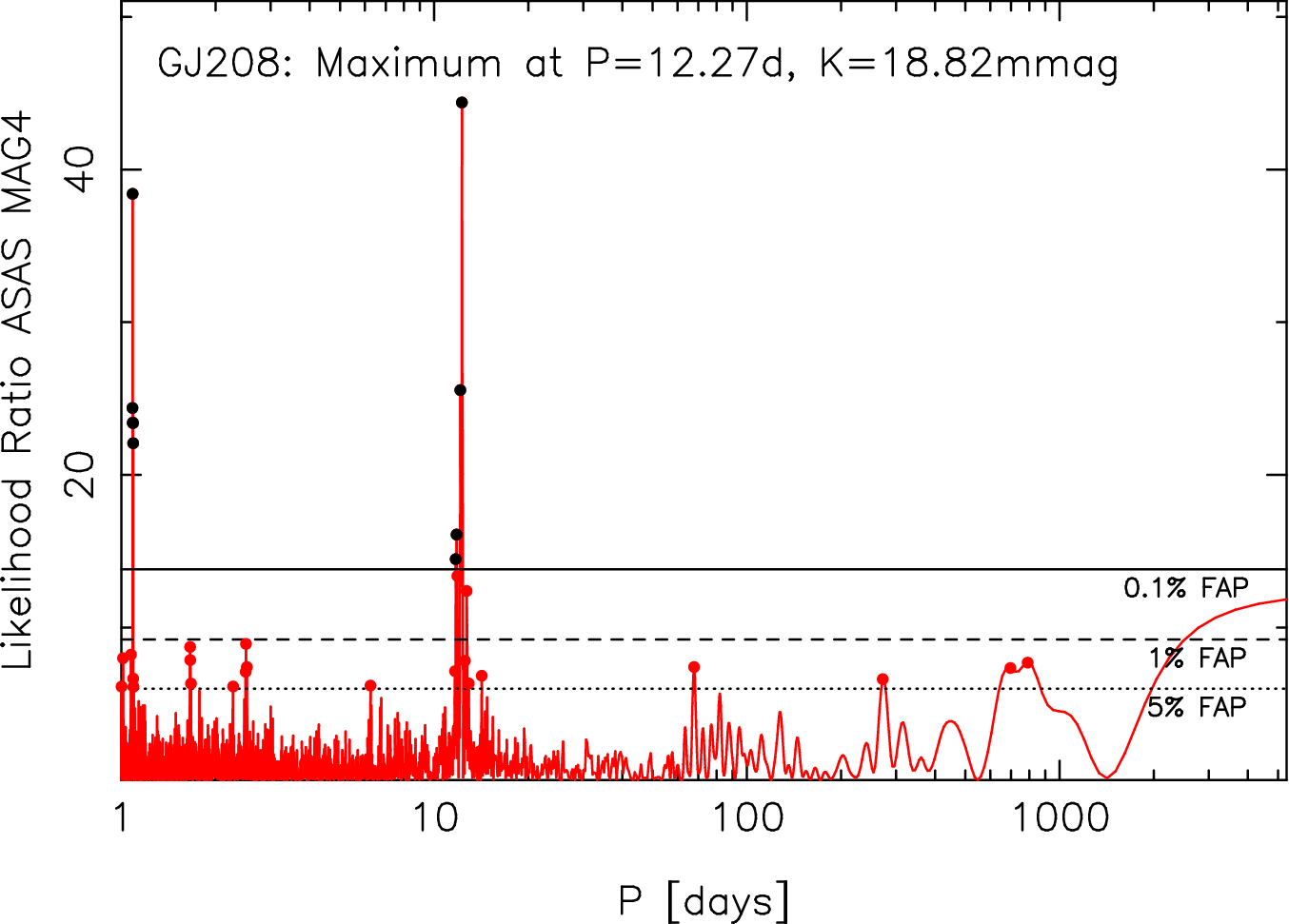}
\includegraphics[angle=270, width=0.19\textwidth,clip]{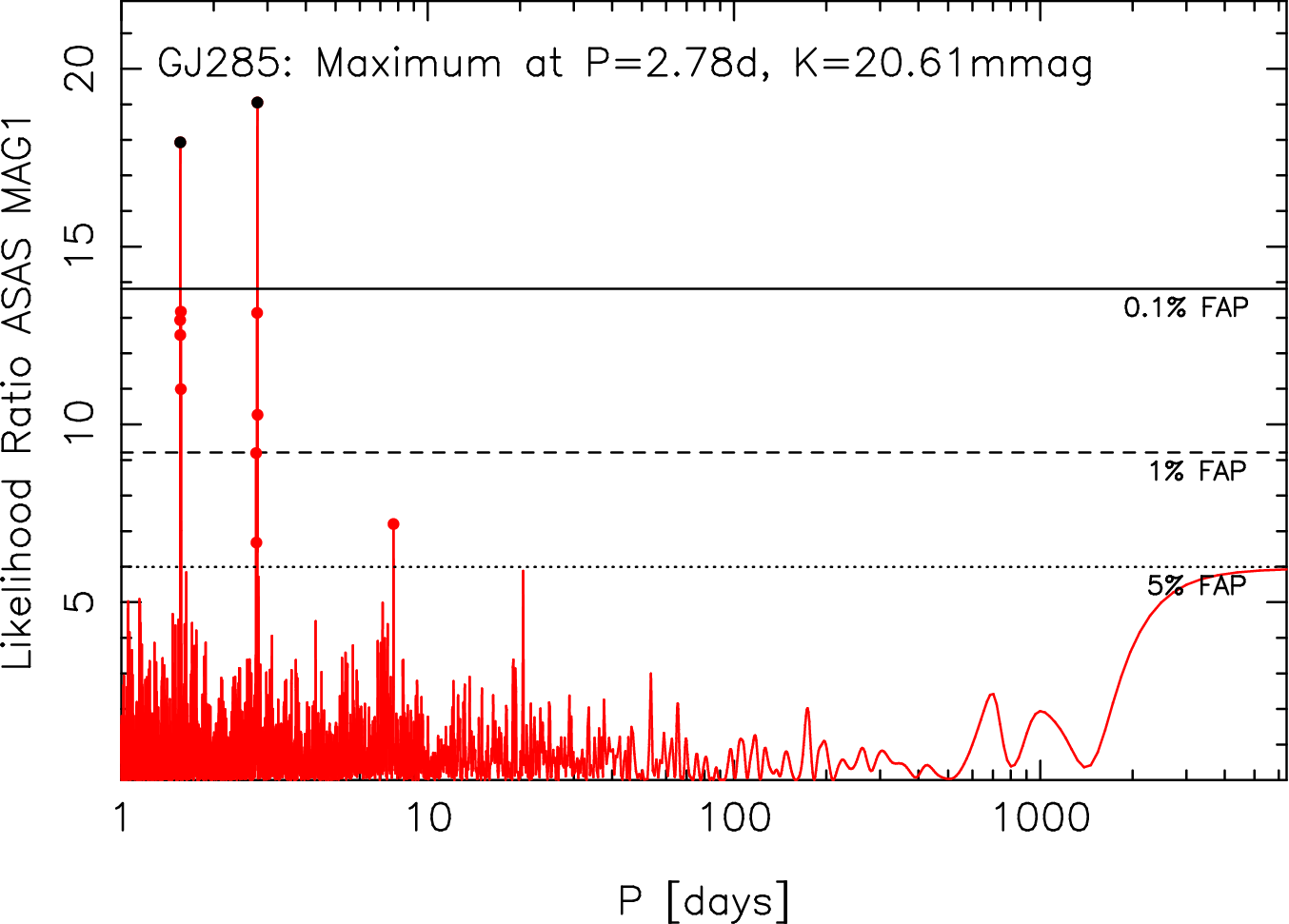}
\includegraphics[angle=270, width=0.19\textwidth,clip]{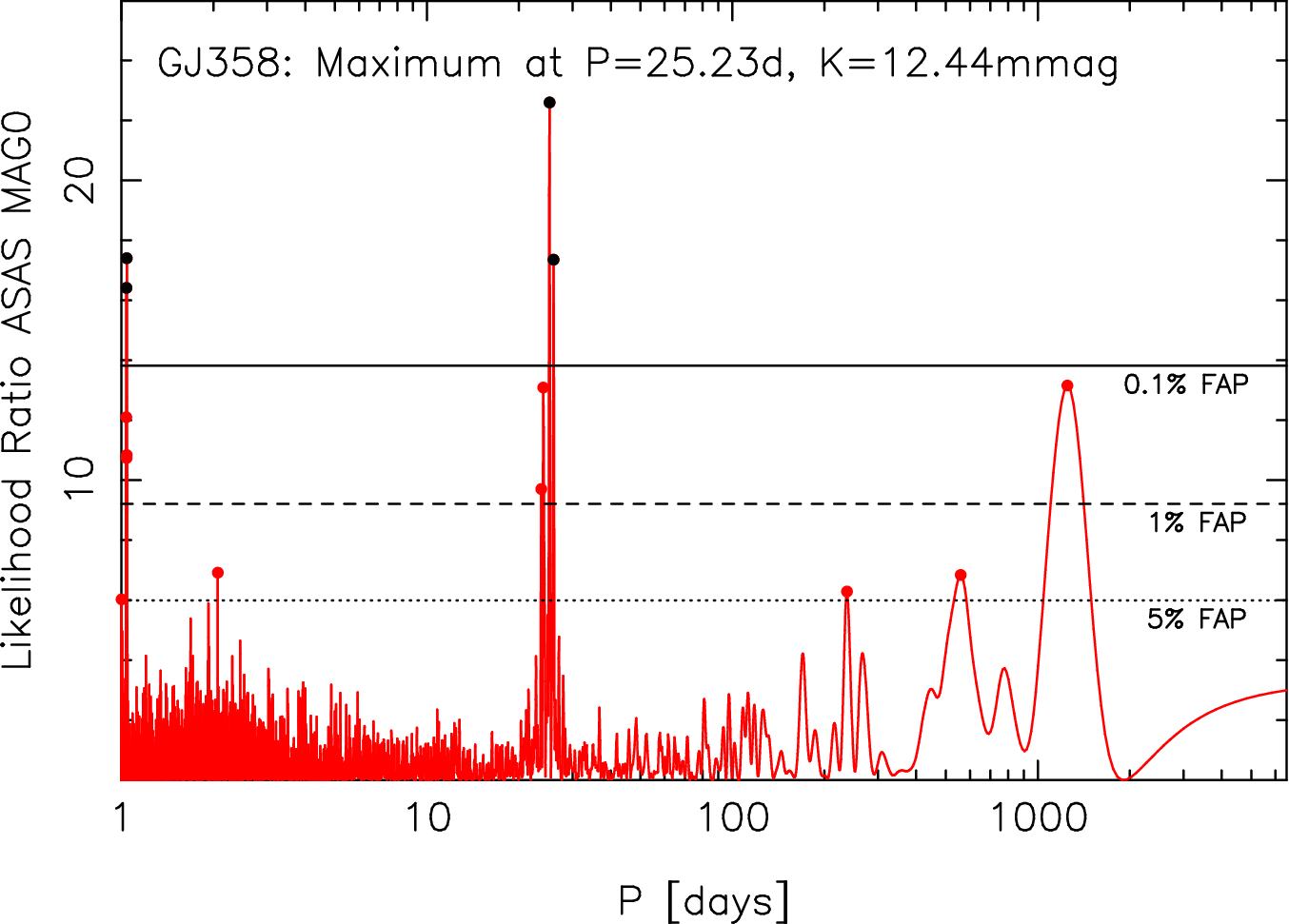}
\includegraphics[angle=270, width=0.19\textwidth,clip]{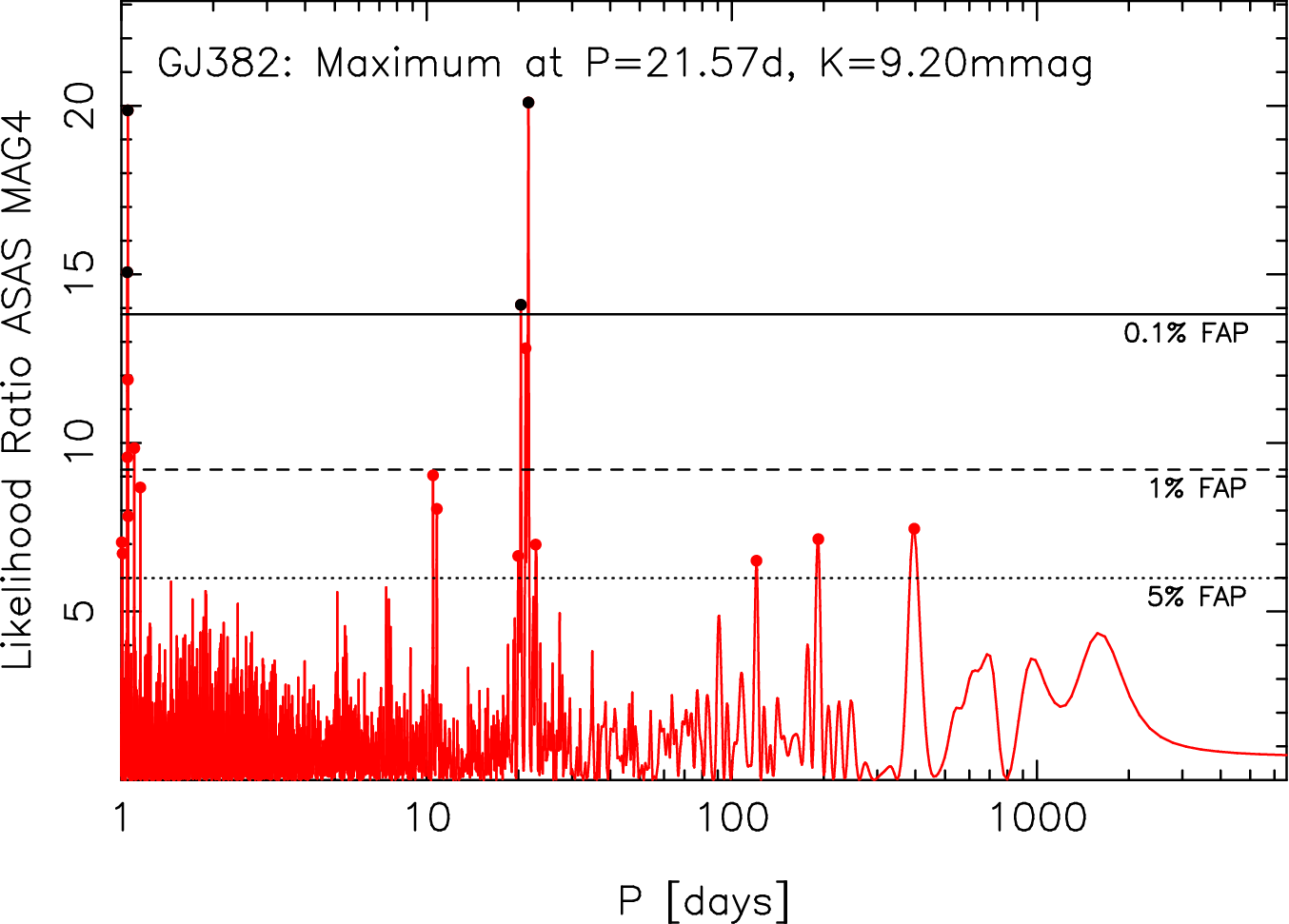}
\includegraphics[angle=270, width=0.19\textwidth,clip]{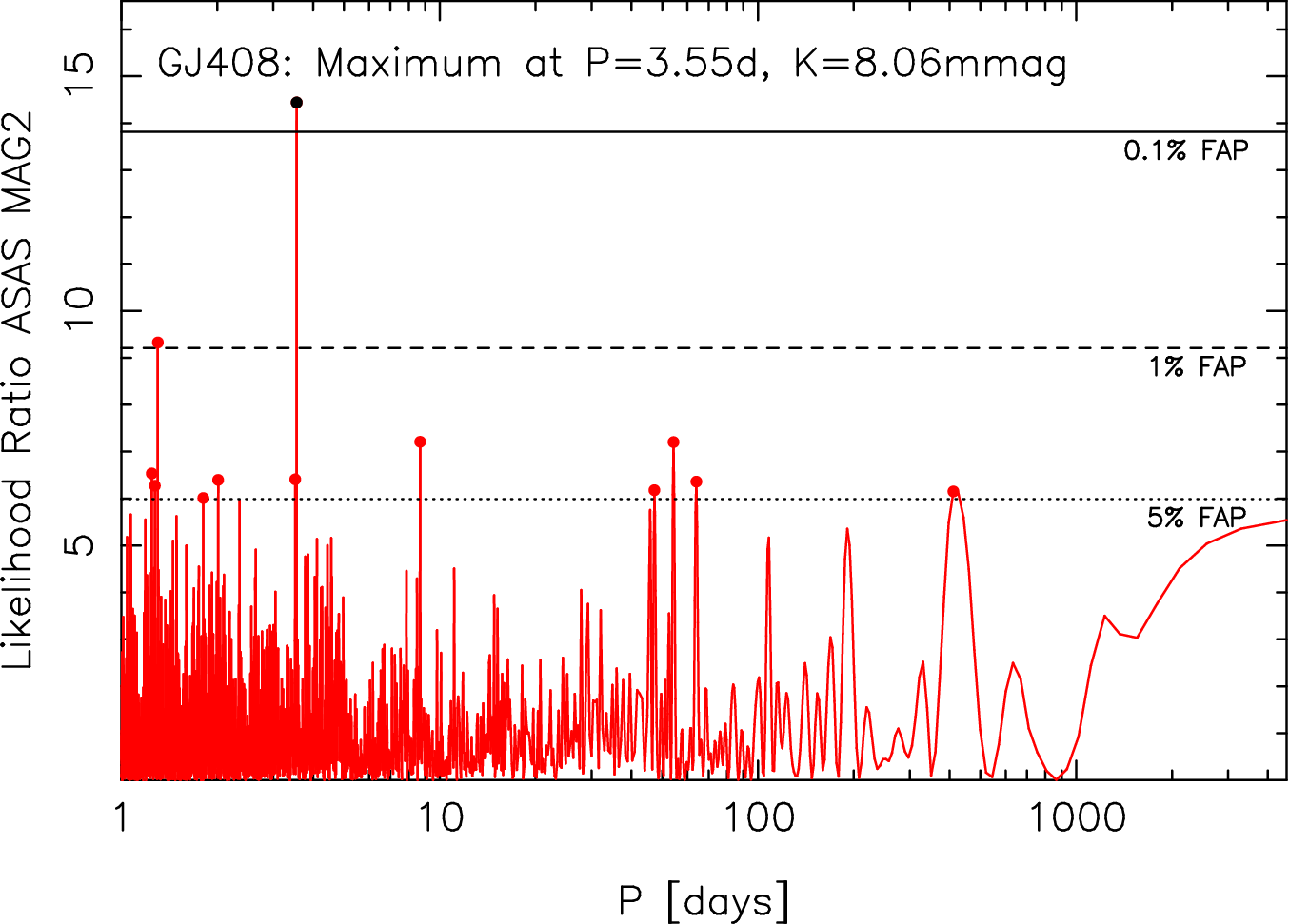}
\includegraphics[angle=270, width=0.19\textwidth,clip]{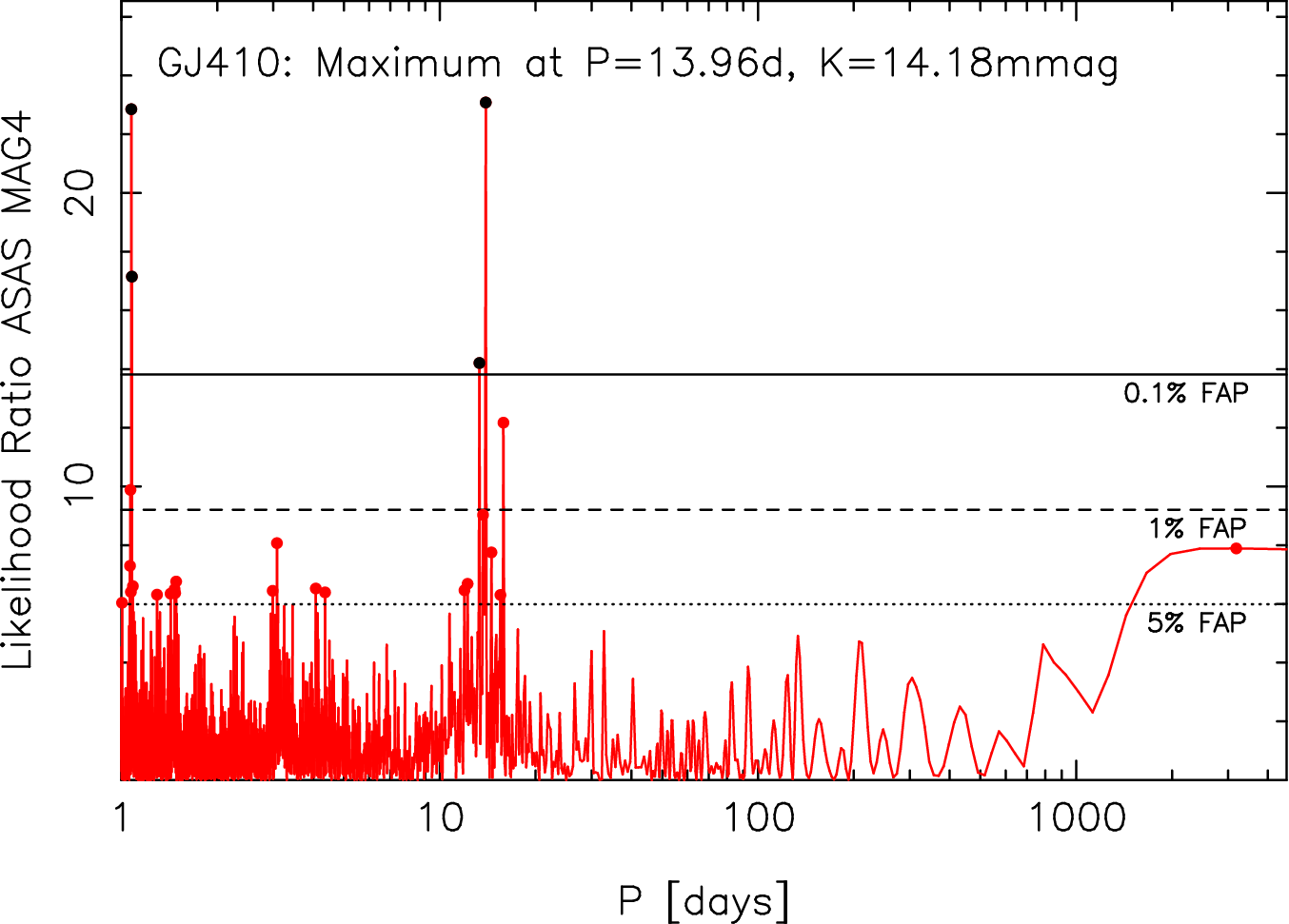}
\includegraphics[angle=270, width=0.19\textwidth,clip]{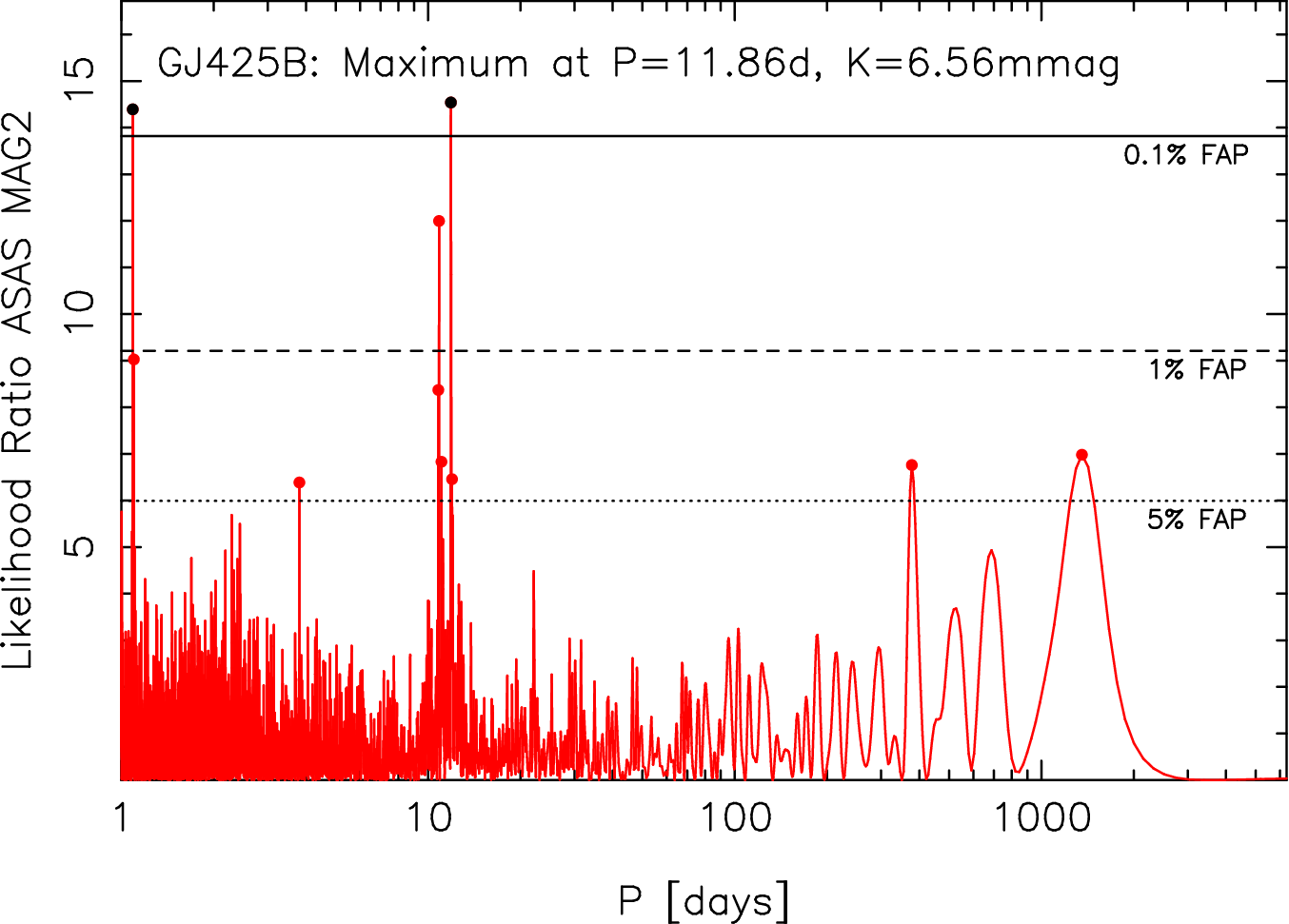}
\includegraphics[angle=270, width=0.19\textwidth,clip]{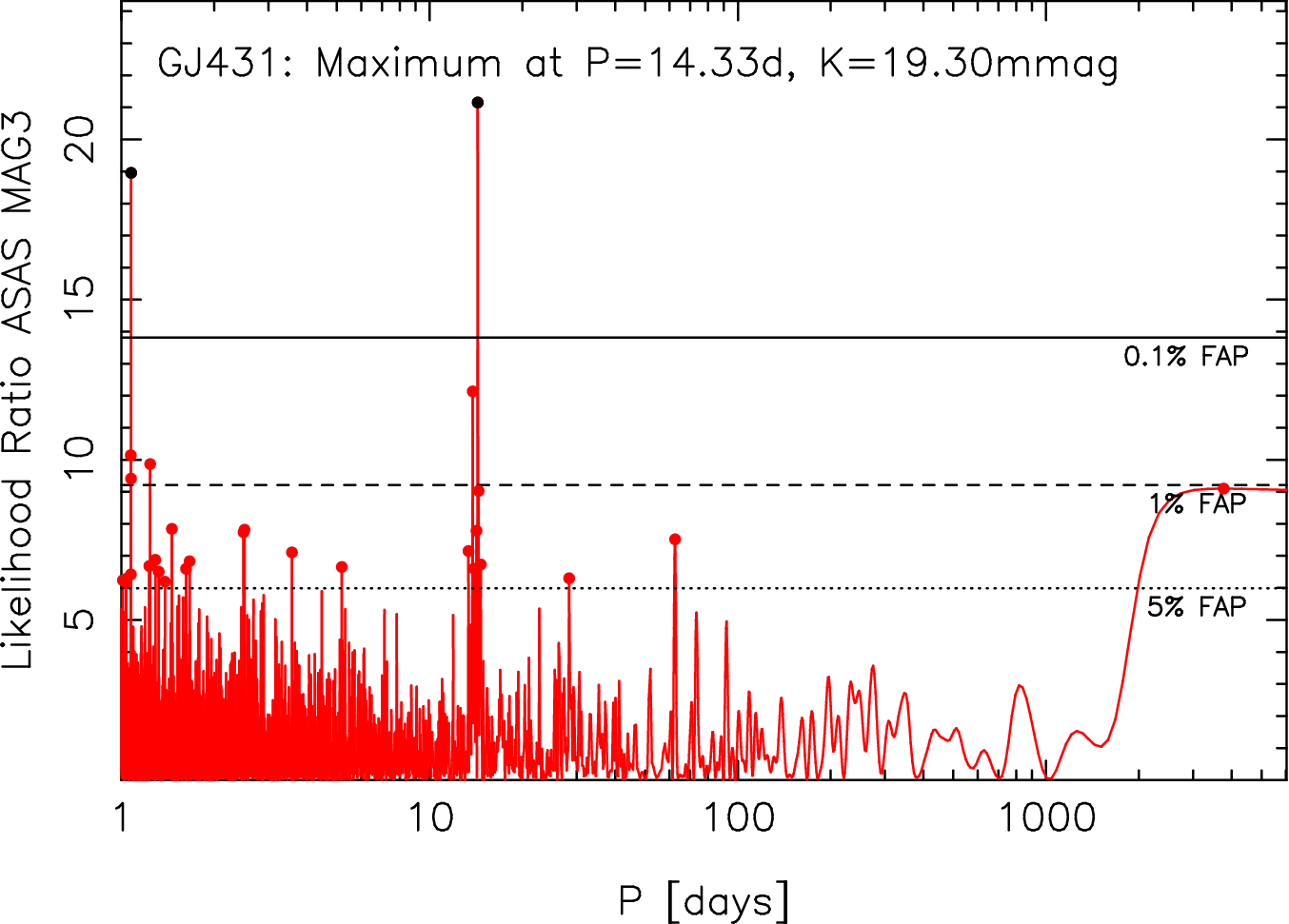}
\includegraphics[angle=270, width=0.19\textwidth,clip]{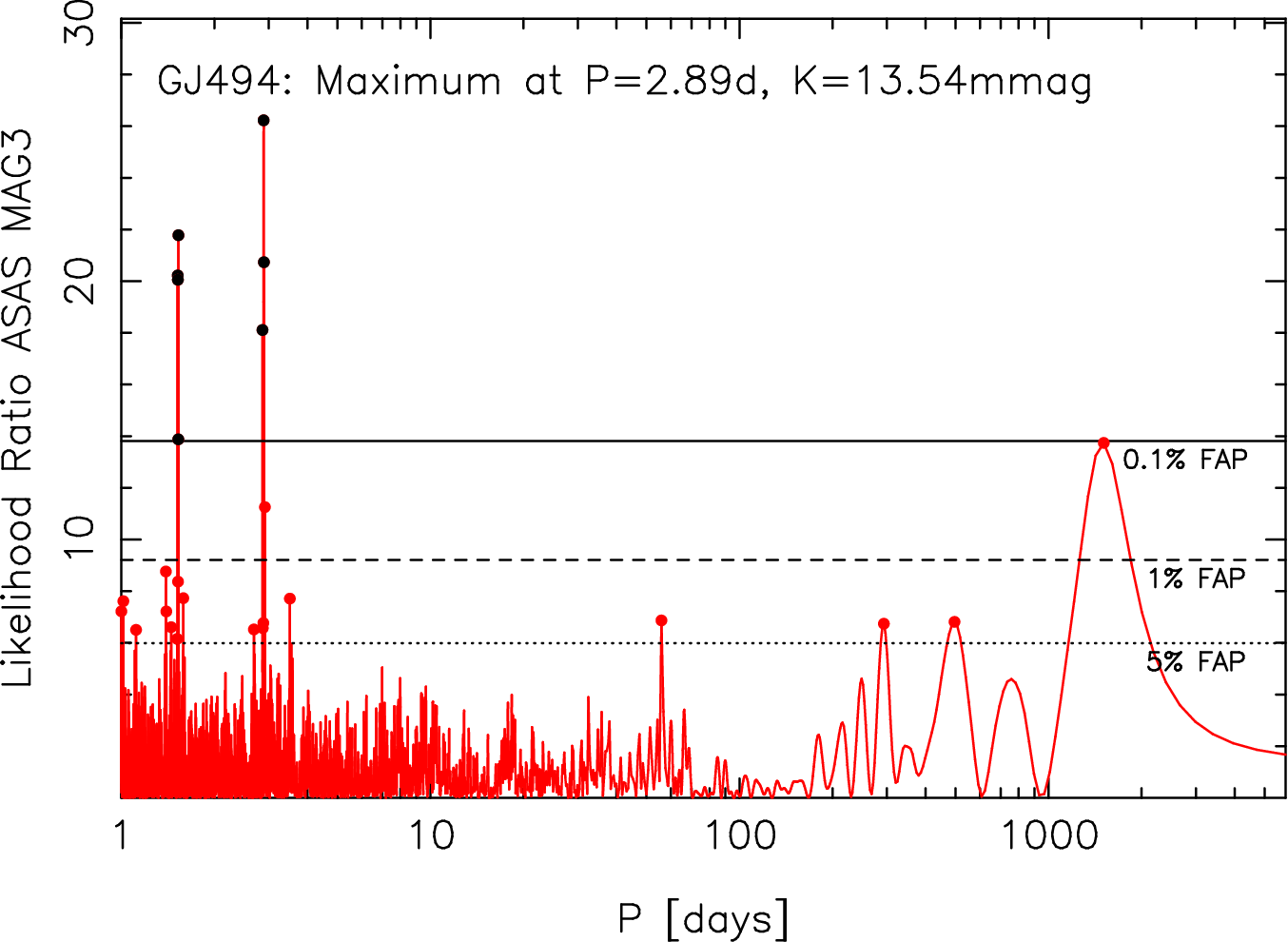}
\includegraphics[angle=270, width=0.19\textwidth,clip]{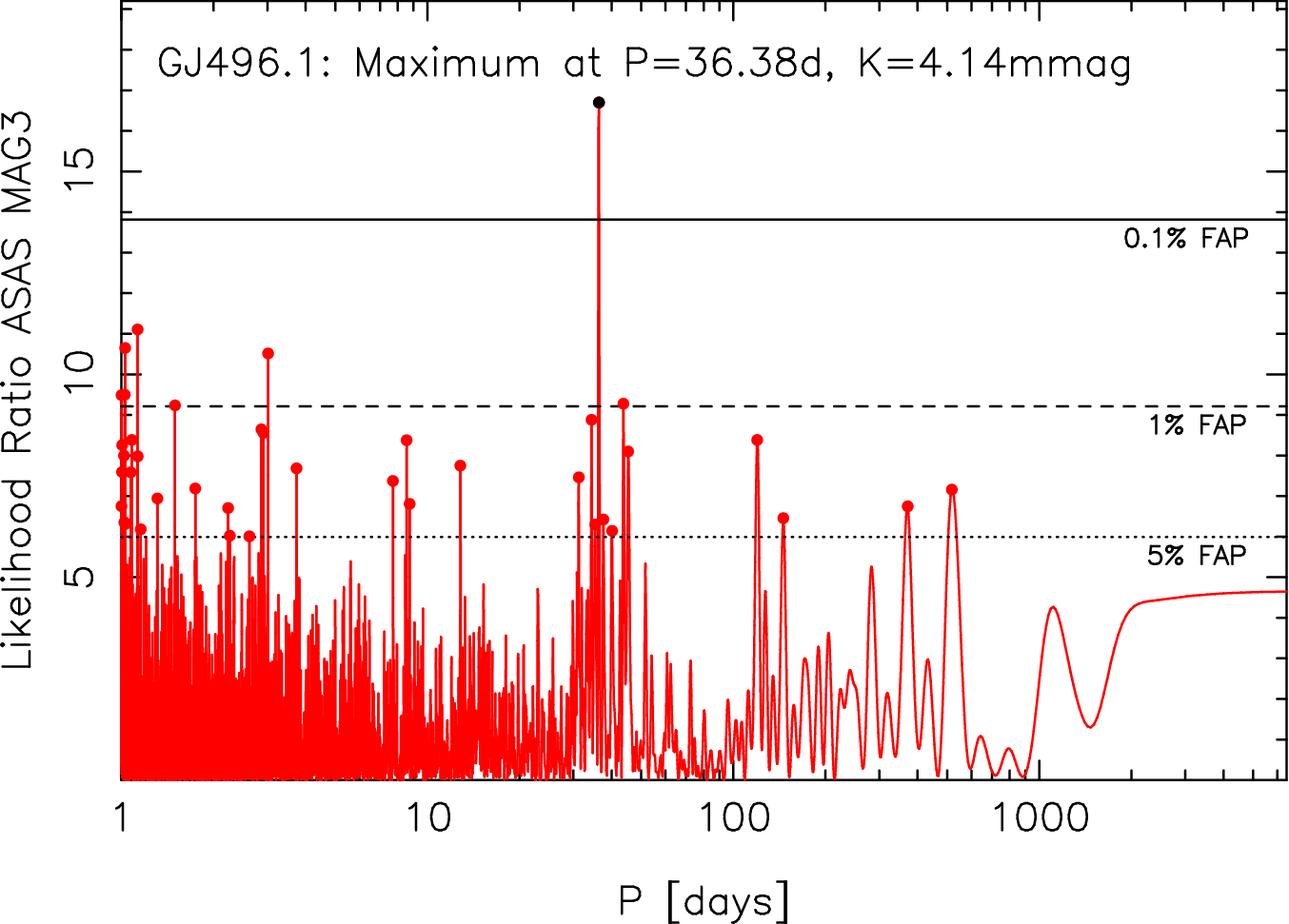}
\includegraphics[angle=270, width=0.19\textwidth,clip]{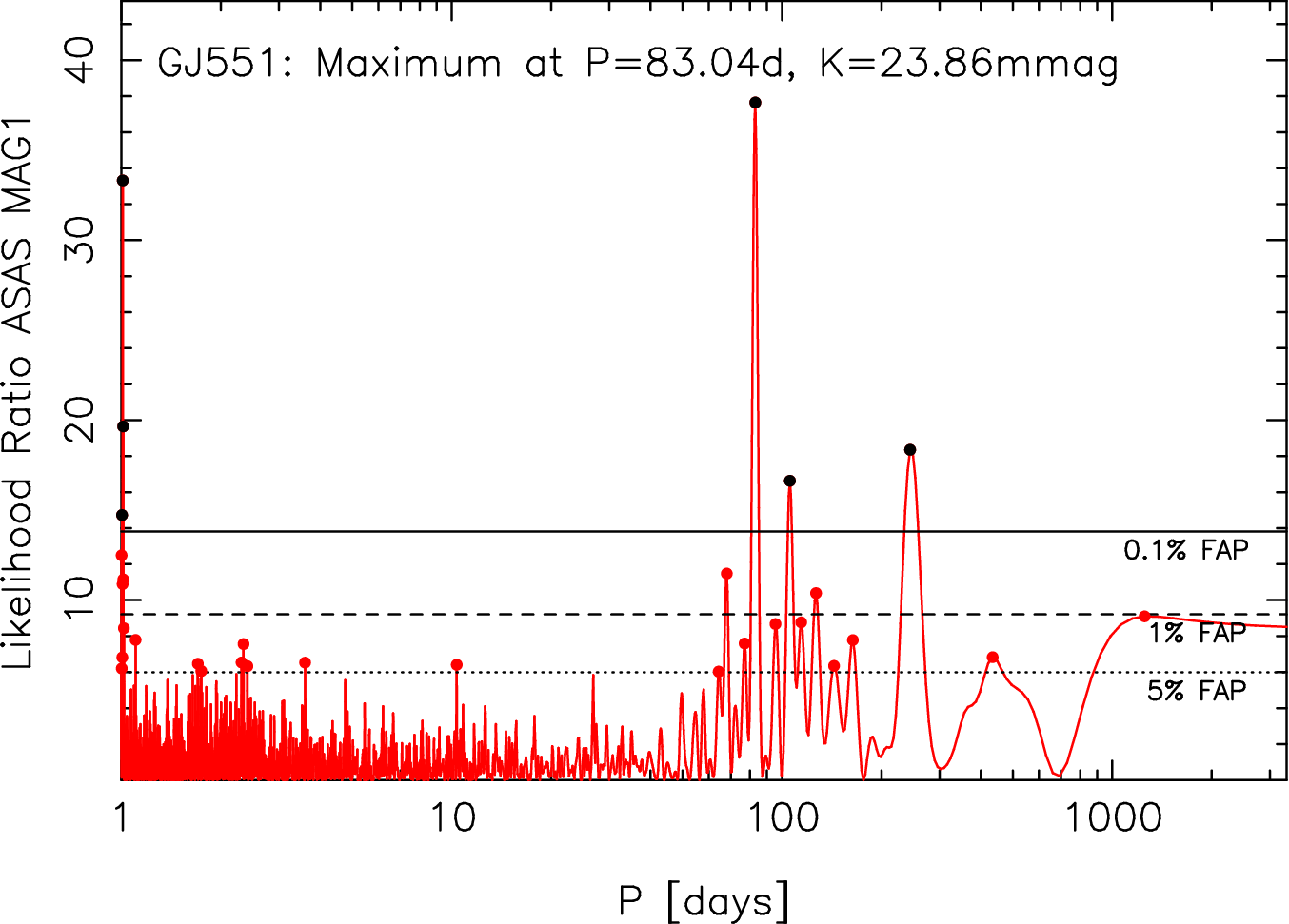}
\includegraphics[angle=270, width=0.19\textwidth,clip]{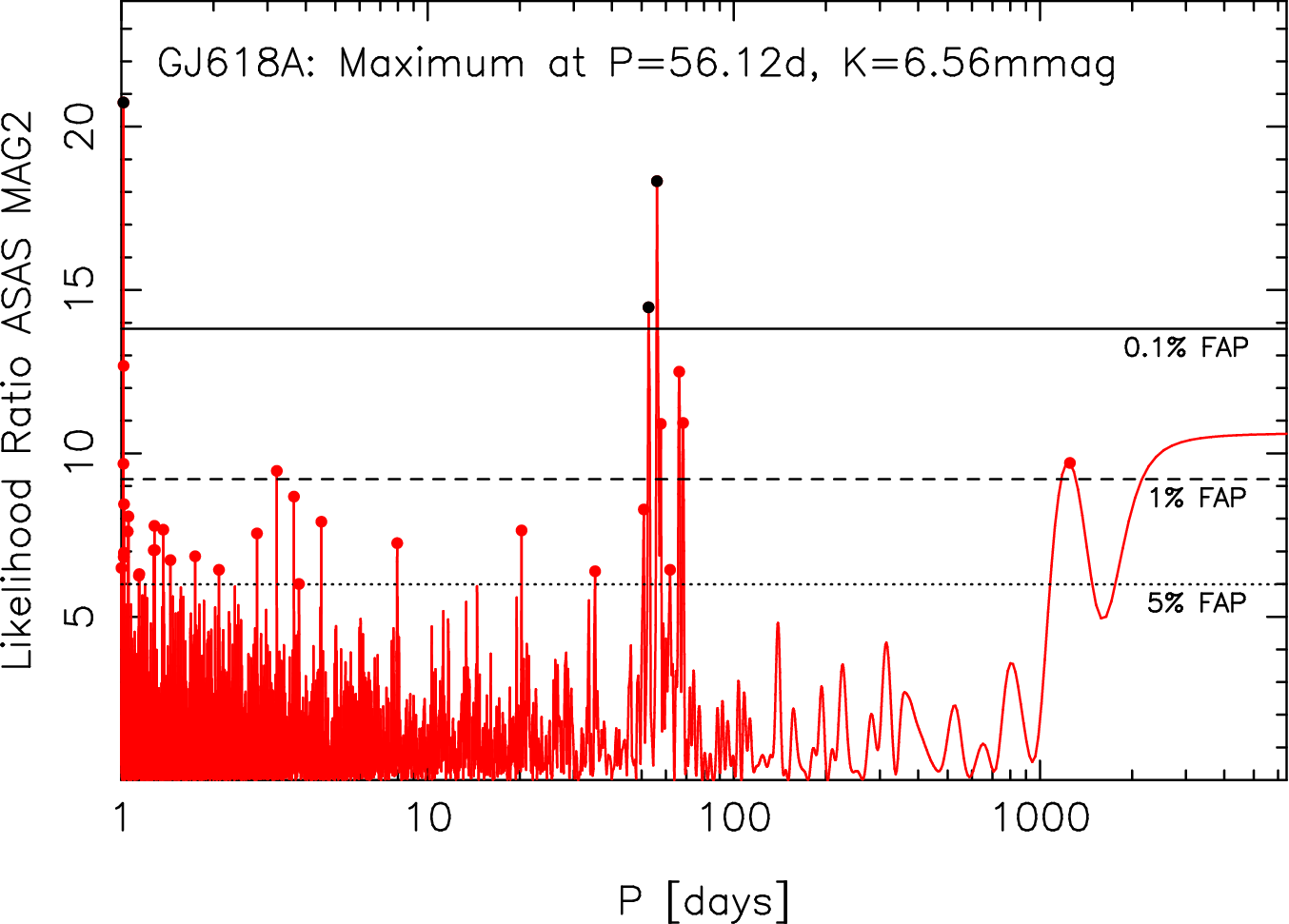}
\includegraphics[angle=270, width=0.19\textwidth,clip]{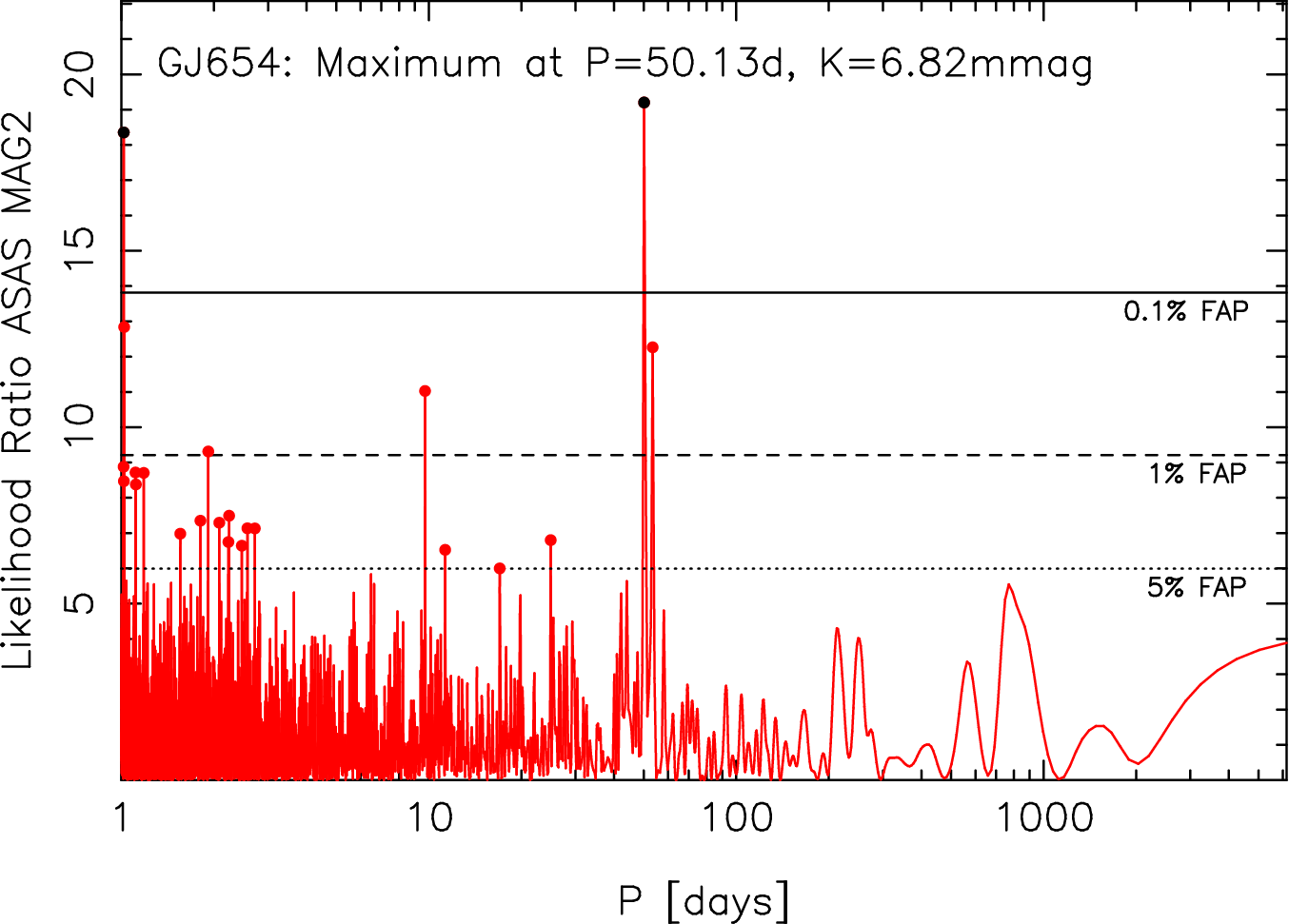}
\includegraphics[angle=270, width=0.19\textwidth,clip]{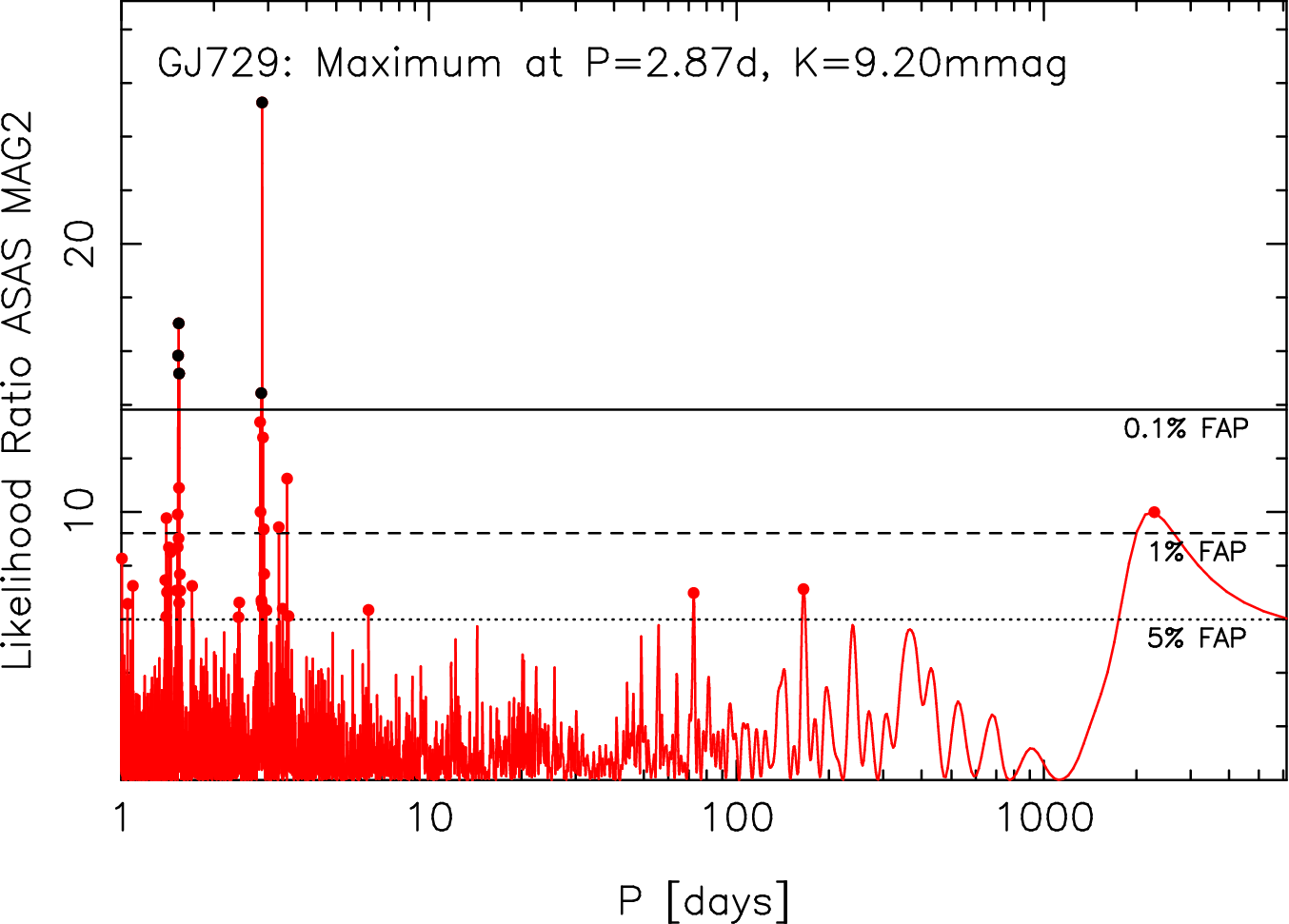}
\includegraphics[angle=270, width=0.19\textwidth,clip]{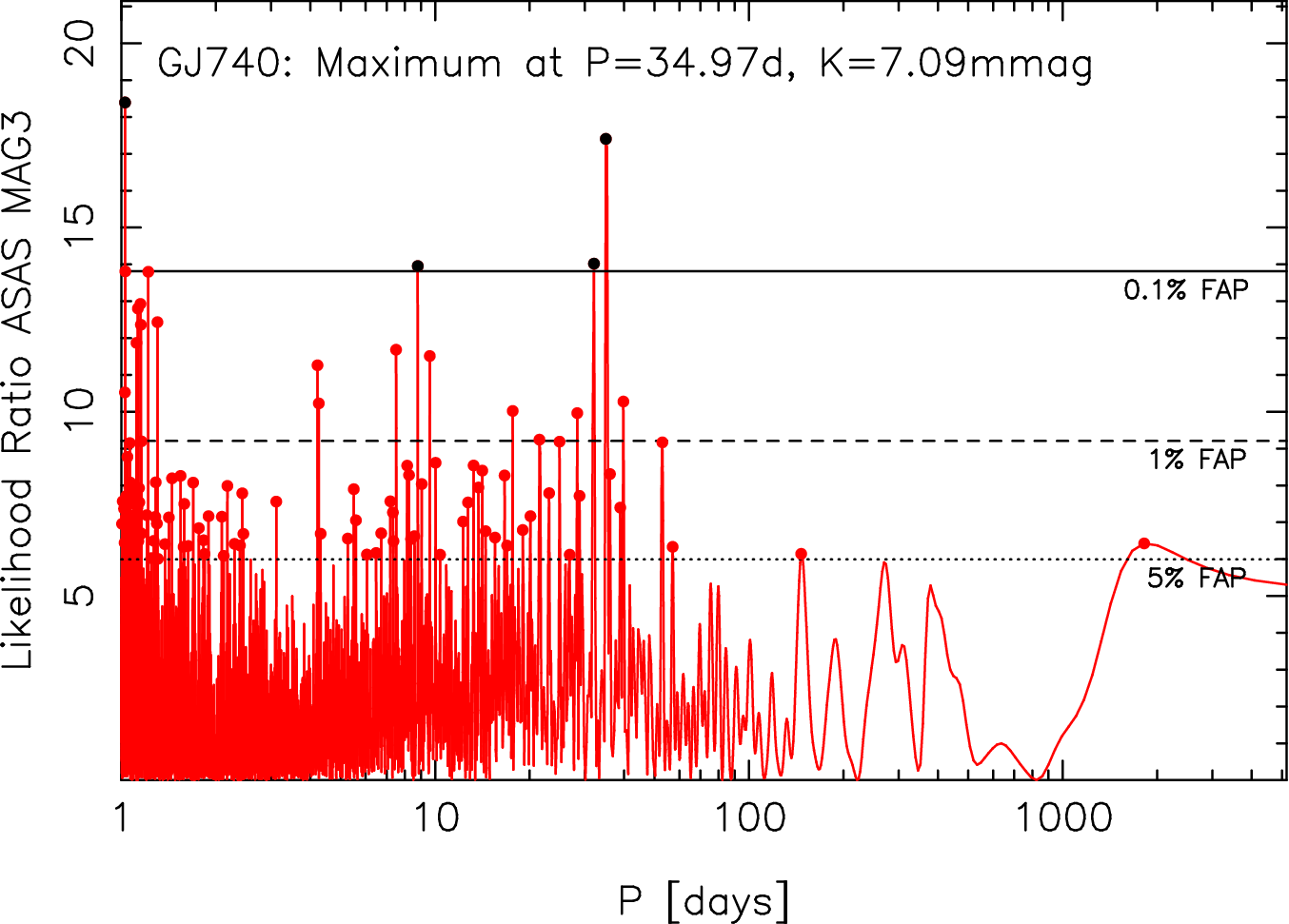}
\includegraphics[angle=270, width=0.19\textwidth,clip]{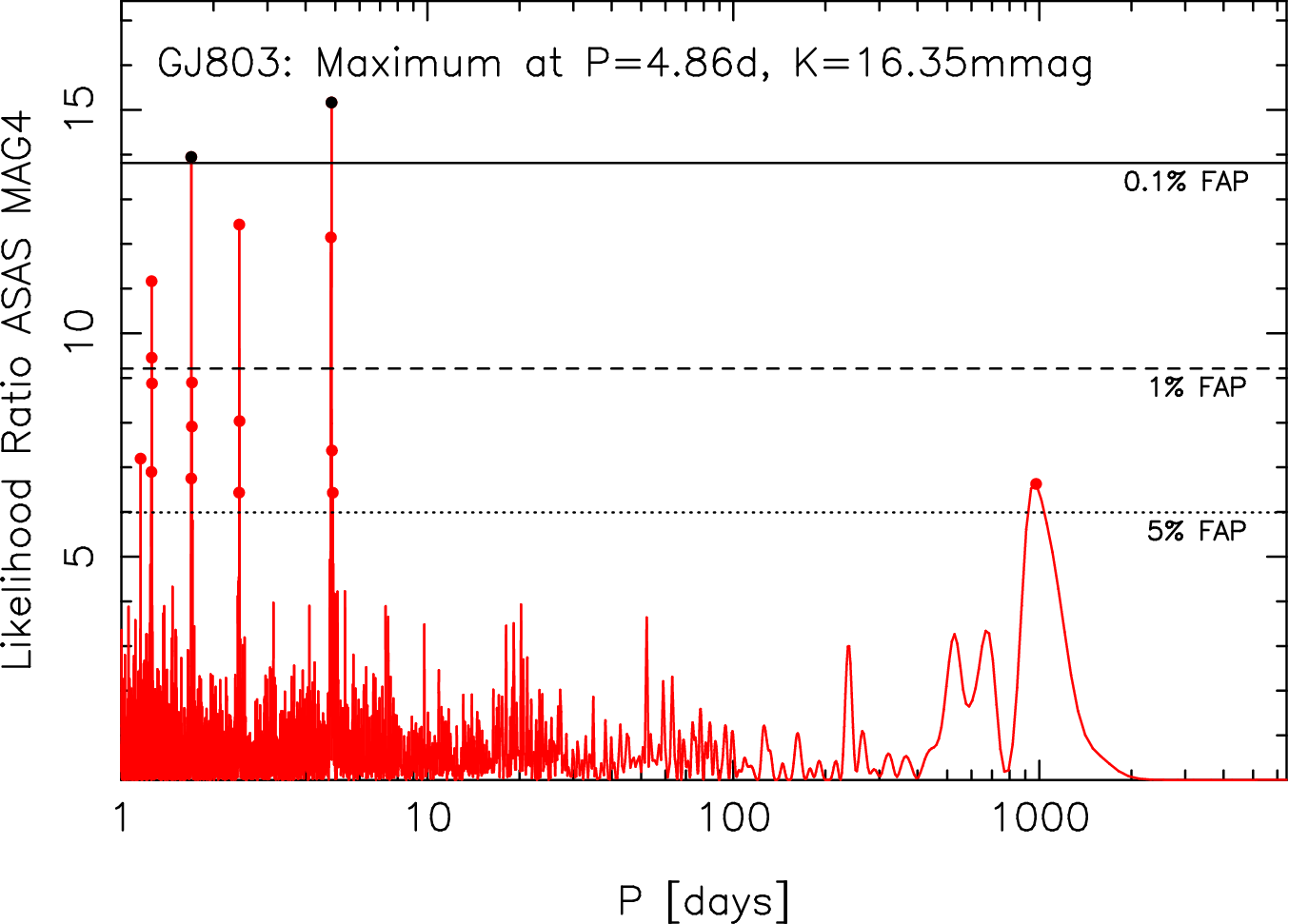}
\includegraphics[angle=270, width=0.19\textwidth,clip]{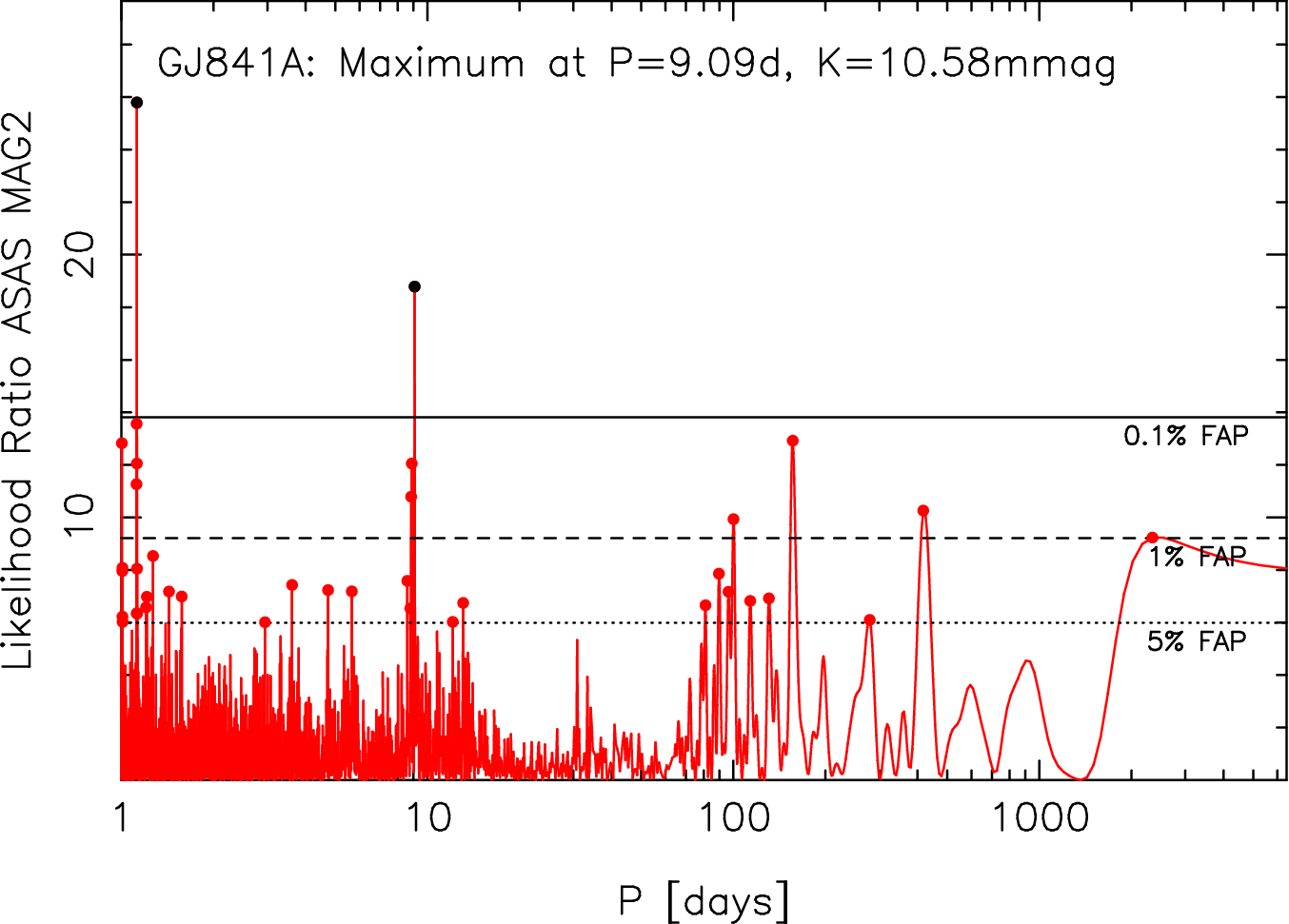}
\includegraphics[angle=270, width=0.19\textwidth,clip]{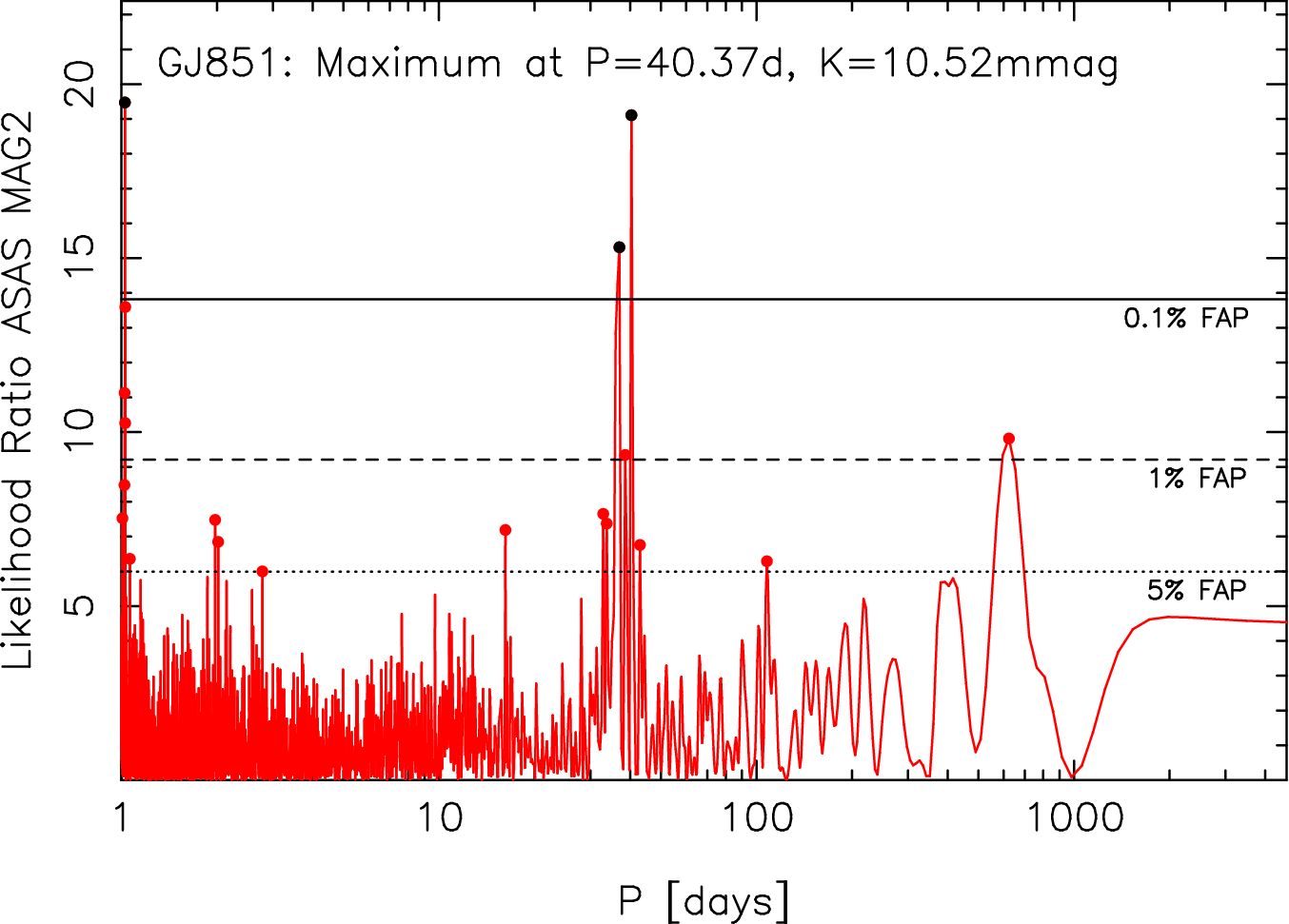}
\includegraphics[angle=270, width=0.19\textwidth,clip]{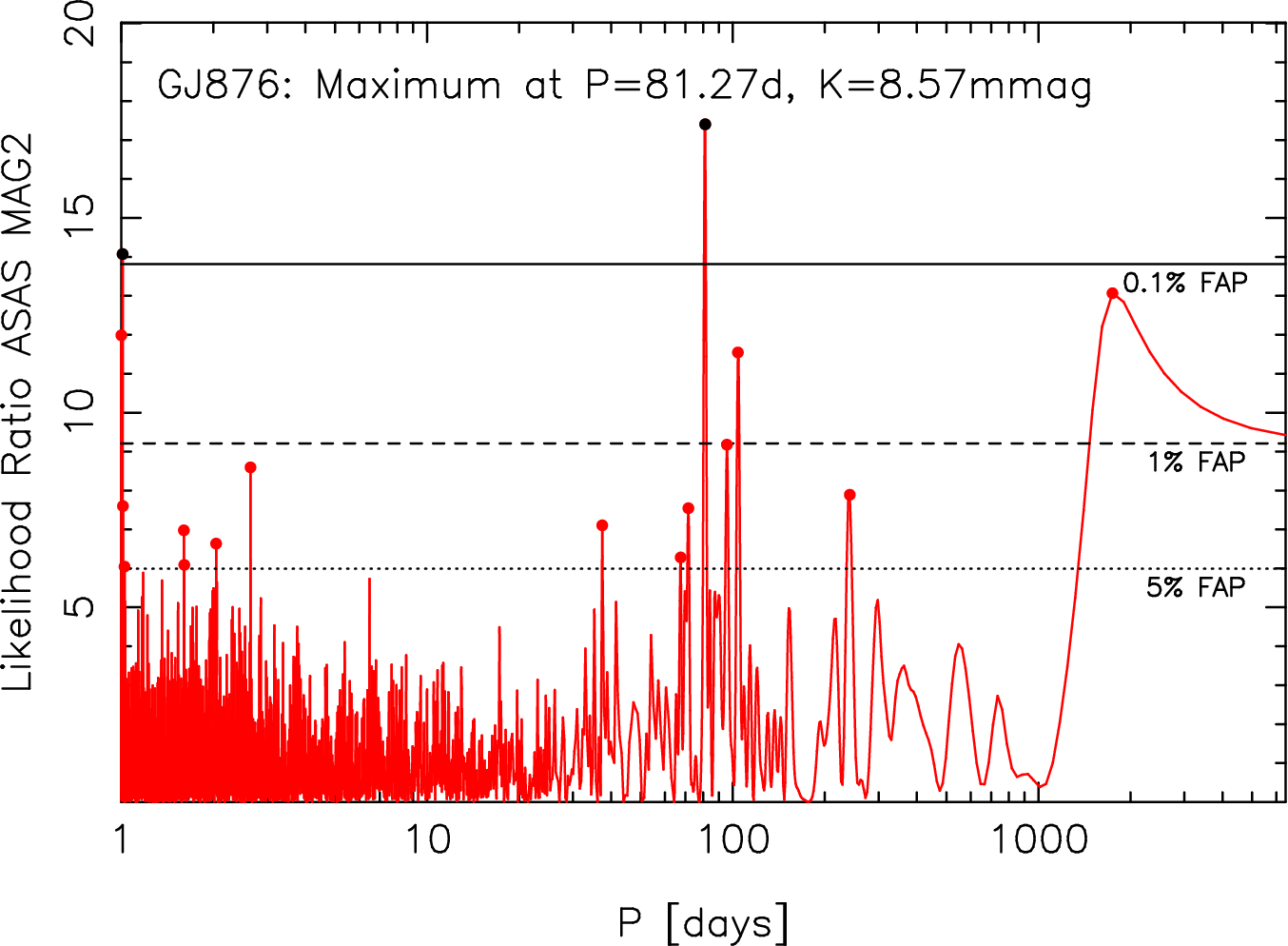}
\includegraphics[angle=270, width=0.19\textwidth,clip]{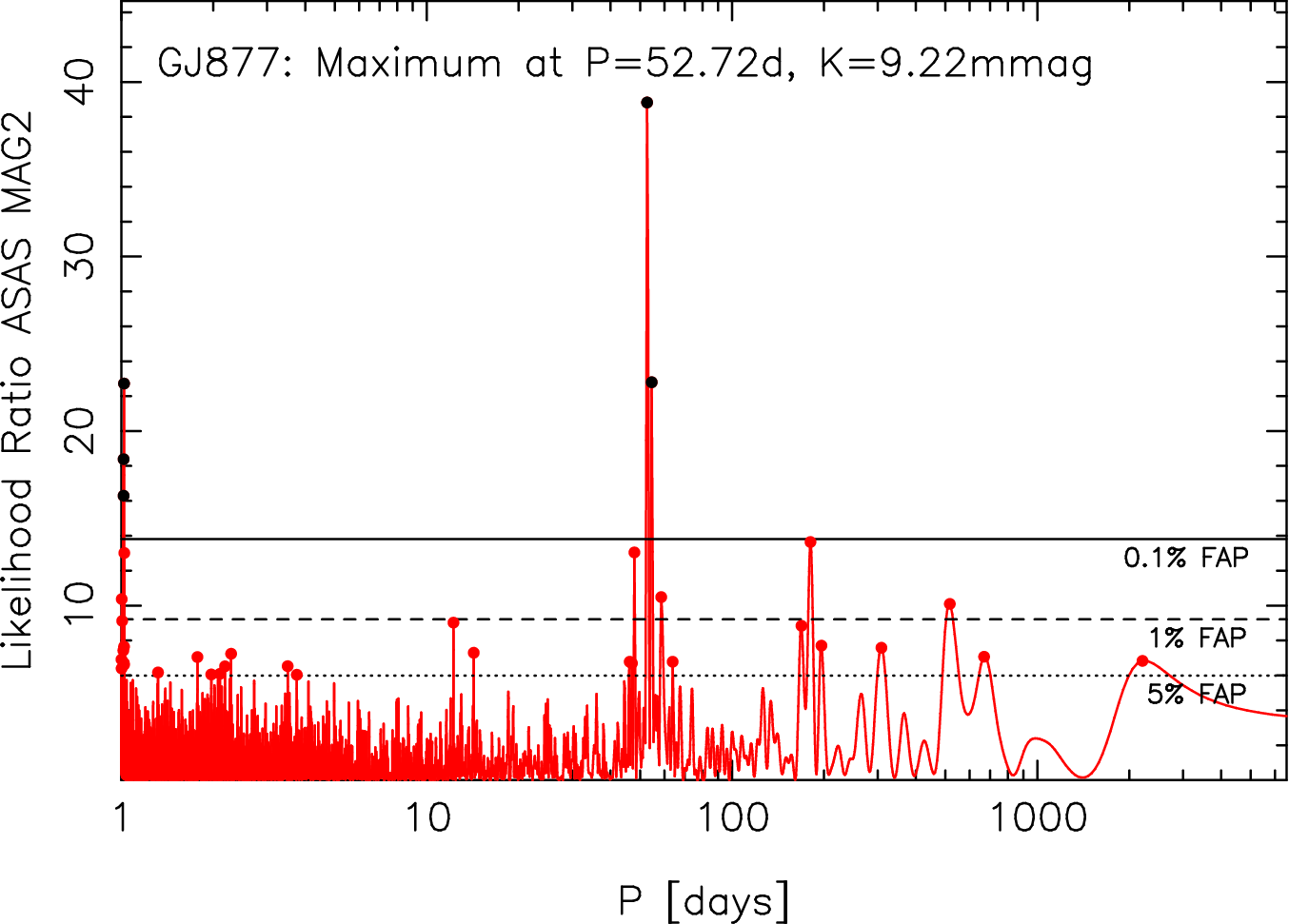}
\includegraphics[angle=270, width=0.19\textwidth,clip]{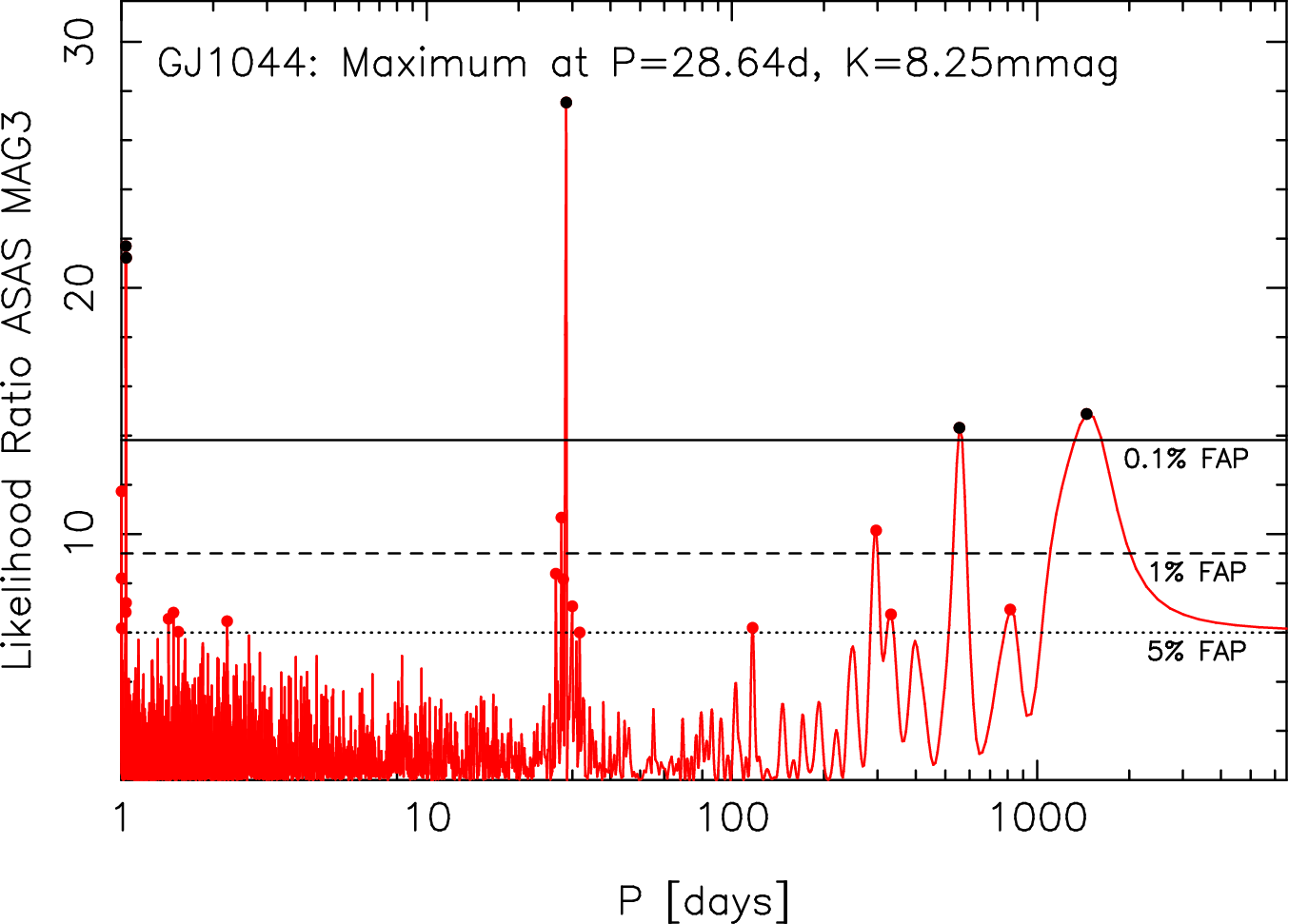}
\includegraphics[angle=270, width=0.19\textwidth,clip]{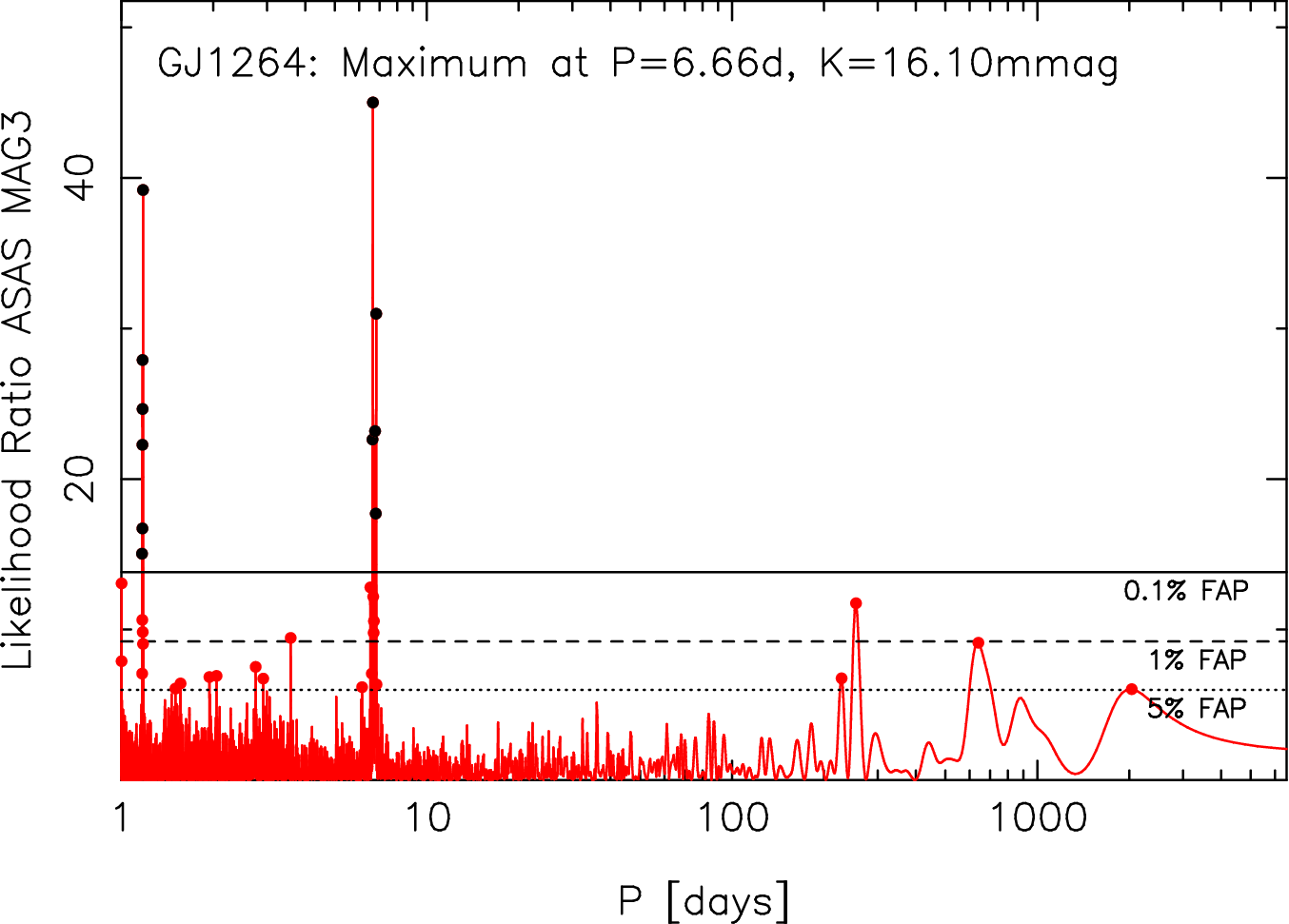}
\includegraphics[angle=270, width=0.19\textwidth,clip]{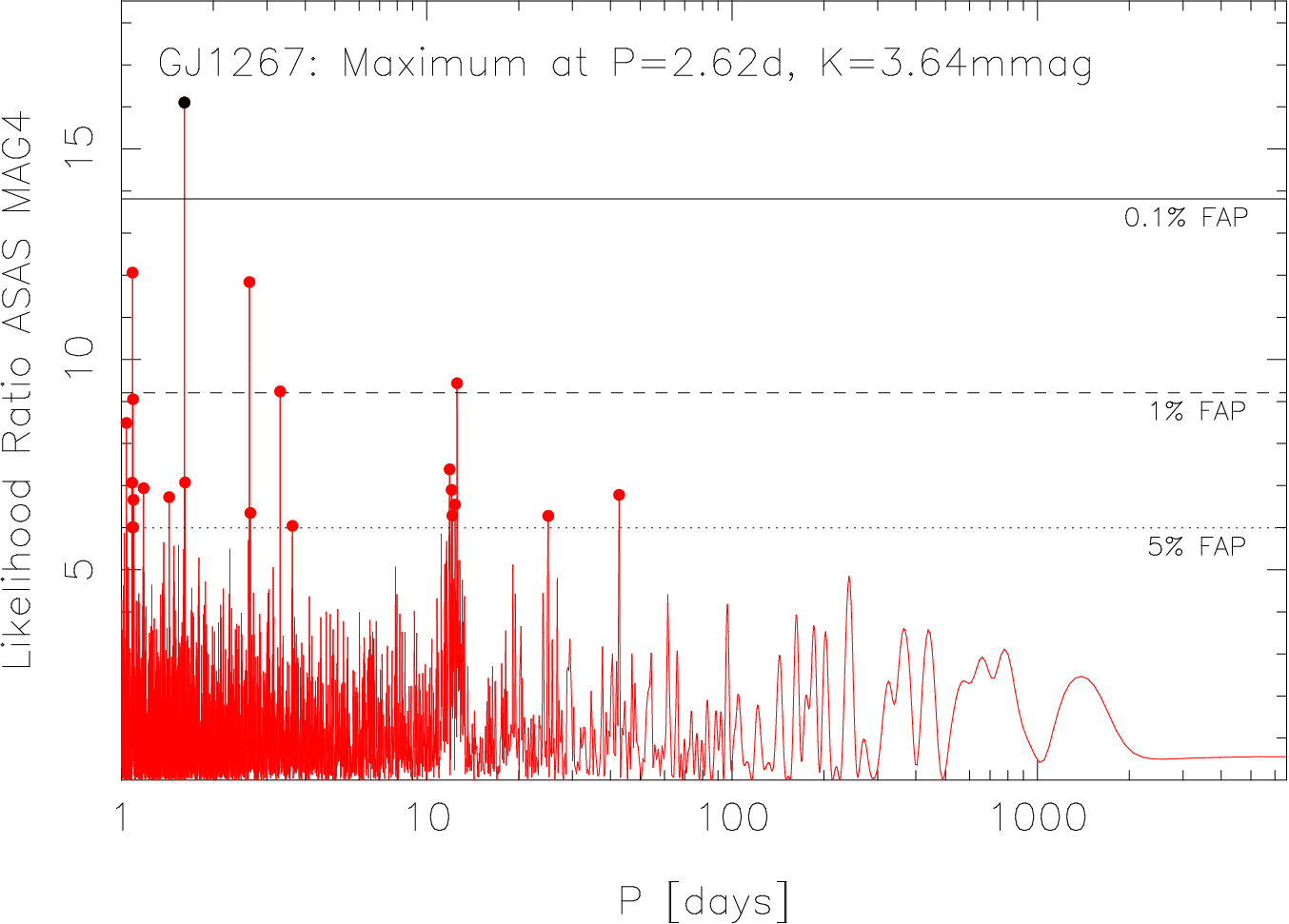}
\includegraphics[angle=270, width=0.19\textwidth,clip]{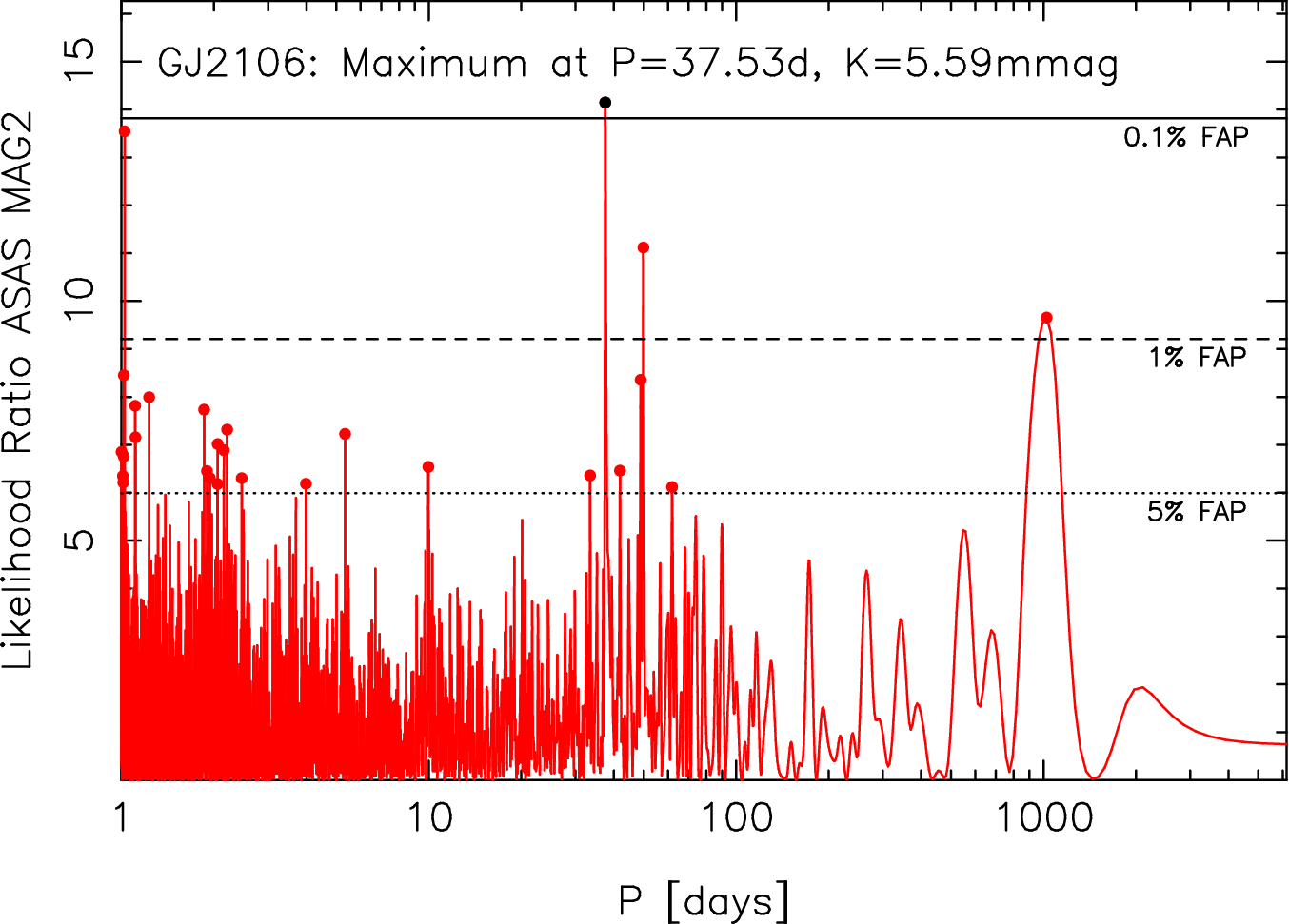}
\includegraphics[angle=270, width=0.19\textwidth,clip]{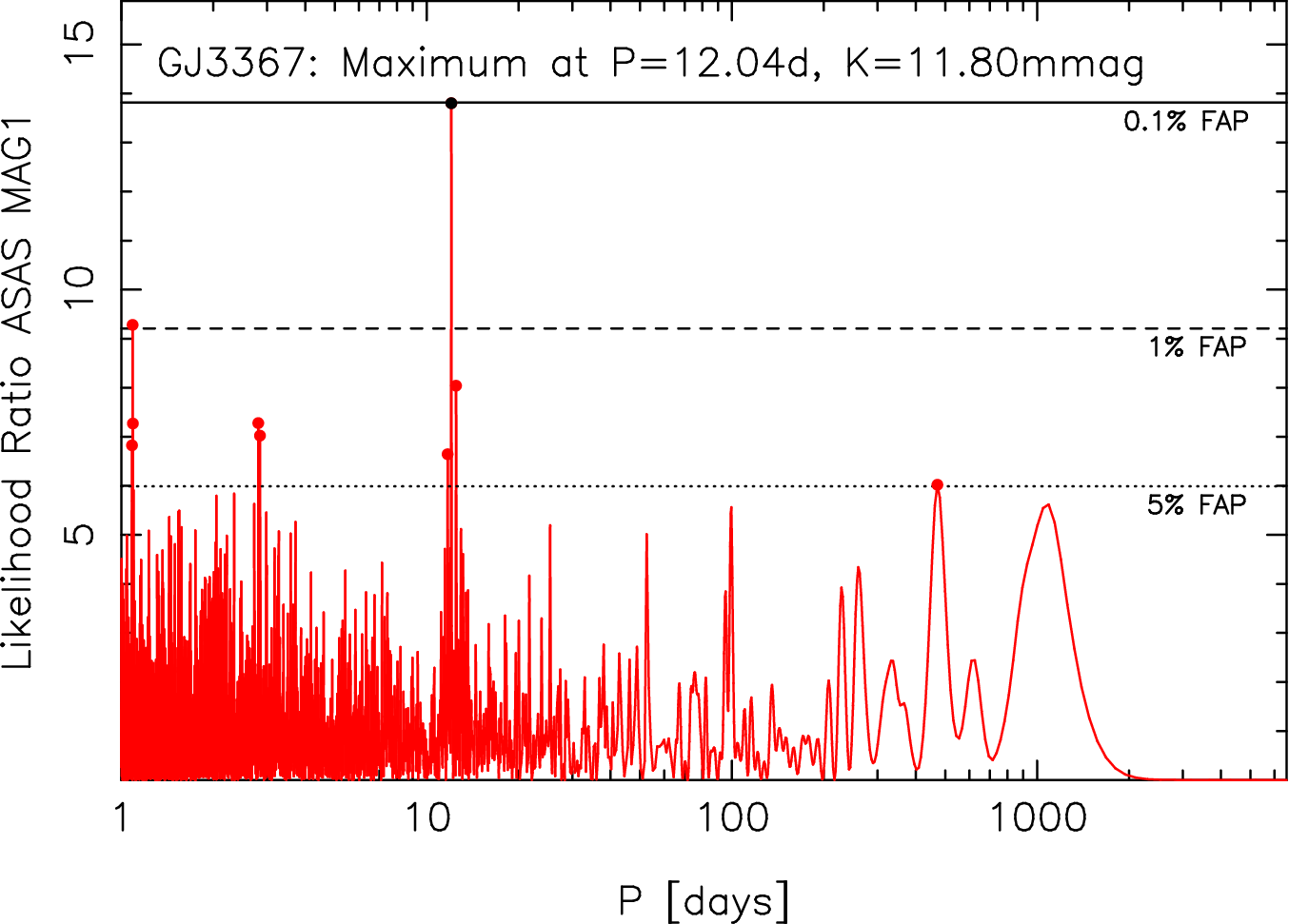}
\includegraphics[angle=270, width=0.19\textwidth,clip]{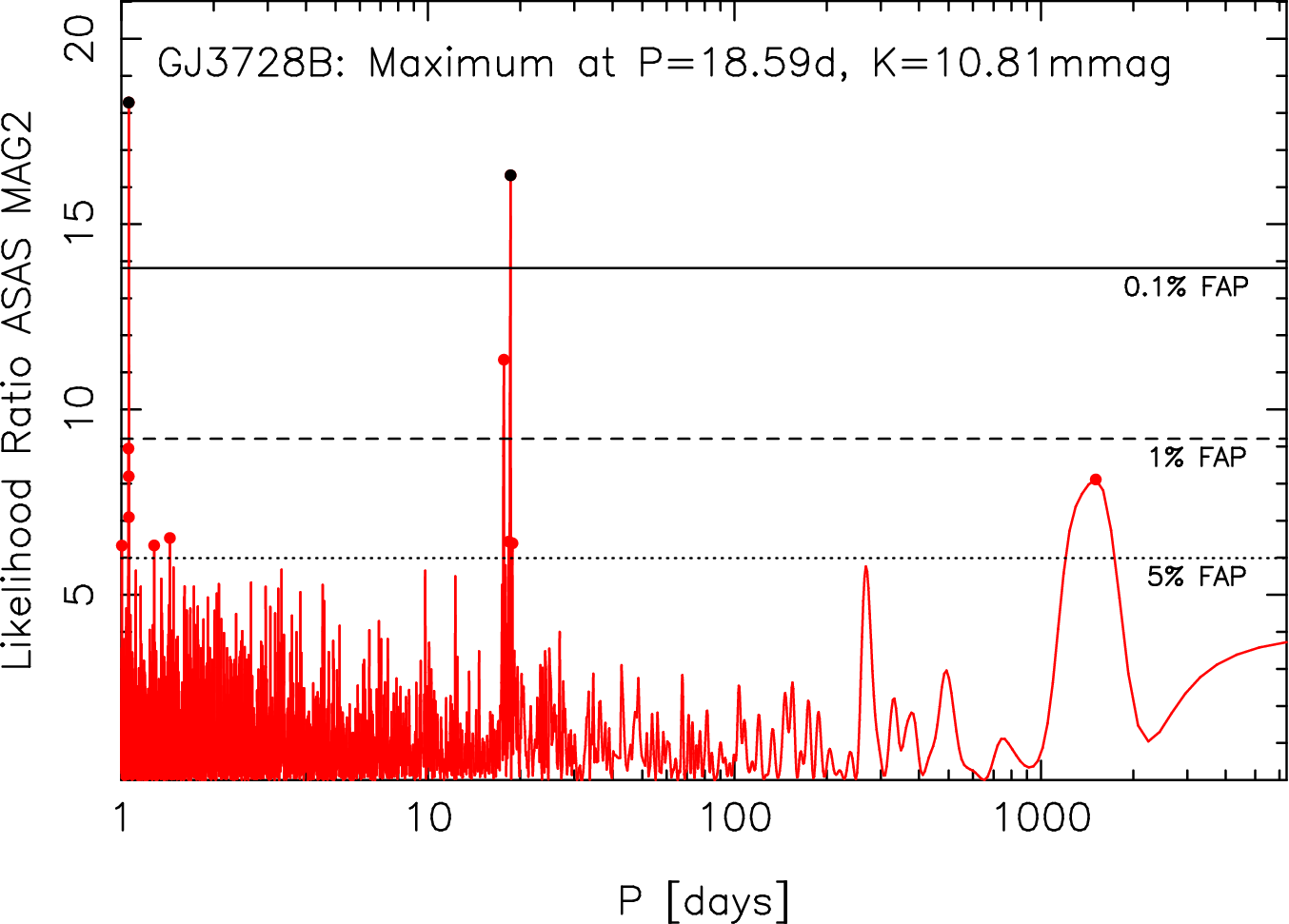}
\includegraphics[angle=270, width=0.19\textwidth,clip]{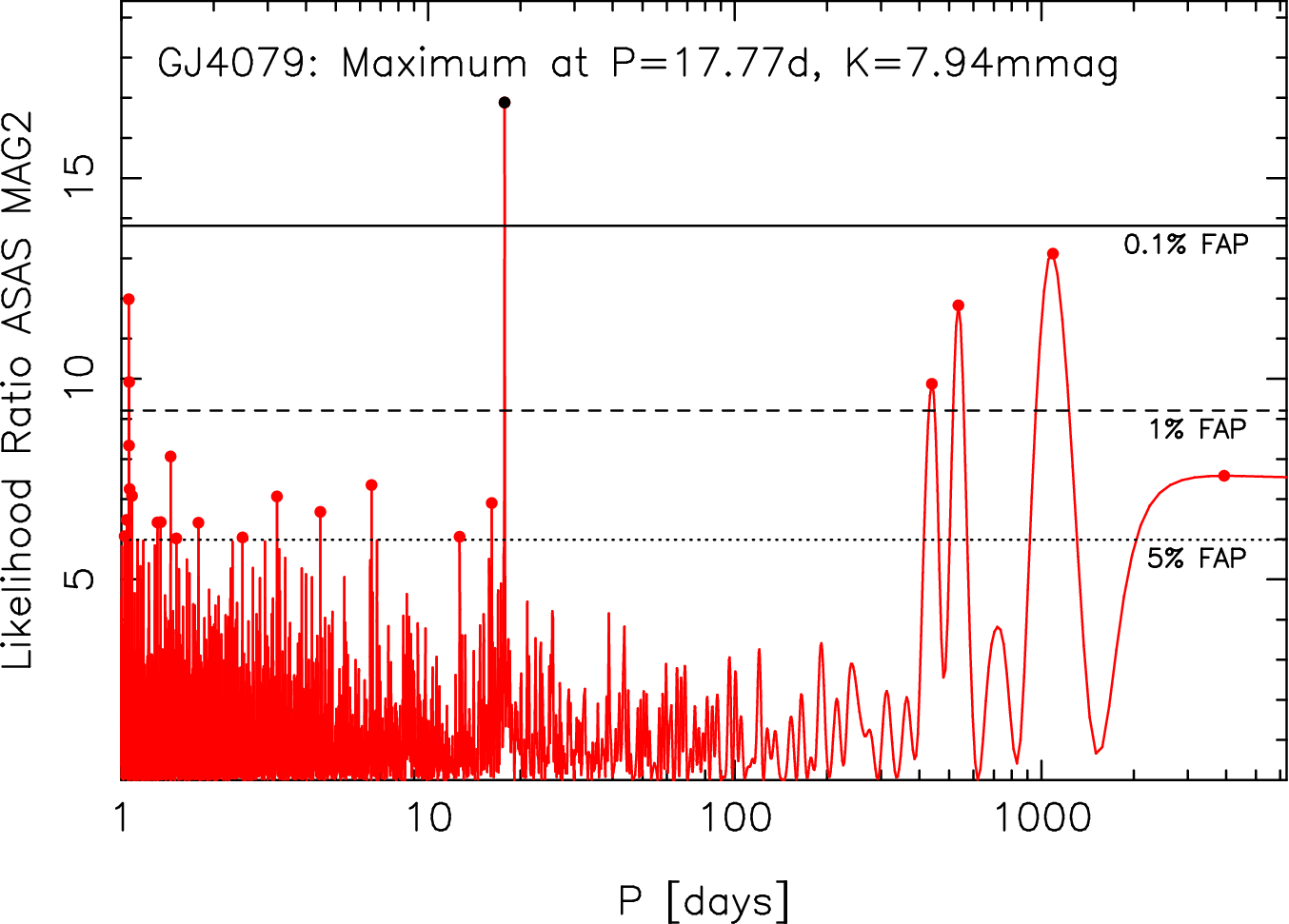}
\includegraphics[angle=270, width=0.19\textwidth,clip]{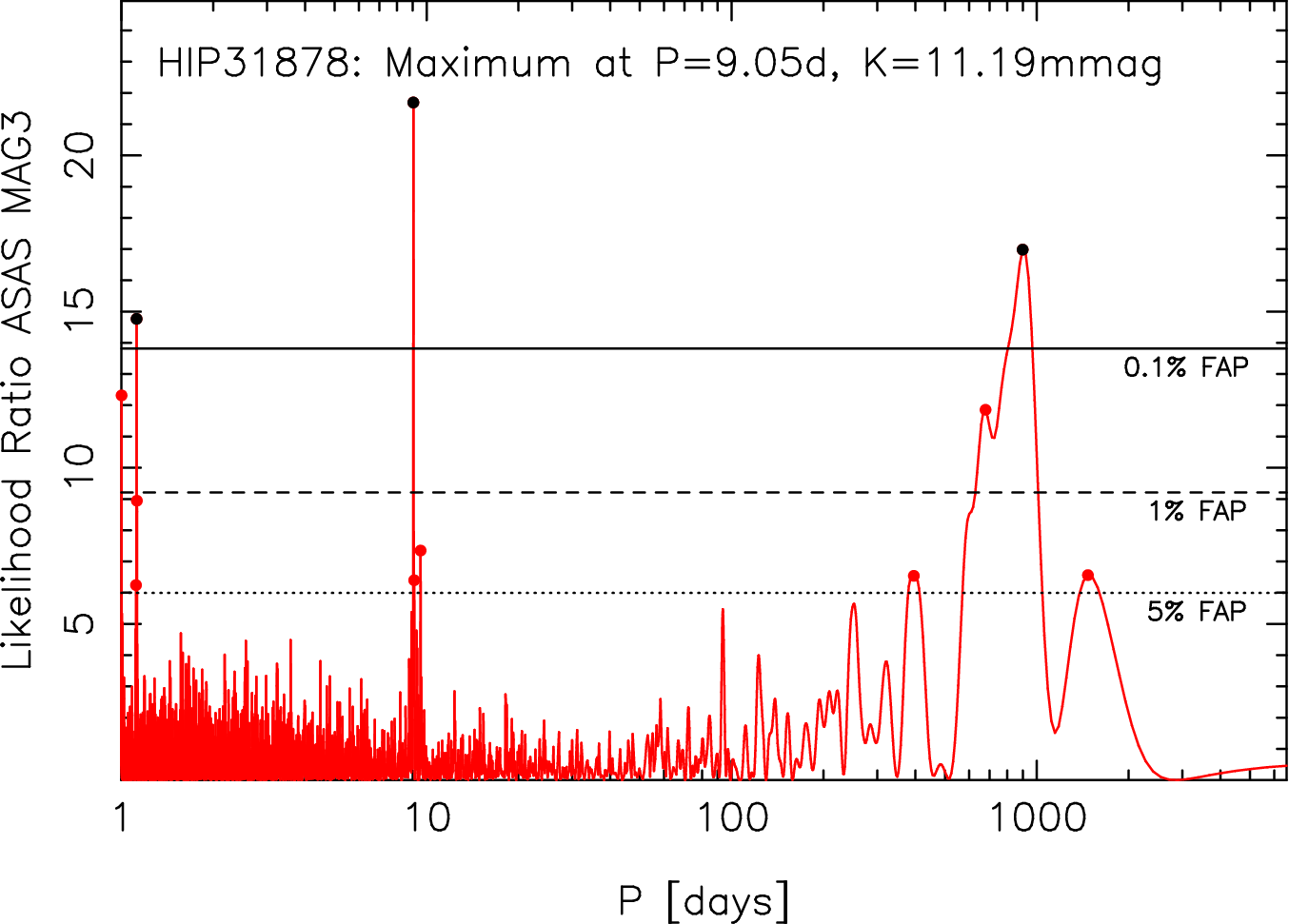}
\includegraphics[angle=270, width=0.19\textwidth,clip]{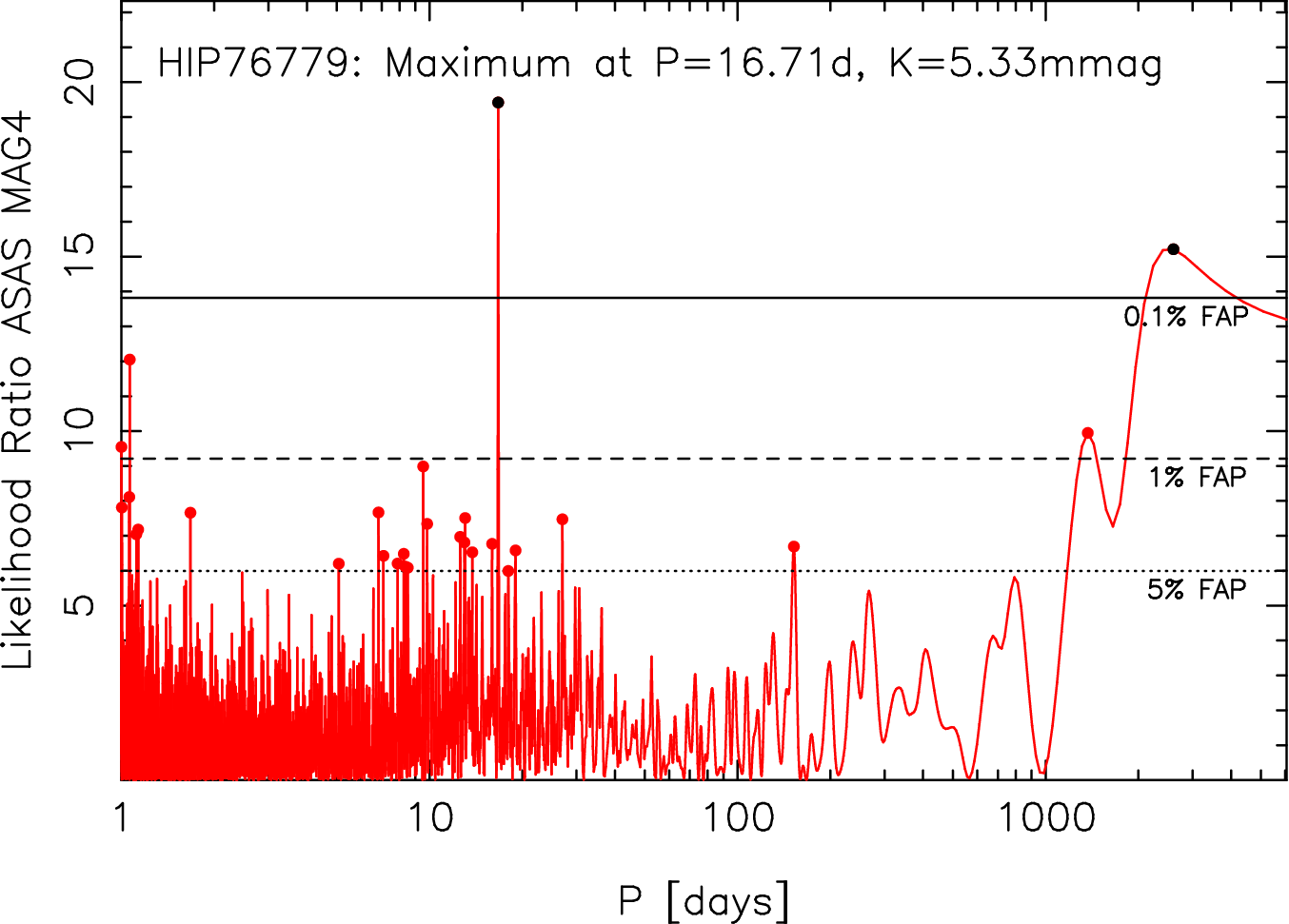}
\includegraphics[angle=270, width=0.19\textwidth,clip]{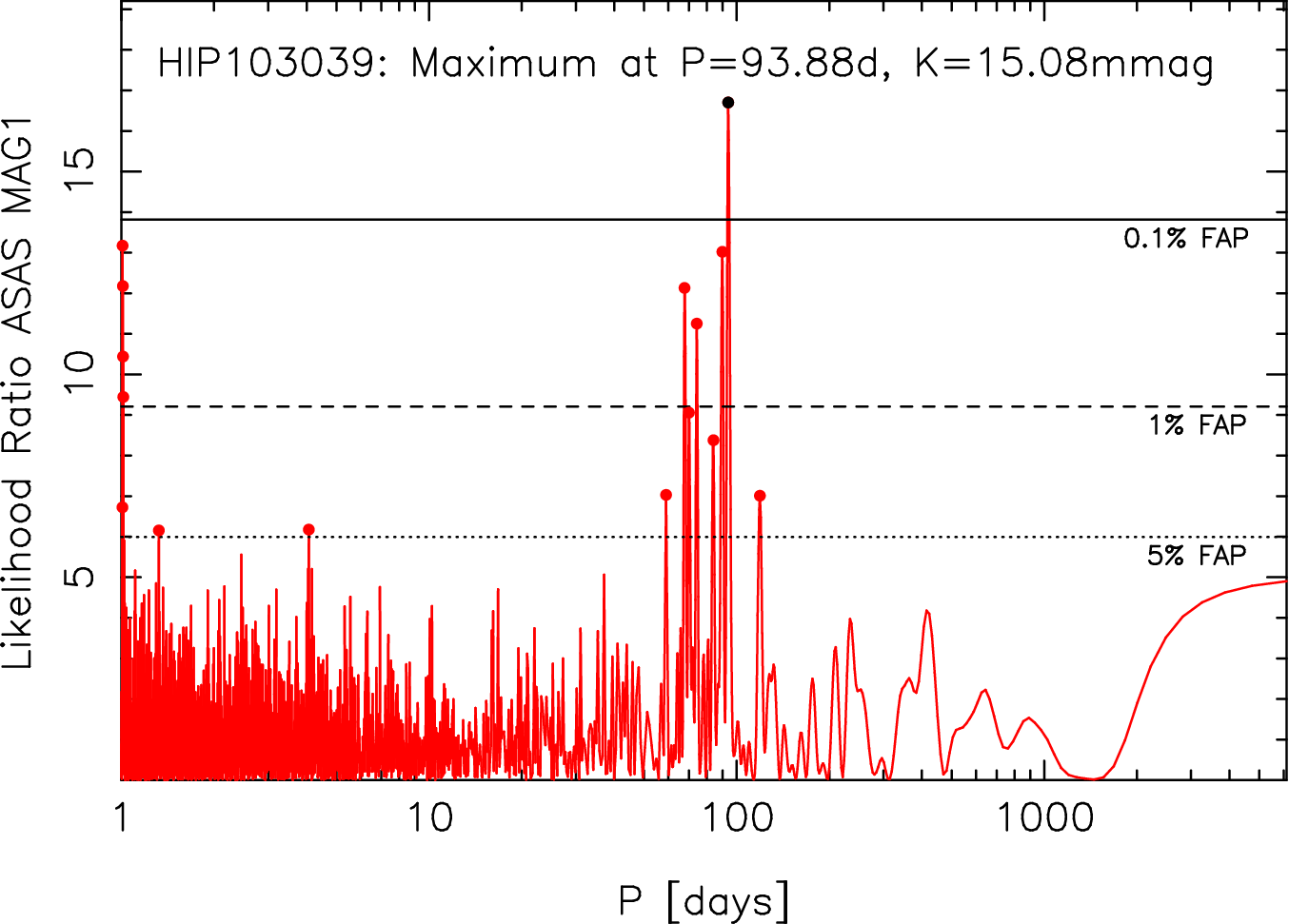}
\caption{Likelihood-ratio periodograms of selected ASAS V-band photometry time-series that show evidence for photometric rotation periods. The black (red) filled circles denote the maxima that exceed the 0.1\% (5\%) FAP threshold ratio.}\label{fig:asas_periodograms}
\end{figure}

In Table \ref{tab:kiraga_rotations}, we have listed the estimated rotation periods of some of the targets in our sample based on analyses in the current work, as well as \citet{kiraga2007} and \citet{mascareno2015} together with some additional studies. This includes the estimates we could find in the literature, as well as our estimates for some of the targets that showed clear photometric variability concentrated around a well-defined periodicity below 150 days with a FAP exceeding 0.1\% in the likelihood-ratio periodogram. We note that we do not report uncertainties to avoid confusion -- none of the photometric periodicities in ASAS data are at a precise period but rather spread around a dominating periodicity. We consider this as expected for the case where differential rotation is significant \citep[e.g.][]{messina2003}. We thus estimate that if one wishes to assess the uncertainties of our estimates, a conservative 10\% uncertainty could be present. We have plotted the periodograms of the corresponding time-series of asas V-band photometry in Fig. \ref{fig:asas_periodograms} to demonstrate the significance of the tabulated photometric rotation periods.

\begin{deluxetable}{lccccc}
\tablewidth{0pt}
\tablecaption{List of M dwarfs in the sample for which we could find estimated rotation periods (in units of days). The estimates from \citet{kiraga2007}, based on ASAS photometry (denoted as K07), and those of \citet{mascareno2015}, based on spectral activity indicators (SM15), are tabulated in separate columns. If no reference is given, the value is based on the analyses of ASAS photometry in the current work. We also tabulated other available estimates found in the literature. UPDATE GJ3543 -- CHECK ASAS VALUES AND MAKE SURE THIS IS CORRECT!\label{tab:kiraga_rotations}}
\tablehead{
Star & K07 & SM15 & This work & Other & Ref.
}
\startdata
GJ1 &  & 60.1$\pm$5.7 & $^{}$ &  & 2 \\
GJ15A &  &  & $^{}$ & 44 & 7 \\
GJ54.1 &  &  & 71.72$^{}$ &  &  \\
GJ91 &  &  & 43.61$^{}$ &  &  \\
GJ163 &  & 61.0$\pm$0.3 & $^{}$ &  & 2 \\
GJ169 &  &  & 24.94$^{}$ &  &  \\
GJ176 & 38.92 & 39.3$\pm$0.1 & 40.85$^{}$ & 39.61 & 1,2,11 \\
GJ182 & 4.41 &  & 4.37$^{}$ &  & 1 \\
GJ190 &  &  & 70.34$^{}$ &  &  \\
GJ205 & 33.61 & 35.0$\pm$0.1 & 33.61$^{}$ &  & 1,2 \\
GJ208 &  &  & 12.27$^{}$ & 12.04 & 13 \\
GJ273 &  & 115.6$\pm$19.4 & $^{}$ &  & 2 \\
GJ285 &  &  & 2.78$^{}$ &  &  \\
GJ357 &  & 74.3$\pm$1.7 & $^{}$ &  & 2 \\
GJ358 & 25.26 & 16.8$\pm$1.6 & 25.20$^{}$ & 24.7 & 1,2,13 \\
GJ382 & 21.56 & 21.7$\pm$0.1 & 21.57$^{}$ & 21.56 & 1,2,13 \\
GJ408 &  &  &  3.55$^{}$ &  &  \\
GJ410 &  &  & 13.96$^{}$ &  &  \\
GJ425B &  &  & 11.86$^{}$ &  &  \\
GJ431 & 14.31 &  & 14.33$^{}$ & 0.9328 & 1,13 \\
GJ433 &  & 73.2$\pm$16.0 & $^{}$ &  & 2 \\
GJ436 &  & 39.9$\pm$0.8 & $^{}$ &  & 2 \\
GJ494 & 2.889 &  & 2.89$^{}$ & 2.886 & 1,13 \\
GJ496.1 &  &  & 36.38$^{}$ &  &  \\
GJ514 &  & 28.0$\pm$2.9 & 25.64$^{b}$ &  & 2 \\
GJ526 &  & 52.3$\pm$1.7 & $^{}$ &  & 2 \\
GJ536 &  & 43.8$\pm$0.1 & $^{}$ &  & 2 \\
GJ551 & 82.53 & 116.6$\pm$0.7 & 83.04$^{}$ & 82.6$\pm$0.1 & 1,2,14 \\
GJ569A & 13.68$^{c}$ &  &  &  & 1 \\
GJ581 &  & 132.5$\pm$6.3 & $^{}$ & 94.1$\pm$1.0; 130$\pm$2 & 2,5,12 \\
GJ588 &  & 61.3$\pm$6.5 & $^{}$ &  & 2 \\
GJ618A & 56.52 &  & 56.11$^{}$ &  & 1 \\
GJ649 &  &  & $^{}$ & 24.8$\pm$1.0 & 8 \\
GJ654 &  &  & 50.13$^{}$ &  &  \\
GJ667C &  & 103.9$\pm$0.7 & $^{}$ &  & 2 \\
GJ674 & 33.29 &  & 33.15$^{a}$ &  & 1 \\
GJ676A &  & 41.2$\pm$3.8 & $^{}$ &  & 2 \\
GJ687 &  &  & $^{}$ & 60.8$\pm$1.0 & 9 \\
GJ699 &  & 148.6$\pm$0.1 & $^{}$ &  & 2 \\
GJ701 &  & 127.8$\pm$3.2 & $^{}$ &  & 2 \\
GJ729 & 2.869 &  & 2.87$^{}$ & 2.869 & 1,13 \\
GJ740 &  &  & 34.97$^{}$ &  &  \\
GJ752A &  & 46.5$\pm$0.3 & $^{}$ &  & 2 \\
GJ754 &  &  & 127.95$^{b}$ &  &  \\
GJ784 &  &  & 20.22$^{a}$ &  &  \\
GJ803 & 4.848 &  & 4.86$^{}$ &  & 1 \\
GJ832 &  & 45.7$\pm$9.3 & $^{}$ &  & 2 \\
GJ841A & 1.124 &  & 9.09$^{d}$ &  & 1 \\
GJ846 &  & 31.0$\pm$0.1 & $^{}$ &  & 2 \\
GJ849 &  & 39.2$\pm$6.3 & $^{}$ &  & 2 \\
GJ851 &  &  & 40.37$^{}$ &  &  \\
GJ867A & 4.233$^{c}$ &  &  &  & 1 \\
GJ871.1A &  &  &  & 2.358 & 13 \\
GJ876 &  & 87.3$\pm$5.7 & 81.27$^{}$ & 95$\pm$1 & 2,4 \\
GJ877 &  & 116.1$\pm$0.7 & 52.72$^{}$ &  & 2 \\
GJ880 &  & 37.5$\pm$0.1 & $^{}$ &  & 2 \\
GJ1044 &  &  & 28.64$^{}$ &  &  \\
GJ1146 &  &  & $^{}$ & 98.1$\pm$0.6 & 6 \\
GJ1264 &  &  & 6.66$^{}$ & 6.67 & 13 \\
GJ1267 &  &  & 2.62$^{}$ &  &  \\
GJ2106 &  &  & 37.53$^{}$ &  &  \\
GJ3293 &  &  & $^{}$ & 41 & 10 \\
GJ3323 &  &  &   & 88.50 & 13 \\
GJ3367 & 12.05 &  & 12.04$^{}$ & 12.485 & 1,13 \\
GJ3470 &  & 21.9$\pm$1.0 & $^{}$ & 20.70$\pm$0.15 & 2,3 \\
GJ3728B &  &  & 18.59$^{}$ &  &  \\
GJ4079 &  &  & 17.77$^{}$ &  &  \\
HIP31878 &  &  & 9.15$^{}$ &  &  \\
HIP76779 &  &  & 16.71$^{}$ &  &  \\
HIP103039 &  &  & 93.88$^{}$ &  &  \\
\enddata
\tablenotetext{a}{The signal in the ASAS photometry does not exceed the 0.1\% FAP and is not considerably higher than other likelihood maxima making the interpretation of this signal as a rotation period rather uncertain.}
\tablenotetext{b}{Estimate based on a periodic signal in the HARPS and HIRES S-indices rather than photometric data.}
\tablenotetext{c}{Although \citet{kiraga2007} reported a photometric rotation based on ASAS photometry, we could not confirm this claim. There was no conclusive evidence for periodic signals in the ASAS V-band photometry measurements that we analysed although the signal reported by \citet{kiraga2007} did correspond to a maximum in our likelihood ratio tests.}
\tablenotetext{d}{This signal at a period of 9.09 days has an alias at the period of 1.12 days reported by \citet{kiraga2007} that we do not accept as our solution because we limit our rotation periods \emph{a priori} above two days.}
\tablerefs{(1) \citet{kiraga2007}; (2) \citet{mascareno2015}; (3) \citet{biddle2014}; (4) \citet{nelson2016}; (5) \citet{robertson2014}; (6) \citet{haghighipour2010}; (7) \citet{howard2014}; (8) \citet{johnson2010}; (9) \citet{burt2014}; (10) \citet{astudillo2015}; (11) \citet{robertson2015}; (12) \citet{vogt2010}; (13) \citet{kiraga2012}; (14) \citet{collins2017}}
\end{deluxetable}

Although our sample of photometric rotation periods is small, the majority of them coincide with the observation of \citet{mcquillan2014b} that the didstribution of M dwarf rotation periods peaks at roughly 19 and 33 days (bimodal distribution) and ranges approximately from 10 to 60 days for stars with 0.30 M$_{\oplus} <$ M$_{\star} < 0.55$ M$_{\oplus}$ albeit with a non-negligible tail towards shorter periods. This is rather unfortunate because this period range corresponds to planetary orbits located inside the HZs of M dwarf stars and it is thus essential to distinguish between rotation-induced radial velocity signals and those caused by planets orbiting the stars. We pay attention to this fact when interpreting the radial velocity signals while acknowledging that some of the signals we have detected and interpreted as candidate planets could in fact have been misinterpreted as stellar rotation periods \citep[e.g.][]{robertson2015,robertson2015b}.

\begin{figure}
\center
\includegraphics[angle=270, width=0.80\textwidth]{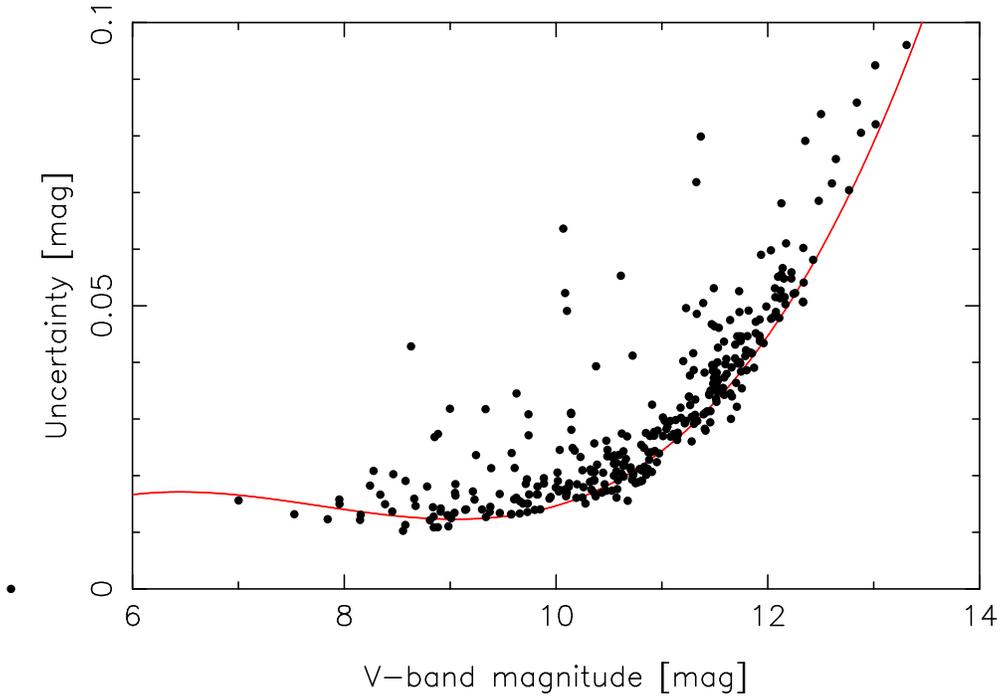}
\caption{The relationship between the ASAS V-band mean magnitude and the corresponding weighted standard deviation of the sample targets. The red solid curve represents a 3rd-order polynomial roughly describing this relationship with parameters $a_{0} = -0.2173, a_{1} = 9.533 \times 10^{-2}, a_{2} = -1.266\times 10^{-2}, a_{3} = 5.443 \times 10^{-4}$. This polynomial has been fitted by weighting the points according to the respective uncertainties in the data sets.}\label{fig:asas_errors}
\end{figure}

It is worth noting that after removing the 5-$\sigma$ outliers from the ASAS photometry data sets, we observed a clear 
relationship between the stellar magnitudes and the corresponding scatter in the data (Fig. \ref{fig:asas_errors}). This relationship was used to assess whether the ASAS data sets were contaminated by a nearby source or by the target star being too bright and thus untrustworthy due to being considerably above the red curve in Fig. \ref{fig:asas_errors}.

\section{Planet candidates orbiting the sample stars}\label{sec:planets}

In this section we tabulate the candidate planets orbiting the sample stars. However, we do not present a rather lenghty discussion of them individually here but show it in Appendix \ref{sec:individual_targets}. In this Appendix, we do not discuss all the targets in detail because of their large number but provide details for a subjectively selected subset whose members were found to contain signals that have not been previously reported, for which we obtain results that differ from those presented in the literature, or provide new insights in terms of e.g. numbers of planet candidates, their parameter estimates and/or the interpretation of the signals. All the planet candidates in the sample have been listed in Table \ref{tab:candidates} together with the maximum \emph{a posteriori} (MAP) estimates and 99\% credibility intervals of the parameters of the Keplerian model. Furthermore, we have tabulated the minimum masses ($m_{p} \sin i$) and semi-major axes ($a$) as estimated by using the stellar masses as shown in Table \ref{tab:targets_physics} and by assuming that the uncertainties of the stellar masses are 10\% of the estimates for all targets.

\begin{deluxetable}{llccccccccl}
\tablewidth{0pt}
\tablecaption{Planet candidates in the sample. Signal amplitudes, periods, eccentricities, and estimated minimum masses and semi-major axes in terms of MAP estimates and 99\% credibility intervals. Estimates for parameters $m_{p} \sin i$ and $a$ are calculated under the assumption that the uncertainties in the stellar masses (see Table \ref{tab:targets_physics}) are 10\% of the estimates. References are given to original discovery papers where the corresponding periodicity was first reported and interpreted as a planet candidate, or if omitted, refer to the current work.\label{tab:candidates}}
\tablehead{
Candidate & & $P$ [days] & $K$ [ms$^{-1}$] & $e$ & $m_{p} \sin i$ [M$_{\oplus}$] & $a$ [AU] & Zone\tablenotemark{a} & Ref.
}
\startdata
GJ1044     & b          &    1721$\pm$87 &    16.68$^{+    3.76}_{-    3.41}$ &     0.07$^{+    0.18}_{-    0.07}$ &    252.4$^{+    80.4}_{-    72.8}$ &    2.526$\pm$0.253 & C   \\
GJ1148     & b          &   41.374$^{+   0.014}_{-   0.018}$ &    36.53$\pm$2.57 &     0.36$\pm$0.05 &     89.8$\pm$18.0 &    0.159$^{+   0.014}_{-   0.017}$ & C & 1 \\
GJ1148     & c          &     532$^{+      8}_{-     11}$ &    10.31$^{+    2.46}_{-    2.75}$ &     0.23$^{+    0.20}_{-    0.23}$ &     58.8$^{+    20.0}_{-    18.1}$ &    0.871$^{+   0.081}_{-   0.090}$ & C   \\
GJ1177B    & b          &    364.7$^{+     2.3}_{-     2.0}$ &   109.41$^{+   16.98}_{-    6.65}$ &     0.01$^{+    0.07}_{-    0.01}$ &    883.5$^{+   218.5}_{-   195.5}$ &    0.851$^{+   0.076}_{-   0.094}$ & C   \\
GJ15A      & b          &  11.4427$^{+  0.0058}_{-  0.0047}$ &     1.95$^{+    0.56}_{-    0.62}$ &     0.06$^{+    0.25}_{-    0.06}$ &      3.6$\pm$1.3 &    0.071$\pm$0.007 & C & 2 \\
GJ160.2    & b          &   9.7471$^{+  0.0043}_{-  0.0050}$ &     3.01$\pm$1.07 &     0.02$^{+    0.22}_{-    0.02}$ &      7.8$\pm$3.1 &    0.079$^{+   0.007}_{-   0.009}$ & H & 3 \\
GJ163      & b          &   8.6312$^{+  0.0023}_{-  0.0021}$ &     5.94$\pm$0.75 &     0.02$^{+    0.12}_{-    0.02}$ &      9.9$\pm$2.3 &    0.060$^{+   0.005}_{-   0.006}$ & H & 4,5 \\
GJ163      & c          &   25.637$\pm$0.042 &     3.22$\pm$0.90 &     0.03$^{+    0.18}_{-    0.03}$ &      7.6$^{+     2.9}_{-     2.3}$ &    0.124$^{+   0.010}_{-   0.013}$ & HZ & 4,5 \\
\enddata
\tablenotetext{a}{HZ: planet inside the estimated stellar liquid-water habitable zone; H (C): planet inside (outside) the inner (outer) edge of the habitable zone. The habitable zones are estimated according to the equations of \citet{kopparapu2013} and listed in Table \ref{tab:targets_physics}.}
\tablenotetext{b}{The estimated eccentricities are lower than reported in \citet{rivera2005} and \citet{rivera2010} because we did not account for planet-planet interactions in our model. We thus obtained the same solution as \citet{jenkins2014b} with additional signals at periods of 10 and 15 days that are likely caused by the fact that the interactions were not modelled.}
\tablerefs{(1) \citet{haghighipour2010}; (2) \citet{howard2014}; (3) \citet{tuomi2014}; (4) \citet{tuomi2013c}; (5) \citet{bonfils2013b}; (6) \citet{forveille2009}; (7) \citet{howard2010}; (8) \citet{anglada2014}; (9) \citet{arriagada2013}; (10) \citet{locurto2013}; (11) \citet{tuomi2014b}; (12) \citet{astudillo2017}; (13) \citet{johnson2007}; (14) \citet{astudillo2015}; (15) \citet{bonfils2012}; (16) \citet{bonfils2011}; (17) \citet{butler2016}; (18) \citet{delfosse2013}; (19) \citet{butler2004}; (20) \citet{moutou2011}; (21) \citet{suarezmascareno2016}; (22) \citet{anglada2016}; (23) \citet{bonfils2005}; (24) \citet{udry2007}; (25) \citet{mayor2009}; (26) \citet{wright2016}; (27) \citet{johnson2010}; (28) \citet{anglada2012}; (29) \citet{anglada2013}; (30) \citet{bonfils2007}; (31) \citet{forveille2011}; (32) \citet{anglada2012c}; (33) \citet{burt2014}; (34) \citet{bailey2009}; (35) \citet{wittenmyer2014}; (36) \citet{butler2009}; (37) \citet{marcy1998}; (38) \citet{delfosse1998}; (39) \citet{marcy2001}; (40) \citet{rivera2005}; (41) \citet{rivera2010}; (42) \citet{ribas2018}; (43) \citet{tuomi2018}}
\end{deluxetable}

To summarise the general properties of the candidates in Table \ref{tab:candidates}, we can say that out of the 118 candidate planets, only seven\footnote{We note that the candidate orbiting GJ 1046, with a minimum mass of 26.1 M$_{\rm Jup}$, is a brown dwarf rather than a planet and is thus omitted from Table \ref{tab:candidates}.} have minimum masses that are greater than one Jupiter-mass indicating that such massive planets are rare around M dwarfs even though they would be rather easy to detect on orbits with periods of up to or slightly more than 4000 days in the current data sets. A total of 88\% of the candidates in the sample have minimum masses below 100 M$_{\oplus}$. Moreover, there are more (54\%) candidates with minimum masses below 10 M$_{\oplus}$ than those with masses above that (46\%) even though the former are much more difficult to detect. This points rather strongly towards planets being more abundant around M dwarfs the smaller they are \citep{bonfils2013,dressing2013,dressing2015,tuomi2014}. We quantify this preliminary result in Section \ref{sec:occurrence} by taking into account the detection probability function of the whole sample.

Another striking feature in Table \ref{tab:candidates} is that roughly 15\% of the candidate planets appear to have orbital distances that place them inside the stellar liquid-water habitable zones, as defined by the equations of \citet{kopparapu2013} and listed in Table \ref{tab:targets_physics}. Again, the detection probability function of the sample has to be taken into account, but it is clear that low-mass habitable-zone planets are very abundant orbiting the M dwarfs in the Solar neighbourhood.

Systems of multiple planets also appear to be rather common. In our sample, there is a total of 39 candidate planets in systems where there is no evidence for additional companions. However, this means that a majority of 79 (67\%) of the candidate planets in the sample are in multiplanet systems. There are thus 28 stars with at least two candidate planets in our sample, which seems to be consistent with the observation of \citet{ballard2016} that half or exoplanet systems around M dwarfs have highly packed systems of at least five planets.

\begin{figure}
\center
\includegraphics[angle=270, width=0.45\textwidth,clip]{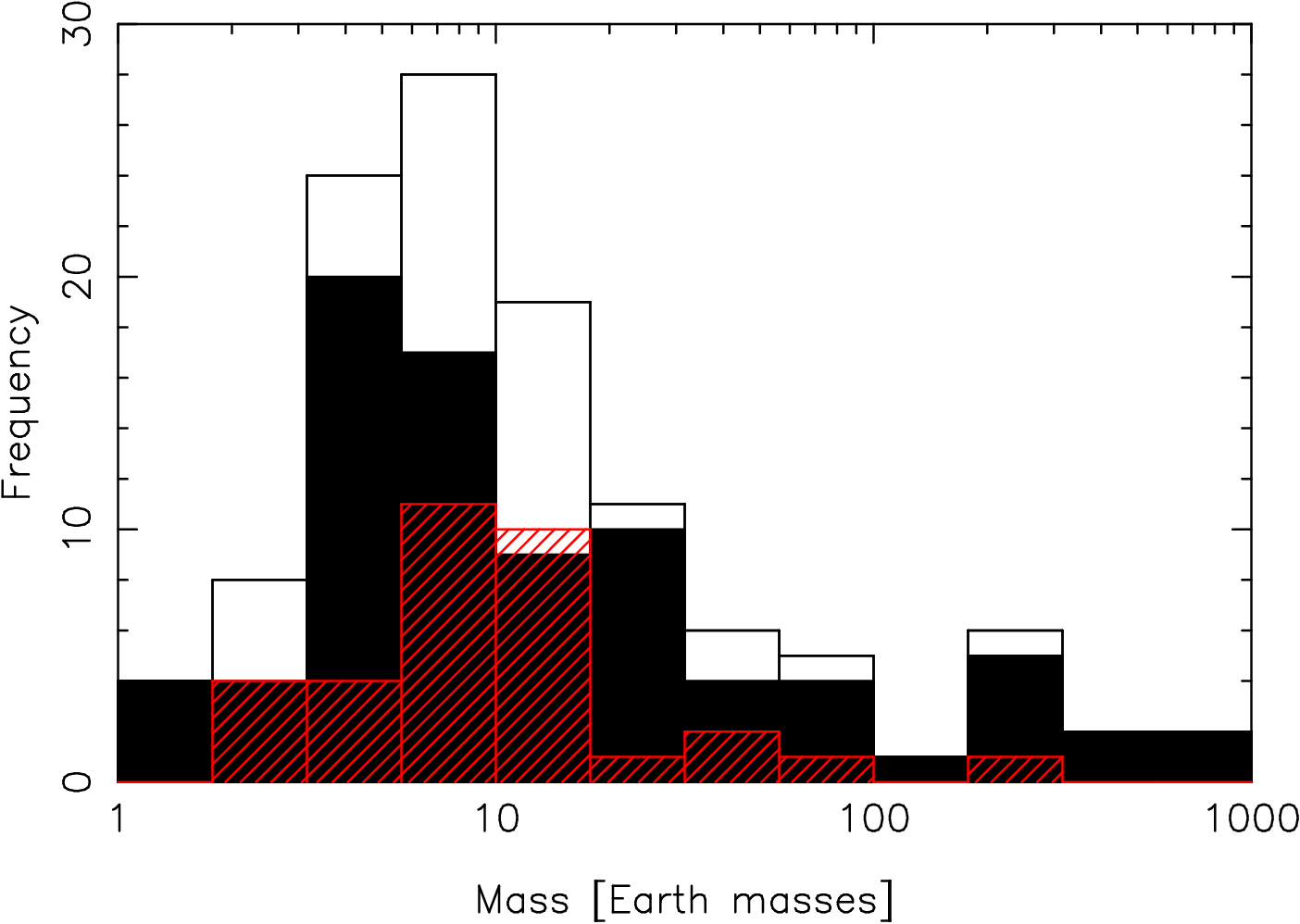}
\includegraphics[angle=270, width=0.45\textwidth,clip]{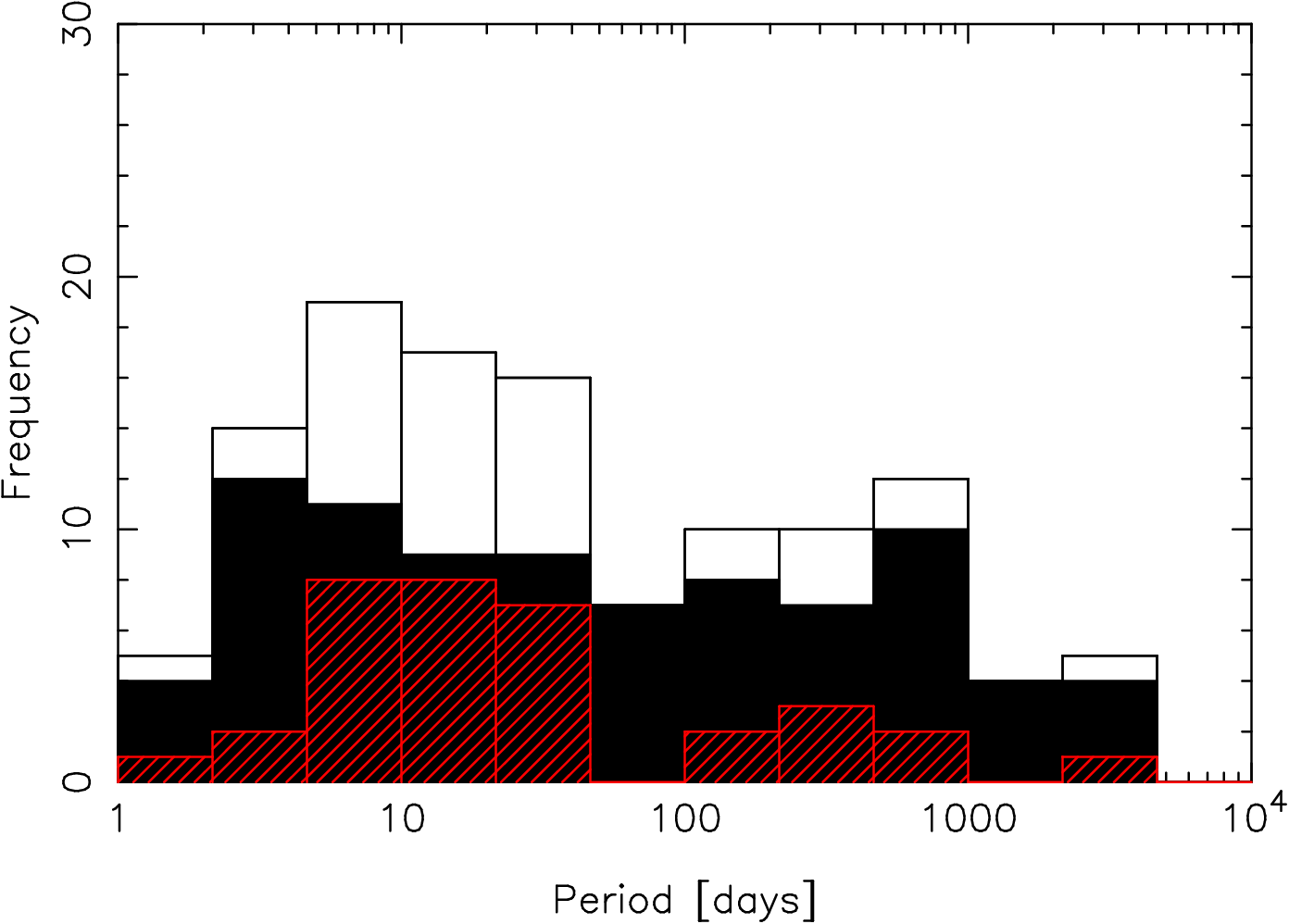}
\caption{Distributions of minimum masses and orbital periods of the candidate planets listed in Table \ref{tab:candidates}. The white histogram lists all candidates while the black (red) distributions denote planets orbiting high-mass (low-mass) subsample of the target stars with m$_{\star} \geq$ 0.43 M$_{\odot}$ (m$_{\star} <$ 0.43 M$_{\odot}$).}\label{fig:mass_period_dist}
\end{figure}

When looking at the distributions of candidate minimum masses and orbital periods (Fig. \ref{fig:mass_period_dist}) it is clear that the minimum masses peak at just below 10 M$_{\oplus}$, which indicates that super-Earths and mini-Neptunes are the most common planets in the sample. However, the detection of planets with the radial velocity method is more difficult the less massive they are, which means the distribution in Fig. \ref{fig:mass_period_dist} (left panel) is affected by detection bias. This is also the case for planets on longer periods but we have still detected dozens of planets with orbital periods in excess of 100 days enabling us to estimate occurrence rates for long-period planets as well.

Finally, looking at Table \ref{tab:candidates} and Fig. \ref{fig:mass_period_dist}, it is clear that, based on the current sample, it is possible to probe parameter space not accessible to previous studies. Unlike with the \emph{Kepler} space telescope whose transiting planet candidates are concentrated to orbits with periods shorter than 50 days due to constrained lifetime of the mission \citep[e.g.][]{howard2012,dressing2013,morton2014}, our sample contains several long-period planet candidates and a total of 32 (27\%) of the candidates have orbital periods longer than 200 days. Some more recent results based on \emph{Kepler} photometry have been reported for up to periods of 200 days \citep{dressing2015,silburt2015}, but this is still an order of magnitude shorter than our typical baselines. Moreover, in comparison to the HARPS and combined UVES and HARPS surveys that were both based on only ten candidates \citep{bonfils2013,tuomi2014}, we can now repeat the occurrence rate computations for a sample some 12 times larger than was previously available.

\subsection{Significances}

\begin{figure}
\center
\includegraphics[angle=270, width=0.80\textwidth,clip]{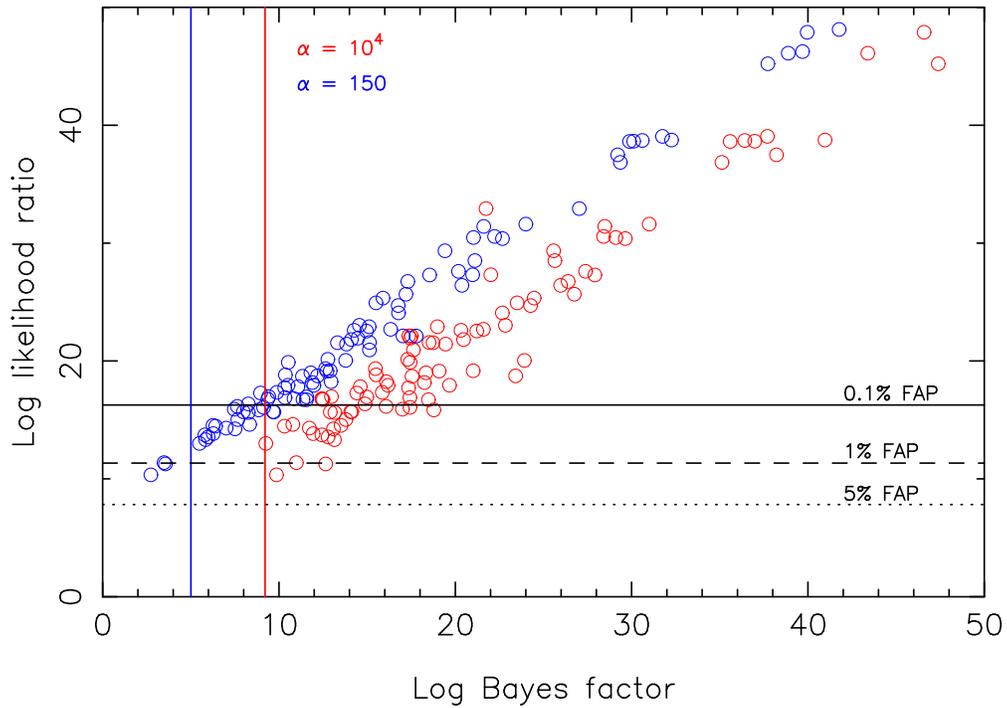}
\caption{Bayes factors (BFs) of models containing the signals with respect to models that do not. The red circles denote the BFs as estimated by using the samples from posterior and prior densities and the blue circles denote the estimates based on BIC values. The red and blue vertical lines denote the corresponding detection thresholds of these methods. The Bayes factors are plotted as a function of the respective likelihood ratios with FAPs based on likelihood-ratio test.}\label{fig:significances}
\end{figure}

The fact that the signals are significantly detected is visualised in Fig. \ref{fig:significances}. As described in Section \ref{sec:significance}, the Bayes factors of all our signals exceed the threshold $\alpha = 10^{4}$ when estimated by using the approximation based on samples from posterior and prior densities (red vertical significance threshold and red circles in Fig. \ref{fig:significances}). However, there are three of them that do not appear significant when applying the Bayes factors based on BIC and a detection threshold of $\alpha = 150$ (blue circles and blue vertical threshold). These three signals correspond to the weakest ones in GJ 163 at a period of 109 days (Section \ref{sec:GJ163}), GJ 221 at a period of 487 days (Section \ref{sec:GJ221}), and GJ 687 at a period of 758 days (Section \ref{sec:GJ687}) data. Their existence thus needs to be verified with future data. The likelihood-ratio test indicates similarly that these three signals are just below the 1\% FAP threshold (Fig. \ref{fig:significances}), which is not surprising considering that both tests are based on maximum-likelihood values. However, since all the signals tabulated in Table \ref{tab:candidates} do satisfy our primary detection criteria and are significant, we assume they are all caused by planets orbiting the respective stars.

While keeping in mind these three exceptions, all our signals are thus significantly present in the data based on all three significance tests. We are thus confident that they have been detected robustly and can be interpreted as planetary signals as discussed in detail in Section \ref{sec:individual_targets}. We note that, as can be seen in Fig. \ref{fig:significances}, the BIC typically provides more conservative Bayes factors becasue it penalises more complex models more heavily. However, it assumes that all the free parameters in the model are equally important for the goodness of the model, which neither can be said to be the case nor can it even be addressed what it means that one parameter is more important for the model. The BIC estimate also assumes that the posterior density is strictly Gaussian \citep[see discussion in][]{kass1995,liddle2007}. This is never the case in practice even though some parameter densities are approximately Gaussian rather frequently. We refer to Table \ref{tab:candidates}, where it can be seen that a majority of parameters have unsymmetric probability densities that give rise to unsymmetric 99\% credibility intervals.

We note that all the Bayes factors and likelihood ratios are not shown in Fig. \ref{fig:significances} because they are outside the maximum threshold of 50 implying that they are even more confidently detected than the signals presented in this figure.

\section{Linear accelerations and noise properties}

The parameters $\theta_{0}$ of the model without Keplerian signals (see Section \ref{sec:rv_model}) should not be dismissed as merely ``nuisance parameters``, although with respect to the planet candidates in the sample their values are typically less important. However, this is not generally so as they form an important part of the statistical model and the solutions we have obtained for the Keplerian parameters are by no means independent of the vector $\theta_{0}$. We discuss these parameters in this section together with the implications of their estimated values.

\subsection{Linear accelerations}

The linear accelerations modelled with the linear acceleration parameter $\dot{\gamma}$ provide information regarding the movement of the star in space on long time-scales. Although we have removed perspective accelerations from the radial velocity data sets, the stars might experience linear acceleration due to having massive, known or otherwise, companions on long-period orbits. Alternatively such apparent acceleration might be caused by magnetic activity cycles but we consider this option less likely as we attempt to account for activity-induced variations by our statistical model.

Having analysed the radial velocity data of such a large number of targets, it is to be expected that some of them indeed show evidence for linear acceleration indicative of the presence of long-period substellar companions and/or known or unknown stellar companions. Known examples of such stars are e.g. CD-44 836A, GJ 105B, GK 166C, GJ 250B, GJ 551, GJ 569A, GJ 618A, GJ 667C, GJ 678.1A, GJ 752A, GJ 3193B, and GJ 3404A, that are all components in stellar binaries or systems of even higher multiplicity. However, we find statistically significant linear trends (at the 99\% creddibility level) in the velocities of 89 of the target stars and only few of these could be caused by known stellar companions, such as the trend in the GJ 667C radial velocity data \citep{anglada2012,anglada2013,delfosse2013}.

We have listed the linear acceleration parameters for all the targets for which it could be quantified\footnote{We did not attempt determining the acceleration for targets with too few velocities or whose data was corrupted by contamination.} in Table \ref{tab:targets_estimates}. The estimated 99\% uncertainty intervals overlap with zero for all but 89 of the targets that we thus consider significantly different from zero. When the acceleration parameter is indeed significantly different from zero, suggesting the presence of a long-period companion, the corresponding stars can be considered good targets for imaging surveys for massive long-period companions to nearby stars.

\subsection{Stellar jitter}

Another important parameter is the estimated excess white noise in the data, sometimes referred to as ''stellar jitter`` \citep{wright2005}. We have denoted this parameter as $\sigma_{l}$ throughout the current work. \citet{wright2005} reported an average jitter level of 5 ms$^{-1}$ for inactive M dwarfs and 5-10 ms$^{-1}$ for active ones, based on the HIRES observations, most of which are included in the current work.

Ideally, $\sigma_{l}$ would quantify the noise originating on the stellar surface -- its inhomogeneities, variations, and other features that produce neither strictly periodic variations nor variations that are correlated in time or connected to the stellar activity and traced by the activity indicators. However, in practice, we know this is not the case and $\sigma_{l}$ in fact contains noise coming not only from the stellar surface but also from the telescopes and instruments used to obtain the measurements. This has been demonstrated by the simple observation that the value of $\sigma_{l}$ is different for different telescope/instrument combinations even when they are targeting the same star \citep[e.g.][]{tuomi2011b}. Therefore, the parameter $\sigma_{l}$ can be considered to quantify the upper limit for the actual stellar jitter arising from stellar surface activity. If there is data from more than one instrument available, it is possible to obtain the least upper limit -- that is, the lowest upper limit of the $\sigma_{l}$ estimates of the different instruments -- to constrain the jitter from above. We define this as the lowest upper limit of the 99\% credibility intervals of the $\sigma_{l}$ parameters of the different instruments.

\begin{figure}
\center
\includegraphics[angle=270, width=0.49\textwidth,clip]{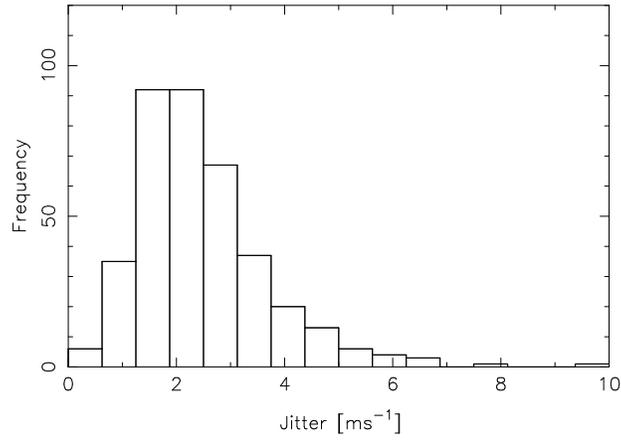}
\caption{Distribution of estimated jitter levels of the sample stars as quantified by parameter $\sigma_{\rm J}$.}\label{fig:jitter_distribution}
\end{figure}

We have listed the estimated jitter levels for all the targets for which this estimation was possible in Table \ref{tab:targets_estimates}. The median value of the estimates is 2.29 ms$^{-1}$ and the obtained estimates range from only 0.41 ms$^{-1}$ for GJ 863 to as much as 44.92 ms$^{-1}$ for HIP 20160 that is a well-known classical T Tauri star \citep[e.g.][]{chen2013}. There are thus stars with hardly any stellar jitter on top of the estimated instrument-induced white noise levels making them excellent radial velocity targets and also stars whose radial velocities are dominated by stellar jitter making it difficult to observe planets orbiting them. The distribution of estimated jitter levels for the sample stars is shown in Fig. \ref{fig:jitter_distribution}. This distribution demonstrates that the selected prior density for the jitter is unlikely to be dominating the obtained estimates because the observed distribution has a much more (less) density on the high (low) value side than the applied Gaussian distribution.

\begin{figure}
\center
\includegraphics[angle=270, width=0.49\textwidth,clip]{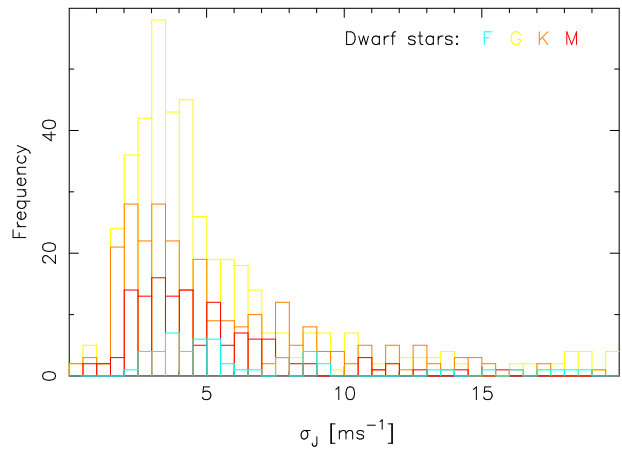}
\caption{Distribution of estimated jitter levels for FGKM stars in the Keck/HIRES exoplanet survey. The estimates have been taken from \citet{butler2016}.}\label{fig:jitter_distribution2}
\end{figure}

This distribution can be contrasted with the values obtained by Keck/HIRES exoplanet survey for FGKM stars \citep{butler2016} where M dwarfs were found to have jitter levels with a median of roughly 3.0 ms$^{-1}$ that is somewhat higher than our estimate of 2.3 ms$^{-1}$ likely because Keck observations have a white noise component arising from the instrument that is not accounted for by the corresponding estimated instrumental uncertainties that are assumed to be photon-noise dominated (Fig. \ref{fig:jitter_distribution2}).

\begin{figure}
\center
\includegraphics[angle=270, width=0.49\textwidth,clip]{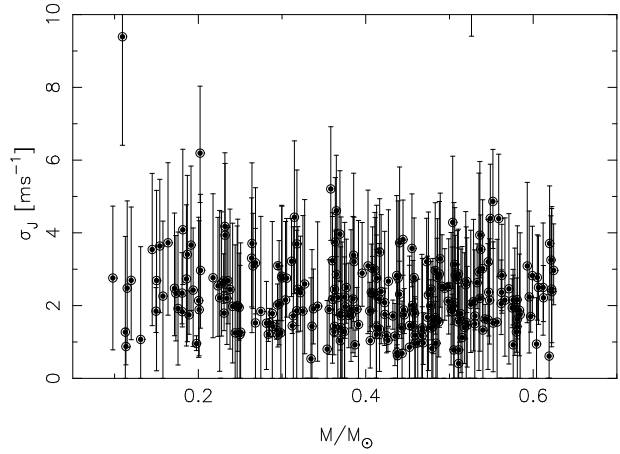}
\caption{Stellar jitter (parameter $\sigma_{\rm J}$) as a function of estimated stellar mass.}\label{fig:jitter_mass}
\end{figure}

However, we did not find any explanatory variables for the varying jitter levels for the targets in the sample. We attempted to find correlations between the jitter and estimated stellar properties but no such correlations were found. We plot the jitter as a function of stellar mass to demonstrate this point in Fig. \ref{fig:jitter_mass}. This indicates that the radial velocity variations of the sample stars are remarkably uniform and modelled well by estimating the uncertainty of each radial velocity measurement by $\sqrt{\sigma_{i}^{2} + \sigma_{\rm J}^{2}}$ where $\sigma_{i}$ is the estimated instrument uncertainty and $\sigma_{\rm J}$ can be estimated to have a value of 2.3 ms$^{-1}$.

This result has two alternative explanations. Either the jitter levels are dominated by instrument noise that naturally does not depend on the stellar properties or M dwarfs show uniform jitter levels over the mass range from 0.1 to 0.6 M$_{\oplus}$. We cannot rule out either one of these explanations. However, in practice this indicates that M dwarfs show very low and uniform levels of radial velocity variations that cannot be explained by intrinsic correlations or activity-induced variability. It is also possible that signals of low-mass planets that are below our detection threshold might be responsible for this apparent uniformity in which case the actual stellar jitter levels would be even lower.

We did not study the wavelength dependence of stellar jitter that is known do decrease towards the red wavelenghts that are less contaminated by stellar activity \citep{anglada2012b}.

\subsection{Activity-induced variability}

\begin{figure}
\center
\includegraphics[angle=270, width=0.49\textwidth,clip]{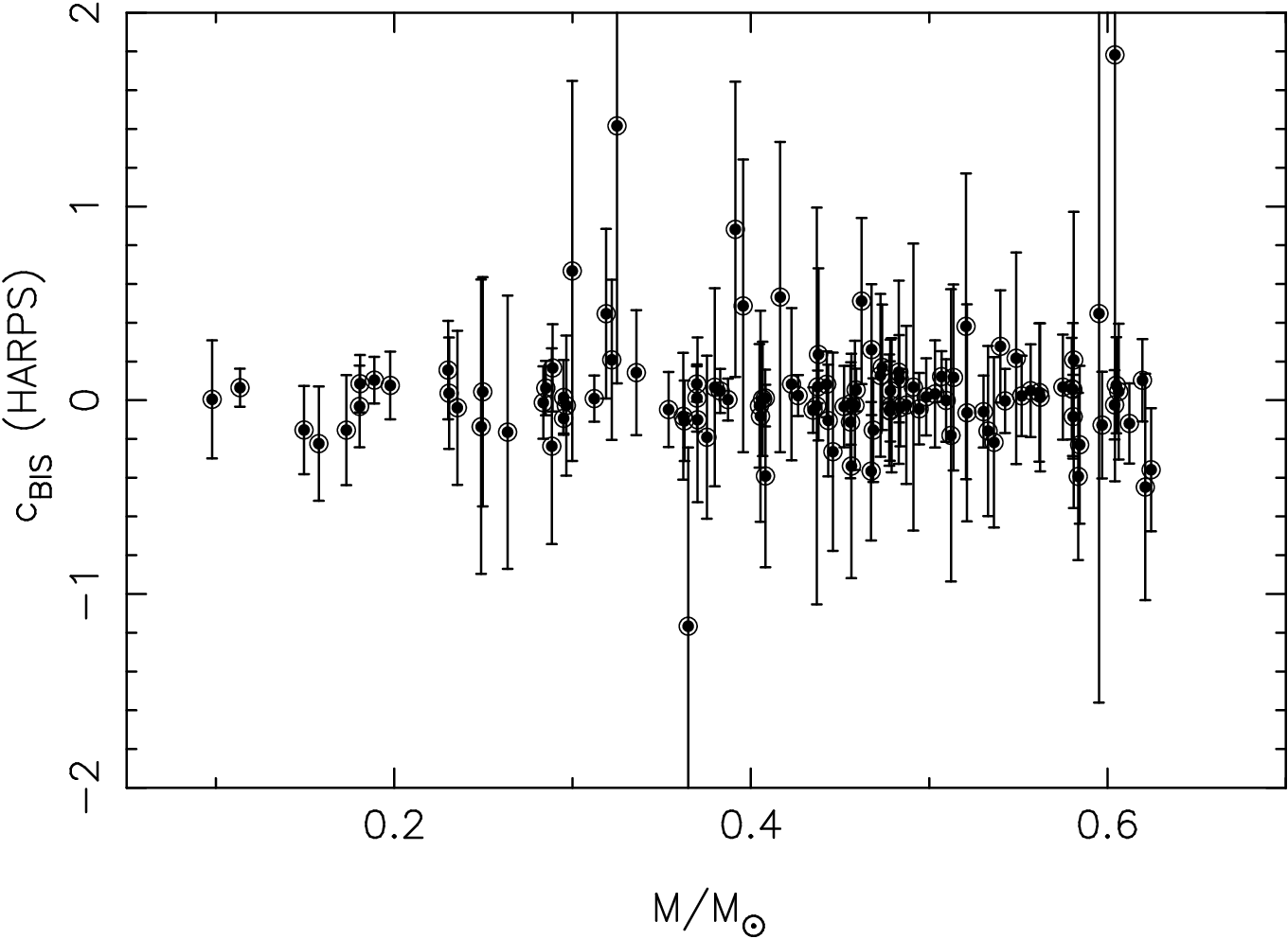}
\includegraphics[angle=270, width=0.49\textwidth,clip]{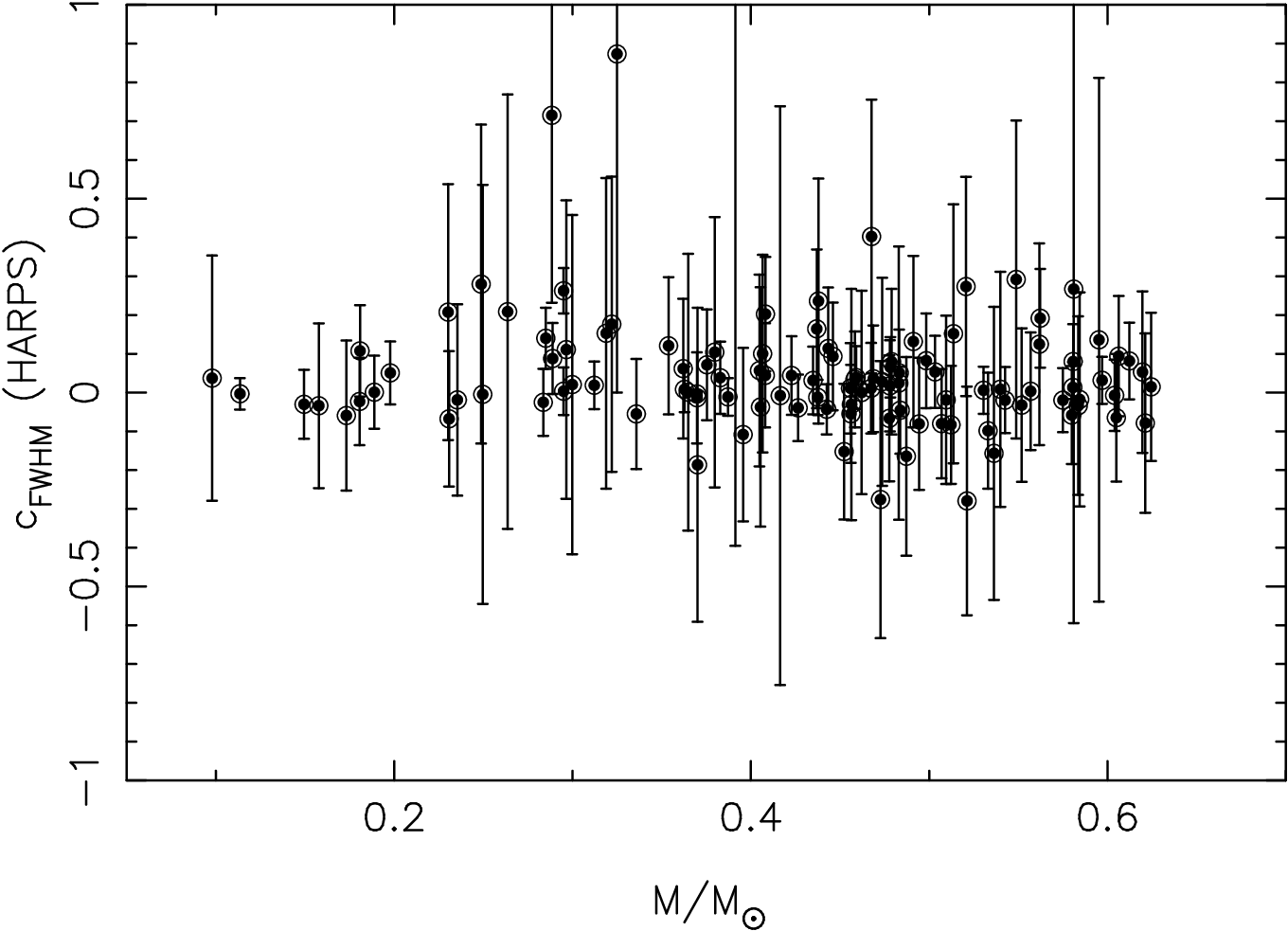}

\includegraphics[angle=270, width=0.49\textwidth,clip]{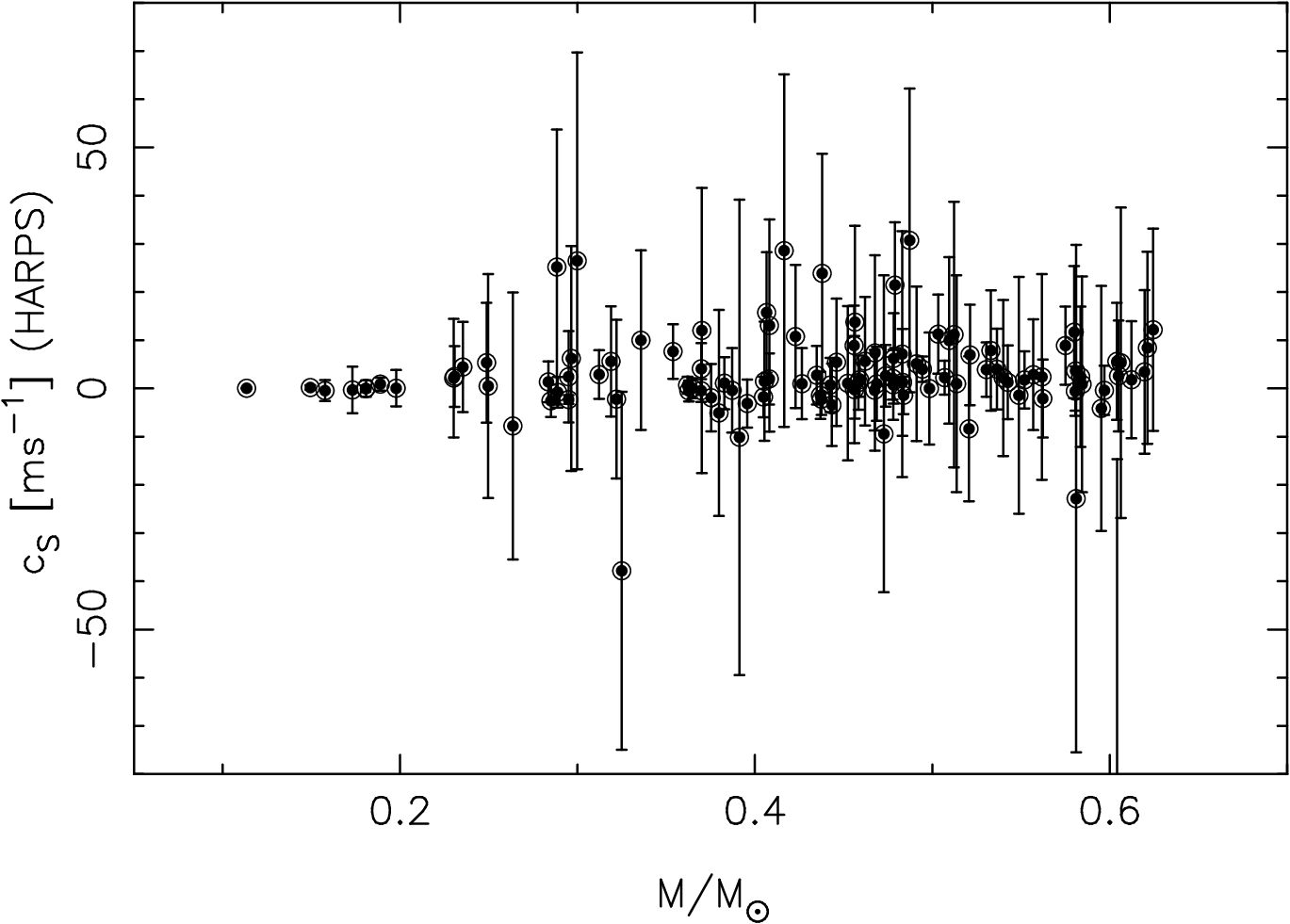}
\includegraphics[angle=270, width=0.49\textwidth,clip]{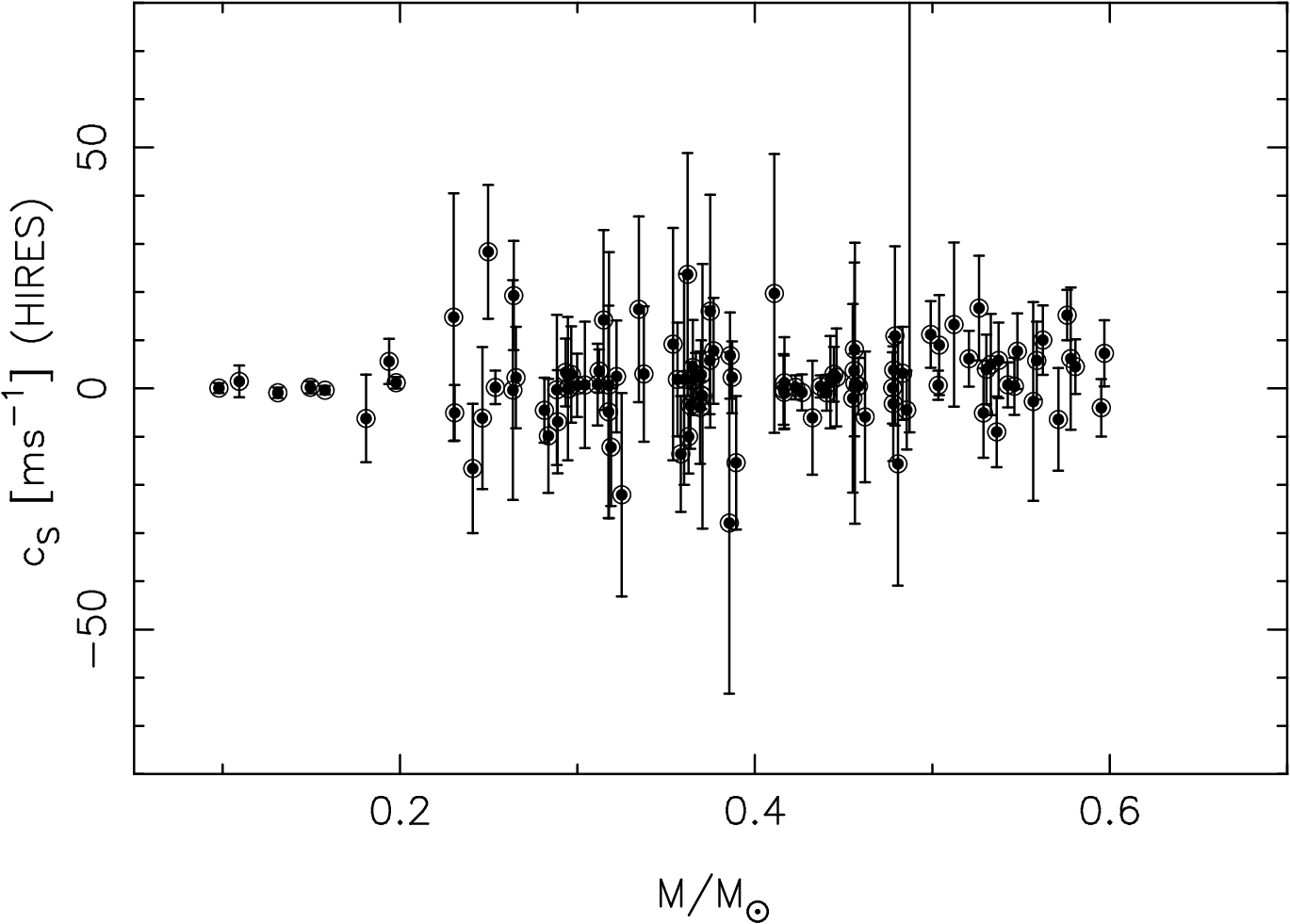}
\caption{MAP parameter estimates quantifying the dependence of radial velocities on the various activity indicators: BIS, FWHM and S-index of HARPS spectra and S-index of HIRES spectra. The parameters are plotted as functions of estimated stellar masses to demonstrate that there are no systematic differences as a function of stellar mass.}\label{fig:activity_relations}
\end{figure}

We also studied the connections between different spectral activity-indicators and radial velocities by quantifying the linear dependence of the latter on the former by using our statistical model. We did not find any evidence that different stellar types in our sample are connected differently to the activity-indices of HARPS and HIRES that were available for us. In fact, we did not obtain any evidence for radial velocities being connected to variations in the activity-indices (Fig. \ref{fig:activity_relations}) but observed that the corresponding parameters were consistent with zero for a majority of the targets implying no statistically significant correlations.\footnote{In fact, we used the significant correlations to spot biases in the results caused by outliers in the activity data that were subsequently removed before re-analysing the data sets.}

Out of a total of 159 data sets, only 32 HIRES data sets, or 20\% of them, showed evidence for a significant connection with a 99\% credibility between the radial velocities and the S-indices that measure the emissions of CaII H\&K lines with respect to the continuum. Due to the lack of correlations between this connection and stellar properties, the physical origin of these connections remains unsure. We also note that \citet{butler2016} observed significant such correlations between radial velocities and S-indices of FGKM stars in the Keck/HIRES sample for 11\% of the stars with 3-$\sigma$ significance. This suggests that radial velocities of M dwarfs might be more likely than FGK dwarfs to be linearly connected to the variations in the S-indices for HIRES.

Similarly, only 20 out of 327 HARPS data sets, or 6\%, showed evidence for a linear connection between the velocities and S-indices whereas this number was 15 (19) for the HARPS BIS (FWHM) values. This suggests that very few M dwarfs have variations that arise demonstrably from the activity-induced variations of the stellar surface or that the linear dependence does not describe these variations adequately accurately. Either way, apart from a handful of exceptions (such as the T Tauri star HIP 20160), we have no evidence in general for such activity-induced origin of radial velocity variability. It is thus possible that the MA component in our statistical model accounts for the corresponding aperiodic and/or quasiperiodic variations disabling the detection of the linear dependence of velocities on activity-indices. Alternatively, the proposed measures of stellar activity (BIS, FWHM and S-index) are not particularly accurate for M dwarfs \citep[see e.g.][who observed correlations between radial velocities and Na I emission]{gomez2012}. 

\section{Occurrence rates of planets around M dwarfs}\label{sec:occurrence}

\begin{figure*}
\center
\includegraphics[angle=270, width=0.95\textwidth]{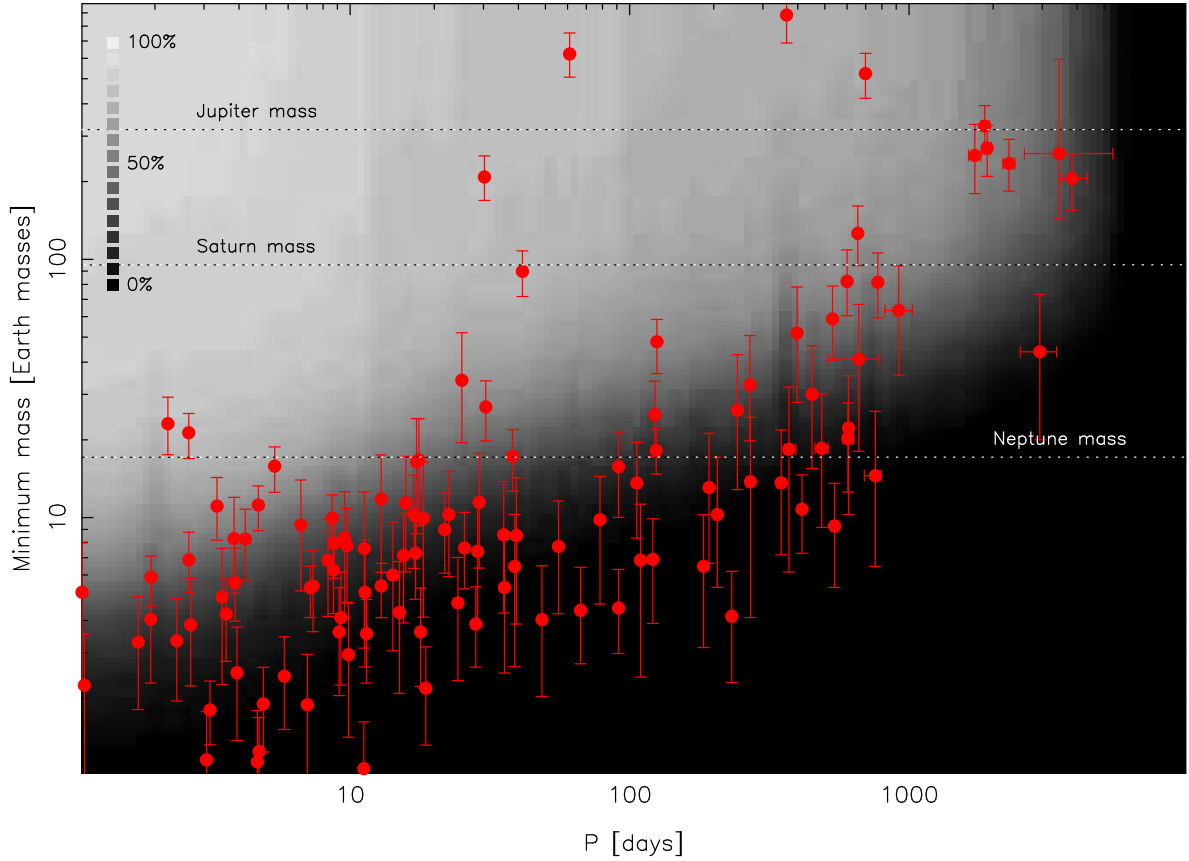}
\caption{Estimated detection probability function of the whole sample as functions of planetary minimum mass and period (gray scale in the background) and candidate planets in the sample (red dots) as listed in Table \ref{tab:candidates}. The rapid decrease in the detection probability function between 3000-5000 days represents the typical baselines of the data sets that effectively prevent the detections of additional planets on longer periods.}\label{fig:population_mass_period}
\end{figure*}

\begin{figure*}
\center
\includegraphics[angle=270, width=0.95\textwidth]{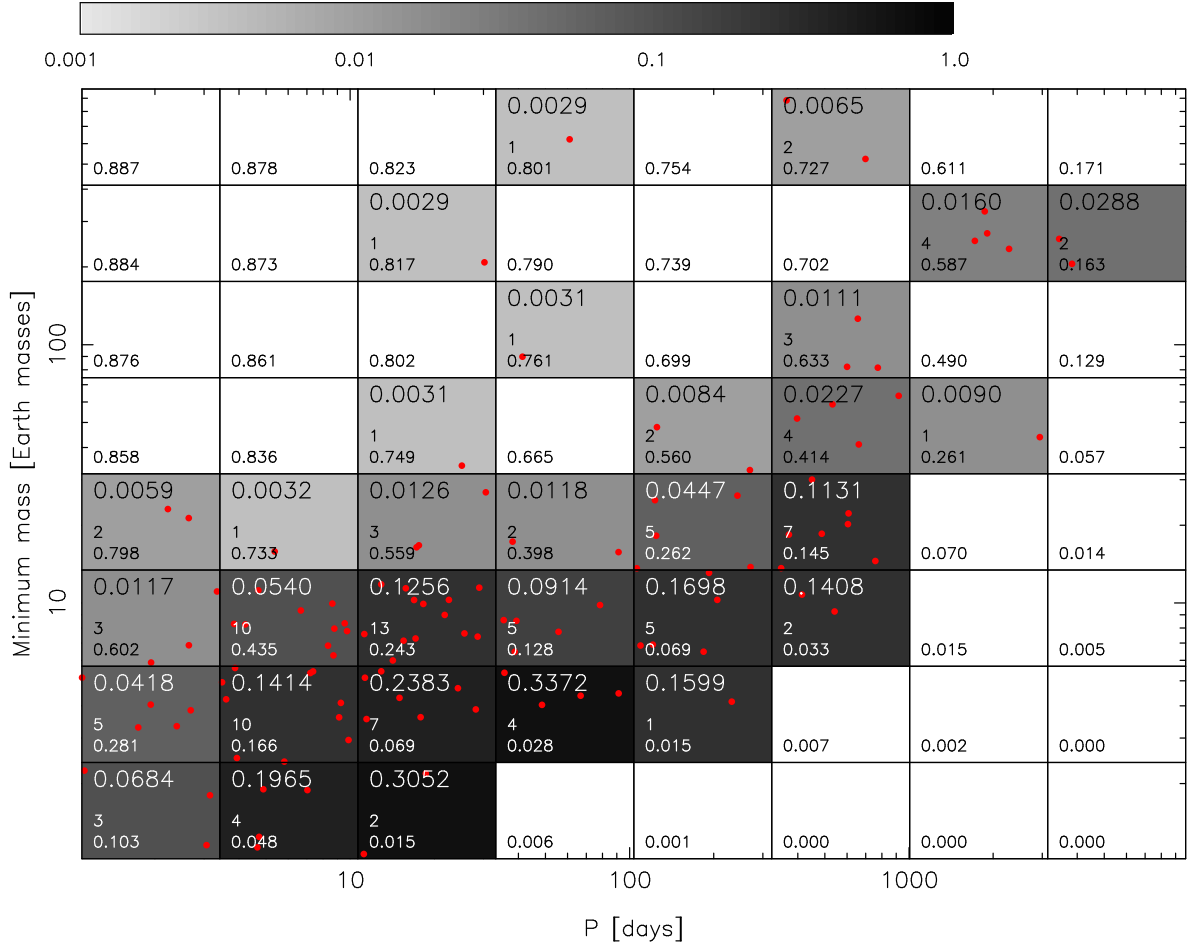}
\caption{Occurrence rate of planets around the sample stars for $m_{p} \sin i \in [1, 1000]$ M$_{\oplus}$ and $P \in [1, 10000]$ days when dividing the parameter space into a 8$\times$8 grid also used in Table \ref{tab:detection_probability}. The grey scale demonstrates the occurrence rate estimates and red dots denote all the candidates tabulated in Table \ref{tab:candidates}. The top number represents the estimated occurrence rate in the grid point, the integer below it denotes the number of candidates and the bottom number represents the average detection probability in the grid.}\label{fig:population_mass_period2}
\end{figure*}

Based on the detection probability function and the detected planet candidates, we calculated the occurrence rates of planets orbiting the sample stars. We have summarised the results in Figs. \ref{fig:population_mass_period} and \ref{fig:population_mass_period2} and tabulated the corresponding numbers in Tables \ref{tab:detection_probability} and \ref{tab:occurrence2}.

\begin{deluxetable}{lcccccccc}
\tablewidth{0pt}
\tablecaption{Detection probability as a function of minimum mass and orbital period based on all the observations of the sample stars.\label{tab:detection_probability}}
\tablehead{
$m_{p} \sin i$ (M$_{\oplus}$) & [1, 3.2] days & [3.2, 10] days & [10, 32] days & [32, 100] days & [100, 320] days & [320, 1000] days & [1000, 3200] days & [3200, 10$^{4}$] days
}
\startdata
 $[420, 1000]$ & 0.887$\pm$0.001 & 0.878$\pm$0.010 & 0.823$\pm$0.043 & 0.801$\pm$0.009 & 0.754$\pm$0.011 & 0.727$\pm$0.025 & 0.611$\pm$0.096 & 0.171$\pm$0.291 \\
 $[180, 420]$  & 0.884$\pm$0.003 & 0.873$\pm$0.012 & 0.817$\pm$0.047 & 0.79$\pm$0.0180 & 0.739$\pm$0.019 & 0.702$\pm$0.035 & 0.587$\pm$0.099 & 0.163$\pm$0.29 \\
 $[75, 180]$   & 0.876$\pm$0.009 & 0.861$\pm$0.019 & 0.802$\pm$0.050 & 0.761$\pm$0.030 & 0.699$\pm$0.041 & 0.633$\pm$0.071 & 0.490$\pm$0.162 & 0.129$\pm$0.279 \\
 $[32, 75]$    & 0.858$\pm$0.017 & 0.836$\pm$0.032 & 0.749$\pm$0.089 & 0.665$\pm$0.100 & 0.560$\pm$0.132 & 0.414$\pm$0.185 & 0.261$\pm$0.244 & 0.057$\pm$0.187 \\
 $[13, 32]$    & 0.798$\pm$0.054 & 0.733$\pm$0.102 & 0.559$\pm$0.211 & 0.398$\pm$0.229 & 0.262$\pm$0.238 & 0.145$\pm$0.153 & 0.070$\pm$0.125 & 0.014$\pm$0.057 \\
 $[5.6, 13]$   & 0.602$\pm$0.172 & 0.435$\pm$0.269 & 0.243$\pm$0.243 & 0.128$\pm$0.160 & 0.069$\pm$0.121 & 0.033$\pm$0.056 & 0.015$\pm$0.030 & 0.005$\pm$0.014 \\
 $[2.4, 5.6]$  & 0.281$\pm$0.226 & 0.166$\pm$0.163 & 0.069$\pm$0.107 & 0.028$\pm$0.045 & 0.015$\pm$0.021 & 0.007$\pm$0.012 & 0.002$\pm$0.008 & 0.000$\pm$0.007 \\
 $[1.0, 2.4]$  & 0.103$\pm$0.099 & 0.048$\pm$0.081 & 0.015$\pm$0.032 & 0.006$\pm$0.015 & 0.001$\pm$0.011 & 0.000$\pm$0.005 & -- & -- \\
\enddata
\end{deluxetable}

\begin{deluxetable}{lcccccccc}
\tablewidth{0pt}
\tablecaption{Occurrence rate in units of ''planets per star`` as a function of minimum mass and orbital period based on all the observations of the sample stars. Two upper limits of the uncertainty intervals could not be estimated for the low-mass planets because the detection probability function falls to zero in the corresponding intervals.\label{tab:occurrence2}}
\tablehead{
$m_{p} \sin i$ (M$_{\oplus}$) & [1, 3.2] days & [3.2, 10] days & [10, 32] days & [32, 100] days & [100, 320] days & [320, 1000] days & [1000, 3200] days & [3200, 10$^{4}$] days
}
\startdata
$[420, 1000]$  & $<$0.0027 & $<$0.0027 & $<$0.0029 & 0.0029$^{+0.0002}_{-0}$ & $<$0.0032 & 0.0065$^{+0.0004}_{-0.0002}$ & $<$0.0050 & -- \\
$[180, 420]$ & $<$0.0027 & $<$0.0027 & 0.0029$^{+0.0001}_{-0.0002}$ & $<$0.0032 & $<$0.0033 & $<$0.0036 & 0.0160$^{+0.0056}_{-0.0023}$ & 0.0288$^{+-0.0288}_{-0.0184}$ \\
$[75, 180]$ & $<$0.0027 & $<$0.0028 & $<$0.0031 & 0.0031$^{+0.0003}_{-0.0001}$ & $<$0.0037 & 0.0111$^{+0.0029}_{-0.0011}$ & $<$0.0083 & -- \\
$[32, 75]$ & $<$0.0028 & $<$0.0030 & 0.0031$^{+0.0007}_{-0.0003}$ & $<$0.0046 & 0.0084$^{+0.0057}_{-0.0016}$ & 0.0227$^{+0.0218}_{-0.0070}$ & 0.0090$^{+0.0196}_{-0.0043}$ & -- \\
$[13, 32]$ & 0.0059$^{+0.0010}_{-0.0004}$ & 0.0032$^{+0.0012}_{-0.0004}$ & 0.0126$^{+0.0105}_{-0.0035}$ & 0.0118$^{+0.0142}_{-0.0043}$ & 0.0447$^{+0.0743}_{-0.0213}$ & 0.1131$^{+0.2553}_{-0.058}$ & -- & -- \\
$[5.6, 13]$ & 0.0117$^{+0.0075}_{-0.0026}$ & 0.054$^{+0.061}_{-0.0206}$ & 0.1256$^{+0.2355}_{-0.0628}$ & 0.0914$^{+0.2027}_{-0.0508}$ & 0.1698$^{+0.5445}_{-0.1081}$ & 0.1408$^{+0.3592}_{-0.0882}$ & -- & -- \\
$[2.4, 5.6]$ & 0.0418$^{+0.0459}_{-0.0187}$ & 0.1414$^{+0.2934}_{-0.0699}$ & 0.2383$^{+0.7617}_{-0.1449}$ & 0.3372$^{+0.9961}_{-0.2082}$ & 0.1599$^{+?}_{-0.0933}$ & -- & -- & -- \\
$[1.0, 2.4]$ & 0.0684$^{+0.1191}_{-0.0335}$ & 0.1965$^{+0.8035}_{-0.1238}$ & 0.3052$^{+?}_{-0.2052}$ & -- & -- & -- & -- & -- \\
\enddata
\end{deluxetable}

We find that there are, on average, 2.39$^{+4.58}_{-1.36}$ planets per star orbiting the sample stars given our detection probability function illustrated in Figs. \ref{fig:population_mass_period} and \ref{fig:population_mass_period2} and tabulated in Table \ref{tab:detection_probability}. This is based on the 118 candidate planets in our sample that satisfy the planet candidate detection criteria. We note that because we could only estimate occurrence rates for the intervals for which there were candidate planets in the sample, the overall occurrence rate is an underestimate. Specifically, because the detection probability function is zero or close to zero such that only a few candidate planets could be detected below the line from roughly (1 M$_{\oplus}$, 10 days) to (50 M$_{\oplus}$, 10$^{4}$ days) we suspect that the real occurrence rate of planets with masses above 1 M$_{\oplus}$ could be considerably higher than the estimated 2.4 planets per star (Fig. \ref{fig:population_mass_period}).

\begin{figure}
\center
\includegraphics[angle=270, width=0.60\textwidth]{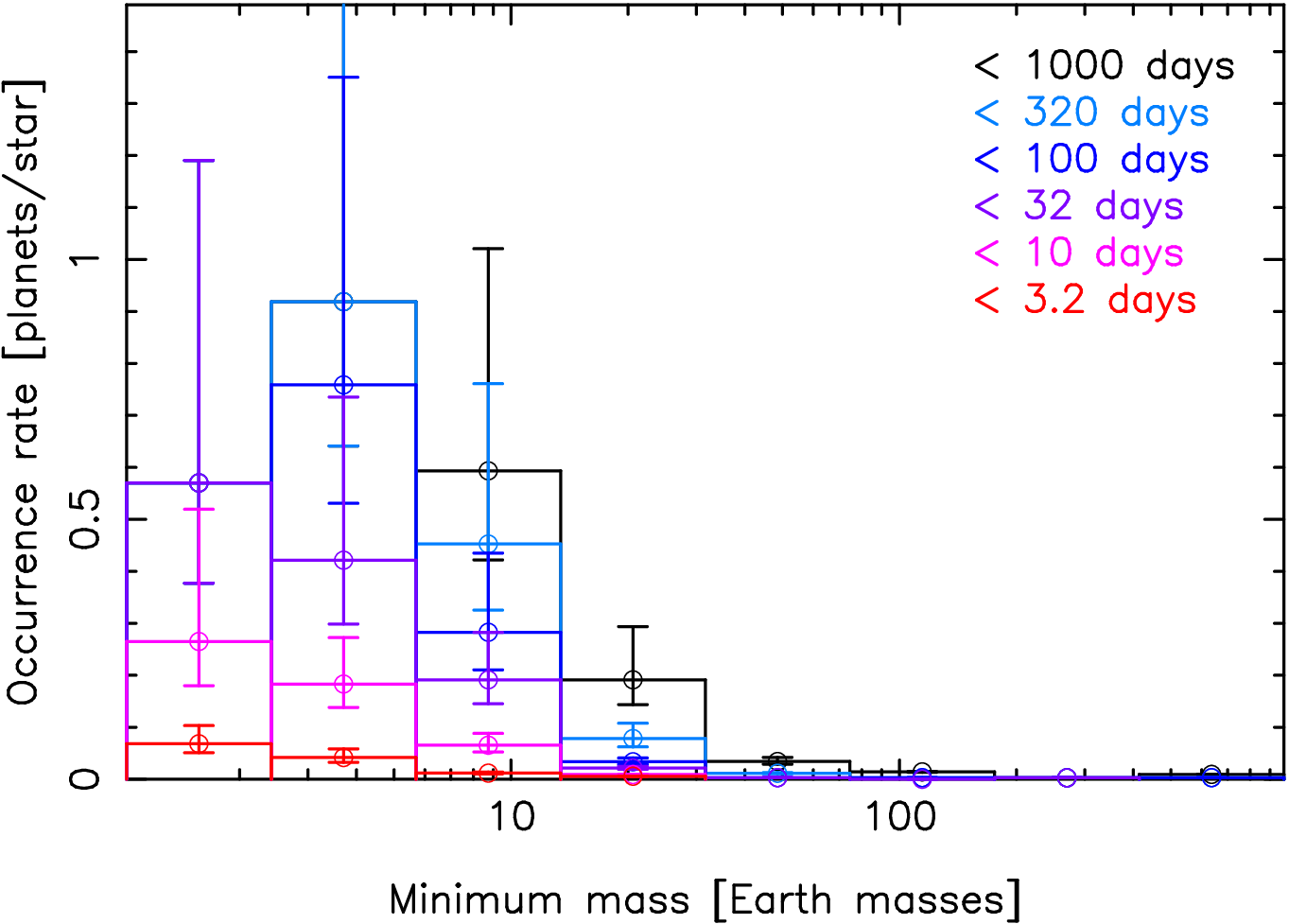}

\includegraphics[angle=270, width=0.60\textwidth]{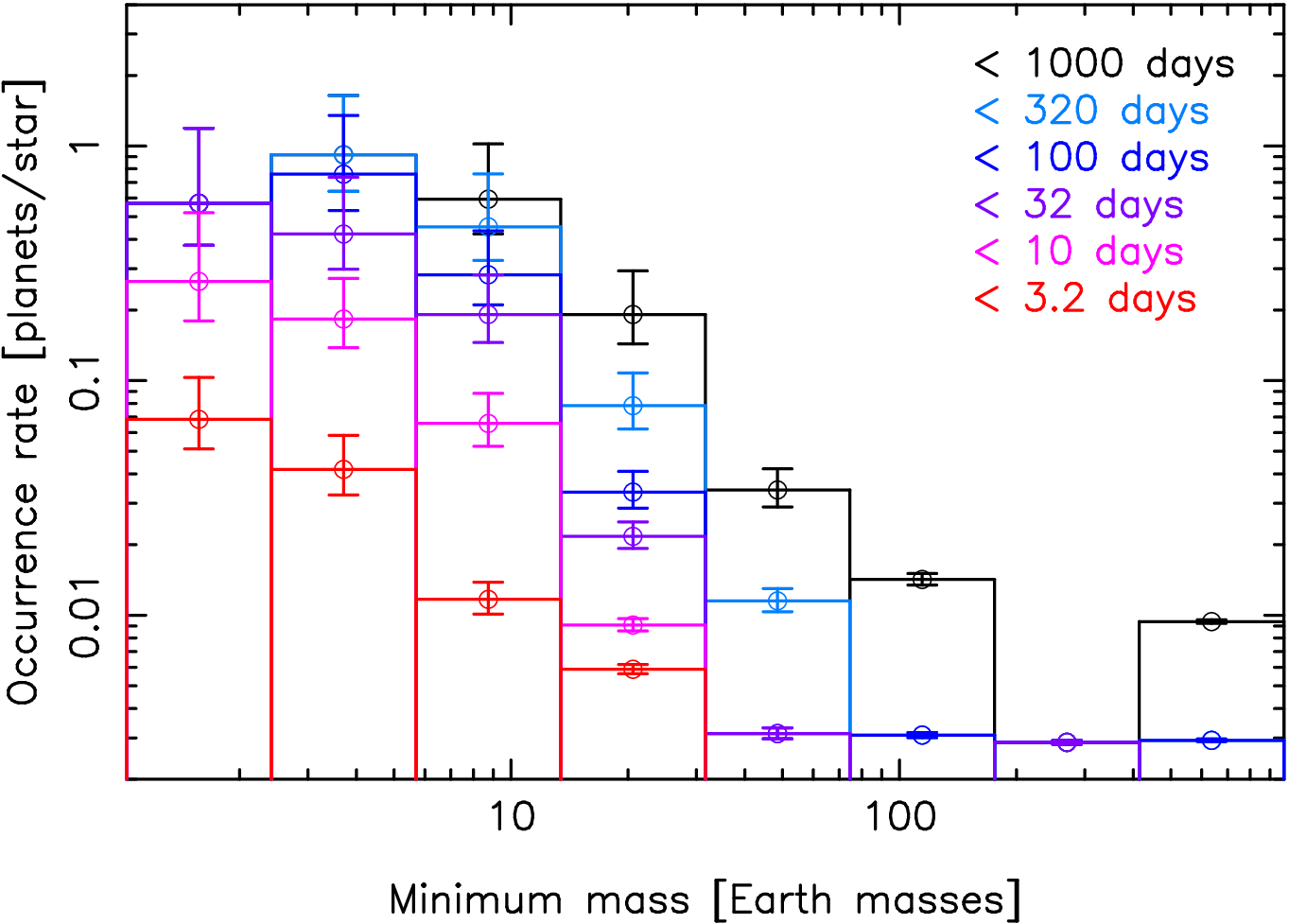}
\caption{Top panel: occurrence rate of planet candidates orbiting the sample stars as a function of their minimum mass. The different colours denote the cumulative occurrence rate up to the given period thresholds. The same plot is shown with linear vertical axis in the bottom panel to highlight the increasing occurrence rate towards smaller masses for virtually all period ranges for which data is available.}\label{fig:population_massfunction}
\end{figure}

We further illustrate the results by plotting the occurrence rate of planets up to a limiting upper period threshold as a function of minimum mass (Fig. \ref{fig:population_massfunction}). It can be seen that the occurrence rate of planets increases as their mass decreases. Moreover, this is the case for all period ranges and indicates that the mass function increases rapidly as a function of decreasing mass. We have used the standard 1-$\sigma$ errors in Fig. \ref{fig:population_massfunction} rather than the obtained minimum and maximum limits to make the plot clearer.

According to these results, it is clear that short-period ($P < 10$ days) planets with minimum masses above 32 M$_{\oplus}$ are very rare -- in the current sample, none were detected and the estimated upper limit for such planets is 0.022 planets per star. In contrast, low-mass planets with 1 M$_{\oplus} < m_{p} \sin i <$ 32 M$_{\oplus}$ are very common on such short-period orbits with an occurrence rate of 0.52$^{+1.33}_{0.27}$. Dividing these low-mass candidates into two sets, planets with 1 M$_{\oplus} < m_{p} \sin i <$ 5.6 M$_{\oplus}$ roughly corresponding to Earths and super-Earths, and planets with 5.6 M$_{\oplus} < m_{p} \sin i <$ 32 M$_{\oplus}$ corresponding to mini-Neptunes and Neptunes, we can see that the latter have an occurrence rate of only 0.075$^{+0.071}_{-0.024}$ planets per star whereas the former are super-abundant around M dwarfs with an occurrence rate of 0.45$^{+1.26}_{-0.25}$ planets per star. It can also be seen that the occurrence rate peaks for super-Earths and mini-Neptunes with minimum masses 2.4 M$_{\oplus} < m_{p} \sin i <$ 13 M$_{\oplus}$ on orbital periods of 10 days $< P <$ 100 days, where the occurrence rate is 0.79$^{+2.20}_{-0.47}$. This is contrasted with the occurrence rate of more massive planets in this period interval of an order of magnitude lower. This rapid increase in occurrence rate thus happens at about 13 M$_{\oplus}$ for such short-period planets.

Although there are not enough planets in the parameter space above periods of 100 days and minimum masses of 32 M$_{\oplus}$, we can still conclude that overall there are at least 0.094$^{+0.038}_{-0.033}$ super-Neptunes and giant planets per star with minimum masses and periods such that 32 M$_{\oplus} < m_{p} \sin i <$ 1000 M$_{\oplus}$ and 320 days $< P < 10^{4}$ days. However, smaller planets with minimum masses 5.6 M$_{\oplus} < m_{p} \sin i <$ 32 M$_{\oplus}$ are much more frequent even for the shorter period interval of 320 days $< P < 10^{3}$ days with an occurrence rate of 0.25$^{+0.61}_{-0.15}$, which indicates the occurrence rate increases towards lower masses even for long-period planets.

Given the detection probability function for the whole sample, we could now repeat the occurrence rate estimation for planets that can be classified either as Earths or super-Earths in the stellar habitable zones. We obtained a value of 0.48$^{+0.46}_{-0.16}$ planets per star, which indicates that, on average, every second M dwarf has an Earth- or super-Earth-type planet in their HZs.

\begin{figure}
\center
\includegraphics[angle=270, width=0.80\textwidth]{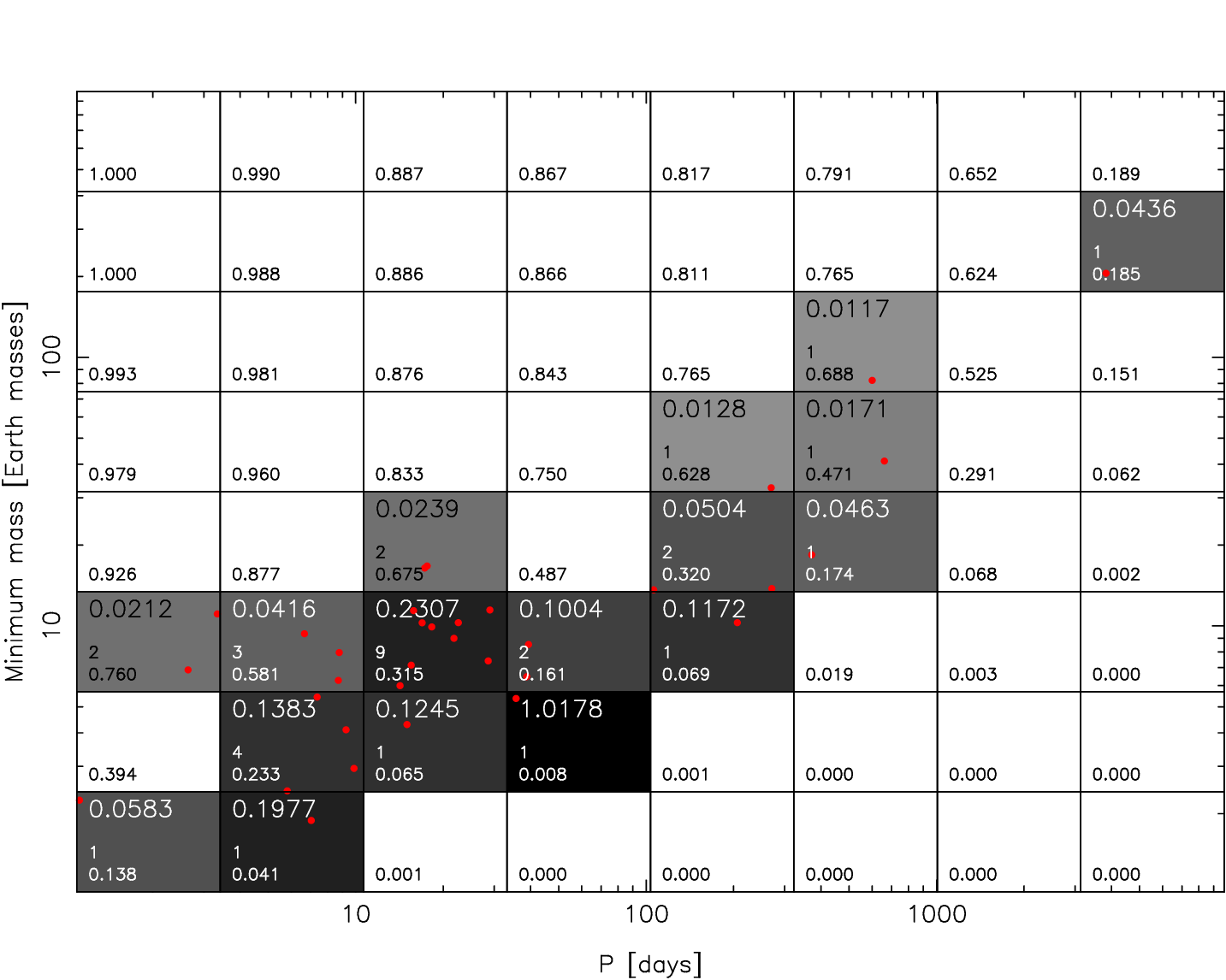}

\includegraphics[angle=270, width=0.80\textwidth]{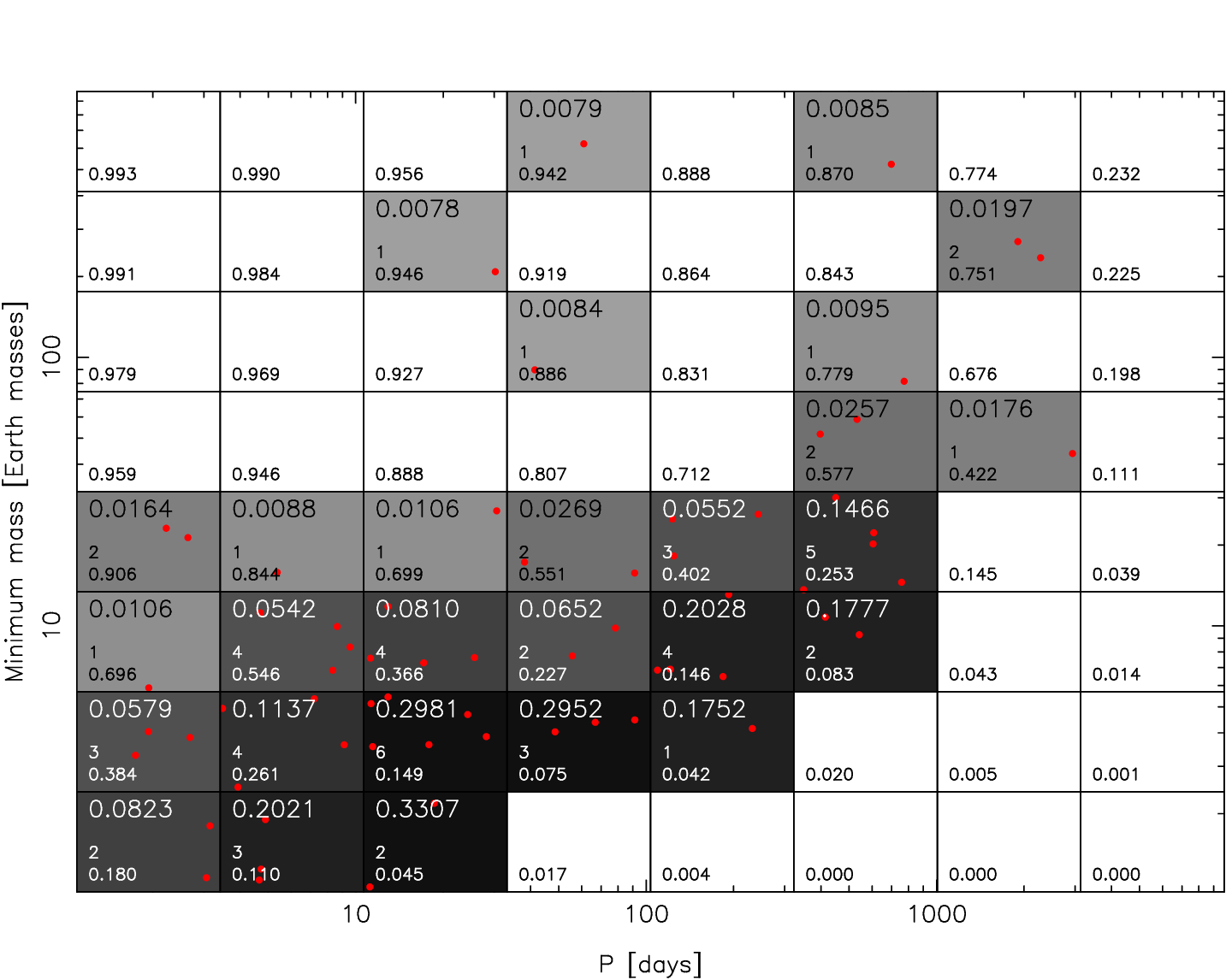}
\caption{As in Fig. \ref{fig:population_mass_period2} but for subsample of targets with masses above (top panel) and below (bottom panel) the median mass of the sample of 0.43 M$_{\oplus}$}\label{fig:occurrence_low_high}
\end{figure}

We could also test whether there was evidence for occurrence rate changing as a function of stellar properties -- the stellar mass and metallicity that are known to affect the occurrence rates of giant planets \citep[e.g.][]{johnson2010b}. We divided the population of stellar targets into high and low mass subsets by adopting the median stellar mass of 0.43 M$_{\oplus}$ as a limit. According to our results, there is only evidence for dependence of occurrence rate on M dwarf mass for the long-period mini-Neptunes and Neptunes with 5.6 M$_{\oplus} < m_{p} \sin i <$ 32 M$_{\oplus}$ and 320 days $< P < 10^{3}$ days. For these planets, the occurrence rate around low-mass M dwarfs is 0.32$^{+0.53}_{-0.16}$ planets per star whereas it is only 0.046$^{+0.287}_{-0.024}$ for such planets around high-mass M dwarfs. This suggests that long-period mini-Neptunes and Neptunes could be an order of magnitude more frequent orbiting low-mass M dwarfs, which is a probable indication that their formation rate is affected by stellar mass. However, although the detection probability function has been accounted for, we note that for the high-mass M dwarfs this result is based on only one detected planet because detection of planets in the high-mass subsample of targets is more difficult due to the respective lower signal amplitudes. We illustrate this result in Fig. \ref{fig:occurrence_low_high} where we have also removed stars for which detection thresholds could not be calculated for simplicity and for which accurate mass estimates were not available.

We similarly searched for differences in the occurrence rate for subsamples with higher or lower metallicity than the median value. No differences could be found in the overall occurrence rate but there was weak evidence for a difference between occurrence rates of mini-Neptunes and Neptunes with 5.6 M$_{\oplus} < m_{p} \sin i <$ 32 M$_{\oplus}$ on periods of 320 days $< P < 10^{3}$ days. Namely, for the low-metallicity sub-sample based on \citet{gaidos2014} estimates (with [Fe/H] $<$ -0.06) we obtained an occurrence rate of 0.83$^{+0.17}_{-0.48}$ planets per star whereas only an upper limit of 0.50 planets per star could be obtained for the high-metallicity sub-sample with \citet{gaidos2014} metallicities below -0.06 because no such planets were detected. This suggests that M dwarfs with sub-Solar metallicities might have many more cool Neptune-type planets orbiting them than those with super-Solar metallicities. However, this result is not very confidently justified by the data due to the low number of available metallicity estimates for the sample stars in \citet{gaidos2014}. When repeating these computations for the \citet{neves2012,neves2013} metallicity estimates, we obtained occurrence rates of 0.25$^{+0.50}_{-0.15}$ and 0.053$^{+0.147}_{-0.028}$ planets per star, respectively. These values also appear to suggest that cool Neptunes are more frequent orbiting low-metallicity M dwarfs than high-metallicity M dwarfs. However, it is very clear that a re-analysis of the current data with uniformly derived spectroscopic metallicities for the whole sample is needed.

It is possible that the sample is still too small to see reliable evidence for potential differences in the occurrence rates as a function of stellar properties. However, our results are likely robust with respect to occurrence rate as a function of planetary properties, such as orbital period and minimum mass as discussed above.

\subsection{Comparison to \emph{Kepler} estimates}

The results obtained based on transit photometry data from the \emph{Kepler} space-telescope provide an independent estimate for the occurrence rate of planets orbiting M dwarfs \citep{dressing2013,dressing2015,morton2014}. However, their comparison with results obtained by radial velocity surveys is not straightforward because the transit data provides occurrence rates as a function of planetary radius whereas radial velocity method can only be used to determine it as a function of the planetary minimum mass. This is especially problematic because the average densities of super-Earths and mini-Neptunes are not well-constrained, which introduces additional uncertainty into the results \citep[e.g.][]{dressing2015b,wolfgang2016,zeng2016}.

Scenarios describing the formation of the population of planets orbiting M dwarfs do not provide very narrow constraints either. Typical planets orbiting M dwarfs with orbital periods ranging from a few to a few dozen days (Figs. \ref{fig:population_mass_period} and \ref{fig:population_mass_period2}) could be anything from evaporated cores of mini-Neptunes \citep{luger2015}, depending on the early X-ray/extreme ultraviolet radiation environment, or desert planets \citep{tian2015}, to super-Earths with thick gaseous envelopes \citep{charbonneau2009} or even ''water worlds`` covered by a hundreds of kilometres deep ocean due to their volatile-rich nature \citep{alibert2016}. This results in a high level of uncertainty and makes it difficult to define a general mass-radius relationship for small exoplanets.

\begin{figure}
\center
\includegraphics[angle=270, width=0.80\textwidth]{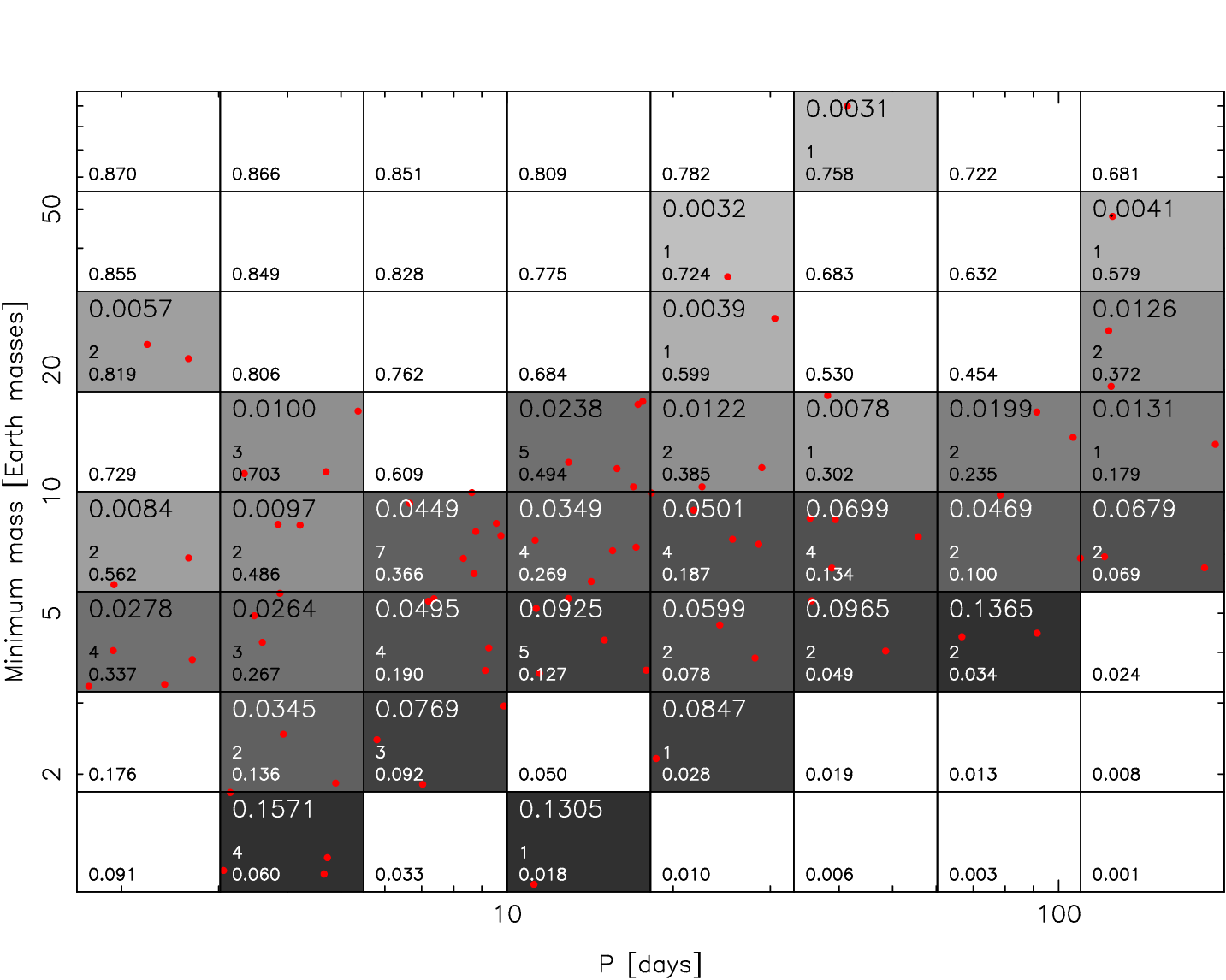}
\caption{As in Fig. \ref{fig:population_mass_period2} but with a scaled grid such that it coincides with the choice of \citet{dressing2015} to better enable the comparisons of the results with those from \emph{Kepler} transit photometry.}\label{fig:occurrence_dressing_comparison}
\end{figure}

The obtained total occurrence rate of 2.39$^{+4.58}_{-1.36}$ planets per star in our sample is remarkably close to the estimate obtained by \citet{dressing2015} with \emph{Kepler} transit photometry results based on 156 planet candidates (Fig. \ref{fig:occurrence_dressing_comparison}). The estimate of \citet{dressing2015} of $2.5\pm0.2$ planets per star is more precise but consistent with our estimate. However, the Kepler transit photometry is sensitive to planets with radii as low as 0.5 R$_{\oplus}$. Because radial velocities are not sensitive enough to detect such small planets with masses likely well below 1 M$_{\oplus}$, it is clear that the 0.60$^{+0.09}_{-0.07}$ planets per star for orbital periods between 0.5 and 200 days and radii between 0.5 and 1.0 R$_{\oplus}$ reported by \citet{dressing2015} are practically all below the detection threshold\footnote{By detection threshold of the sample we mean the area in mass-period space where detection probability falls to zero.} of our sample that is limited to about 1 M$_{\oplus}$ even for short-period planets. This implies that accounting for the existence of this population of small planets detected by \emph{Kepler} transit photometry, i.e. assuming our sample of targets also has such small planets orbiting them with an occurrence rate of 0.6 planets per star, means that the actual occurrence rate of planets around M dwarfs is at least 3.0 planets per star. Conversely, planets on orbits with periods longer than 200 days are not included in the \citet{dressing2015} sample. Because we find that the occurrence rate of planets with orbital periods above 200 days is 0.54$^{+1.06}_{-0.29}$ planets per star, this also implies a total occurrence rate of at least 3.0 planets per star.

For planets on orbits with periods below 100 days, the observed increase in occurrence rate by an order of magnitude has been reported e.g. by \citet{morton2014} and \citet{dressing2015} when moving below roughly 3 R$_{\oplus}$ in planet radius. Our results thus agree with those based on \emph{Kepler} transit photometry given that this increase in occurrence rate at 3 R$_{\oplus}$ approximately corresponds to a minimum mass of 13 M$_{\oplus}$ and corresponds to an order of magnitude higher frequency of super-Earths and mini-Neptunes than Neptunes on short-period orbits.

Similarly, our results agree with those of \citet{dressing2015} with respect to the occurrence rate of Earths and super-Earths in the stellar habitable zones of M dwarfs. While \citet{dressing2015} obtained an estimate of 0.43$^{+0.14}_{-0.09}$ potentially habitable planets with 1 R$_{\oplus} < R <$ 2 R$_{\oplus}$ per star, our results indicate a consistent estimate of 0.48$^{+0.46}_{-0.16}$ planets classified as Earths or super-Earths per star such that their minimum masses satisfy 1 M$_{\oplus} < m_{p} \sin i <$ 10 M$_{\oplus}$ and the upper 99\% credibility boundary of the minimum mass estimate is below 10 M$_{\oplus}$ in accordance with their classification as super-Earths (Table \ref{tab:mass_class}).

\section{Discussion}\label{sec:discussion}

Over the recent years, it has become gradually evident that virtually all M dwarf stars have planets orbiting them; that these planets are more abundant the smaller they are; and there is a population of Earths and super-Earths around M dwarfs with orbital periods ranging from a few days to few months \citep[e.g.][]{bonfils2013,dressing2013,dressing2015,morton2014,tuomi2014}. Because M dwarfs are the most common stars in the Solar neighbourhood and the Galaxy, the occurrence rate of planets around such stars thus dominates the general estimates for occurrence rates of planets. Early attempts to estimate the occurrence rate of M dwarf planets \citep{butler2004,endl2006,johnson2007,cumming2008,zechmeister2009} based on radial velocity surveys only succeeded in calculating limits for the occurrence rate of giant planets around M dwarfs. More recent studies have managed to quantify the occurrence rates of planets around stars in the Solar neighbourhood for a range of masses above 1 M$_{\oplus}$ and periods of up to 4000 days \citep{bonfils2013,tuomi2014}. This has also been possible for an independent sample of M dwarfs in the \emph{Kepler} space-telescope's field \citep{dressing2013,dressing2015} and has yielded broadly consistent results. Although the comparison of occurrence rate estimates based on two different detection techniques, radial velocity and transit photometry (that can only be used to determine minimum masses and radii, respectively), is difficult without reliable models predicting planetary compositions for various sized planetary-mass objects, some conclusions are evident based on the aforementioned studies and current work.

First, according to our results, there are at least 2.39$^{+4.58}_{-1.36}$ planets per M dwarf and very probably at least 3.0 when accounting for the fact that we observe a different parameter space from that accessible by \emph{Kepler} space-telescope transit photometry \citep{dressing2015}. Moreover, it appears evident that the rapid increase in occurrence rate observed for periods below roughly 100 days that happens for planets with radii below 3 R$_{\oplus}$ corresponds to a similar increase for minimum masses below $\sim$ 13 M$_{\oplus}$ (see Fig. \ref{fig:occurrence_dressing_comparison}). Moreover, this increased occurrence rate of smaller planets applies to longer orbital periods as well but is shifted towards higher minimum masses of roughly 13-30 M$_{\oplus}$, which might be supported by recent analyses of microlensing  data \citep{suzuki2016}. Moreover, we have presented evidence that the occurrence rate of such mini-Neptunes and Neptunes (see Table \ref{tab:mass_class}) is an order of magnitude higher around low-mass M dwarfs with masses below 0.43 M$_{\odot}$, which suggests that their formation is affected by stellar mass.

The population of planets around M dwarfs thus consists of super-Earths and mini-Neptunes on orbits with periods ranging from a few days to a hundred days and Neptunes and super-Neptunes with orbital periods longer than a few hundred days. Although planets with minimum masses consistent with, or slightly larger than, that of the Earth are the most abundant in all period ranges (Fig. \ref{fig:population_massfunction}) we cannot establish this for orbital periods longer than 32 days because that would require observational precision below $\sim$ 1 ms$^{-1}$ that is approximately the limiting precision of current radial velocity data.

With results that appear to be in agreement with those based on transit photometry, we confirm that M dwarfs have very rich planetary systems around them even in the immediate Solar neighbourhood. This indicates that M dwarfs are the primary targets for detections of nearby Earth-like planets \citep[e.g][]{anglada2016} that could be considered candidate habitable planets.

\acknowledgements

MT and HRAJ are supported by grants from the Leverhulme Trust (RPG-2014-281) and the Science and Technology Facilities Council (ST/M001008/1). MT and JSJ acknowledge funding by Fondecyt through grant 1161218 and JSS acknowledges partial support from CATA-Basal (PB06 Conicyt). SSV gratefully acknowledges support from NSF grants AST-0307493 and AST-0908870. The authors acknowledge the significant efforts of the HARPS-ESO team in improving the instrument and its data reduction pipelines. We also acknowledge the efforts of all the individuals that are not included in this work but have been involved in observing the target stars with the spectroscopic instruments. The work herein is partially based on observations obtained at the W. M. Keck Observatory, which is operated jointly by the University of California and the California Institute of Technology, and we thank the UC-Keck and NASA-Keck Time Assignment Committees for their support. We also wish to extend our special thanks to those of Hawaiian ancestry on whose sacred mountain of Mauna Kea we are privileged to be guests. Without their generous hospitality, the Keck observations presented herein would not have been possible. This paper includes data gathered with the 6.5 meter Magellan Telescopes located at Las Campanas Observatory, Chile. The work herein was also based on observations obtained at the Automated Planet Finder (APF) facility and its Levy Spectrometer at Lick Observatory. This research has made use of the SIMBAD database, operated at CDS, Strasbourg, France.

\appendix

\clearpage
\newpage

\section{Selected targets}\label{sec:individual_targets}

\subsection{GJ 15A}

GJ 15A (HD 1326, HIP 1475) is a nearby M dwarf that has recently been reported to host a super-Earth orbiting it \citep{howard2014} and subsequently confirmed by \citep{butler2016}. We analysed an updated set of 340 HIRES velocities of the star in an attempt to verify the results of \citet{howard2014} and \citep{butler2016} and to search for signals corresponding to additional planet candidates orbiting the star.

\begin{figure}
\center
\includegraphics[angle=270, width=0.49\textwidth,clip]{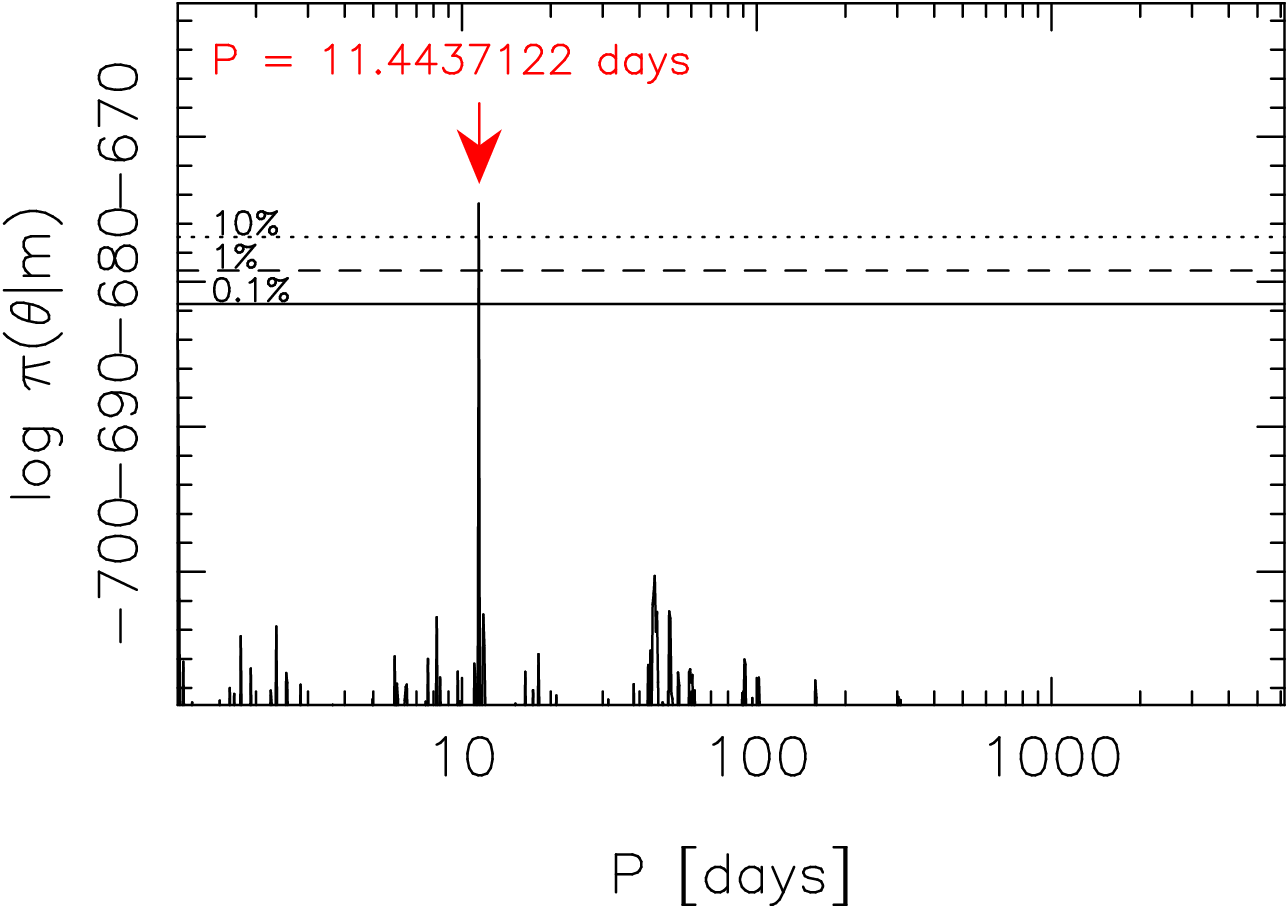}

\includegraphics[angle=270, width=0.49\textwidth,clip]{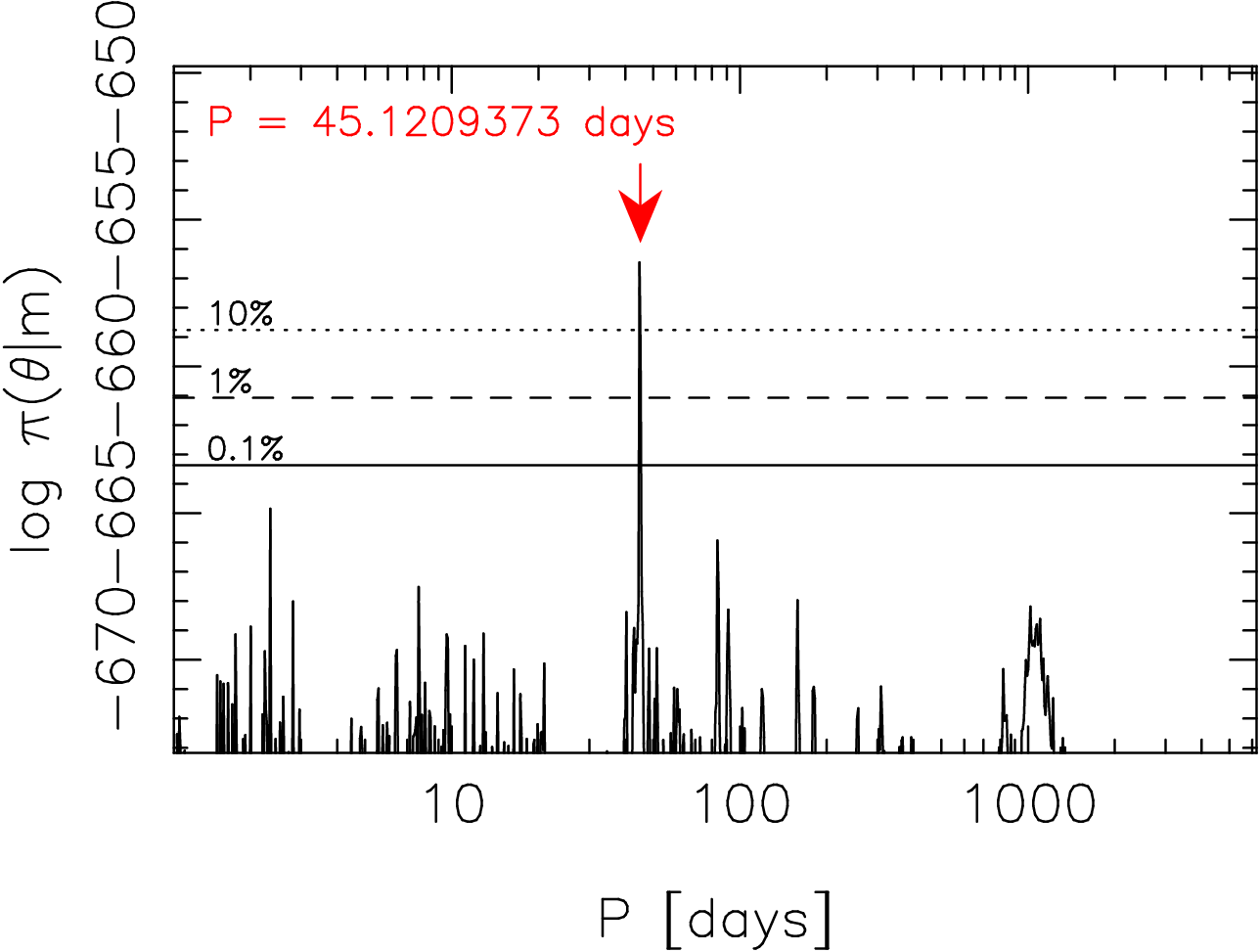}
\caption{Estimater posterior probability density of the period parameter of the signal in a one-Keplerian model (top panel) and the second signal in a two-Keplerian model (bottom) given HIRES data of GJ 15A. The red arrow indicates the global maximum and the horizontal lines denote the 10\% (dotted), 1\% (dashed), and 0.1\% (solid) equiprobability thresholds with respect to this maximum.}\label{fig:GJ15A_period_search}
\end{figure}

The signal corresponding to GJ 15A b was easy to find with the DRAM samplings of the posterior probability density (Fig. \ref{fig:GJ15A_period_search}, top panel). The signal at a period of 11.4427 [11.4379, 11.4485] days has an amplitude of 1.95 [1.32, 2.52] ms$^{-1}$ and it thus corresponds to a 3.6 [2.3, 4.9] M$_{\oplus}$ cool super-Earth. Our estimate of the stellar luminosity of GJ 15A \citep{boyajian2012} suggests that the stellar habitable zone as estimated according to the equations of \citet{kopparapu2013} is closer to the stellar surface than reported by \citet{howard2014} placing the planet at the outer edge of the HZ.

\begin{figure}
\center
\includegraphics[angle=270, width=0.49\textwidth,clip]{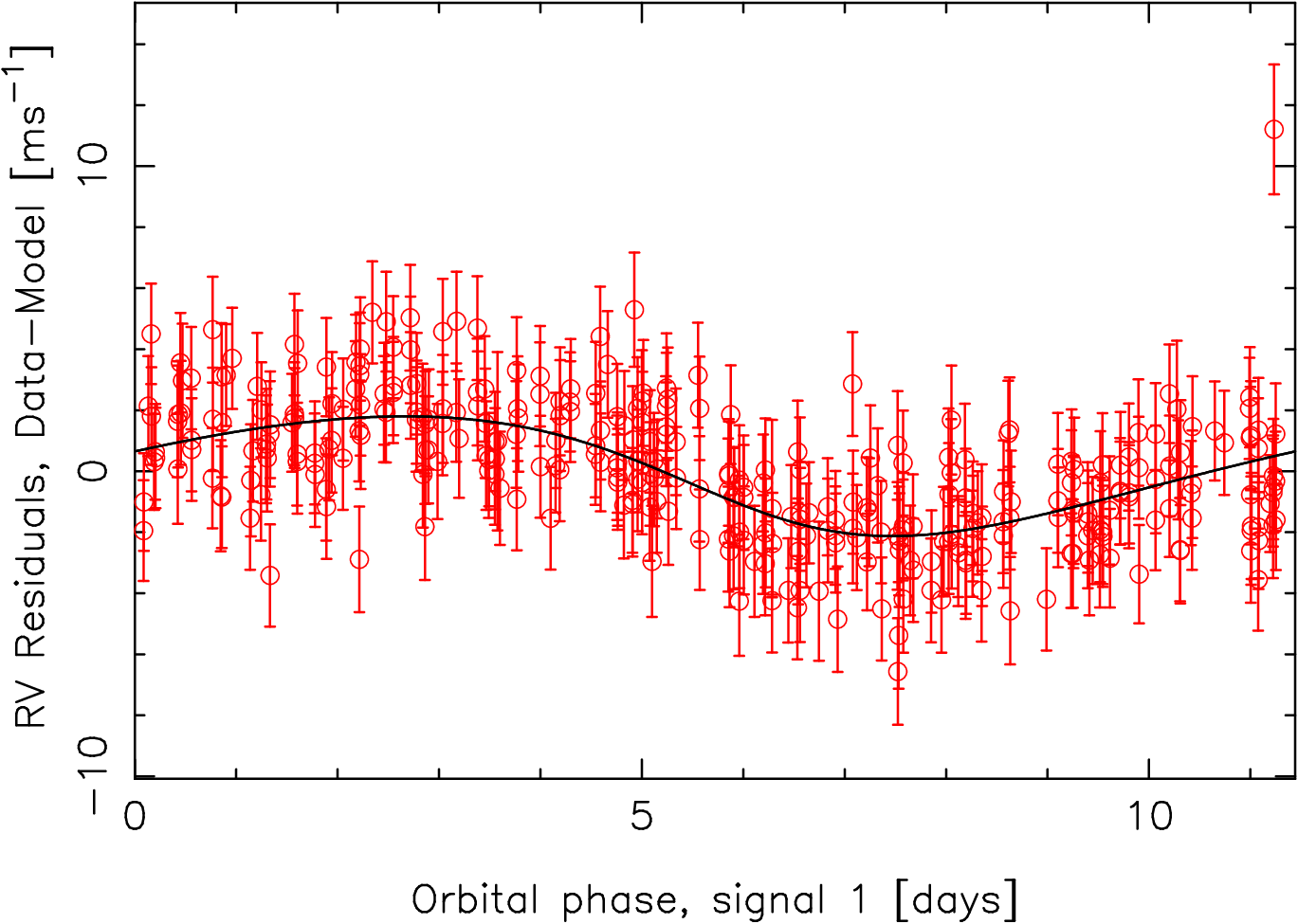}

\includegraphics[angle=270, width=0.49\textwidth,clip]{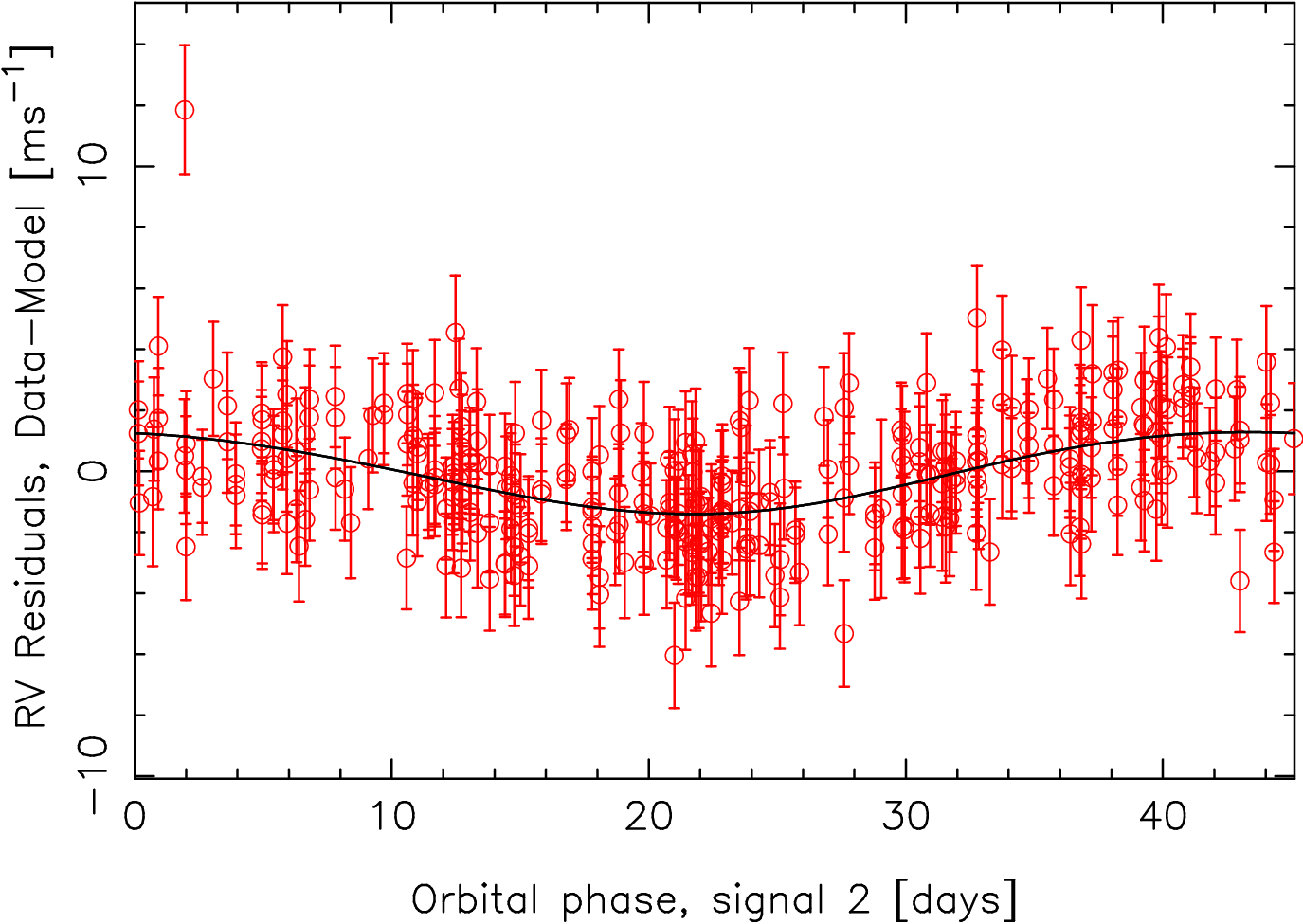}
\caption{HIRES radial velocities of GJ 15A folded of the phases of the signals with the other signal and all deterministic components of the model subtracted from the two panels. The solid curve denotes the MAP signal.}\label{fig:GJ15A_phased}
\end{figure}

However, there appeared to be an additional signal in the HIRES radial velocities of GJ 15A. As can be seen in Fig.\ref{fig:GJ15A_period_search} (bottom panel), the period parameter of the second signal has a clear maximum at a period of 45 days. However, \citet{howard2014} also reported a signal at a period of 44.8 days in the HIRES S-indices. We had no difficulties in identifying the periodicity in the S-indices and it indeed appears likely, as also pointed out by \citet{howard2014} and \citep{butler2016}, that the radial velocity signal that is present in the HIRES data at a period of 45.128 [44.993, 45.293] days is a genuine signal in the data produced by stellar rotation rather than a planet. The signals are shown in Fig. \ref{fig:GJ15A_phased} by folding the radial velocity residuals on their phases after subtracting the other signal an the deterministic components of the model.

There were no ASAS photometry measurements available for GJ 15A disabling the possibility to photometrically verify that the star rotates with a period of roughly 45 days. Neverheless, we accept that this is likely to be the rotation period of the star based on the counterpart signal in HIRES S-indices.

\clearpage

\subsection{GJ 27.1}

\begin{figure}
\center
\includegraphics[angle=270, width=0.49\textwidth,clip]{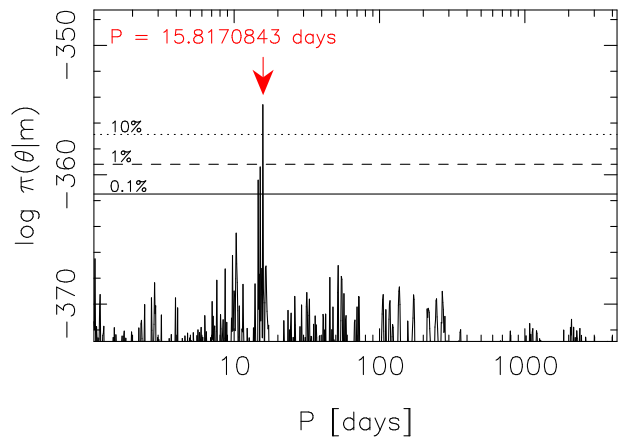}
\caption{Estimated posterior probability density of the period parameter of a signal given combined GJ 27.1 data from UVES, HARPS, and HIRES spectrographs. The red arrow indicates the global maximum and the horizontal lines denote the 10\%, 1\%, and 0.1\% equiprobability thresholds with respect to this maximum. The maximum has been split into several nearby maxima due to aliasing caused by the sampling of the data (see Fig. \ref{fig:GJ27.1_data}).}\label{fig:GJ27.1_period_search}
\end{figure}

\begin{figure}
\center
\includegraphics[angle=270, width=0.49\textwidth,clip]{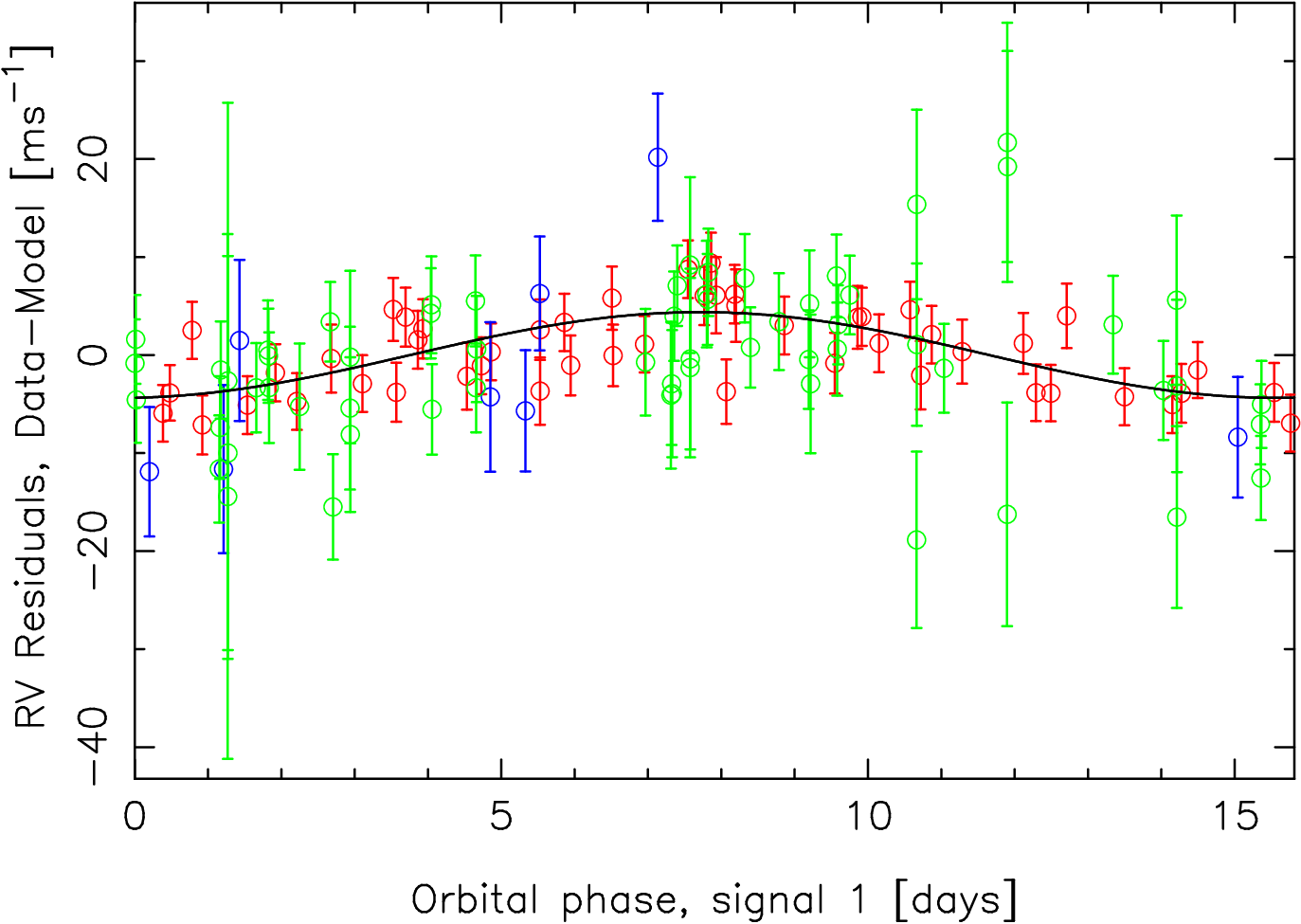}

\includegraphics[angle=270, width=0.49\textwidth,clip]{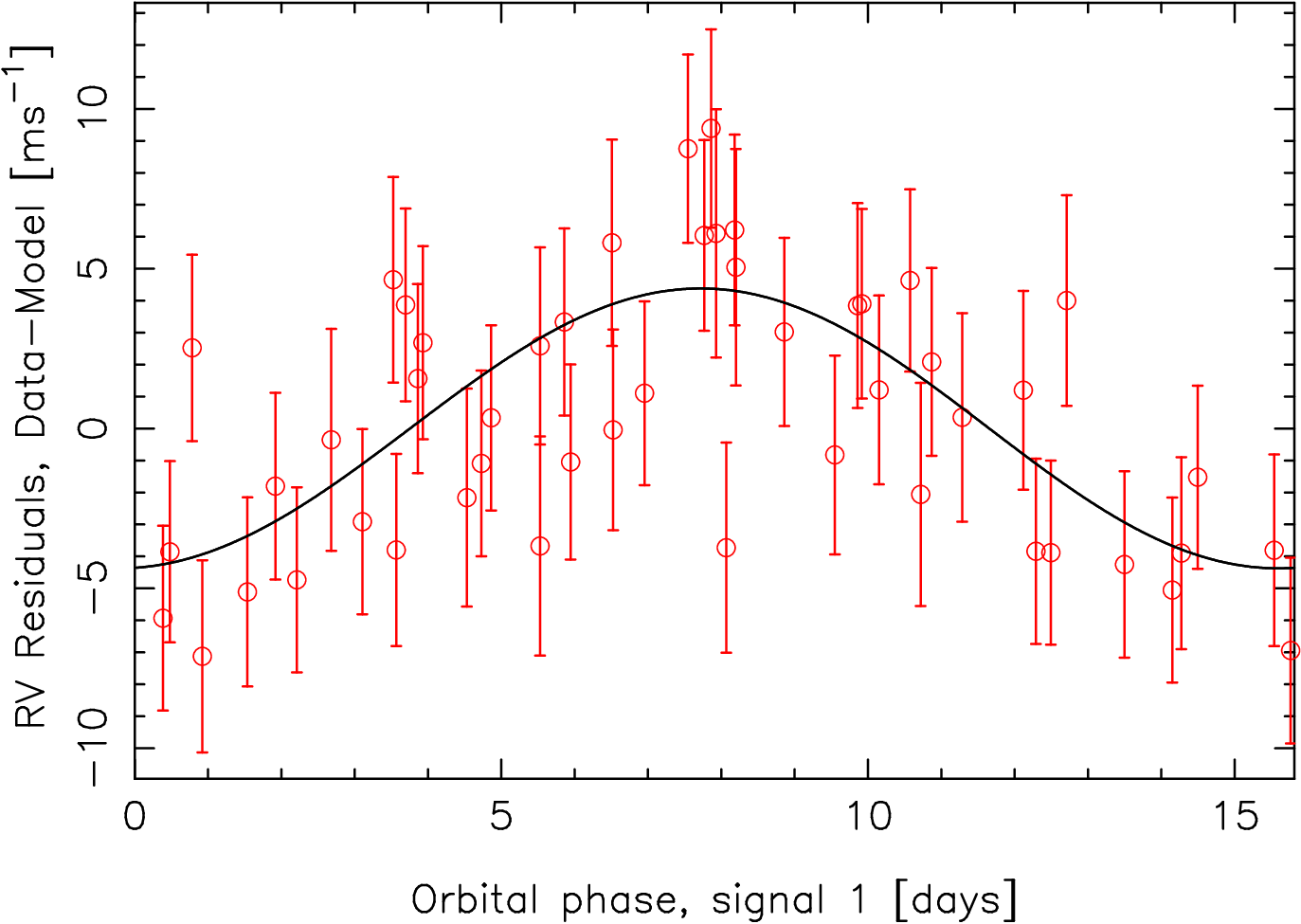}
\caption{Top panel: Phase-folded radial velocities of GJ 27.1 with respect to HARPS (red), HIRES (blue), and UVES (green) data. The solid line indicates the estimated Keplerian curve. Bottom panel: Phase-folded HARPS data alone.}\label{fig:GJ27.1_signal}
\end{figure}

\begin{figure}
\center
\includegraphics[angle=270, width=0.49\textwidth,clip]{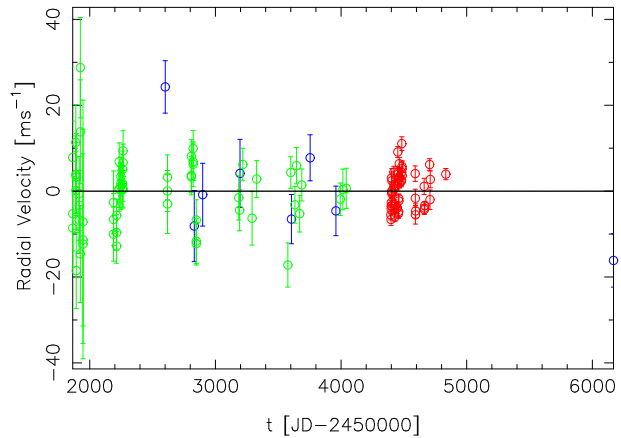}
\caption{HARPS (red), HIRES (blue), and UVES (green) radial velocity data residuals of GJ 27.1 after subtracting linear trend.}\label{fig:GJ27.1_data}
\end{figure}

The candidate planet orbiting GJ 27.1 (HIP 3143) was reported by \citet{tuomi2014} based on combined HARPS ($N =$ 50) and UVES ($N =$ 62) data. We re-analysed the combined data set together with 8 new velocities from HIRES and again observed a signal at a period of 15.819 [15.794, 15.841] days (Figs. \ref{fig:GJ27.1_period_search} and \ref{fig:GJ27.1_signal}). Although this signal is split into several maxima in the period space, the splitting is caused by the rather pathological data sampling that provides no overlap between HARPS and UVES data sets (Fig. \ref{fig:GJ27.1_data}). We did not find other significant maxima in the period space exceeding the 0.1\% probability threshold that were independent of the global maximum and could thus correspond to an independent periodicity.

Despite the fact that the new velocities from HIRES cannot be expected to change the results much due to there being 62 and 50 UVES and HARPS velocities, respectively, we report the results of the analysis of the combined HARPS, HIRES, and UVES data. The HARPS activity indices (BIS, FWHM, S-index) were not statistically connected to the radial velocities. The same was found for HIRES velocities and S-indices. Moreover, we could not find any periodic signals in any of these indices. There is thus no evidence for the signal we observe in the radial velocities at a period of 15.8 days being caused by variations traced by these indices and thus resulting from stellar activity.

We obtained 417 ASAS V-band photometry measurements covering a baseline of 3173.3 days. However, this photometric data was not found to contain any periodic signals indicative of, perhaps, stellar rotation or activity cycles. The target GJ 27.1 thus does not appear to have variations in brightness that could be connected to the observed periodicity in the radial velocity data at a period of 15.8 days. Therefore, we interpret the variations in the radial velocities as indication of a candidate planet orbiting the star. This candidate has a minimum mass of 11.4 [6.1, 17.2] M$_{\oplus}$ and semi-major axis of 0.101 [0.089, 0.109] corresponding to a hot mini-Neptune-type candidate.

\clearpage

\subsection{GJ 49}

\begin{figure}
\center
\includegraphics[angle=270, width=0.49\textwidth,clip]{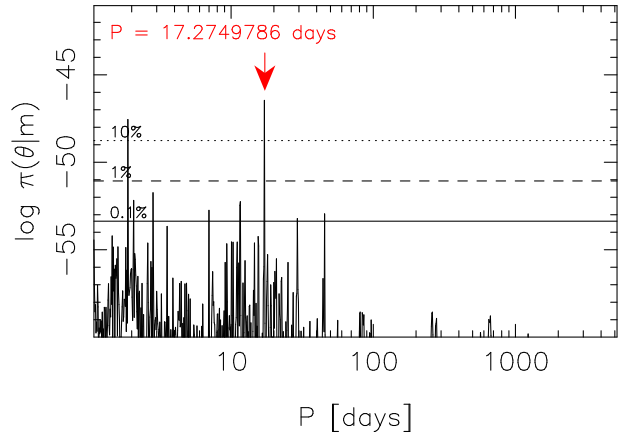}
\caption{As in Fig. \ref{fig:GJ27.1_period_search} but for the HIRES data of GJ 49.}\label{fig:GJ49_period_search}
\end{figure}

\begin{figure}
\center
\includegraphics[angle=270, width=0.49\textwidth,clip]{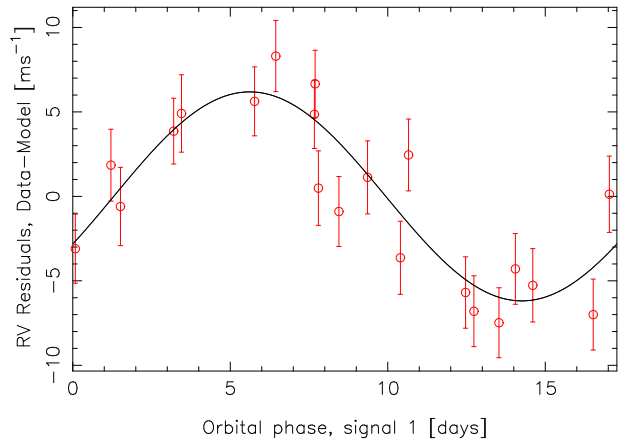}
\caption{HIRES radial velocities of GJ 49 folded on the phase of the detected signal (solid curve).}\label{fig:GJ49_curve}
\end{figure}

GJ 49 (HIP 4872) has been observed by HIRES a total of 21 times over a period of 5200 d. The HIRES data of this target was also analysed in \citet{butler2016} but no signals were reported based on their likelihood-ratio criterion. We observed a signal satisfying our detection criteria in the corresponding radial velocity data at a period of 17.272 [17.259, 17.286] d with an amplitude of 6.18 [3.60, 9.11] ms$^{-1}$. The probability maximum corresponding to this signal was reasonably unique in the period space (Fig. \ref{fig:GJ49_period_search}) with a local maximum at a period of 1.8 d. However, these two maxima are not independent as we could not fit a two-Keplerian model to the data in a meaningful way. Moreover, the local maximum coincides with a 2-day alias of the global maximum, which shows that the two are indeed related via aliasing. We have plotted the Keplerian curve together with the radial velocities in Fig. \ref{fig:GJ49_curve}

No ASAS data was available for GJ 49 and the HIRES S-indices showed no evidence for periodicities. We thus interpret the signal as a planet candidate orbiting the star with a minimum mass of 16.4 [8.6, 24.2] M$_{\oplus}$ and classify it as a hot Neptune.

\clearpage

\subsection{GJ 54.1}

GJ 54.1 (HIP 5643) has recently been observed intensively by HARPS and we obtained a set of 114 radial velocities by processing the HARPS data products that were available in the ESO archive. We combined these HARPS velocities with a set of 21 HIRES velocities and obtained evidence for two significant periodicities.

\begin{figure}
\center
\includegraphics[angle=270, width=0.49\textwidth,clip]{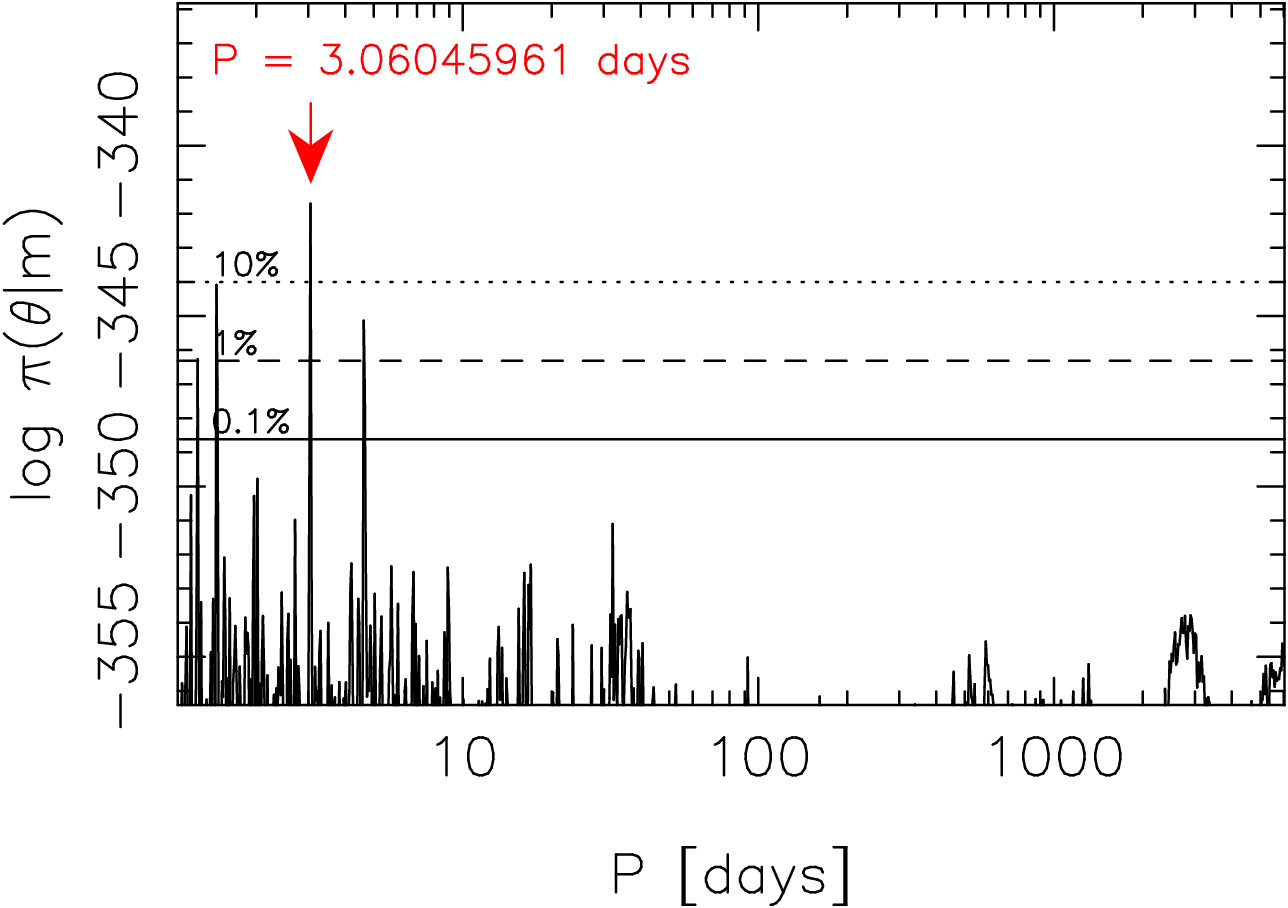}

\includegraphics[angle=270, width=0.49\textwidth,clip]{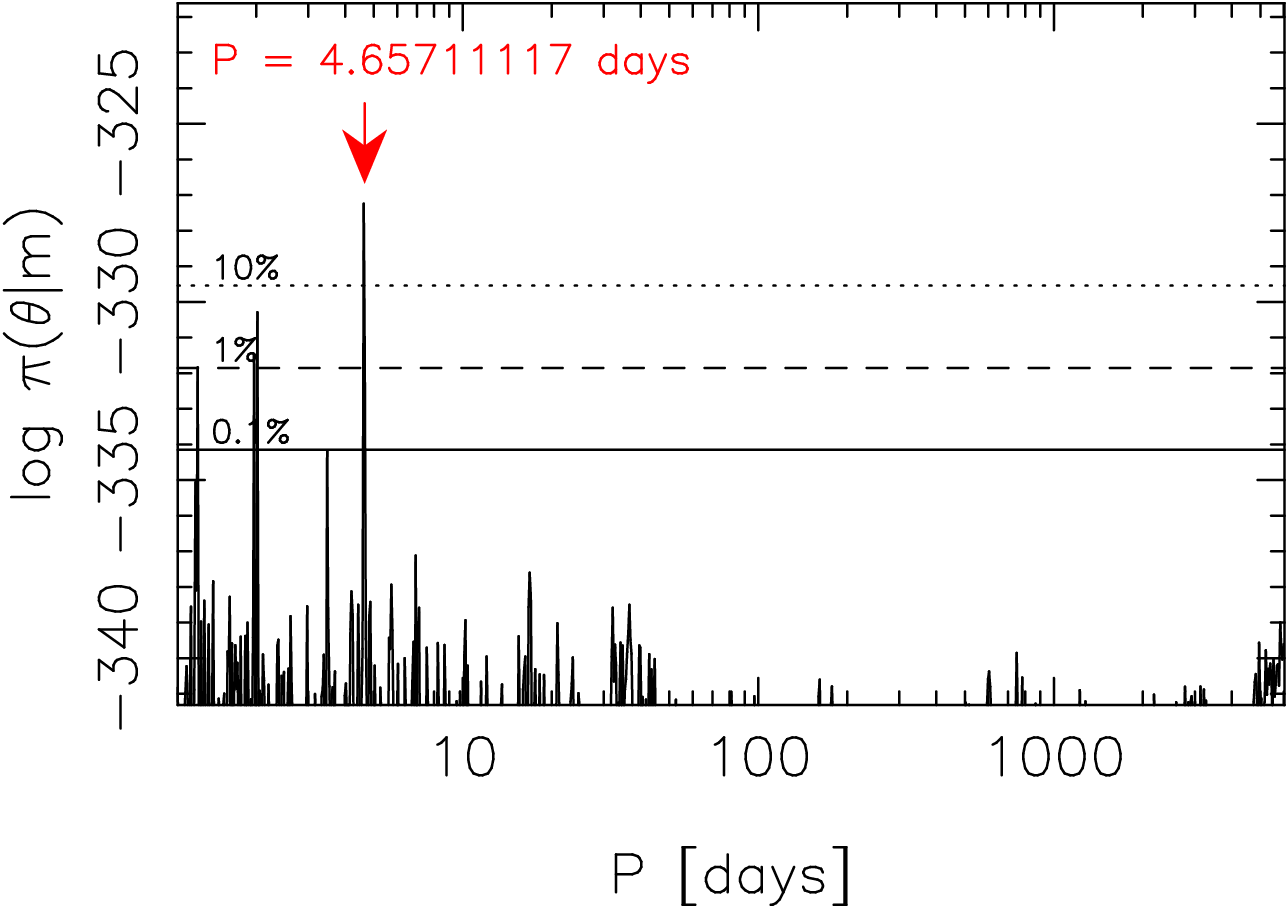}
\caption{Top (bottom) panel shows the estimated posterior probability density as a function of the period of the signal (second signal) in a one-Keplerian (two-Keplerian) model given the combined HARPS and HIRES radial velocities of GJ 54.1.}\label{fig:GJ54.1_period_search}
\end{figure}

\begin{figure}
\center
\includegraphics[angle=270, width=0.49\textwidth,clip]{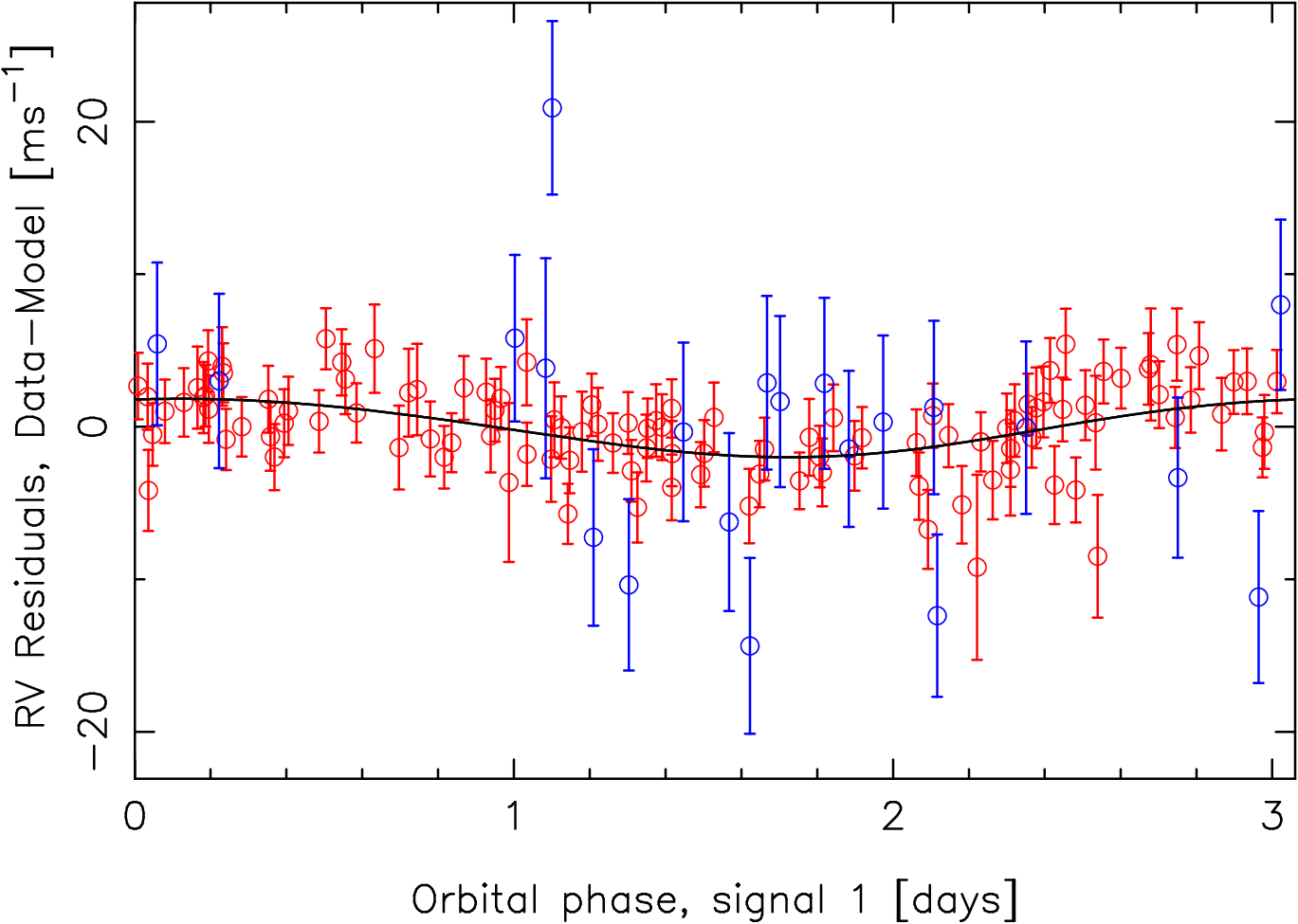}

\includegraphics[angle=270, width=0.49\textwidth,clip]{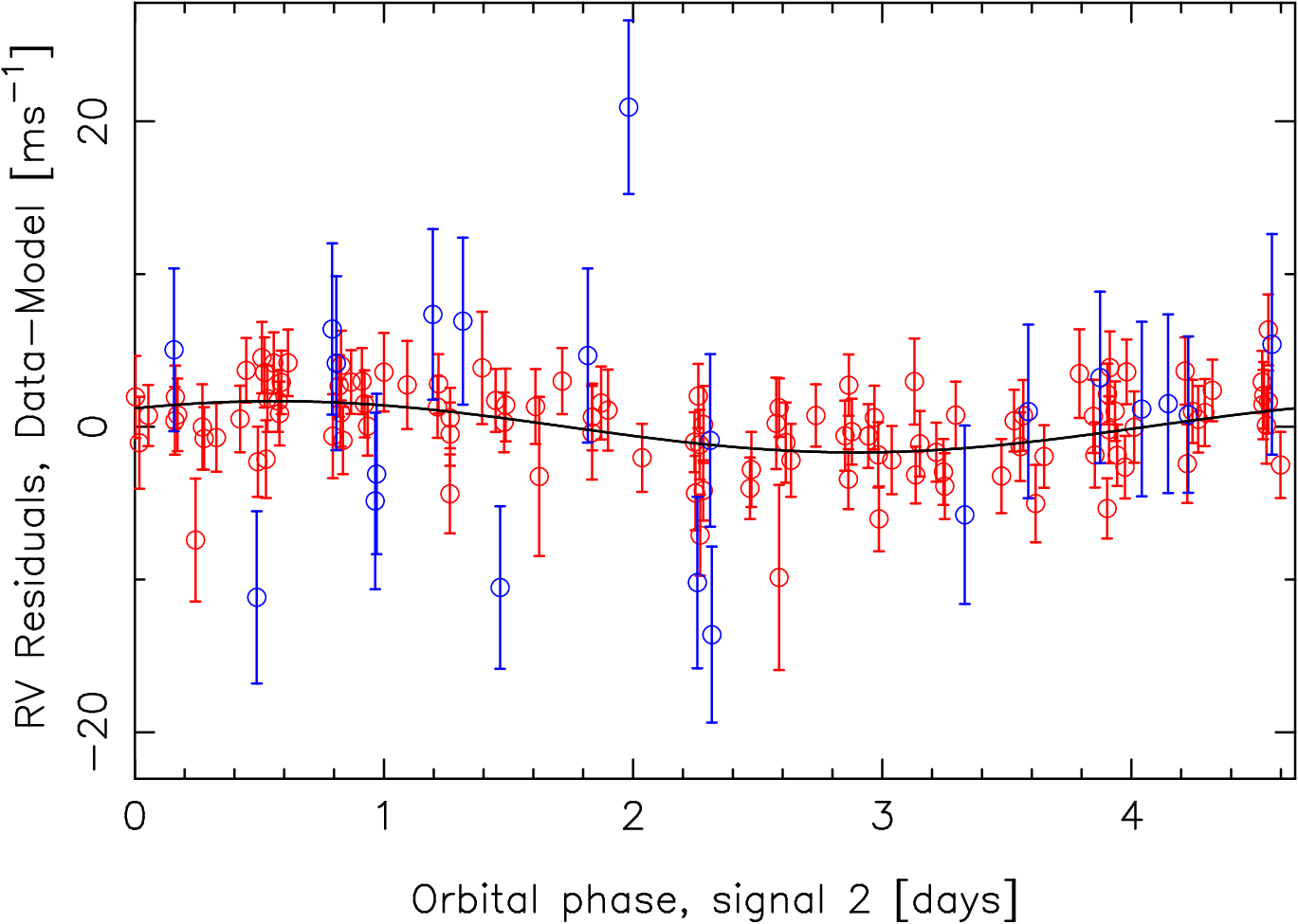}
\caption{HARPS (red) and HIRES (blue) radial velocities of GJ 54.1 folded on the phases of the two signals with the other signal subtracted from each panel.}\label{fig:GJ54.1_signals}
\end{figure}

The estimated posterior probability density of the one-Keplerian model (Fig. \ref{fig:GJ54.1_period_search}, top panel) revealed four maxima out of which we identified the global maximum at a period of 3.0604 [3.0597, 3.0613] days. The second most significant maximum was found at a period of 1.485 days and corresponds to the primary daily alias of the global maximum. The search for a second signal (Fig. \ref{fig:GJ54.1_period_search}, bottom panel) confirmed the presence of a second signal and its daily alias as a significant solution to the data. This second signal was found at a period of 4.6572 [4.6556, 4.6586] days and its alias at a period of 1.273 days. We performed additional searches for periodicities with one-, two- and three-Keplerian models by limiting the upper boundary of the period space to 12 days for better resolution. The two signals were detected uniquely and we also obtained hints for another potential signal at a period of 1.969 days together with its daily alias at a nearby period of 2.032 days. However, this third periodicity did not satisfy our detection criteria. We have plotted the radial velocities of GJ 54.1 folded on the phases of the signals in Fig. \ref{fig:GJ54.1_signals} for visual inspection.

\begin{figure}
\center
\includegraphics[angle=270, width=0.49\textwidth,clip]{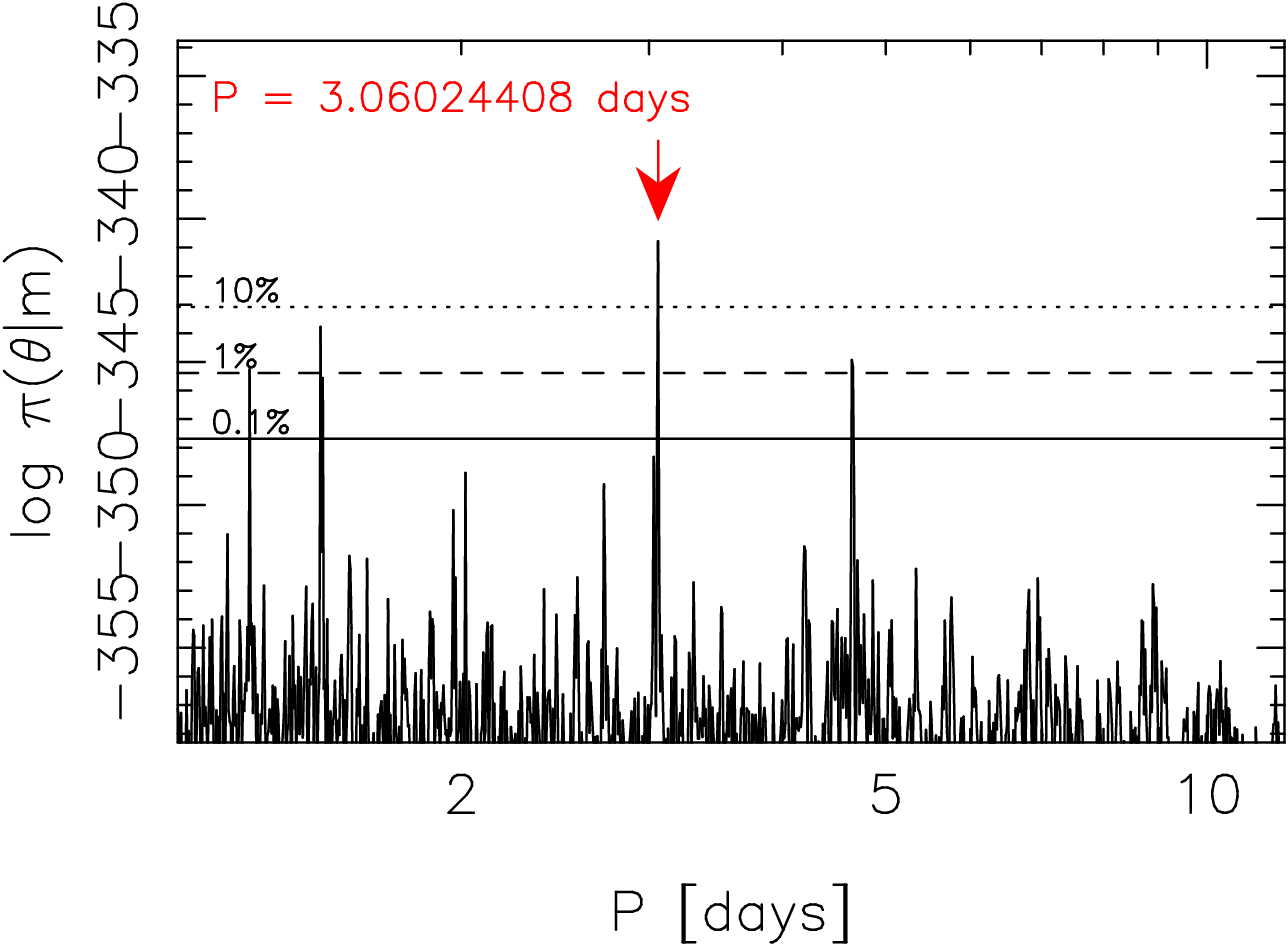}

\includegraphics[angle=270, width=0.49\textwidth,clip]{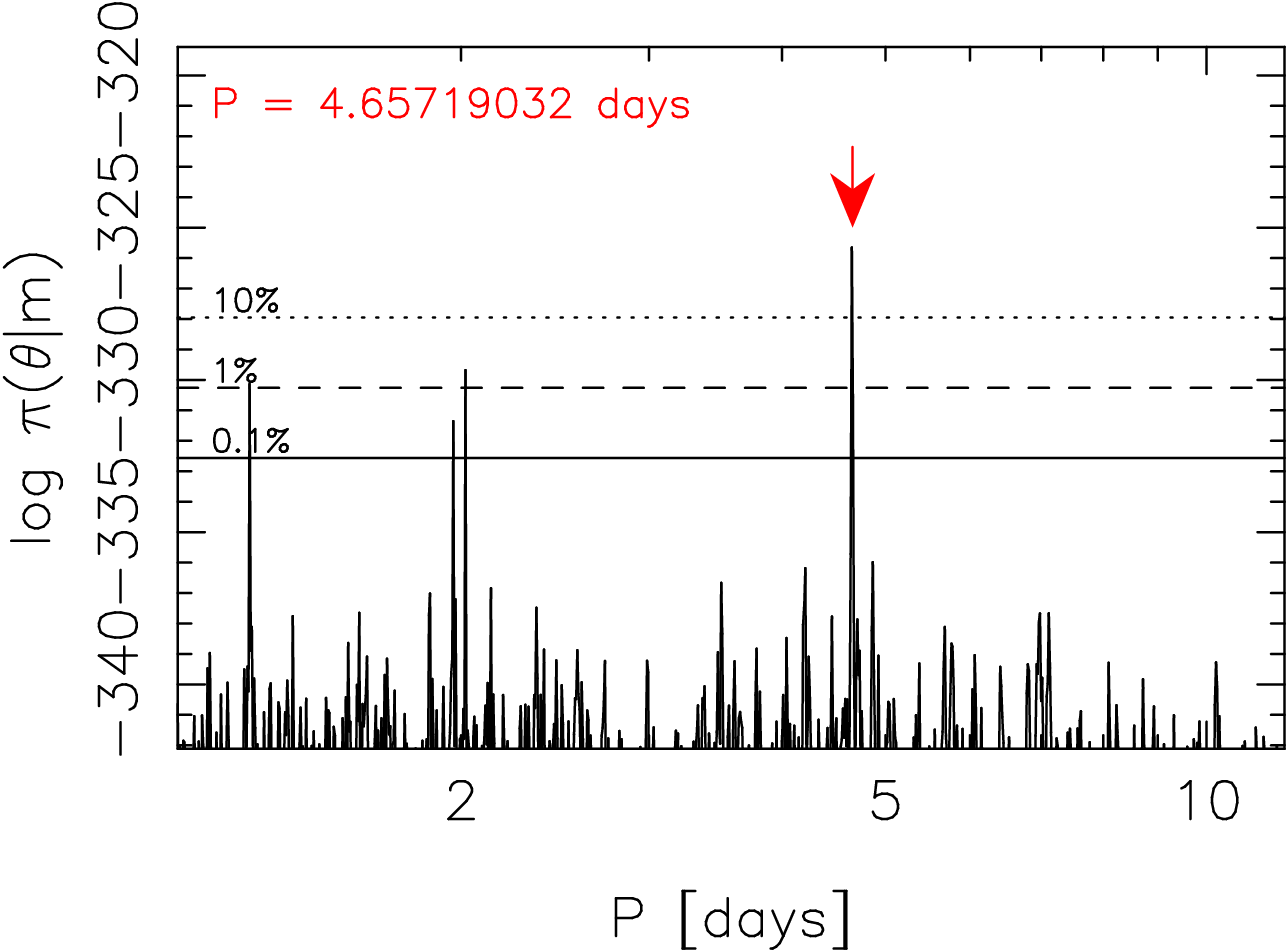}

\includegraphics[angle=270, width=0.49\textwidth,clip]{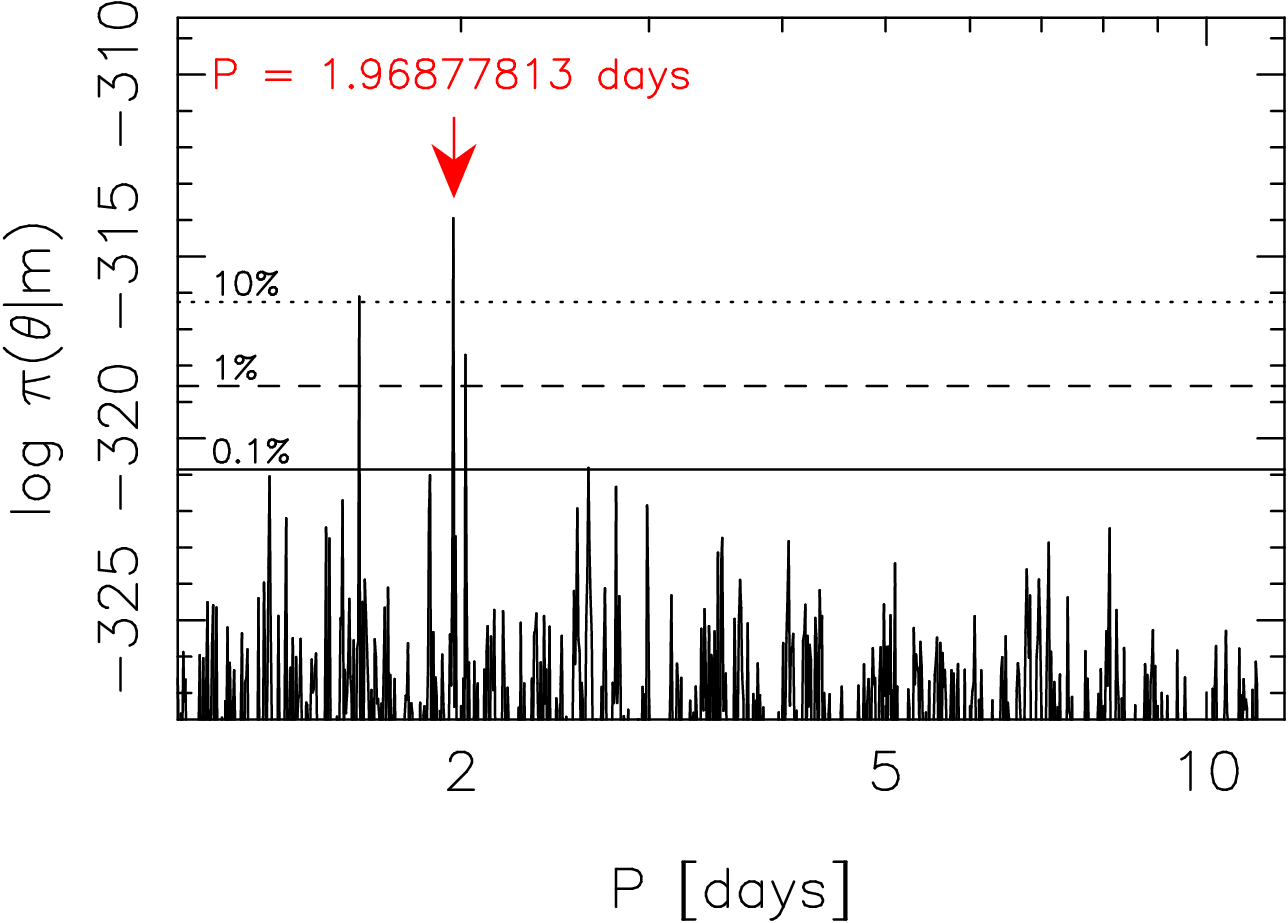}
\caption{As in Fig. \ref{fig:GJ54.1_period_search} but for models with 1-3 Keplerian signals with period space limited by an upper bound of 12 days.}\label{fig:GJ54.1_period_search2}
\end{figure}

\begin{figure}
\center
\includegraphics[angle=270, width=0.49\textwidth,clip]{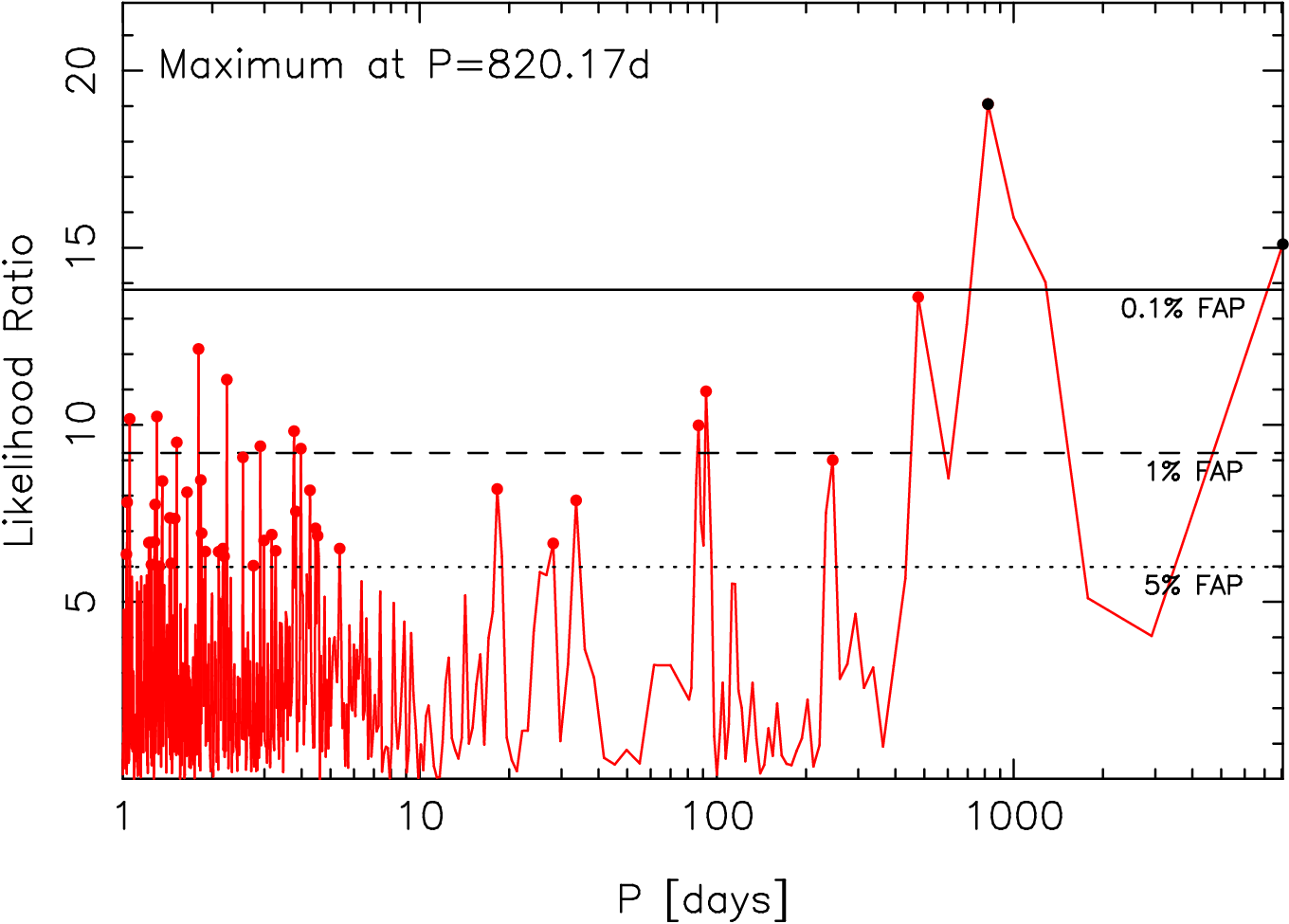}

\includegraphics[angle=270, width=0.49\textwidth,clip]{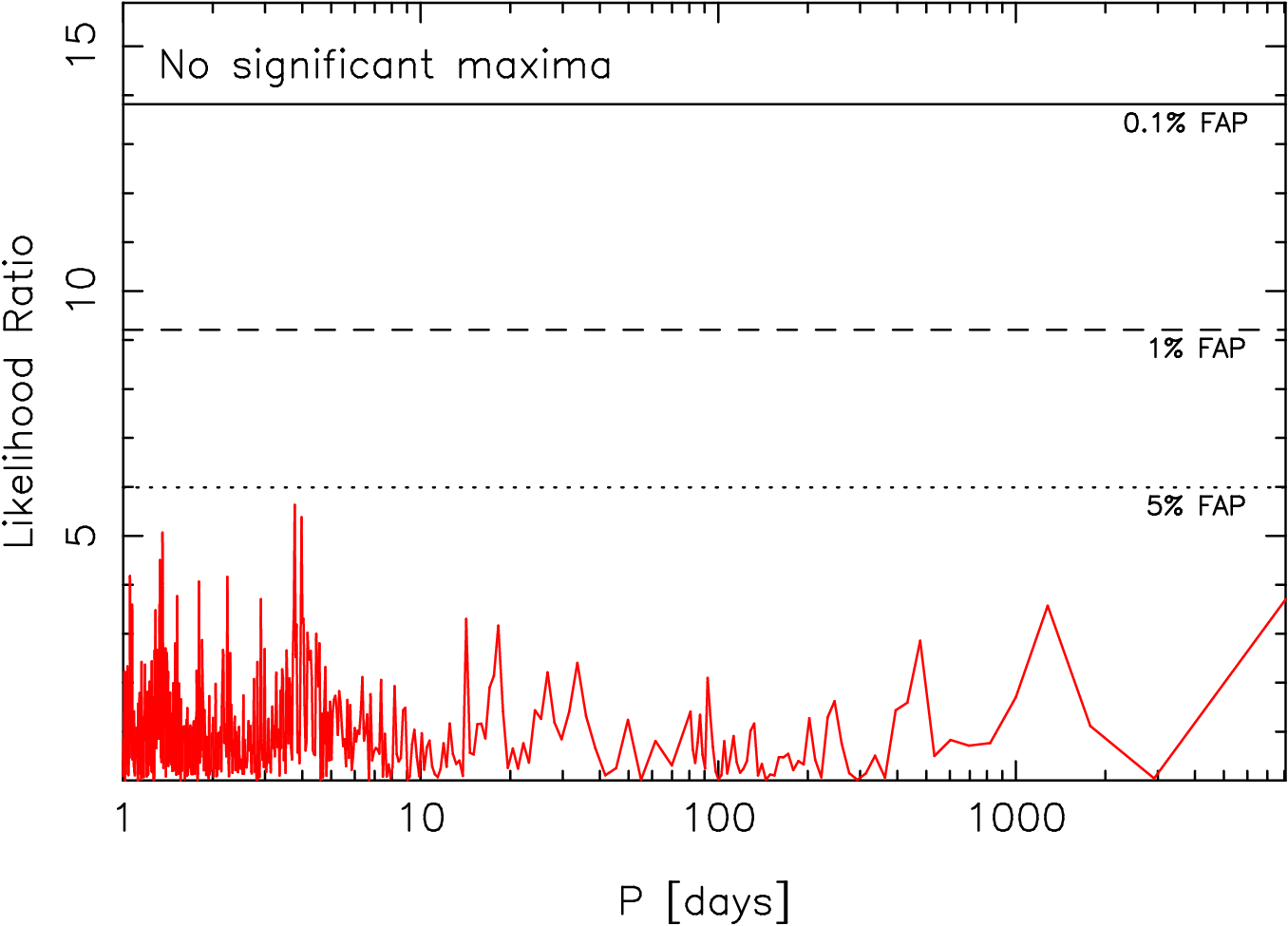}
\caption{Likelihood-ratio periodogram of the HARPS S-indices of GJ 54.1 (top panel) and the periodogram of residuals after subtracting the signal corresponding to the highest ratio (bottom panel).}\label{fig:GJ54.1_S}
\end{figure}

We searched for periodicities in the HARPS activity indicators of GJ 54.1. The only index to provide evidence for periodicities in excess of the 0.1\% FAP was the HARPS S-index that was found to have a signal at a period of 820 days (Fig. \ref{fig:GJ54.1_S}). However, there were no signals at short periods and we could thus not find any evidence for counterparts of the signals in the radial velocities.

\begin{figure}
\center
\includegraphics[angle=270, width=0.49\textwidth,clip]{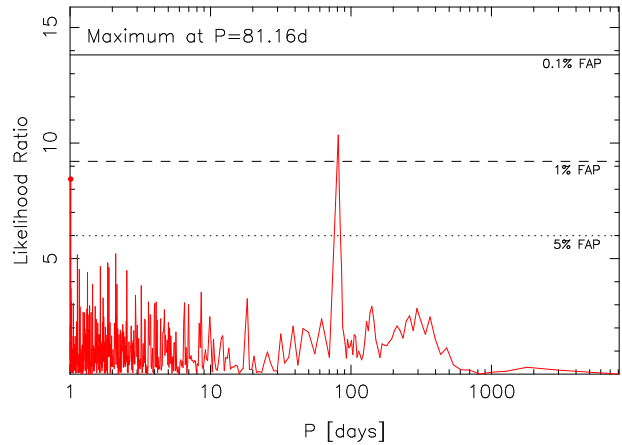}
\caption{Likelihood-ratio periodogram of the HARPS FWHM values of GJ 54.1.}\label{fig:GJ54.1_FWHM}
\end{figure}

We also found a signal in the HARPS FWHM values at a period of 81.16 days but this signal is only suggestive as it did not exceed the 0.1\% FAP threshold (Fig. \ref{fig:GJ54.1_FWHM}). However, given that there is a photometric periodicity in the ASAS data of GJ 54.1 at a reasonably close period of 71.72 days (Fig. \ref{fig:asas_periodograms}) with nearby powers from 70-90 days, we interpret the suggestive signal in the FWHM as a signature of the stellar (differential) rotation and conclude that the short period signals in the radial velocities are unlikely to be connected to rotation or other periodic phenomena of the stellar surface.

The two unique signals in the radial velocities of GJ 54.1 at periods of 3.06 and 4.66 days have amplitudes of 1.99 [1.04, 2.94] and 1.66 [0.74, 2.66] ms$^{-1}$, respectively. They thus correspond to Keplerian signals of candidate planets with minimum masses of 1.2 [0.5, 1.8] and 1.1 [0.4, 1.8] M$_{\oplus}$ that are classified as hot Earths -- they orbit the star, an M4 dwarf, just inside the estimated inner boundary of the liquid-water habitable zone.

\clearpage

\subsection{GJ 69}

\begin{figure}
\center
\includegraphics[angle=270, width=0.49\textwidth,clip]{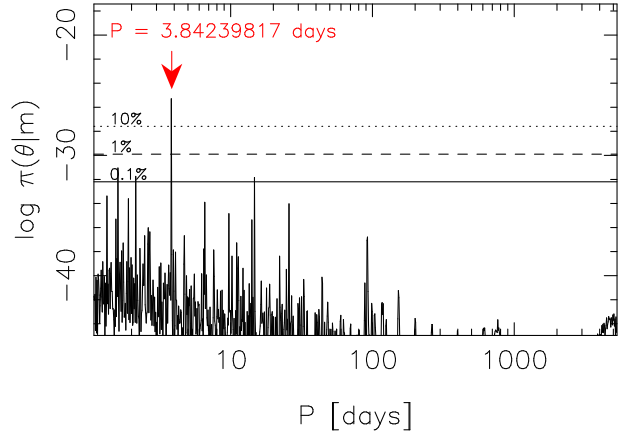}
\caption{Estimated posterior probability density as a function of the period parameter given the radial velocities of GJ 69. The red arrow indicates the global maximum at a period of 3.84 days and the horizontal lines denote the 10\% (dotted), 1\% (dashed), and 0.1\% (solid) equiprobability thresholds with respect to the global maximum.}\label{fig:GJ69_period_search}
\end{figure}

\begin{figure}
\center
\includegraphics[angle=270, width=0.49\textwidth,clip]{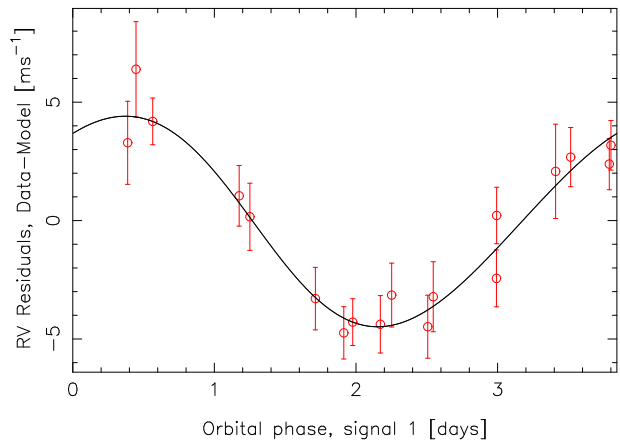}
\caption{HIRES radial velocities of GJ 69 folded on the phase of the signal.}\label{fig:GJ69_phased}
\end{figure}

We observed a short-period signal in the HIRES velocities of GJ 69 (HD 10436, HIP 8070). This signal, at a period of 3.84237 [3.84183, 3.84322] days and with an amplitude of 4.41 [2.92, 5.89] ms$^{-1}$, was found according to our signal detection criteria despite the fact that there were only 18 HIRES radial velocities available. Yet, the probability maximum corresponding to the signal in the period space was unique without local maxima exceeding the 1\% probability threshold of the global maximum (Fig. \ref{fig:GJ69_period_search}). We have plotted the HIRES radial velocities in Fig. \ref{fig:GJ69_phased} after folding them on the phase of the signal. Additional signals were not identified in the data.

The radial velocity signal did not have counterparts in the HIRES S-indices. Although ASAS photometry data was not available for GJ 69, we have no evidence suggesting that the signal would be caused by stellar activity cycles or rotation rather than a candidate planet. We thus interpret the signal as evidence in favour of a hot mini-Neptune with a minimum mass of 8.3 [5.0, 12.0] M$_{\oplus}$ orbiting the star.

\newpage

\subsection{GJ 83.1}

\begin{figure}
\center
\includegraphics[angle=270, width=0.49\textwidth,clip]{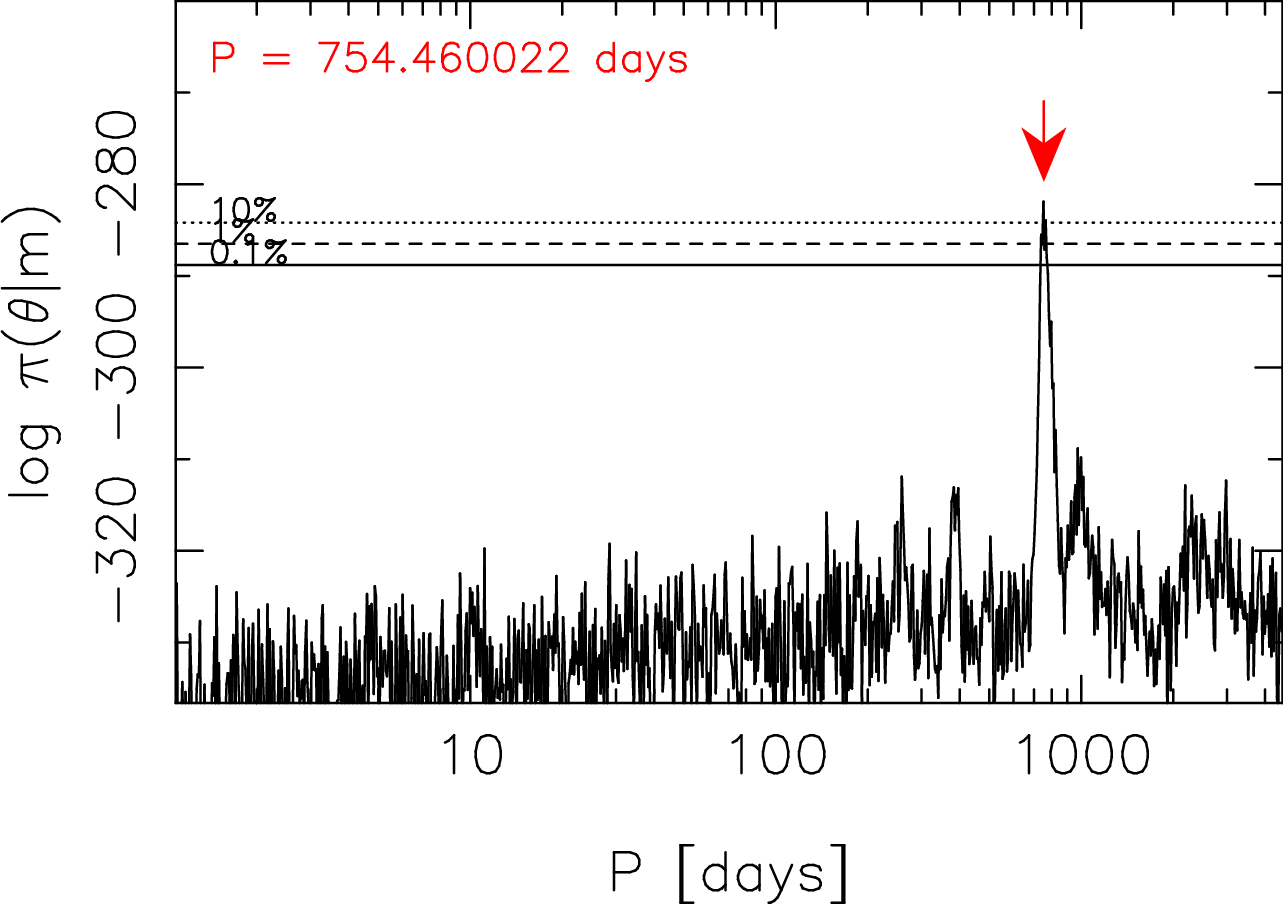}

\includegraphics[angle=270, width=0.49\textwidth,clip]{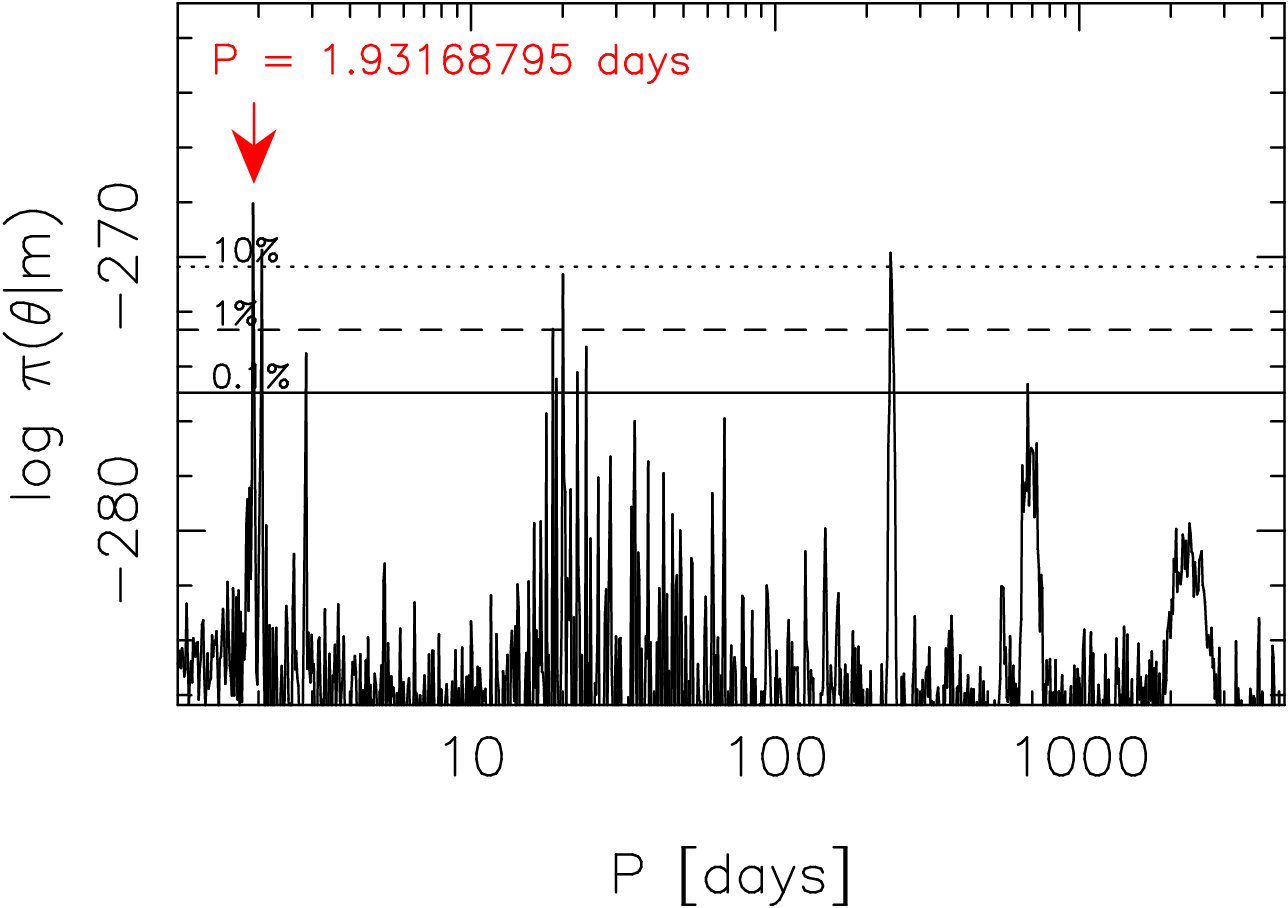}

\includegraphics[angle=270, width=0.49\textwidth,clip]{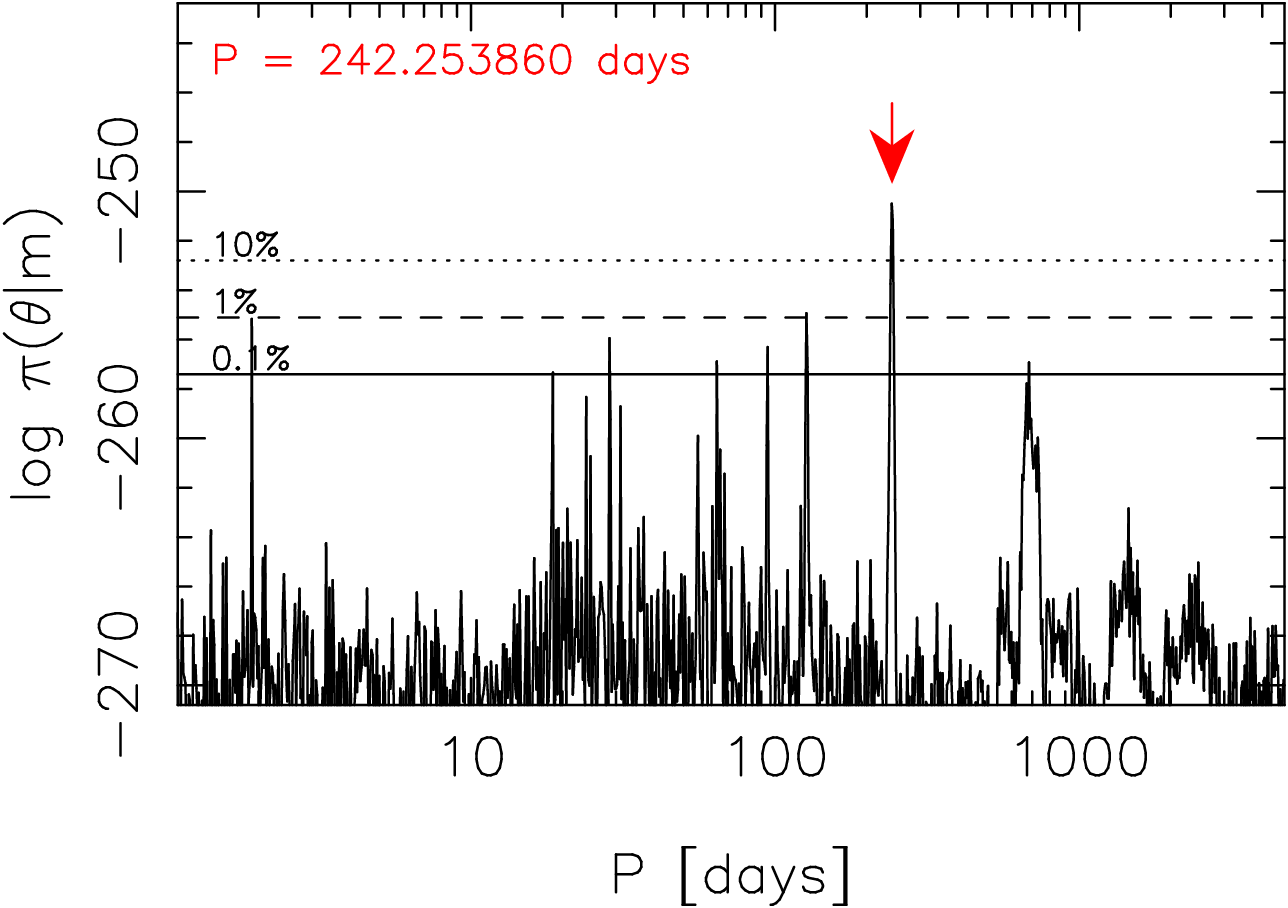}
\caption{Estimated posterior probability densities of models with $k$ Keplerian signals as a function of the period of the $k$th signal for $k=1$, $k=2$, and $k=3$ (from top to bottom). The red arrows denote the global maxima in the period space and the horizontal lines show the 10\% (dotted), 1\% (dashed), and 0.1\% (solid) equiprobability thresholds with respect to the maxima.}\label{fig:GJ83.1_psearch}
\end{figure}

\begin{figure}
\center
\includegraphics[angle=270, width=0.49\textwidth,clip]{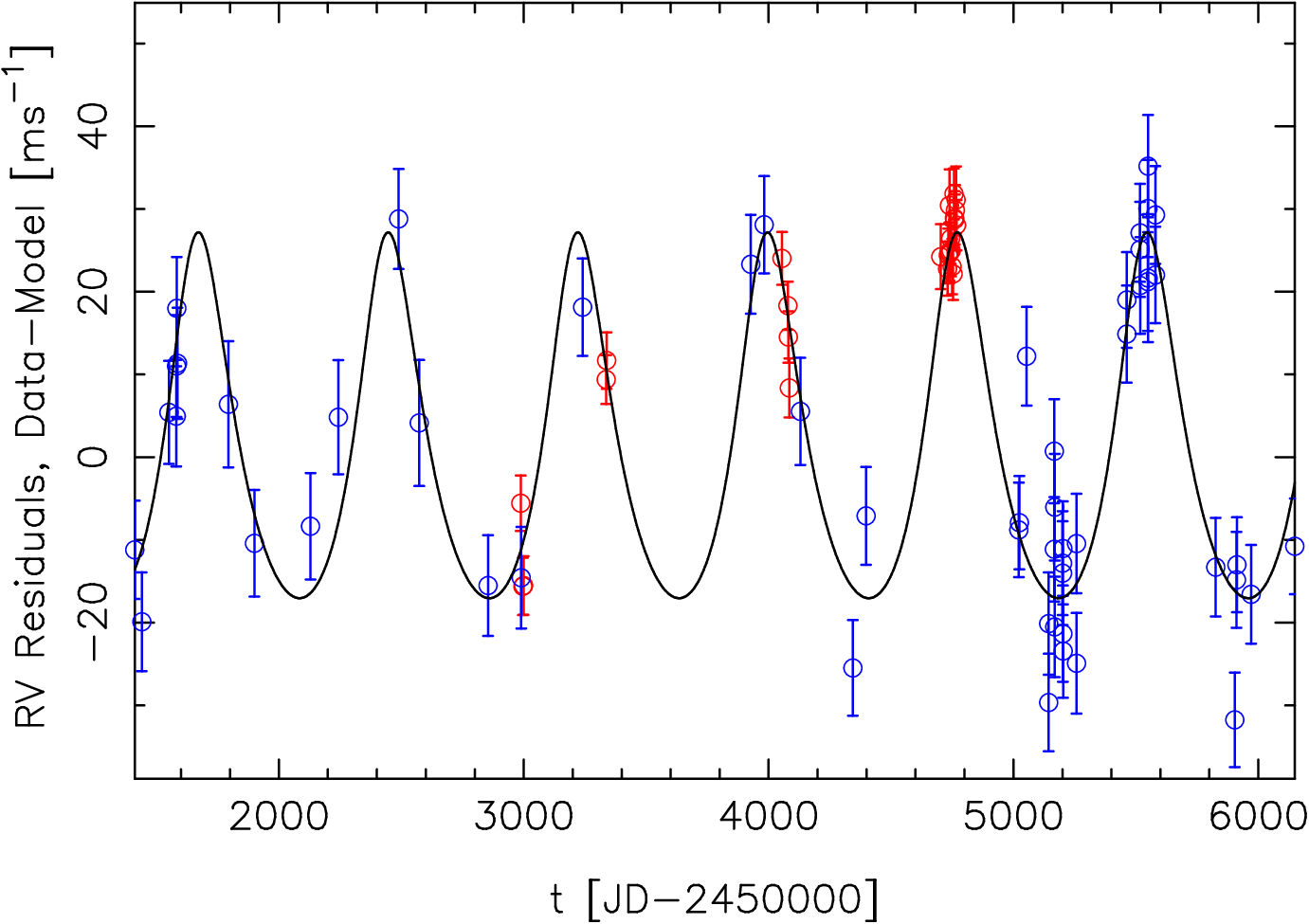}

\includegraphics[angle=270, width=0.49\textwidth,clip]{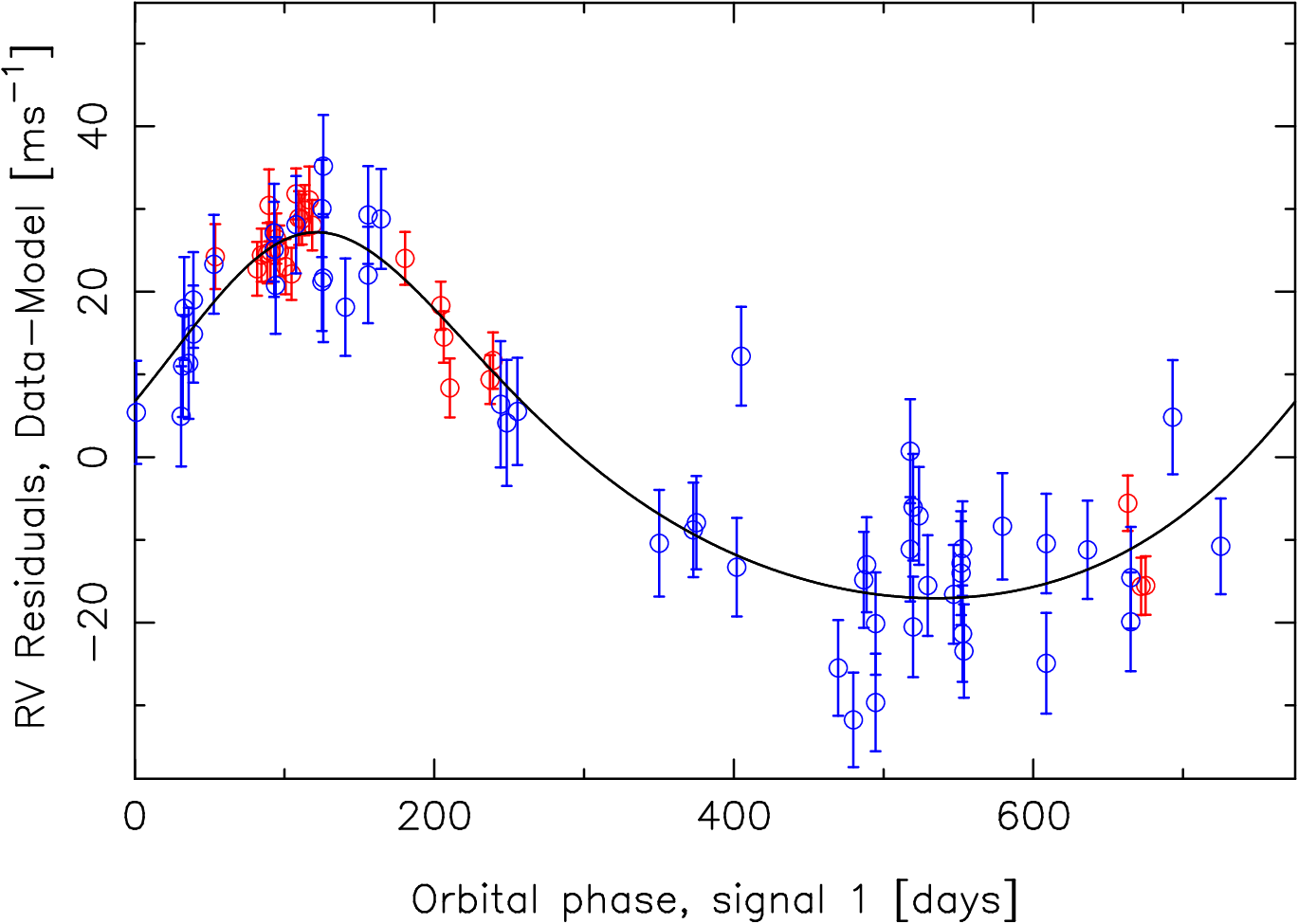}
\caption{HARPS (red) and HIRES (blue) radial velocities of GJ 83.1 modelled with a Keplerian function corresponding to the 773-day candidate planet (solid curve, top panel). Bottom panel shows the phase-folded velocities of this candidate.}\label{fig:GJ83.1_curve}
\end{figure}

As also suggested by \citet{butler2016} based on HIRES data alone, the combined HARPS and HIRES radial velocities of GJ 83.1 support strongly the existence of a giant planet with a minimum mass of 135 [96, 174] M$_{\oplus}$ orbiting the star with a period of 773.4 [756.9, 789.9] days. The corresponding signal, with an amplitude of 20.28 [15.99, 24.57] ms$^{-1}$ was easily detected as a probability maximum in the period space by using our DRAM samplings of the parameter space (Fig. \ref{fig:GJ83.1_psearch}). It is clear that the signal is supported mainly by the HIRES velocities because HARPS data has rather poor phase-coverage as well as much more limited baseline (Fig. \ref{fig:GJ83.1_curve}). However, we also obtained evidence for additional signals in the combined HARPS and HIRES data of GJ 83.1.

\begin{figure}
\center
\includegraphics[angle=270, width=0.49\textwidth,clip]{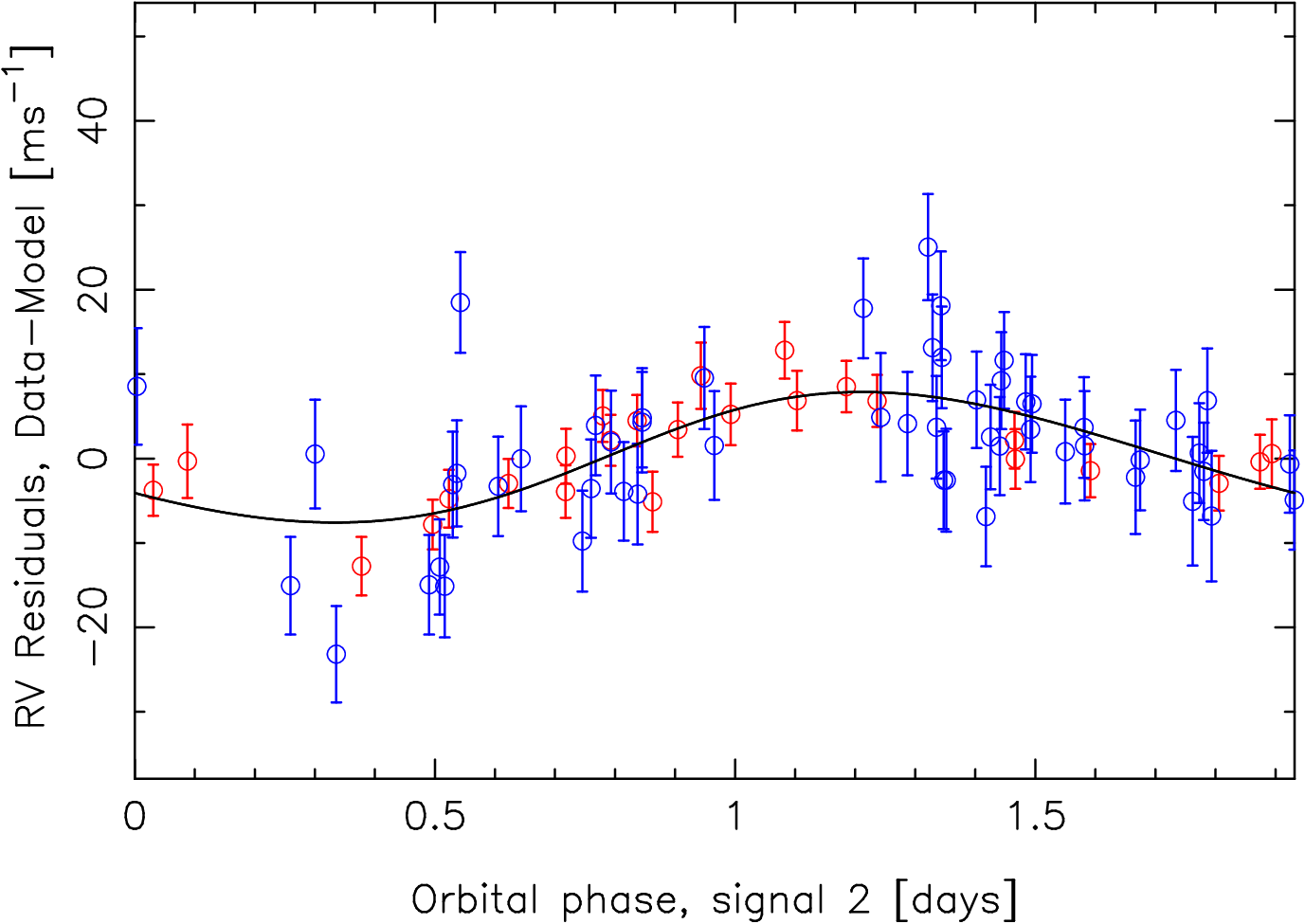}

\includegraphics[angle=270, width=0.49\textwidth,clip]{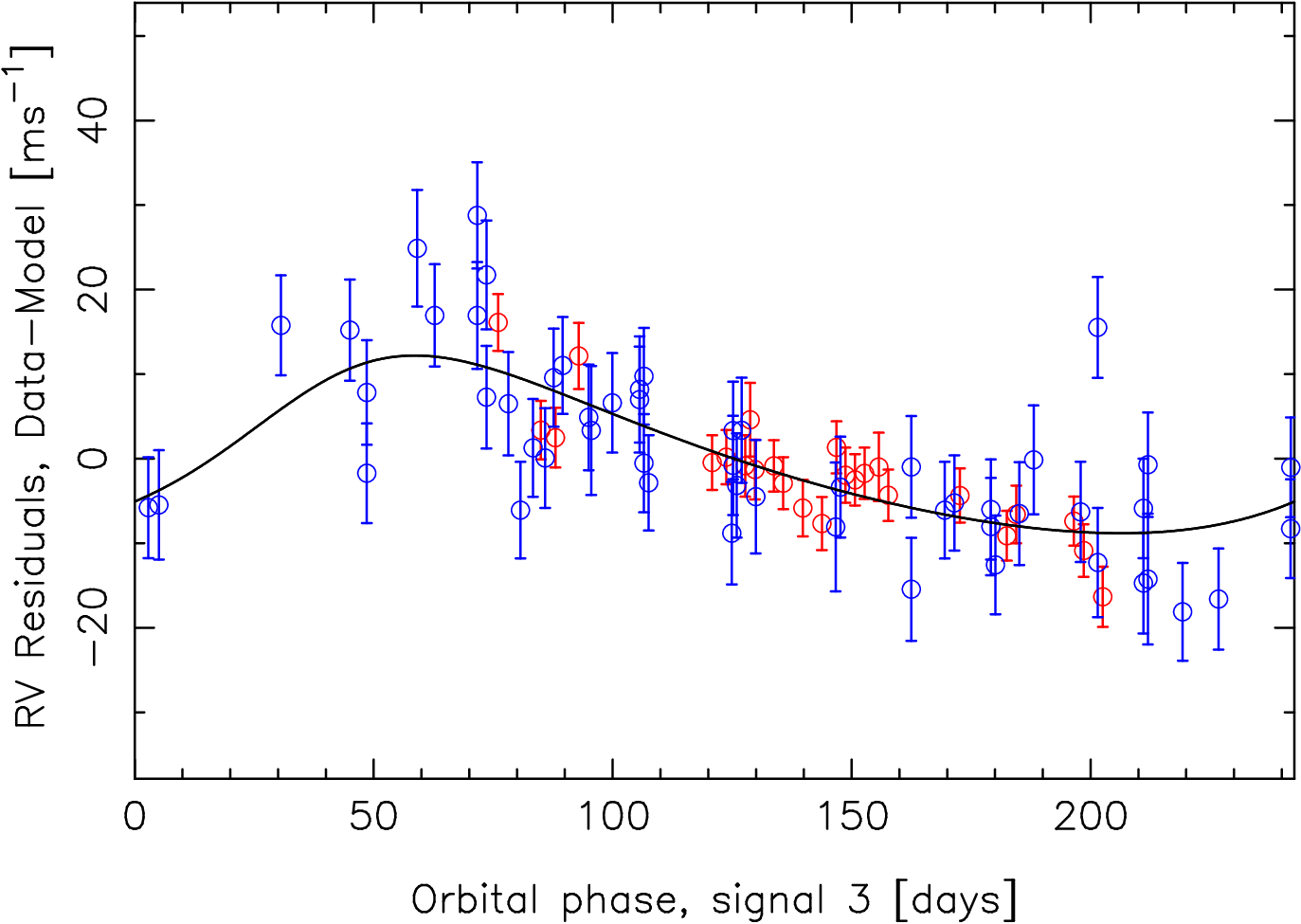}
\caption{HARPS (red) and HIRES (blue) radial velocities of GJ83.1 folded on the phases of the two weaker signals with the other signals subtracted.}\label{fig:GJ83.1_signals}
\end{figure}

As can be seen in Fig. \ref{fig:GJ83.1_psearch} (middle panel), searches for a second periodic signal enabled us to identify three prominent maxima in the period space. The most significant of these, at a period of 1.93 days, was found to satisfy our signal detection criteria. As can be seen in Fig. \ref{fig:GJ83.1_psearch} (middle panel), there is also a clearly visible daily alias of this signal at a period of 2.07 days. Moreover, a third signal at a period of 240 days was significant according to our criteria as well. Conversely, probability maxima corresponding to a signal with a period of roughly 20 days was not significant as it could not be constrained in the period and amplitude spaces due to the existence of several nearby local maxima. This was not the case for the signals at periods of 1.93 and 240 days that were rather isolated in the period space (Fig. \ref{fig:GJ83.1_psearch}) and thus unique as Keplerian signals should be. It is thus our interpretation that there are additional signals at periods of 1.93177 [1.93156, 1.93194] and 243.1 [239.1, 248.3] days (Fig. \ref{fig:GJ83.1_signals}) that correspond to a hot super-Earth and a cool Neptune with minimum masses of 4.0 [2.2, 6.0] and 26.0 [12.8, 42.8] M$_{\oplus}$, respectively.

We did not observe any periodicities in either the ASAS photometry or the activity indices of HARPS or HIRES. We thus interpret the results such that GJ83.1 is a host to a diverse system of three candidate planets.

\clearpage

\subsection{GJ 160.2}

GJ 160.2 (HIP 19165) is also a target that was reported to have a signal by \citet{tuomi2014} and interpreted as a planet candidate with an orbital period of 5.24 d. This result was based on 100 UVES and 7 HARPS radial velocities. Considering that UVES data might be prone to biases (see Section \ref{sec:UVES}), this result is therefore subject to doubt as the combined data set is dominated by UVES.

\begin{figure}
\center
\includegraphics[angle=270, width=0.49\textwidth,clip]{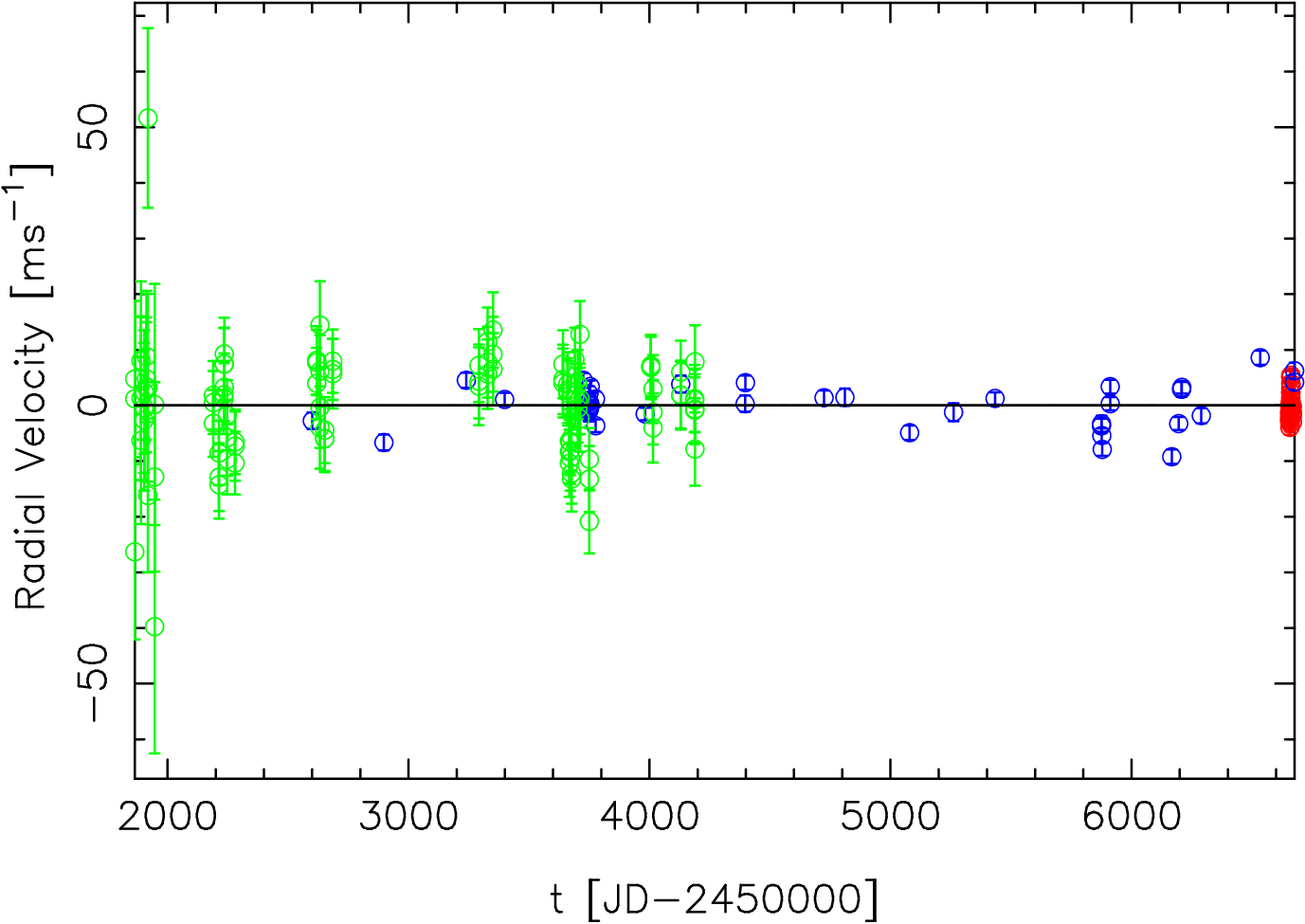}

\includegraphics[angle=270, width=0.49\textwidth,clip]{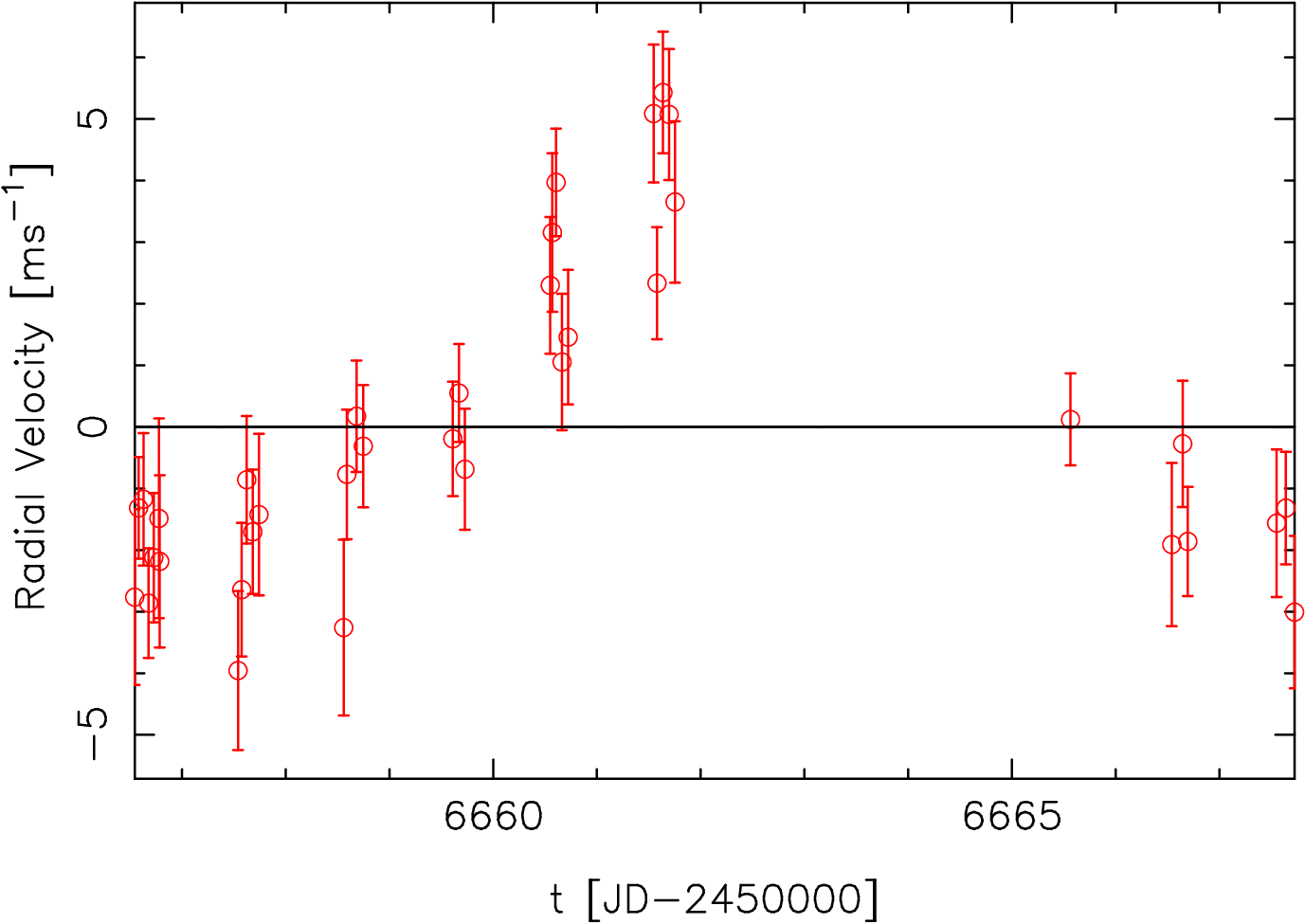}
\caption{Top panel: HARPS (red), HIRES (blue), and UVES (green) radial velocity data of GJ 160.2. Bottom panel shows the HARPS data only.}\label{fig:GJ160.2_data}
\end{figure}

To verify the existence of the signal reported by \citet{tuomi2014}, we obtained 36 new HARPS velocities and also 38 HIRES radial velocities, although the majority of the latter velocities had been observed prior to the discovery of the signal by \citet{tuomi2014}. Moreover, as we accounted for the linear dependence of the radial velocities on the activity indices, we had to remove the 7 early HARPS velocities from the analyses as the corresponding activity indices suggested spectral contamination and/or low S/N ratios due to having FWHM 2.0 kms$^{-1}$ in excess of the data median of 4.2 kms$^{-1}$ despite the deviation of only 0.018 kms$^{-1}$ for the new HARPS velocities. Moreover, the corresponding BIS values were similarly indicative of being drawn from a different population making the first seven HARPS observations of suspect. The combined data and the resulting HARPS velocities with a baseline of 11 d are shown in Fig. \ref{fig:GJ160.2_data}. Even rudimentary visual inspection of the HARPS velocities suggests that there are variations in the data that cannot be explained as pure noise.

\begin{figure}
\center
\includegraphics[angle=270, width=0.49\textwidth,clip]{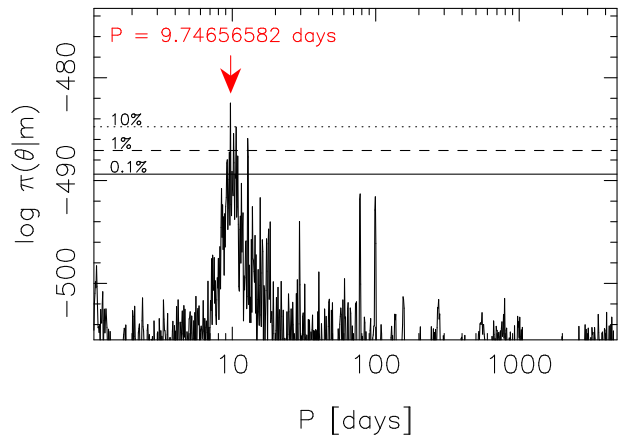}
\caption{As in Fig. \ref{fig:GJ27.1_period_search} but for the signal in the combined HARPS, HIRES, and UVES velocities of GJ 160.2.}\label{fig:GJ160.2_period_search}
\end{figure}

This suggestion can also be verified by analysing the data and observing that there is a strong, albeit rather broad and splintered, probability maximum at a period of 9.7 days with an almost equally high local maximum at a period of 10.4 days (Fig. \ref{fig:GJ160.2_period_search}). There are also several local maxima next to these two maxima because the HARPS data only cover roughly the phase of the signal making the estimation of the exact period rather difficult. We thus adopt the global maximum as our solution while keeping in mind that this is a subjective choice. The splicing of the maximum is again caused by the less-than-optimal sampling of the three data sets (Fig. \ref{fig:GJ160.2_data}, top panel). 

\begin{figure}
\center
\includegraphics[angle=270, width=0.49\textwidth,clip]{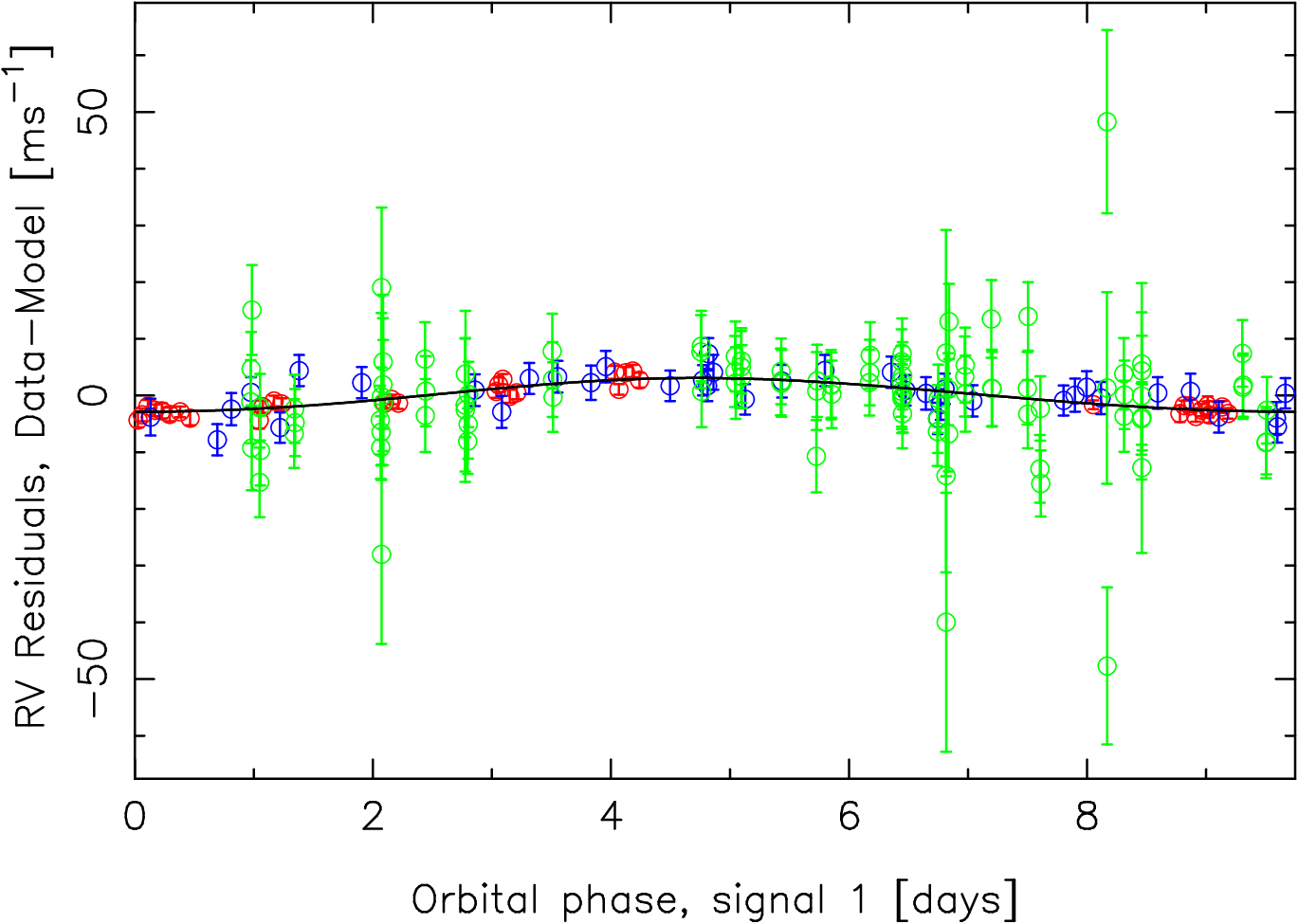}

\includegraphics[angle=270, width=0.49\textwidth,clip]{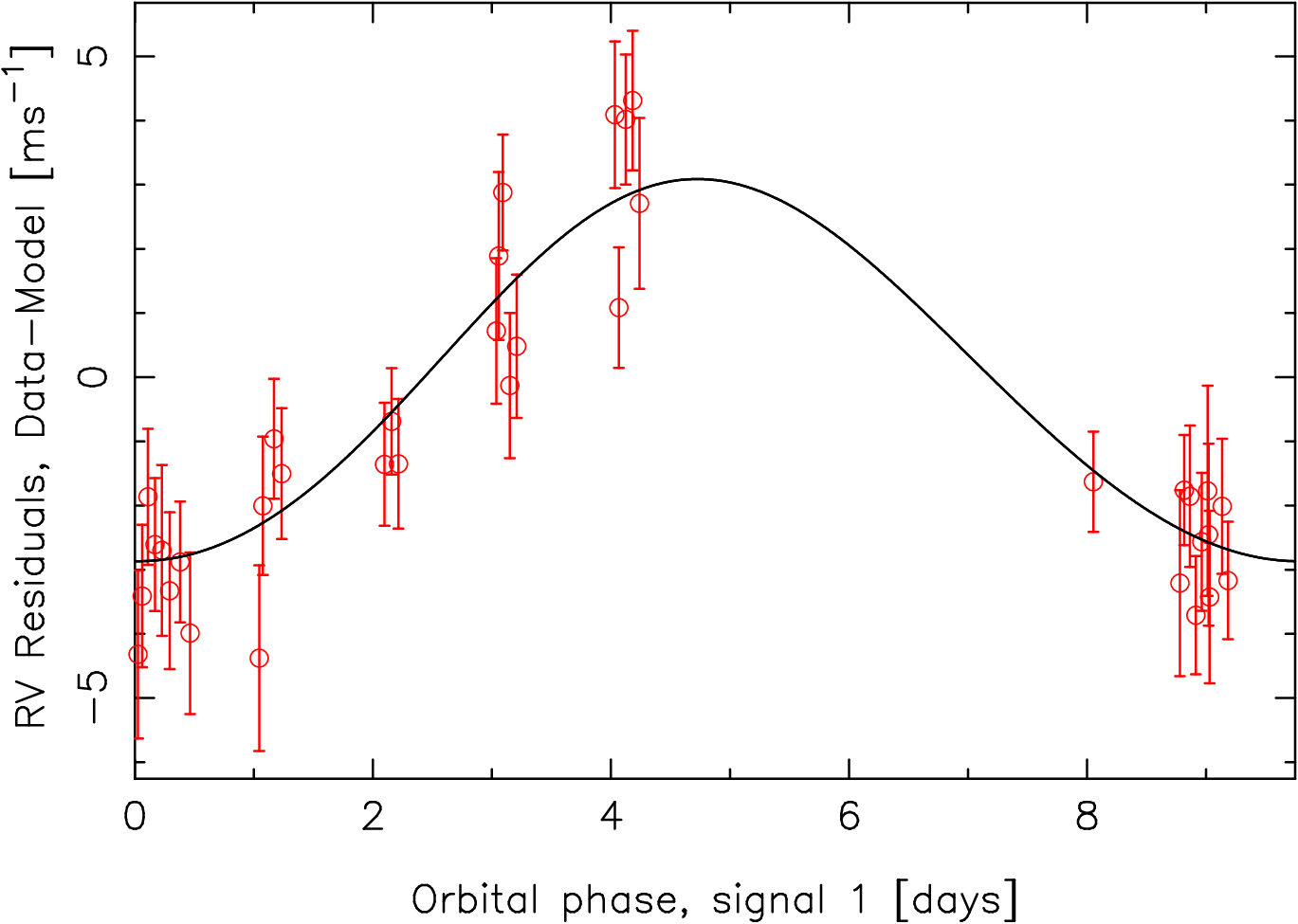}
\caption{Top panel: HARPS (red), HIRES (blue), and UVES (green) radial velocity data of GJ 160.2 folded on the period of the signal. Bottom panels shows the HARPS data alone.}\label{fig:GJ160.2_signal}
\end{figure}

However, the existence of the signal is well established by the available data and detected in accordance with our criteria. We have plotted the phase-folded signal in Fig. \ref{fig:GJ160.2_signal}. Moreover, although the presence of the signal is most obvious visually in HARPS data, all three data sets support its existence when comparing the maximum likelihood values of each data set given a model with a Keplerian signal and a reference model without it. The natural logarithms of the maximum likelihood values increase from -67.6, -93.4 and -342.3 to -50.2, -90.6 and -339.7 for HARPS, HIRES, and UVES, respectively, when including the signal in the model.

\begin{figure}
\center
\includegraphics[angle=270, width=0.49\textwidth,clip]{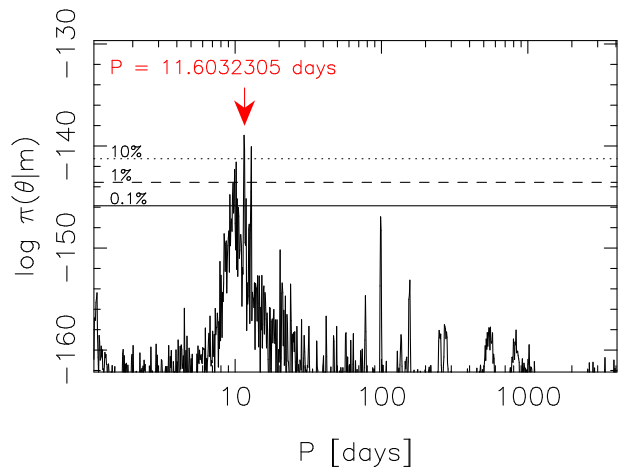}
\caption{As in Fig. \ref{fig:GJ160.2_period_search} but for the signal in the combined HARPS and HIRES radial velocities of GJ 160.2.}\label{fig:GJ160.2_period_search2}
\end{figure}

Finally, due to possible biases in the UVES data, we analysed the combined HARPS and HIRES data in order to see whether the UVES velocities altered the solution. This time the highest probability maximum was found at a period of 11.6 days but the obtained result was consistent with the presence of a signal due to the presence of almost equally high local maxima at a period of 9.7 days (Fig. \ref{fig:GJ160.2_period_search2}). This indicates that even though we do not know the exact period of the signal, it is present in the data between 9 and 12 days.

We obtained 551 ASAS V-band measurements and analysed them in order to see whether the signal at a period of 9.7 days has a photometric counterpart. These measurements were found to contain suggestive evidence in favour of a signal at a period of 89.46 days that might correspond to the stellar rotation, although the signal is not significant enough to qualify as photometric rotation period of  the star and is thus omitted from Table \ref{tab:kiraga_rotations}. We failed to detect any signals at or near the period of the radial velocity signal. We also failed to detect any signals in the activity indices of the HARPS data disabling us to find a connection between the radial velocity variability interpreted as a signal and variations in the activity indicators. Although HIRES velocities and S-indices were found to be linearly connected with 99\% credibility, whether we accounted for this connection or not did not change the results qualitatively and only slightly quantitatively. We thus interpret the signal with a period of 9.7471 [9.7420, 9.7514] days and an amplitude of 3.01 [1.93, 4.08] ms$^{-1}$ as being caused by a candidate planet with a minimum mass of 7.8 [4.7, 10.9] M$_{\oplus}$ orbiting the star that we thus classify as a hot mini-Neptune. The aforementioned uncertainty regarding the exact orbital period is unlikely to affect the conclusions regarding the occurrence rate of planets around M dwarfs, which is the main subject of the current work.

We note that the solution we obtained based on the new HARPS and HIRES radial velocities is different from that reported by \citet{tuomi2014}. The estimated orbital period is almost twice the one reported by \citet{tuomi2014}, which can result from poor data sampling and lack of continuous phase-coverage of the HARPS data that we have obtained with a high-cadence observing run. It can also be caused by the problems in the first 7 HARPS measurements that we removed from the full HARPS data set due to the anomalies in the FWHM and BIS values.

\clearpage

\subsection{GJ 163}\label{sec:GJ163}

GJ 163 (HIP 19394) has been reported to be a host to a system of ``up to four'' planets \citep{tuomi2013c} based on a set of 55 HARPS radial velocities, although a three-Keplerian solution was the more robust solution. This result was subsequently confirmed with roughly three times more HARPS velocities ($N = 150$) by \citet{bonfils2013b}. This example thus serves to demonstrate the higher sensitivity of HARPS-TERRA data reduction \citep{anglada2012b} and the posterior samplings as signal search technique applied in the current work in comparison to the traditional HARPS Cross Correlation Function (CCF) reduction and simple periodogram analyses of the velocities. We could now obtain a set of 170 HARPS radial velocities from the ESO arhive advanced data products with the TERRA algoritm. The evidence for the planets orbiting GJ 163 is thus ripe for a revisit.

\begin{figure}
\center
\includegraphics[angle=270, width=0.49\textwidth,clip]{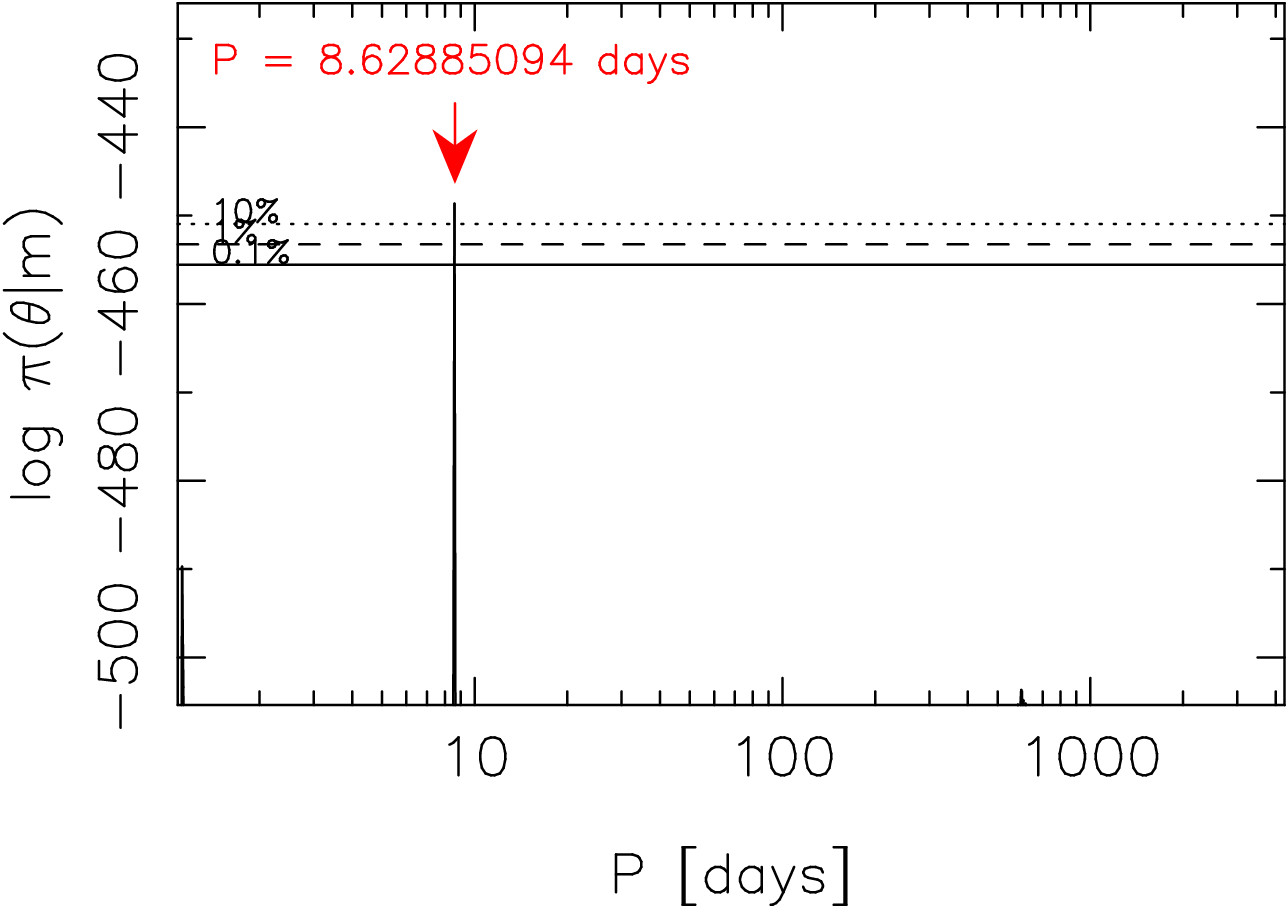}

\includegraphics[angle=270, width=0.49\textwidth,clip]{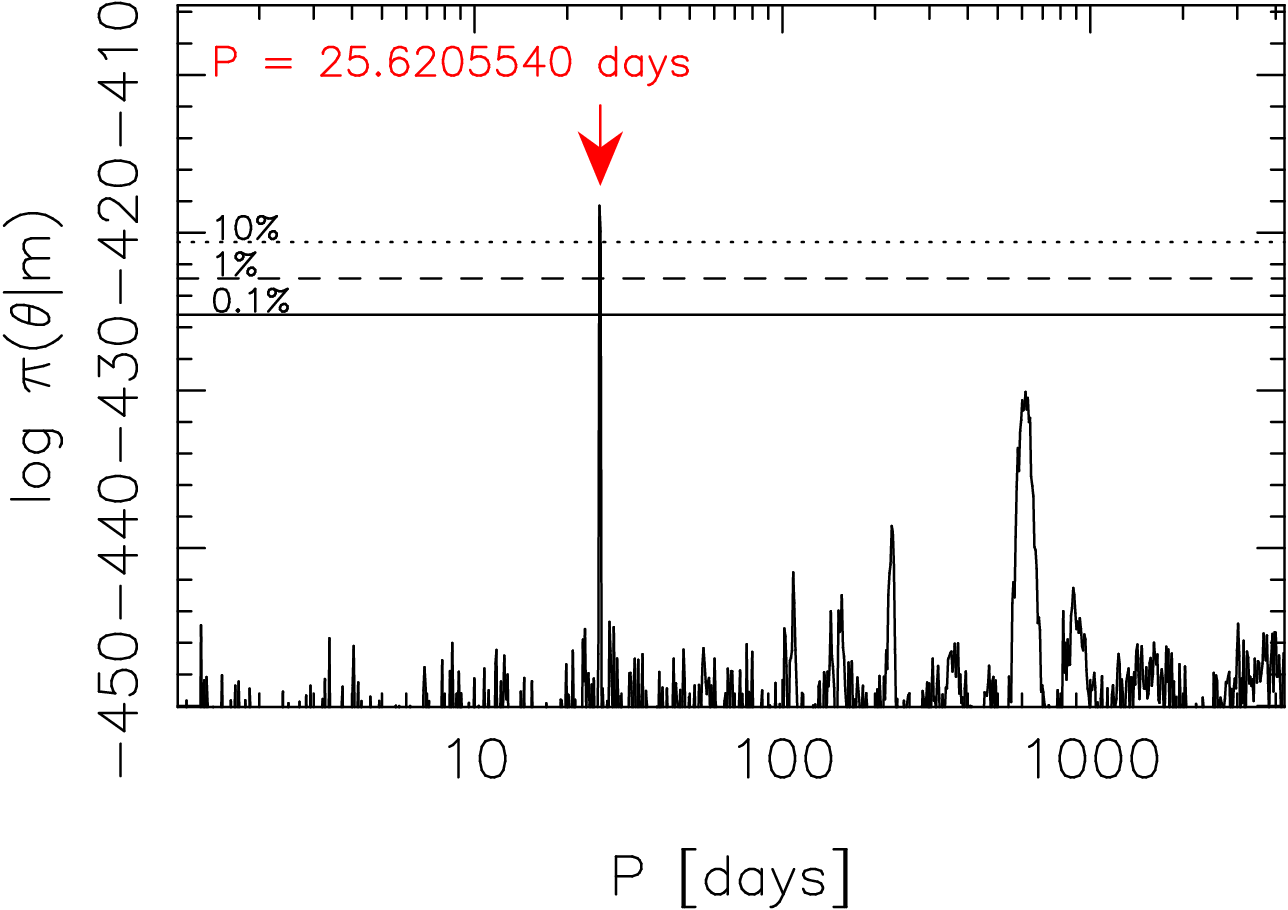}

\includegraphics[angle=270, width=0.49\textwidth,clip]{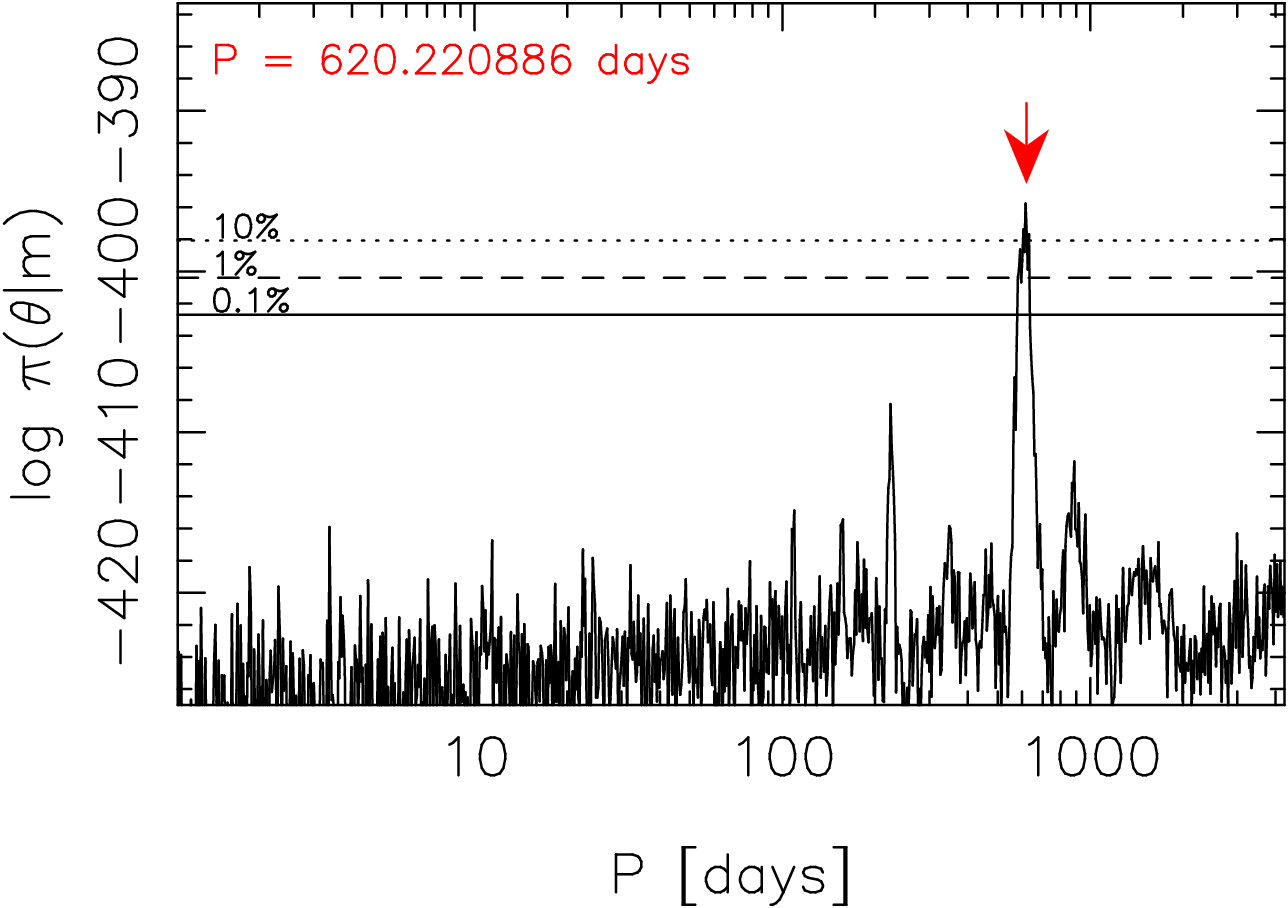}
\caption{Estimated posterior probability densities of models with $k$ Keplerian signals given DJ 163 data as a function of the period of the $k$th signal for $k=1$, $k=2$, and $k=3$ (from top to bottom). The red arrows denote the positions of the global maxima in the period space and the horizontal lines show the 10\% (dotted), 1\% (dashed), and 0.1\% (solid) equiprobability thresholds with respect to the maxima.}\label{fig:GJ163_psearch}
\end{figure}

We could easily replicate the results of \citet{tuomi2013c} with the available larger set of HARPS radial velocities. We found easily the three signals (Fig. \ref{fig:GJ163_psearch}) discussed by \citet{tuomi2013c} and \citet{bonfils2013b} at periods of 8.6311 [8.6291, 8.6335], 25.637 [25.594, 25.680], and 604.3 [580.2, 633.4] corresponding to planets orbiting the star with minimum masses of 9.9 [7.6, 12.3], 7.6 [5.2, 10.5], and 20.2 [12.5, 27.9] M$_{\oplus}$. These values are consistent with the estimates obtained by \citet{bonfils2013b} and \citet{tuomi2013c}.

\citet{tuomi2013c} also discussed the possibility that there could be a third candidate planet orbiting the star. Although they could not make this claim confidently, they obtained evidence for a periodicity of 125.0 [123.3, 128.0] days in the HARPS data. Similarly, \citet{bonfils2013b} discussed two additional signals at periods of 19.46$\pm$0.02 and 108.4$\pm$0.5 days. We thus searched for additional signals in the HARPS data in order to verify these results with our HARPS-TERRA data and more flexible statistical model.

\begin{figure}
\center
\includegraphics[angle=270, width=0.49\textwidth,clip]{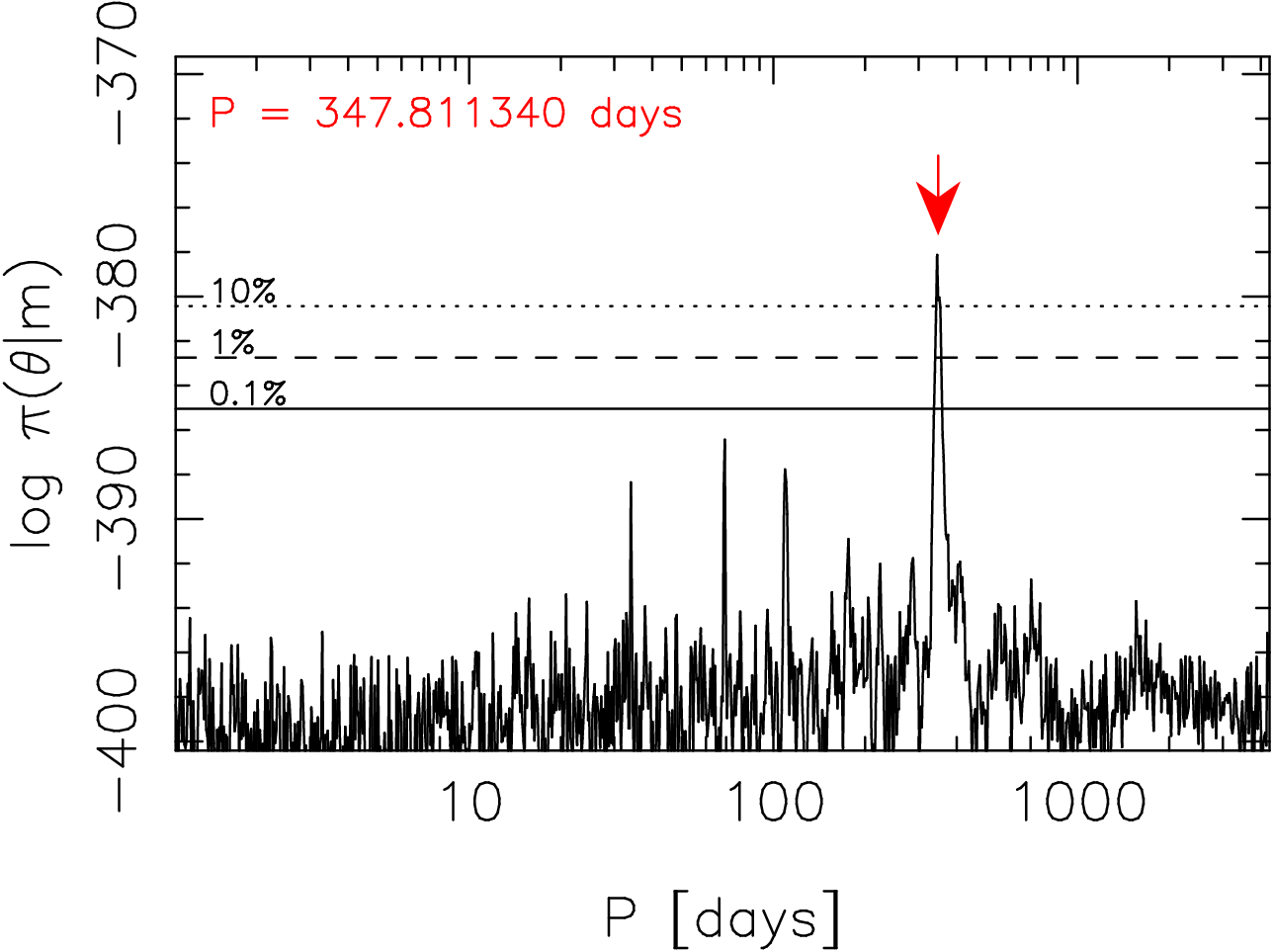}

\includegraphics[angle=270, width=0.49\textwidth,clip]{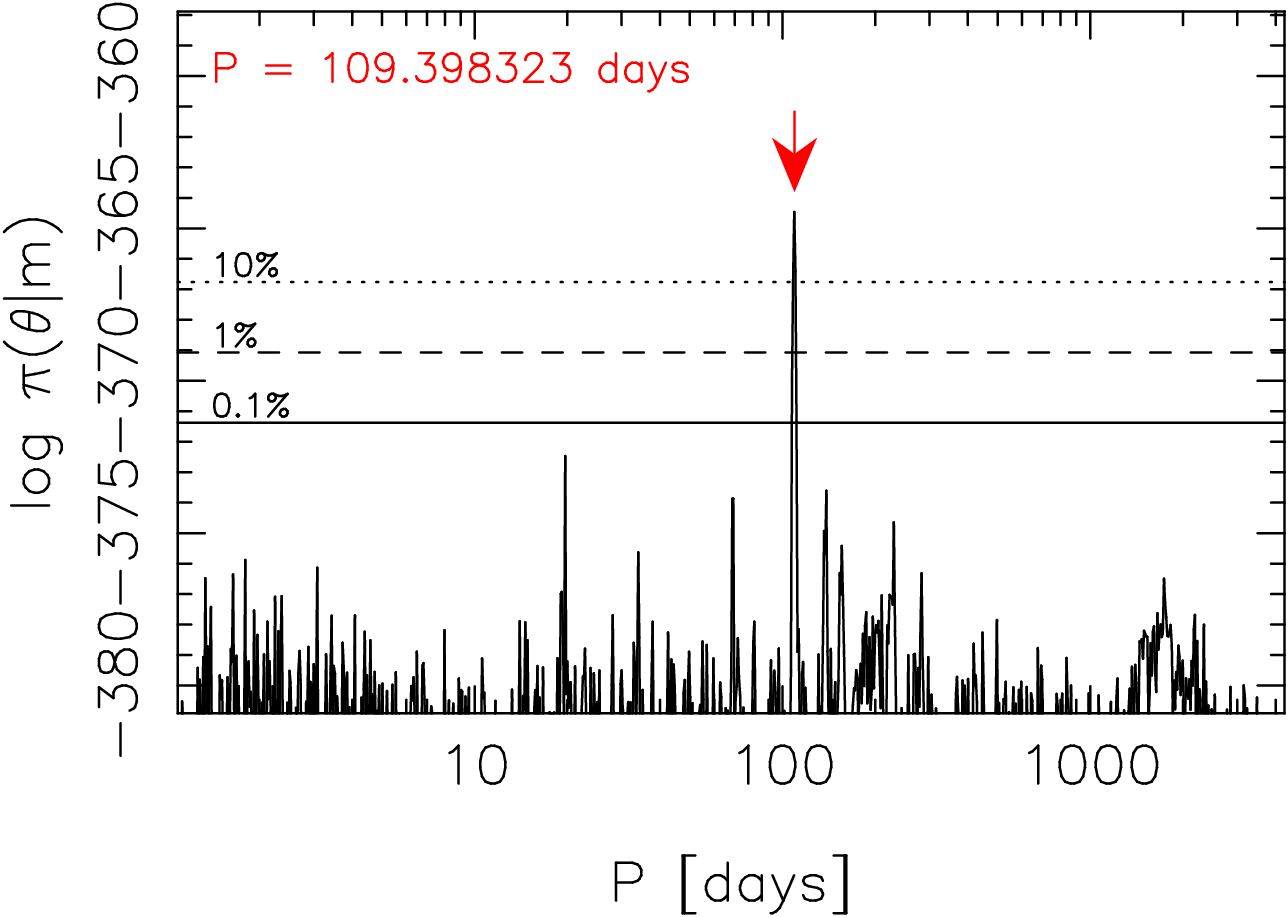}
\caption{As in Fig. \ref{fig:GJ163_psearch} but for $k=4$ and $k=5$.}\label{fig:GJ163_psearch2}
\end{figure}

\begin{figure}
\center
\includegraphics[angle=270, width=0.49\textwidth,clip]{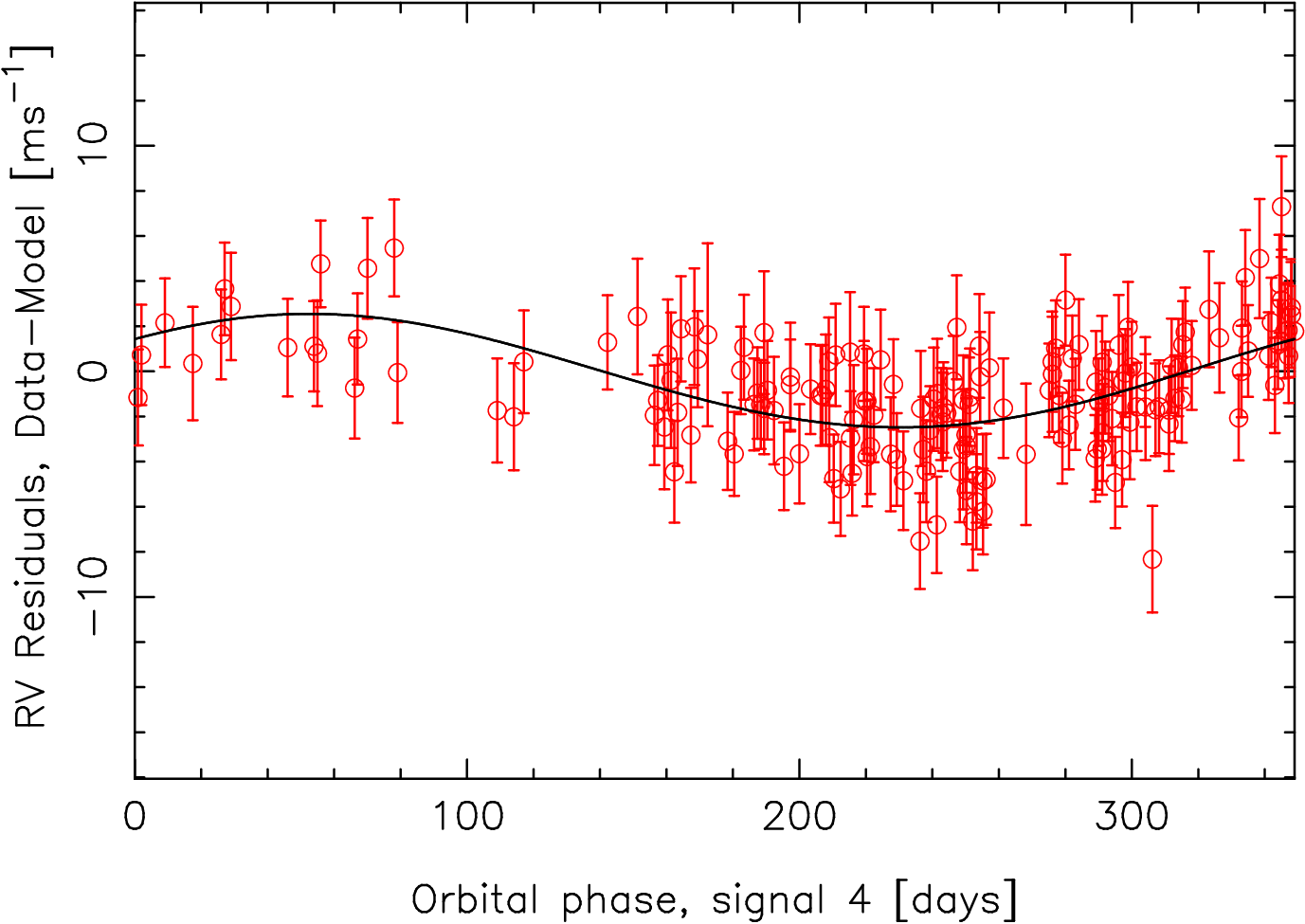}
\includegraphics[angle=270, width=0.49\textwidth,clip]{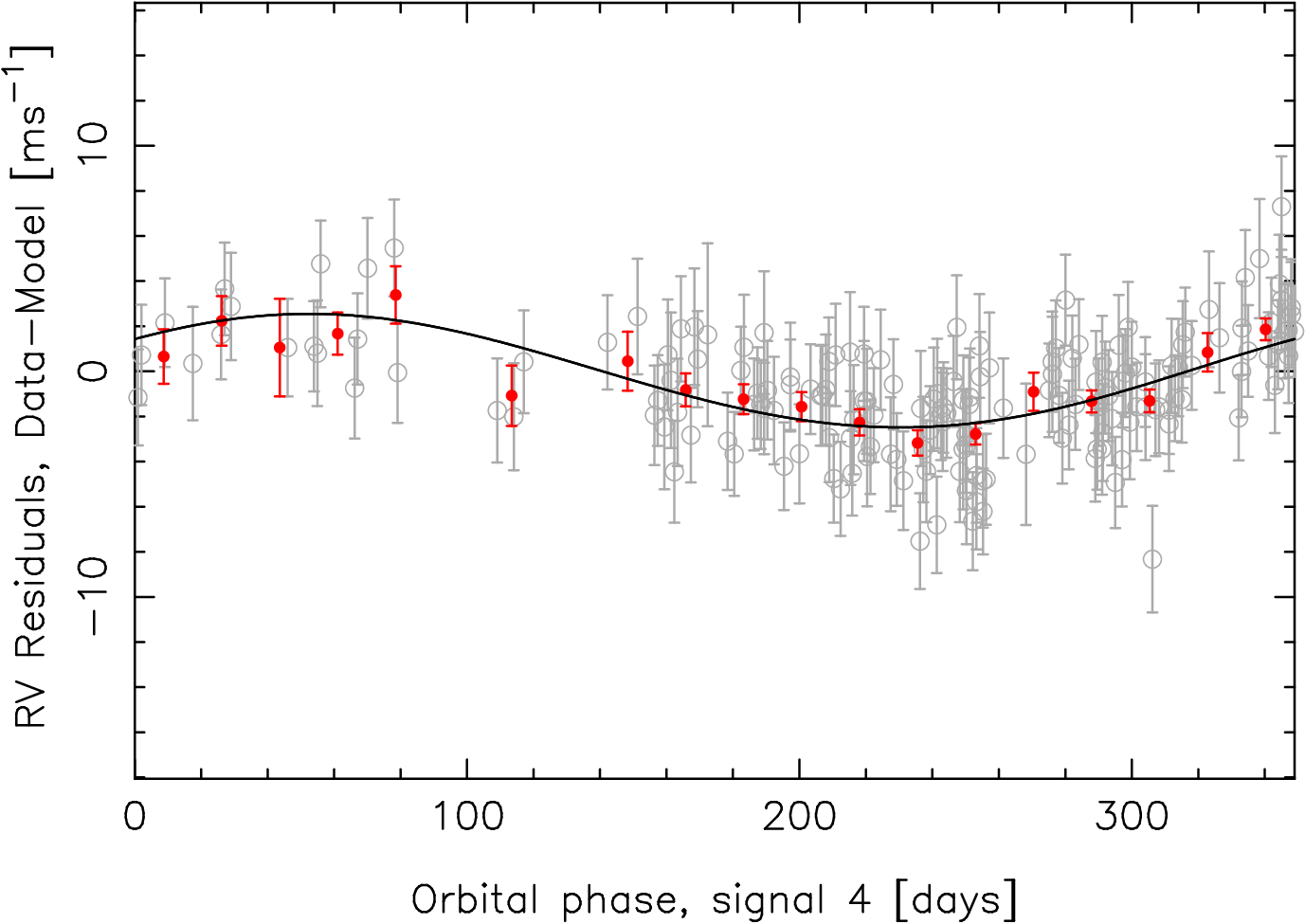}

\includegraphics[angle=270, width=0.49\textwidth,clip]{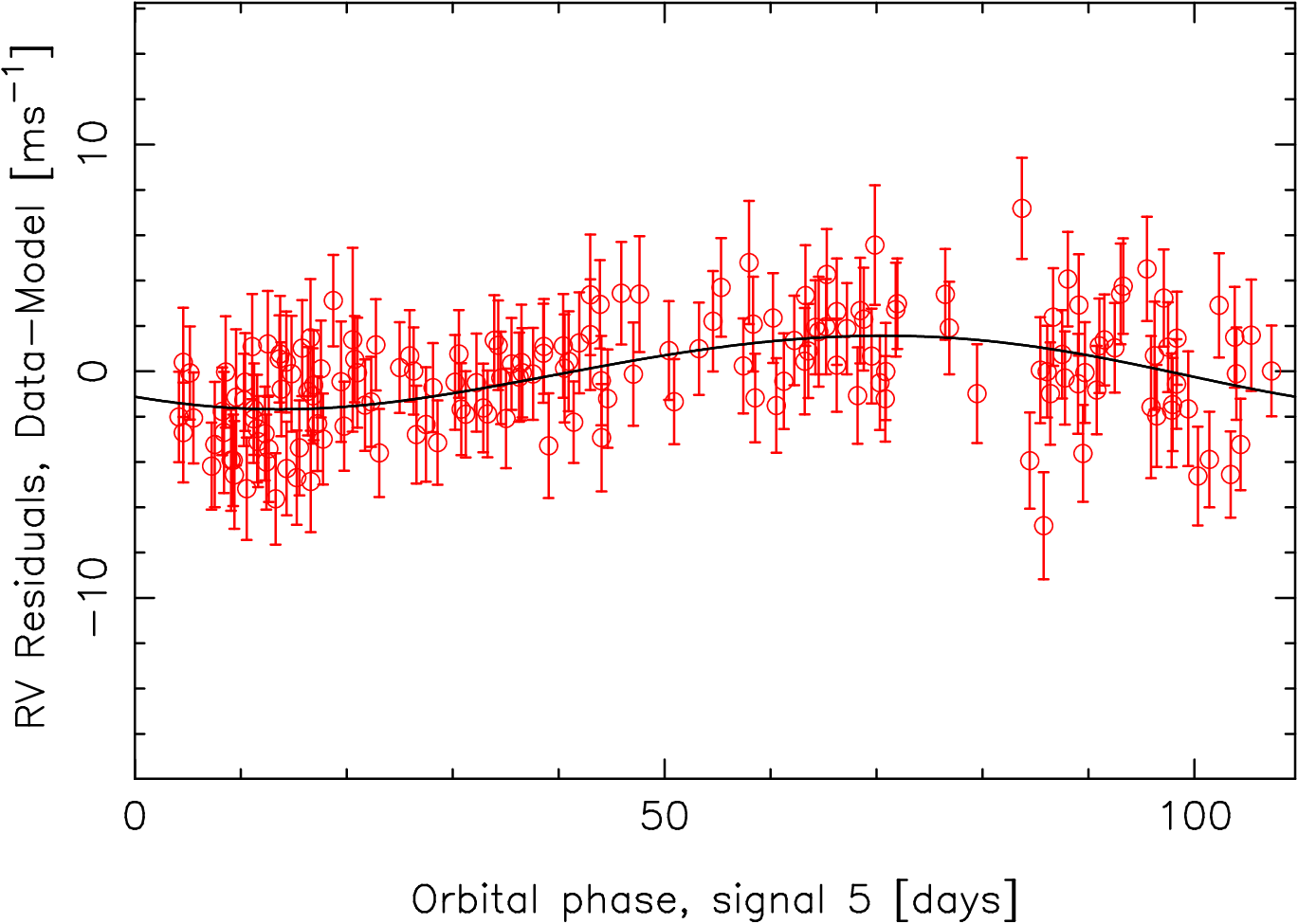}
\includegraphics[angle=270, width=0.49\textwidth,clip]{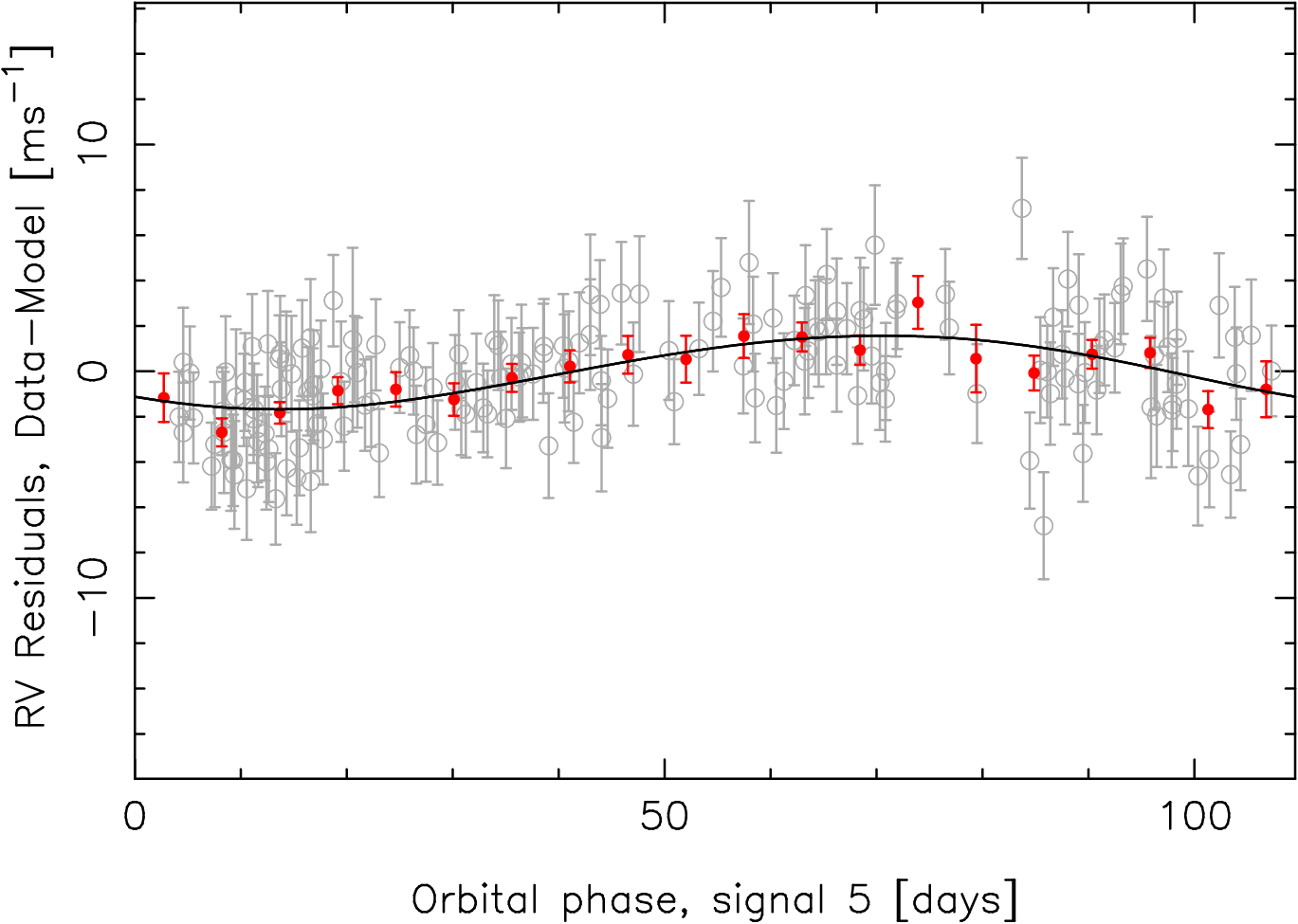}
\caption{HARPS radial velocities of GJ 163 folded on the phases of the 4th and 5th Keplerian signals detected in the data with the other signals subtracted.}\label{fig:GJ163_phased}
\end{figure}

We observed unique probability maxima at periods of 348.6 [338.1, 360.3] and 109.47 [108.06, 111.03] days that corresponded to signals satisfying our detection criteria. These signals were found to have amplitudes of 2.47 [1.30, 3.64] and 1.66 [0.69, 2.85] ms$^{-1}$, respectively, making them higher in amplitude than the typical instrument noise of HARPS of roughly 1.0 ms$^{-1}$ and of GJ 163 data in particular that has a mean value of 1.26 ms$^{-1}$. We demonstrate the uniqueness of these signals in Fig. \ref{fig:GJ163_psearch2} and show the phase-folded HARPS radial velocities in Fig. \ref{fig:GJ163_phased}. The latter one of these signals likely corresponds to the variability discussed by \citet{bonfils2013b} and \citet{tuomi2013c} at periods of 108 and 125 days, respectively, whereas the former has not been discussed earlier likely due to its rather poor phase-coverage. It is noteworthy that the 19.5-day signal discussed in detail by \citet{bonfils2013b} cannot be detected in the updated data set with our statistical models and signal search techniques making it likely that it was a spurious signal related to stellar activity rather than a planet.

Neither the HARPS activity indicators nor the ASAS V-band photometry data showed any evidence for periodicities. We thus interpret the five signals in the data as evidence for five planets orbiting the star in a dynamically packed configuration. Although we did not perform dynamical analyses of the system, it appears possible that the corresponding 5-planet system could be stable due to sufficient orbital spacing. However, we leave full-scale dynamical analyses of the system for a future publication. 

\clearpage

\subsection{GJ 176}\label{sec:GJ176}

\begin{figure}
\center
\includegraphics[angle=270, width=0.49\textwidth,clip]{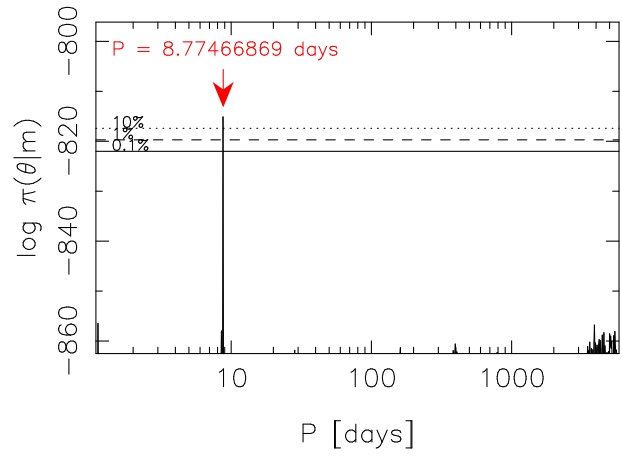}
\caption{As in Fig. \ref{fig:GJ27.1_period_search} but for the signal in the combined HARPS, HIRES, and HET velocities of GJ 176.}\label{fig:GJ176_period_search}
\end{figure}

The candidate planet orbiting GJ 176 (HD 285968, HIP21932), with an orbital period of 8.78 days and a minimum mass of 8.4 M$_{\oplus}$, was reported originally by \citet{forveille2009} in a study that also disputed, together with the independent work of \citet{butler2009}, the existence of a previously proposed candidate with an orbital period of 10.24 days and a minimum mass of 24 M$_{\oplus}$ \citep{endl2008}. We could easily detect the signal at a period of 8.7748 [8.7721, 8.7778] days and with an amplitude of 3.94 [3.14, 4.75] ms$^{-1}$ in the combined set of 71 HARPS, 103 HIRES \citep[most of which are new but some have already been published by][]{butler2009}, and 99 HET velocities \citep{robertson2015}, with a combined baseline of 5836 days. The signal is shown as a unique probability maximum in the period space in Fig. \ref{fig:GJ176_period_search}. This signal could also be detected in the HIRES data alone \citep{butler2016}.

\begin{figure}
\center
\includegraphics[angle=270, width=0.49\textwidth,clip]{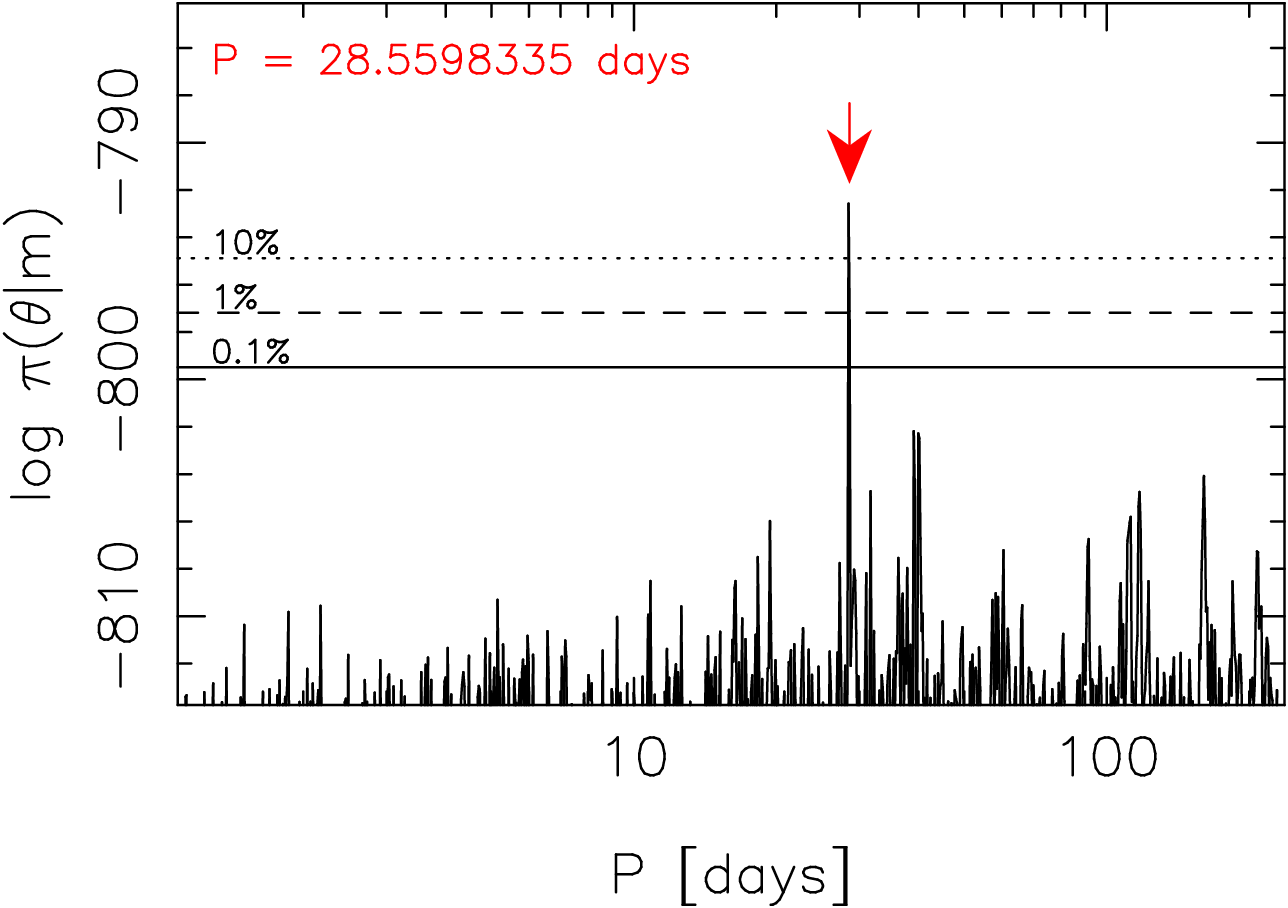}

\includegraphics[angle=270, width=0.49\textwidth,clip]{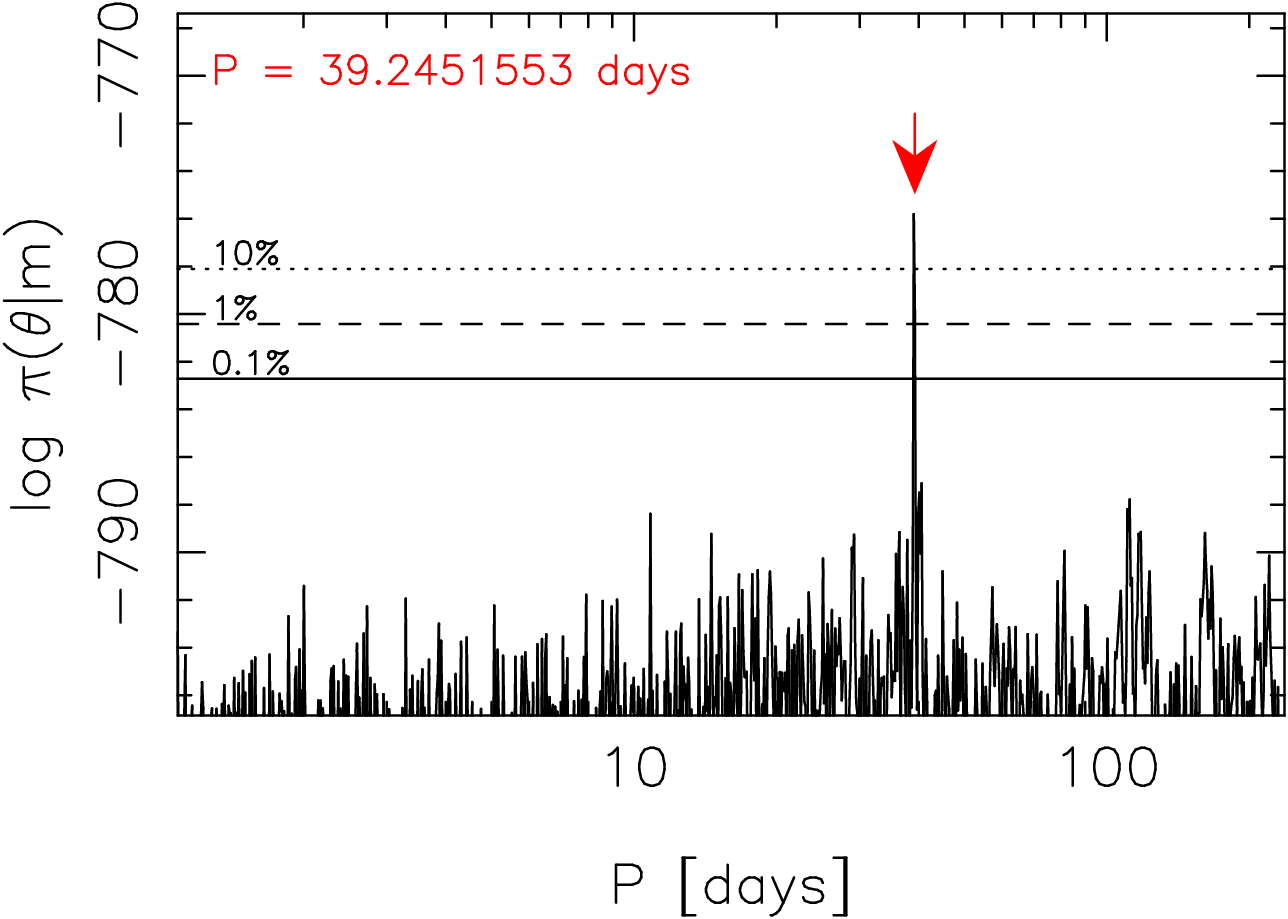}
\caption{Estimated posterior probability densities of the two- (top panel) and three-Keplerian (bottom panel) models given GJ 176 radial velocity data as functions of the period parameters of the second and third Keplerian signals, respectively. The red arrows indicate the positions of the global maxima in the period space and the horizontal lines denote the 10\% (dotted), 1\% (dashed), and 0.1\% (solid) equiprobability thresholds with respect to the maxima.}\label{fig:GJ176_period_search2}
\end{figure}

\begin{figure}
\center
\includegraphics[angle=270, width=0.49\textwidth,clip]{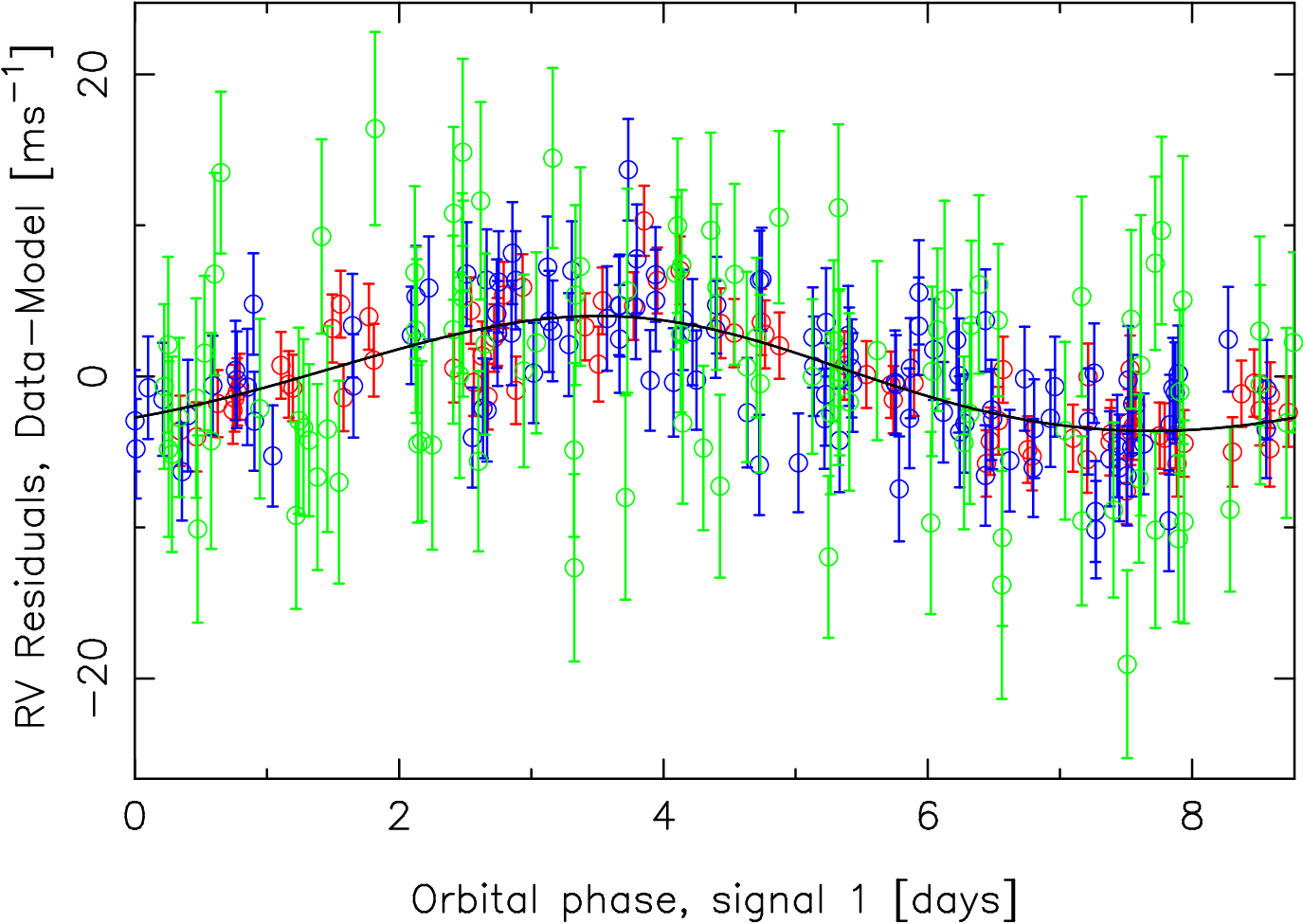}

\includegraphics[angle=270, width=0.49\textwidth,clip]{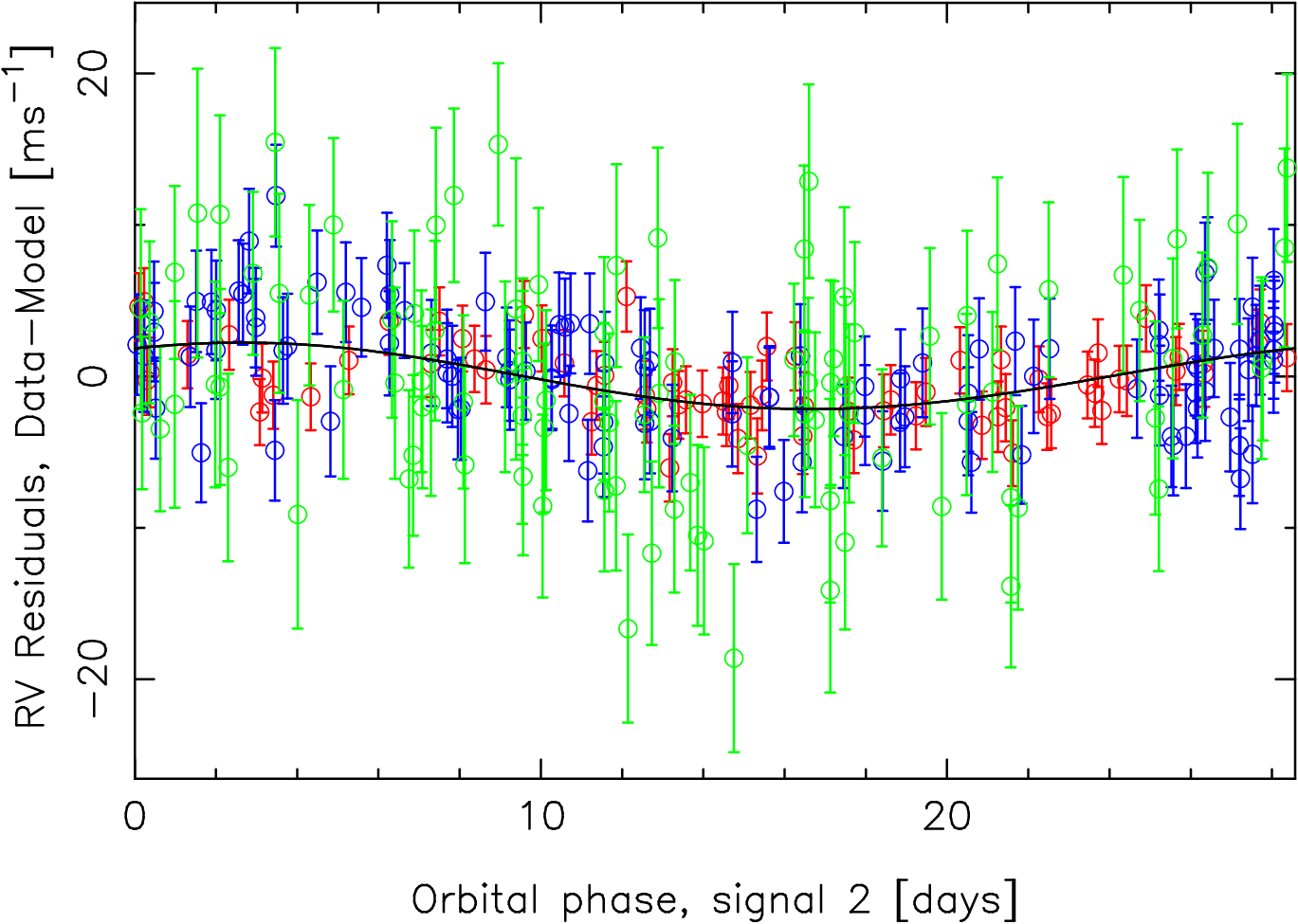}
\caption{HARPS (red), HIRES (blue), and HET (green) velocities of GJ 176 folded of the phases of the two signals of candidate planets detected in the combined data. The solid curve denotes the preferred Keplerian signal. The other signals and the second order polynomial have been subtracted from each panel.}\label{fig:GJ176_curve}
\end{figure}

However, the signal at a period of 8.77 days was not the only one we identified in the combined HARPS, HIRES, and HET radial velocities. Due to the presence of long-period variations likely connected to stellar activity, as well as apparent acceleration modelled by using a second order polynomial rather than only a linear trend, we first limited the period space to 220 days to avoid spurious solutions caused by long-period variability and its aliasing due to annual sampling.  Most notably, we obtained evidence for two independent and unique signals at periods of 28.586 [28.533, 28.633] and 39.233 [39.144, 29.334] days, respectively (Fig. \ref{fig:GJ176_period_search2}) out of which the latter likely corresponds to the rotation period of the star also observed by \citet{forveille2009}. However, the former signal is previously unknown and given its independence of the 40-d signal, it is likely Keplerian signal of another candidate planet orbiting the star. We show the radial velocities folded of the phases of the signals corrresponding to the candidate planets in Fig. \ref{fig:GJ176_curve}.

\begin{figure}
\center
\includegraphics[angle=270, width=0.49\textwidth,clip]{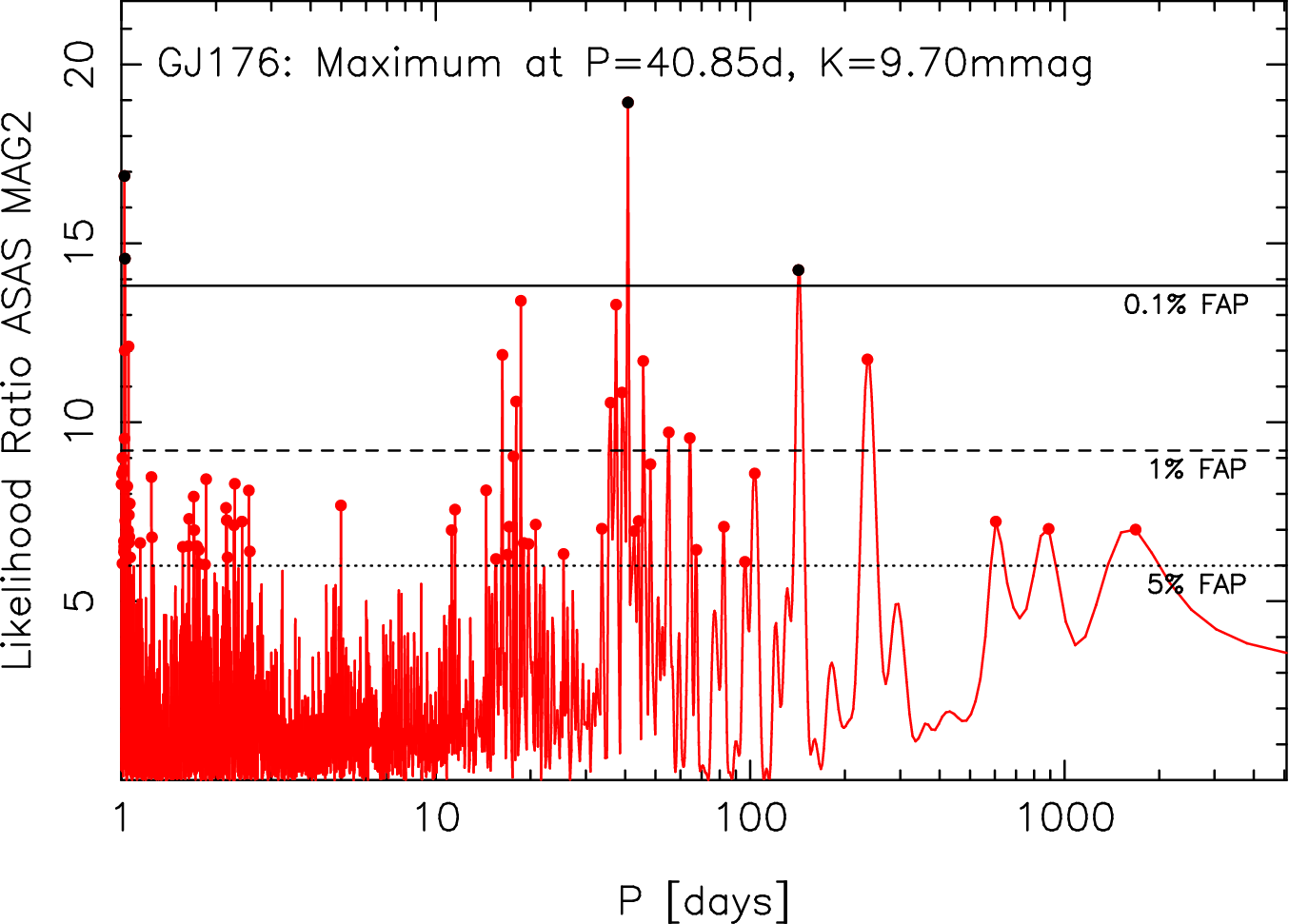}
\caption{Likelihood-ratio periodogram of the ASAS V-band photometry of GJ 176.}\label{fig:GJ176_asas}
\end{figure}

When looking at the set of 301 ASAS V-band photometry measurements we discovered a significant periodicity at a period of 40.85 days (Fig. \ref{fig:GJ176_asas}) -- also reported by \citet{kiraga2007}. We classify this as the photometric rotation period of the star. However, the 8.77 and 28.6-day signals have no counterparts in activity indicators or photometry and are thus classified as a planetary signals corresponding to a hot mini-Neptune and a warm mini-Neptune with minimum masses of 8.0 [5.8, 10.4] and 7.4 [3.8, 11.4]  M$_{\oplus}$, respectively.

We note that we accounted for the correlations between the HET radial velocities and the four activity indicators discussed in \citet{robertson2015}, namely, the two different sodium indices, H$_\alpha$ index, and Ca I line index. Out of these, only the last one appeared to trace activity-induced radial velocity variations whereas the first three were uncorrelated with the HET velocity data. This contradicts the claim in \citet{robertson2015} that ``... sodium resonance lines [...] correlate with RV ...''. We found that the estimates of the parameters describing this correlation, $c_{\rm Na1}$ and $c_{\rm Na2}$, that quantify the dependence of radial velocities on the corresponding indices published in \citet{robertson2015}, were 0.49 [-0.52, 1.41] and -0.30 [-1.30, 0.69] $10^{3}$ ms$^{-1}$, respectively, which implies that they are not statistically significantly different from zero and that the correlations thus cannot be said to exist with more than roughly 1-$\sigma$ credibility level. In contrast, the value for a similar correlation between velocities and Ca I index was -14.5 [-23.1, -5.1] $10^{3}$ ms$^{-1}$, which is significantly different from zero with a 99\% credibility implying a linear negative dependence of the velocities on the Ca I index.

\clearpage

\subsection{GJ 179}

\begin{figure}
\center
\includegraphics[angle=270, width=0.49\textwidth,clip]{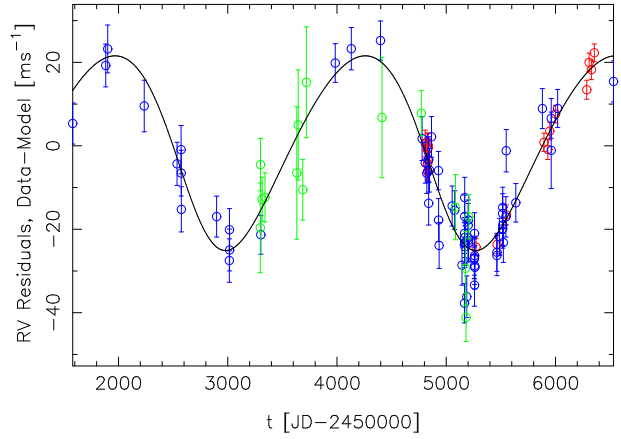}
\caption{HARPS (red), HIRES (blue), and HET (green) radial velocities of GJ 179 and the MAP Keplerian function of GJ 179 b (solid curve).}\label{fig:GJ179_curve}
\end{figure}

The combined HARPS, HIRES, and HET radial velocities of GJ 179 (HIP 22627) were consistent with the giant planet orbiting the star (Fig. \ref{fig:GJ179_curve}) reported by \citet{howard2010} based on HIRES and HET radial velocities. We estimate that this planet with a minimum mass of 250.1 [188.5, 311.7] M$_{\oplus}$ or 0.787 [0.592, 0.981] M$_{\rm Jup}$ has an orbital period of 2300 [2180, 2450] days implying a semi-major axis of 2.44 [2.17, 2.69] AU. We also observed a positive connection between HARPS velocities and BIS values and a negative one between HIRES velocities and S-indices with 95\% but not with 99\% credibility but find that the Keplerian signal corresponding to GJ 179 b cannot be shown to be affected by these connections implying that it is likely independent of the activity-induced variability and thus interpreted as a planet candidate orbiting the star.

\begin{figure}
\center
\includegraphics[angle=270, width=0.49\textwidth,clip]{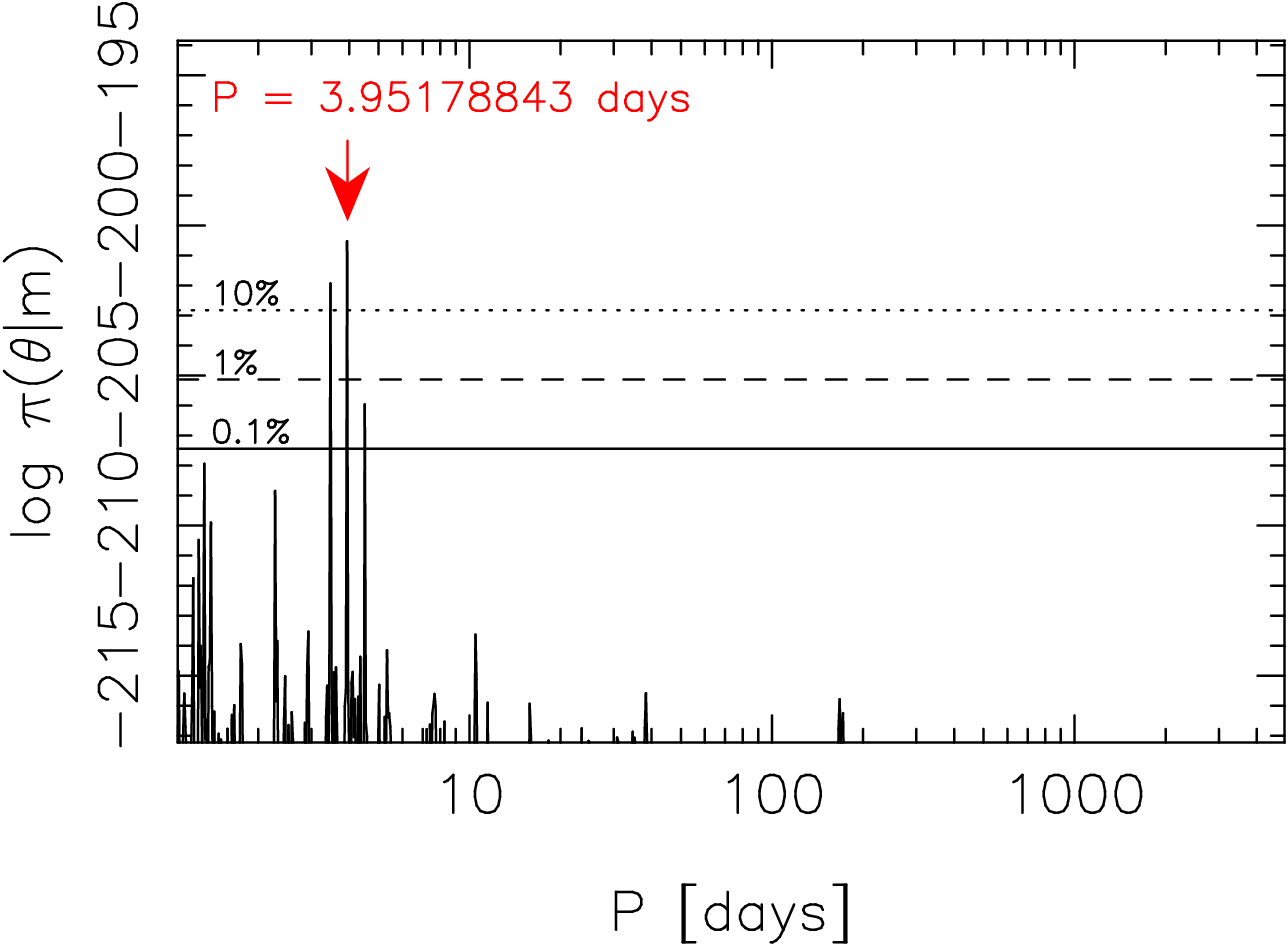}

\includegraphics[angle=270, width=0.49\textwidth,clip]{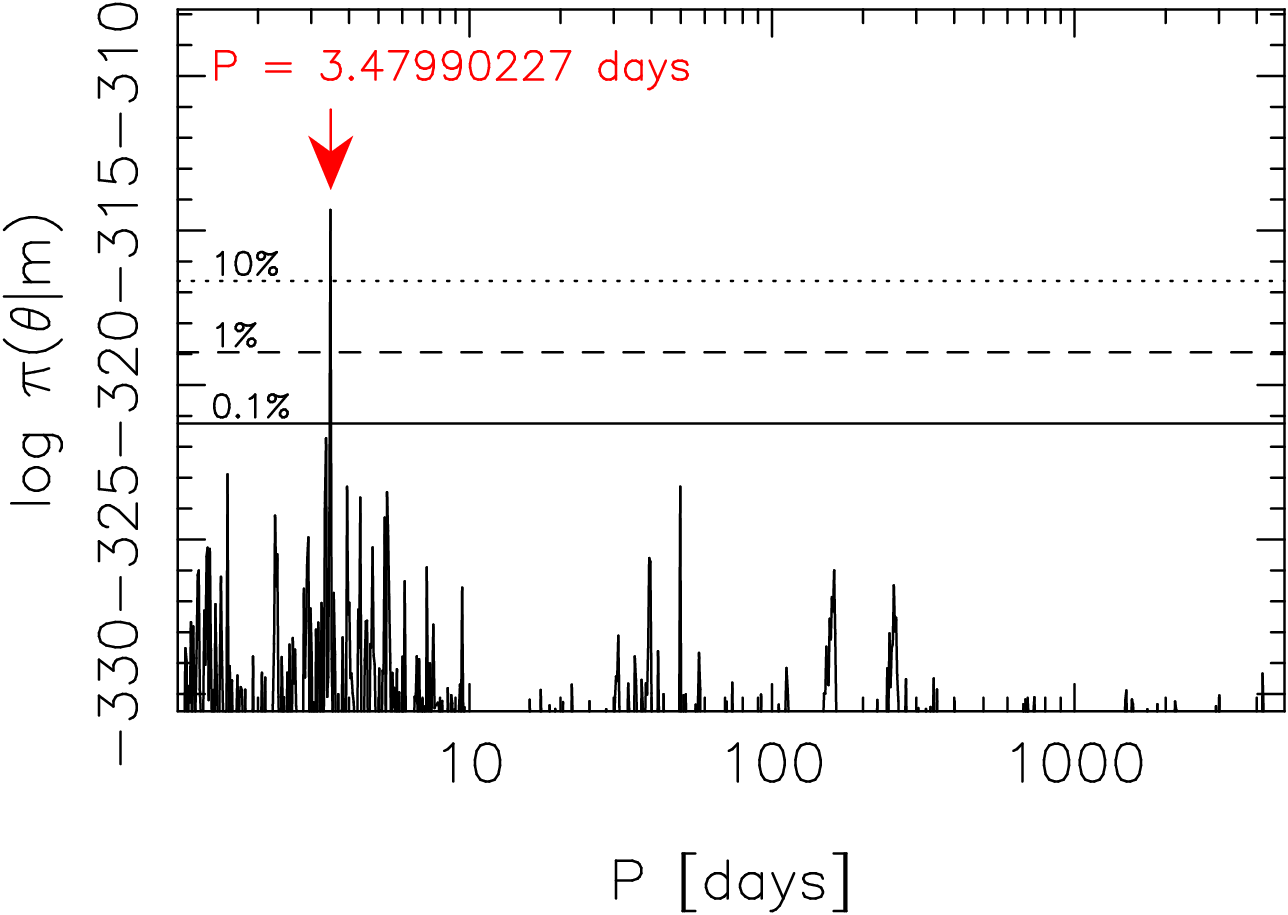}
\caption{Top panel: posterior probability density as a function of the period parameter of the second Keplerian signal in the GJ 179 data as obtained by \citet{butler2016}. Bottom panel shows the same given the combined data as analysed in the current work. The red arrow indicates the position of the global maximum in the period space and the horizontal lines denote the 10\% (dotted), 1\% (dashed), and 0.1\% (solid) equiprobability contours with respect to the maximum.}\label{fig:GJ179_psearch}
\end{figure}

\citet{butler2016} reported a ``signal requiring confirmation'' at a period of 3.95148$\pm$0.00037 days based on their analysis of the HIRES data because the signal did not satisfy their detection threshold for a candidate planet. We have reproduced here the period search of \citet{butler2016} for the HIRES data of GJ 179. The signal they reported can be seen together with two nearby local maxima in the period space (Fig. \ref{fig:GJ179_psearch}, top panel). When analysing the combined data we spotted the second signal at a period of 3.4798 [3.4787, 3.4812] days (Fig. \ref{fig:GJ179_psearch}, bottom panel) that coincides with one of the distinguishable local maxima in the posterior based on HIRES data alone. It thus seems evident that while HIRES data allows three different periodicities as solutions, the HARPS and HET data are able to rule out two of these maxima leaving only one unique probability maximum corresponding to a periodic signal that satisfies our detection criteria.

\begin{figure}
\center
\includegraphics[angle=270, width=0.49\textwidth,clip]{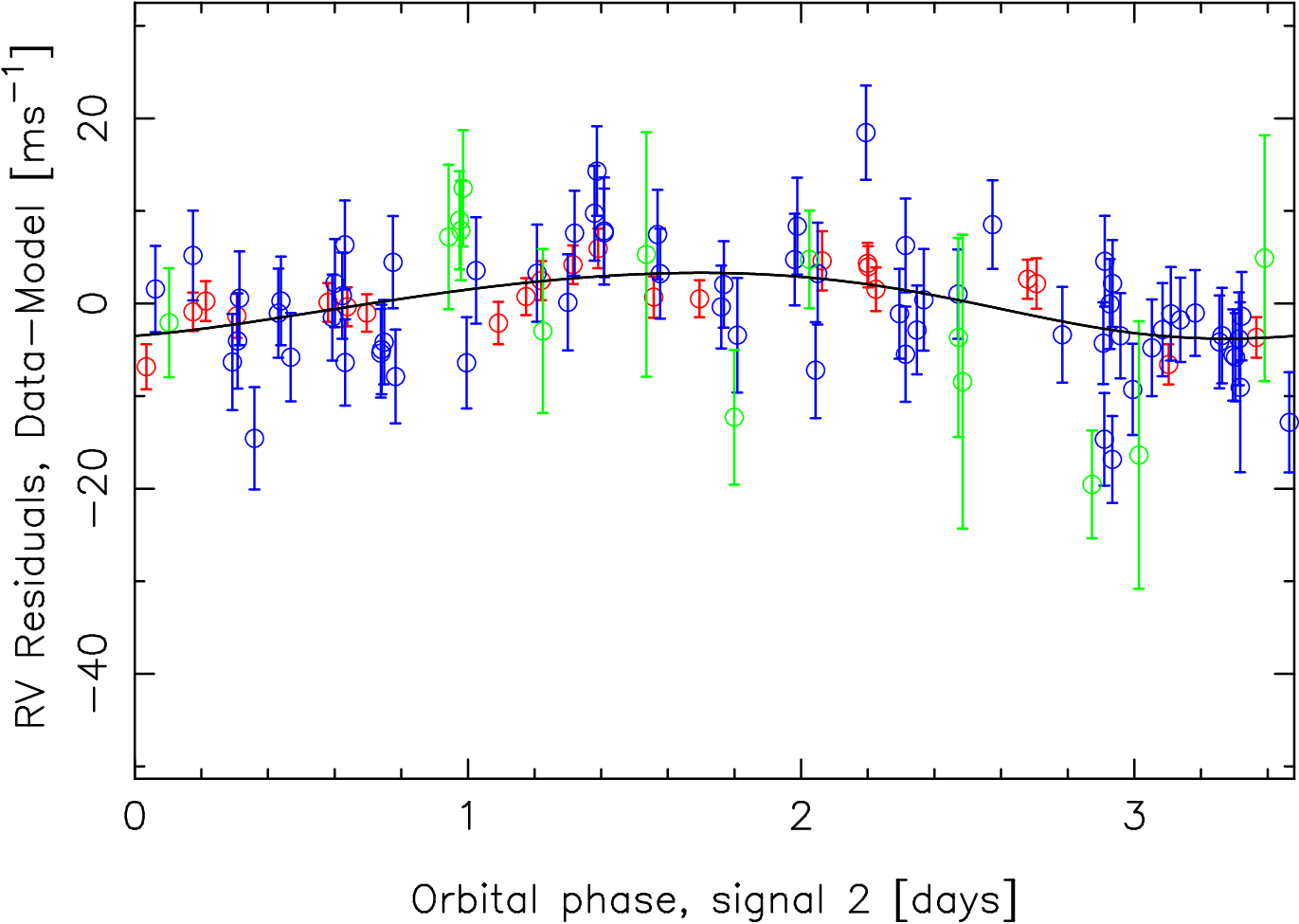}

\includegraphics[angle=270, width=0.49\textwidth,clip]{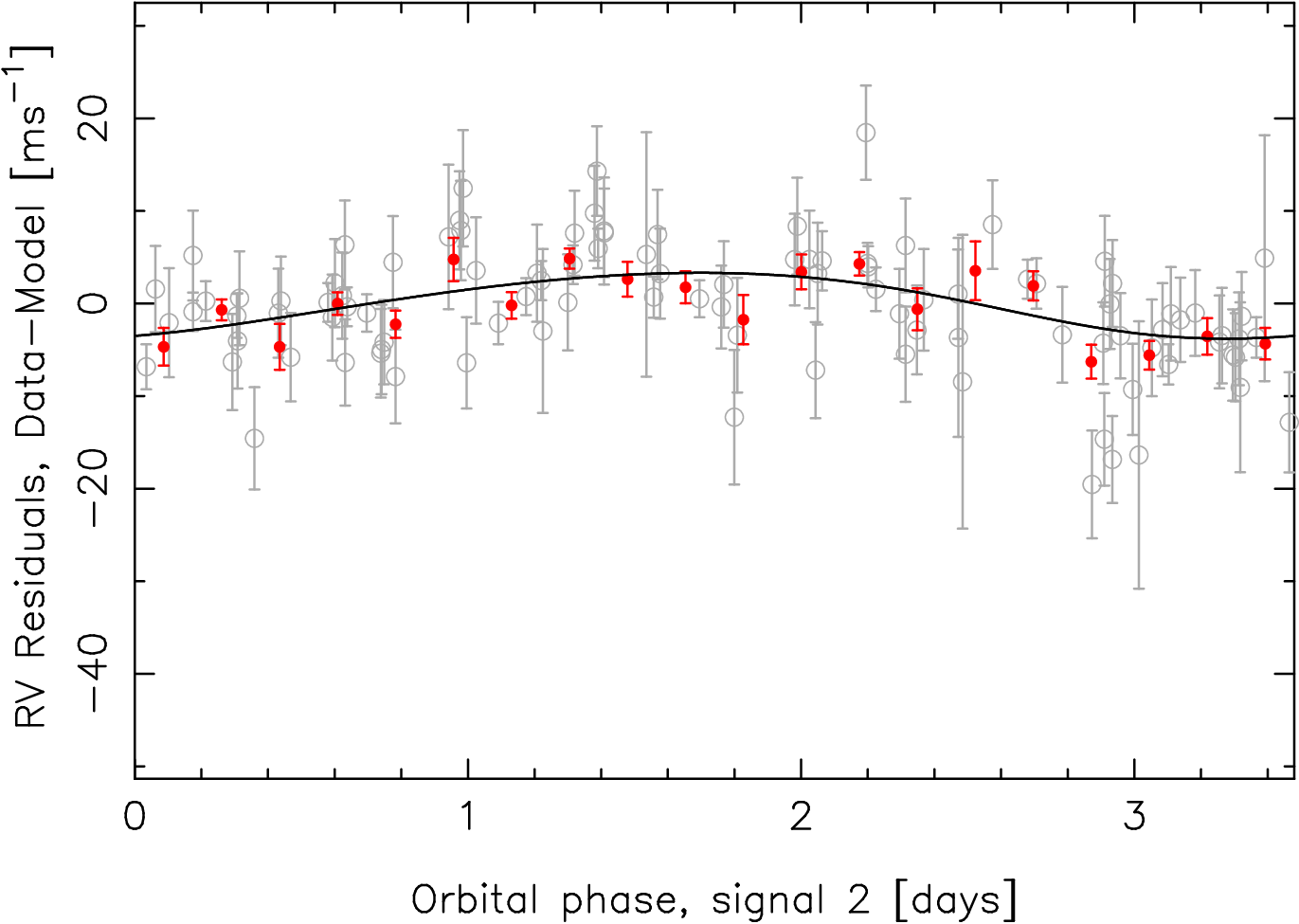}
\caption{Top panel: HARPS (red), HIRES (blue), and HET (green) radial velocities of GJ 179 folded on the phase of the second signal in the data. The signal of GJ 179 b has been subtracted. Bottom panel shows the same when the phase is divided into 20 bins (red filled circles).}\label{fig:GJ179_phased}
\end{figure}

We have thus obtained evidence for a second periodic signal in the radial velocities of GJ 179. This signal, with an amplitude of 4.04 [1.95, 6.13] ms$^{-1}$, corresponds to a hot super-Earth with a minimum mass of 4.9 [2.2, 7.7] M$_{\oplus}$. The corresponding radial velocity signal is shown in Fig. \ref{fig:GJ179_phased}. As the radial velocity component connected to the activity indicators is independent of the two Keplerian signals, we interpret this signal as a candidate planet orbiting the star. We have also obtained evidence for linear acceleration in the combined data of -1.16 [-1.78, 0.53] ms$^{-1}$year$^{-1}$.

Although 193 ASAS photometry measurements of GJ 179 were available, the time-series contained variations with semi-amplitude of 500 mmag. This variability appears unlikely to be caused by the star because \citet{koen2010} reported a V magnitude of 12.018 with an uncertainty of 110 mmag that, although rather large, cannot explain the variations in the ASAS data. We thus consider the ASAS data to be contaminated and not representative of the actual stellar brightness of GJ 179.

\clearpage

\subsection{GJ 205}\label{sec:GJ205}

\begin{figure}
\center
\includegraphics[angle=270, width=0.49\textwidth,clip]{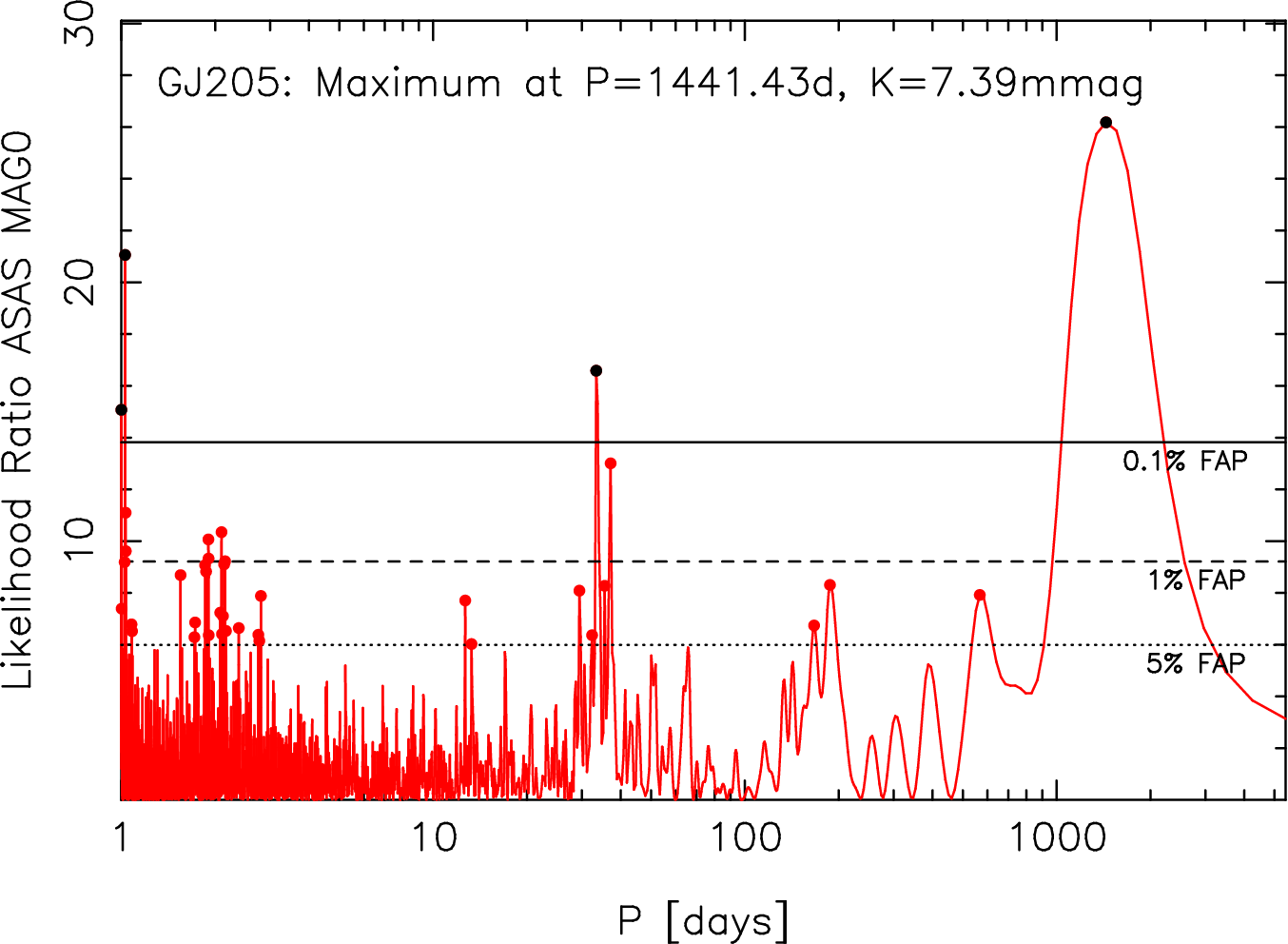}

\includegraphics[angle=270, width=0.49\textwidth,clip]{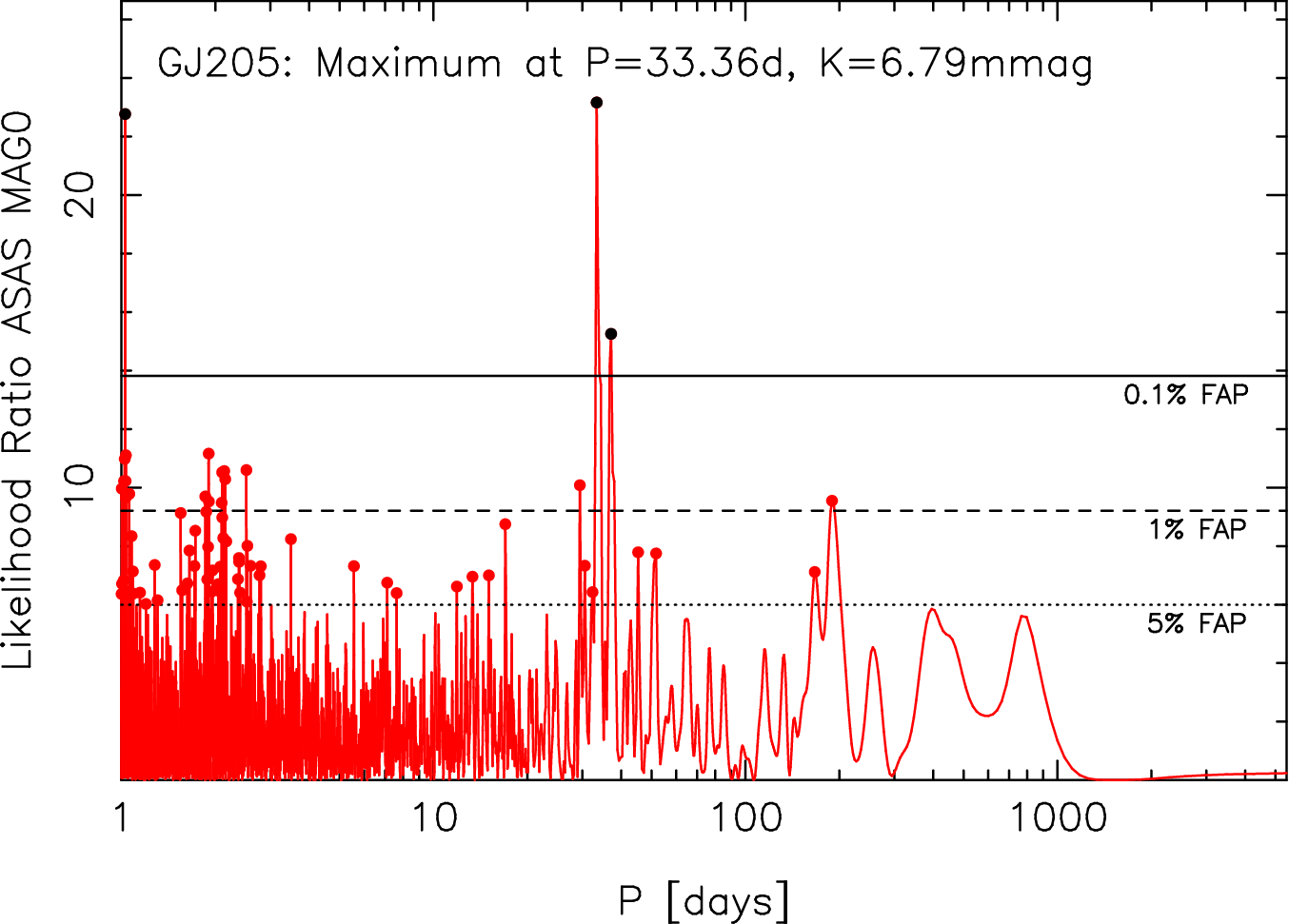}
\caption{Likelihood-periodogram of the ASAS V-band photometry of GJ 205 with a signal likely corresponding to stellar rotation at a period of 33.36 days together with a long-period photometric cycle. The horizontal lines denote the 5\% (dotted), 1\% (dashed), and 0.1\% (solid) FAPs, respectively. The red (black) filled circles highlight the maxima exceeding the 5\% (0.1\%) FAP. Botom panel shows the residual periodogram after subtracting the long-period cycle.}\label{fig:GJ205_photom}
\end{figure}

Although \citet{bonfils2013} reported a signal in the HARPS radial velocities of GJ 205 (HD 36395, HIP 35878) at 32.7 days, they interpreted it as being caused by stellar activity because they also found periodogram powers at the period of 33 days in the HARPS activity indices. This interpretation was reinforced by the results of \citet{kiraga2007}, who observed a 33.61-day periodicity in the photometric data of the target. Indeed, we could verify the existence of such a periodic signal in the ASAS V-band photometry. As shown in Fig. \ref{fig:GJ205_photom}, this photometric periodicity at 33.36 days is very strong, exceeding the 0.1\% FAP threshold. Following \citet{kiraga2007} we interpret this signal as an indication of the rotation period of the star. The stronger periodicity of 1440 days is interpreted as the stellar magnetic cycle. There were no indications of other periodic signals in the ASAS photometry data. We also found a signal at a period of roughly 35 d in the HARPS S-index reinforcing the interpretation that the stellar rotation period is indeed somewhere around 33 days.

\begin{figure}
\center
\includegraphics[angle=270, width=0.49\textwidth]{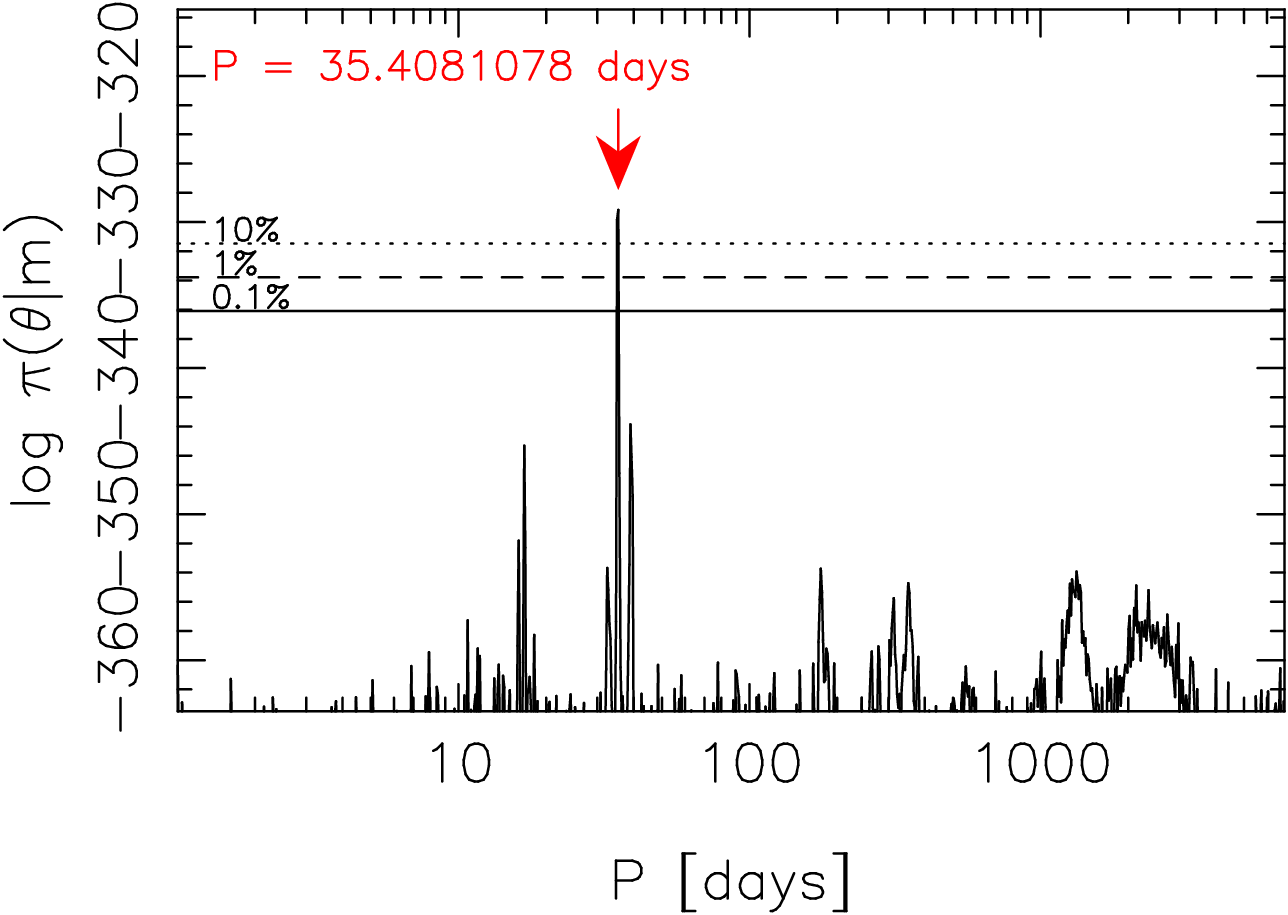}

\includegraphics[angle=270, width=0.49\textwidth]{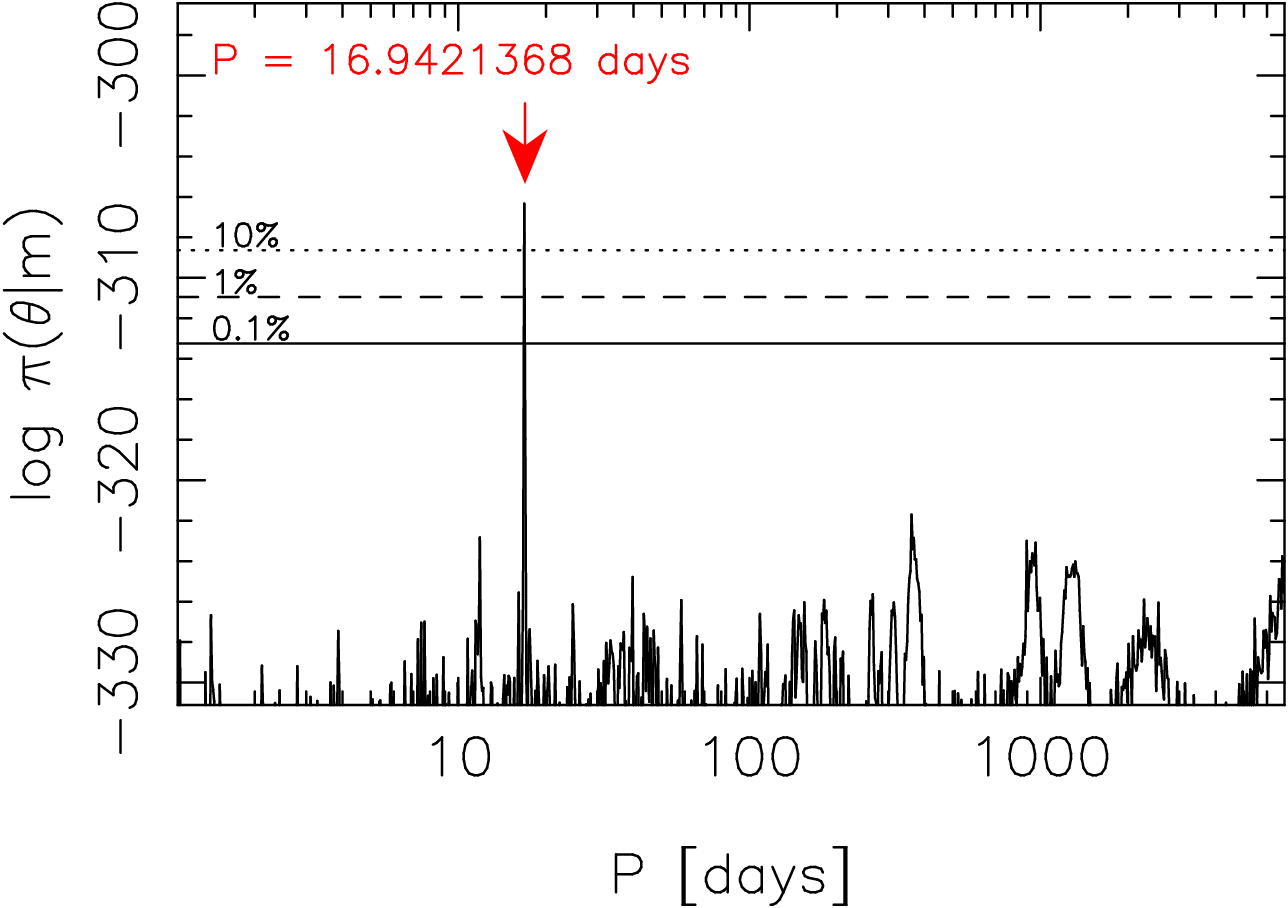}

\includegraphics[angle=270, width=0.49\textwidth]{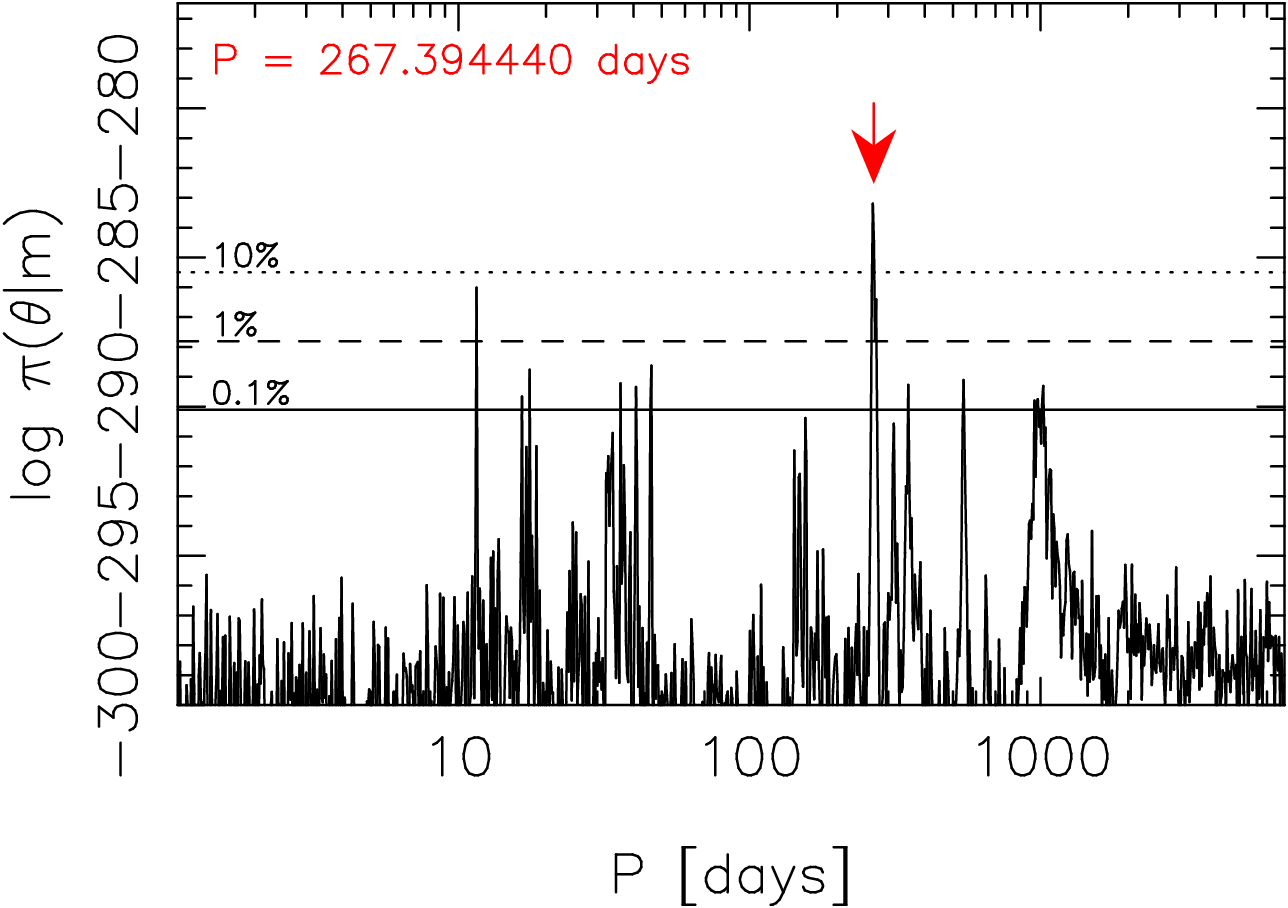}
\caption{As in Fig. \ref{fig:GJ27.1_period_search} but for the signal of a one-Keplerian model (top), the second signal in the two-Keplerian model (middle), and the third signal in a three-Keplerian model given the combined HARPS, HIRES, and PFS velocities of GJ 205.}\label{fig:GJ205_signal_search}
\end{figure}

\begin{figure}
\center
\includegraphics[angle=270, width=0.49\textwidth]{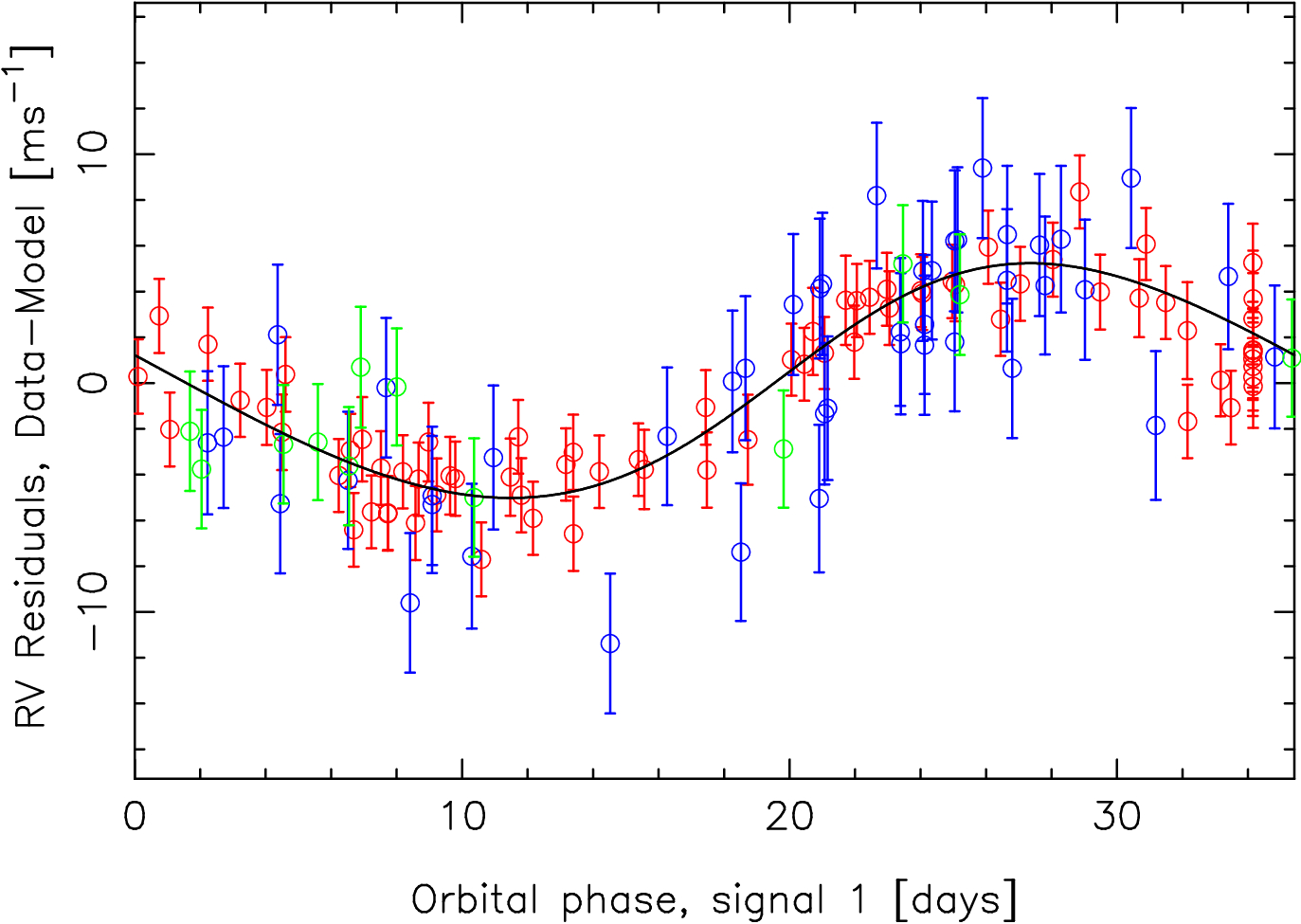}

\includegraphics[angle=270, width=0.49\textwidth]{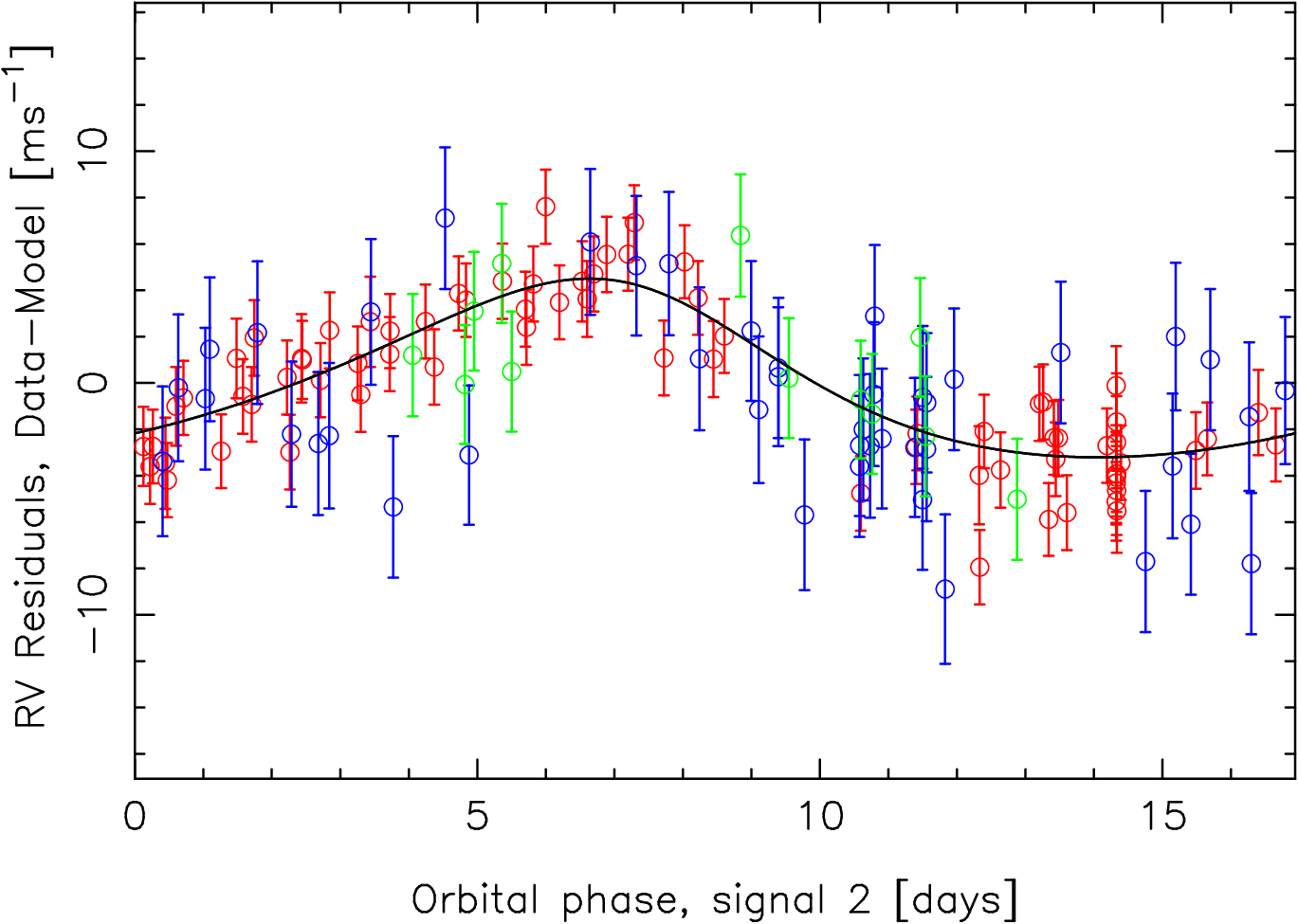}

\includegraphics[angle=270, width=0.49\textwidth]{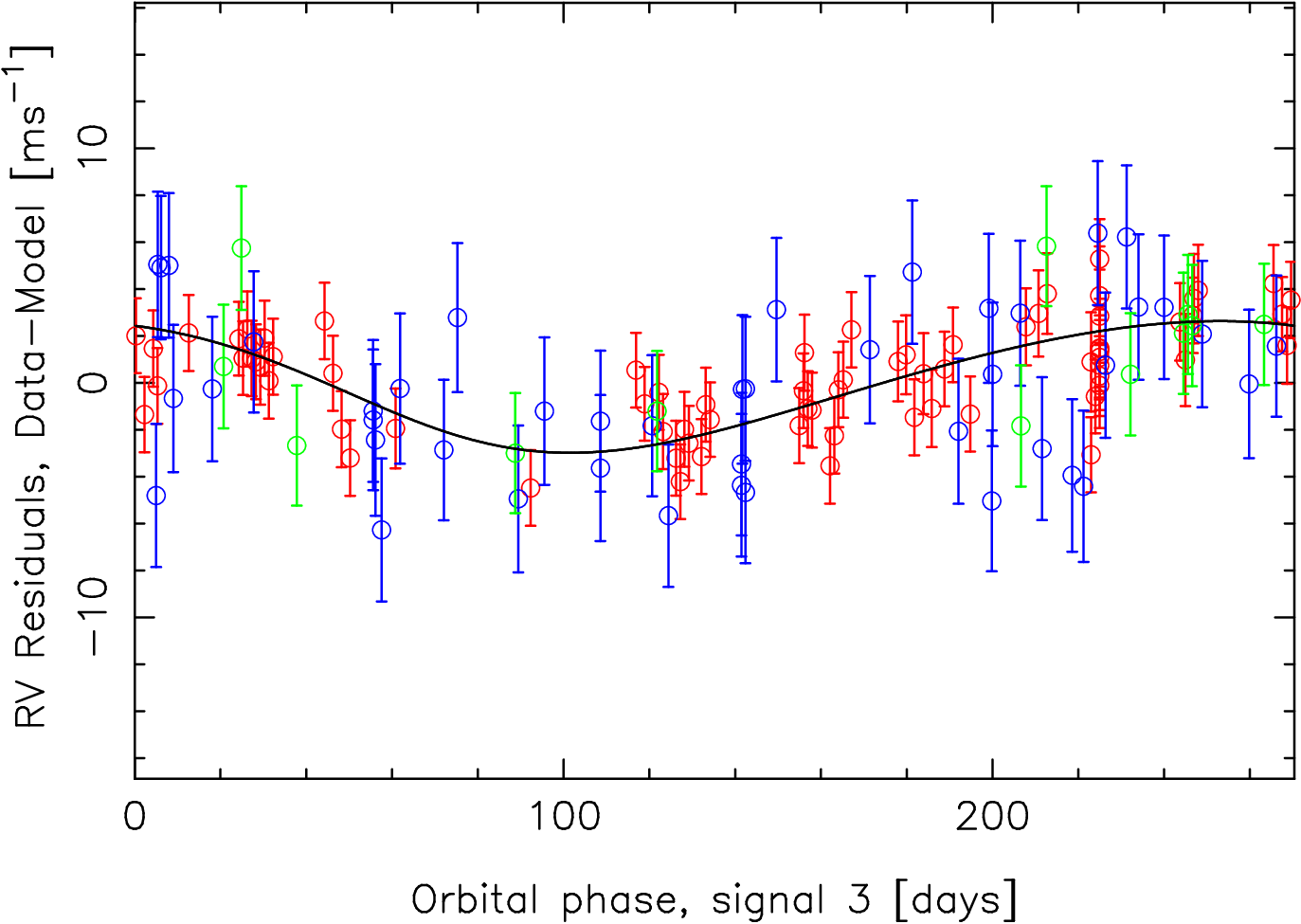}
\caption{Phase-folded HARPS (red), HIRES (blue), and PFS (green) radial velocities given the signals present in the GJ 205 data (solid curves) when the two other signals and the deterministic components of the statistical model have been removed from each panel.}\label{fig:GJ205_signals}
\end{figure}

The analysis of the corresponding set of 74 HARPS, 44 HIRES, and 12 PFS radial velocities enabled us to detect a signal at a period of 35.419 [35.378, 35.464] days likely corresponding to the photometric periodicity. We demonstrate the presence of this signal in the velocities in Fig. \ref{fig:GJ205_signal_search} (top panel). However, this was not the only signal we detected in the GJ 205 velocity data. We also discovered another signal at a period of 16.937 [16.916, 16.950] days with an amplitude of 3.60 [2.32, 4.87] ms$^{-1}$ in the combined data that has no photometric counterparts (Fig. \ref{fig:GJ205_signal_search}, middle panel). The search for a third signal in the combined data revealed a significant probability maximum at a period of 270.8 [261.9, 276.5] d with a local maximum at roughly 11 days (Fig. \ref{fig:GJ205_signal_search}). This third signal, with an amplitude of 1.99 [0.68, 3.30] ms$^{-1}$, satisfies our signal detection criteria and does not appear to have counterparts in the photometry of the activity indicators either. We have plotted the radial velocities folded on the phases of the signals in Fig. \ref{fig:GJ205_signals} for visual inspection.

Although the third signal, at a period of 270 days, is not entirely unique with a local maximum at a period of 11 d exceeding the 1\% probability threshold of the global maximum, we interpret it as a candidate planet orbiting the star. We note that additional observations are required to confirm this candidate but as the signal satisfies our detection criteria, and as it has no counterparts in the activity indicators or photometry, its interpretation as a planetary signal is justified. We thus conclude that there is evidence for two candidate planets orbiting GJ 205. With minimum masses of 10.3 [6.3, 14.7] and 13.8 [4.1, 24.5] M$_{\oplus}$, respectively. These two candidates are classified as a hot Neptune an a cool Neptune.

\clearpage

\subsection{GJ 208}\label{sec:GJ208}

\begin{figure}
\center
\includegraphics[angle=270, width=0.49\textwidth,clip]{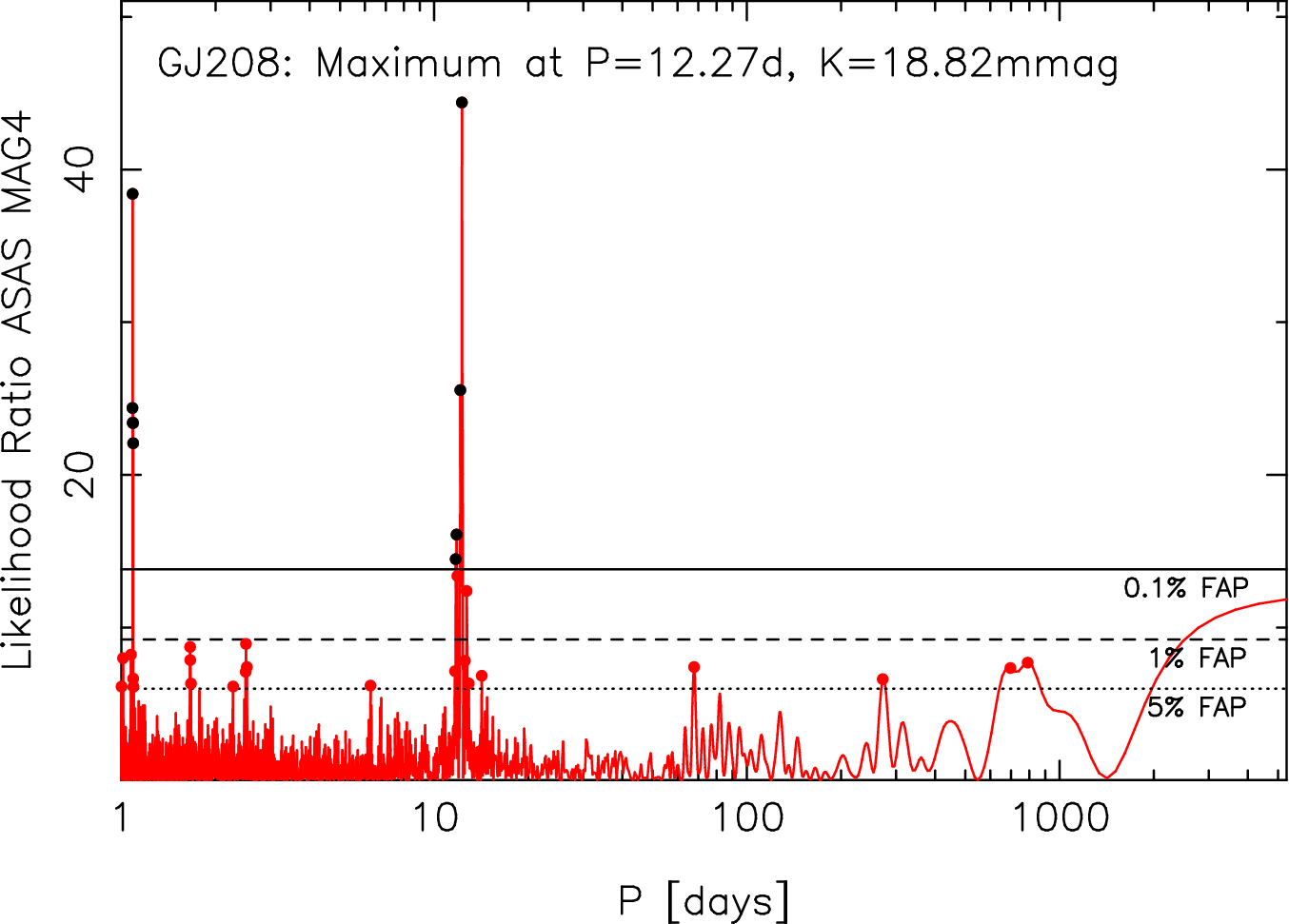}
\caption{Likelihood-ratio periodogram of the ASAS V-band photometry of GJ 208.}\label{fig:GJ208_asas}
\end{figure}

\begin{figure}
\center
\includegraphics[angle=270, width=0.49\textwidth,clip]{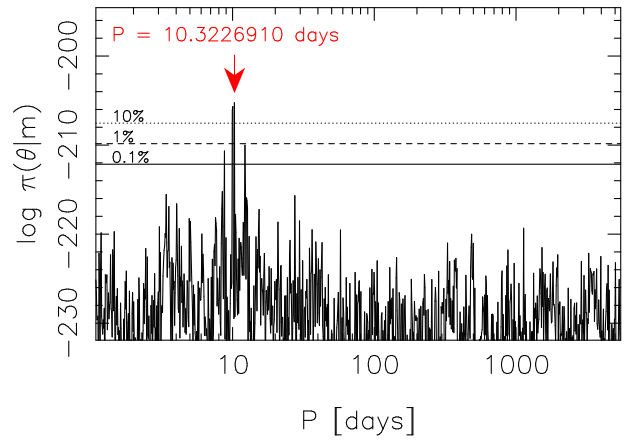}
\caption{As in Fig. \ref{fig:GJ27.1_period_search} but for the signal in the combined HARPS and HIRES velocities of GJ 208.}\label{fig:GJ208_psearch}
\end{figure}

The ASAS V-band photometry data ($N=$437) of GJ 208 (HD 245409, HIP 26335, V2689 Ori) contains a strong signal indicative of stellar photometric rotation at a period of 12.28 days (Fig. \ref{fig:GJ208_asas}). \citet{kiraga2012} report a consistent periodicity of 12.04 d based on a larger set of 450 (619) ASAS V-band (I-band) observations. We could also observe a counterpart of this likely rotation-induced signal in the combined HARPS and HIRES radial velocities (Fig. \ref{fig:GJ208_psearch}). This signal was detected amid elevated white noise levels in the data -- we estimated the jitter to have a value of 5.05 [3.18, 6.92] ms$^{-1}$, which is significantly larger than the median radial velocity jitter of the sample stars of 2.26 ms$^{-1}$. It is thus our interpretation that these periodicities indeed imply that the star has a rotation period at or near the photometric periodicity.

GJ 208 provides one of the clearest examples of a periodic signal in the radial velocity data for which the best explanation is that it is caused by the co-rotation of active and/or inactive regions on the stellar surface and has a substantial photometric counterpart.

\clearpage

\subsection{GJ 221}\label{sec:GJ221}

One of the most interesting stars in our sample is the K7V or M0V dwarf GJ 221 (HIP 27803). It has been reported to be a host to two planets with orbital periods of 125.06$\pm$1.10 and 3.8731$\pm$0.0007 days in two independent studies based on HARPS data \citep{locurto2013} and combined HARPS and PFS data \citep{arriagada2013}. A third candidate planet was later reported by \citet{tuomi2014b} with an orbital period of 496 [474, 525] days. While we had no difficulties in identifying these signals in the data, our analyses show that three Keplerian signals are not sufficient in modelling the radial velocity variations of the combined HARPS ($N=109$) and PFS ($N=38$) data.

\begin{figure}
\center
\includegraphics[angle=270, width=0.49\textwidth,clip]{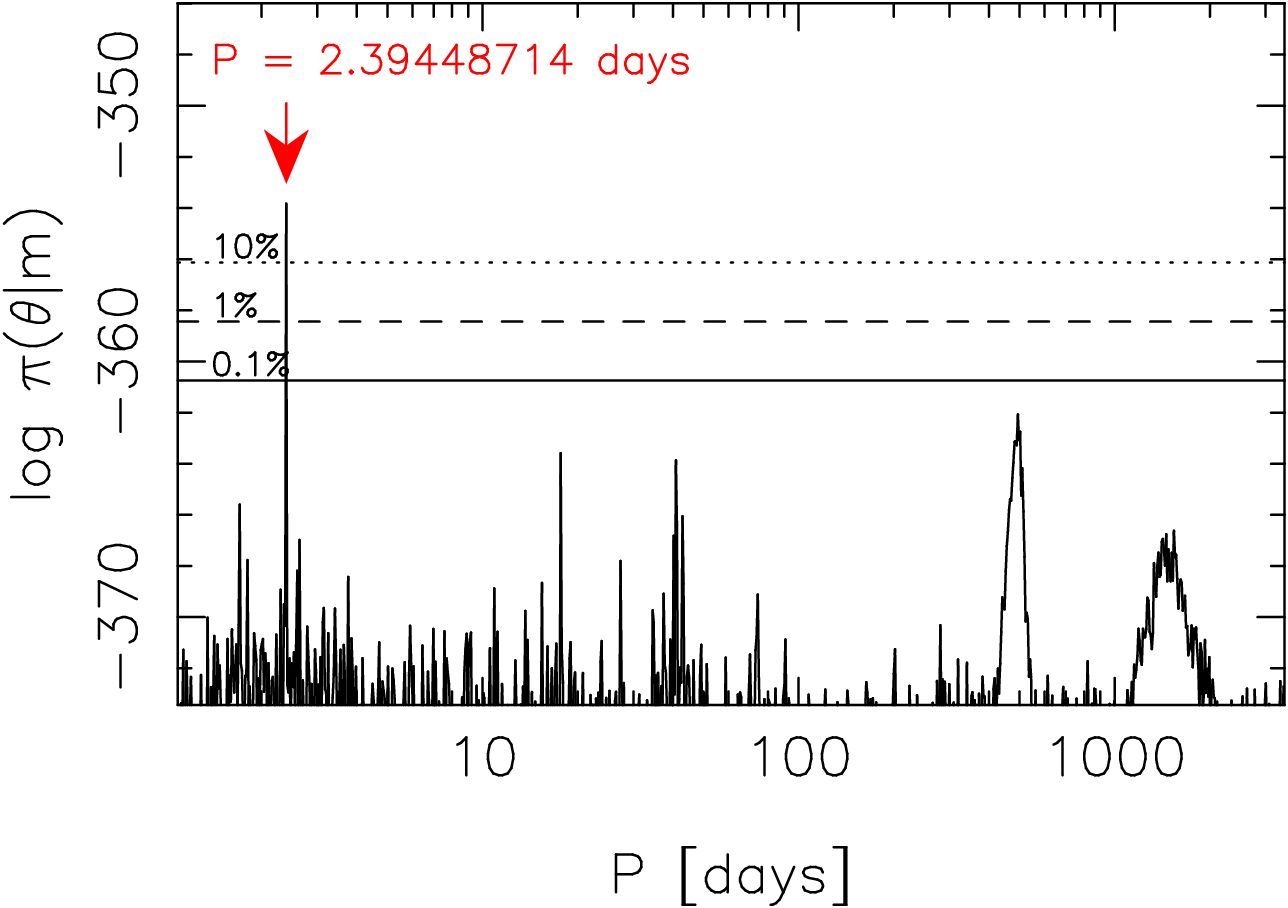}

\includegraphics[angle=270, width=0.49\textwidth,clip]{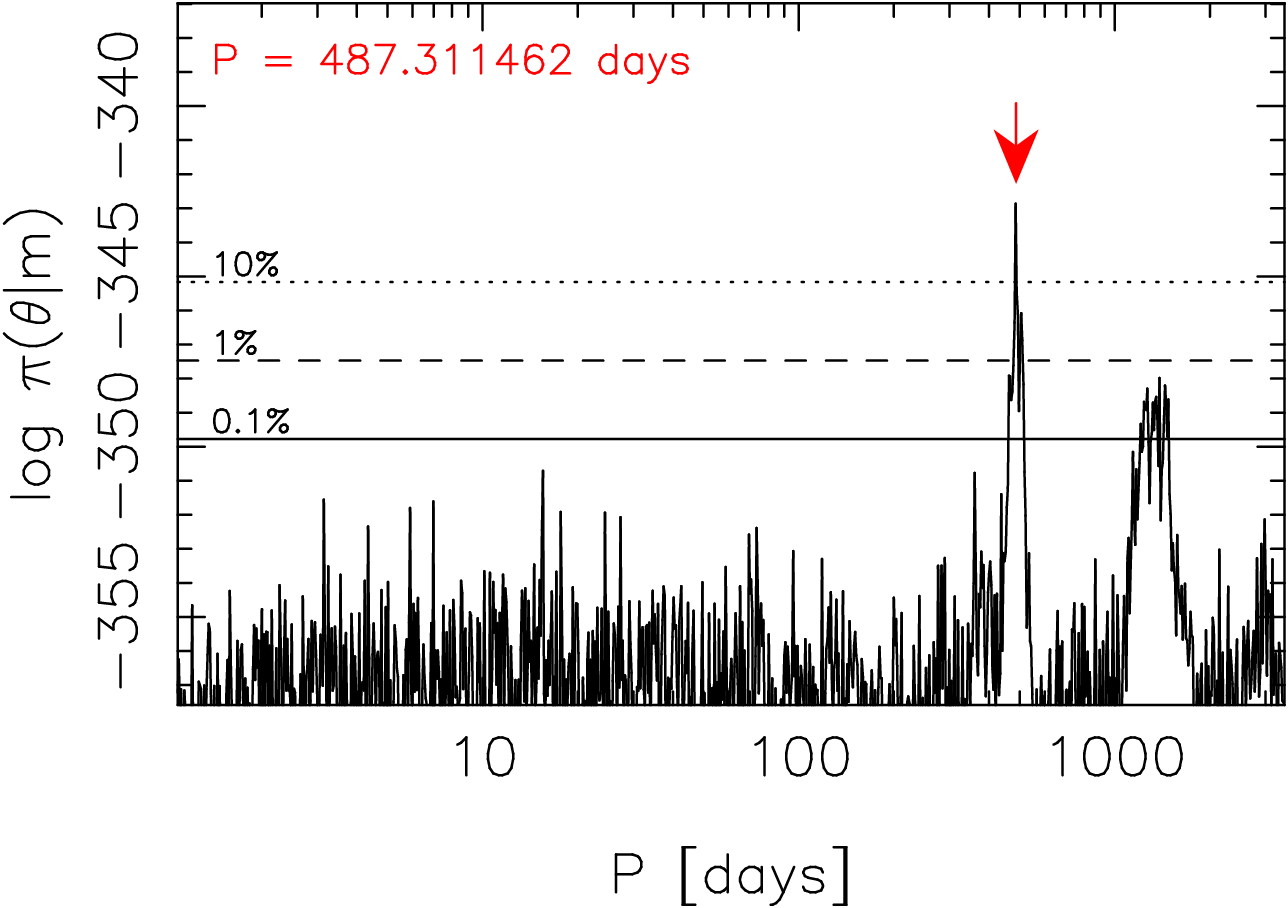}
\caption{Estimated posterior probability density for GJ 221 as a function of the period parameter of the $k$th signal for models with three (top panel) and four (bottom panel) Keplerian signals. The horizontal lines denote the 10\% (dotted), 1\% (dashed), and 0.1\% (solid) equiprobability thresholds with respect to the maxima highlighted with the red arrows.}\label{fig:GJ221_psearch}
\end{figure}

\begin{figure}
\center
\includegraphics[angle=270, width=0.49\textwidth,clip]{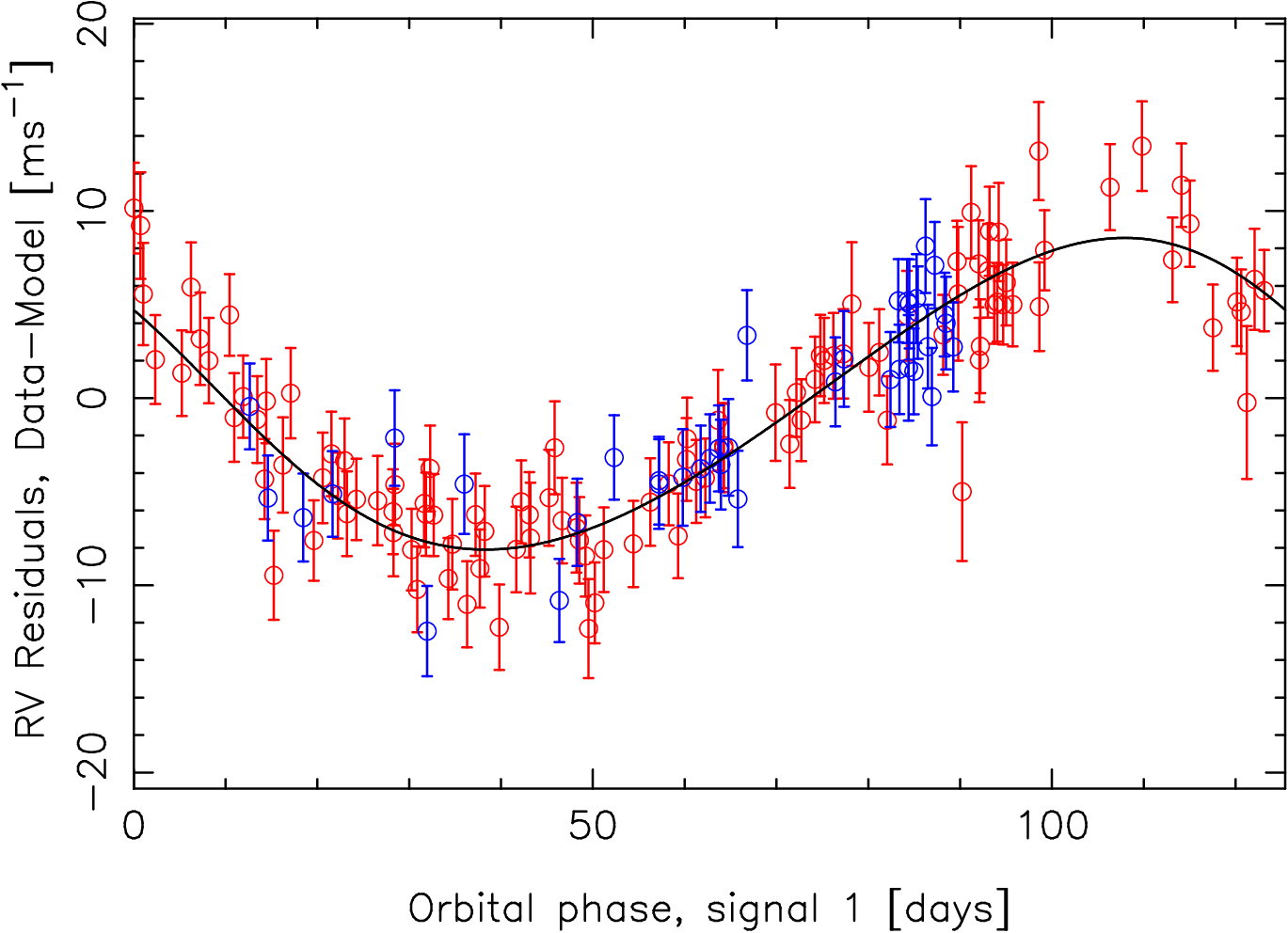}
\includegraphics[angle=270, width=0.49\textwidth,clip]{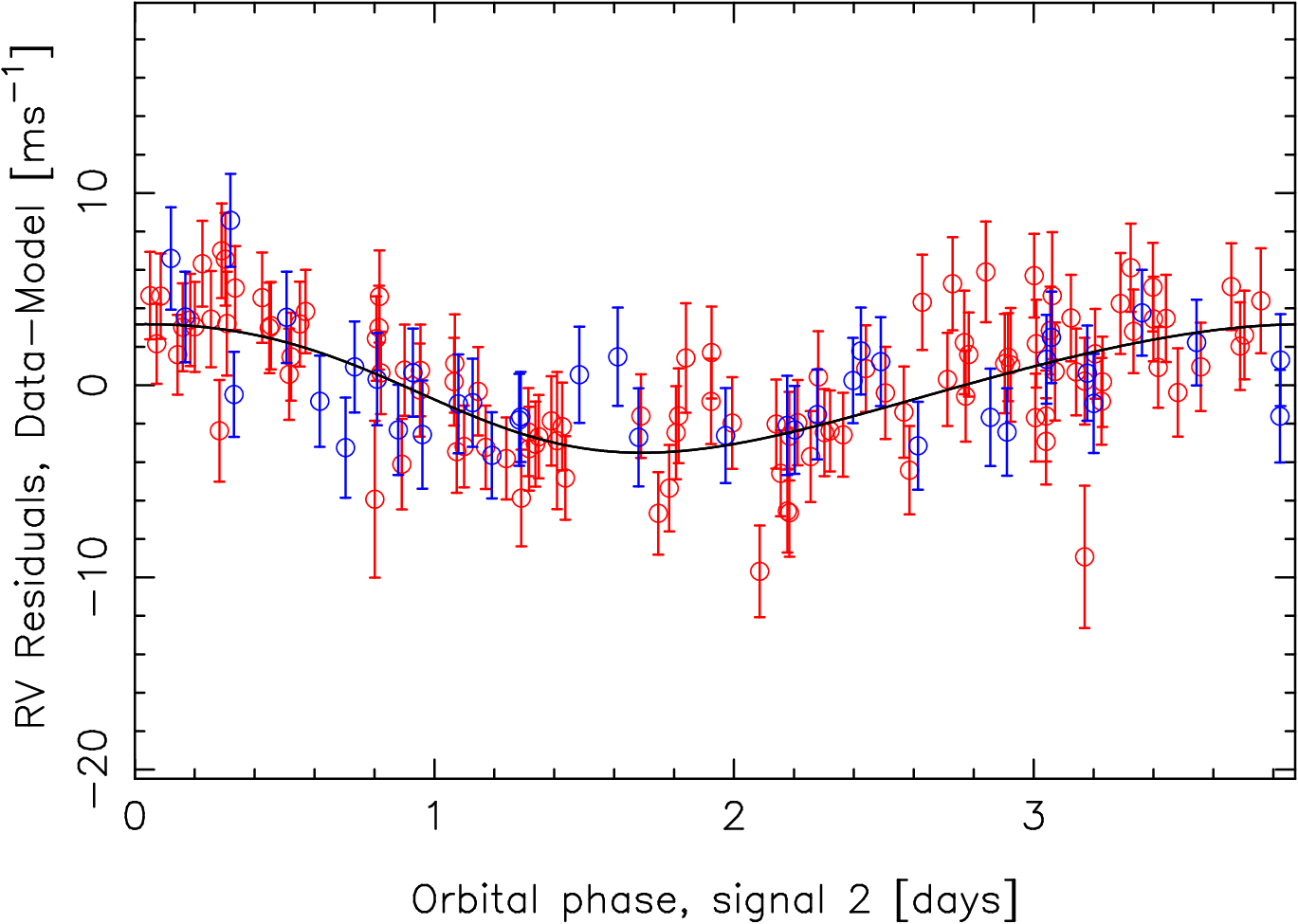}

\includegraphics[angle=270, width=0.49\textwidth,clip]{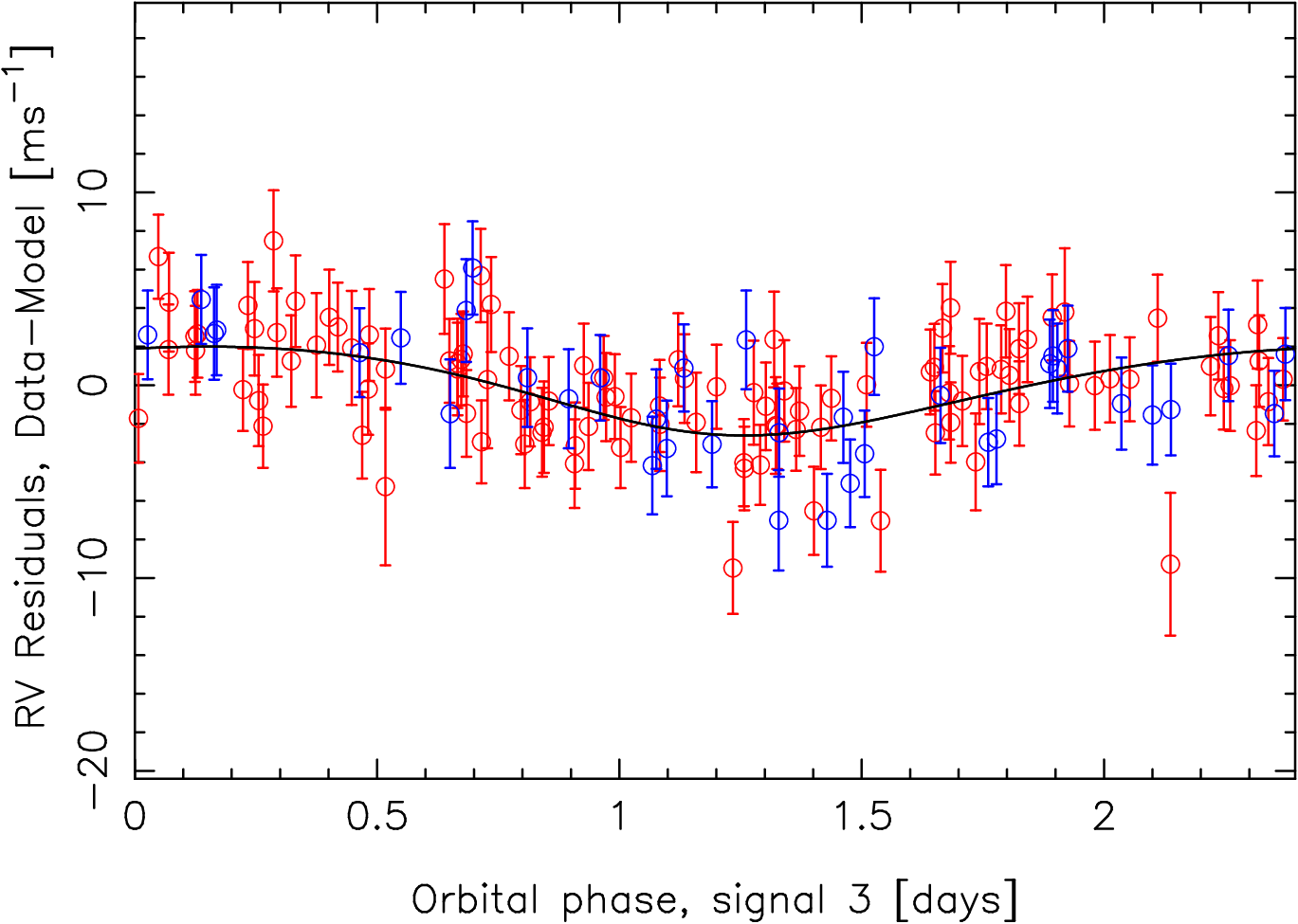}
\includegraphics[angle=270, width=0.49\textwidth,clip]{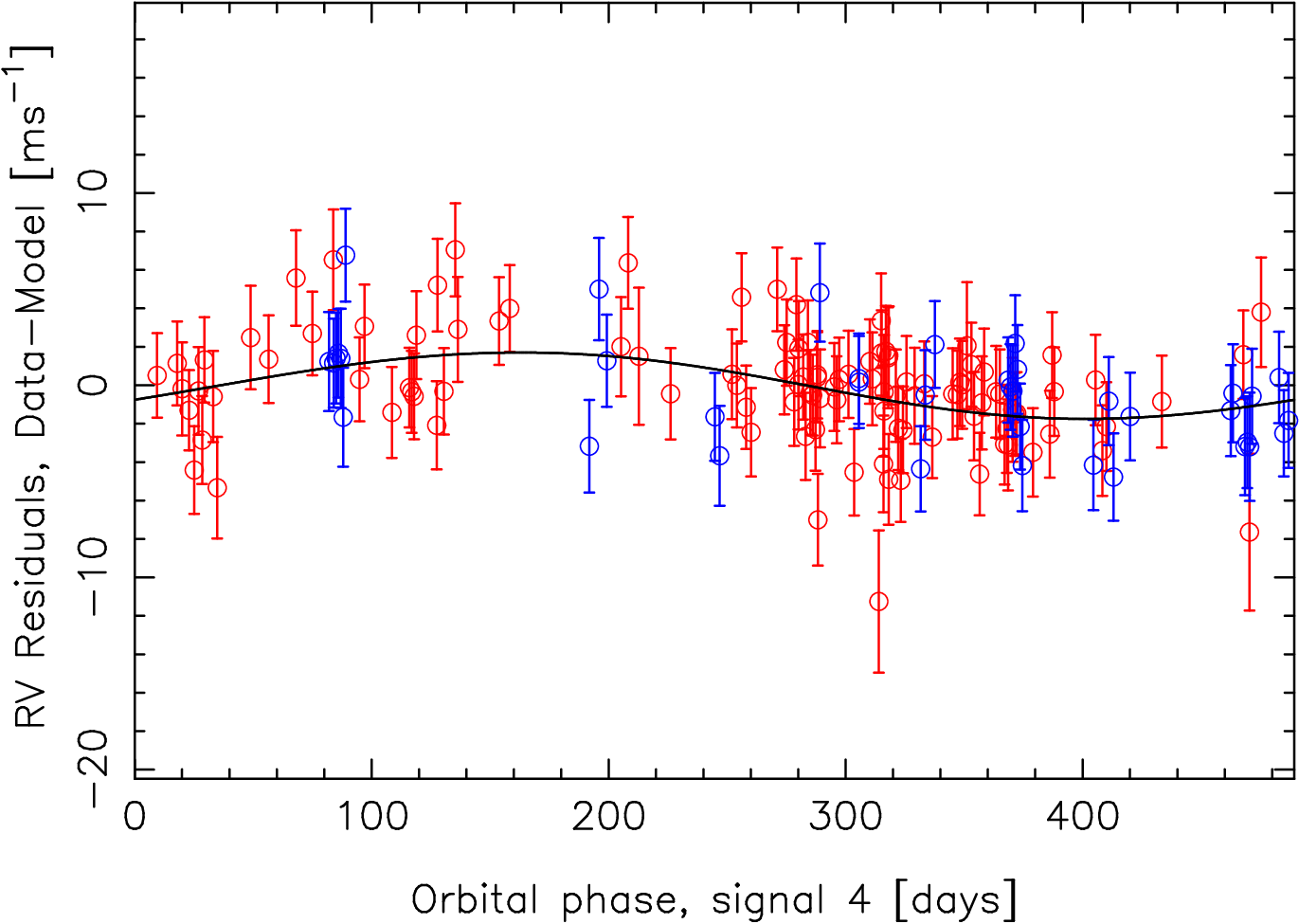}
\caption{HARPS (red) and PFS (blue) radial velocities of GJ 221 folded on the periods of the four signals in the combined data with the other signals subtracted from each panel.}\label{fig:GJ221_curves}
\end{figure}

The 125-day and 3.9-day signals were easily detected in the data, and because they were already presented by \citet{arriagada2013} and \citet{locurto2013} and verified by \citet{tuomi2014b} by using the planet detection criteria also applied in the current work, we omit the corresponding searches for periodicity in this brief re-visit of the star's radial velocities. We present the DRAM samplings in a search for additional signals in Fig. \ref{fig:GJ221_psearch}. We detect two additional signals in the combined HARPS and PFS data at periods of 2.39456 [2.39415, 2.39498] and 485.4 [460.6, 512.9] days, respectively. The latter one of these was reported by \citet{tuomi2014b} but their sampling technique was unable to spot very narrow probability maxima such as that one corresponding to the signal at the period of 2.39 days presented in Fig. \ref{fig:GJ221_psearch} (top panel). As all these four signals are independently present in the data, it is our interpretation that they all correspond to Keplerian signals of candidate planets orbiting the star. We visualise the four signals in the data by showing the phase-folded radial velocities in Fig. \ref{fig:GJ221_curves}. We did not detect any additional signals in the data, although there was an isolated probability maximum in the 5-Keplerian model at a period of 17 days that did not quite satisfy our detection criteria.

To increase the credibility of the four signals in the GJ 221 data as Keplerian signals of candidate planets, we calculated the likelihood-ratio periodograms of the HARPS activity indices. There were no significant periodicities in these indices suggesting that stellar activity does not contribute strongly to any of the observed radial velocity periodicities. We also obtained 475 ASAS V-band photometry measurements but failed to find any periodic signals in them either. This strenghtens the interpretation that the four signals indeed correspond to planet candidates. We thus interpret the signals as one 47.6 [36.6, 58.6] M$_{\oplus}$ minimum mass warm super-Neptune, two hot super-Earths with minimum masses of 5.7 [3.7, 7.7] and 3.3 [1.9, 4.9] M$_{\oplus}$ (at the periods of 3.9 and 2.4 days, respectively), and one cold Neptune with a minimum mass of 19.2 [9.2, 30.4] M$_{\oplus}$.

\clearpage

\subsection{GJ 229}

\begin{figure}
\center
\includegraphics[angle=270, width=0.49\textwidth,clip]{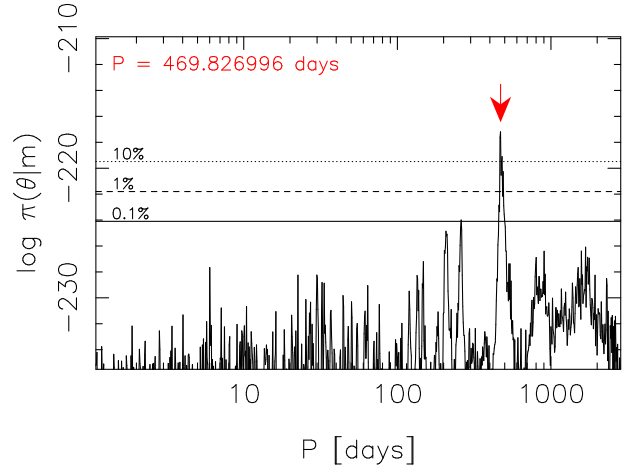}
\caption{As in Fig. \ref{fig:GJ27.1_period_search} but for the signal in the combined HARPS and UVES velocities of GJ 229 as discussed in \citet{tuomi2014}.}\label{fig:GJ229_period_search}
\end{figure}

\citet{tuomi2014} reported a signal in the HARPS and UVES radial velocities of GJ 229 (HD 42581, HIP 2929) and interpreted it as a cool super-Neptune orbiting the star, if classified according to the classification system presented in the current work. We re-analysed the data available to \citet{tuomi2014} by accounting for the correlations between the HARPS velocities and the three activity indicators -- BIS, FWHM, and S-index. This revealed the same signal that \citet{tuomi2014} reported at a period of 471 days (Fig. \ref{fig:GJ229_period_search}). However, when updating the data set with additional HARPS data and HIRES data set, the same signal could no longer be obtained as a solution, casting doubt on its origin.

In fact, we suspect that the mistake in the UVES barycentric correction (see Section \ref{sec:UVES}) might have been responsible for the signal reported by \citet{tuomi2014}. We thus assume that the UVES data of GJ 229 is biased and base our conclusions on only HARPS and HIRES radial velocities of the star.

\begin{figure}
\center
\includegraphics[angle=270, width=0.49\textwidth,clip]{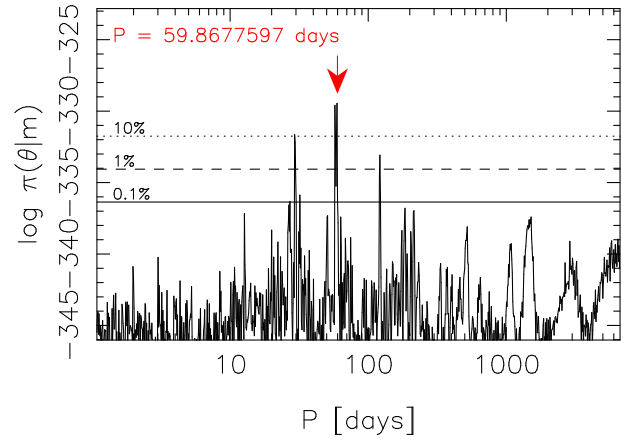}
\caption{As in Fig. \ref{fig:GJ27.1_period_search} but for the signal in the combined HARPS and HIRES velocities of GJ 229.}\label{fig:GJ229_period_search2}
\end{figure}

The analysis of the combined HARPS and HIRES radial velocities resulted in conclusions differing from those reported by \citet{tuomi2014}. We could not obtain any evidence in favour of the signal at a period of 471 days, which strongly suggests that it was indeed primarily caused by a bias in the UVES velocities or, possibly, stellar activity-induced cycle. Instead, we obtained hints of signals at periods of approximately 30, 60, and 120 days (Fig. \ref{fig:GJ229_period_search2}). However, none of these probability maxima were strong enough to comply with our signal detection criteria.

We analysed the set of 323 ASAS V-band photometry measurements but could not identify any periodic signals that could link the suggestive signals in radial velocities to photometric variations. Similarly, we analysed the HARPS and HIRES activity indicators and failed to find any signs of periodicities in them. Furthermore, although the HARPS velocities appear to be independent of the respective activity indicators, the HIRES radial velocities are weakly connected to the corresponding S-indices and we estimate that the parameter $c_{\rm S}$ has a value of 4.5 [-1.9, 10.9] ms$^{-1}$, which indicates that it is significantly different from zero with 95\% but not with 99\% credibility. We thus conclude that there is no credible evidence in favour of candidate planets orbiting GJ 229. If the signal reported by \citet{tuomi2014} was indeed caused by biases in the UVES data, e.g. due to mistakes in the barycentric correction, it should not be there when re-processing the UVES spectra and calculating the radial velocities from scratch.

\clearpage

\subsection{GJ 251}\label{sec:GJ251}

\begin{figure}
\center
\includegraphics[angle=270, width=0.49\textwidth]{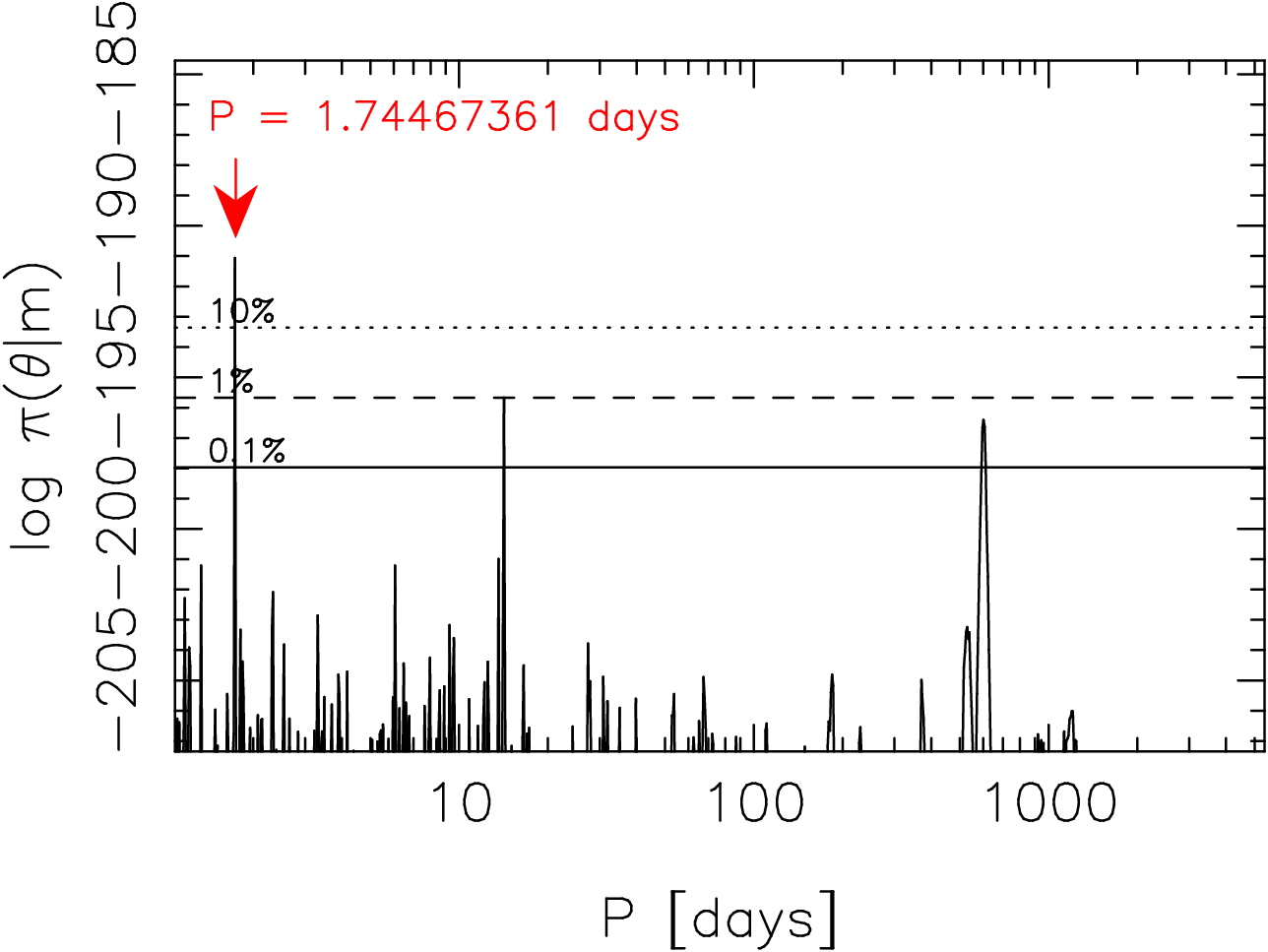}

\includegraphics[angle=270, width=0.49\textwidth]{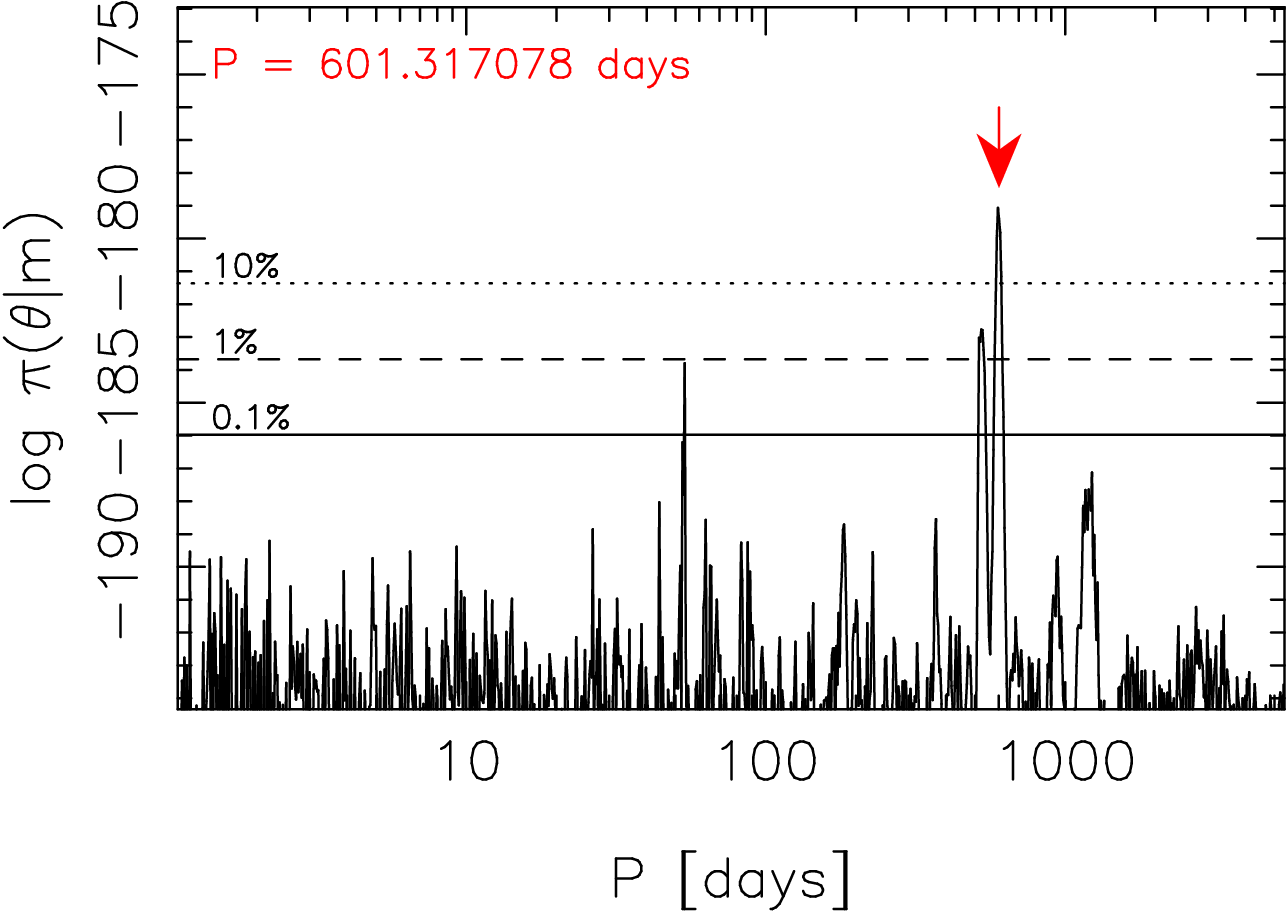}
\caption{Estimated posterior probability densities given GJ 251 data as functions the period parameter of a model with one Keplerian signal (top panel) and the period parameter of the second signal of the two-Keplerian model. The horizontal lined denote the 10\% (dotted), 1\% (dashed) and 0.1\% (solid) equiprobability thresholds with respect to the global maxima indicated by red arrows.}\label{fig:GJ251_signal_search}
\end{figure}

\begin{figure}
\center
\includegraphics[angle=270, width=0.49\textwidth]{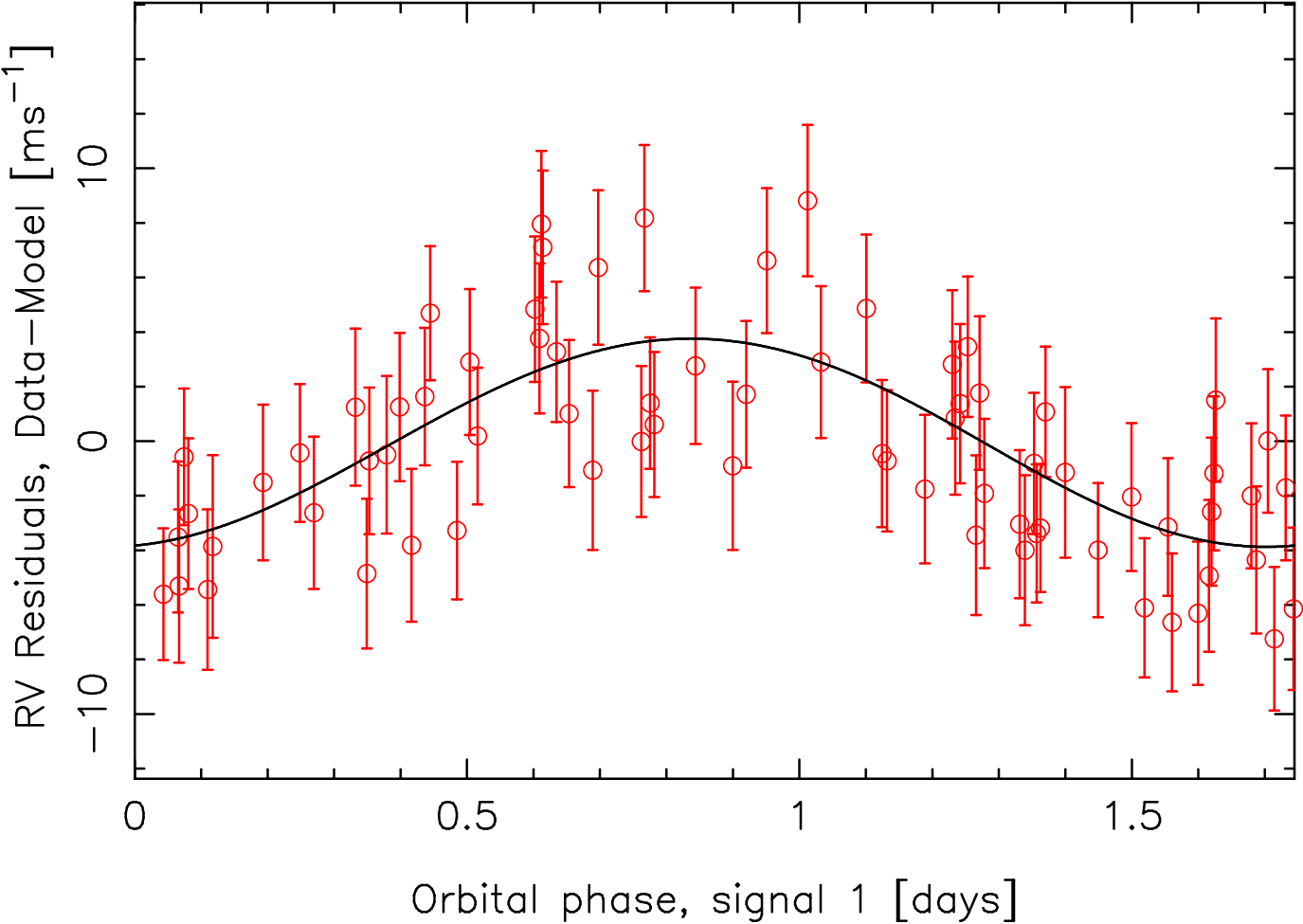}

\includegraphics[angle=270, width=0.49\textwidth]{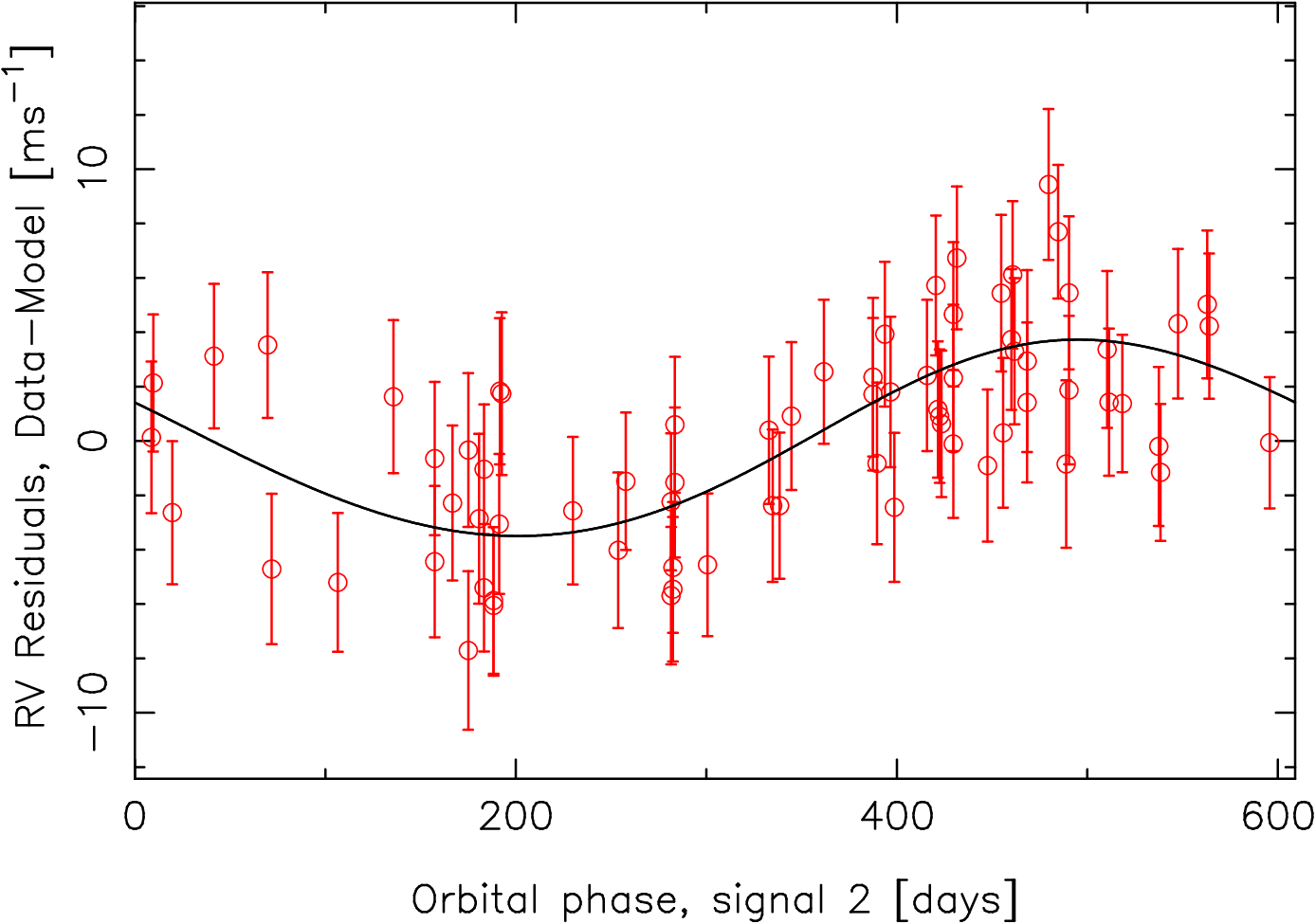}
\caption{HIRES radial velocities folded on the phases of the two signals detected in the GJ 251 data. The other signal has been subtracted from each panel.}\label{fig:GJ251_signals}
\end{figure}

We obtained evidence for two signals in the HIRES radial velocities of GJ 251 (HD265866, HIP 33226) at periods of 1.74476 [1.74459, 1.74498] and 607 [588, 623] days \citep[see also][]{butler2016}. These two signals, with amplitudes of 3.66 [2.22, 5.12] and 3.49 [1.74, 5.23] ms$^{-1}$, respectively, were identified as unique posterior probability maxima in the period spaces of the one- and two-Keplerian models (Fig. \ref{fig:GJ251_signal_search}). There were three prominent maxima in the period space of the one-Keplerian model (Fig. \ref{fig:GJ251_signal_search}, top panel) corresponding to the two independent signals and a third maximum at a period of 13.69 days corresponding to an alias caused by daily sampling in the data (it occurs in the frequency space at $f = f_{s} - 2f_{a}$, where $f_{s}$ is the short-period signal and $f_{a}$ corresponds to the daily aliasing frequency). Similarly, the signal at a period of 607 days also has a local maximum at a period of 520 days due to aliasing -- the two maxima are thus representative of the same periodic phenomenon. We show the HIRES radial velocities folded on the phases of the signals in Fig. \ref{fig:GJ251_signals}.

The HIRES S-indices did not contain any periodicities that could have been interpreted as counterparts of the radial velocity signals. This suggests that the signals correspond to Doppler periodicities of Keplerian origin. Although we could not study the photometric lightcurve of the star because no ASAS photometry was available, we interpret the two signals as candidate planets with minimum masses of 3.3 [1.8, 5.0] and 22.2 [10.2, 35.5] M$_{\oplus}$ orbiting the star.

\clearpage

\subsection{GJ 273}\label{sec:GJ273}

\begin{figure}
\center
\includegraphics[angle=270, width=0.49\textwidth,clip]{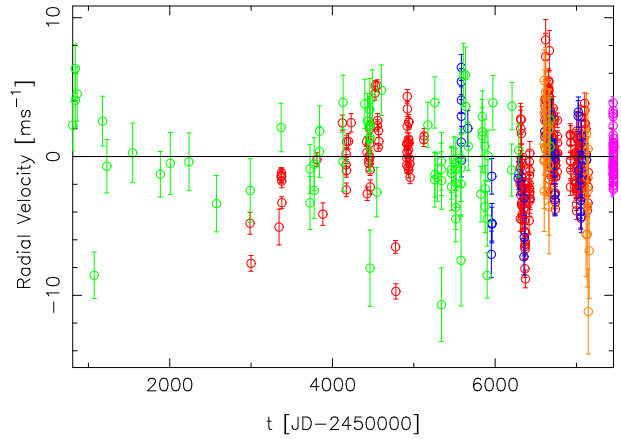}
\caption{Radial velocities of GJ 273 from HARPS (red), HIRES (green), PFS (blue), APF (orange), and HARPN (purple).}\label{fig:GJ273_data}
\end{figure}

\begin{figure}
\center
\includegraphics[angle=270, width=0.49\textwidth]{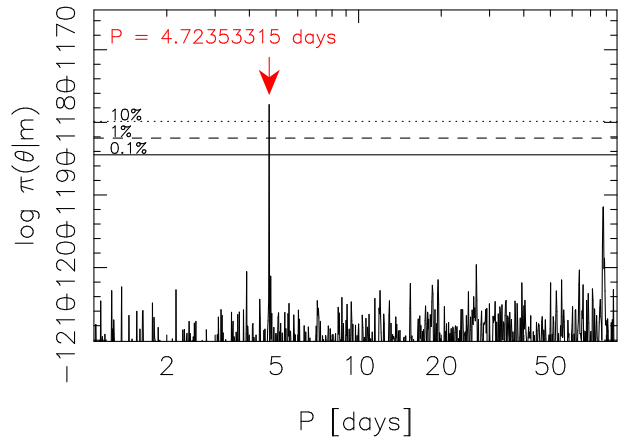}
\caption{Estimated posterior probability density given GJ 273 data as a function of the period of the Keplerian signal in a one-Keplerian model. The red arrow indicates the global probability maximum and the horizontal lines denote the 10\% (dotted), 1\% (dashed), and 0.1\% (solid) equiprobability thresholds with respect to the maximum.}\label{fig:GJ273_psearch1}
\end{figure}

GJ 273, also known as Luyten's star, has been intensively observed by HARPS ($N=277$), HIRES ($N=75$), and PFS ($N=36$). Moreover, we also obtained 27 velocities from the APF from two observing runs approximately one year apart as well as another 51 velocities from HARPS-N covering a single 5-day oberving run. The resulting set of 466 radial velocities (Fig. \ref{fig:GJ273_data}) showed evidence for several periodic signals, as also reported by \citet{astudillo2017}. The most significant was a signal at a period of 4.72322 [4.72240, 4.72403] days with an amplitude of 1.10 [0.78, 1.41] ms$^{-1}$ (Fig. \ref{fig:GJ273_psearch1}).

\begin{figure}
\center
\includegraphics[angle=270, width=0.49\textwidth]{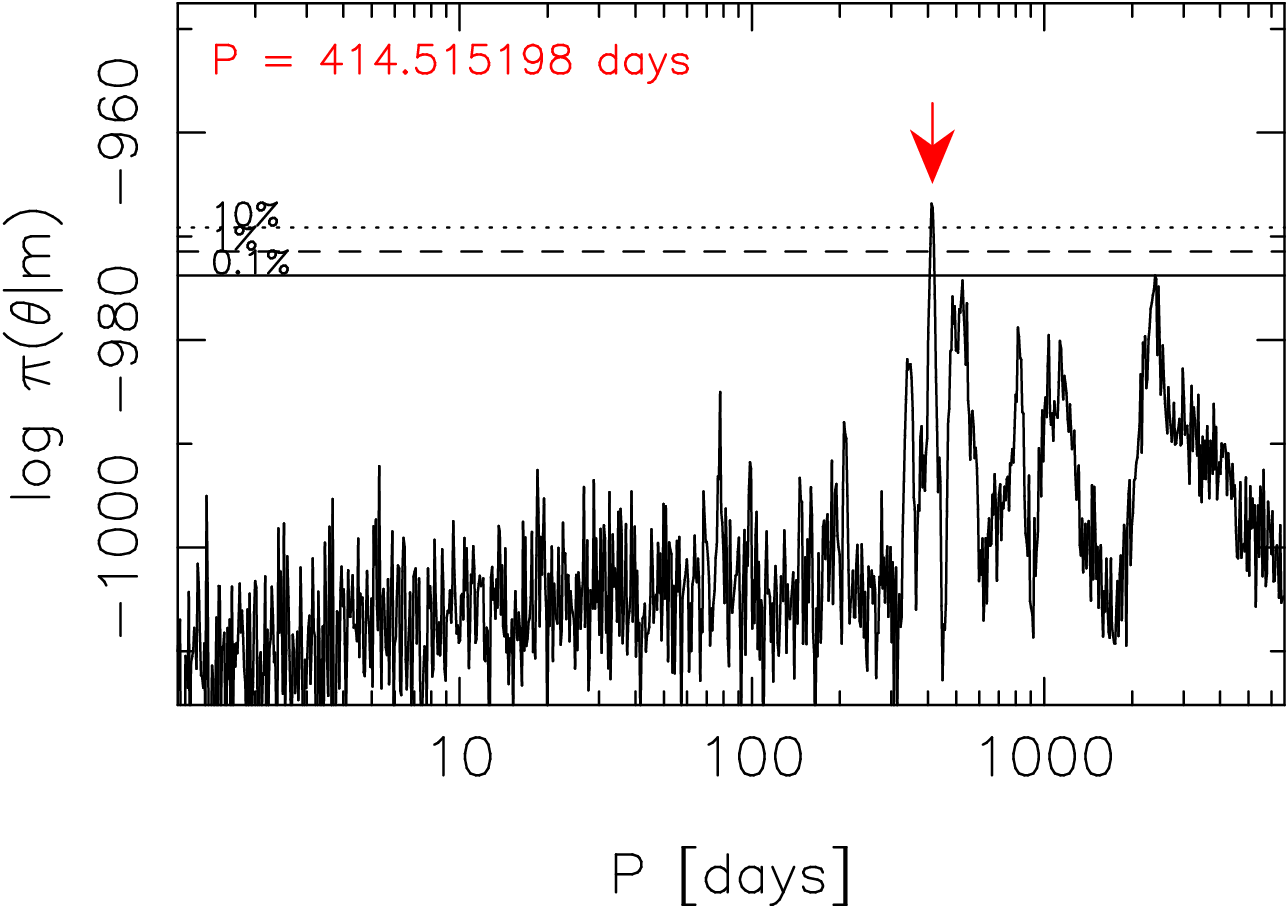}

\includegraphics[angle=270, width=0.49\textwidth]{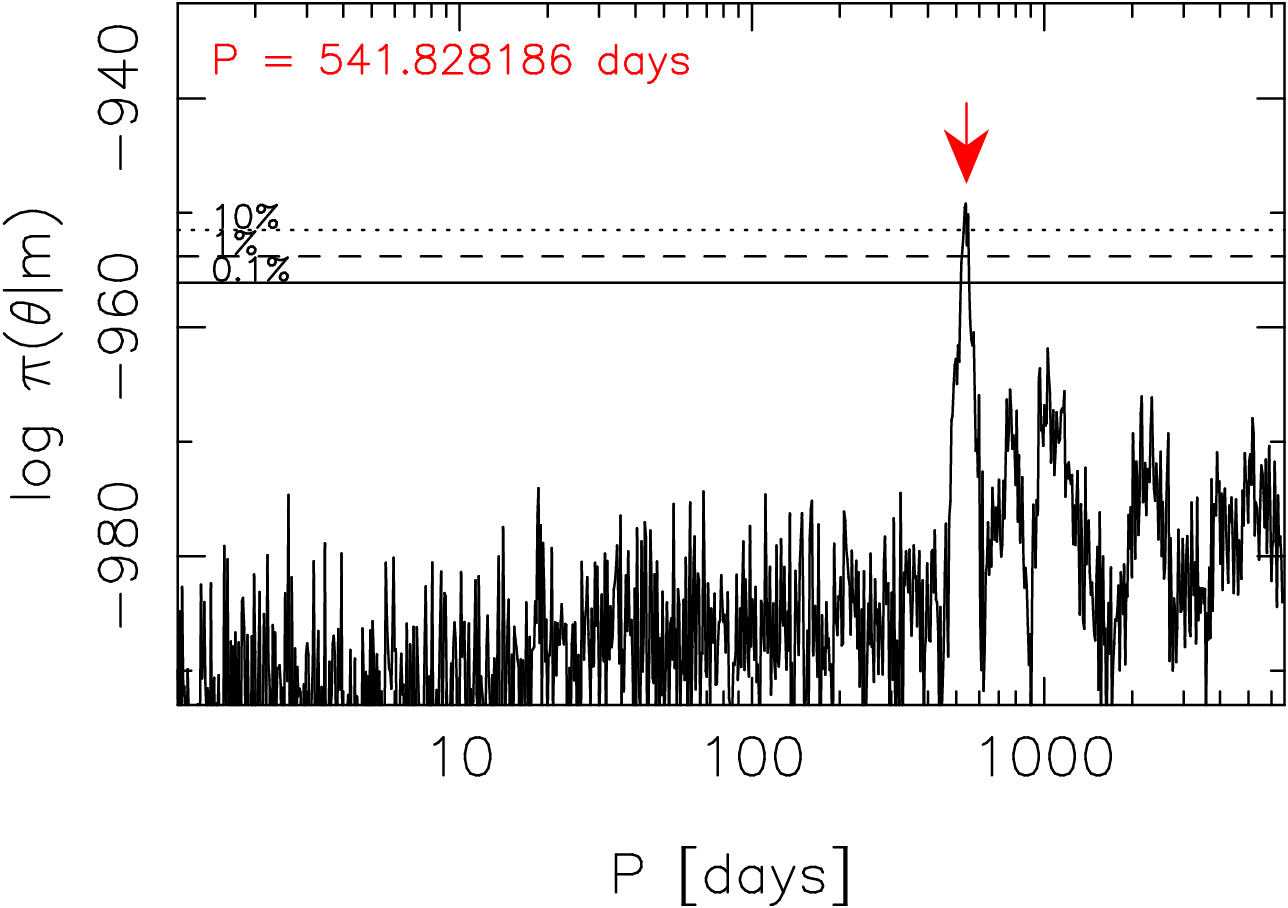}
\caption{As in Fig. \ref{fig:GJ273_psearch1} but for the second (top panel) and third (bottom panel) signals in models containing two and three Keplerian signals, respectively.}\label{fig:GJ273_psearch2}
\end{figure}

\begin{figure}
\center
\includegraphics[angle=270, width=0.49\textwidth]{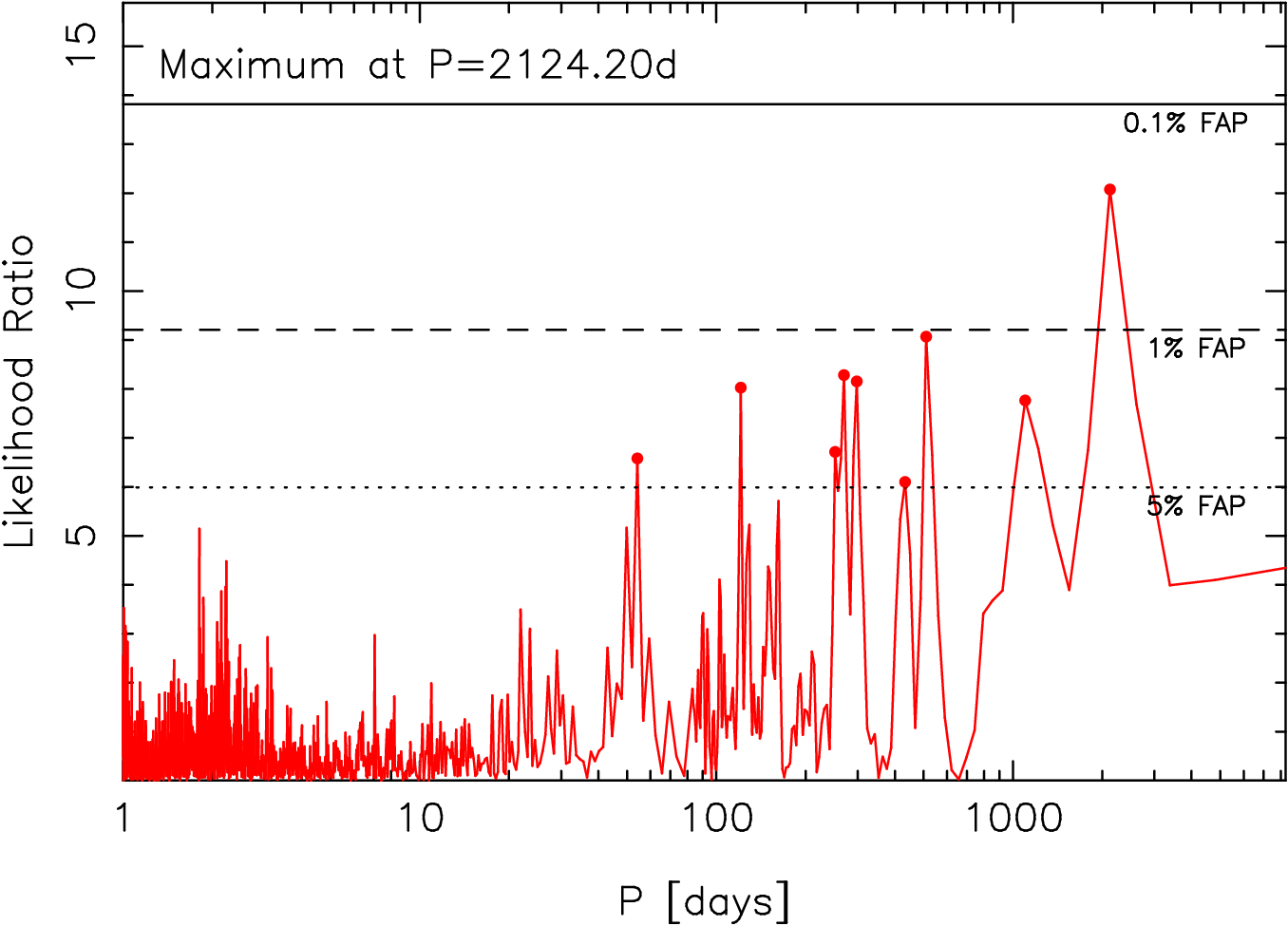}
\includegraphics[angle=270, width=0.49\textwidth]{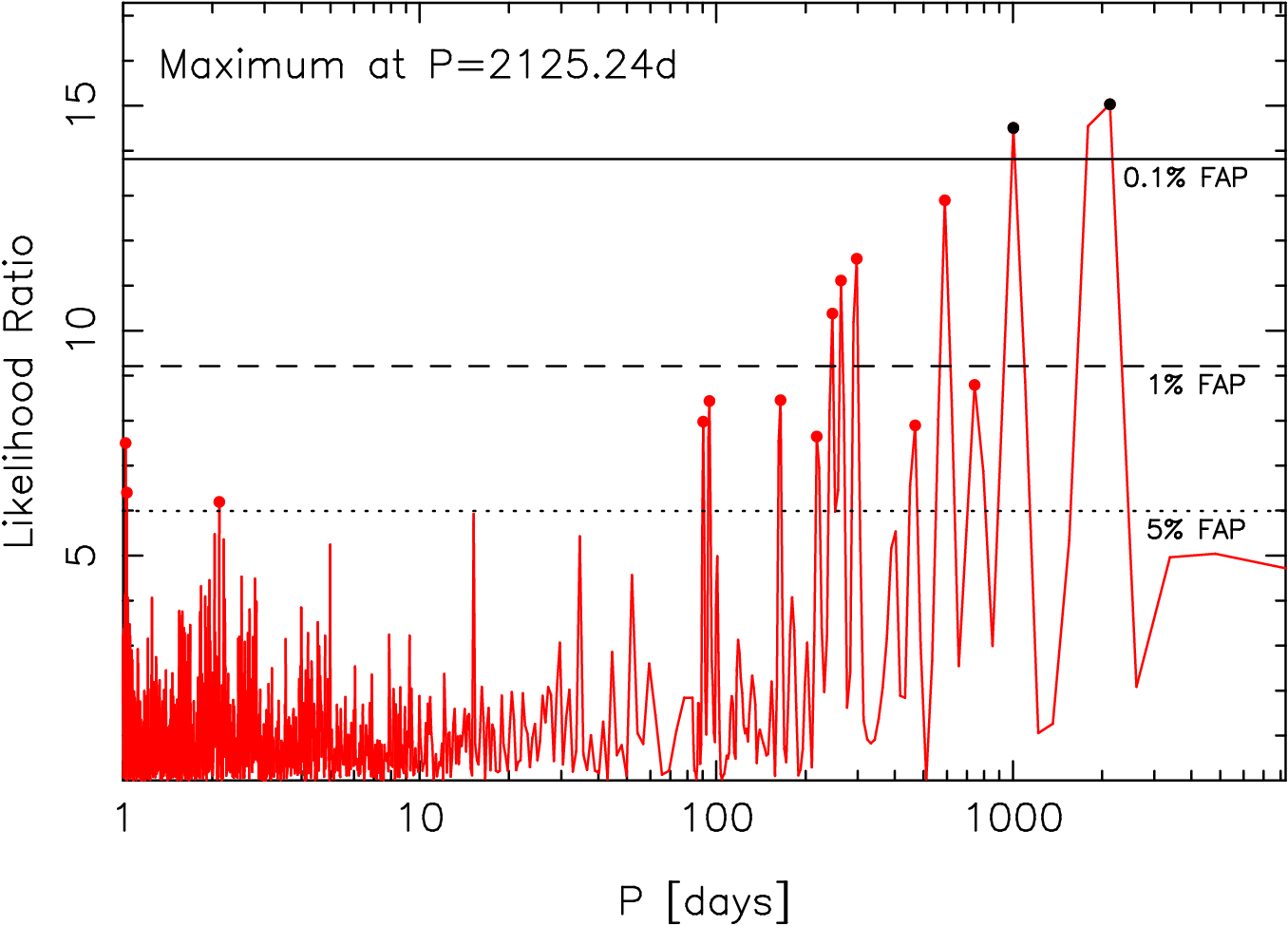}

\includegraphics[angle=270, width=0.49\textwidth]{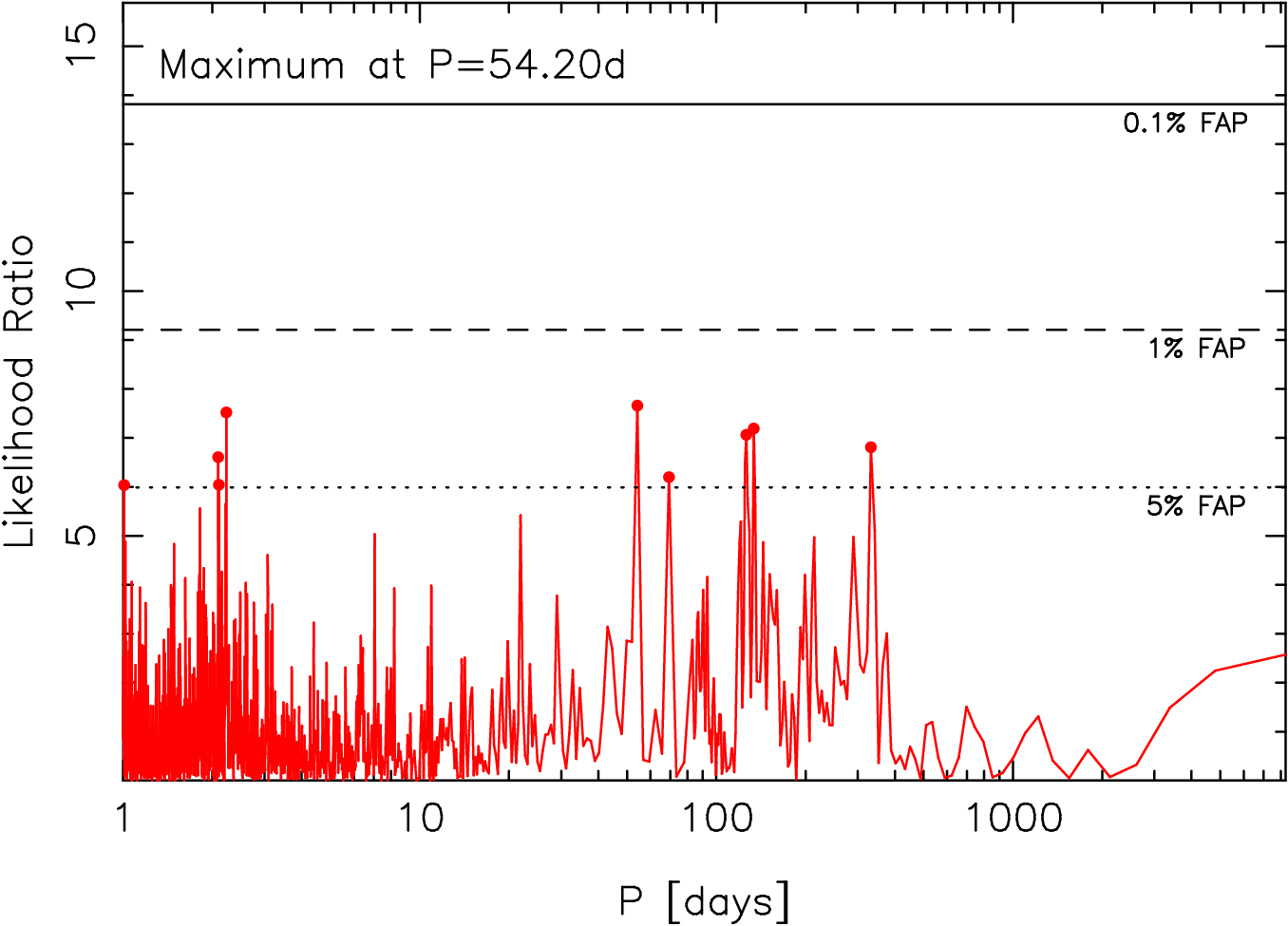}
\includegraphics[angle=270, width=0.49\textwidth]{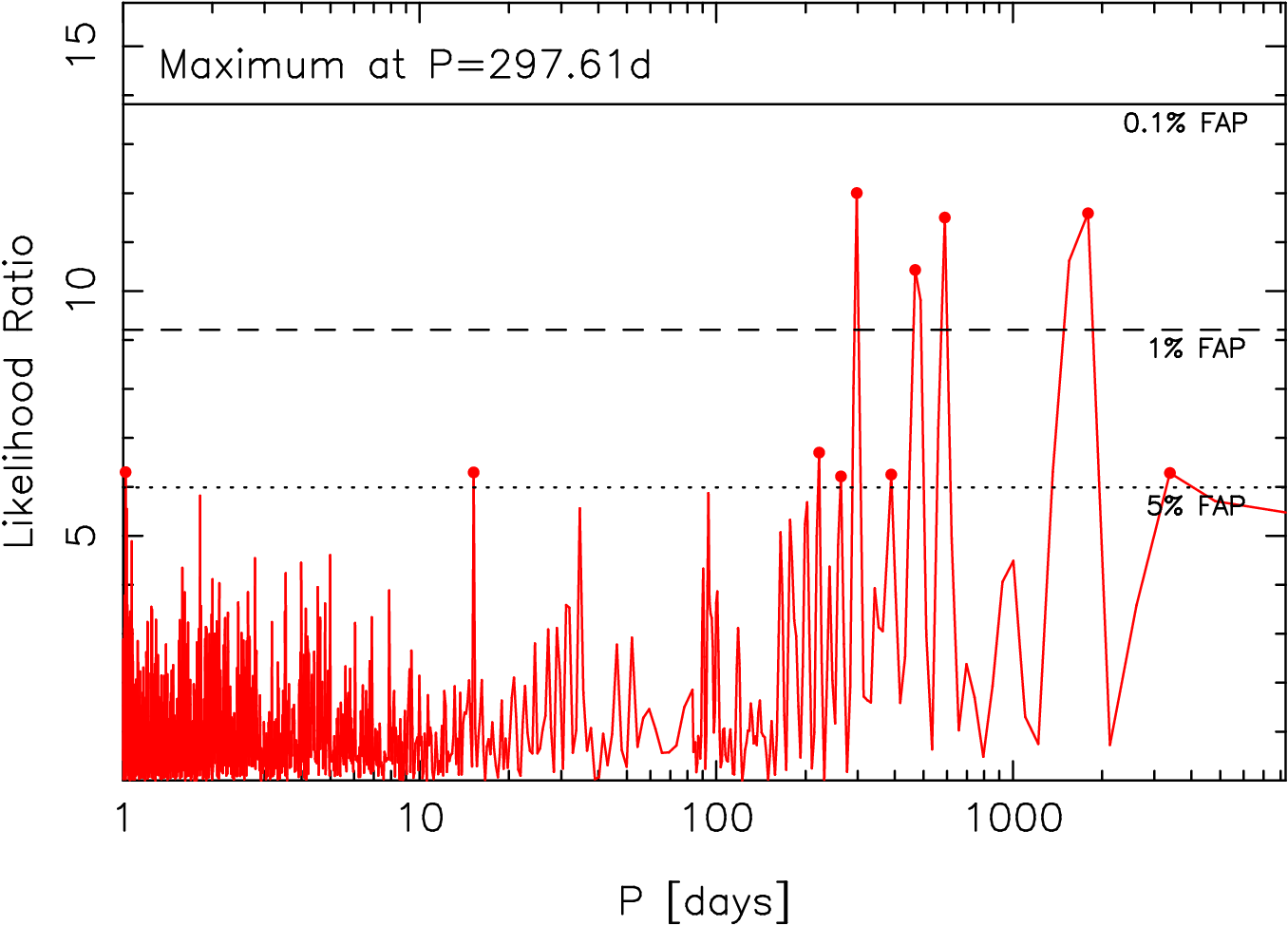}
\caption{Likelihood-ratio periodograms of HARPS FWHM (left panels) and S-index (right panels). The bottom panels show residual periodograms after subtracting the most significant periodicities.}\label{fig:GJ273_activity}
\end{figure}

The combined data set also showed evidence for longer periodicities at periods of 413.9 [408.4, 418.3] and 541.5 [525.7, 557.3] days with amplitudes of 2.15 [1.54, 2.76] and 1.73 [1.07, 2.40] ms$^{-1}$, respectively (Fig. \ref{fig:GJ273_psearch2}). Should these signals correspond to planets orbiting the star, such a system could be stable due to mean motion resonances -- a potential 4:3 resonance in this particular case with a period ratio of $1.31\pm0.06$. However, such dynamical analyses are beyond the current work and we did not analyse the dynamical interactions of such a system but leave it for future work. Furthermore, we also observe weak evidence for activity-induced variability at nearby periods making it possible that these two radial velocity signals are caused by stellar activity. We show the likelihood ratio periodograms of GJ 273 HARPS activity indices in Fig. \ref{fig:GJ273_activity}. These periodograms demonstrate that there might be activity-induced variability at periods ranging from 100 to 2000 days but it is difficult to tell whether the radial velocity signals actually have counterparts in the activity data.

\begin{figure}
\center
\includegraphics[angle=270, width=0.49\textwidth]{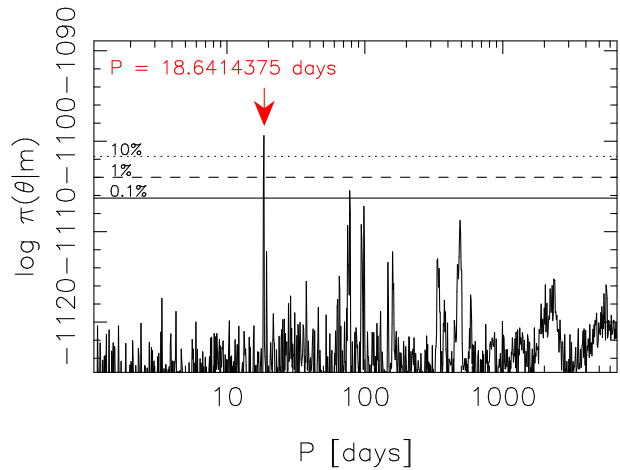}
\caption{As in Fig. \ref{fig:GJ273_psearch1} but for the fourth signal in a model containing four Keplerian signals.}\label{fig:GJ273_psearch3}
\end{figure}

\begin{figure}
\center
\includegraphics[angle=270, width=0.49\textwidth]{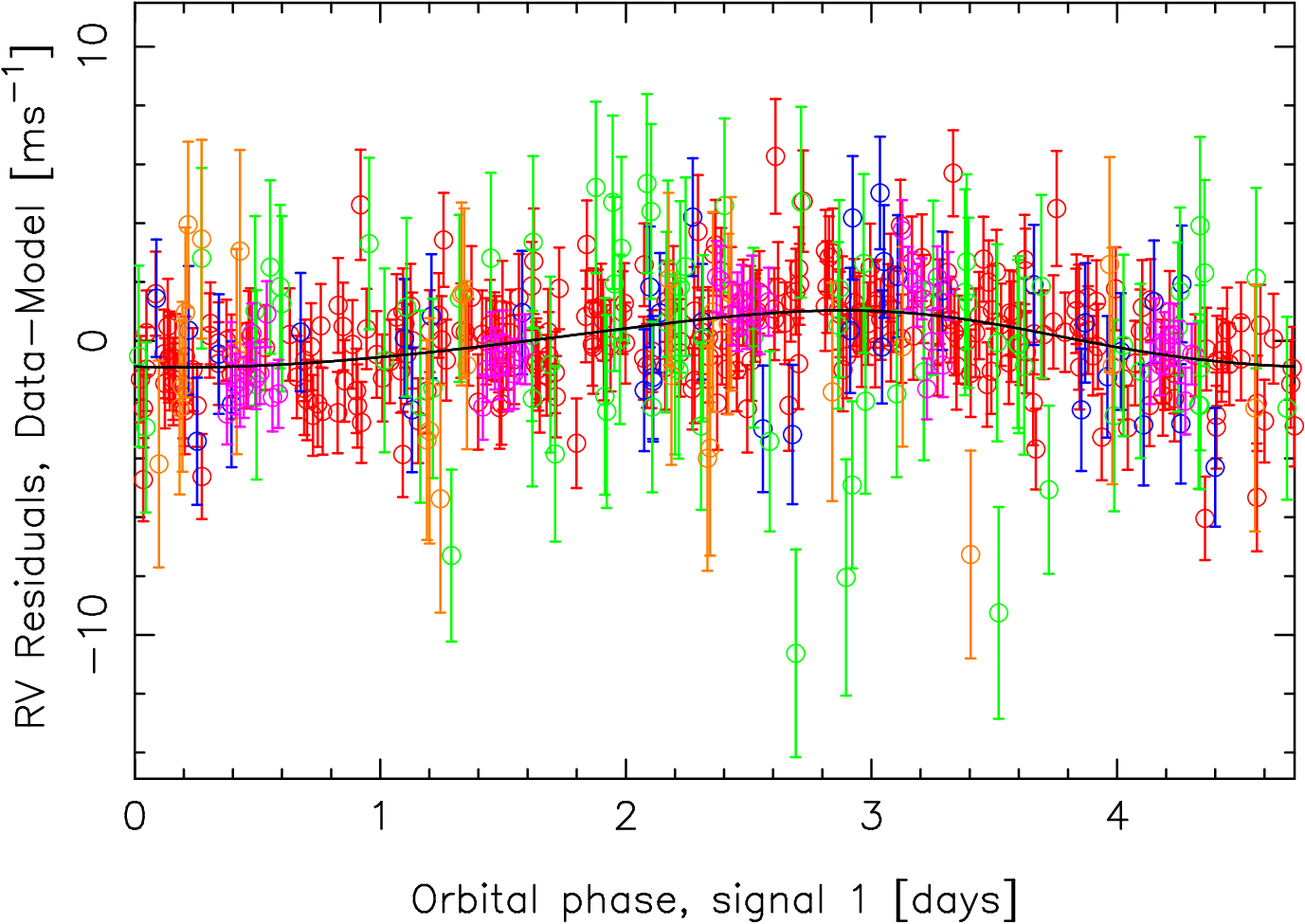}
\includegraphics[angle=270, width=0.49\textwidth]{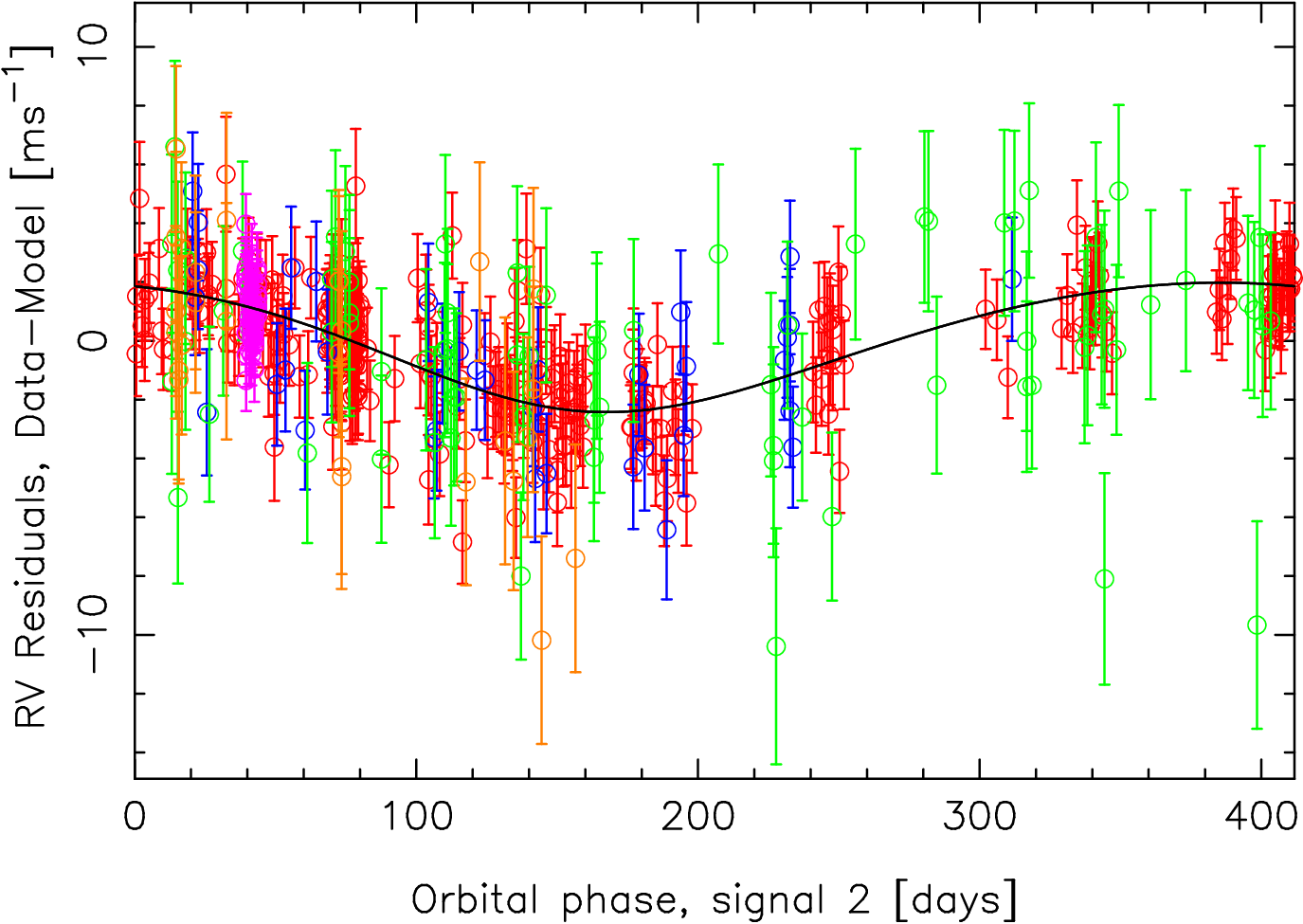}

\includegraphics[angle=270, width=0.49\textwidth]{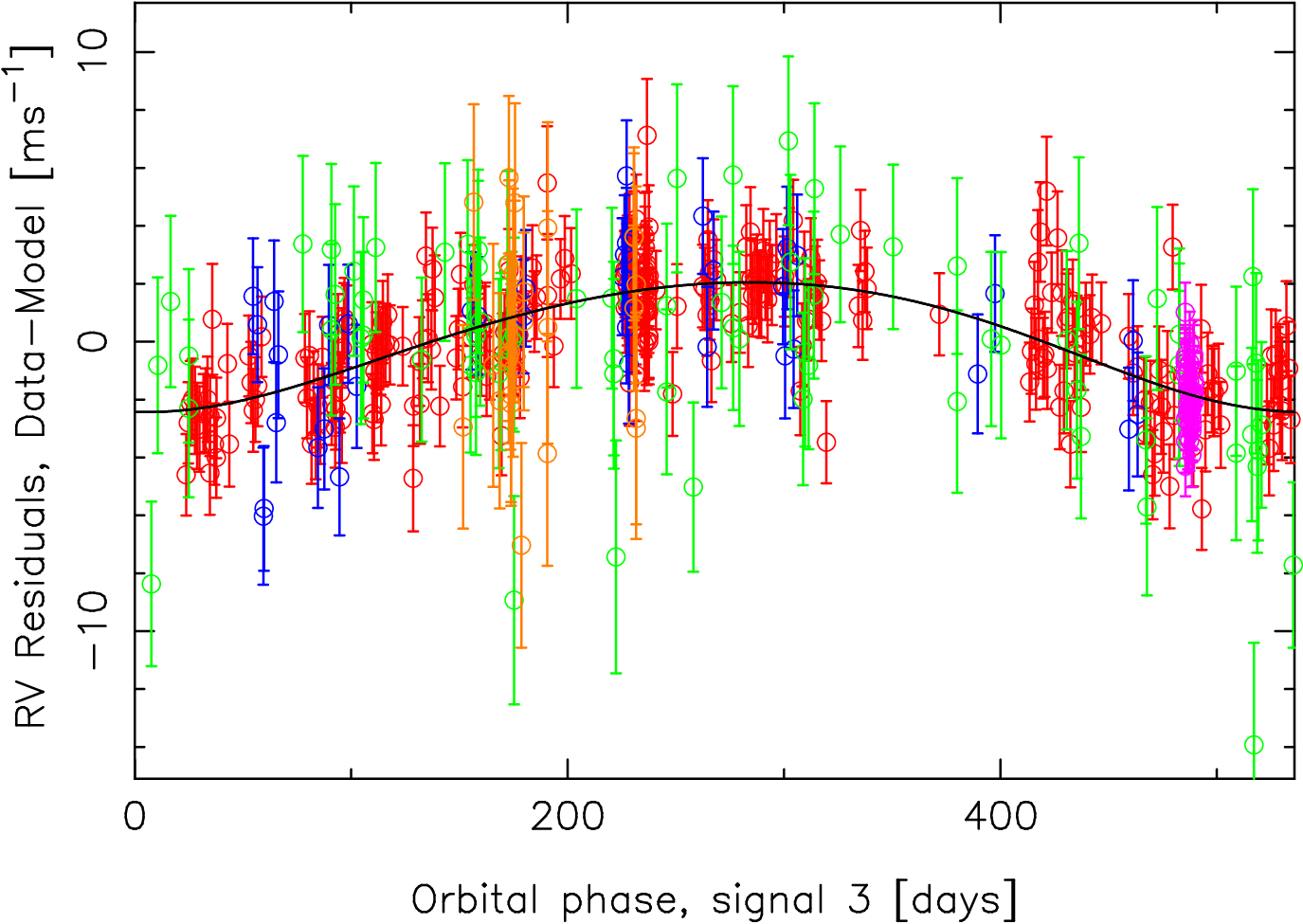}
\includegraphics[angle=270, width=0.49\textwidth]{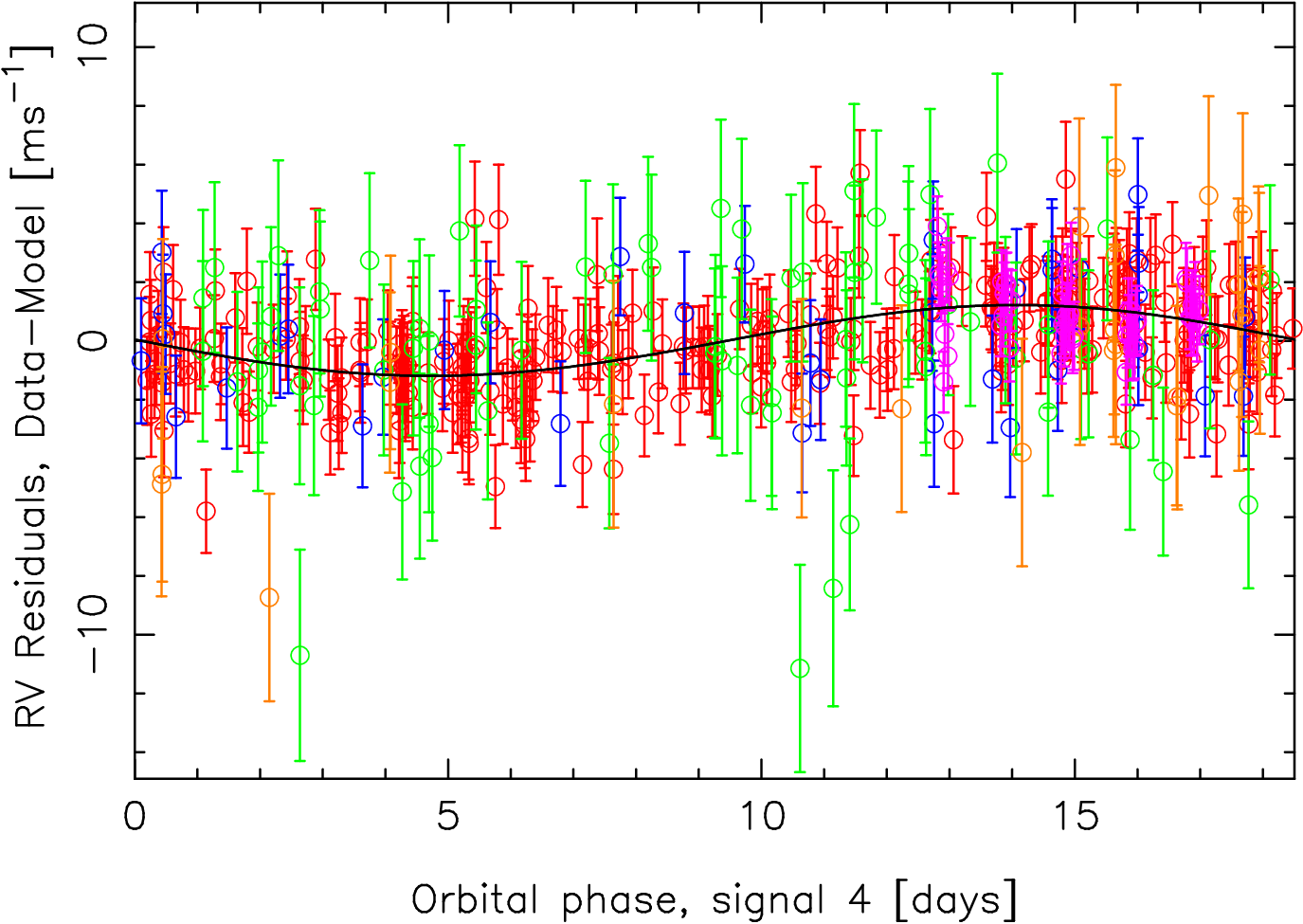}
\caption{HARPS (red), PFS (blue), HIRES (green), APF (orange), and HARPN (purple) radial velocities of GJ 273 folded on the periods of the four signals with the other signals subtracted from each panel.}\label{fig:GJ273_phased_curves}
\end{figure}

\begin{figure}
\center
\includegraphics[angle=270, width=0.49\textwidth,clip]{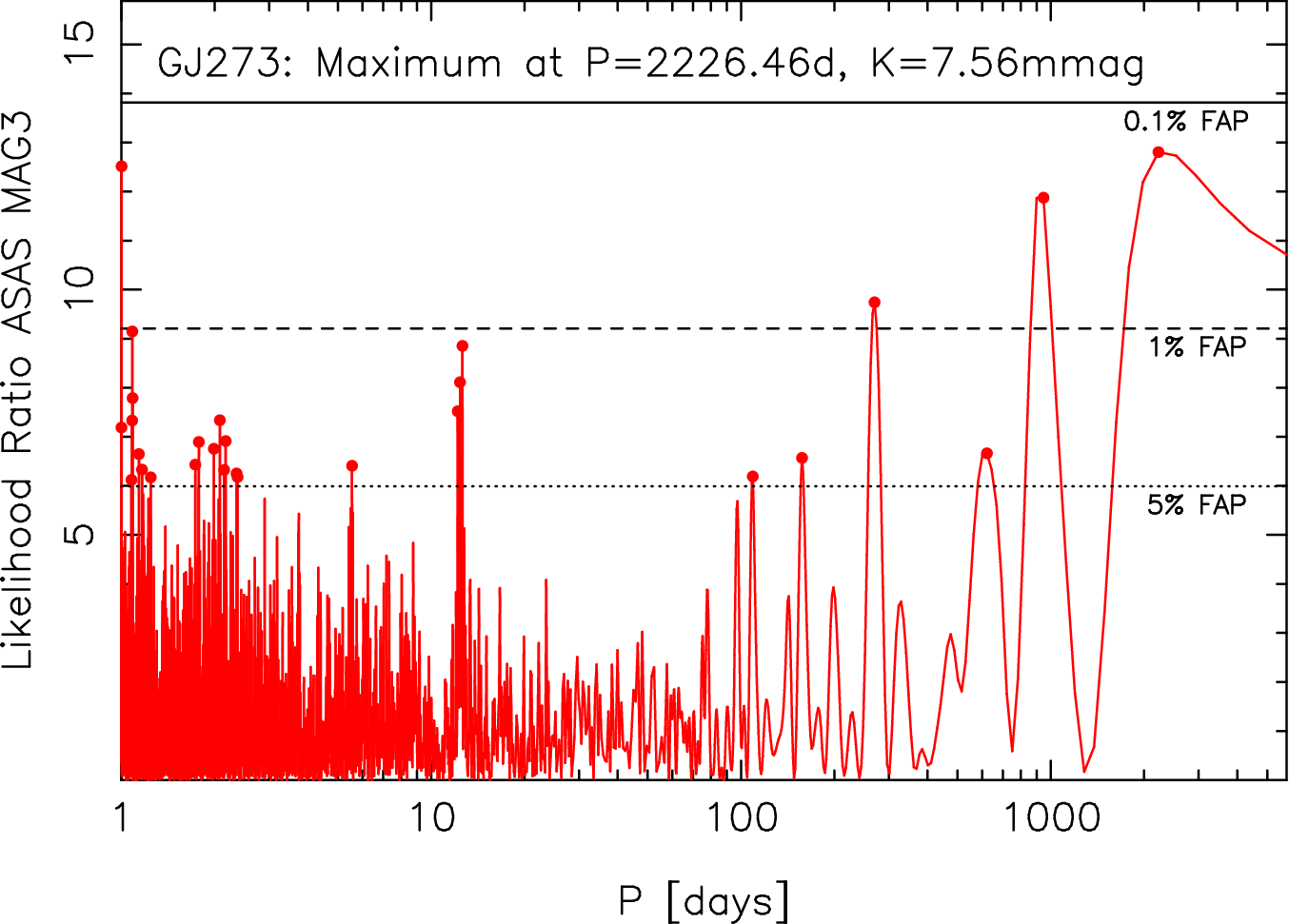}

\includegraphics[angle=270, width=0.49\textwidth,clip]{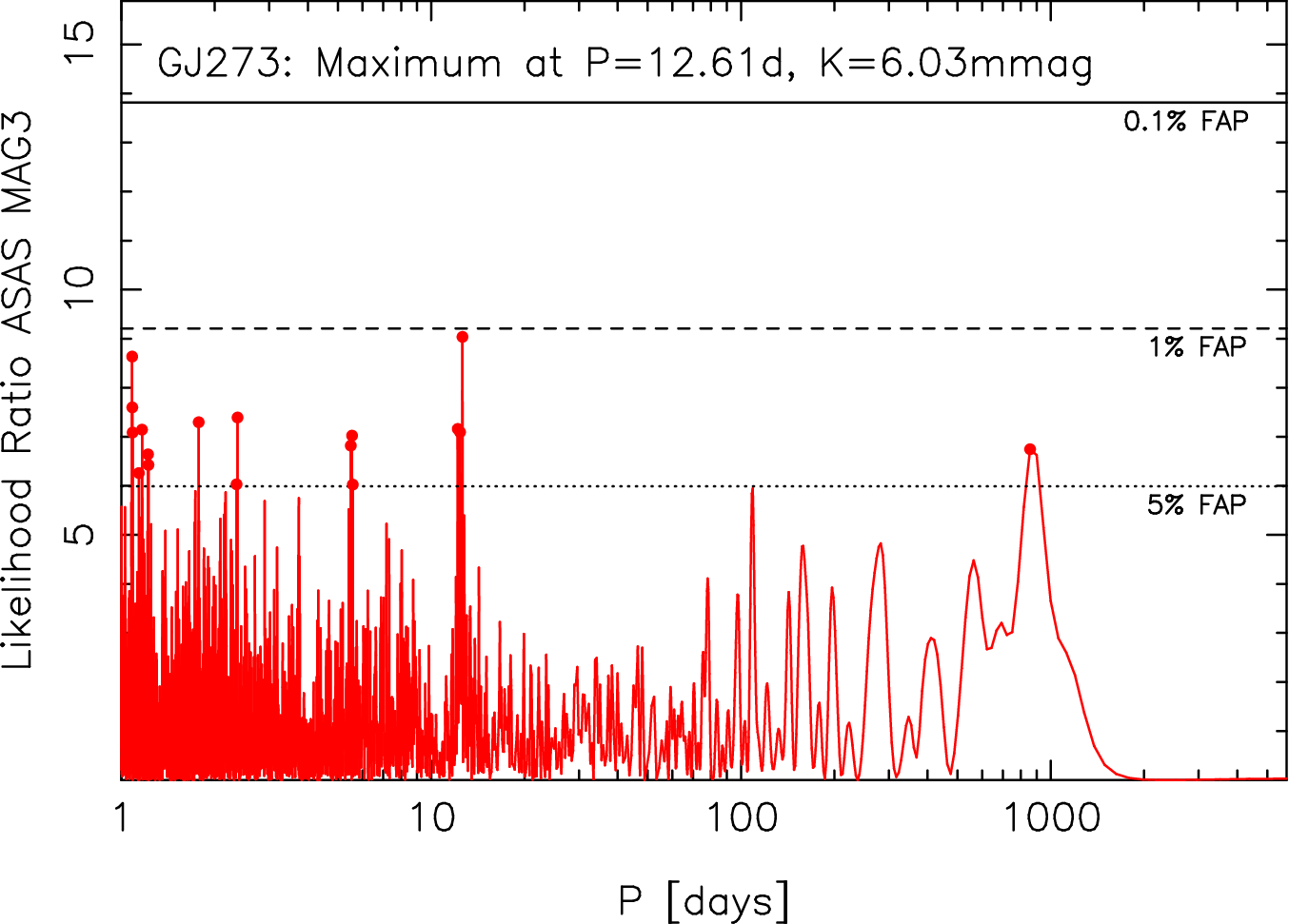}
\caption{Likelihood periodogram of the ASAS V-band photometry data of GJ 273 (top panel) and the residual periodogram after subtracting the dominant frequency (bottom panel).}\label{fig:GJ273_photom}
\end{figure}

Analysing the combined data with a model containing four Keplerian signals revealed yet another significant signal at a period of 18.640 [18.626, 18.656] days that was found to be independent of the variations in the activity indices and did not have periodic counterparts in the activity data either. We have plotted the estimated posterior densitiy as a function of the period of this signal to demonstrate its uniqueness in Fig. \ref{fig:GJ273_psearch3}. We also show the radial velocities folded on the phases of the signals (all four of them) in Fig. \ref{fig:GJ273_phased_curves}. It is our interpretation that the two short-period signals correspond to candidate planets because the available 321 ASAS V-band photometry measurements indicate the presence of photometric periodicities neither at the period of 4.72 days nor 18.64 days. However, the photometric data provides suggestive evidence for a periodicity of 2200 days that together with the S-index and FWHM values enables us to conclude that there is a probable magnetic activity cycle at that period (Fig. \ref{fig:GJ273_activity}). There is also even more weakly suggestive evidence for a photometric periodicity of 12.61 days (Fig. \ref{fig:GJ273_photom}, bottom panel), although the corresponding periodogram power does not even exceed the 1\% FAP. This potential periodicity does not coincide with any of the signals in radial velocities and might suggest that the rotation period of the star is 12.61 days. However, we do not tabulate such weak and suggestive signals in Table \ref{tab:kiraga_rotations} listing the known rotation periods of the stars in our sample.

Although less certain, it is also our interpretation that other two signals in the combined data at period of 414 and 542 days correspond to candidate planets orbiting the star because they, too, satisfy all the detection criteria and do not appear to have clear counterparts in the activity data. We have thus presented evidence for four candidate planets orbiting GJ 273 with orbital periods of 4.72, 414, 542, and 18.64 days and minimum masses of 1.2 [0.8, 1.7], 10.8 [7.2, 14.7], 9.3 [5.3, 13.6], and  1.2 [1.3, 3.2] M$_{\oplus}$, respectively. They are thus classified as a hot Earth, two cool mini-Neptunes, and a warm super-Earth. Given the fact that according to the estimates of \citet{kopparapu2013}, the liquid water habitable-zone of GJ 273 is located between 0.06 and 0.12 AU and that the candidate planet with orbital period of 18.64 days lies squarely in the middle of this zone, we have also obtained evidence for another \citep[e.g. Proxima b;][]{anglada2016} potentially habitable likely rocky candidate planet orbiting a very nearby star.

We note that \citet{bonfils2013} reported that they detected a signal at a period of 440 days but that the preferred solution corresponded to a high-eccentricity orbit due to poor phase-coverage and they could not draw conclusions regarding the nature of the signal. This likely corresponds to the signal we detect at a period of 410 days. \citet{bonfils2013} also speculated that the apparent periodogram power at a period of 440 days might be a yearly alias of the long term trend. We did not obtain any evidence in favour of this being the case, in particular, as the signal is present in the data even when taking into account the possibility that there indeed is a linear trend in the data (significant trend of 0.308$\pm$0.050 ms$^{-1}$year$^{-1}$). Although we call the signals with periods of 414 and 542 days candidate planets in the current work, more data is needed to verify this interpretation.

\clearpage

\subsection{GJ 300}\label{sec:GJ300}

\begin{figure}
\center
\includegraphics[angle=270, width=0.49\textwidth,clip]{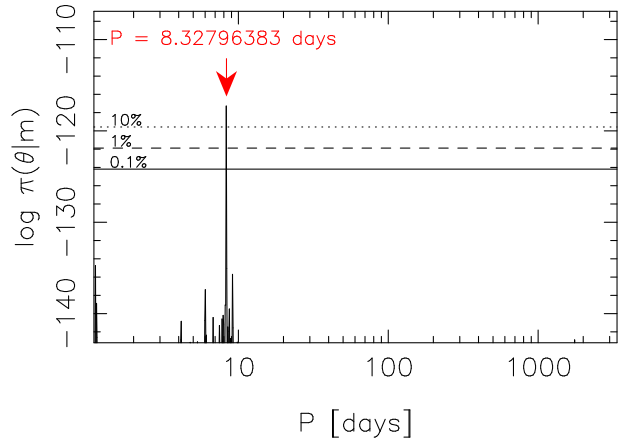}
\caption{Estimated posterior probability density as a function of the period parameter of a Keplerian signal given the HARPS and HARPN radial velocities of GJ 300. The red arrow denotes the position of the global maximum and the horizontal lines indicate the 10\% (dotted), 1\% (dashed), and 0.1\% (solid) equiprobability thresholds with respect to the maximum.}\label{fig:GJ300_period_search}
\end{figure}

\begin{figure}
\center
\includegraphics[angle=270, width=0.49\textwidth]{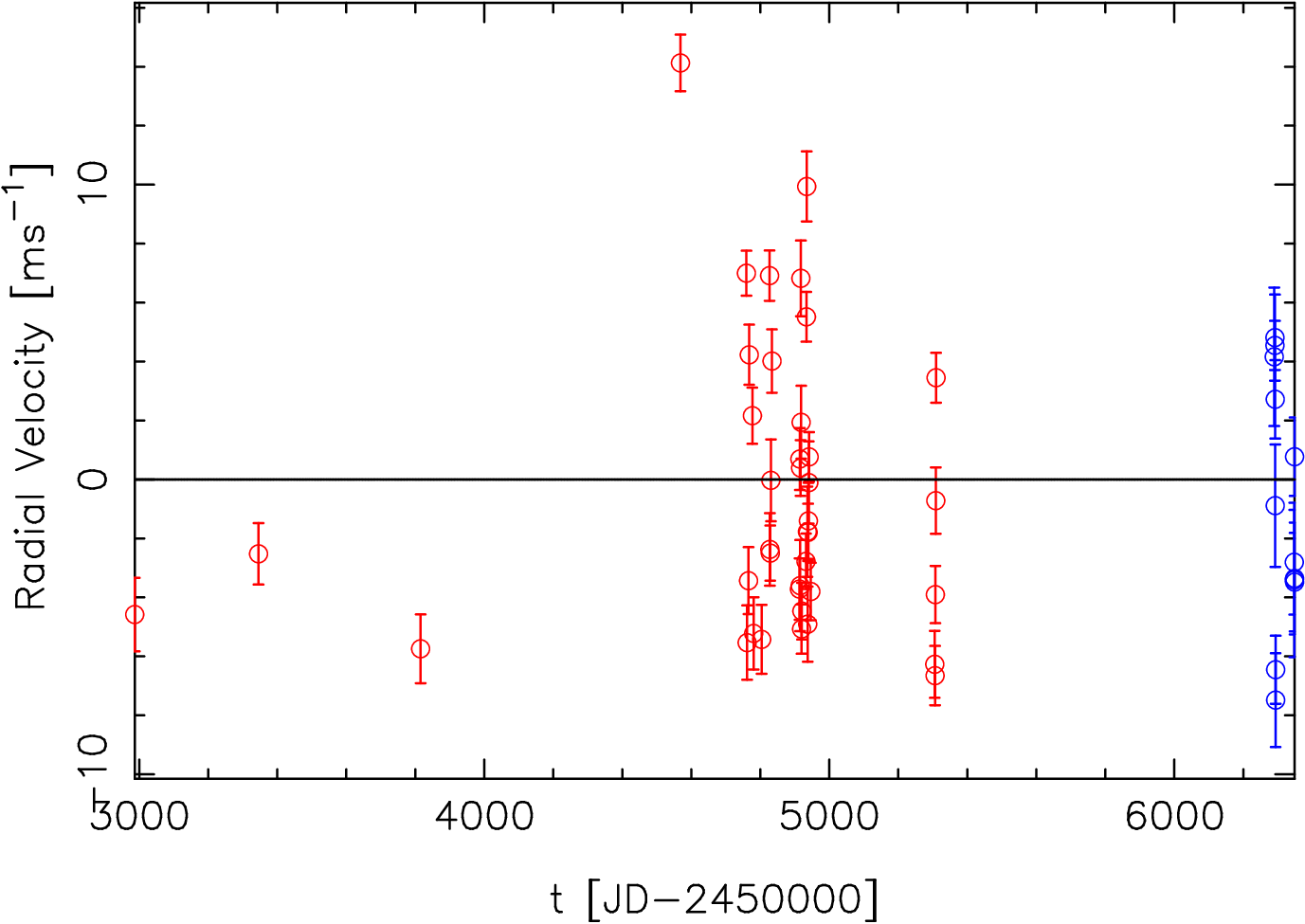}

\includegraphics[angle=270, width=0.49\textwidth]{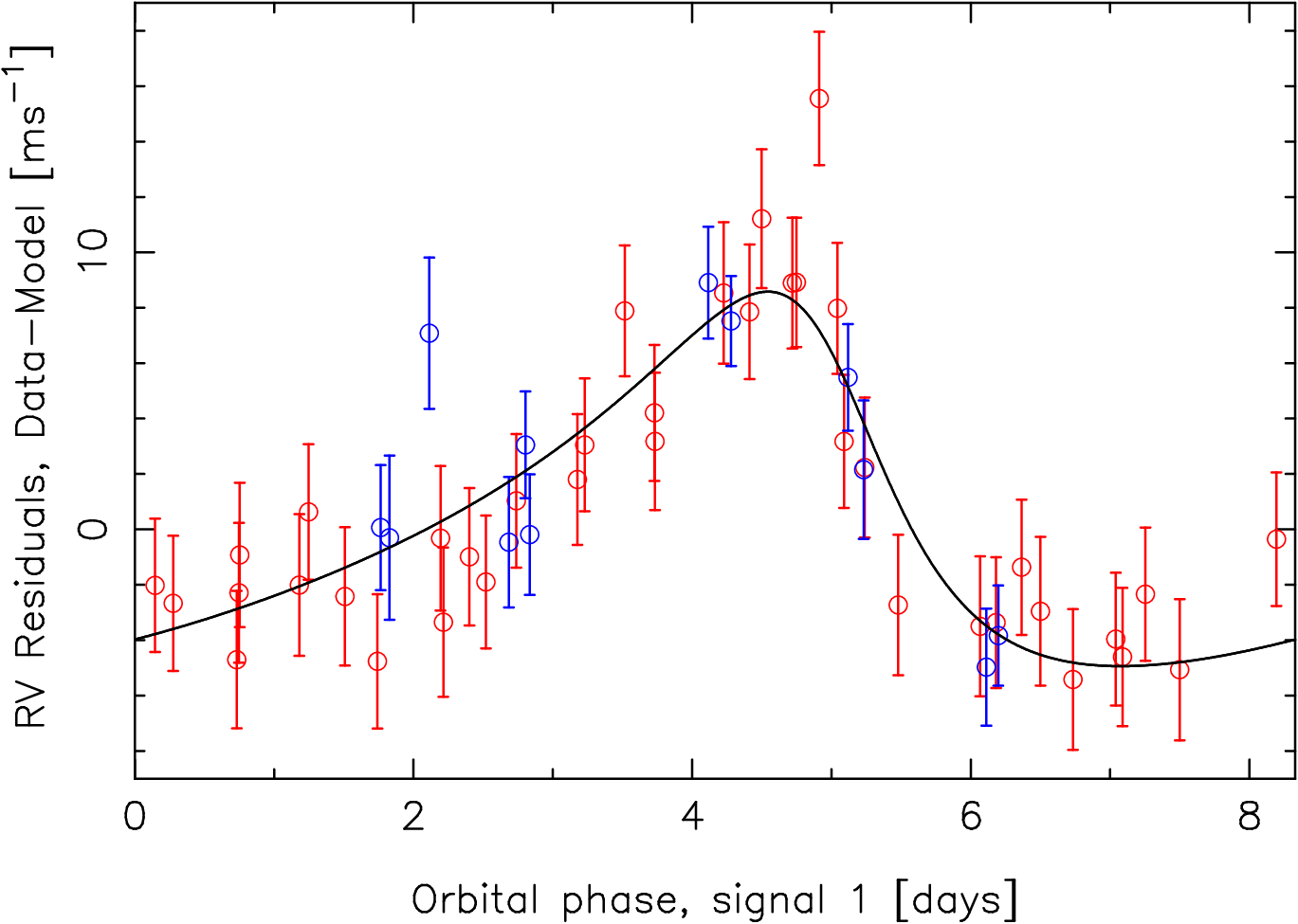}
\caption{HARPS (red) and HARPN (blue) radial velocities of GJ 300 (top panel). Bottom panel shows the velocities folded on the phase of the signal of the candidate planet.}\label{fig:GJ300_phase}
\end{figure}

The combined 39 HARPS and 12 HARPN radial velocities were found to contain a strong and unique signal with a period of 8.3279 [8.3189, 8.3337] days and an amplitude of 5.80 [3.62, 7.73] ms$^{-1}$ (Figs. \ref{fig:GJ300_period_search} and \ref{fig:GJ300_phase}). We consider this to be a candidate planet orbiting the star because the signal has no counterparts in either the ASAS photometry or HARPS activity indicators.

The liquid-water HZ of the star, according to the equations of \citep{kopparapu2013}, is between 0.04 and 0.06 AU. Therefore, the semi-major axis of GJ 300 b of 0.050 [0.045, 0.055] AU places the planet inside the stellar HZ. With a minimum mass of 6.8 [4.1, 9.9] M$_{\oplus}$ we thus classify the candidate planet orbiting GJ 300 as a warm super-Earth. We note that, as seen in Fig. \ref{fig:GJ300_phase} (top panel), the eccentricity of GJ 300 b appears considerable -- we obtained an estimate of 0.29 [0, 0.58]. However, this estimate is not statistically significantly different from zero with 99\% credibility and is thus likely affected by the ``pathological'' data sampling -- the HARPS data is found mainly in a cluster covering one observing season and the data sets from the two instruments do not overlap (see Fig. \ref{fig:GJ300_phase}, top panel). Moreover, the bias towards high eccentricities observed in radial velocity searches for planets likely also plays a role in this case \citep{zakamska2011}.

\clearpage

\subsection{GJ 310}

\begin{figure}
\center
\includegraphics[angle=270, width=0.49\textwidth]{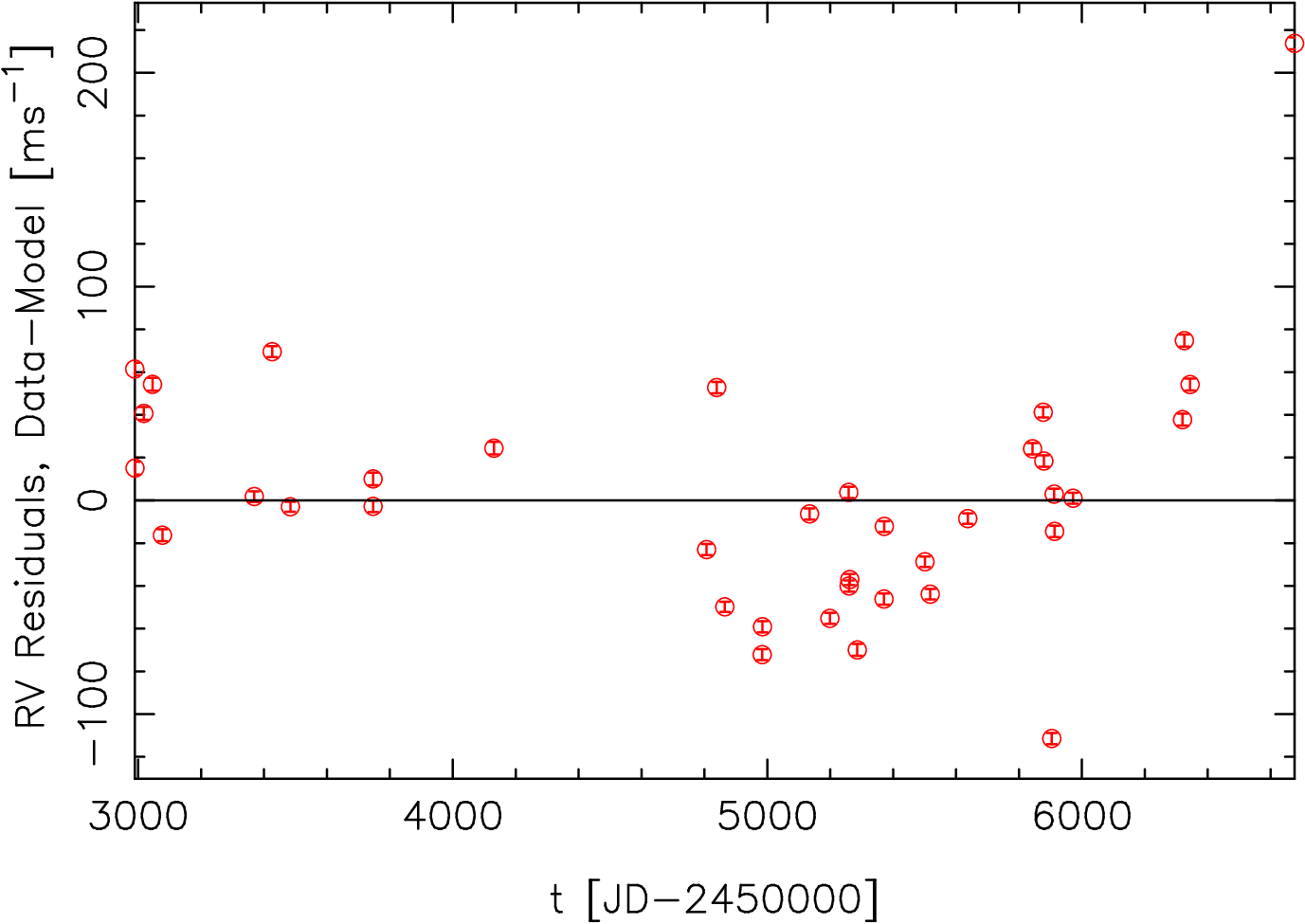}

\includegraphics[angle=270, width=0.49\textwidth]{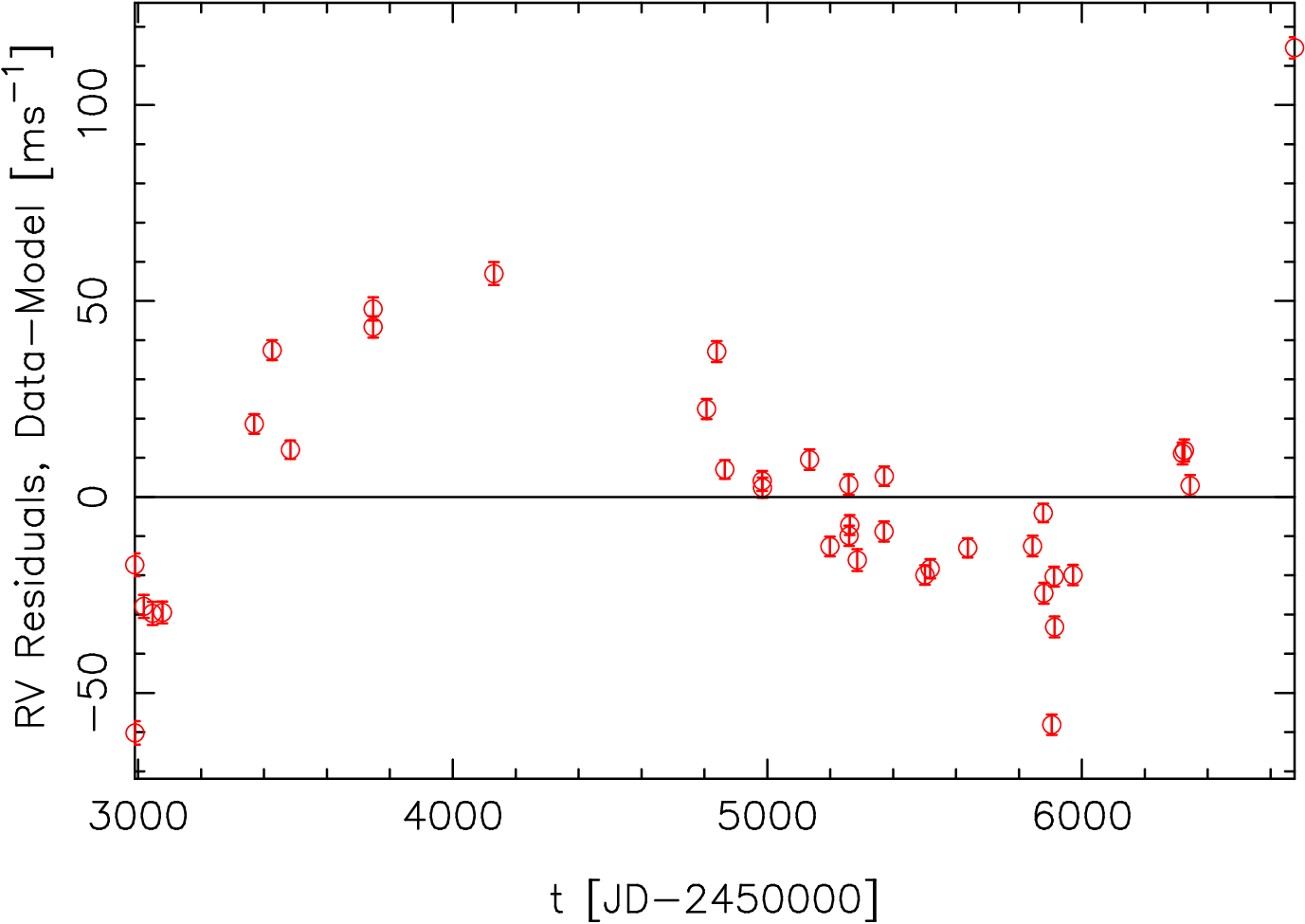}

\includegraphics[angle=270, width=0.49\textwidth]{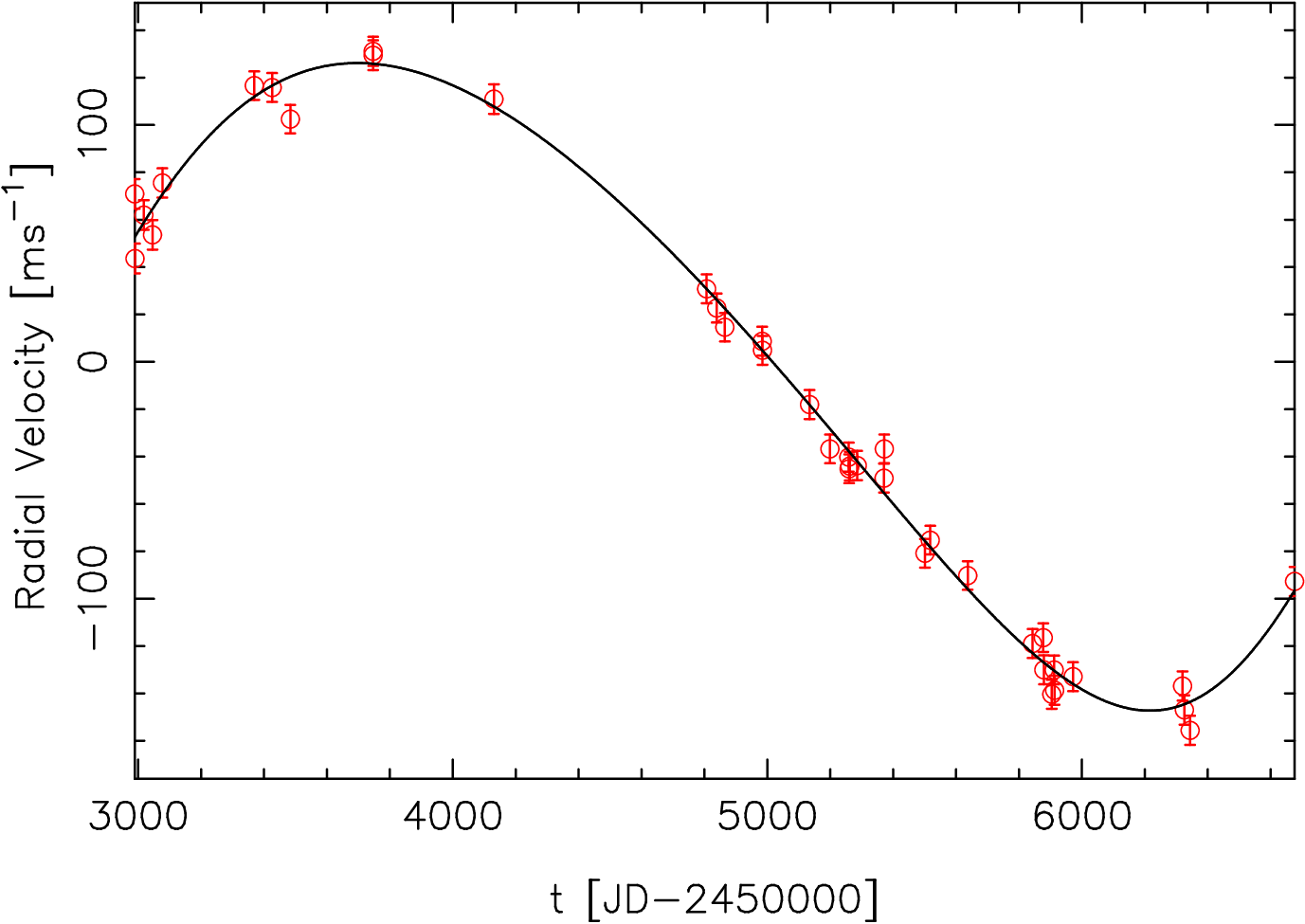}
\caption{HIRES data residuals of GJ 310 with linear trend subtracted (top panel), with second-order polynomial subtracted (middle panel), and when the data has been modelled with a second-order polynomial and a Keplerian signal (solid curve; bottom panel).}\label{fig:GJ310_residual}
\end{figure}

There was a clearly identifiable long-period signal in the set of 38 HIRES radial velocities of GJ 310 (HIP 42220) corresponding to a candidate planet classified as a Jupiter-analog. Although this signal had a period of 4283 [4161, 4524] days that was longer than the data baseline of 3687 days, we include the corresponding candidate planet in our sample because the orbit was well-constrained from above in accordance to our detection criteria. This signal was missed by \citet{butler2016} because they did not use second-order polynomials in their model for accelerating movement. The existence of the signal is apparent from Fig. \ref{fig:GJ310_residual} where we have plotted the residuals of the standard model with linear acceleration subtracted (top panel) and the same residuals after subtracting the second-order polynomial (middle panel). The signal of the planet candidate orbiting GJ 310 is shown in the bottom panel where we have used the full model to describe the data with the MAP parameters of a one-Keplerian model.

Because there is evidence for a second-order polynomial acceleration in the data, it seems evident that there is also another long-period substellar companion to the star in addition to the giant planet (or possibly a brown dwarf) with a minimum mass of 8.18 [6.40, 10.40] M$_{\rm Jup}$. We did not observe any significant signals in the HIRES S-indices. Moreover, ASAS photometry was not available for this star.

\clearpage

\subsection{GJ 317}

GJ 317 is a well-known planet-host and has been reported to have a giant planet with a minimum mass of 1.2 M$_{\rm Jup}$ orbiting with an orbital period of 692.9$\pm$4 days \citep{johnson2007}. It was also observed by \citet{johnson2007} that the star accelerates in the radial direction which indicates that there is another massive long-period companion to the star. The minimum mass was later revised to 1.81 M$_{\rm Jup}$ by \citet{anglada2012d} who also attempted to constrain the orbit of the outer companion with astrometric observations. We had no problems in identifying the strong signal corresponding to the candidate planet GJ 317 b. We modelled the combined HARPS and HIRES radial velocities with a statistical model containing a second-order polynomial trend that enabled us to account for the changing acceleration cause by the long-period companion \citep{anglada2012d}. However, we do not classify it as a planet candidate in the current work because its orbit cannot be constrained above in the period space, which violates our signal detection criteria.

\begin{figure}
\center
\includegraphics[angle=270, width=0.49\textwidth]{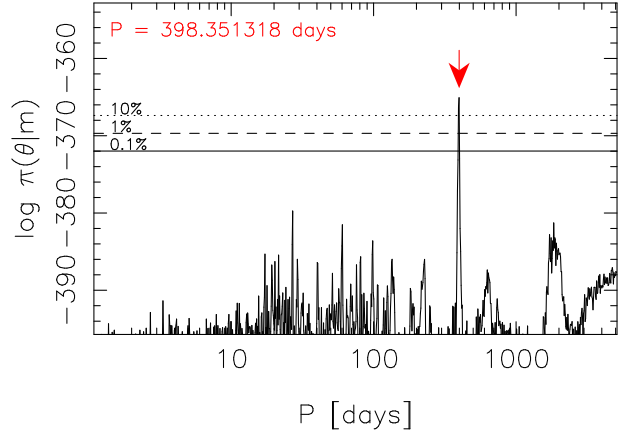}
\caption{Estimated posterior probability density given GJ 317 radial velocities as a function of the period parameter of the second Keplerian signal. The red arrow shows the global maximum and the horizontal lines denote the 10\% (dotted), 1\% (dashed), and 0.1\% (solid) equiprobability thresholds with respect to the identified maximum.}\label{fig:GJ317_psearch}
\end{figure}

However, we did identify a second signal in the GJ 317 radial velocities. This signal, with an amplitude of 8.98 [4.86, 12.70] ms$^{-1}$, was found as a unique probability maximum in the period space at a period of 397.7 [390.3, 401.3] days (Fig. \ref{fig:GJ317_psearch}). We interpret this signal as evidence for a candidate planet orbiting the star with a minimum mass of 51.8 [27.9, 78.2] M$_{\oplus}$ and classify it as a cool super-Neptune. This is because we could not obtain any evidence suggesting that the two signals, the second one in particular, are not caused by planets orbiting the star. There were no significant periodicities in the HARPS or HIRES activity indicators. Moreover, ASAS V-band photometry data did not provide evidence for photometric periodicities.

\begin{figure}
\center
\includegraphics[angle=270, width=0.49\textwidth]{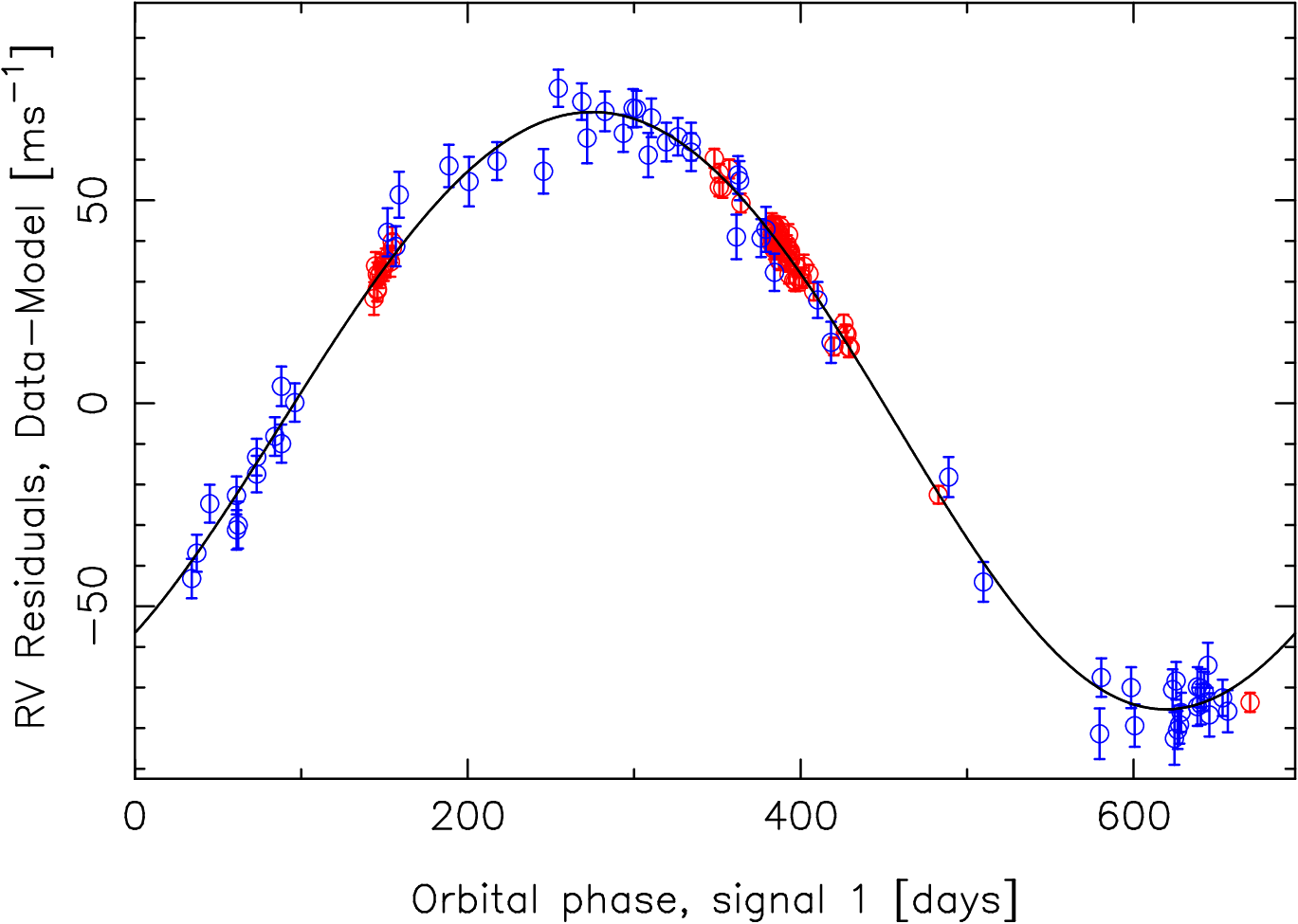}

\includegraphics[angle=270, width=0.49\textwidth]{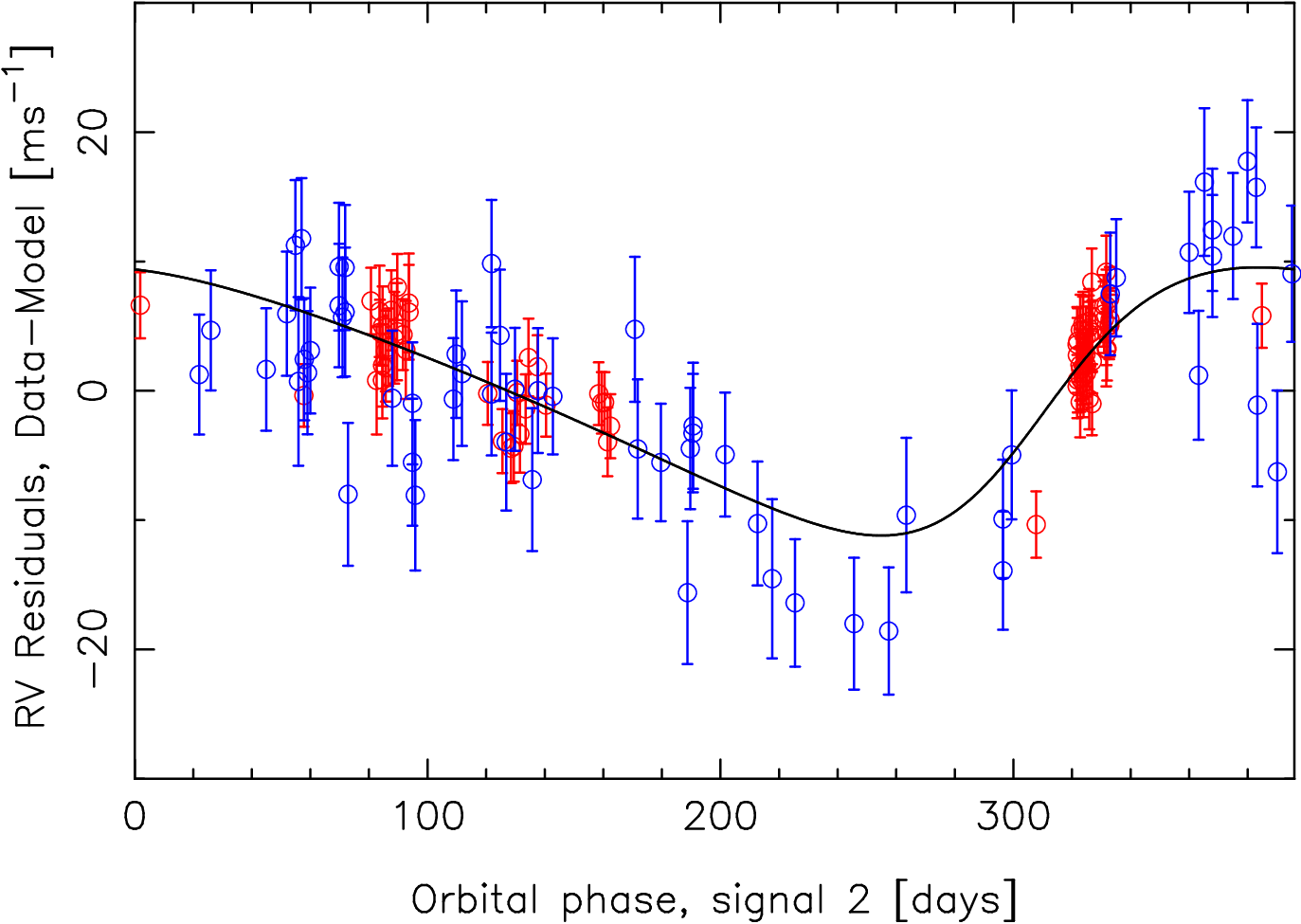}
\caption{HARPS (red) and HIRES (blue) radial velocities of GJ317 folded on the phases of the two significant signals.}\label{fig:GJ317_phased}
\end{figure}

We have plotted the combined radial velocities folded on the phases of the signals in Fig \ref{fig:GJ317_phased}. Although the second signal appears to have an elevated eccentricity of 0.36, we note that the 99\% credibility interval is [0.04, 0.63], which means that low eccentricities cannot be ruled out. Moreover, due to a combination of poor phase-coverage and the overall bias of radial velocity data towards higher eccentricities \citep{zakamska2011}, we estimate that the actual eccentricity of this signal is lower than the MAP estimate of 0.36 if it is indeed caused by a planet orbiting the star. Dynamical analyses of the proposed two-planet system, also possibly accounting for the long-period signal we could not constrain from above in the period space, are needed to better constrain the eccentricity of the second candidate planet and whether the system can be dynamically stable.

\clearpage

\subsection{GJ 357}

We analysed the combined set of HARPS, HIRES, and UVES velocities of GJ 357 (HIP 47103). This star was included in the samples of \citet{zechmeister2009} and \citet{bonfils2013}, and consequently in \citet{tuomi2014}, but no planets have been reported orbiting the star. In addition to the 70 UVES and 5 HARPS velocities analysed in \citet{tuomi2014}, we obtained 44 new HARPS velocities from the data products in the ESO archive together with 35 HIRES velocities. This yielded a set of 156 velocities for GJ 357 after excluding two HIRES velocities and four HARPS velocities corresponding to clear 3-$\sigma$ outliers in the activity indicators suggestive of stellar flares.

\begin{figure}
\center
\includegraphics[angle=270, width=0.49\textwidth,clip]{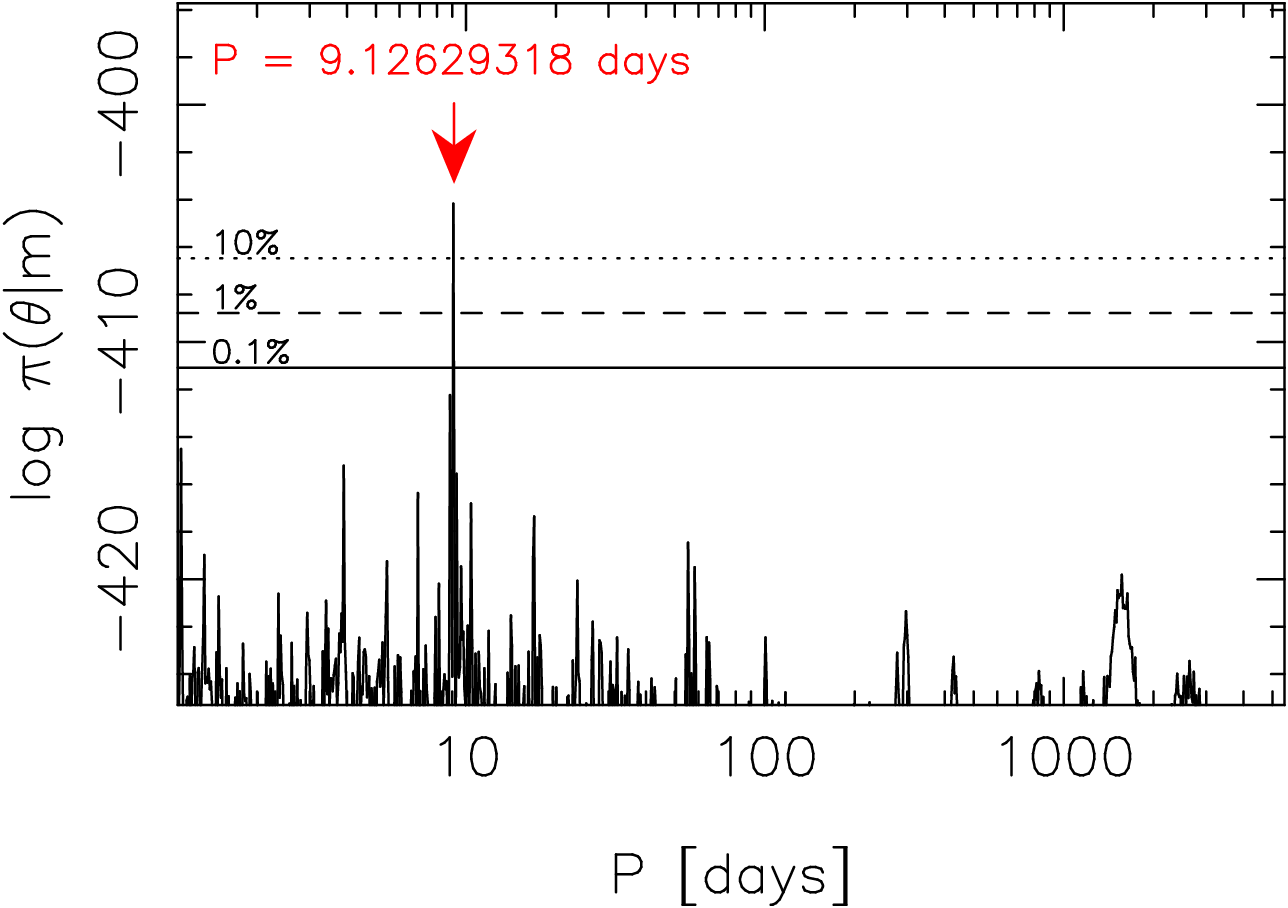}

\includegraphics[angle=270, width=0.49\textwidth,clip]{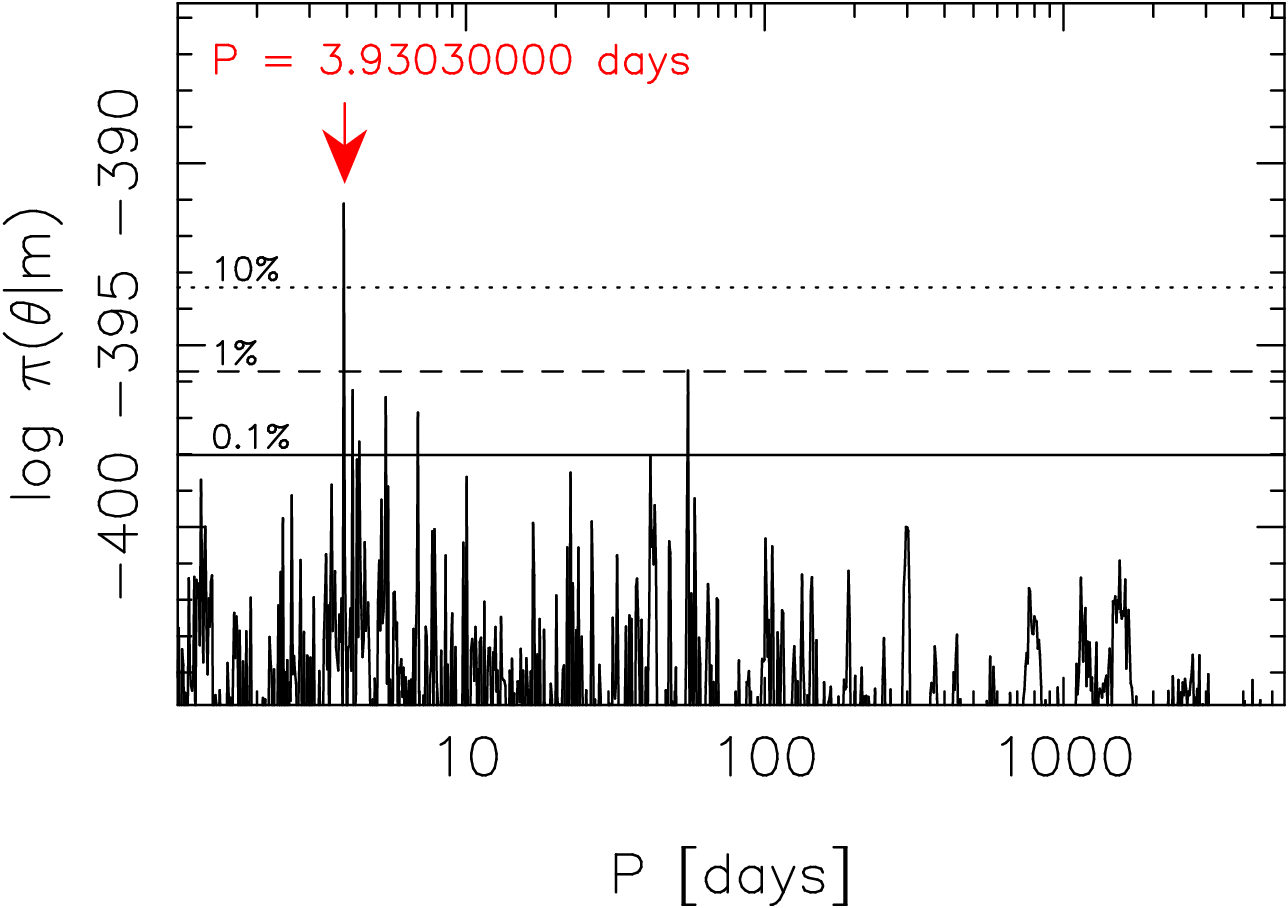}

\includegraphics[angle=270, width=0.49\textwidth,clip]{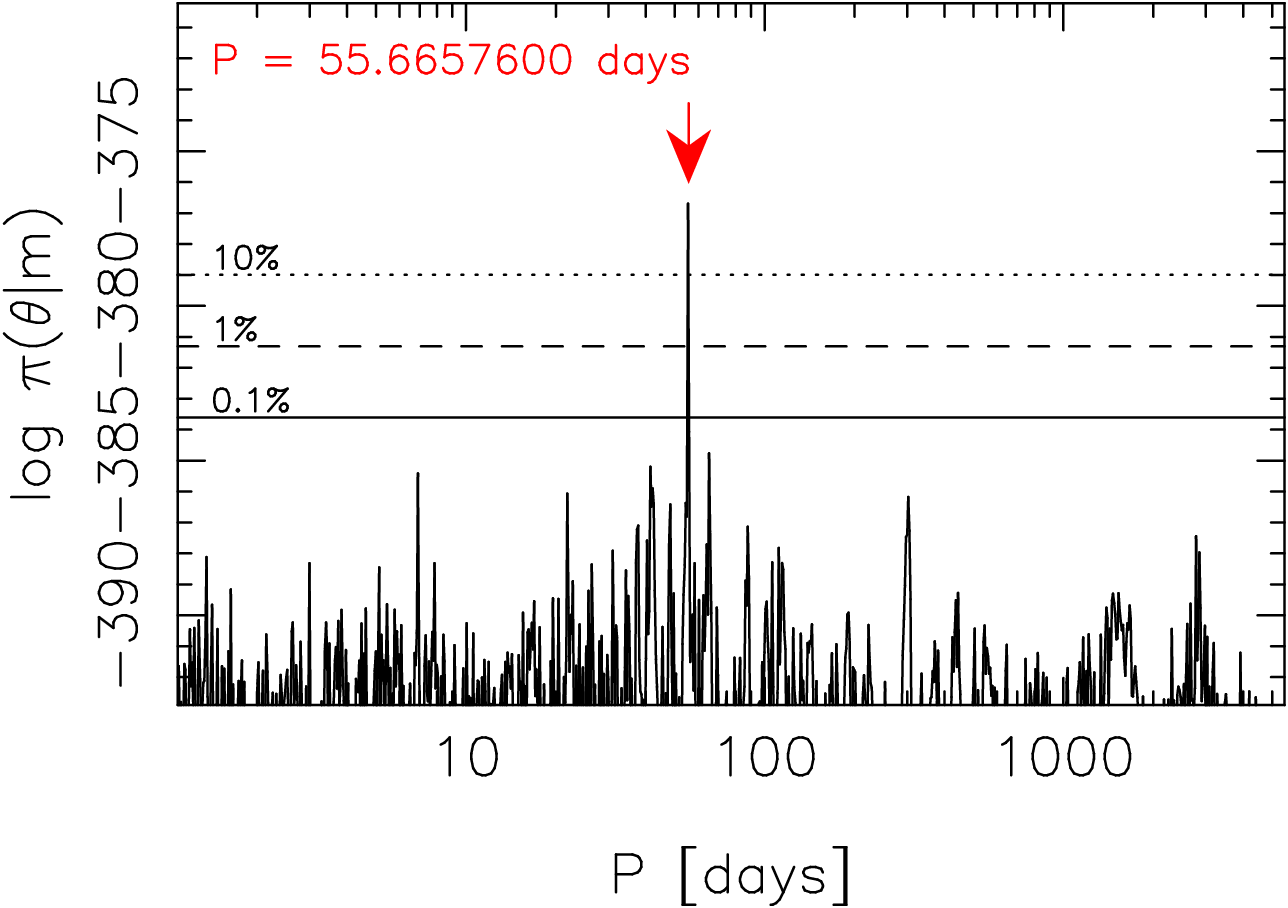}
\caption{Estimated posterior probability density as a function of the period of the $k$th Keplerian signal given the combined HARPS, HIRES, and UVES velocities of GJ 357 for models with $k=1$ (top), $k=2$ (middle), and $k=3$ (bottom). The red arrows indicate the global maxima and the horizontal lines denote the 10\% (dotted), 1\% (dashed), and 0.1\% (solid) equiprobability thresholds with respect to the maxima.}\label{fig:GJ357_period_search}
\end{figure}

When analysing the combined data set, we obtained evidence for three signals in the combined radial velocities. The first signal was detected confidently at a period of 9.1265 [9.1221, 9.1302] days with an amplitude of 2.21 [1.32, 3.22] ms$^{-1}$ (Fig. \ref{fig:GJ357_period_search}, top left panel). The second signal was found at a period of 3.9303 [3.9293, 3.9311] days with an amplitude of 2.03 [1.17, 2.89] ms$^{-1}$ rather less uniquely (Fig. \ref{fig:GJ357_period_search}, top right panel). However, the only local maximum exceeding the 1\% equiprobability threshold with respect to the posterior maximum was found at the position of the third signal and we thus accept the global maximum as a signal as it is detected according to our criteria.

\begin{figure}
\center
\includegraphics[angle=270, width=0.49\textwidth]{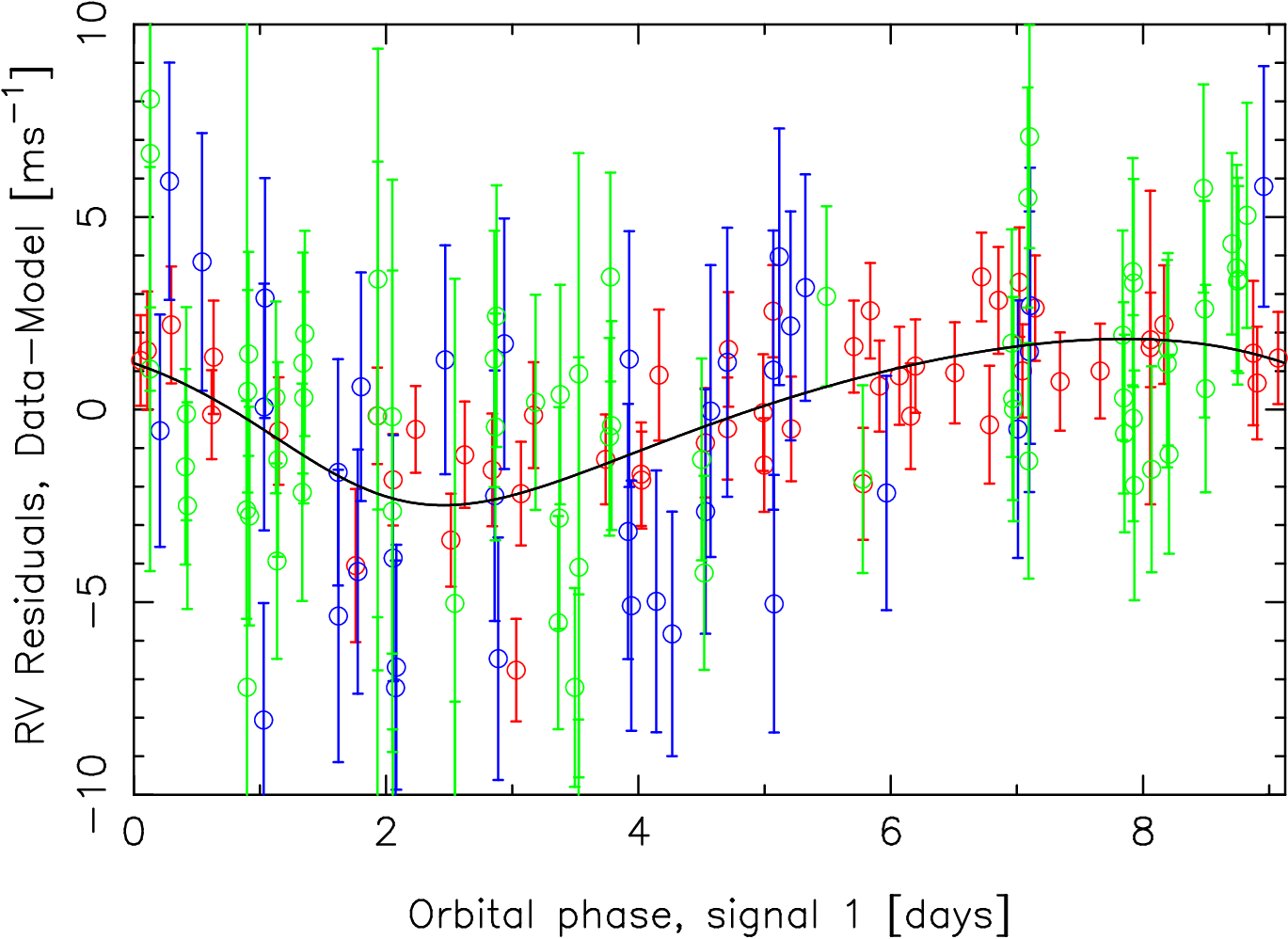}
\includegraphics[angle=270, width=0.49\textwidth]{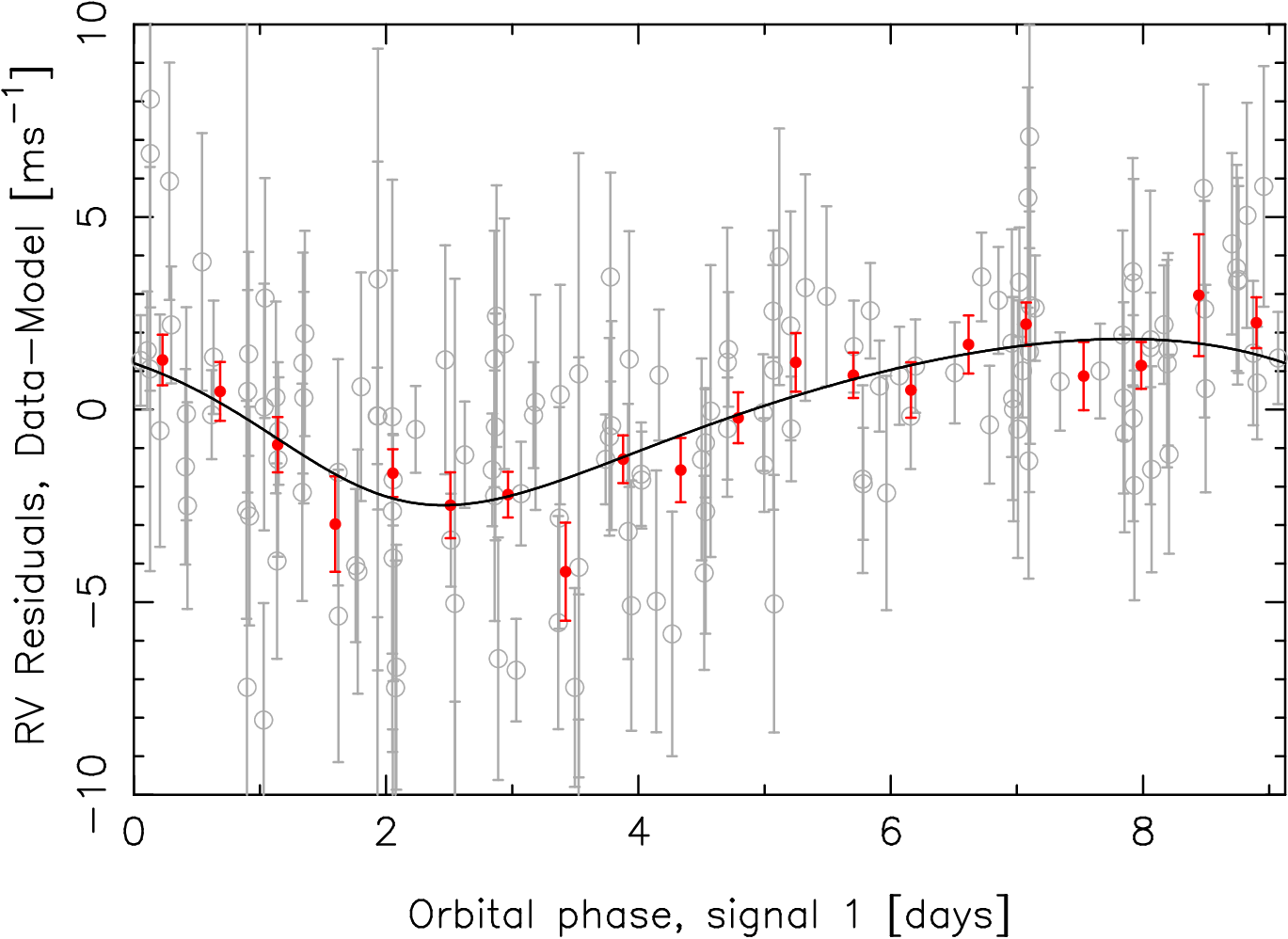}

\includegraphics[angle=270, width=0.49\textwidth]{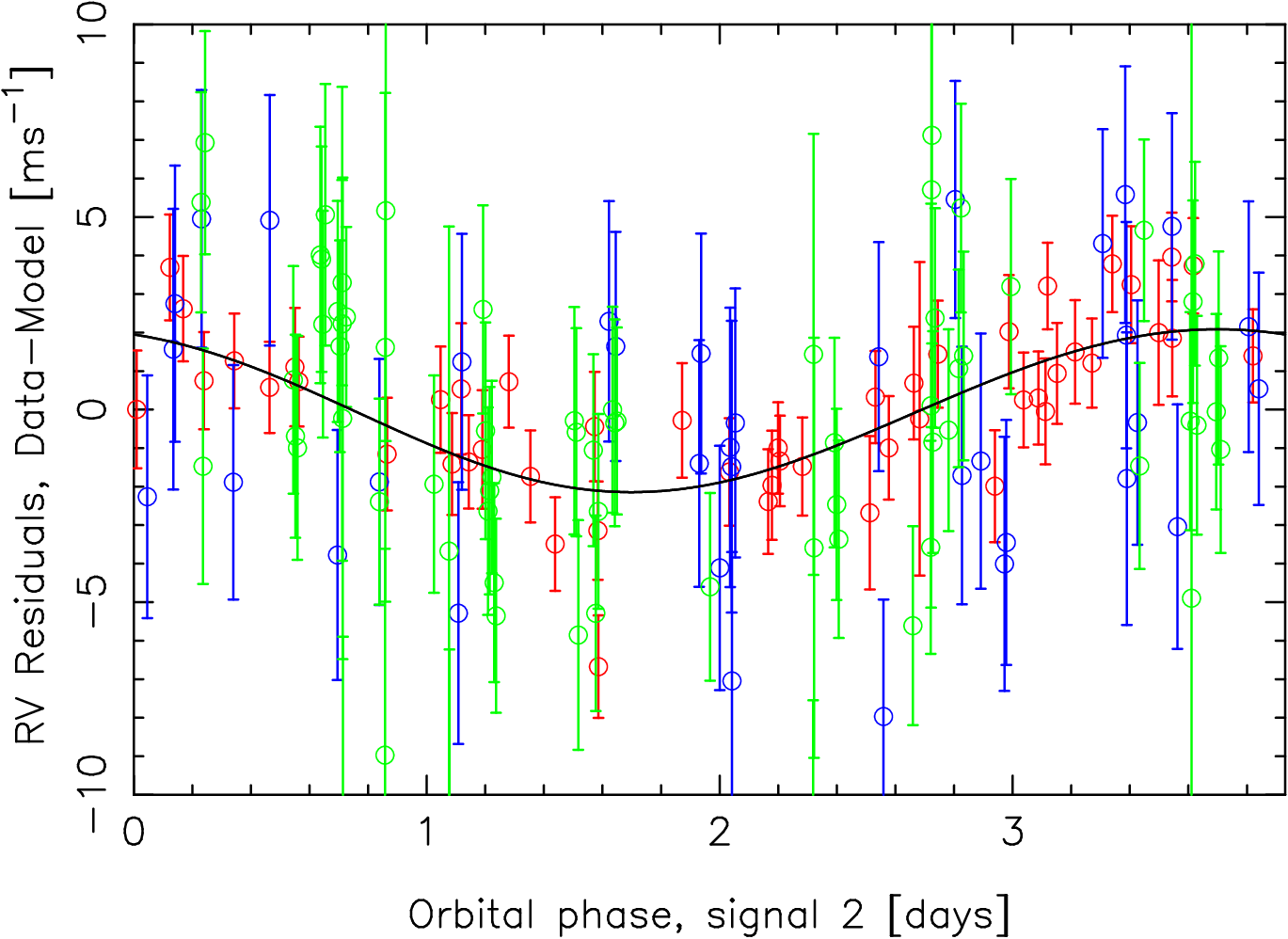}
\includegraphics[angle=270, width=0.49\textwidth]{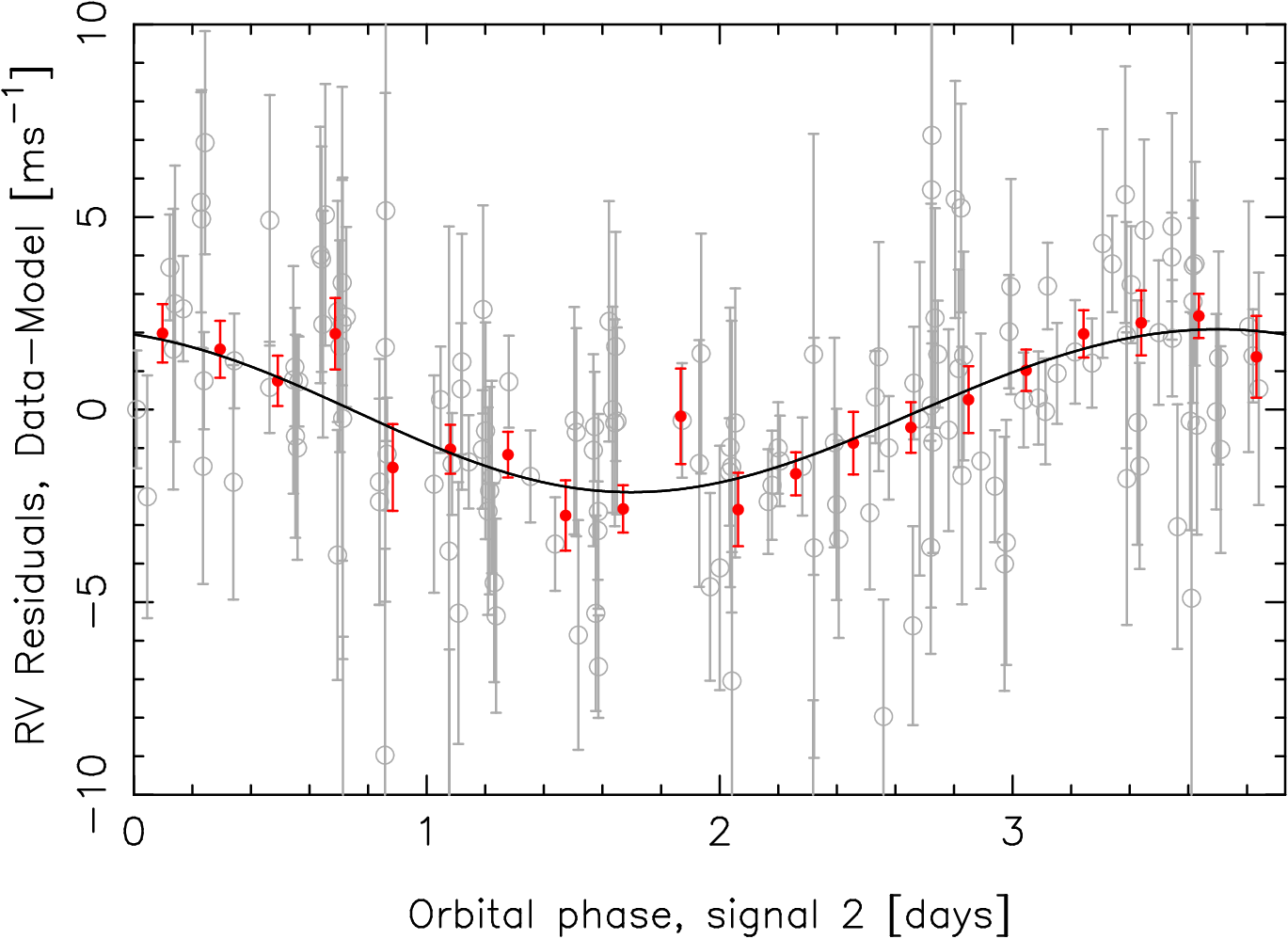}

\includegraphics[angle=270, width=0.49\textwidth]{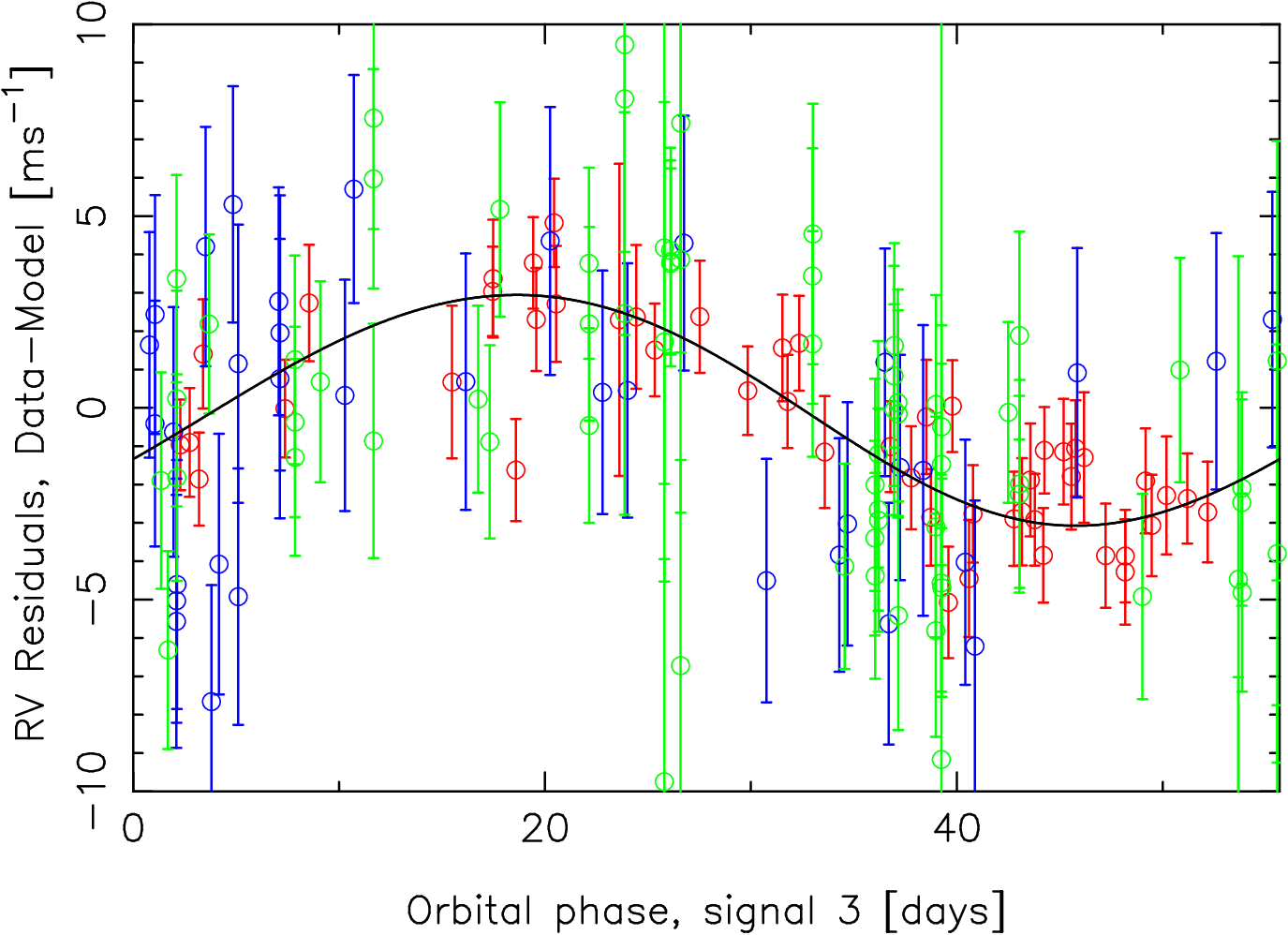}
\includegraphics[angle=270, width=0.49\textwidth]{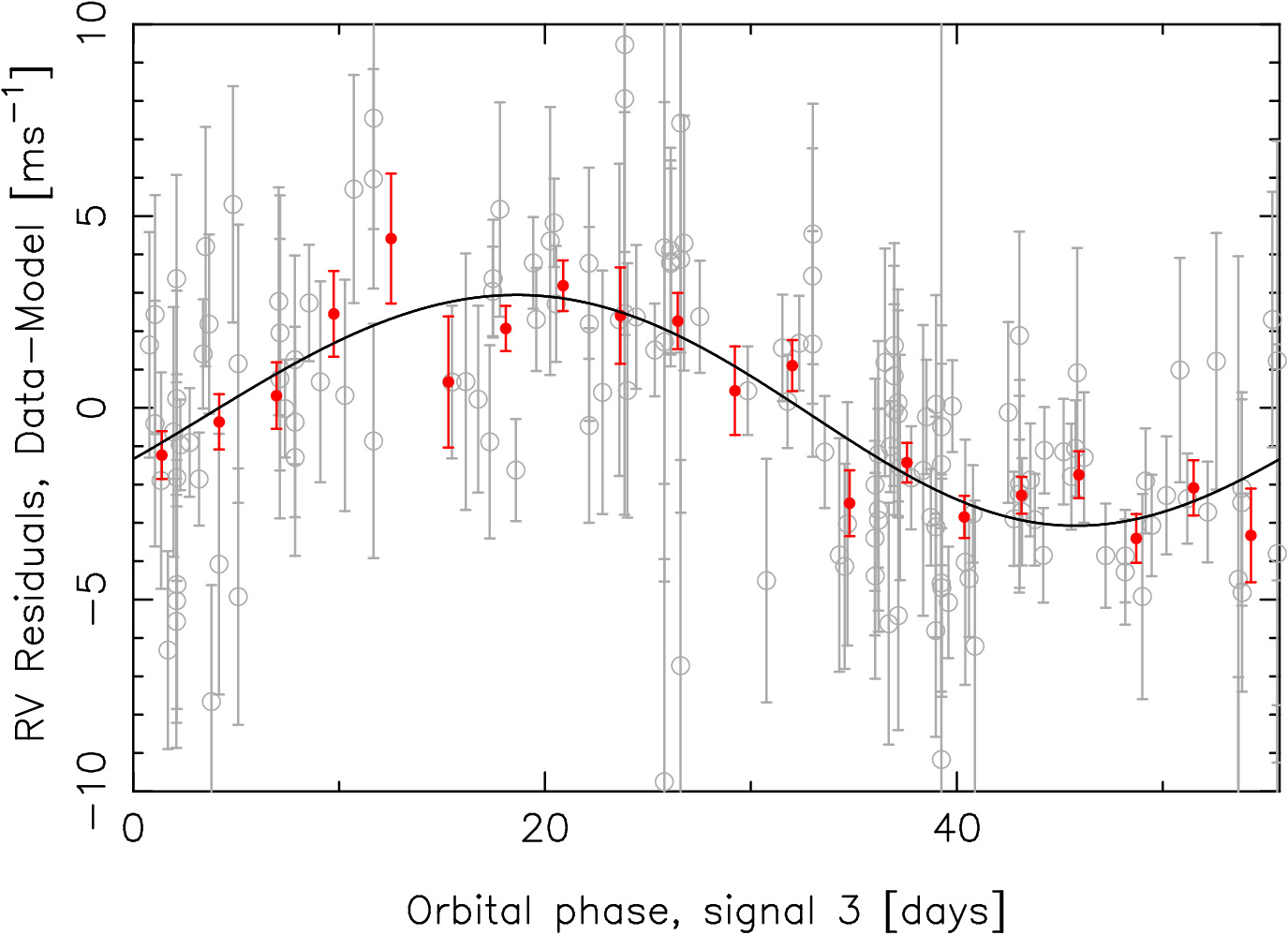}
\caption{Combined HARPS (red), HIRES (blue), and UVES (green) radial velocities of GJ 357 folded on the phases of the four signals detected in the data (left panels). The right panels show the same but with the velocities binned into 30 bins (red dots).}\label{fig:GJ357_phased}
\end{figure}

The third signal was detected as a unique probability maximum. This third signal, at a period of 55.664 [55.520, 55.808] days was also detected according to our criteria. In all, the Bayes factors in favour of the three signals were $1.6 \times 10^{9}$, $1.5 \times 10^{5}$, and $3.6 \times 10^{5}$, which indicates that they indeed exceed the detection threshold of $10^{4}$. Moreover, apart from the smallest data set from HIRES that only supported the existence of the strongest signal at a period of 9.1 days in the data, HARPS and UVES data sets provided consistently evidence in favour of the two weaker signals enabling us to conclude that there are indeed three periodic signals in the combined data of GJ 357. We have plotted the phase-folded radial velocities corresponding to all three signals in Fig. \ref{fig:GJ357_phased} for visual inspection.

The HARPS and HIRES activity indicators did not show any evidence in favour of activity-induced periodicities. Similarly, we could not identify any periodicities in the ASAS V-band photometry of the star. We thus interpret the signals in the radial velocities as evidence in favour of a system of three planet candidates orbiting the star. These candidates with orbital periods of 9.1, 3.9, and 56 days have minimum masses of 3.6 [2.0, 5.4], 2.5 [1.3, 3.8], and 7.7 [4.2, 11.6] M$_{\oplus}$, respectively, making it possible to classify them as two hot super-Earths and a warm mini-Neptune orbiting the star.

\clearpage

\subsection{GJ 358}\label{sec:GJ358}

\begin{figure}
\center
\includegraphics[angle=270, width=0.49\textwidth,clip]{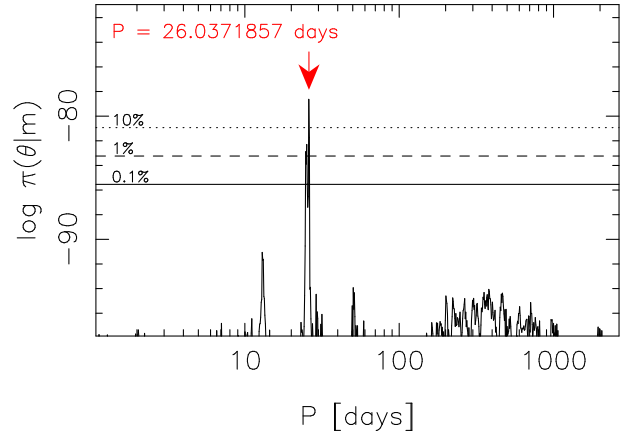}
\caption{As in Fig. \ref{fig:GJ357_period_search} but for the signal in the HARPS data of GJ 358.}\label{fig:GJ358_period_search}
\end{figure}

\begin{figure}
\center
\includegraphics[angle=270, width=0.49\textwidth]{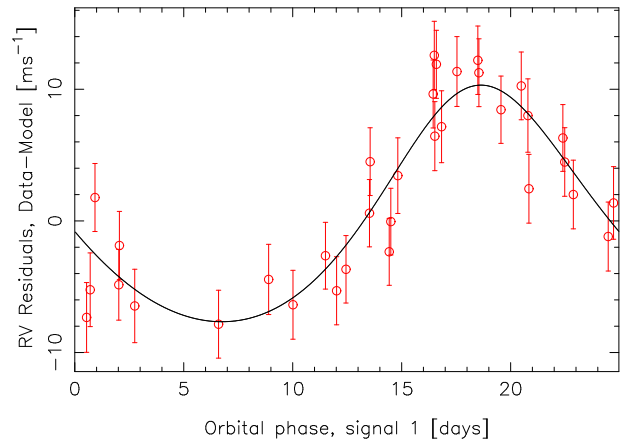}
\caption{HARPS radial velocity data of GJ 358 folded on the phase of the detected signal.}\label{fig:GJ358_phase}
\end{figure}

\begin{figure}
\center
\includegraphics[angle=270, width=0.49\textwidth]{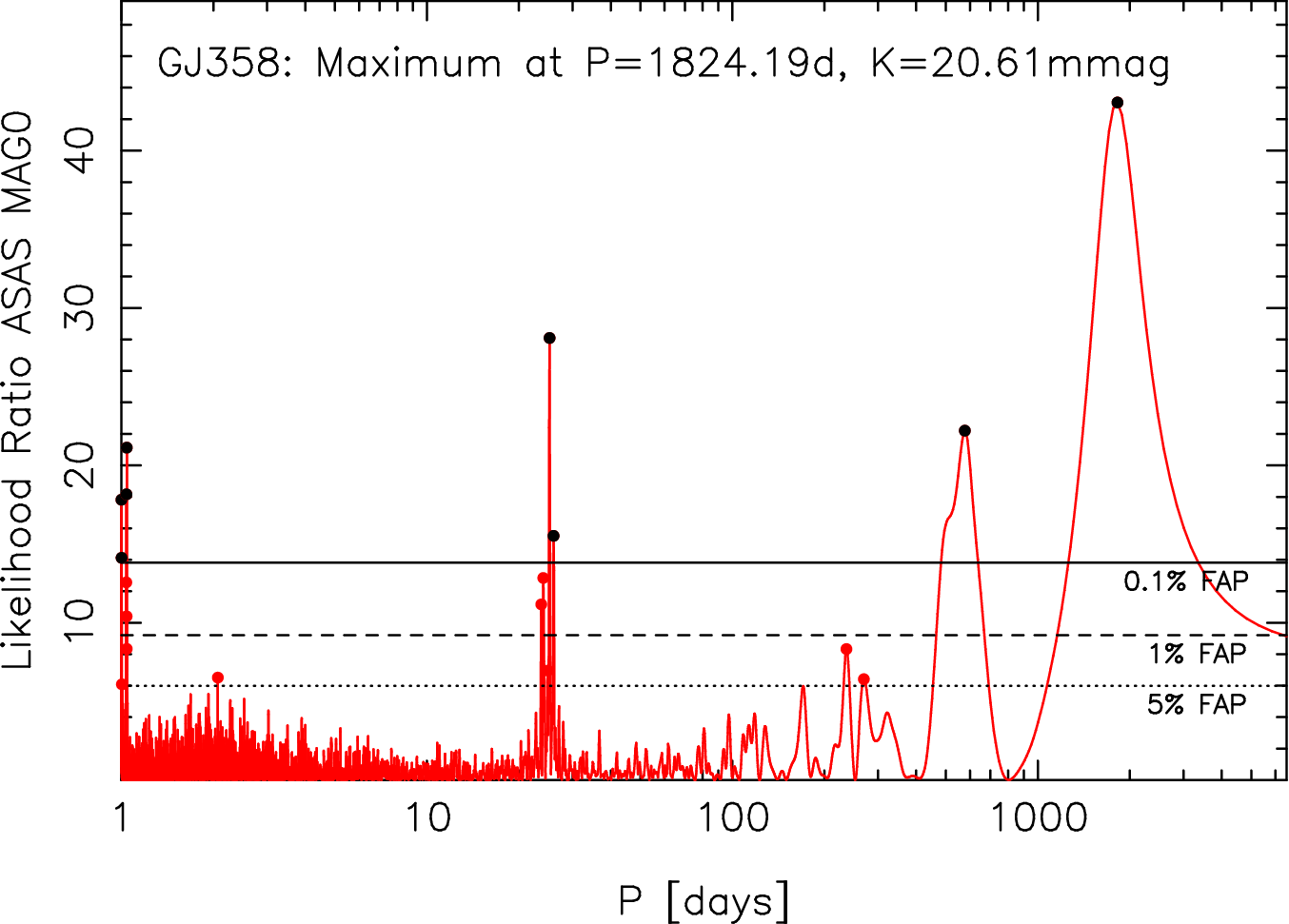}

\includegraphics[angle=270, width=0.49\textwidth]{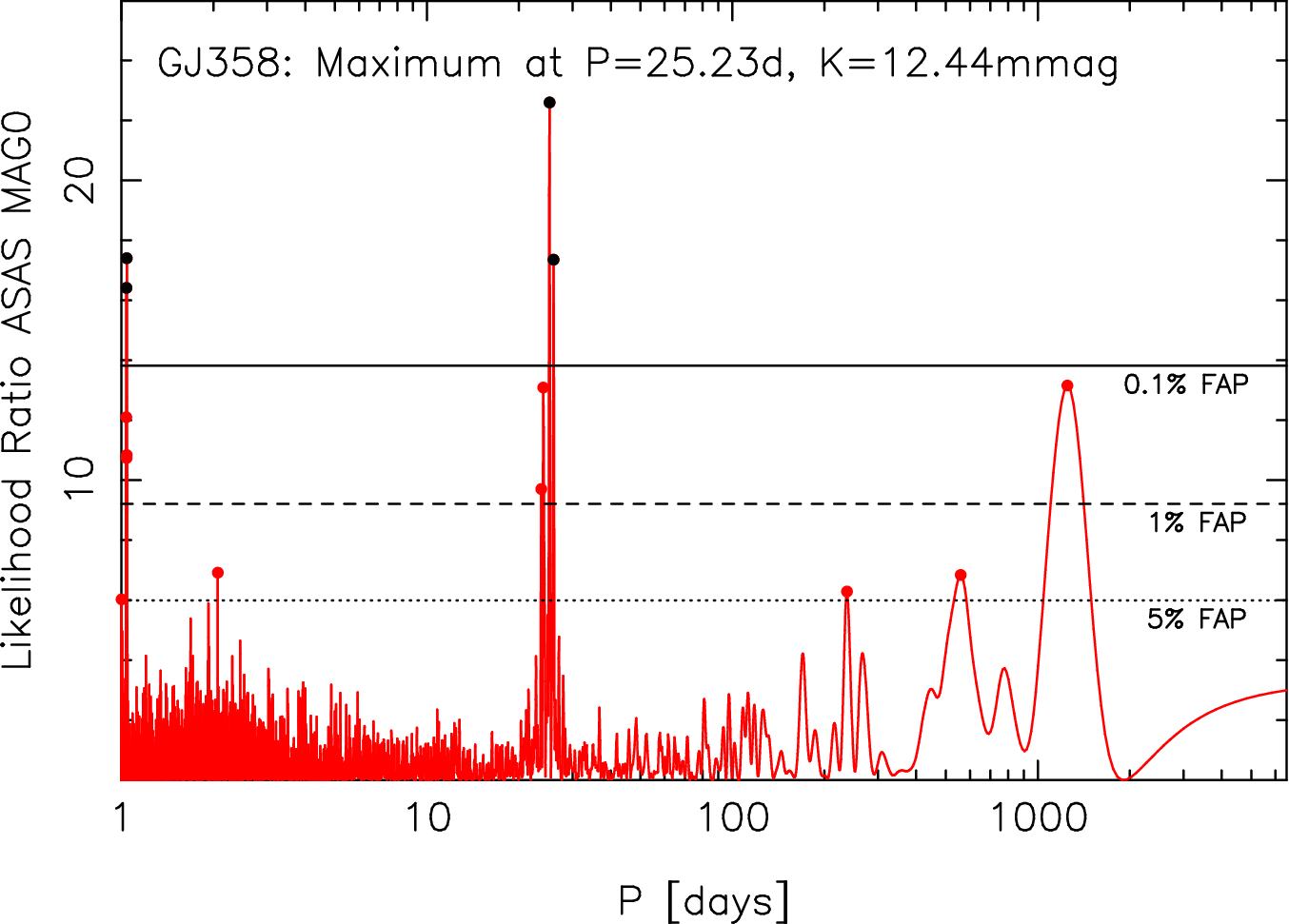}
\caption{Likelihood periodogram of ASAS V-band photometry of GJ 358 (top panel) and the residual periodogram after subtracting the long-period signal (bottom panel).}\label{fig:GJ358_asas}
\end{figure}

The 34 HARPS velocities of GJ 358 (HIP 47425) were found to contain a unique and well-constrained periodic signal at a period of 24.966 [24.932, 25.000] days with an amplitude of 8.58 [4.84, 12.31] ms$^{-1}$ (Figs. \ref{fig:GJ358_period_search} and \ref{fig:GJ358_phase}). However, as also reported by \citet{kiraga2012}, this signal has a photometric counterpart in the set of 907 ASAS V-band photometric measurements. It is thus rather clear, as also discussed by \citet{bonfils2013}, that the radial velocity signal corresponds to stellar rotation that has a period of roughly 25.23 d based on ASAS photometry (Fig. \ref{fig:GJ358_asas}). The photometric rotation period was originally determined by \citet{kiraga2007} based on ASAS data. We note that the dominant feature in the ASAS photometry is the periodicity at a period of 1800 days likely corresponding to a stellar magnetic cycle.

We have mainly included GJ 358 in this section because it serves as an example of a detection of the rotation period of an M dwarf. Although the photometric and radial velocity signals do not coincide exactly, we consider it to be the most probable explanation in such a case when they have a similar period.

Unlike the claim by \citet{bonfils2013}, we did not find linear relationships between the velocities and any of the HARPS activity indicators. This was apparent because the parameters quantifying the (linear) relationship were not found to be statistically significantly different from zero with a 99\% credibility.

\clearpage

\subsection{GJ 361}

\begin{figure}
\center
\includegraphics[angle=270, width=0.49\textwidth]{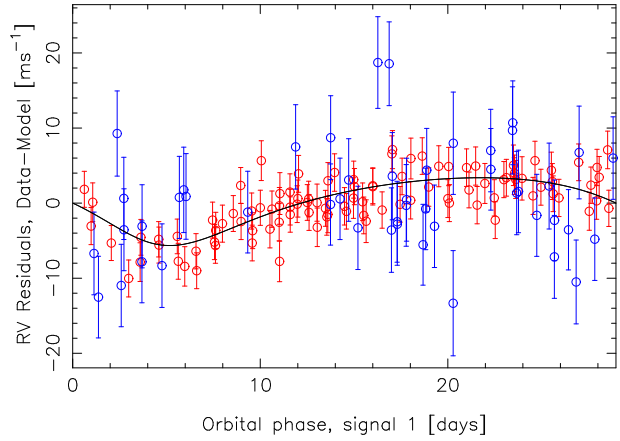}
\caption{HARPS (red) and HIRES (blue) radial velocities of GJ 361 folded on the phase of the radial velocity signal.}\label{fig:GJ361_phased}
\end{figure}

\begin{figure}
\center
\includegraphics[angle=270, width=0.49\textwidth,clip]{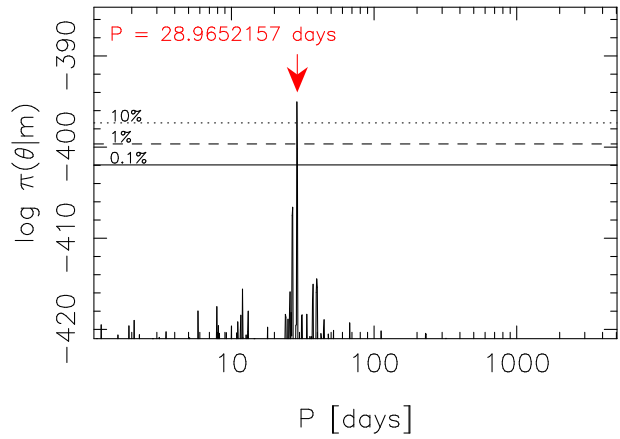}
\caption{Posterior probability density given the radial velocities of GJ 361 as a function of the period of the Keplerian signal. The red arrow denotes the position of the global maximum in the period space and the horizontal lines denote the 10\% (dotted), 1\% (dashed), and 0.1\% (solid) equiprobability thresholds with respect to the maximum.}\label{fig:GJ361_period_search}
\end{figure}

GJ 361 (HIP 47513) has been intensively observed with HARPS. With 101 radial velocities available in the ESO archive -- together with 42 HIRES data that we obtained with a baseline of more than 4000 days -- we expected to be able to detect the signals of super-Earths in the stellar liquid-water habitable zone of GJ 361 estimated to be between 0.13 and 0.22 AU based on the equations of \citet{kopparapu2013}. Indeed, we observed a significant signal at a period of 28.958 [28.886, 29.010] days with an amplitude of 3.81 [2.10, 5.52] ms$^{-1}$ (Figs. \ref{fig:GJ361_phased} and \ref{fig:GJ361_period_search}). Although the phase-folded Keplerian curve in Fig. \ref{fig:GJ361_phased} appears to correspond to an eccentric orbit, the eccentricity parameter is not statistically significantly different from zero.

We could not identify any periodicities in the available HARPS and HIRES spectral activity indicators. Moreover, we could not find any significant periodicities in the ASAS V-band photometric measurements of the star. It therefore appears the most likely explanation that the radial velocity signal is caused by a warm Neptune with a minimum mass of 11.5 [5.8, 17.1] M$_{\oplus}$ orbiting the star in the stellar liquid-water habitable zone.

\clearpage

\subsection{GJ 388}\label{sec:GJ388}

GJ 388 (AD Leo) is a photometrically variable star and has been reported to have a photometric periodicity of 2.7$\pm$0.05 d with an amplitude of $\Delta m_{\rm V} = 24\pm2$ mmag \citep{spiesman1986}. Although they do not discuss the statistical significance of this periodicity, \citet{spiesman1986} speculate that this periodicity might be caused by starspots co-rotating on the stellar surface and that it thus, essentially, corresponds to the stellar rotation period. A similar period was later observed by \citet{morin2008} based on spectropolarimetry. \citet{morin2008} reported a rotation period of 2.2399$\pm$0.0006 d with alternative local solutions at periods of 2.2264 and 2.2537 days. This periodicity was also readily detected in MOST photometry at a period of 2.23$^{+0.36}_{-0.27}$ days \citep{hunt2012}. We note that \citet{engle2009} also reported a photometric periodicity of 2.23 d for GJ 388.

\begin{figure}
\center
\includegraphics[angle=270, width=0.49\textwidth,clip]{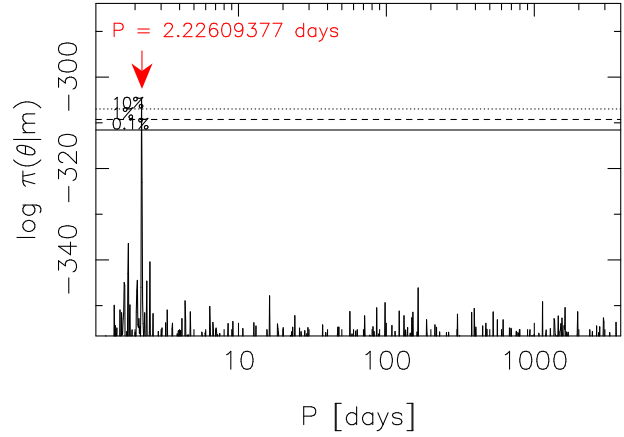}
\caption{Estimated posterior probability density as a function of the signal period given the combined HARPS and HIRES data of GJ 388. The red arrow indicates the position of the global probability maximum and the horizontal lines denote the 10\% (dotted), 1\% (dashed), and 0.1\% (solid) equiprobability thresholds with respect to the maximum.}\label{fig:GJ388_period_search}
\end{figure}

\begin{figure}
\center
\includegraphics[angle=270, width=0.49\textwidth,clip]{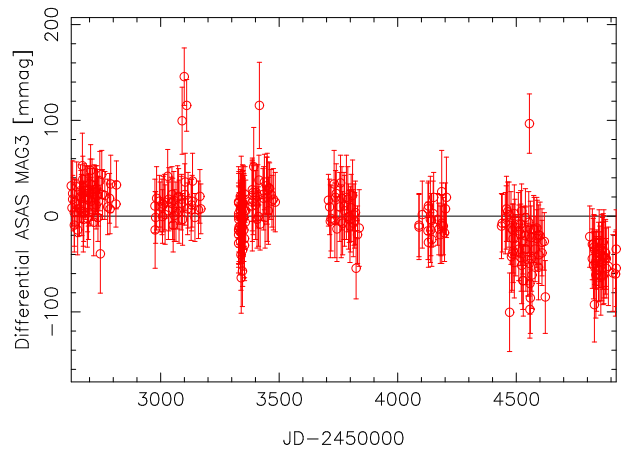}
\caption{ASAS V-band photometry data of GJ 388.}\label{fig:GJ388_asas_data}
\end{figure}

\begin{figure}
\center
\includegraphics[angle=270, width=0.49\textwidth]{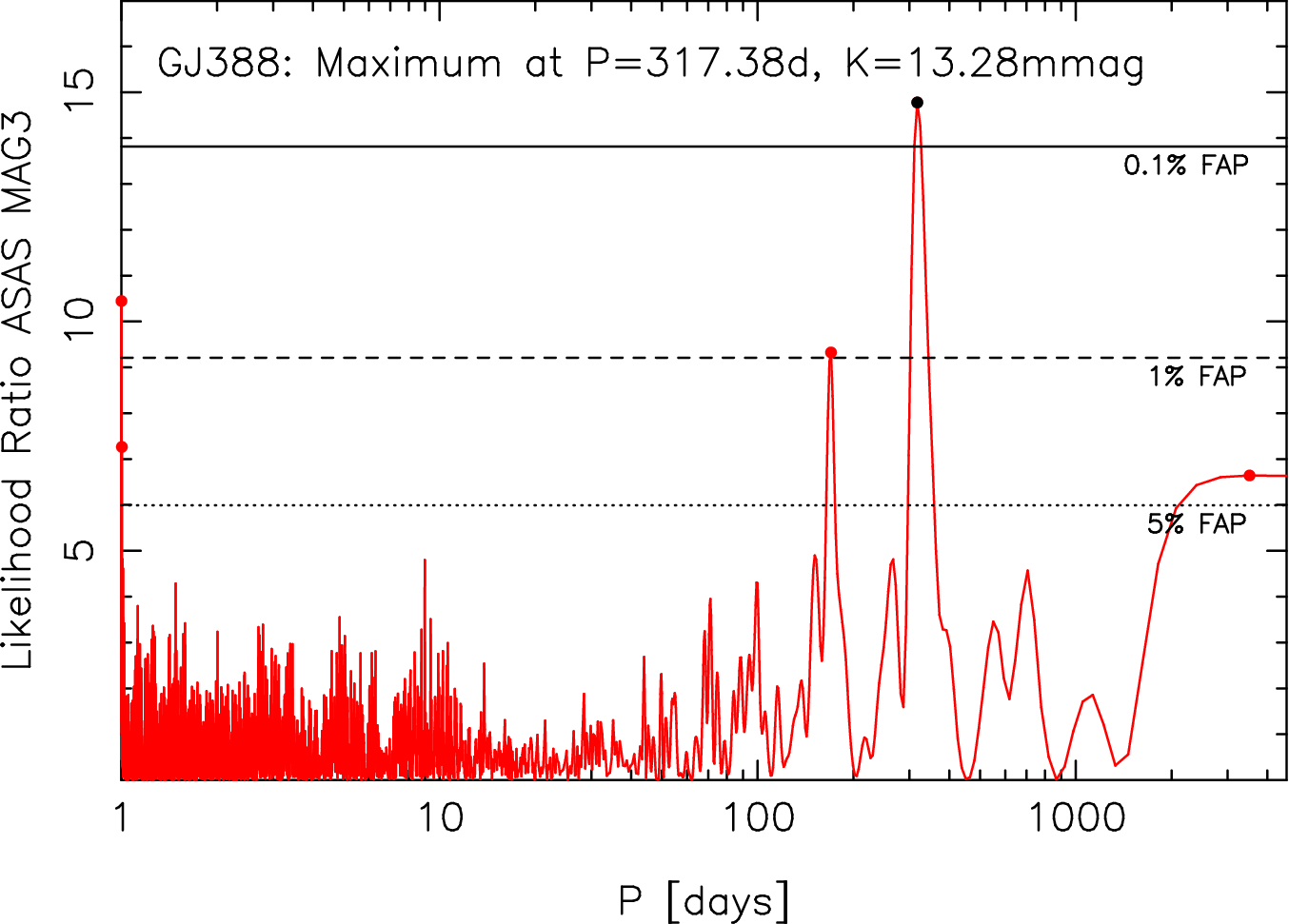}

\includegraphics[angle=270, width=0.49\textwidth]{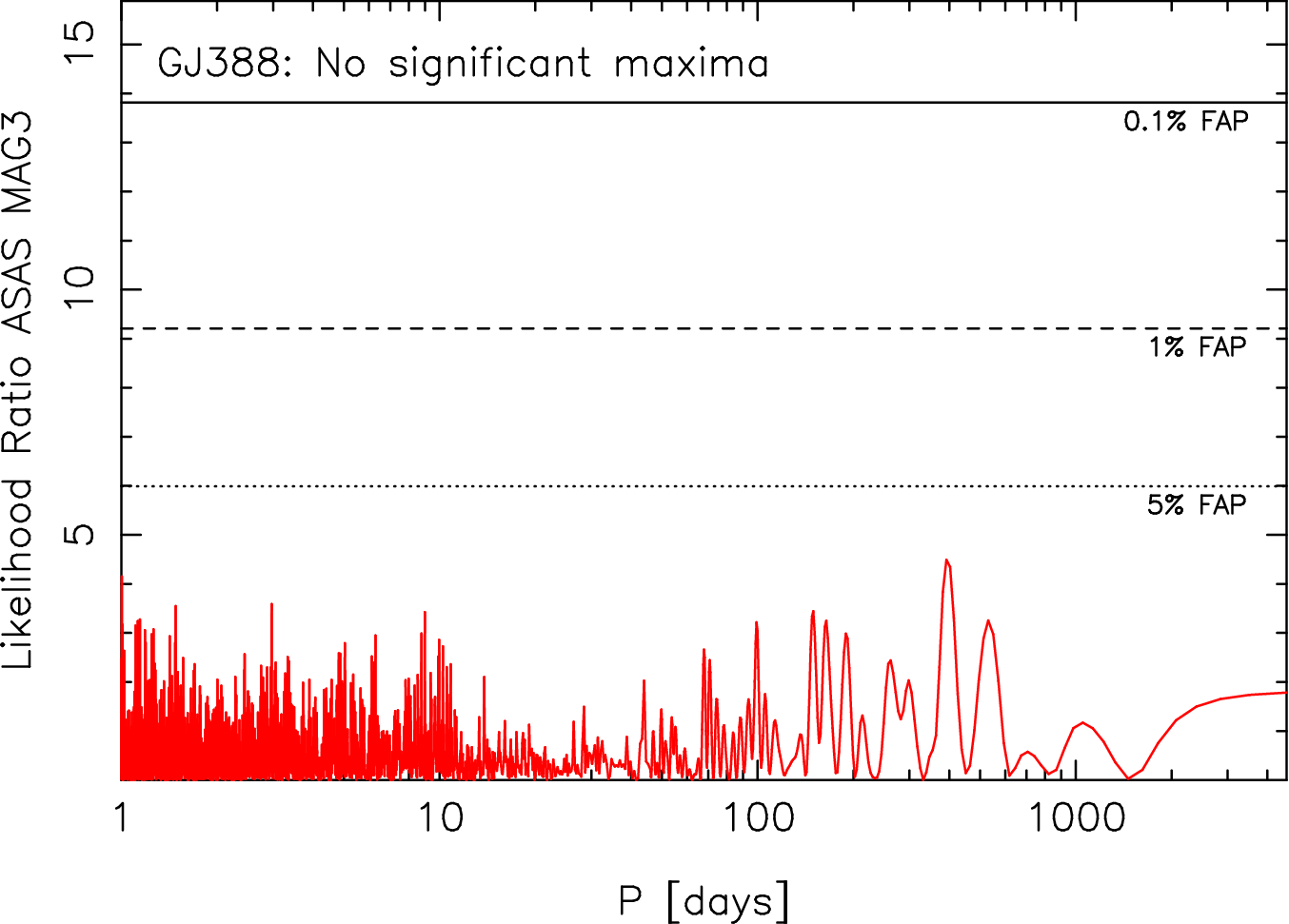}
\caption{Likelihood periodogram of ASAS V-band photometry of GJ 388 (top panel) and the residual periodogram after subtracting the long-period signal (bottom panel).}\label{fig:GJ388_asas}
\end{figure}

We observed a strong periodicity in the combined HARPS and HIRES radial velocities of GJ 388 at a period of 2.22599 [2.22557, 2.22687] days corresponding to the rotation period or planetary signal identified by \citet{morin2008}, \citet{hunt2012} and \citet{tuomi2018} as demonstrated in Fig. \ref{fig:GJ388_period_search}. A periodicity of 2.22 d, together with another one at 1.8 d, was also reported by \citet{bonfils2013} and \citet{reiners2013} based on HARPS radial velocities. However, we could not confirm the existence of the photometric periodicity when analysing the set of 326 ASAS V-band photometry measurements (Fig. \ref{fig:GJ388_asas_data}). In fact apart from a periodicity of 317 days, we did not see any hints of photometric periodicities at or near 2.2 days although signals with amplitudes of 24 mmag would have been easily observed in the available ASAS V-band data at such a period (the 317-day signal has an amplitude of only 13.3 mmag). This is demonstrated by showing the likelihood periodogram of the 326 ASAS photometry measurements in Fig. \ref{fig:GJ388_asas} where the highest likelihood ratios, after subtracting the significant signal at a longer period, are all well below the 5\% FAP threshold.

\begin{figure}
\center
\includegraphics[angle=270, width=0.24\textwidth]{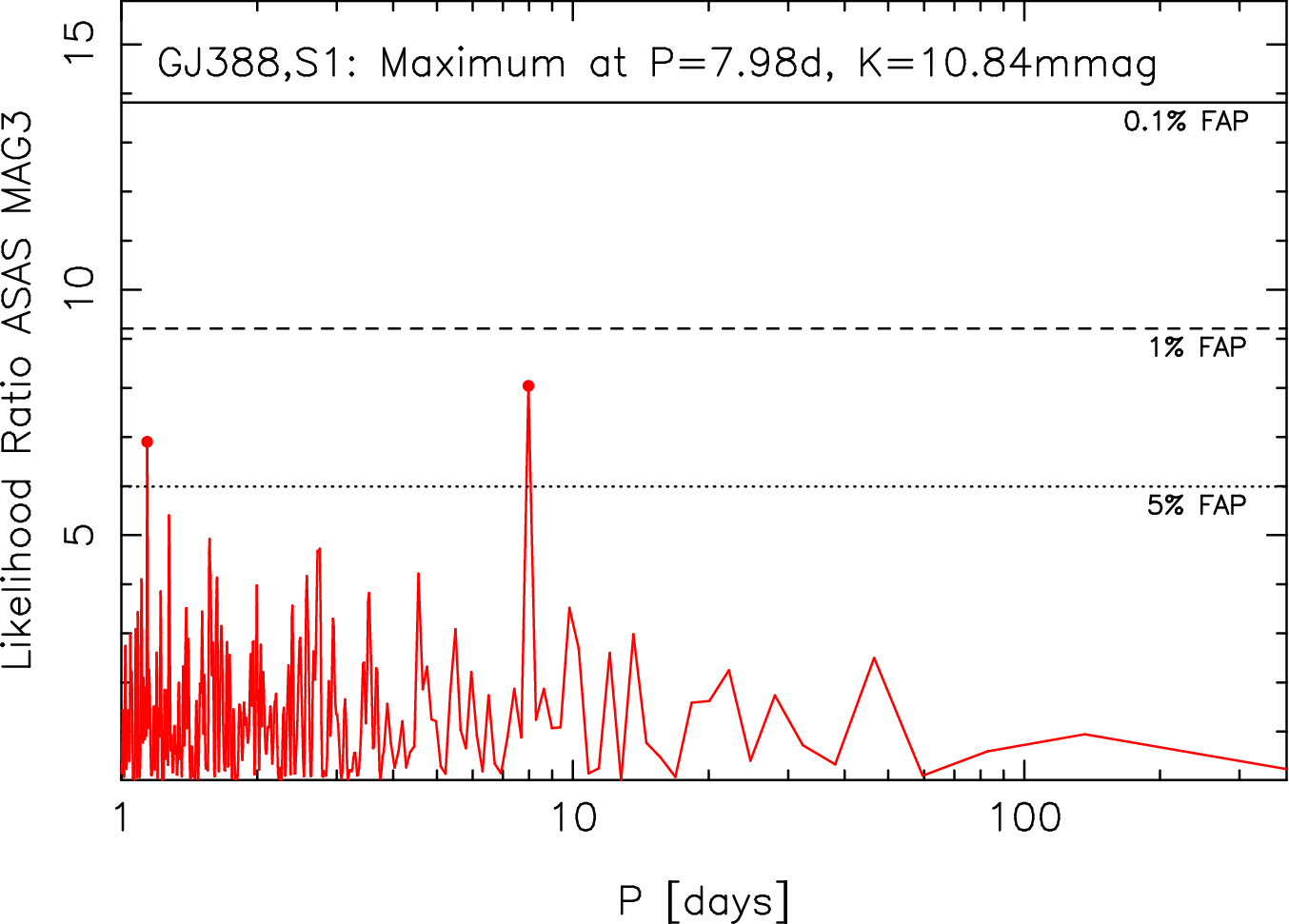}
\includegraphics[angle=270, width=0.24\textwidth]{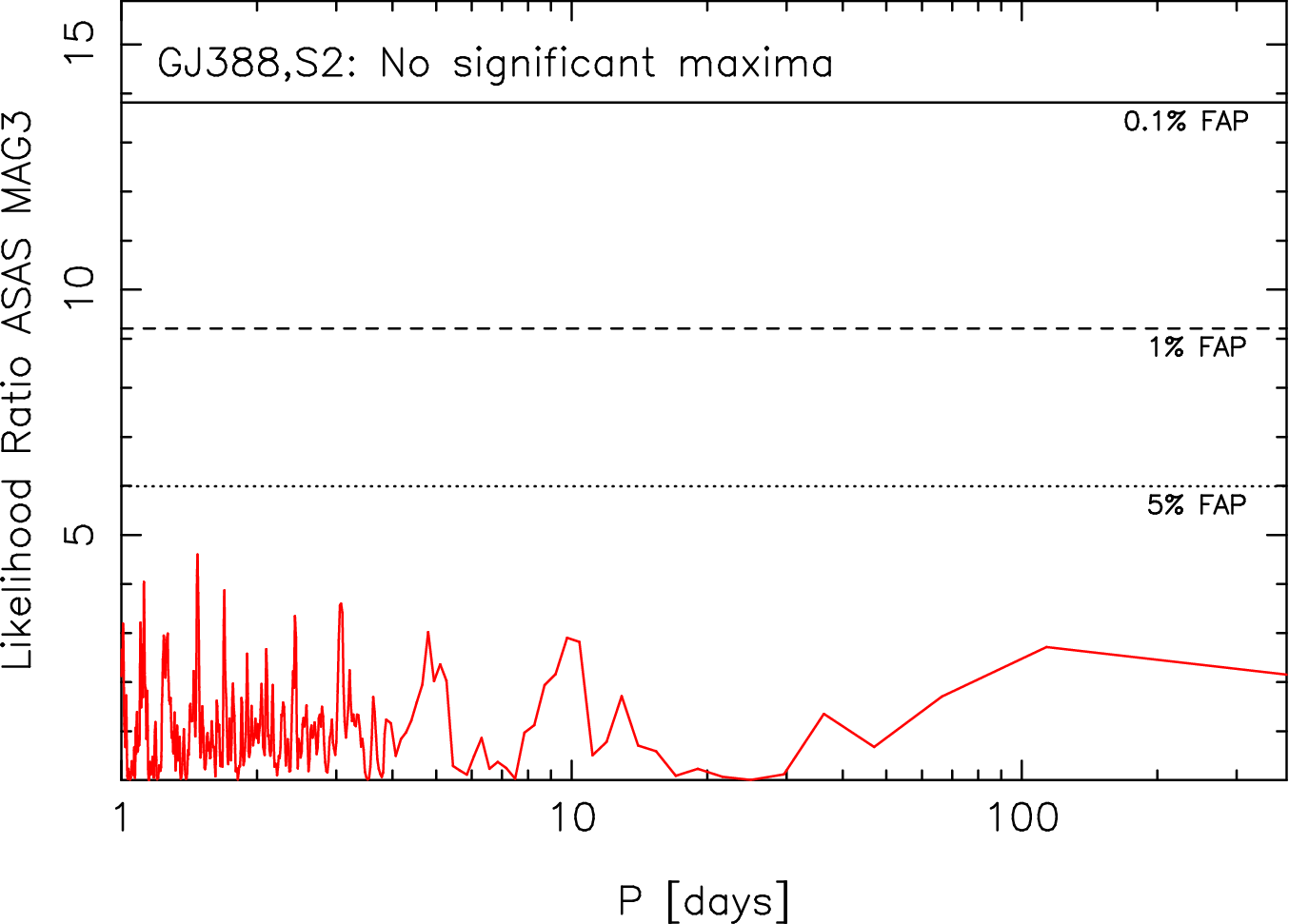}
\includegraphics[angle=270, width=0.24\textwidth]{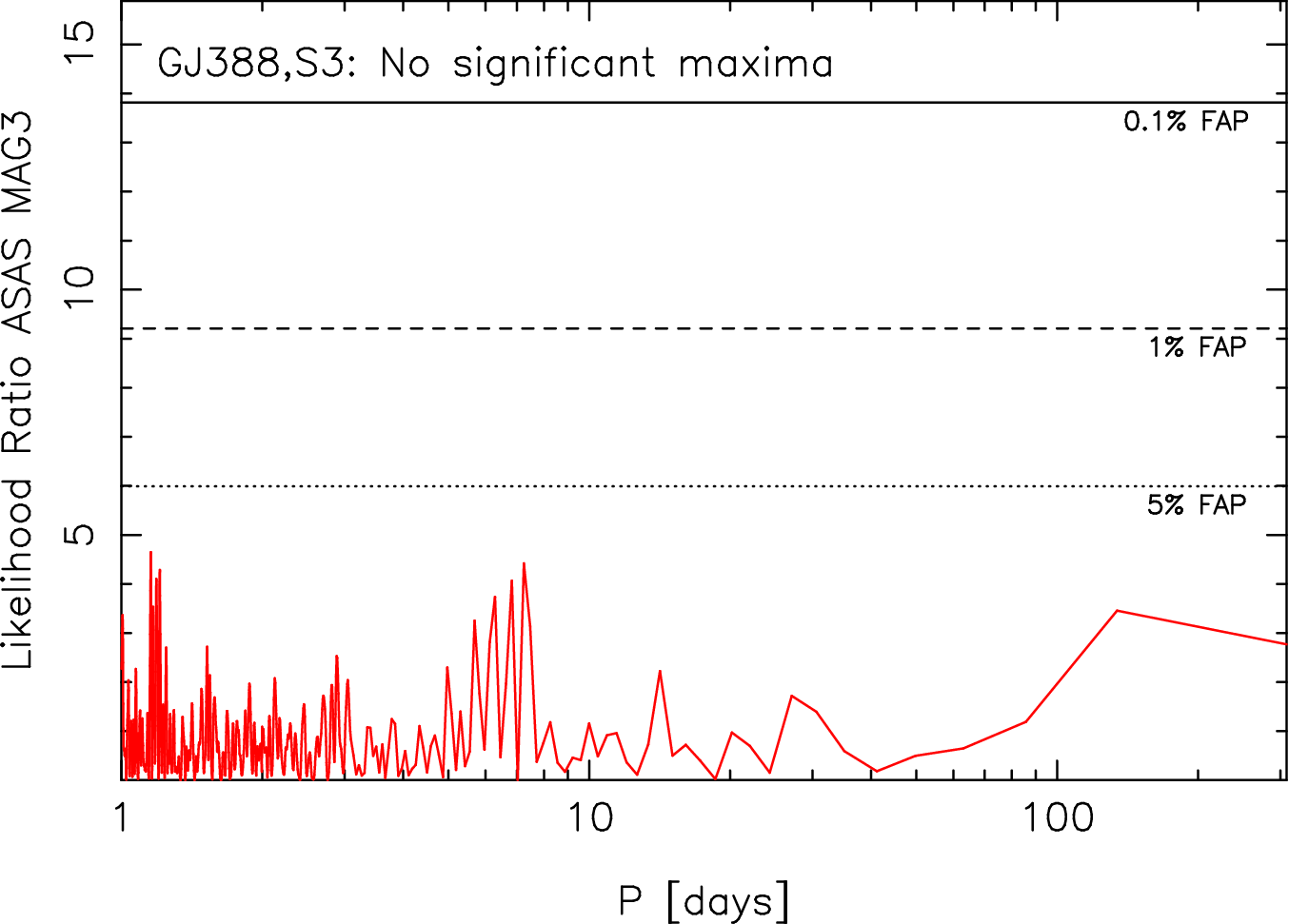}

\includegraphics[angle=270, width=0.24\textwidth]{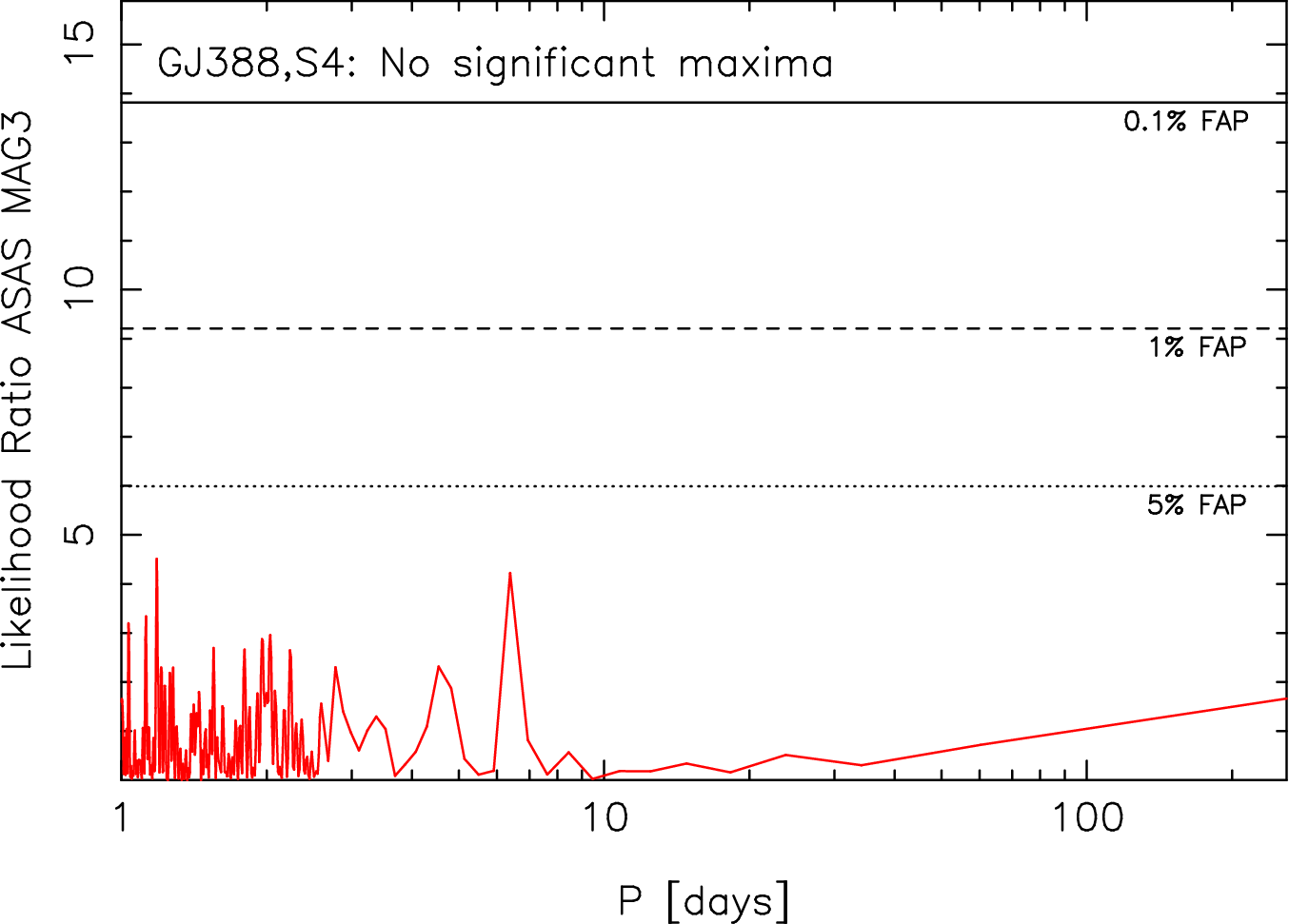}
\includegraphics[angle=270, width=0.24\textwidth]{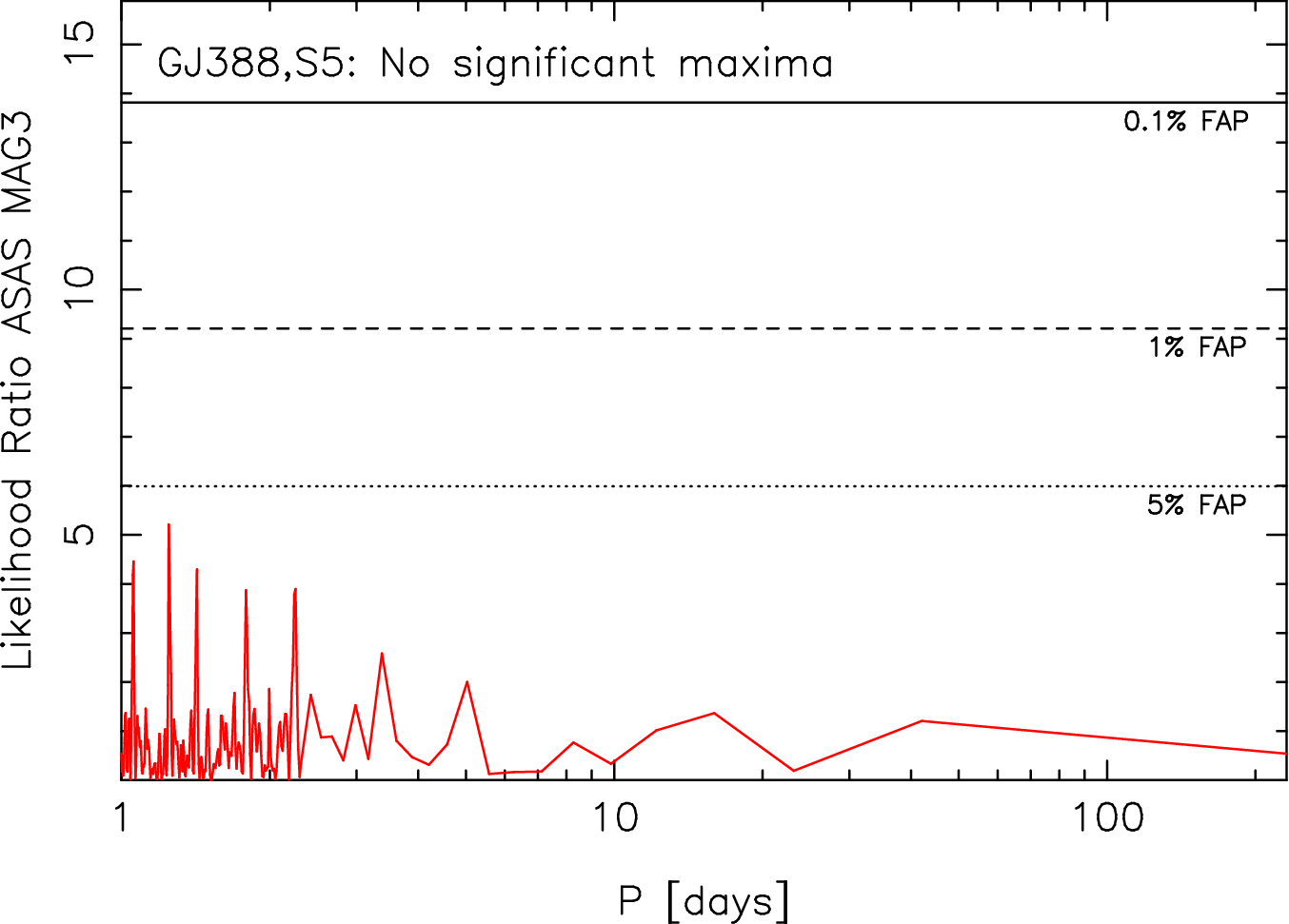}
\includegraphics[angle=270, width=0.24\textwidth]{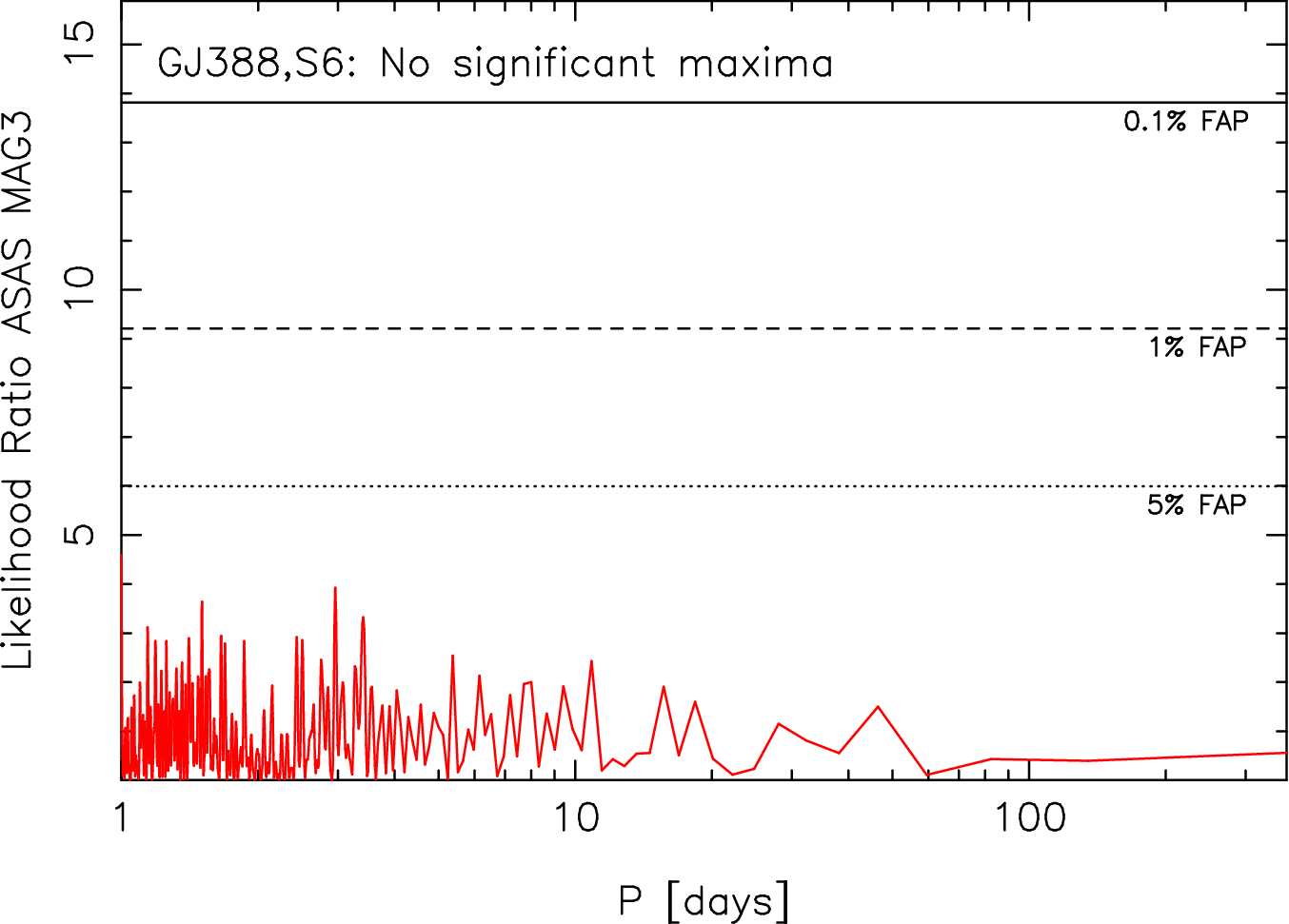}

\includegraphics[angle=270, width=0.24\textwidth]{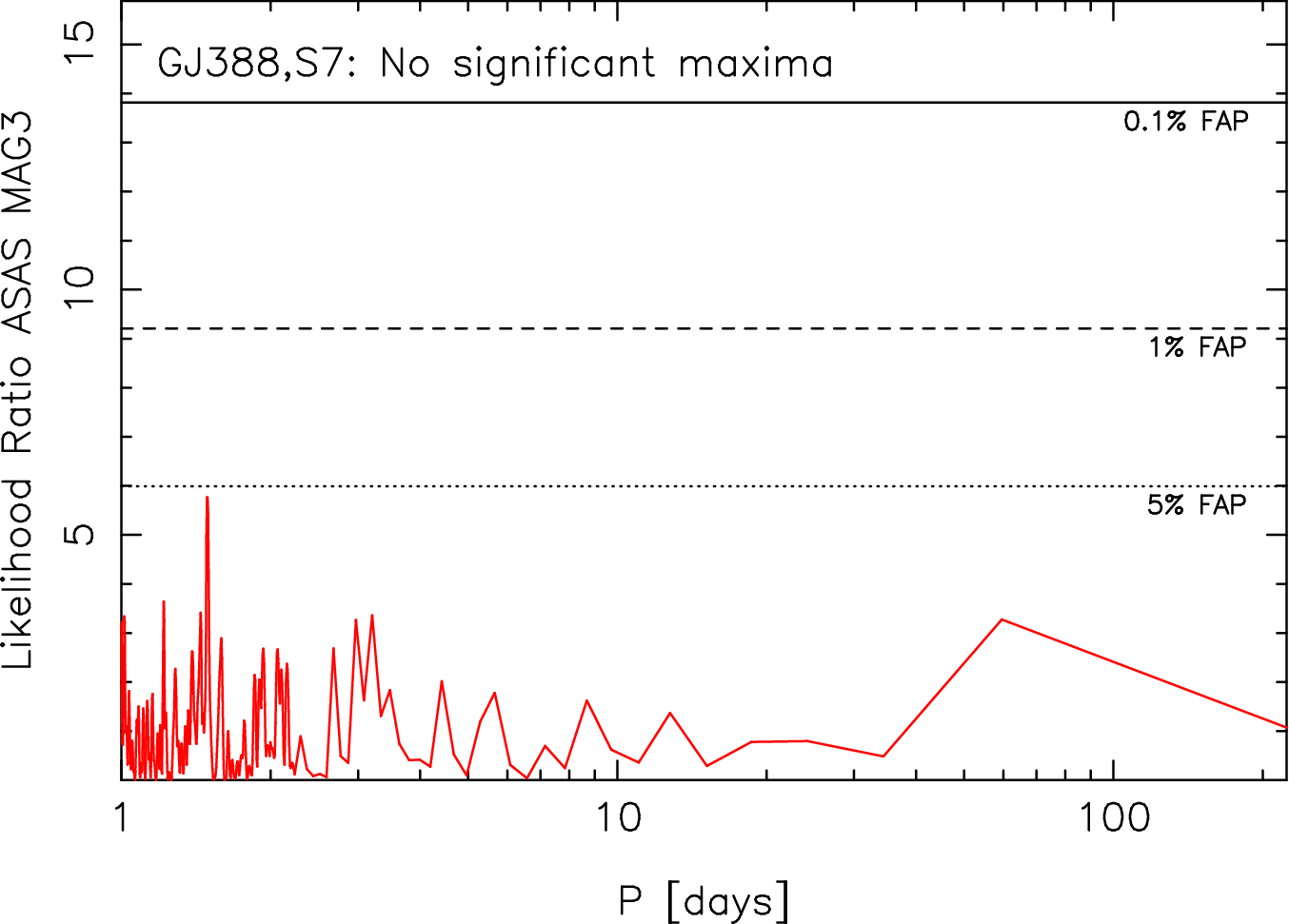}
\caption{Likelihood-ratio periodogram of ASAS photometry of GJ 388 for each observing season (see Fig. \ref{fig:GJ388_asas_data}. The seven observing seasons are denoted by S1, ..., S7.}\label{fig:GJ388_asas_seasons}
\end{figure}

We also calculated the likelihood-ratio periodograms of each of the ASAS observing seasons separately (Fig. \ref{fig:GJ388_asas_seasons}). Apart from the first season (S1), that showed weak evidence for a periodicity of 8.0 days, we could not identify periodic short-term signals in any of them. It is thus apparent that there is no evidence for photometric rotation period in the data of any of the observing seasons either. This, in turn, implies that unlike the RV signal \citep{tuomi2018}, the photometric signal detected by \citet{hunt2012} in MOST photometry is likely not stable on time-scale of several months.

\begin{figure}
\center
\includegraphics[angle=270, width=0.49\textwidth]{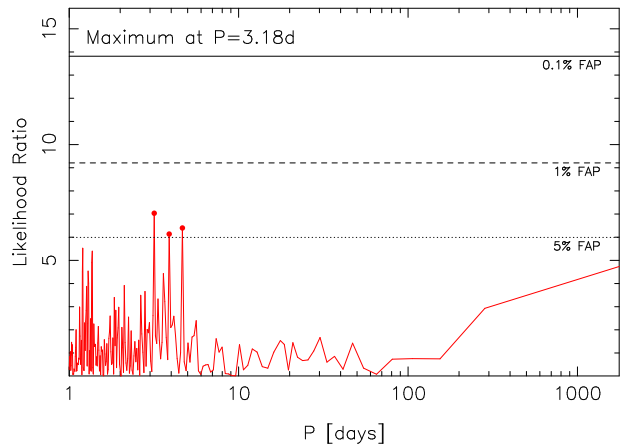}
\caption{Likelihood periodogram of the HARPS BIS values of GJ 388.}\label{fig:GJ388_bis}
\end{figure}

Similarly, we did not observe any periodic signals in the HARPS and HIRES activity indicators at or around a period of 2 days. \citet{bonfils2013} stated that ``... the bisector span demonstrates that stellar activity is responsible for the [radial velocity] variation.'' However, when analysing the corresponding BIS values of HARPS, we did not see any significant periodicities in the data and, in contrast, observed a clear lack of periodogram power around 2.2 days (Fig. \ref{fig:GJ388_bis}). A consistent result was obtained by \citet{reiners2013} who found weak, albeit inconclusive, evidence for a periodicity in the HARPS S-indices. We did find that the HARPS radial velocities and BIS values are correlated and that this correlation is statistically significant (with 99\% credibility) but whether we accounted for this correlation in the statistical model or not, the signal at a period of 2.22 d remained unchanged suggesting that it is independent of the variations in the BIS values. We thus have evidence for a strong radial velocity signal without corresponding counterparts in activity-indices or photometry and that also appears independent of the variations in the activity data.

The analysis results of GJ 388 radial velocities and photometry are in stark contrast with those of e.g. GJ 176 (Section \ref{sec:GJ176}), GJ 205 (Section \ref{sec:GJ205}), GJ 208 (Section \ref{sec:GJ208}) and GJ 358 (Section \ref{sec:GJ358}) that show evidence for photometric periodicities at or very close to the periodicities in radial velocities. We thus expect that if a rotationally induced signal arising from the co-rotation of starspots and other active and inactive regions on the stellar surface had an amplitude as high as is observed for the signal in GJ 388 radial velocities, there would certainly be at least a weak photometric counterpart that was stable in long-baseline photometry. However, although ASAS data is precise enough for the detection of variations with amplitudes in the range 10-20 mmag, we could not find such signals in the data. We thus see no evidence for the 2.2 d period being caused by stellar activity and rotation rather than a close-in planet.

Although \citet{reiners2013} suggested that the radial velocity signal is wavelength-dependent, \citet{tuomi2018} showed that this is not the case. It is thus probable, as concluded by \citet{tuomi2018} in their detailed analysis, that the signal is caused by a planet orbiting the star.

According to our criteria, we thus classify the signal in the GJ 388 data as a candidate planet. It is then possible that, with an estimated minimum mass of 23.1 [17.5, 29.3] M$_{\oplus}$, such a short-period hot super-Neptune planet -- possibly much more massive than that as \citet{morin2008} speculate that the star is seen almost pole-on -- affects the stellar surface to the extent that \citet{morin2008} detected spectropolarimetric variations with a period corresponding to its orbit. In addition to the signal at a period of 2.2 d, we did not find any other significant periodicities in the combined HARPS and HIRES radial velocities of GJ 388.

\clearpage

\subsection{GJ 393}\label{sec:GJ393}

\begin{figure}
\center
\includegraphics[angle=270, width=0.49\textwidth,clip]{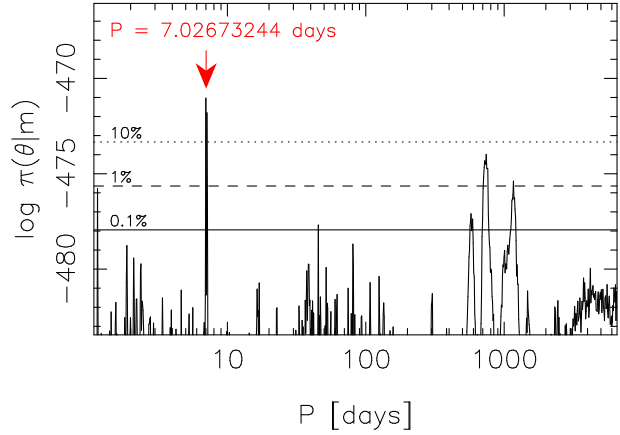}
\caption{Posterior probability density as a function of the period of the Keplerian signal given GJ 393 HARPS and HIRES velocities. The red arrow denotes the global maximum and the horizontal lines indicate the 10\% (dotted), 1\% (dashed), and 0.1\% equiprobability thresholds with respect to the maximum.}\label{fig:GJ393_period_search}
\end{figure}

GJ 393 (HIP 51317) has been intensively targeted by both HARPS and HIRES but no candidate planets have been reported to date. When searching for periodic signals in the combined data, we obtained evidence for a reasonably unique probability maximum in the period space at a period of 7.0269 [7.0228, 7.0304] days (Fig. \ref{fig:GJ393_period_search}). This global maximum was clearly the dominant feature in the posterior probability density but the other two, at periods of 700 and 1200 were exceeding or approaching the 1\% probability threshold of the global maximum and appeared to represent maxima clearly above the background noise levels.

\begin{figure}
\center
\includegraphics[angle=270, width=0.49\textwidth,clip]{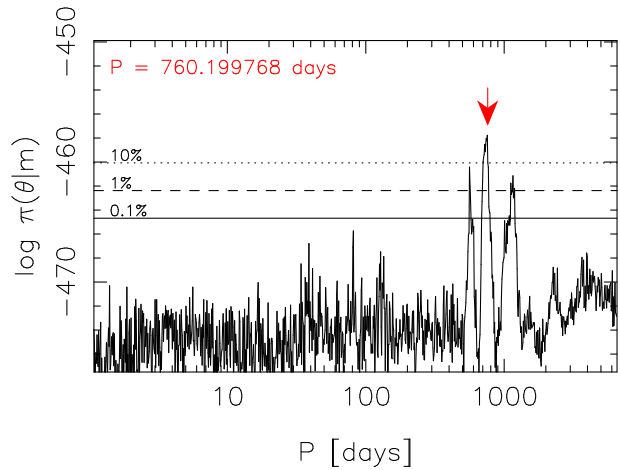}
\caption{As in Fig. \ref{fig:GJ393_period_search} but for the period parameter of the second Keplerian signal in a two-Keplerian model.}\label{fig:GJ393_psearch2}
\end{figure}

\begin{figure}
\center
\includegraphics[angle=270, width=0.49\textwidth,clip]{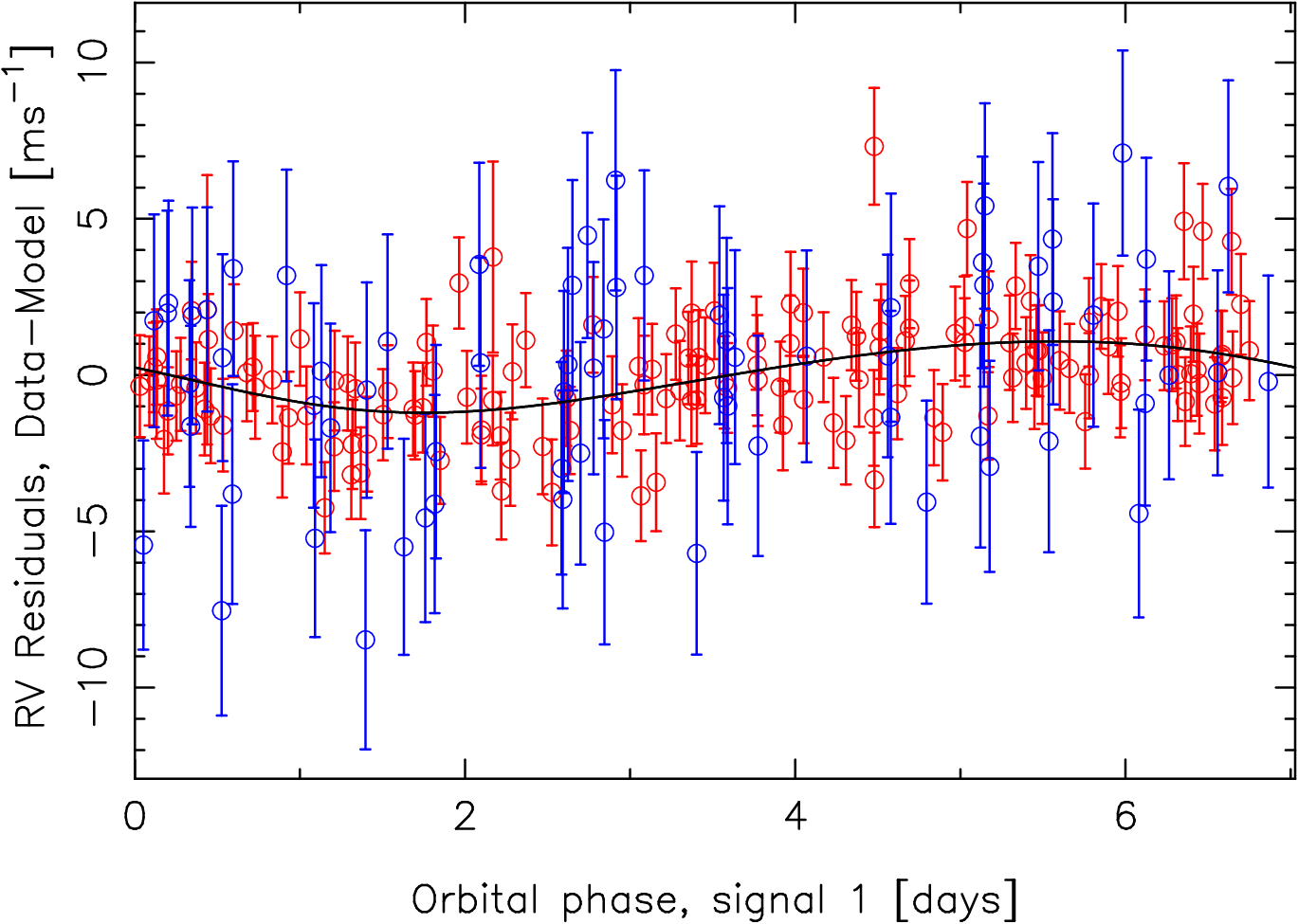}=

\includegraphics[angle=270, width=0.49\textwidth,clip]{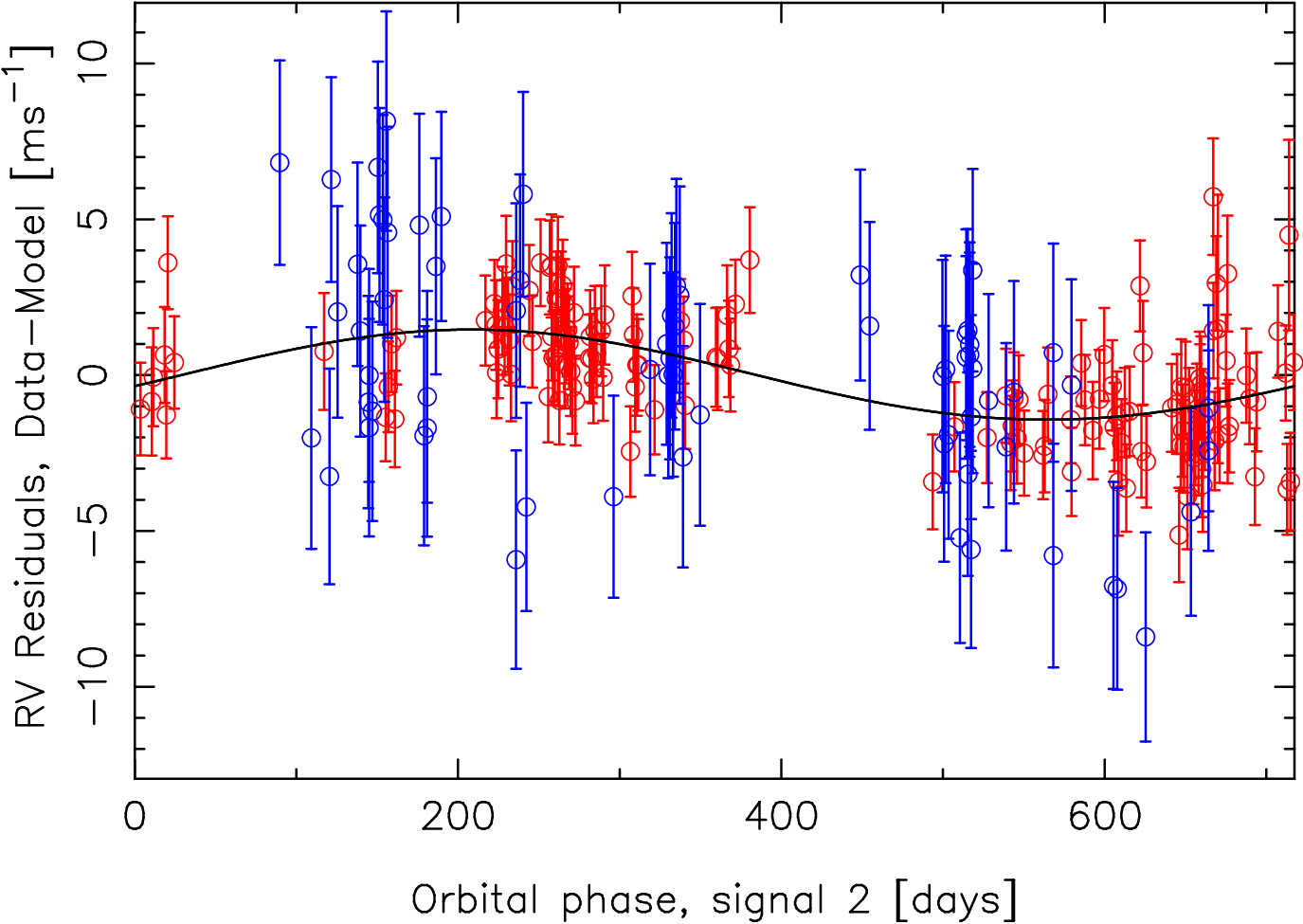}
\caption{Combined HARPS (red) and HIRES (blue) radial velocities of GJ 393 folded on the phases of the signals.}\label{fig:GJ393_phased}
\end{figure}

When searching for additional signals, we could identify a significant signal at a period of 731 [693, 783] days, together with its aliases on both sides (Fig. \ref{fig:GJ393_psearch2}). This aliasing is caused by a long gap in the HARPS data of 1550 days that yields a strong sampling frequency corresponding to a period of 2500 days. It is thus clear that all three maxima in the posterior probability density are actually caused by a single underlying periodicity of 730 days in the data. In Fig. \ref{fig:GJ393_phased}, we show the combined radial velocities folded on the periods of the signals for visual inspection.

\begin{figure}
\center
\includegraphics[angle=270, width=0.49\textwidth,clip]{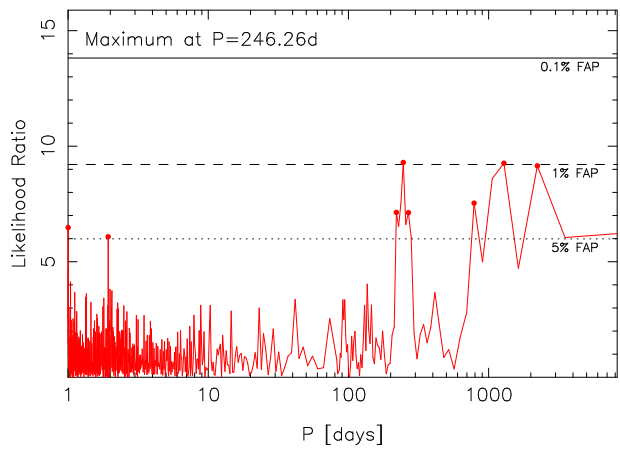}
\caption{Likelihood periodogram of the HARPS S-indices of GJ 393.}\label{fig:GJ393_S_periodogram}
\end{figure}

\begin{figure}
\center
\includegraphics[angle=270, width=0.49\textwidth,clip]{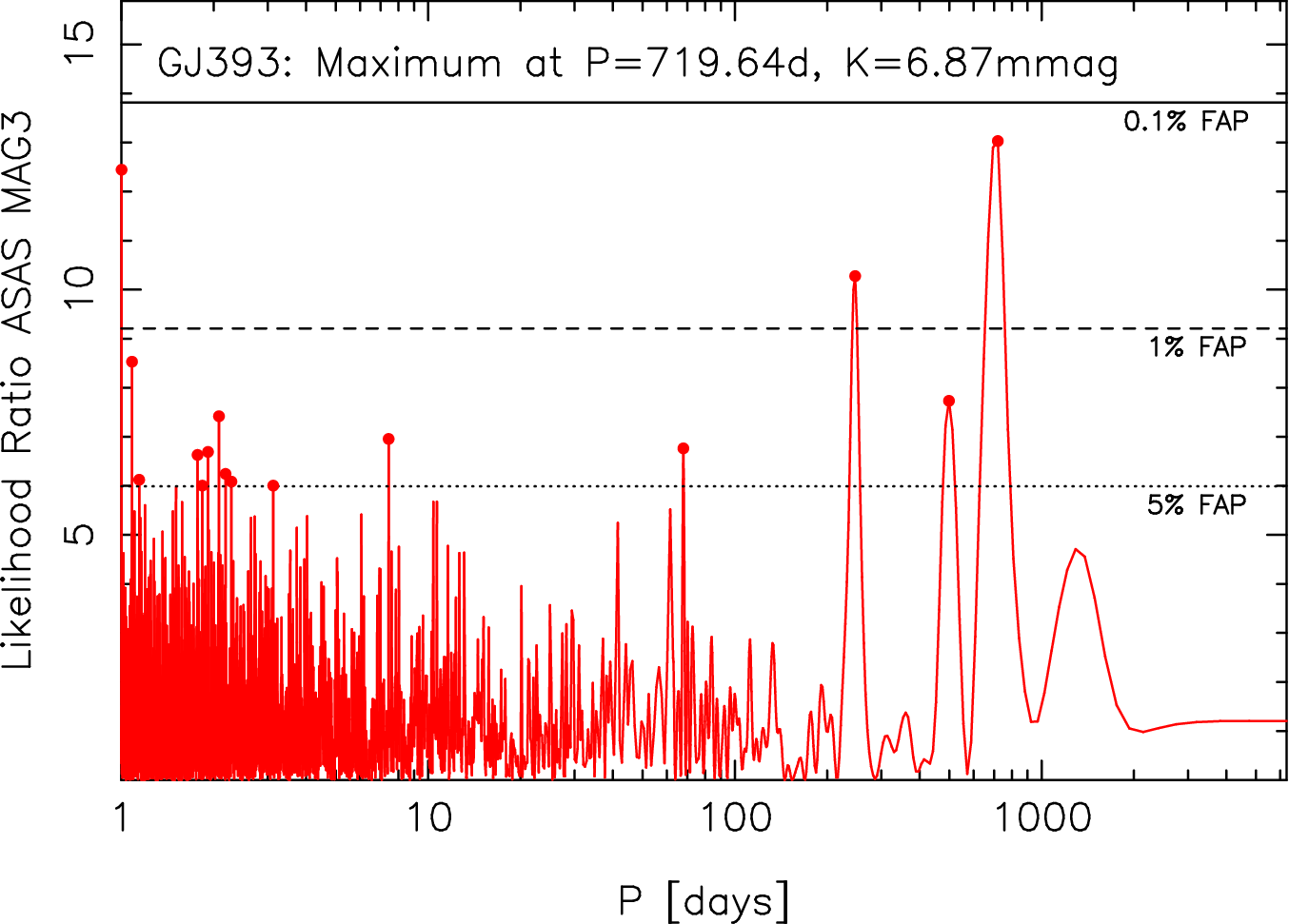}
\caption{Likelihood-ratio periodogram of the ASAS V-band photometry of GJ 393.}\label{fig:GJ393_ASAS}
\end{figure}

The HARPS S-indices suggested the presence of signals at a period of 246, 1300 or 2200 days (Fig. \ref{fig:GJ393_S_periodogram}) but evidence in favour of these periodicities is weak at most. We could not identify other significant signals in the HARPS and HIRES activity indicators. The set of 321 ASAS V-band photometry measurements, however, showed evidence for a periodicity of 720 days. Although this signal did not exceed the 0.1\% FAP threshold in (Fig. \ref{fig:GJ393_ASAS}), it coincides with the second periodicity in the radial velocities. It thus appears likely that the 730-day periodicity in the radial velocities is caused by stellar activity, possibly magnetic cycle, rather than a planet orbiting the star. We note that the suggestive 250-day signal in the HARPS S-indices appears to have a counterpart in the ASAS photometry as well.

\begin{figure}
\center
\includegraphics[angle=270, width=0.49\textwidth,clip]{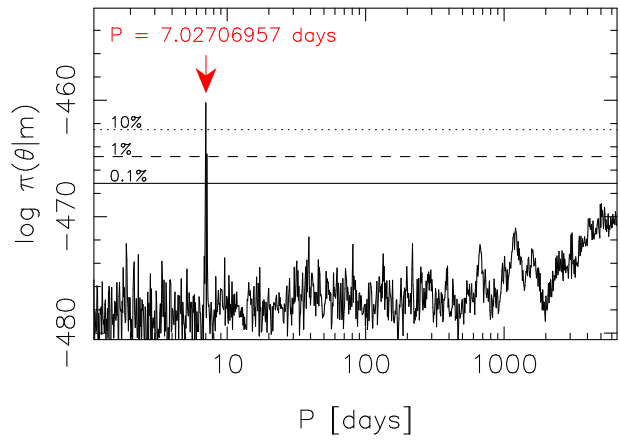}
\caption{As in Fig. \ref{fig:GJ393_psearch2} but when modelling the 730-day signal with the first Keplerian function when sampling the posterior density .}\label{fig:GJ393_psearch3}
\end{figure}

We could increase the uniqueness of the shorter signal by modelling the long-period one at a period of 730 days first, and then searching for a second periodicity. The corresponding posterior density is shown in Fig. \ref{fig:GJ393_psearch3} and demonstrates rather clearly that the 7.03-day signal is strongly present in the data.

The longer one of the radial velocity signals has a counterpart in the ASAS photometry and also likely in the HARPS S-indices (Figs. \ref{fig:GJ393_S_periodogram} and \ref{fig:GJ393_ASAS}). We thus interpret it as evidence for stellar activity cycle rather than a planet orbiting GJ 393. This is the only example in the current work where a photometric signal, potentially indicative of a stellar magnetic cycle at that period, has a counterpart in the radial velocity data. However, the shorter signal in the radial velocities does not have counterparts in the activity indices or photometry and we thus interpret it as a signal caused by a candidate planet orbiting the star. With an estimated minimum mass of 1.9 [0.9, 3.0] M$_{\oplus}$ and a semi-major axis of 0.055 [0.049, 0.060] AU, this planet is classified as a hot super-Earth.

\clearpage

\subsection{GJ 397}

\begin{figure}
\center
\includegraphics[angle=270, width=0.49\textwidth,clip]{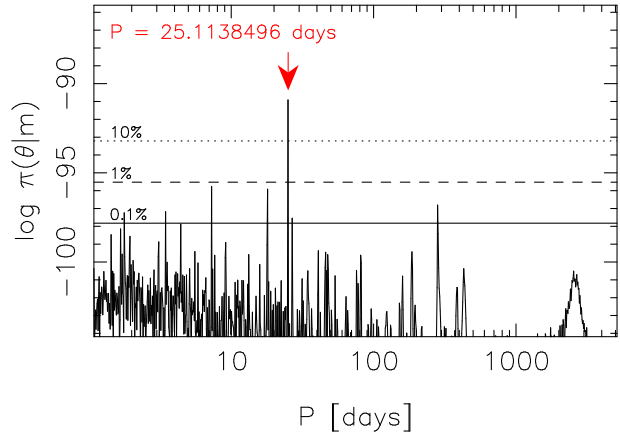}
\caption{As in Fig. \ref{fig:GJ357_period_search} but for the signal in the HIRES data of GJ 397.}\label{fig:GJ397_period_search}
\end{figure}

\begin{figure}
\center
\includegraphics[angle=270, width=0.49\textwidth,clip]{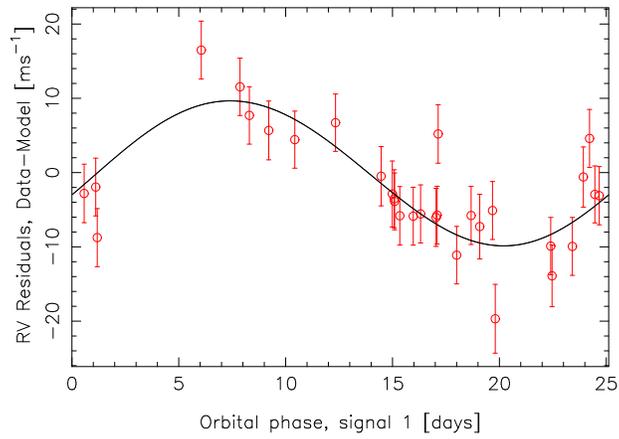}
\caption{Phase-folded HIRES radial velocities of GJ 397 and the Keplerian signal (solid curve).}\label{fig:GJ397_curve}
\end{figure}

With only 30 radial velocities from HIRES, we did not expect to be able to constrain the possible planetary system around GJ 397 (HIP 51525) very well. However, we obtained evidence for a hot Neptune candidate, when classified according to our definitions in Section \ref{sec:planet_classification}. We obtain a period of 25.110 [25.087, 25.140] days and an amplitude of 9.41 [5.32, 13.50] ms$^{-1}$ for this signal, which already indicates that it is strong enough to be expected to be seen in such a HIRES data set (Figs. \ref{fig:GJ397_period_search} and \ref{fig:GJ397_curve}). This signal did not exceed the detection threshold of \citet{butler2016} and was therefore not tabulated by them.

\begin{figure}
\center
\includegraphics[angle=270, width=0.49\textwidth,clip]{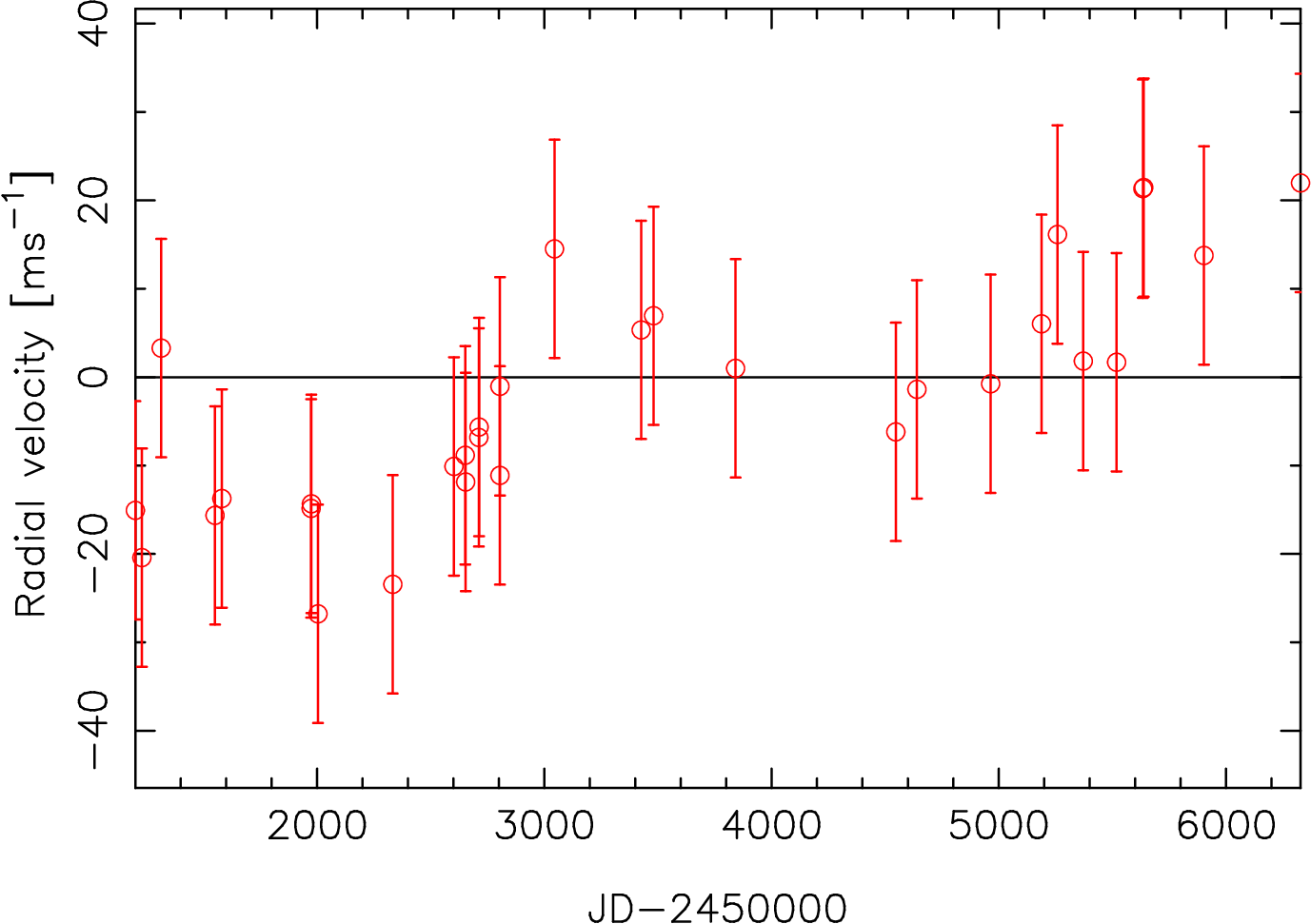}

\includegraphics[angle=270, width=0.49\textwidth,clip]{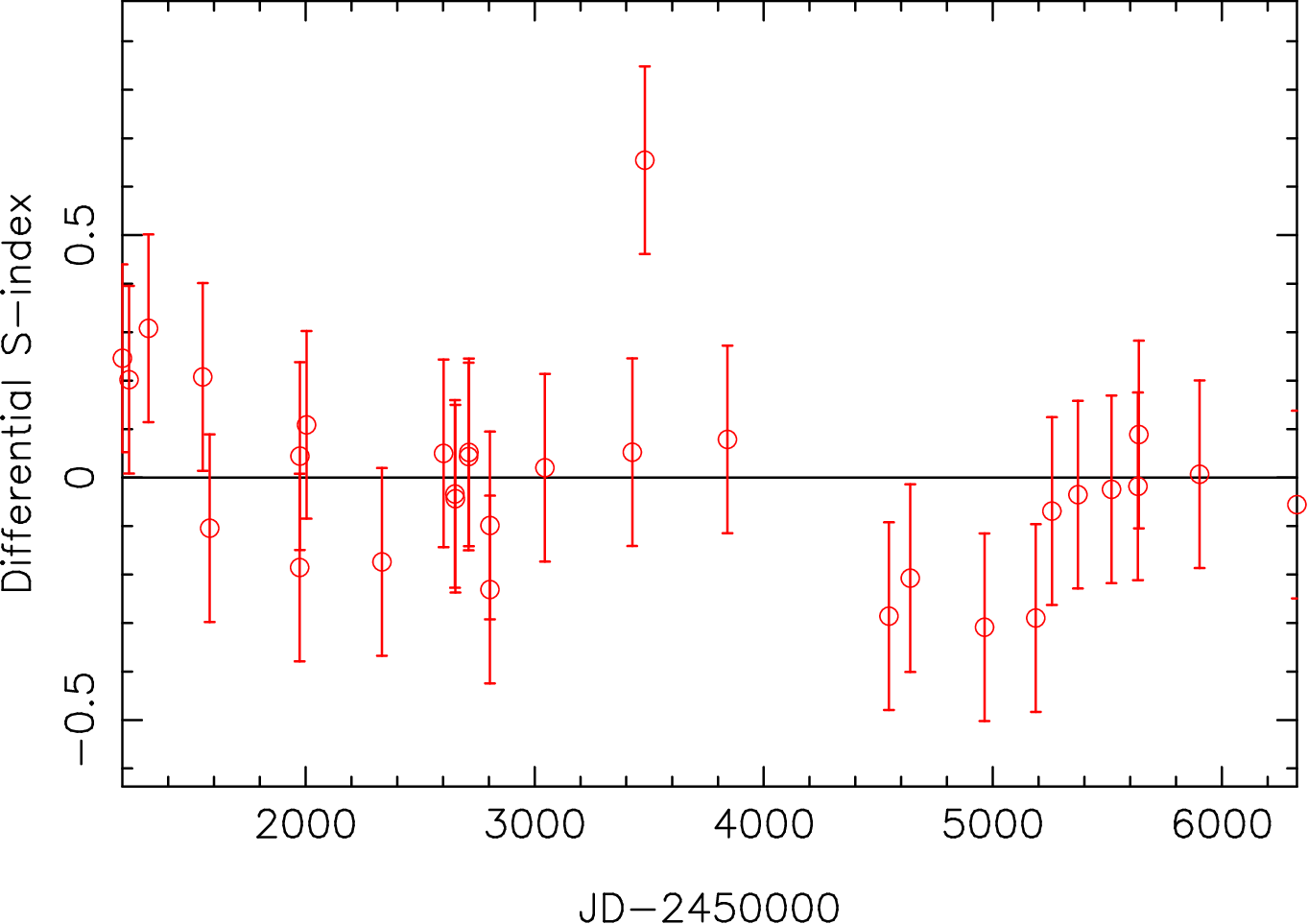}
\caption{HIRES radial velocities of GJ 397 (top) and the corresponding S-indices (bottom). The means have been subtracted from both time-series and the estimated excess white noise has been added in quadrature to the uncertainty estimates.}\label{fig:GJ397_data}
\end{figure}

The radial velocity signal did not have counterparts in the HIRES S-indices and we thus interpret it as a planet candidate. However, some of the variations in the radial velocities were connected to the corresponding S-indices (Fig. \ref{fig:GJ397_data}). The parameter $c_{\rm S}$ was found to have an estimate of 14.7 [1.8, 27.6], which indicates that it is statistically significantly different from zero with 99\% credibility implying a positive connection between the radial velocities and S-indices. We observed a weak periodicity exceeding a 5\% FAP in the S-indices suggestive of a stellar magnetic cycle at a period of 2400 days. There was no ASAS photometry data available for GJ 397.

\clearpage

\subsection{GJ 406}

\begin{figure}
\center
\includegraphics[angle=270, width=0.49\textwidth,clip]{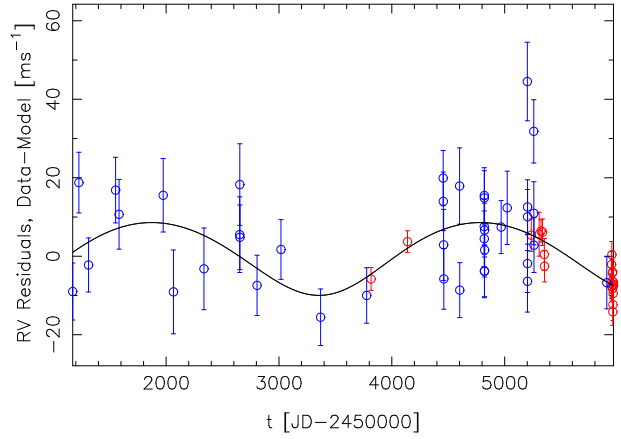}
\caption{Long-period signal in the combined HARPS (red) and HIRES (blue) radial velocities of GJ 406.}\label{fig:GJ406_long}
\end{figure}

\begin{figure}
\center
\includegraphics[angle=270, width=0.49\textwidth,clip]{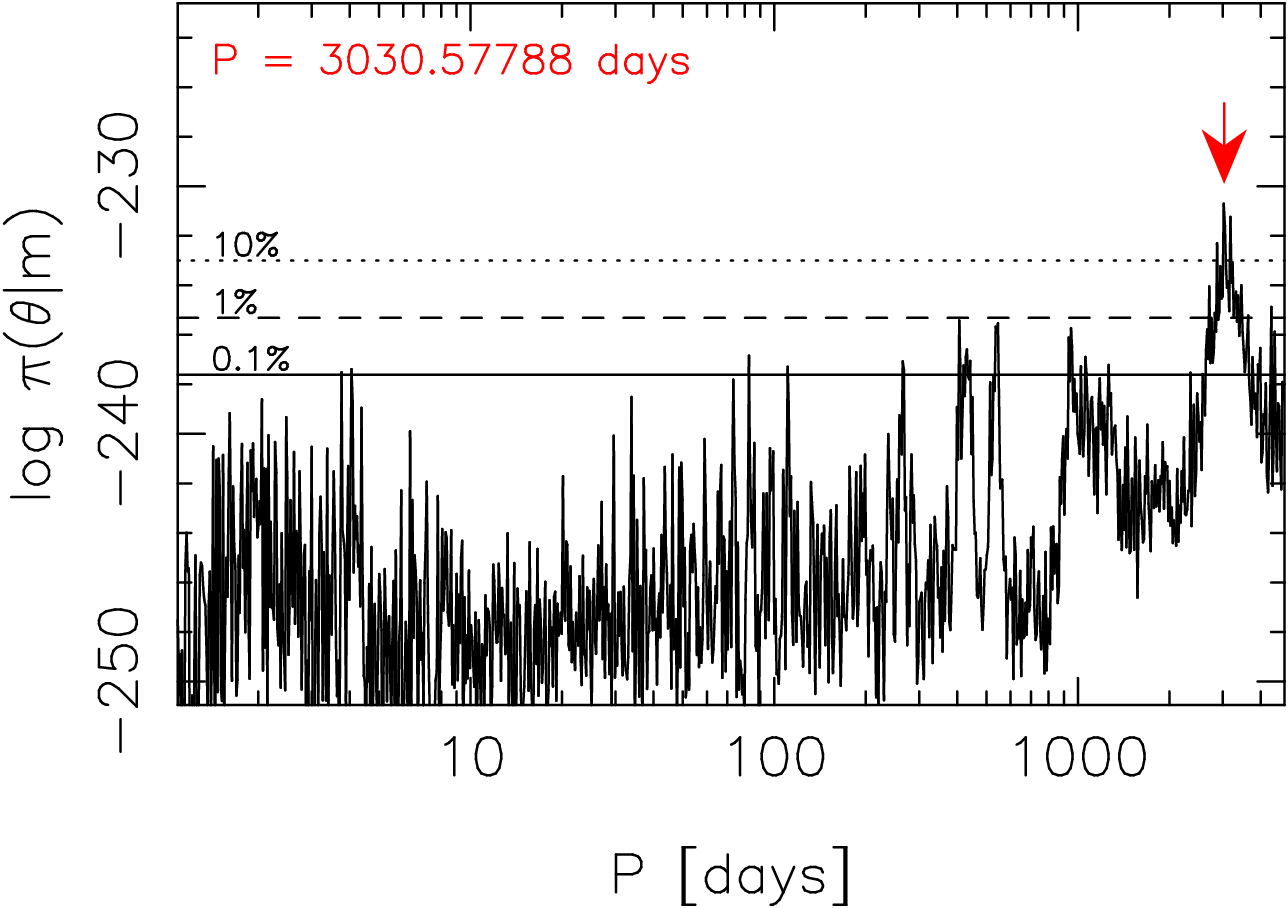}

\includegraphics[angle=270, width=0.49\textwidth,clip]{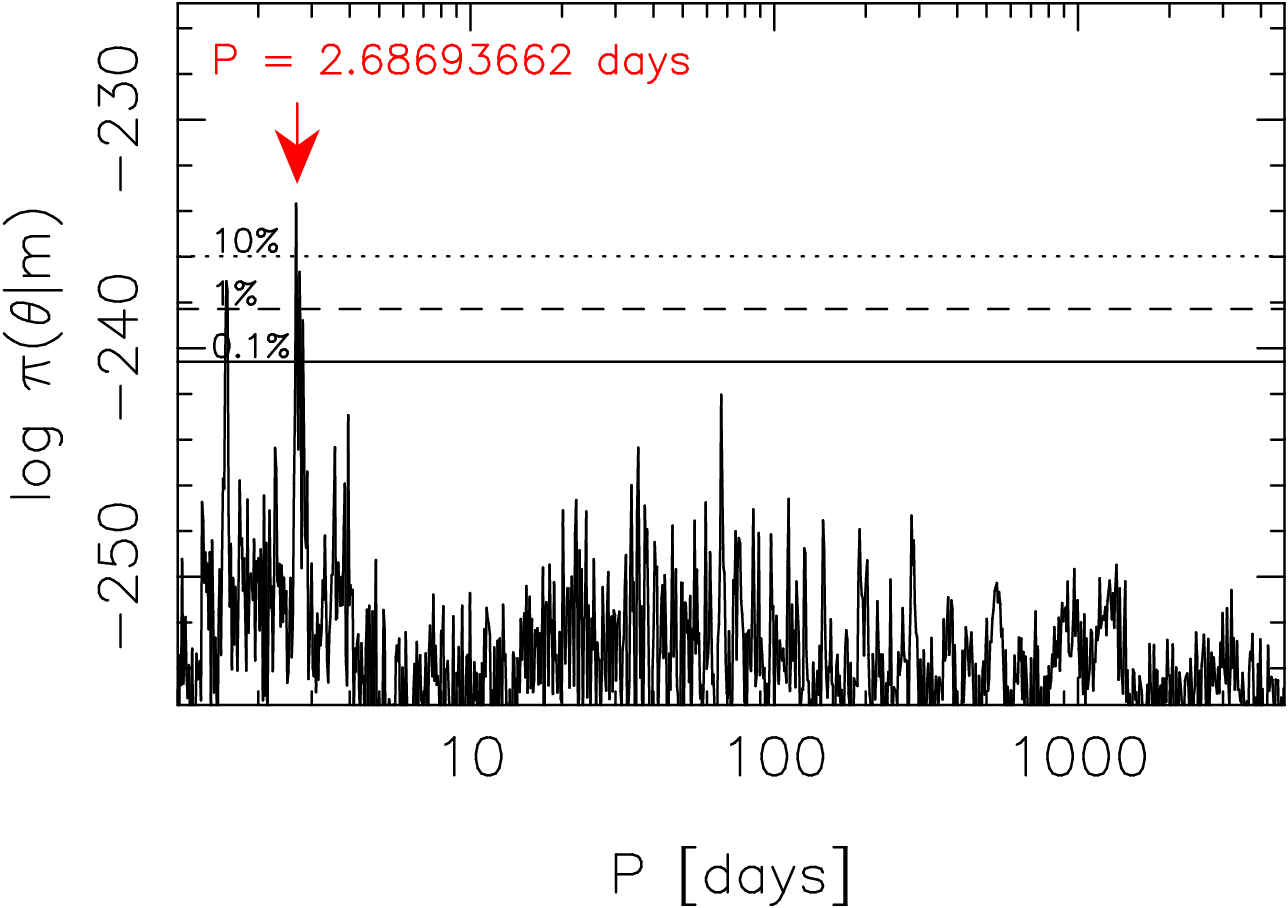}
\caption{Estimated posterior probability densities given GJ 406 radial velocity data as functions of the period of the signal in a one-Keplerian model (top panel) and the period of the second signal in a two-Keplerian model (bottom panel).}\label{fig:GJ406_period_search}
\end{figure}

GJ 406 has been observed with HARPS and HIRES and we obtained a data set containing a total of 63 radial velocities. We observed evidence for long-period variation in the data that satisfied the detection criteria (Fig. \ref{fig:GJ406_long}). Although mostly supported by HARPS data, this signal with a period of 2900 [2300, 3700] days has an amplitude of 8.68 [4.13, 13.77] ms$^{-1}$, which renders it unlikely to be caused by random variations. However, we also detected evidence for a shorter periodicity of 2.68689 [2.68657, 2.68736] days with a roughly equal amplitude of 7.65 [4.12, 11.18] ms$^{-1}$ that also satisfied the signal detection criteria (Fig. \ref{fig:GJ406_period_search}).

\begin{figure}
\center
\includegraphics[angle=270, width=0.49\textwidth,clip]{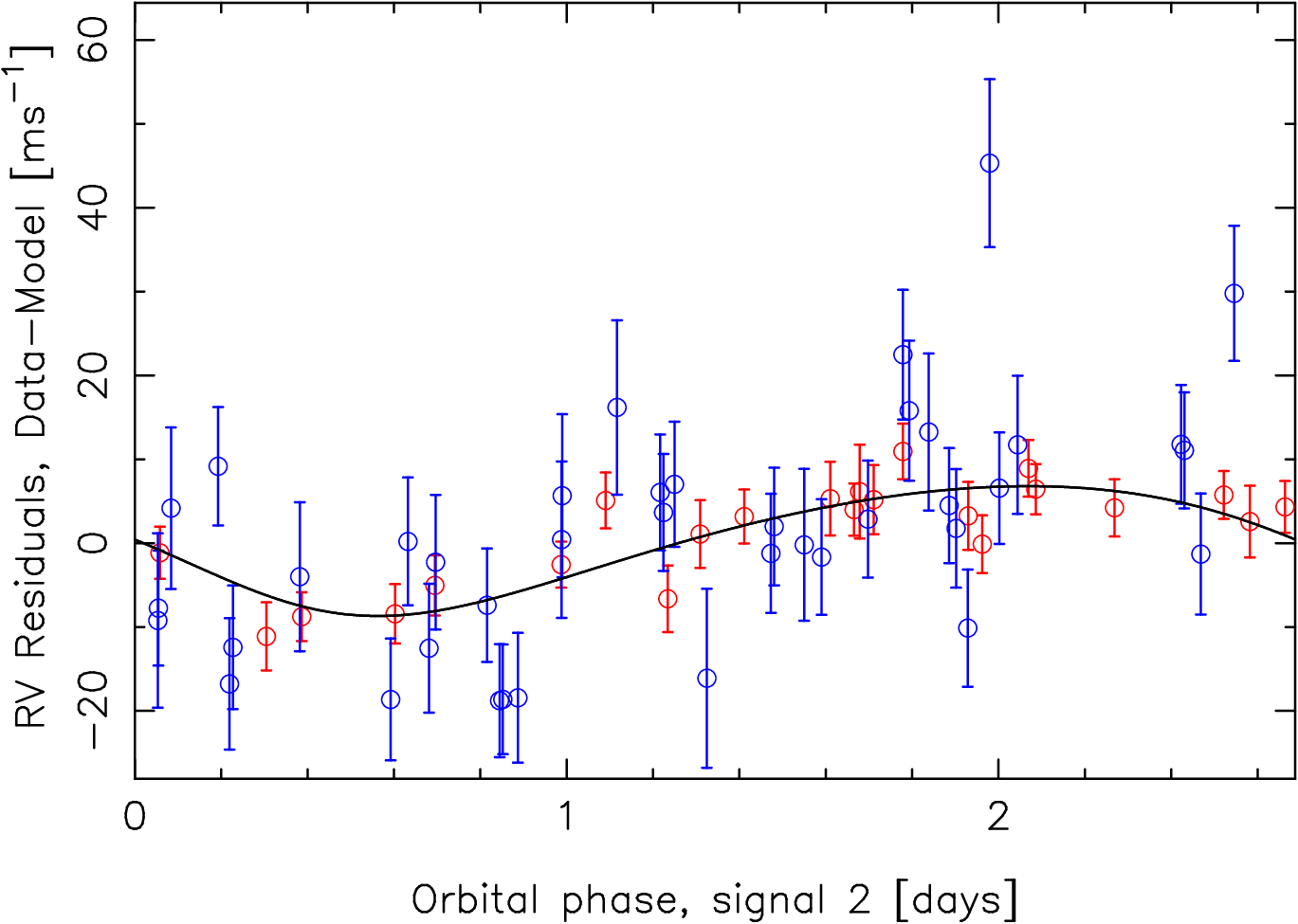}

\includegraphics[angle=270, width=0.49\textwidth,clip]{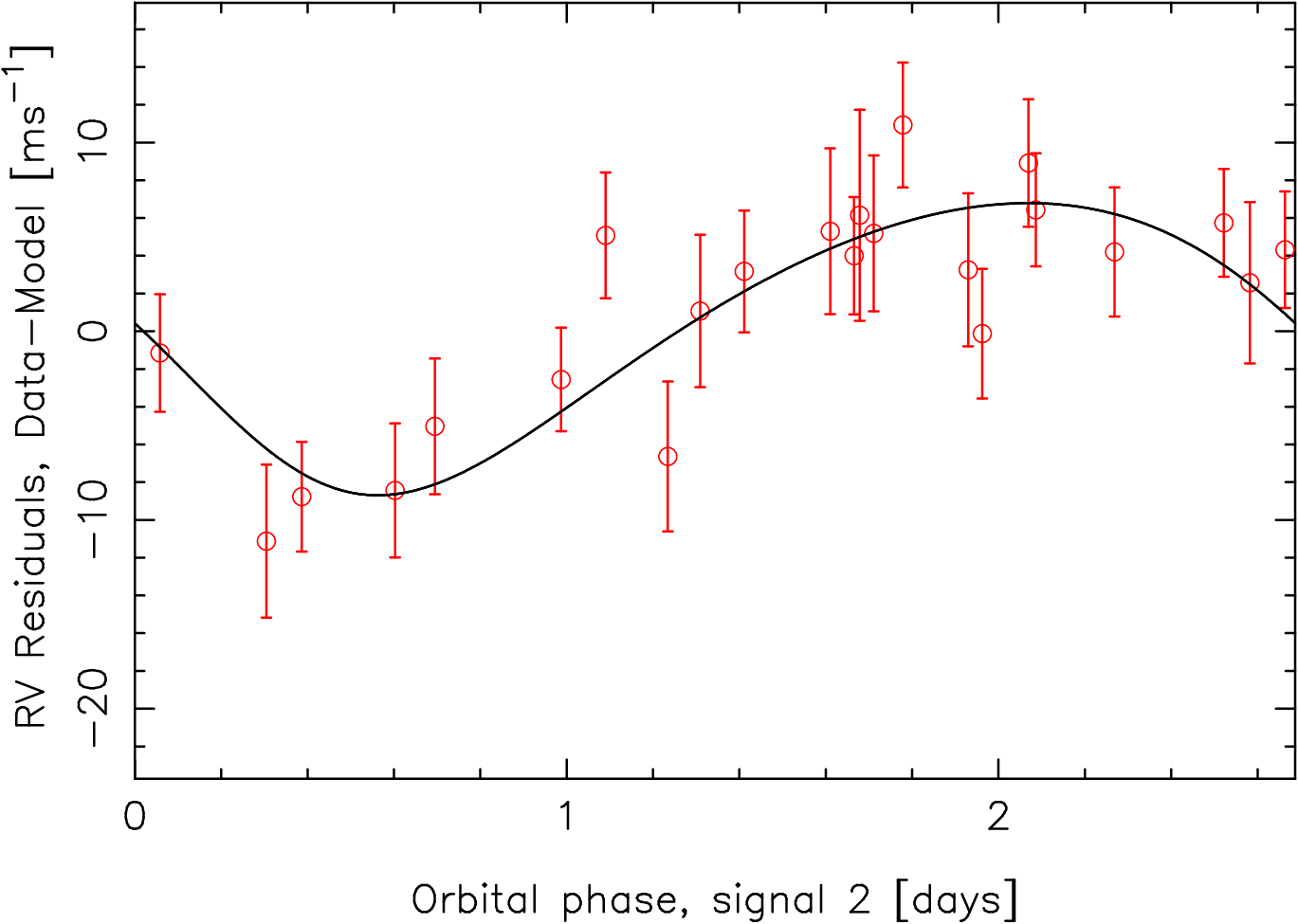}
\caption{Residuals of the HARPS (red) and HIRES (blue) radial velocities of GJ 406 after subtracting the long-period signal folded on the phase of the short period signal at a period of 2.69 days (top panel). The bottom panel shows the HARPS velocities alone.}\label{fig:GJ406_short}
\end{figure}

The signal at a period of 2.69 days was accompanied by its daily alias at a period of 1.59 days (Fig. \ref{fig:GJ406_period_search}). It is therefore rather clear that the corresponding variations in the data are explained by one signal that contributes two maxima to the posterior density in the period space due to aliasing. We have plotted this signal together with the data residuals in Fig. \ref{fig:GJ406_short}. We could not identify any additional signals in the combined data.

We did not find any significant periodicities in the HARPS or HIRES activity indicators. We also analysed the available 120 ASAS V-band photometry measurements but failed to find any significant periodicities suggestive of stellar rotation and/or activity cycles. We thus interpret the two signals as candidate planets orbiting the star. These planets are classified as a cool super-Neptune and a hot super-Earth, respectively.

\clearpage

\subsection{GJ 411}\label{sec:GJ411}

\begin{figure}
\center
\includegraphics[angle=270, width=0.49\textwidth,clip]{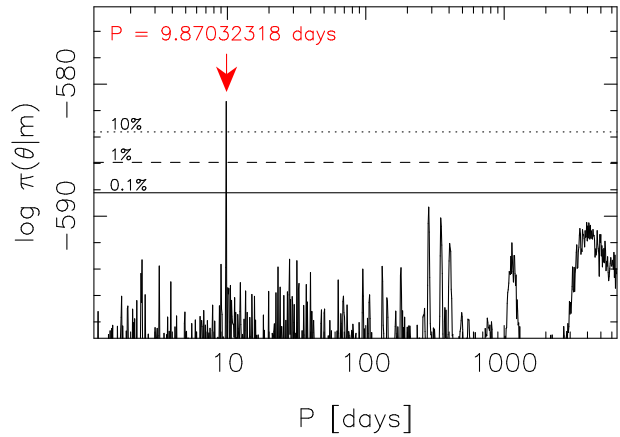}
\caption{Posterior probability density given GJ 411 radial velocities as a function of the period parameter of a Keplerian signal. The red arrow denotes the position of the global maximum and the horizontal lines show the 10\% (dotted), 1\% (dashed), and 0.1\% (solid) equiprobability contours with respect to the maximum.}\label{fig:GJ411_period_search}
\end{figure}

\begin{figure}
\center
\includegraphics[angle=270, width=0.49\textwidth,clip]{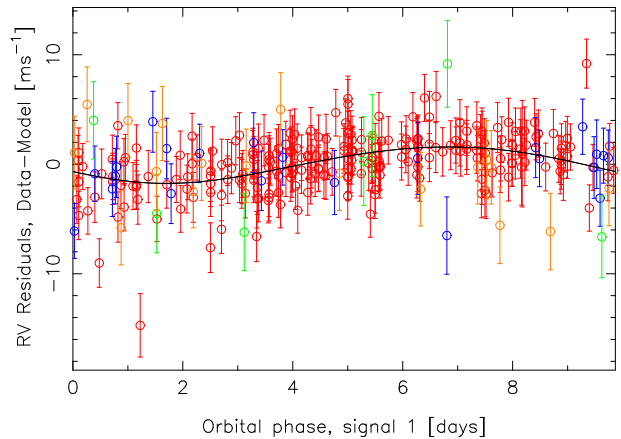}
\caption{Combined HIRES (red), APF (blue), SOPHIE (first 7, green; last 18, orange) radial velocities of GJ 411 folded on the phase of the signal.}\label{fig:GJ411_curve}
\end{figure}

The HIRES radial velocities of the very nearby star GJ 411 (Lalande 21185, HD95735, HIP 54035) have been found to contain evidence in favour of a planet candidate with minimum mass of roughly 3.8 M$_{\oplus}$ orbiting the star with an orbital period of 9.9 days \citep{butler2016}. Based on a combined velocity data set of HIRES, APF, and SOPHIE radial velocities we could also detect this hot super-Earth and obtained an estimate of 2.9 [1.4, 4.7] M$_{\oplus}$ for its minimum mass and an orbital period of 9.8684 [9.8620, 9.8733] days. We demonstrate the existence of the corresponding signal in Figs. \ref{fig:GJ411_period_search} and \ref{fig:GJ411_curve}. Because of a gap of 1800 days between the two SOPHIE observing runs, we treated the first 7 and last 18 SOPHIE velocities as independent data sets in our analyses.

Although we observed a periodogram power in the HIRES S-indices at a period of 174 days \citep{butler2016}, the signal in the radial velocities of GJ 411 has no counterparts in the activity indicators. No ASAS photometry data was available for the star. We could thus not search for photometric counterparts of the radial velocity signal. Therefore, we interpret the signal in the radial velocities of GJ 411 as evidence for a hot super-Earth orbiting the star.

\clearpage

\subsection{GJ 422}

\begin{figure}
\center
\includegraphics[angle=270, width=0.49\textwidth,clip]{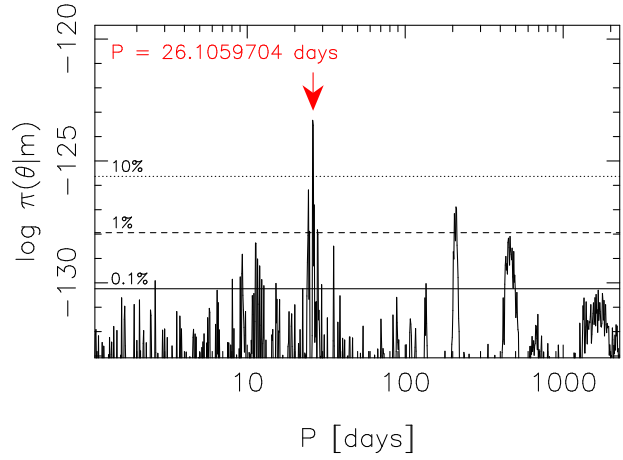}
\caption{Estimated posterior probability density as a function of the period of the Keplerian signal in the model given GJ 422 radial velocities of \citet{tuomi2014}. The red arrows indicates the maximum of the density and the horizontal lines denote the 10\% (dotted), 1\% (dashed), and 0.1\% (solid) equiprobability thresholds with respect to the maximum.}\label{fig:GJ422_period_search}
\end{figure}

Based on UVES and HARPS radial velocities, \citet{tuomi2014} reported a signal at a period of 26.121 [26.063, 26.243] days that they interpreted as a candidate planet orbiting GJ 422 (HD 304043, HIP 55042). We re-analysed the data used in \citet{tuomi2014} with the DRAM samplings and could detect the signal rather easily as the global maximum in the period space (Fig. \ref{fig:GJ422_period_search}, top panel) with only its aliases, shown as ``sidelobes'' of the global maximum, and another local maximum at a period of 205 d exceeding the 1\% probability threshold.

We obtained 19 additional HARPS velocities from the publicly available data products in the ESO archive and analysed the updated HARPS data set in combination with the UVES velocities of \citet{zechmeister2009}. However, to be able to account for the linear correlations between the HARPS velocities and activity indices, we had to remove some suspicious outliers from the data set. There were three spectra where the activity indicators (all BIS, FWHM, and S-index) showed suspiciously large or small values that were considerably off of the respective mean values enabling us to classify them very clearly as outliers. Two of the corresponding velocities were included in the analyses of \citet{tuomi2014}, which means they could have affected the solution.

\begin{figure}
\center
\includegraphics[angle=270, width=0.49\textwidth,clip]{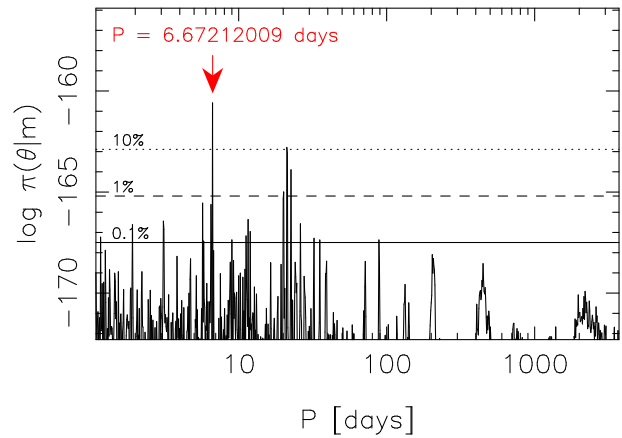}
\caption{As in Fig. \ref{fig:GJ422_period_search} but when analysing the full HARPS and UVES data set.}\label{fig:GJ422_period_search2}
\end{figure}

When analysing the full HARPS data of 41 velocities (after removing outliers) in combination with the UVES data, we observed a signal at a period of 6.6710 [6.6655, 6.6752] days together with a local maximum at a period of 21.3 days. However, although both of these signals were detected according to our criteria, they were not significant enough to make the cut as planet candidates. It is thus our conclusion that there is no clear evidence for candidate planets orbiting GJ 422 although the variations in the velocities centainly warrant future Doppler monitoring of the star (Fig. \ref{fig:GJ422_period_search2}). We thus conclude that there is only suggestive evidence for signals in the GJ 422 data and that the signal detected by \citet{tuomi2014} was not stationary as a Keplerian signal should be and is thus not supported by the new velocities.

\begin{figure}
\center
\includegraphics[angle=270, width=0.49\textwidth]{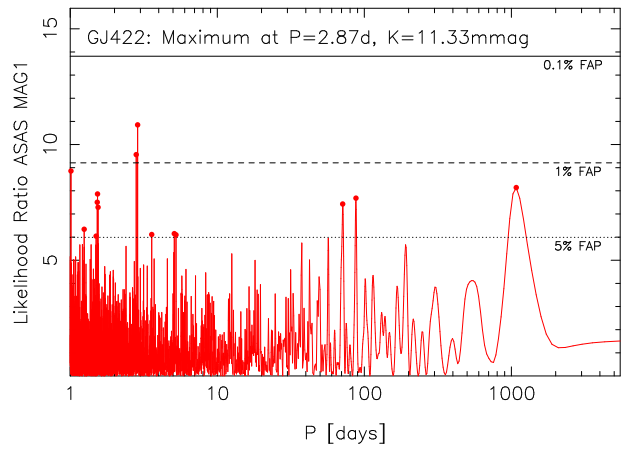}
\caption{Likelihood-ratio periodogram of the ASAS V-band photometry of GJ 422.}\label{fig:GJ422_asas}
\end{figure}

We obtained 365 ASAS V-band photomety measurements from the ASAS archive. The likelihood periodogram in Fig. \ref{fig:GJ422_asas} suggests that there is a photometric periodicity with a period of 2.87 days in the data. However, it seems unlikely that this periodicity is connected to the signal at the period of 26.1 d reported by \citet{tuomi2014}. Moreover, it is not significant enough to qualify as evidence for a photometric rotation period of the star.

\clearpage

\subsection{GJ 433}\label{sec:GJ433}

According to \citet{delfosse2013}, GJ 433 (HIP 56528) is a host to a system of two planetary-mass companions. Based on the analyses of the same HARPS and UVES data, \citet{tuomi2014} reached a similar conclusion and assumed that the two signals with periods of 7.3697 [7.3661, 7.3731] and 3400 [1900, --] days, respectively, were indeed caused by planets orbiting the star although the orbit of the outer companion could not be constrained from above.

We have obtained HIRES and PFS data and analysed the corresponding combined data set and can conclude that the signal at a period of 7.37 days is indeed very significantly present and supported by all four data sets, which makes it a planet candidate according to our criteria. However, the orbit of the proposed outer candidate still cannot be constrained from above, which casts doubt on its periodic nature. Moreover, because this long-period signal is mostly supported by UVES data due to the fact that the maximum likelihood values between one- and two-Keplerian models only change due to an increase in the likelihood of UVES data, we decided to analyse the combined HARPS, HIRES, and PFS data sets and neglect UVES data as published by \citet{zechmeister2009} because it is potentially biased due to a mistake in the barycentric correction.

\begin{figure}
\center
\includegraphics[angle=270, width=0.49\textwidth,clip]{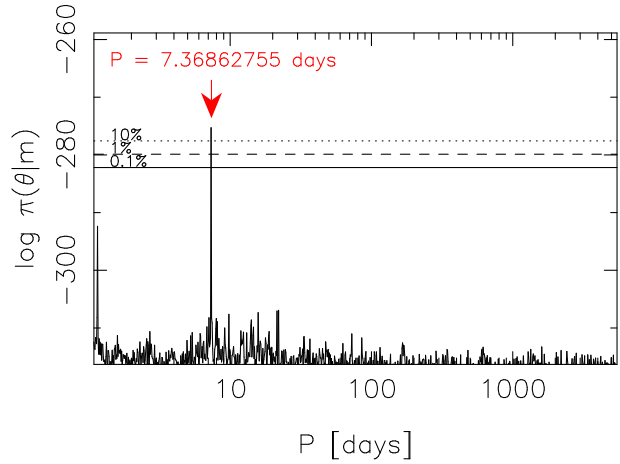}
\caption{Estimated posterior density of the period parameter of the signal in a one-Keplerian model given GJ 433 data. Red arrow denotes the global probability maximum and the horizontal lines correspond to equiprobability contours at 10\% (dotted), 1\% (dashed), and 0.1\% (solid) of the maximum.}\label{fig:GJ433_period_search}
\end{figure}

\begin{figure}
\center
\includegraphics[angle=270, width=0.49\textwidth,clip]{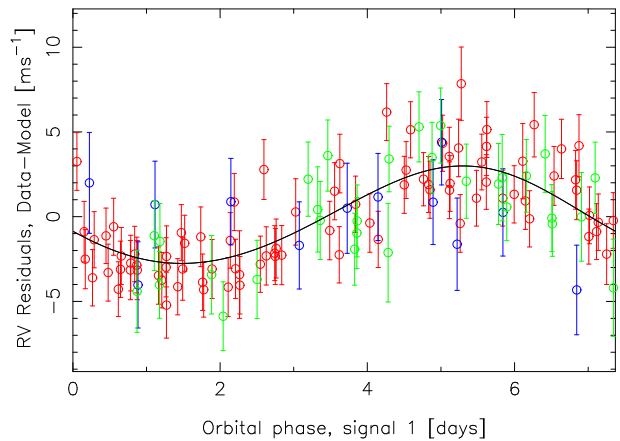}
\caption{Combined HARPS (red), PFS (blue), and HIRES (green) radial velocity data of GJ 433 folded on the phase of the signal. The black solid curve indicates the MAP Keplerian solution.}\label{fig:GJ433_phased}
\end{figure}

Without the UVES velocities, the results indeed appear different. Our samplings of the parameter space yielded evidence for only one signal (Figs. \ref{fig:GJ433_period_search} and \ref{fig:GJ433_phased}) at the familiar period of 7.37064 [7.36933, 7.37190] days. This signal was found to have an amplitude of 2.92 [2.14, 3.71]. It is noteworthy that although the baseline of the combined data, even without UVES, is 5476 days, we could not find any evidence for the signal at a period of 3400 days interpreted as a planet candidate by \citet{delfosse2013} and \citet{tuomi2014}. It thus seems rather clear that this signal is caused by a bias in the UVES velocities.

\begin{figure}
\center
\includegraphics[angle=270, width=0.49\textwidth,clip]{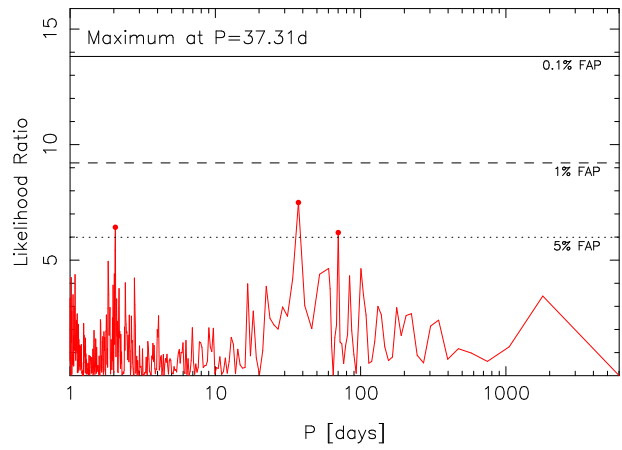}
\caption{Likelihood periodogram of HARPS S-indices of GJ 433.}\label{fig:GJ433_harps_s}
\end{figure}

The activity indicators of HARPS and HIRES did not show strong evidence for periodic signals. However, the HARPS S-indices suggest that there are periodicisies at 37.31 days, although the corresponding likelihood ratio did not exceed the 1\% FAPs (Fig. \ref{fig:GJ433_harps_s}). We also analysed the 683 ASAS V-band photometry measurements. There were no strong periodicities in the ASAS photometry either.

We thus conclude that there is evidence a periodic signal in the radial velocities of GJ 433 that correspond to a planet with a minimum masse of 5.4 [3.6, 7.5] M$_{\oplus}$. This candidate planet is classified as a hot super-Earth.

\clearpage

\subsection{GJ 479}\label{sec:GJ479}

\citet{bonfils2013} reported significant power excesses in the periodogram of the GJ 479 (HIP 61629) HARPS data at period of 11 and 23-24 days, respectively. However, according to them, ``Modeling that RV variability with Keplerians converges toward 2 planets with very similar periods (23.23 and 23.40 d)'', which we interpret as an indication that the apparent 11-day signal was not significant after all according to their periodogram analyses. \citet{bonfils2013} also point out that the activity indices show no periodic features at or near the potential radial velocity periodicities but that photometric data from the Euler telescope shows a maximum periodogram power excess at a period of 23.75 days.

\begin{figure}
\center
\includegraphics[angle=270, width=0.49\textwidth,clip]{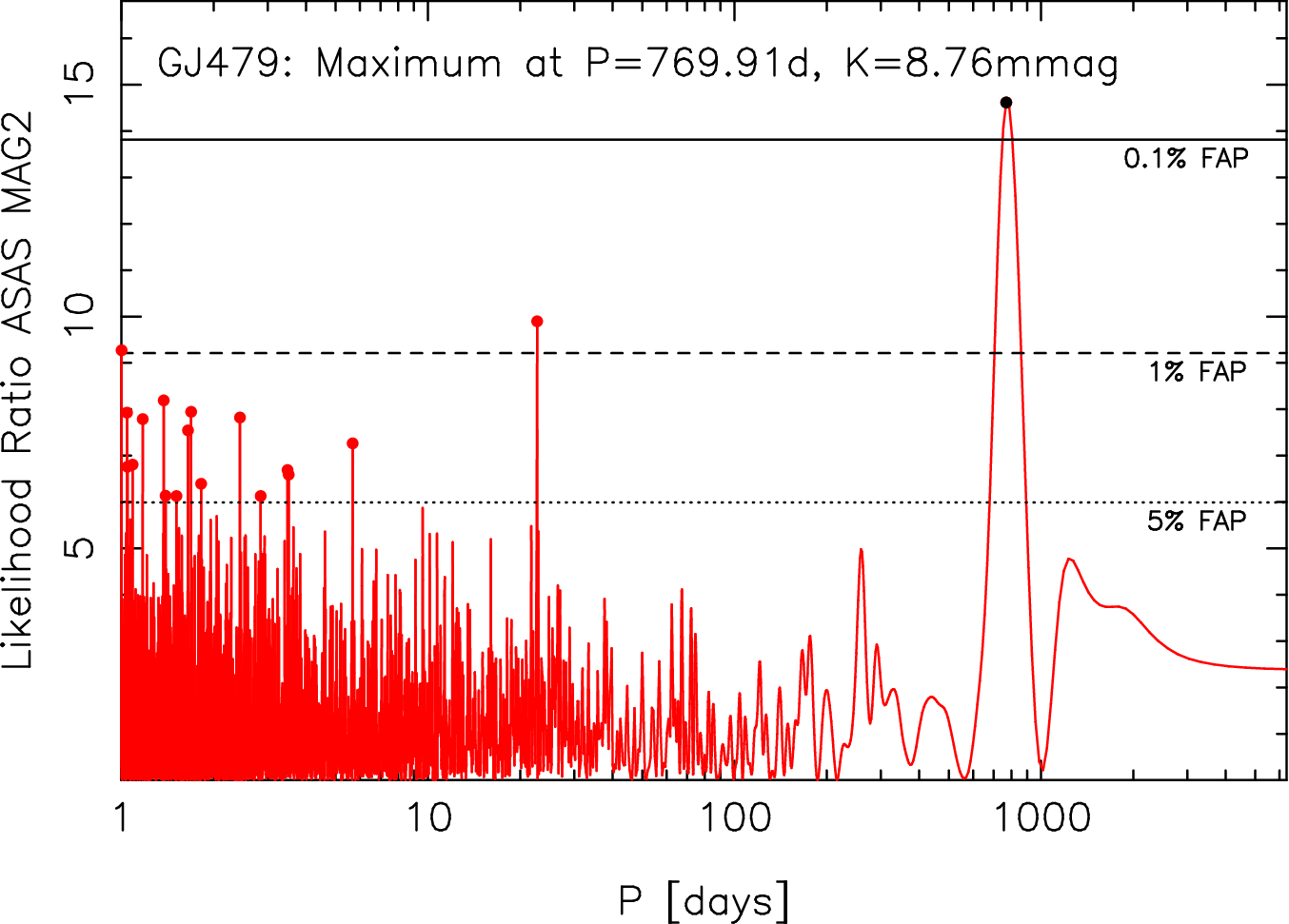}

\includegraphics[angle=270, width=0.49\textwidth,clip]{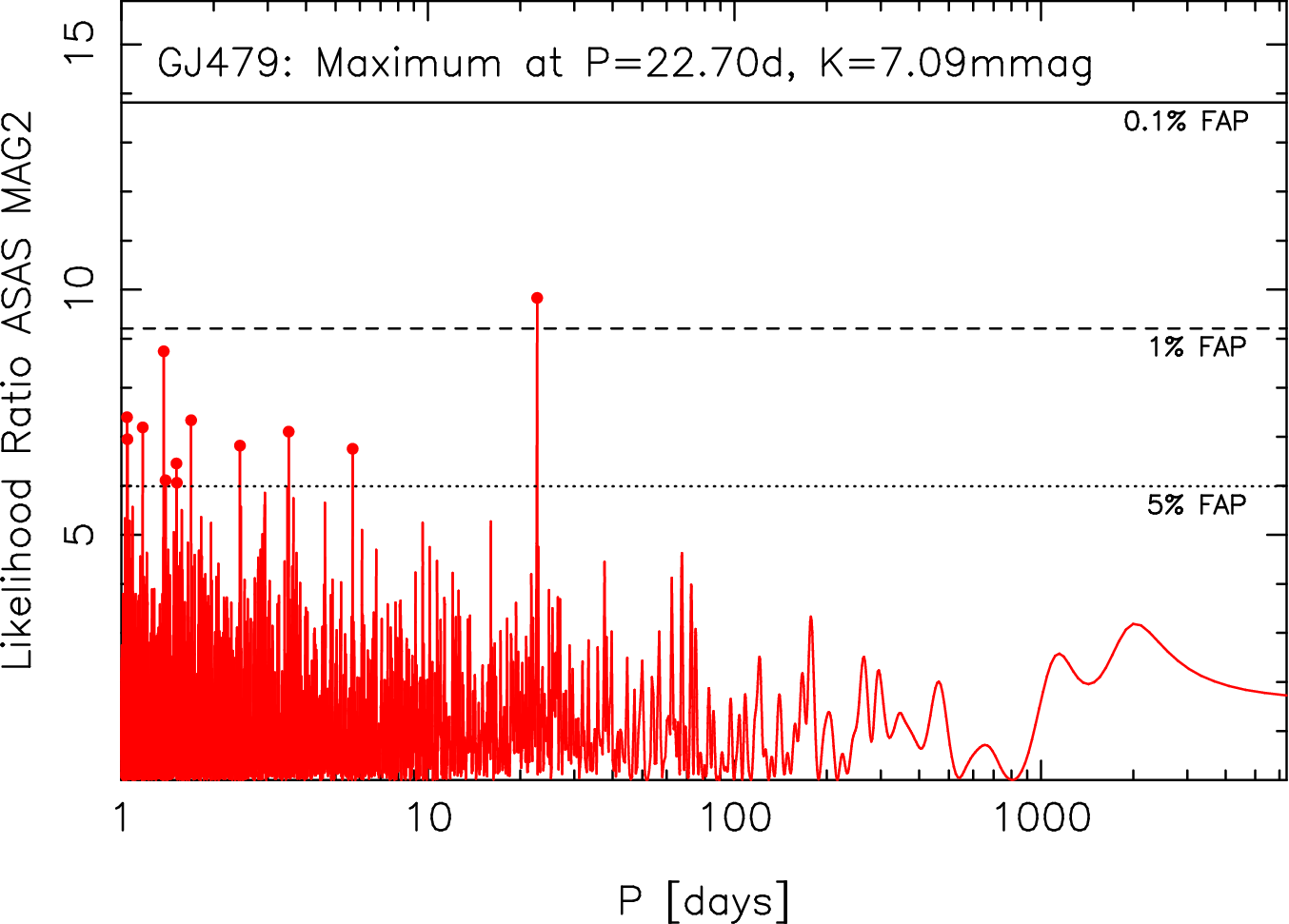}
\caption{Likelihood-ratio periodogram of ASAS V band photometry data of GJ 479 (top panel) and after subtracting the most significant periodicity (bottom panel).}\label{fig:GJ479_asas}
\end{figure}

We observed a strong photometric signal at a period of 770 days and a suggestive one at a period of 22.70 days in a set of 568 ASAS V-band measurements (Fig. \ref{fig:GJ479_asas}). The latter is reasonably close to the signal reported by \citet{bonfils2013} and thus likely has the same origin. Following \citet{bonfils2013}, it is our interpretation that it is indeed caused by stellar rotation. However, we note that this signal in ASAS data does not exceed 0.1\% FAP and is thus uncertain on statistical grounds and does not qualify as a confidently detected photometric rotation period of the star. The HARPS activity indicators do not show any hints of periodicities.

\begin{figure}
\center
\includegraphics[angle=270, width=0.49\textwidth,clip]{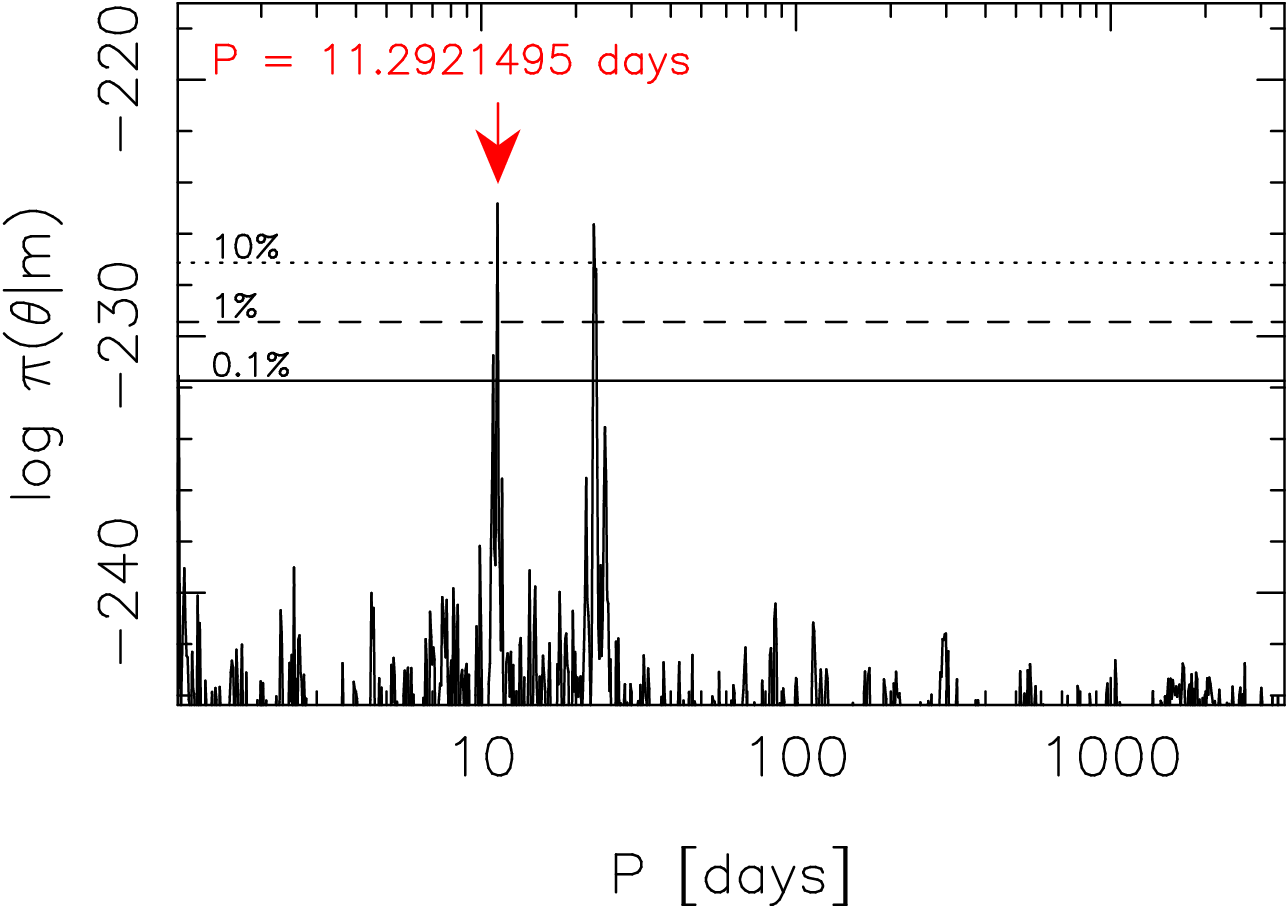}

\includegraphics[angle=270, width=0.49\textwidth,clip]{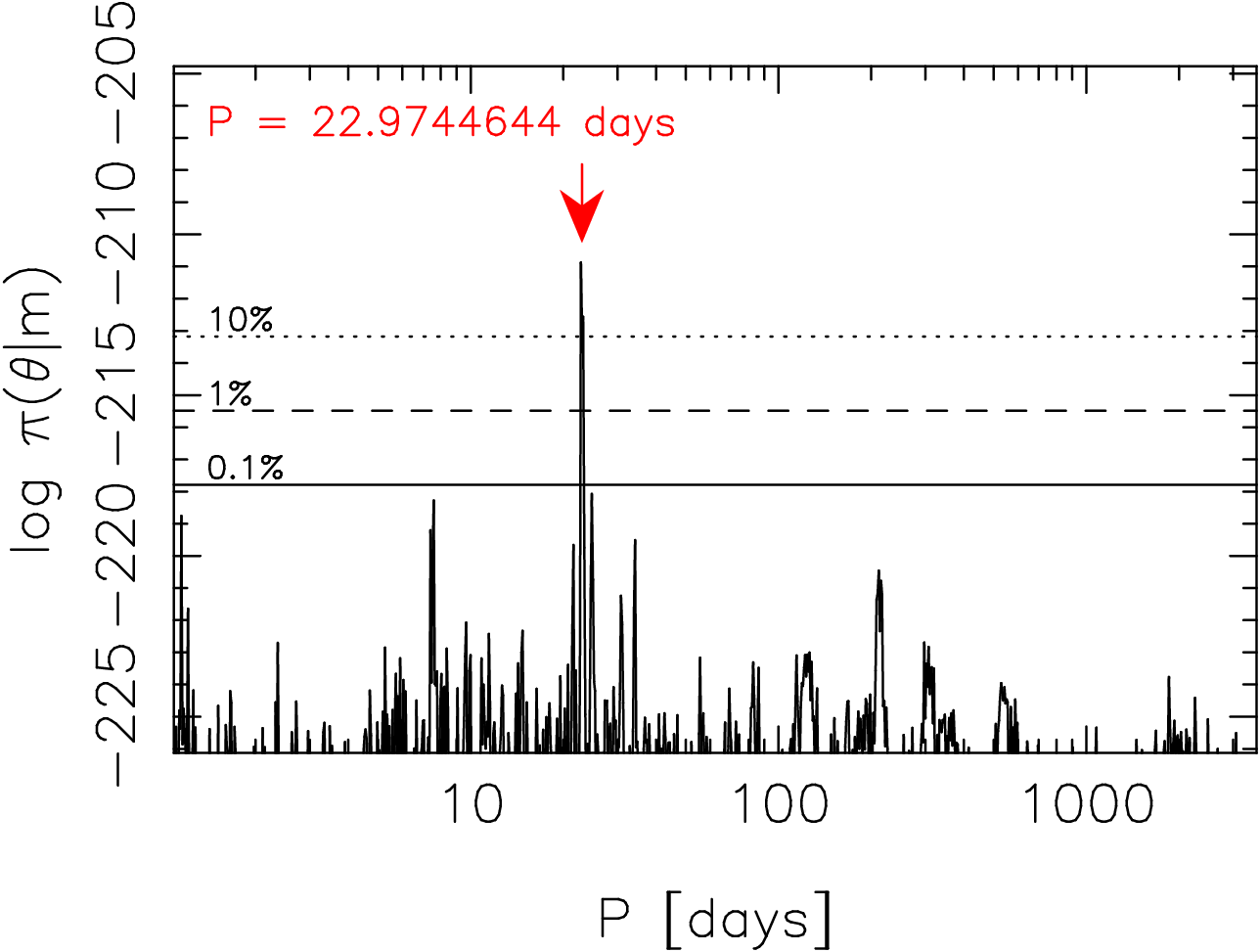}

\includegraphics[angle=270, width=0.49\textwidth,clip]{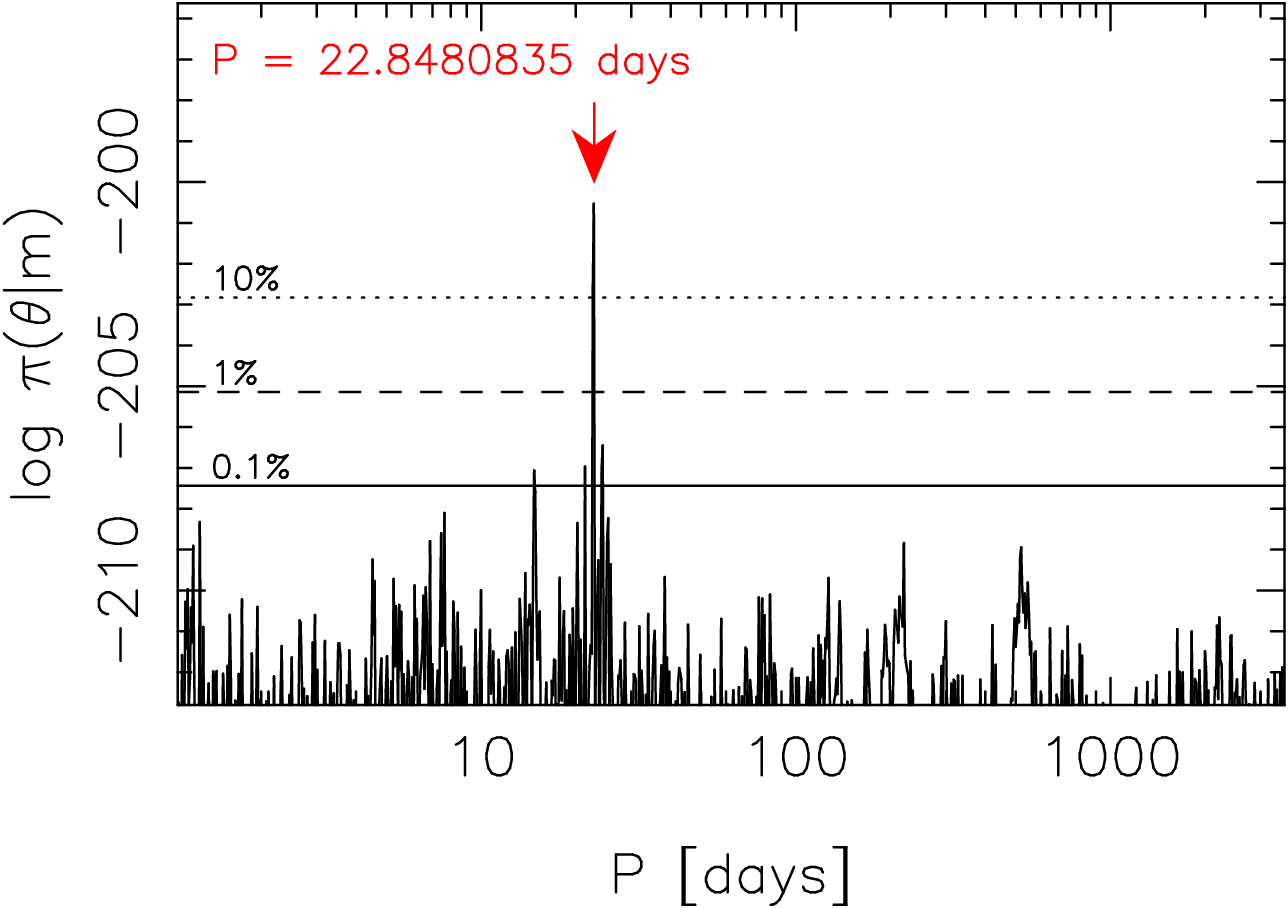}
\caption{Estimated posterior probability density of a model with $k$ Keplerian signals as a function of the period of the $k$th signal given the HARPS and PFS data of GJ 479. The panels (top to bottom) show cases with $k=1$, $k=2$, and $k=3$. Red arrows indicate the positions of the global maima in the period space and the horizontal lines denote the 10\% (dotted), 1\% (dashed), and 0.1\% (solid) equiprobability thresholds with respect to the maxima.}\label{fig:GJ479_period_search}
\end{figure}

\begin{figure}
\center
\includegraphics[angle=270, width=0.49\textwidth,clip]{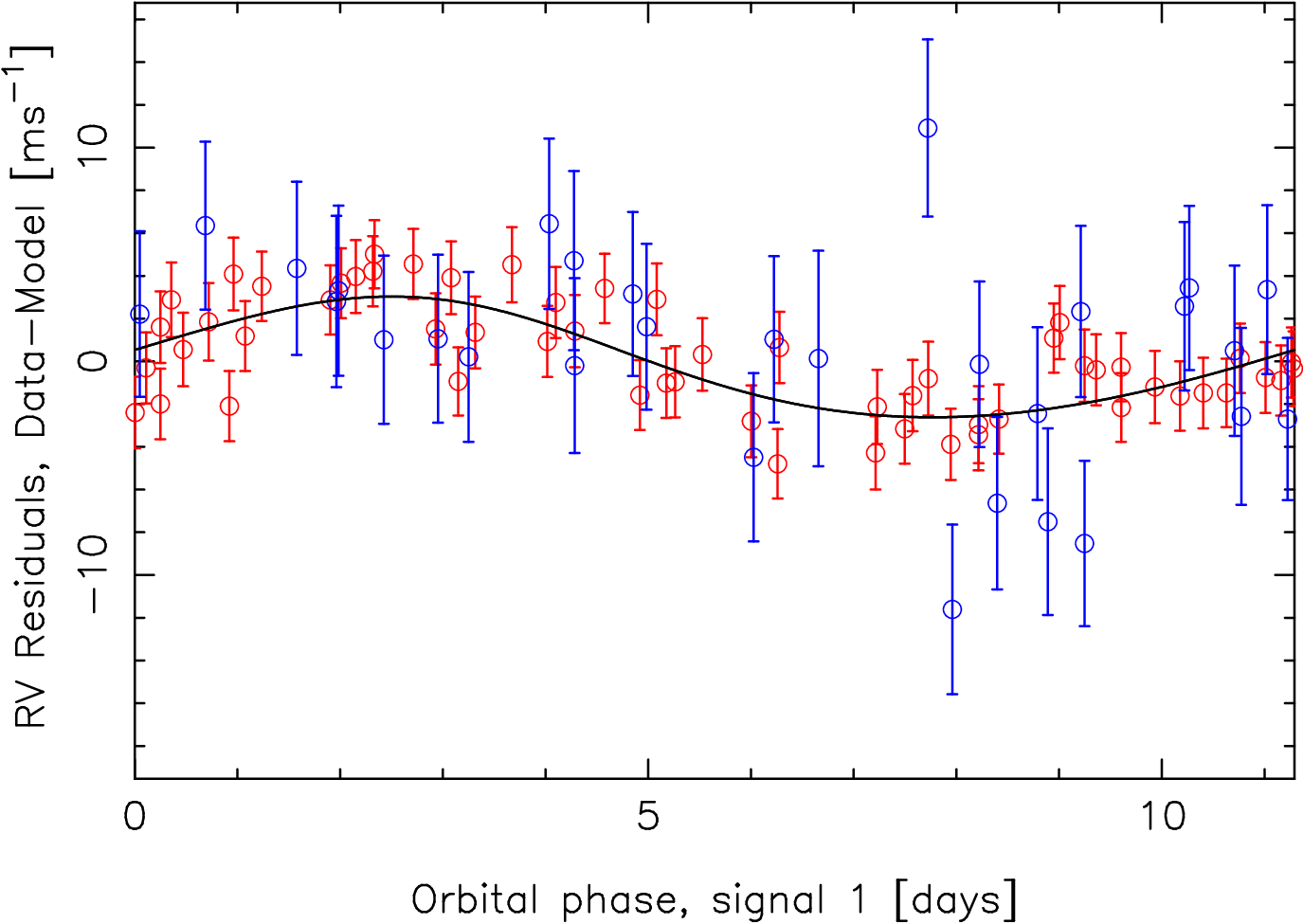}

\includegraphics[angle=270, width=0.49\textwidth,clip]{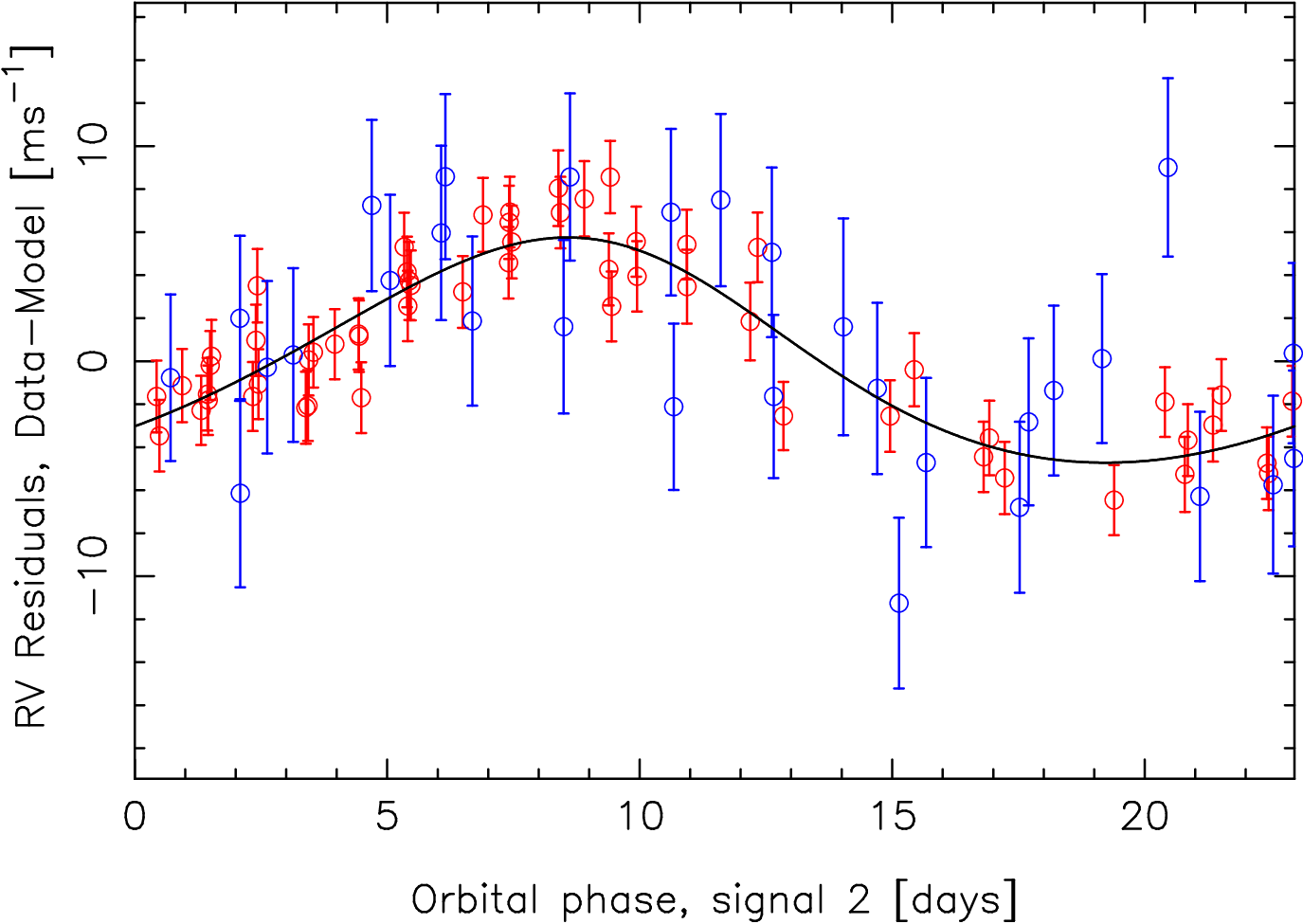}

\includegraphics[angle=270, width=0.49\textwidth,clip]{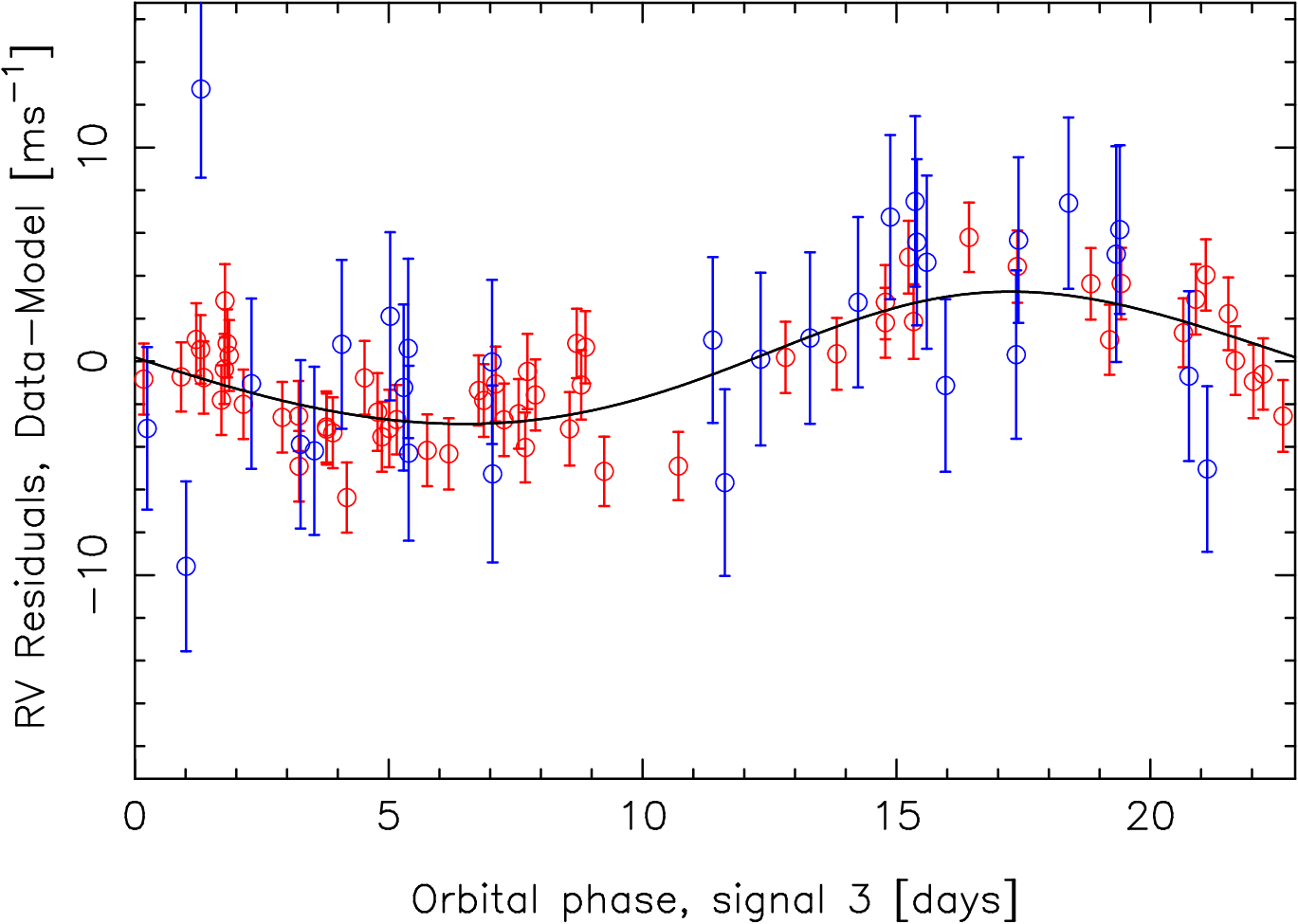}
\caption{HARPS (red) and PFS (blue) radial velocities of GJ 479 folded on the phases of the signals.}\label{fig:GJ479_phased}
\end{figure}

The combined HARPS and PFS radial velocities of GJ 479 showed evidence for three significant signals at periods of 11.292 [11.279, 11.302], 22.958 [22.913, 23.003], and 22.842 [22.781, 22.951] days (Figs. \ref{fig:GJ479_period_search} and \ref{fig:GJ479_phased}). These signals were found to be independent of variations in activity indices as well as independent of one another but the latter two that are also suspiciously close to one another in the period space appear to have a photometric counterpart. We thus interpret the first signal as evidence for a planet candidate with a minimum mass of 5.1 [2.6, 8.2] M$_{\oplus}$ corresponding to a hot super-Earth, whereas the latter signals likely represent stellar rotation.

We note that it appears very likely that the rotation period of the star is $\sim$ 23 days. Although weakly supported by ASAS photometry, this is our interpretation based on the fact that there are two very nearby periodicities in the combined HARPS and PFS radial velocities that could not correspond to planetary signals. Moreover, although the 11.3-day signal is clearly present in the HARPS data alone, the second signal observed in the HARPS data is actually at a period of 23.14 days whereas the corresponding periodicity in the PFS data is at a period of 22.83 days. This means that the periodic variability at and near 23 days does not show the time-invariance and stability that should be expected from a planetary signal. Rather, it shows variations in the period depending on whether it was detected in HARPS data between JD 2453158-2454571 or in PFS data between 2455253-2456735. This suggests that the radial velocity signal at and near 23 days might be caused by starspots co-rotating on the stellar surface at different latitudes and this serve as indication of differential rotation.

Yet, the signal at a period of 11.3 days is supported by both instruments and appears to be invariably present in HARPS and PFS data justifying its interpretation as a signal caused by a candidate planet.

\clearpage

\subsection{GJ 480}

\begin{figure}
\center
\includegraphics[angle=270, width=0.49\textwidth,clip]{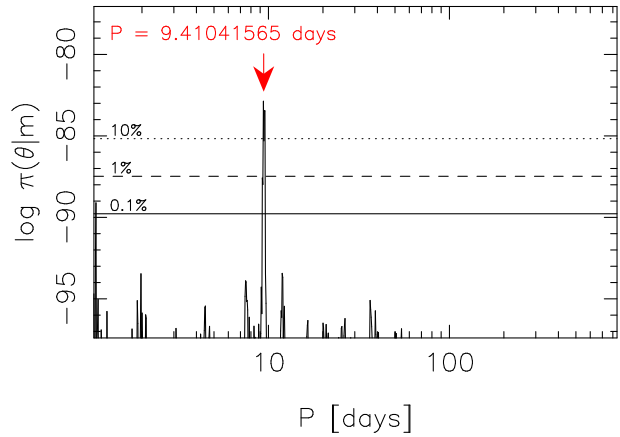}
\caption{Posterior probability density given GJ 480 radial velocity data as a function of the period parameter of the model with one Keplerian signal. The red arrow indicates the location of the global maximum in the period space and the horizontal lines denote the 10\% (dotted line), 1\% (dashed line), and 0.1\% (solid line) probability thresholds with respect to the maximum.}\label{fig:GJ480_psearch}
\end{figure}

\begin{figure}
\center
\includegraphics[angle=270, width=0.49\textwidth,clip]{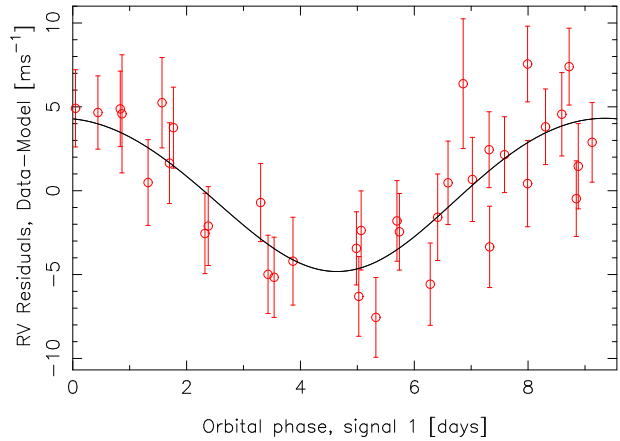}
\caption{Phase-folded HARPS radial velocities of GJ 480 with the modelled radial velocity curve corresponding to GJ 480 b overplotted on top of the data.}\label{fig:GJ480_curve}
\end{figure}

\begin{figure}
\center
\includegraphics[angle=270, width=0.49\textwidth,clip]{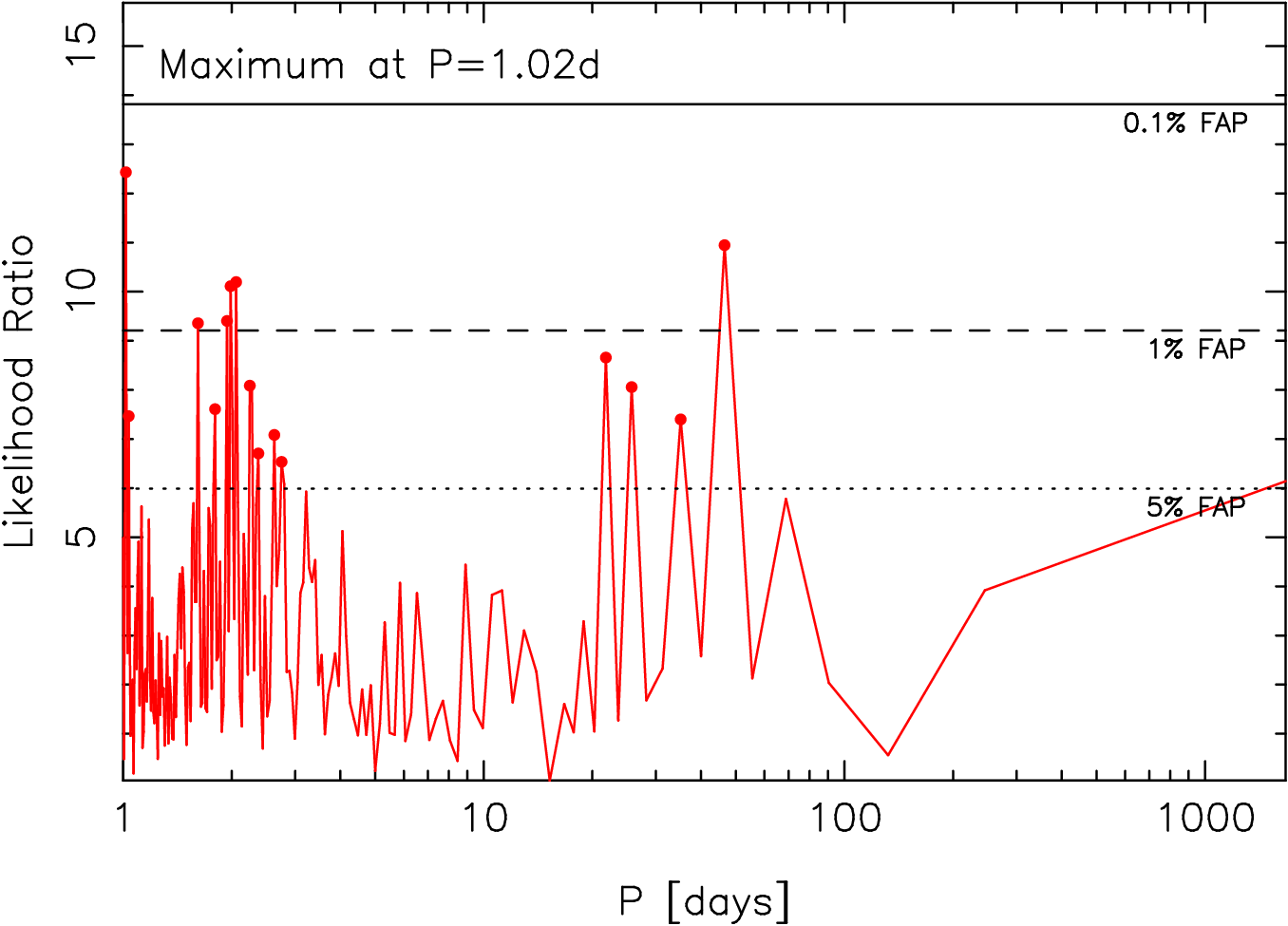}
\caption{Likelihood-ratio periodogram of the HARPS S-indices of GJ 480.}\label{fig:GJ480_HARPS_S}
\end{figure}

We observed a signal at a period of 9.5595 [9.5156, 9.5964] days in the set of 36 HARPS radial velocities of GJ 480 (Wolf 433, HIP 61706) (Figs. \ref{fig:GJ480_psearch} and \ref{fig:GJ480_curve}). Moreover, this signal, with an amplitude of 4.57 [2.18, 6.72] ms$^{-1}$, does not appear to have counterparts in HARPS activity indices or ASAS photometry data. Although the HARPS S-indices show suggestive evidence, in excess of 1\% FAP but not 0.1\% FAP, for periodic behaviour at periods of 47 and 2 days (Fig. \ref{fig:GJ480_HARPS_S}), there is no evidence for signals at or near the location of the radial velocity signal in the period space.

It is noteworthy that the HARPS radial velocities are connected to the corresponding S-indices. There is a positive correlation and the estimate of $c_{\rm S}$ is 14.9 [2.7, 29.9] ms$^{-1}$ indicating that the correlation between radial velocities and S-indices is significant with a 99\% credibility. Yet, the signal in the radial velocities can be detected regardless of whether this connection is accounted for by the model or not and the parameter estimates of the signal remain unaffected. This implies that the activity-induced variations in the radial velocities, traced by the S-indices, are independent of the periodic signal. This, in turn, indicates that the signal likely corresponds to gravitational disturbance caused by a planet with a minimum mass of 8.3 [4.0, 12.6] M$_{\oplus}$. We classify this planet candidate as a hot mini-Neptune orbiting the star.

\clearpage

\subsection{GJ 496.1}

\begin{figure}
\center
\includegraphics[angle=270, width=0.49\textwidth,clip]{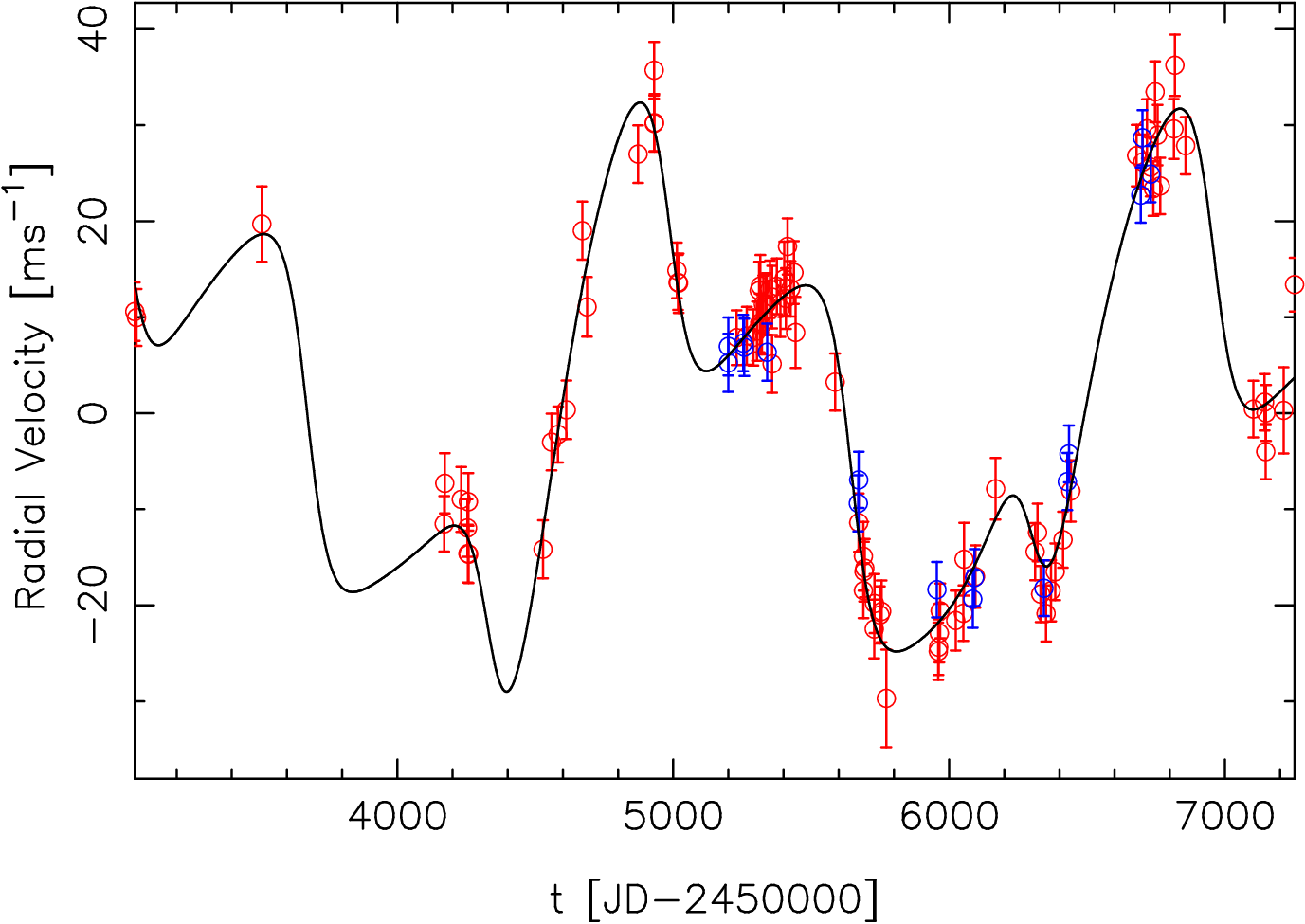}

\includegraphics[angle=270, width=0.49\textwidth,clip]{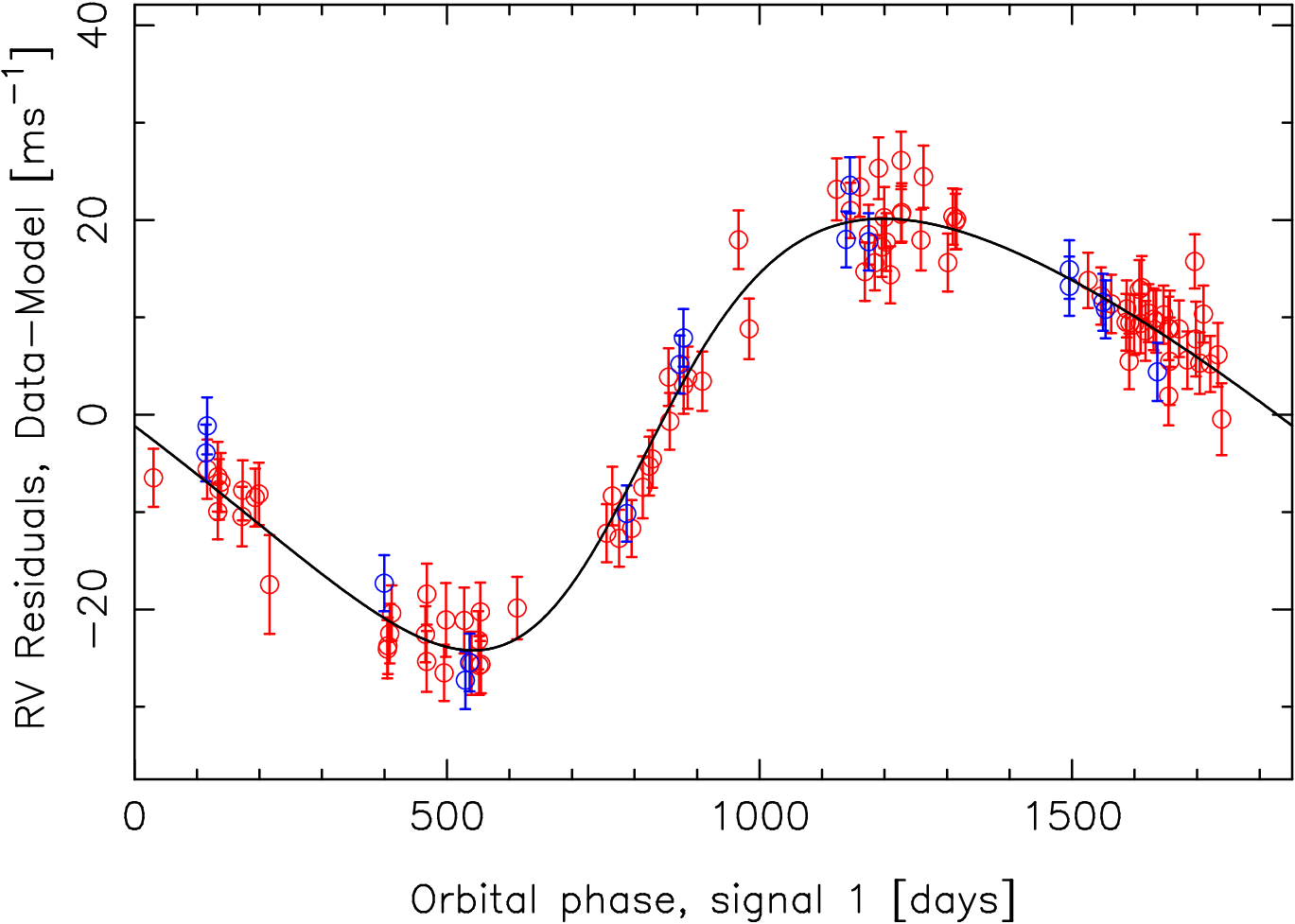}

\includegraphics[angle=270, width=0.49\textwidth,clip]{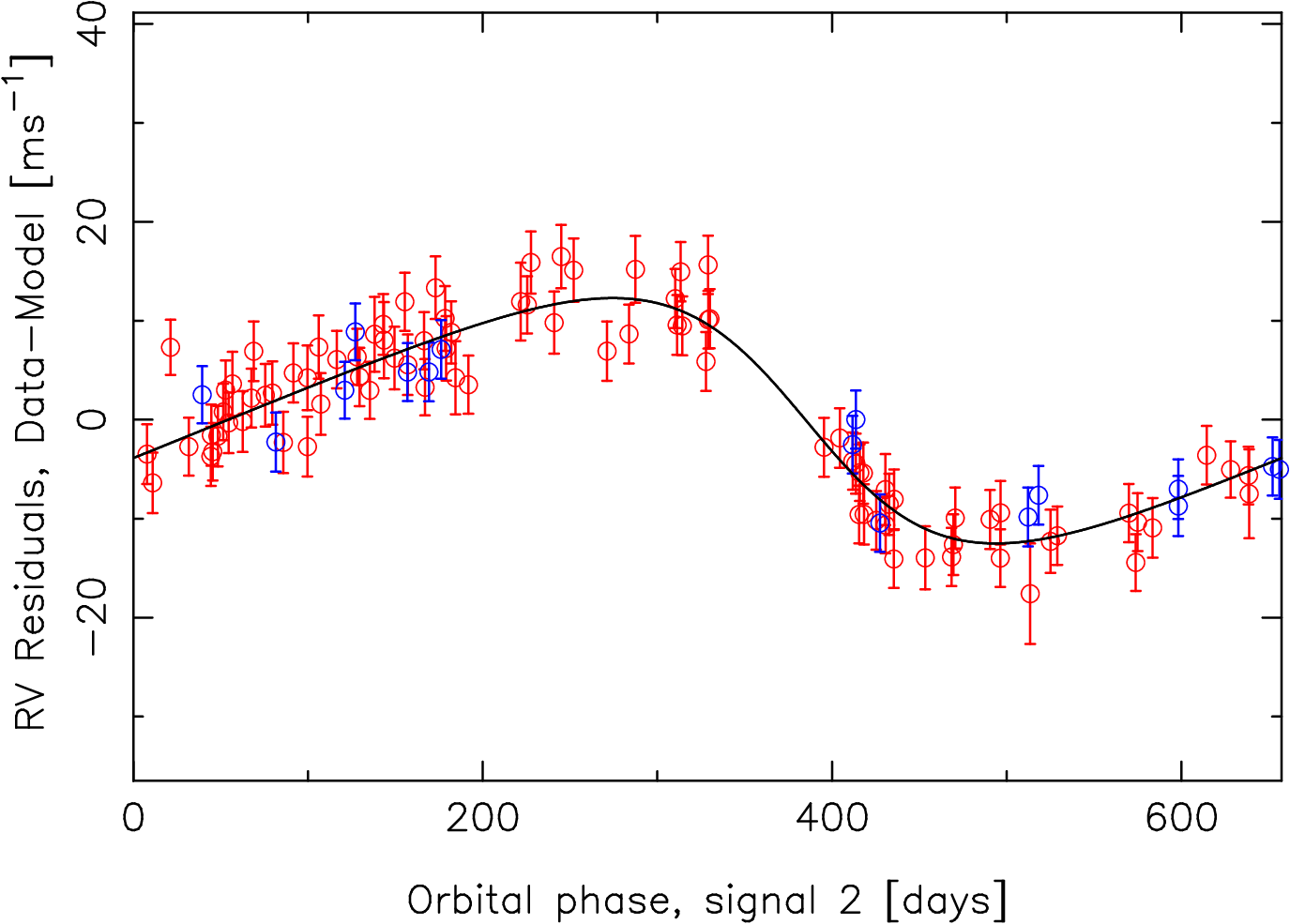}
\caption{Combined radial velocities of GJ 496.1 from HARPS (red) and PFS (blue) together with the modelled two Keplerian functions (top panel). Middle (bottom) panel shows the phase-folded Keplerian curve corresponding to GJ 496.1 b (c) with the signal of the other candidate subtracted.}\label{fig:GJ496.1_signals}
\end{figure}

We observed the two radial velocity signals in the combined HARPS and PFS velocities of GJ 496.1 (HD 113538, HIP 63833) that were reported by \citet{moutou2011}. However, \citet{moutou2011} noted that there is ``a marginal possibility for the radial velocity series for this star to be caused by activity'' as they also described the star as a significantly active one. We obtained no evidence for such activity-induced variability as the excess white noise in the larger HARPS data set was 2.88 [2.26, 3.71] ms$^{-1}$ indicating that the star is a common member of the population of radial velocity inactive M dwarfs. Moreover, we did not obtain any evidence for activity-induced variations in the radial velocities as the HARPS velocities were not connected to the corresponding activity indicators. The signals of the planet candidates in the radial velocities are demonstrated in Fig. \ref{fig:GJ496.1_signals}. We note that our estimates for the orbital eccentricities are much closer to zero than those of \citet{moutou2011}. This might be due to the fact that we penalised high eccentricities \emph{a priori} but also because our results are based on a larger data set minimising the bias \citep{zakamska2011} of radial velocity estimates towards higher eccentricities.

\begin{figure}
\center
\includegraphics[angle=270, width=0.49\textwidth,clip]{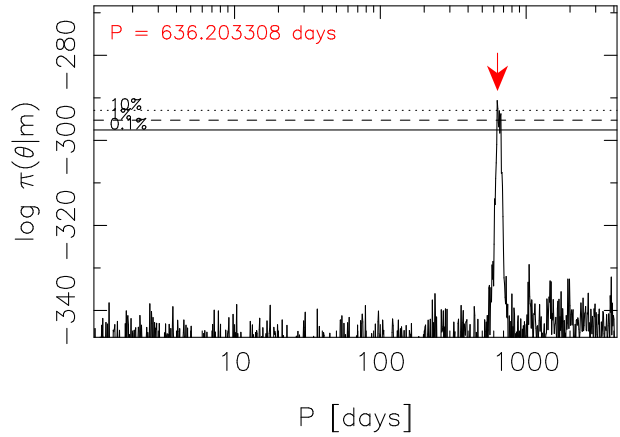}
\caption{Estimated posterior probability density as a function of the period parameter of the second Keplerian signal in a two-Keplerian model of GJ 496.1 radial velocities. The period space has been limited to planets inside the immediate neighbourhood of the giant planet orbiting the star with an orbital period of 1850 days.}\label{fig:GJ496.1_psearch}
\end{figure}

There is also another, more significant, difference between our results and those of \citet{moutou2011}. They reported that the orbital period of the inner candidate planet was 263.3$\pm$2.3 days. However, we obtained a unique solution for the corresponding period of the Keplerian signal at a period of 655.5 [643.0, 669.5] days (Fig. \ref{fig:GJ496.1_psearch}). This is likely due to aliases caused by annual gaps in the HARPS data and it appears rather clear that the apparent contradiction is caused by the fact that \citet{moutou2011} simply observed the yearly alias of the signal that we report as a unique solution in the current work. We could not find any additional significant signals in the radial velocities of GJ 496.1. 

\begin{figure}
\center
\includegraphics[angle=270, width=0.49\textwidth,clip]{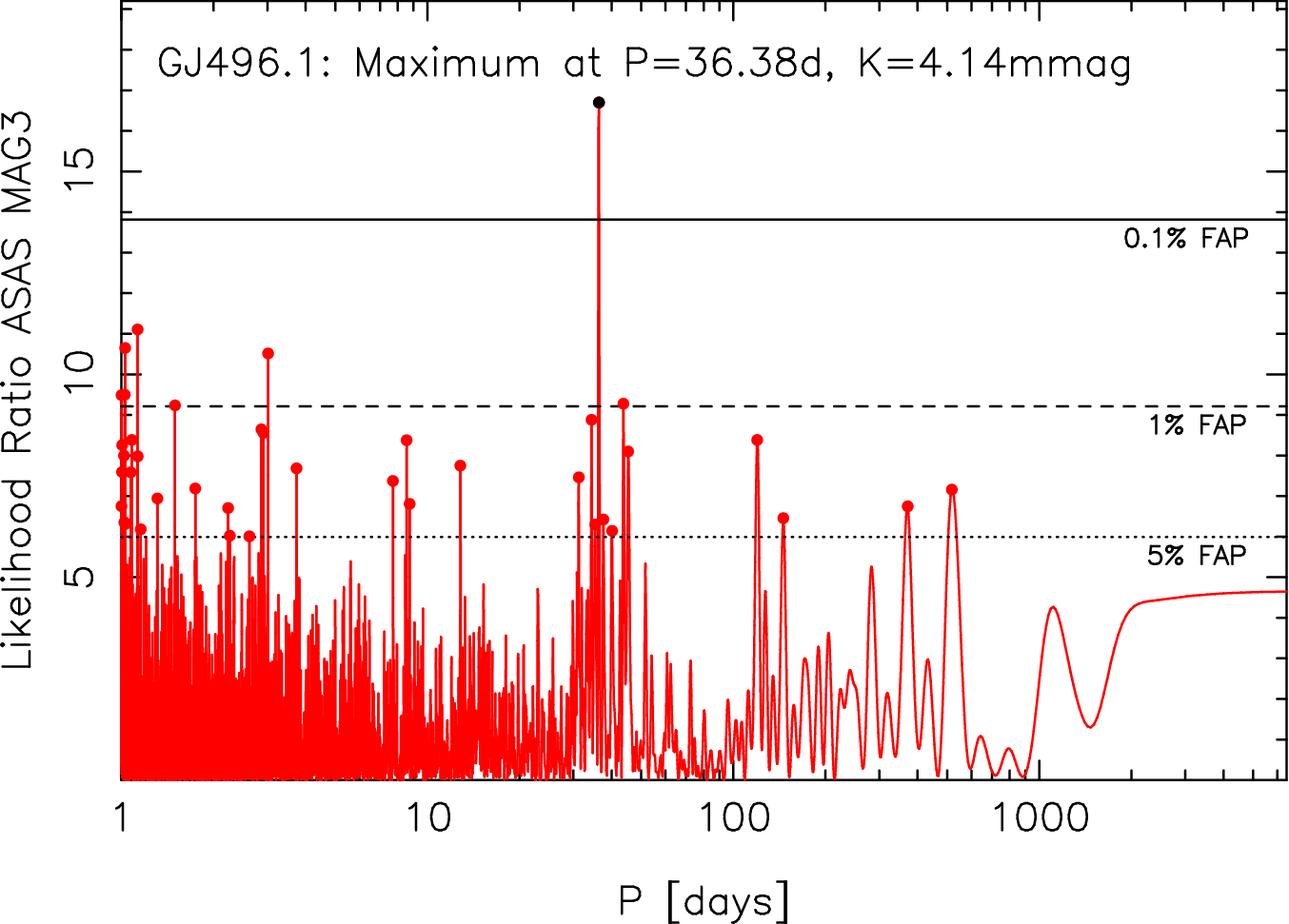}
\caption{Likelihood-ratio periodogram of the ASAS V-band photometry measurements of GJ 496.1. The red (black) dots indicate likelihood ratios exceeding the 5\% (0.1\%) FAP threshold.}\label{fig:GJ496.1_asas}
\end{figure}

We obtained an extensive set of 986 ASAS V-band photometry measurements. This data set indicates that there is a photometric periodicity at a period of 36.38 days that qualifies as an estimate for the photometric rotation period of the star (Fig. \ref{fig:GJ496.1_asas}). We could not identify significant periodicities in the HARPS activity indicators. This also means that it is unlikely that the two radial velocity signals are connected to stellar activity and are thus likely the signals of giant planets orbiting the star.

We note that we modelled the radial velocities by using a model with a 2nd order polynomial. The parameter of the 2nd order term is estimated to be 0.25 [0.02, 0.44] ms$^{-1}$year$^{-1}$ implying that it is statistically significantly different from zero with 99\% credibility. There is thus evidence for increasing acceleration in the data likely indicative of another long-period candidate planet orbiting the star.

\clearpage

\subsection{GJ 514}

\begin{figure}
\center
\includegraphics[angle=270, width=0.49\textwidth,clip]{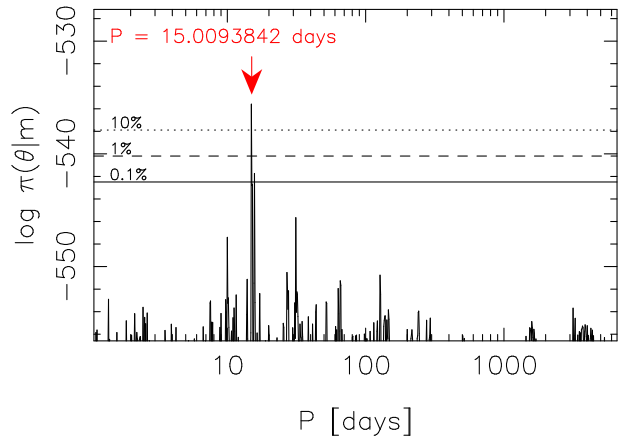}
\caption{Estimated posterior probability density as a function of the period parameter of the Keplerian signals given the GJ 514 radial velocities. The red arrow indicates the position of the global maximum in the period space and the horizontal lines denote the 10\% (dotted), 1\% (dashed), and 0.1\% (solid) equiprobability contours with respect to the global maximum.}\label{fig:GJ514_psearch}
\end{figure}

\begin{figure}
\center
\includegraphics[angle=270, width=0.49\textwidth,clip]{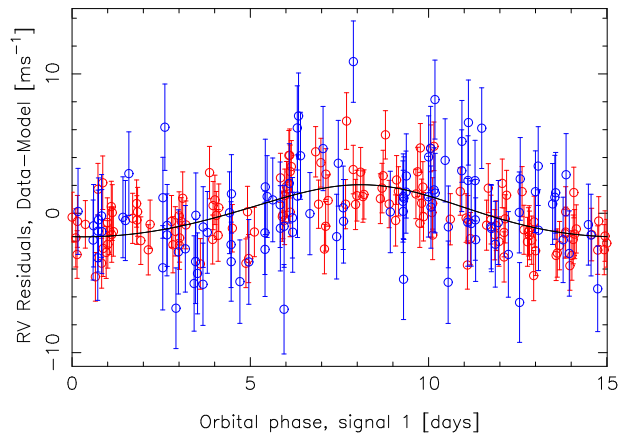}
\caption{Phase-folded HARPS and HIRES radial velocities of GJ 514 with the overplotted Keplerian curve (solid line).}\label{fig:GJ514_curve}
\end{figure}

The combined HARPS ($N = 139$) and HIRES ($N = 103$) radial velocities of GJ 514 (HIP 65859) contain evidence for a periodic signal that can be modelled as a Keplerian function (Figs. \ref{fig:GJ514_psearch} and \ref{fig:GJ514_curve}). This signal, at a period of 15.010 [14.970, 15.022] satisfies all our signal detection criteria and is uniquely present in the combined data set.

\begin{figure}
\center
\includegraphics[angle=270, width=0.49\textwidth,clip]{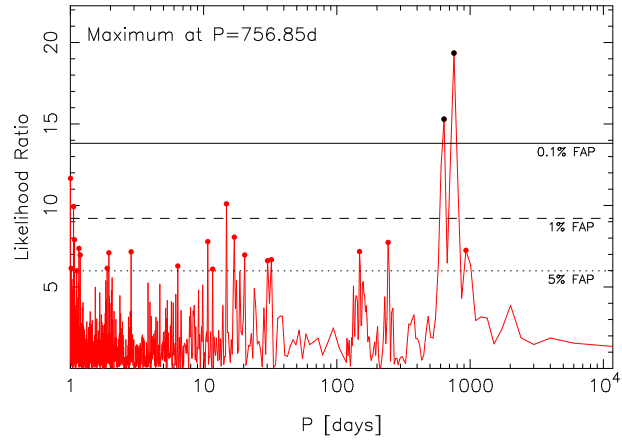}
\caption{Likelihood periodograms of HIRES S-indices of GJ 514. Red (black) dots denote the maxima exceeding the 5\% (0.1\%) FAP thresholds.}\label{fig:GJ514_S}
\end{figure}

The HARPS and HIRES radial velocities of GJ 514 do not appear to show variations that could be interpreted as activity-induced ones. There is no evidence for activity-induced cycles in the HARPS and HIRES velocities and their variability is not connected to the variations in the activity indicators statistically significantly. As did \citet{butler2016}, we observed a significant signal in the likelihood-ratio periodogram of the HIRES S-indices at a period of 760 days (Fig. \ref{fig:GJ514_S}) but this signal did not have counterparts in the radial velocities and we could not identify periodicities corresponding to the signal observed in the radial velocity data. We did not find significant periodicities in the ASAS V-band photometry data. We thus conclude that there is evidence for a hot super-Earth with a minimum mass of 4.3 [2.0 ,6.8] M$_{\oplus}$ orbiting GJ 514 with an orbital period of 15.0 days.

\clearpage

\subsection{GJ 529}

\begin{figure}
\center
\includegraphics[angle=270, width=0.49\textwidth,clip]{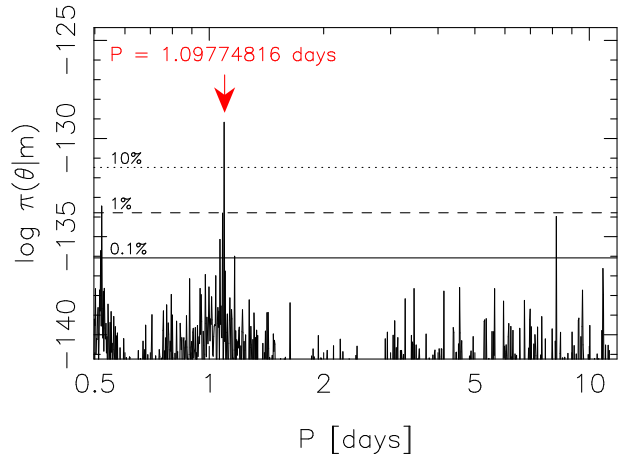}
\caption{Estimated posterior probability density as a function of the period parameter of the Keplerian signal in GJ 592 radial velocities. The red arrow indicates the position of the global maximum in the period space and the horizontal lines denote the 10\% (dotted), 1\% (dashed), and 0.1\% (solid) equiprobability contours with respect to the global maximum.}\label{fig:GJ529_psearch}
\end{figure}

\begin{figure}
\center
\includegraphics[angle=270, width=0.49\textwidth,clip]{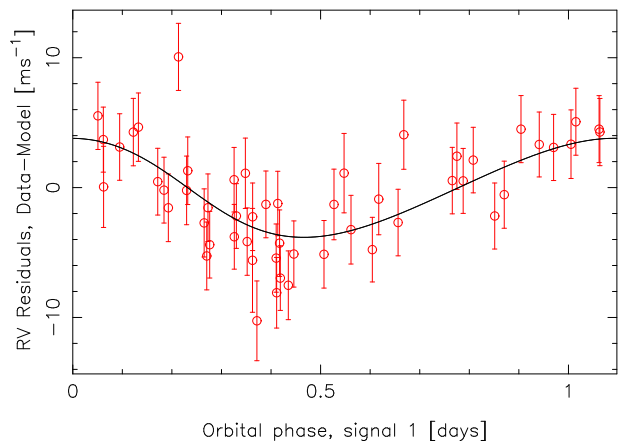}
\caption{HIRES radial velocities of GJ 529 folded on the phase of the signal in the data.}\label{fig:GJ529_phased}
\end{figure}

We found evidence for a periodic signal in the set of 53 KECK velocities of GJ 529 (HD 120467, HIP 67487) at a period of 1.09776 [1.09769, 1.09782] days. This signal is estimated to have an amplitude of 3.77 [1.74, 5.80] ms$^{-1}$ and it thus corresponds to a hot super-Earth with a minimum mass of 5.2 [2.2, 8.1] M$_{\oplus}$ orbiting the star. We demonstrate the uniqueness of the signal in Fig. \ref{fig:GJ529_psearch} where we highlight the period space from 0.5 to 12 days, and have plotted the phase-folded radial velocities in Fig. \ref{fig:GJ529_phased}. This signal did not exceed the detection threshold of \citet{butler2016} and was thus not reported by them.

The interpretation of the radial velocity signal as a candidate planet is enabled by the fact that we did not see any counterpart periodicities in the HIRES S-indices. We also failed to find any periodicities in the set of 391 ASAS V-band photometric observations of the star.

\clearpage

\subsection{GJ 536}

\begin{figure}
\center
\includegraphics[angle=270, width=0.49\textwidth,clip]{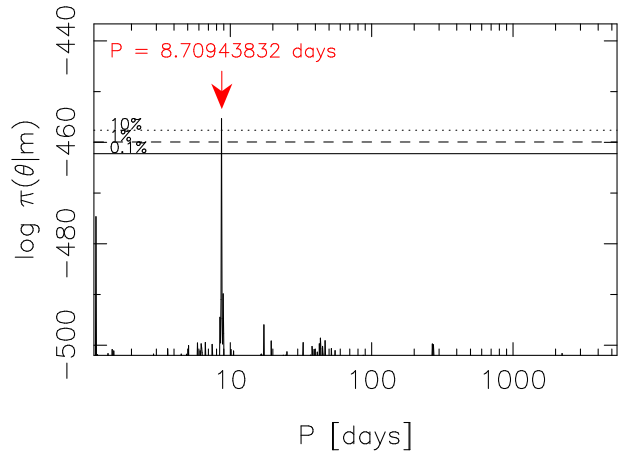}
\caption{Estimated posterior probability density as a function of the period parameter of the Keplerian signal in the GJ 536 radial velocities.}\label{fig:GJ536_psearch}
\end{figure}

With a combined set of HARPS ($N = 122$), HIRES ($N = 70$), and PFS ($N = 6$) radial velocities, we could identify a candidate planet orbiting GJ 536 (HD 122303, HIP 68469) with an orbital period of 8.7085 [8.7057, 8.7115] days. The corresponding radial velocity signal is shown as a significant probability maximum in the period space in Fig. \ref{fig:GJ536_psearch}. As this signal was clearly the dominating feature in the radial velocities, there is little doubt about the existence of the corresponding hot super-Earth with a minimum mass of 6.3 [4.2, 8.3] M$_{\oplus}$. The signal was discovered independently and reported by \citet{suarezmascareno2016}.

\begin{figure}
\center
\includegraphics[angle=270, width=0.49\textwidth,clip]{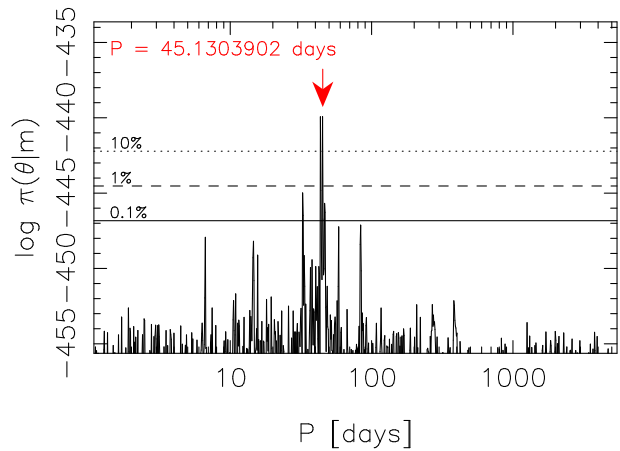}
\caption{Estimated posterior probability density of a model containing two Keplerian signals as a function of the period parameter of the second Keplerian signal in the GJ 536 radial velocities.}\label{fig:GJ536_psearch2}
\end{figure}

\begin{figure}
\center
\includegraphics[angle=270, width=0.49\textwidth,clip]{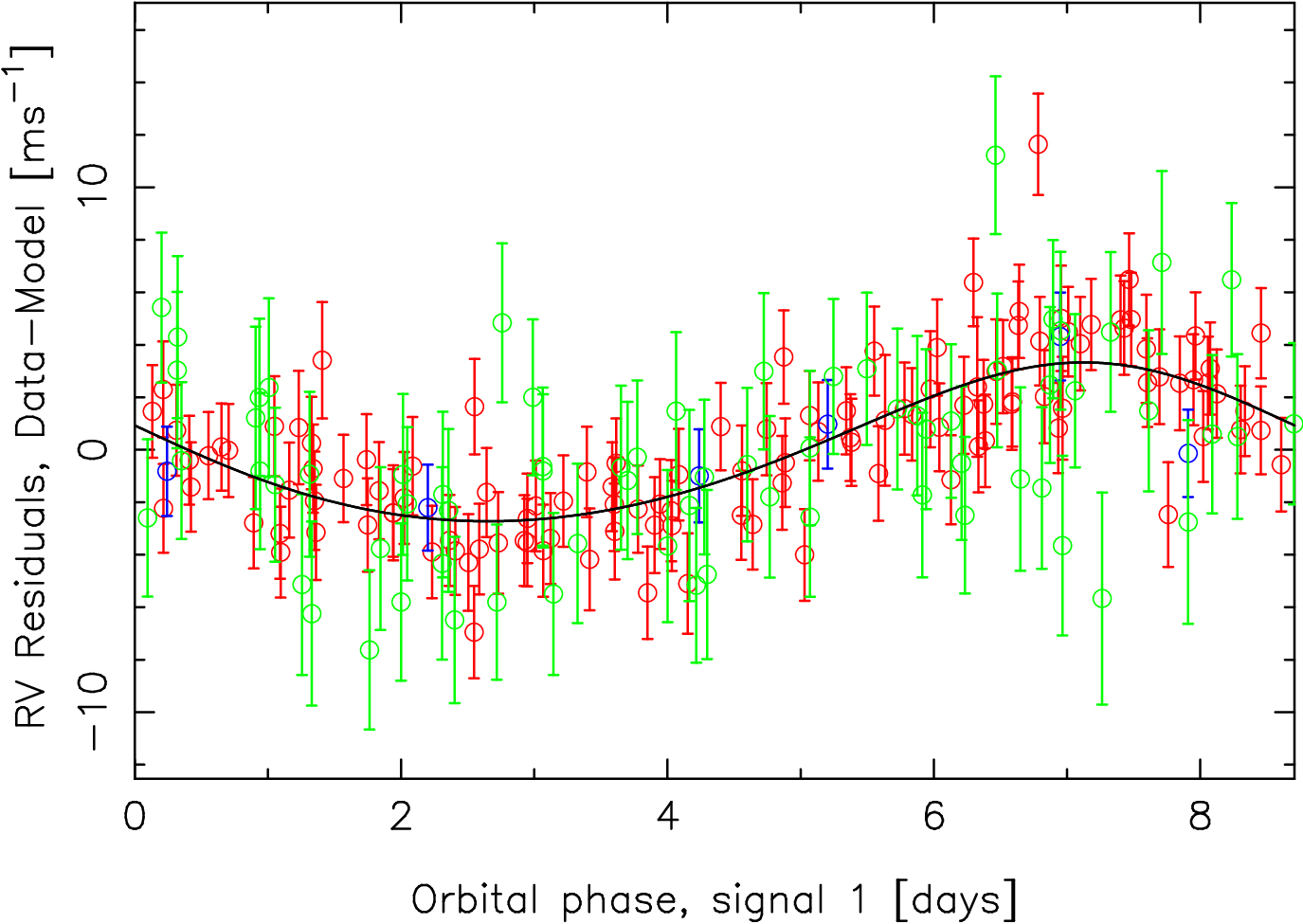}

\includegraphics[angle=270, width=0.49\textwidth,clip]{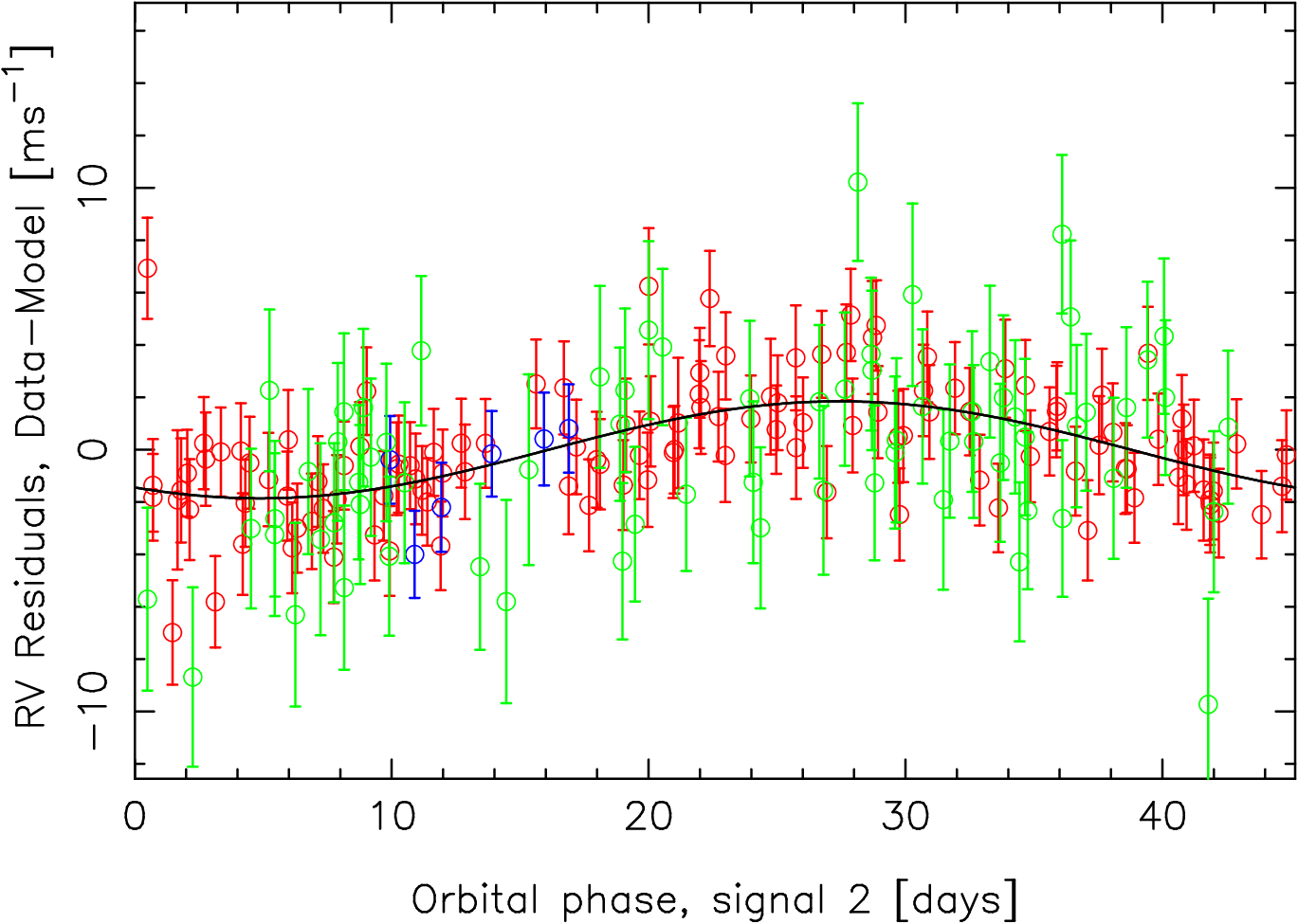}
\caption{Combined HARPS (red), HIRES (green), and PFS (blue) radial velocities of GJ 536 folded on the phases of the two signals with the other signal subtracted from each panel.}\label{fig:GJ536_curve}
\end{figure}

We also found evidence in favour of a second signal in the data at a period of 45.155 [45.017, 45.293] days (Fig. \ref{fig:GJ536_psearch2}). Although the signal is split into two maxima due to aliasing caused by gaps in the data, it still satisfies the detection criteria. We also note that there is a local maximum at a period of 30 days but this maximum does not even exceed the 1\% threshold with respect to the global maximum (Fig. \ref{fig:GJ536_psearch2}). The presence of the two signals in the data is further illustrated by folding the radial velocities on the phases of the signals in Fig. \ref{fig:GJ536_curve}.

Apart from weak evidence in favour of a 130-day periodicity in the HARPS S-indices and an equally weak periodicity of 250 days in the HIRES S-indices \citep{butler2016}, we could not identify any periodic signals in the HARPS and HIRES activity indicators that could be interpreted as counterparts of the radial velocity signals. This can be contrasted by the results of \citet{mascareno2015} and \citet{suarezmascareno2016} who report signals between 42 and 45 days for HARPS S-index and FWHM. However, because they also report a photometric periodicity of 43.33 days that can be considered a counterpart to the radial velocity signal, we interpret that the signal at a period of 45 days is caused by stellar rotation rather than a planet. We thus agree with the interpretation of \citet{suarezmascareno2016} that there is evidence for one planet with an orbital period of 8.7 days corresponding to a hot super-Earth orbiting the star.

\clearpage

\subsection{GJ 553.1}

\begin{figure}
\center
\includegraphics[angle=270, width=0.49\textwidth,clip]{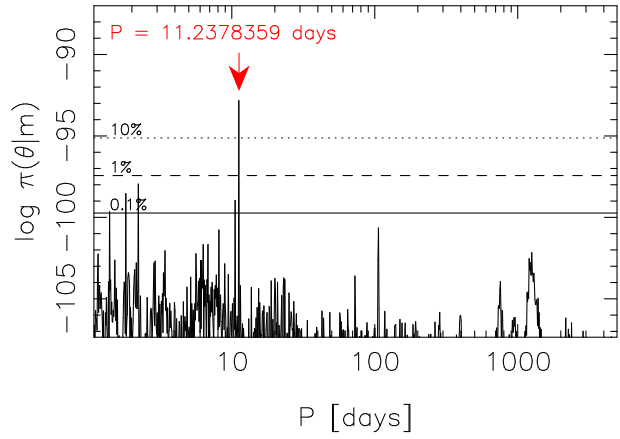}
\caption{As in Fig. \ref{fig:GJ536_psearch} but for the HIRES and HARPS radial velocities of GJ 553.1.}\label{fig:GJ553.1_psearch}
\end{figure}

\begin{figure}
\center
\includegraphics[angle=270, width=0.49\textwidth,clip]{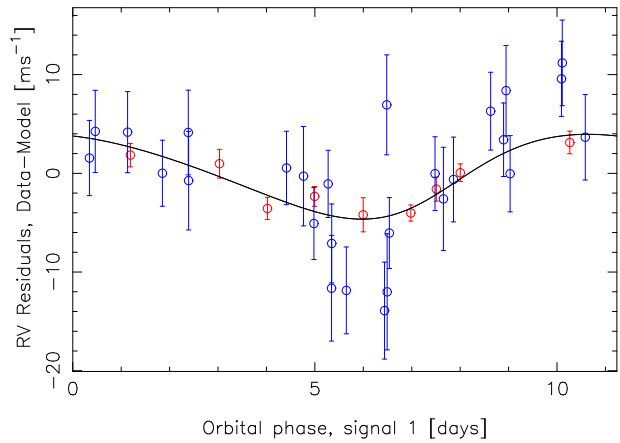}
\caption{HARPS (red) and HIRES (blue) radial velocities of GJ 553.1 folded on the phase of the signal detected in the combined data (solid curve).}\label{fig:GJ553.1_phased}
\end{figure}

We detected a unique periodic signal in the combined HARPS ($N = 9$) and HIRES ($N = 27$) radial velocities of GJ 553.1 (HIP 70975). This signal, at a period of 11.2381 [11.2304, 11.2476] days and with an amplitude of 4.95 [1.92, 7.66] ms$^{-1}$, was rather uniquely present in the data as a probability maximum (Fig. \ref{fig:GJ553.1_psearch}) but we acknowledge that it was one of the weakest significantly detected signals discussed in the current work. We have plotted the phase-folded radial velocities of GJ 553.1 in Fig. \ref{fig:GJ553.1_phased}.

We could not identify any periodic signals in the HARPS and HIRES activity indicators. Similarly, we found no evidence for periodicities in the ASAS V-band photometry of GJ 553.1. We thus interpret the radial velocity signal as evidence for a candidate planet classified as a hot mini-Neptune with a minimum mass of 7.6 [3.1, 12.6] M$_{\oplus}$.

\clearpage

\subsection{GJ 555}

\begin{figure}
\center
\includegraphics[angle=270, width=0.49\textwidth,clip]{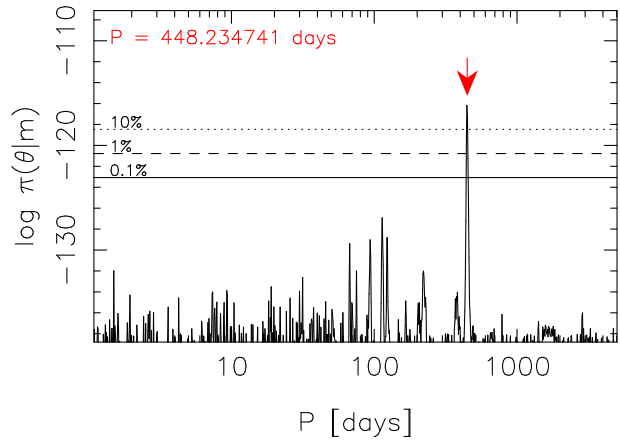}
\caption{Estimated posterior probability density as a function of the period parameter of the Keplerian signal in the GJ 555 radial velocities.}\label{fig:GJ555_psearch}
\end{figure}

\begin{figure}
\center
\includegraphics[angle=270, width=0.49\textwidth,clip]{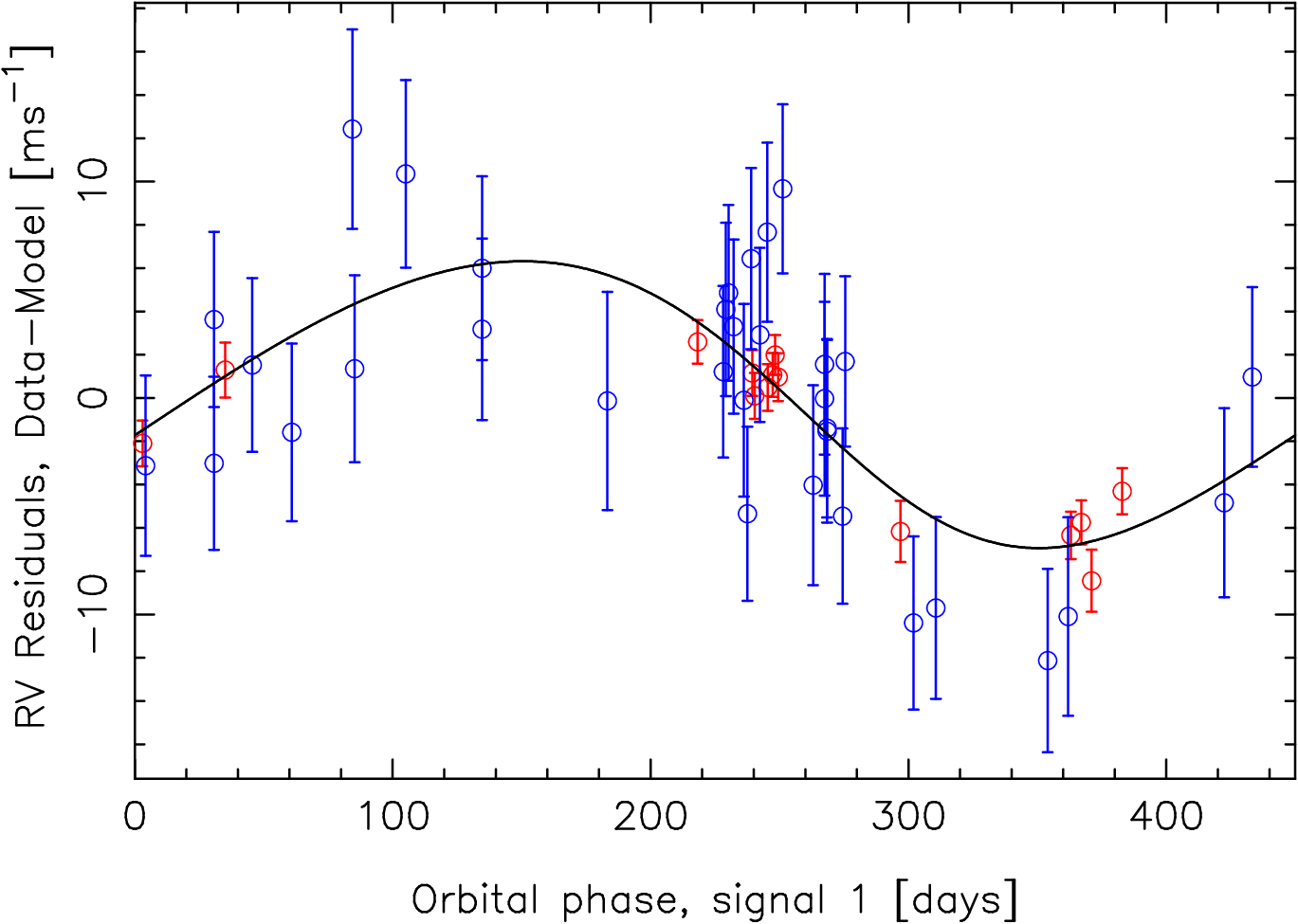}

\includegraphics[angle=270, width=0.49\textwidth,clip]{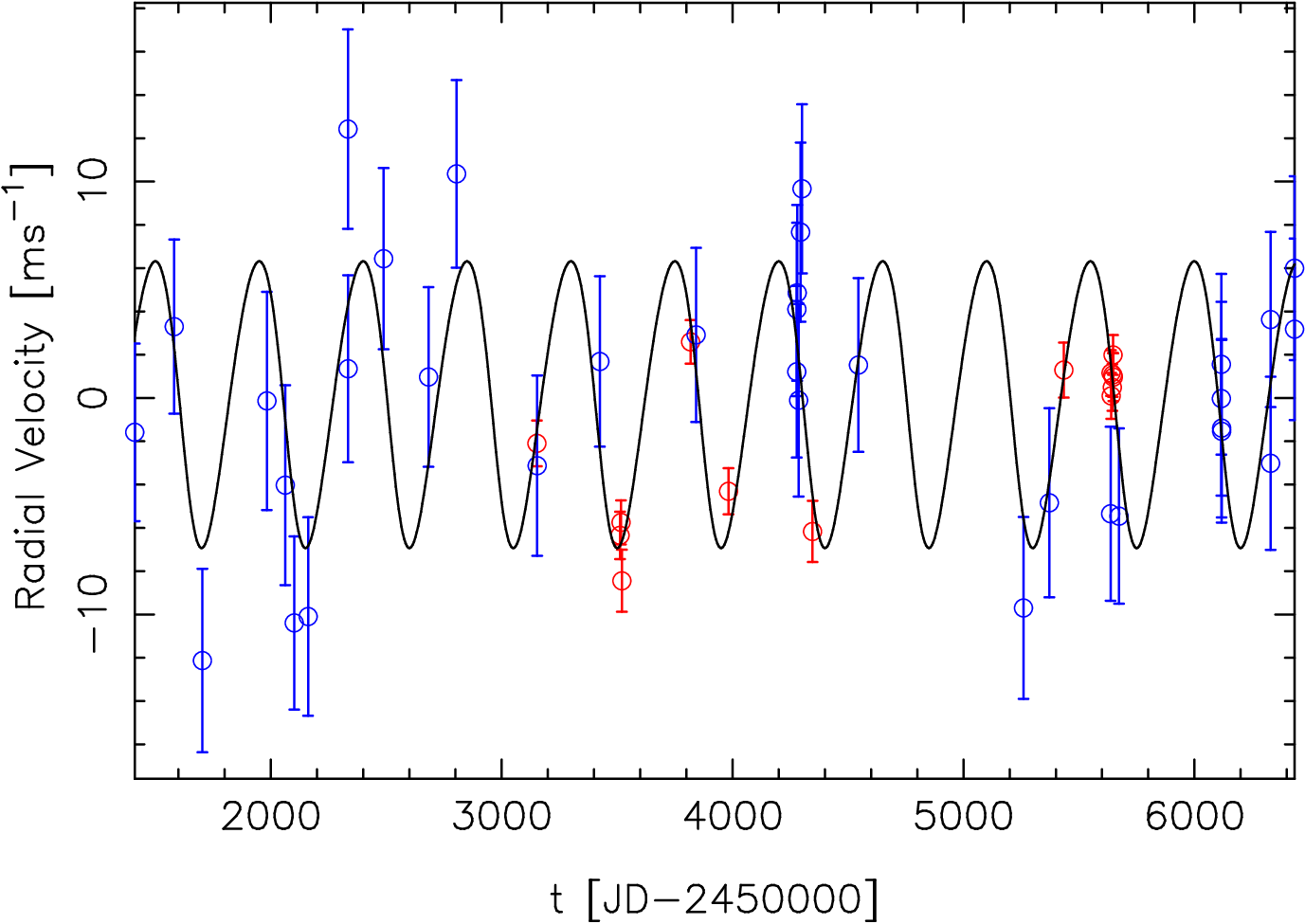}
\caption{The Keplerian signal in the combined HARPS (red) and HIRES (blue) radial velocities of GJ 555.}\label{fig:GJ555_curve}
\end{figure}

GJ 555 (HIP 71253) has been observed with both HARPS and HIRES spectrographs. Although the total number of data is below the average for both instruments, we could identify a signal in the combined velocities at a period of 449.6 [443.7, 456.5] days with an amplitude of 6.33 [3.49, 9.50] ms$^{-1}$ (Figs. \ref{fig:GJ555_psearch} and \ref{fig:GJ555_curve}). This signal satisfied the detection criteria and is present in the data as a unique probability maximum in the period space. We thus interpret the signal as evidence for a planet candidate orbiting the star. However, the phase-coverage of the signal is far from optimal as can be seen in Fig. \ref{fig:GJ555_curve} (bottom panel), which makes further Doppler spectroscopic monitoring of this target more than welcome in order to constrain the signal better.

Even with a rather poor phase-coverage, the signal is detected according to our criteria and is thus classified as a planet candidate that we classify as a cool Neptune with a minimum mass of 30.1 [15.4, 46.4] M$_{\oplus}$. This is because we identified counterparts neither in the HARPS or HIRES activity indicators nor in the ASAS V-band photometry of the star.

\clearpage

\subsection{GJ 588}

\begin{figure}
\center
\includegraphics[angle=270, width=0.49\textwidth,clip]{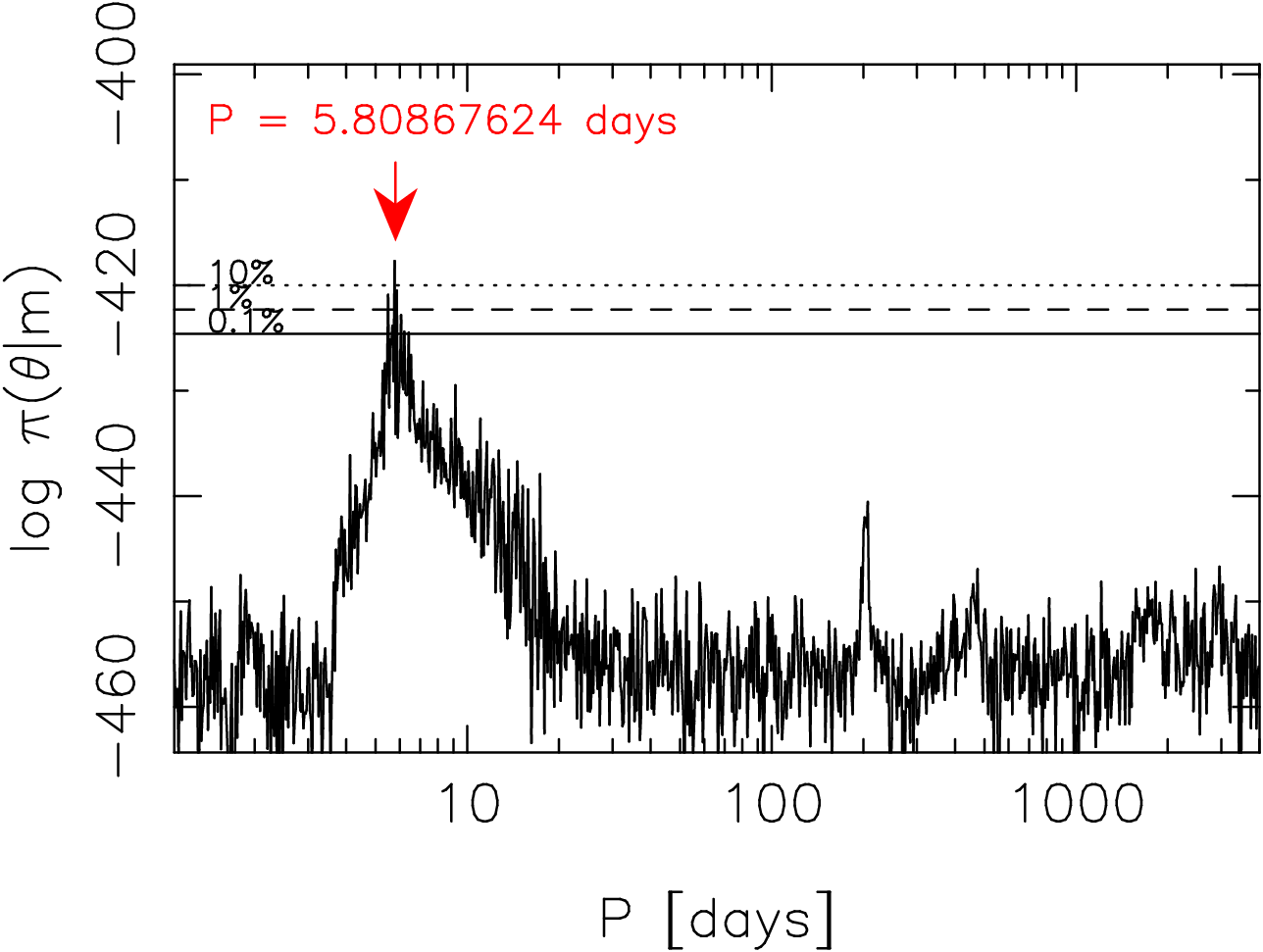}

\includegraphics[angle=270, width=0.49\textwidth,clip]{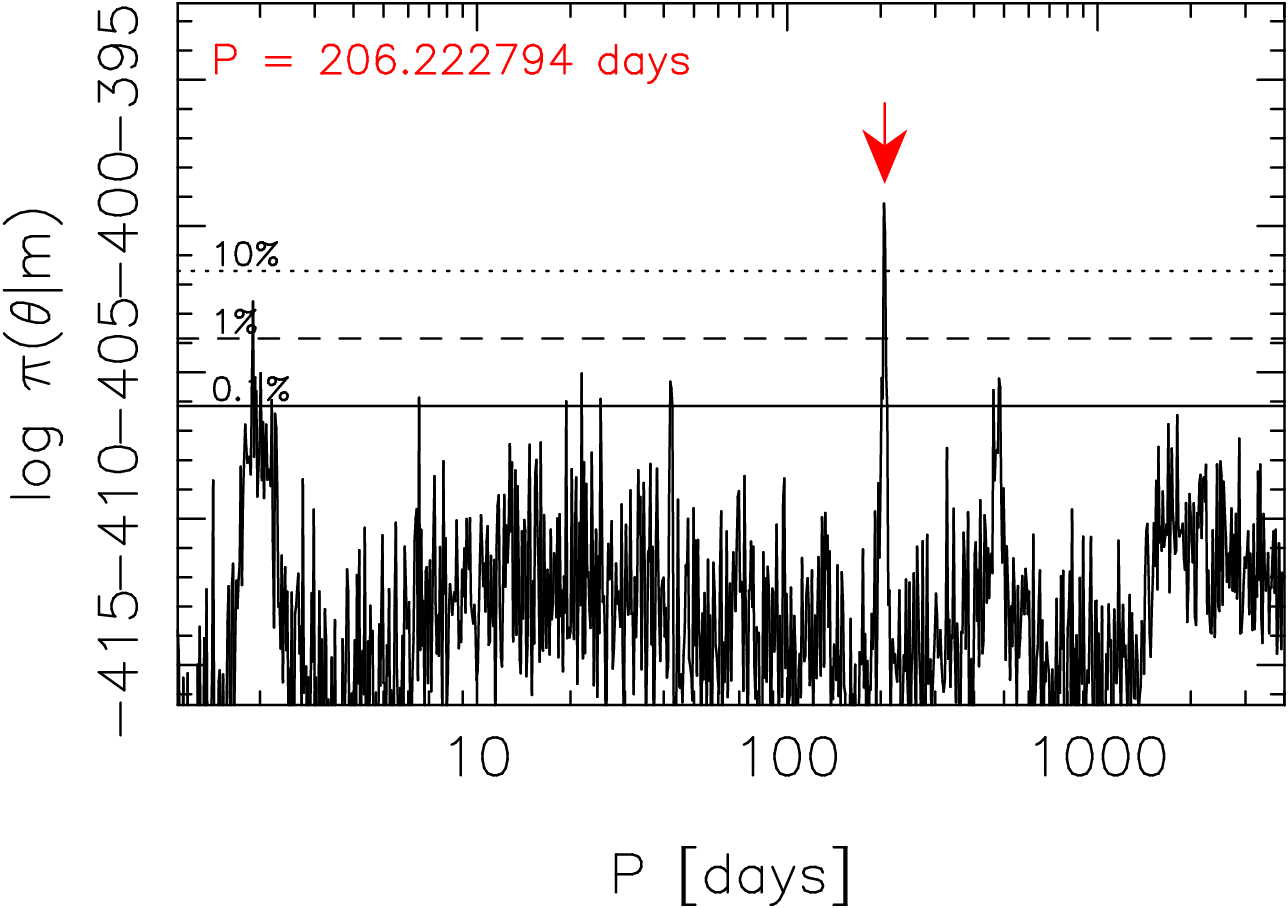}
\caption{Estimated posterior probability densities as a function of the period parameters of the first (top) and second (bottom panel) Keplerian signals in the GJ 588 radial velocities. The global maxima are denoted by red arrows and the horizontal lines indicate the 10\% (dotted), 1\% (dashed), and 0.1\% (solid) probability thresholds with respect to the maxima.}\label{fig:GJ588_psearch}
\end{figure}

\begin{figure}
\center
\includegraphics[angle=270, width=0.49\textwidth,clip]{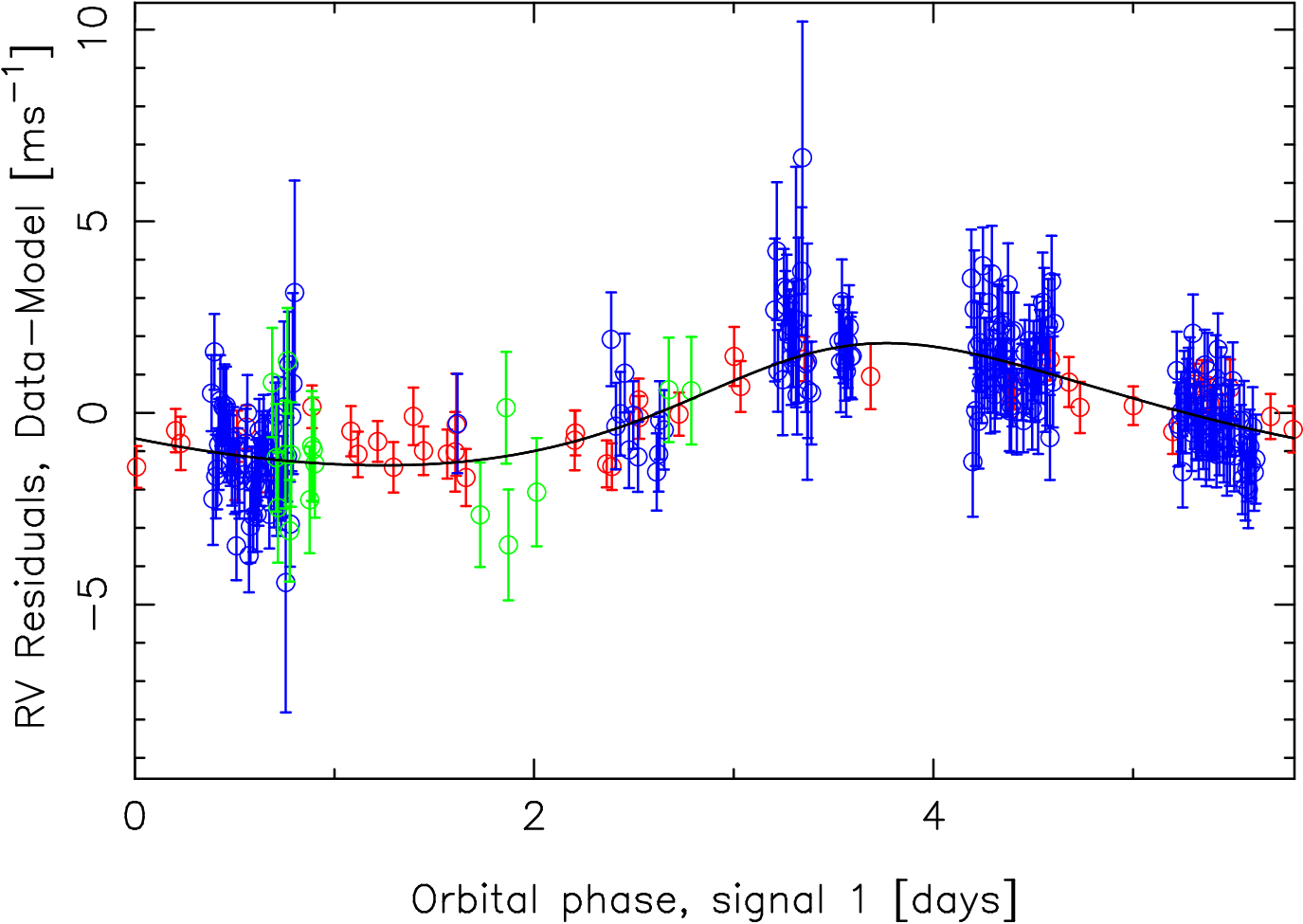}
\includegraphics[angle=270, width=0.49\textwidth,clip]{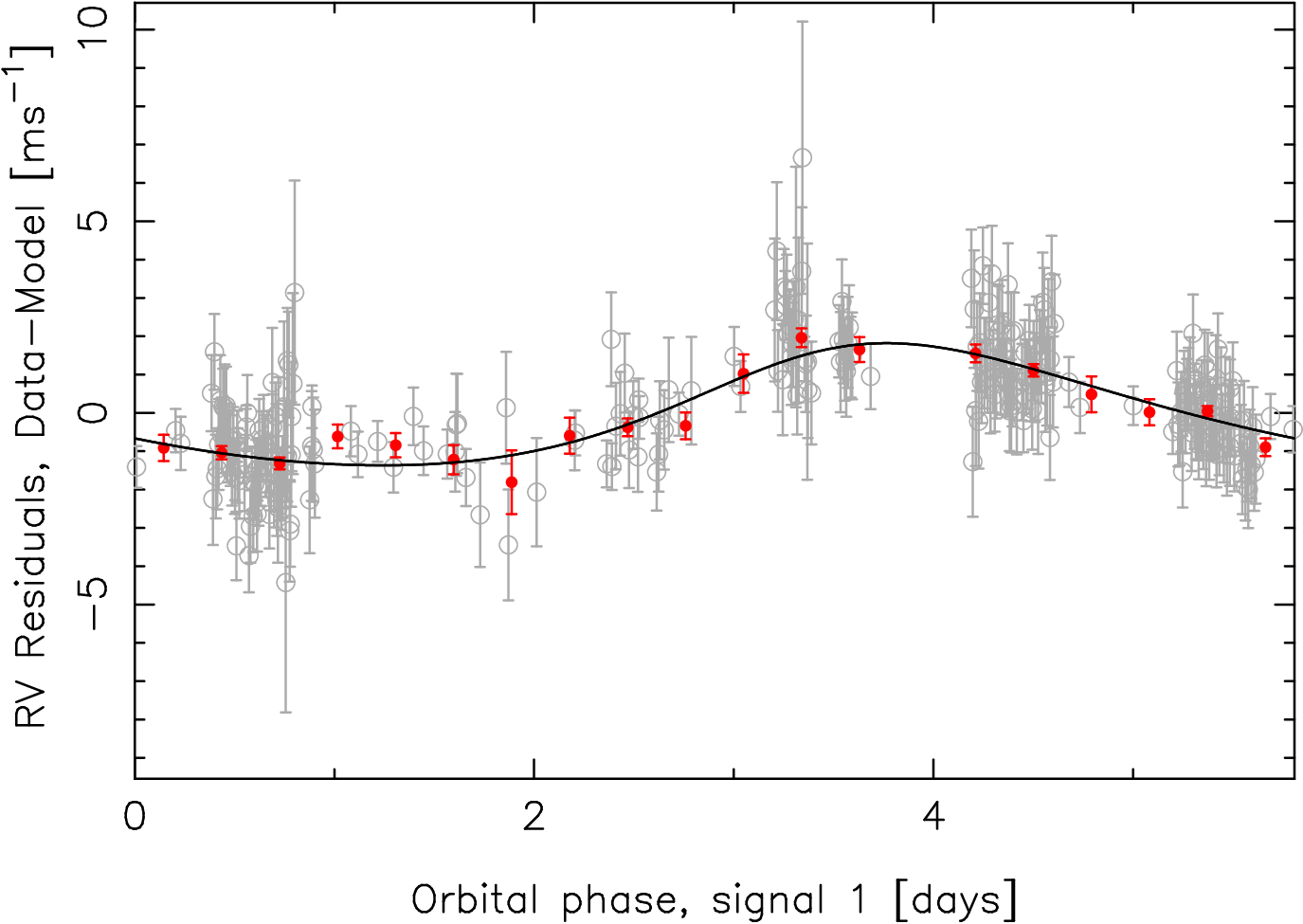}

\includegraphics[angle=270, width=0.49\textwidth,clip]{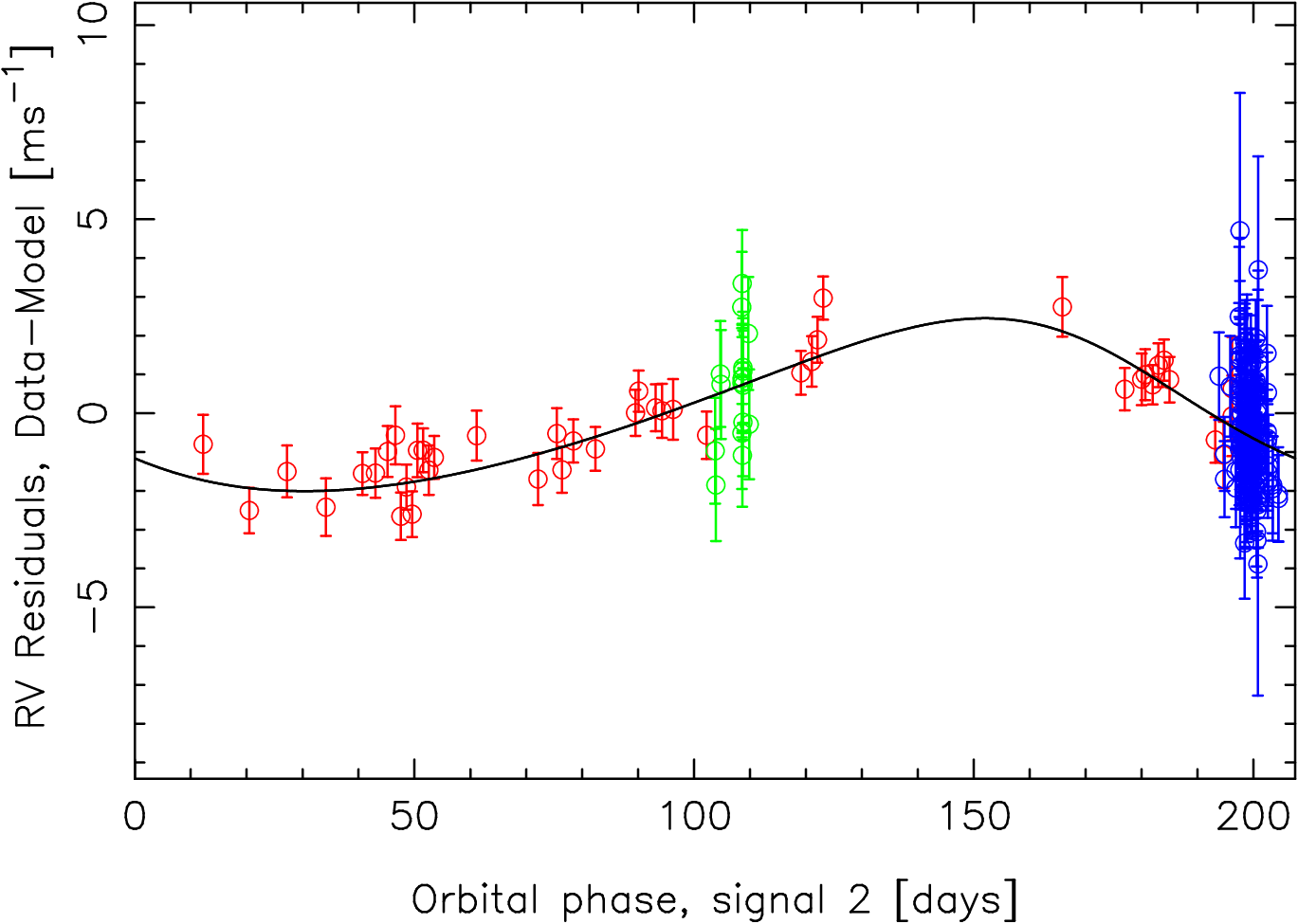}
\includegraphics[angle=270, width=0.49\textwidth,clip]{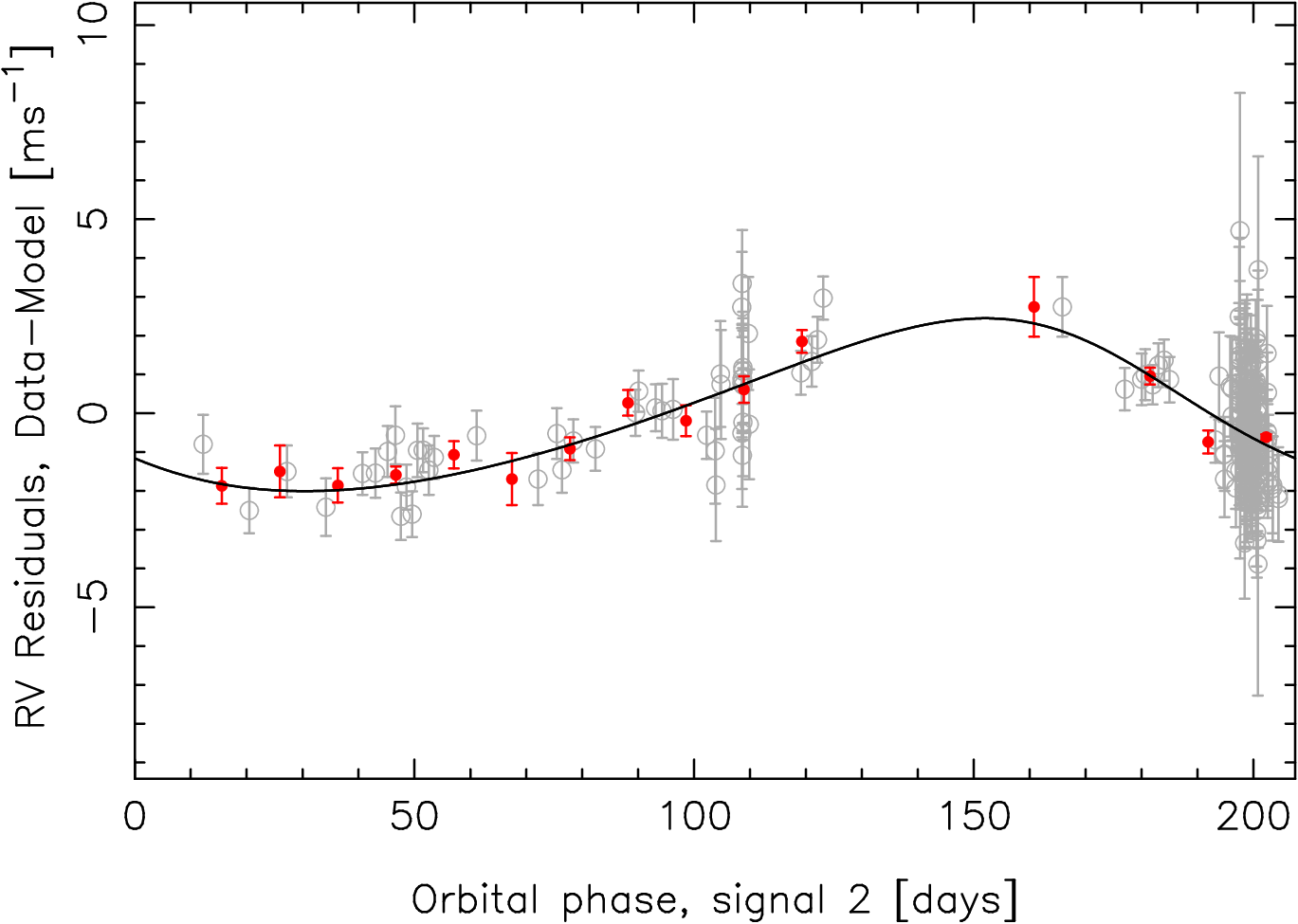}
\caption{HARPS radial velocities of GJ 588 folded to the phases of the two signals detected in the data. The red, blue, and green colours denote the three different observing campaigns. Blue circles correspond to the CTB observing run.}\label{fig:GJ588_curve}
\end{figure}

\begin{figure}
\center
\includegraphics[angle=270, width=0.49\textwidth,clip]{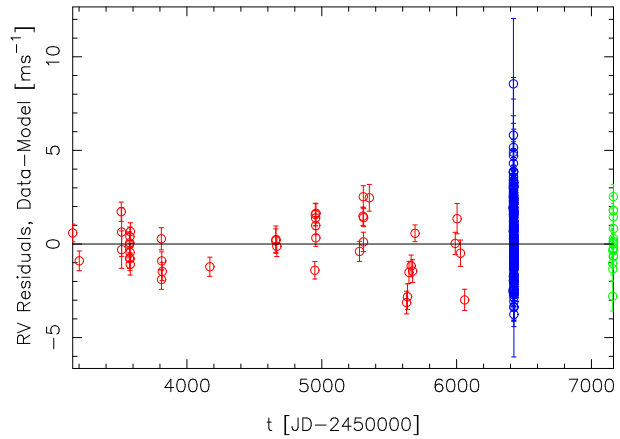}
\caption{HARPS dadial velocities of GJ 588. The CTB observing run is denoted by blue circles.}\label{fig:GJ588_data}
\end{figure}

GJ 588 (HIP 76074) is one of the targets of the \emph{Cool Tiny Beats} (CTB) observing campaign\footnote{Proxima Centauri (GJ 551) was also one of the CTB targets \citep{anglada2016}.}, conducted with the HARPS, whose purpose was to detect small planets on short-period orbits. With a total of 268 HARPS radial velocities, we were able to find evidence for two periodic signals that appeared unique in the period space and satisfied our signal detection criteria (Figs. \ref{fig:GJ588_psearch} and \ref{fig:GJ588_curve}). These signals, at periods of 5.8084 [5.8066, 5.8101] and 206.0 [202.7, 208.1] days have amplitudes below 2 ms$^{-1}$ but are still detected robustly due to HARPS data taken in three observing campaigns that have completely different cadences, enabling detections of low-amplitude signals ranging from a few to a few hundred days (Fig. \ref{fig:GJ588_data}).

\begin{figure}
\center
\includegraphics[angle=270, width=0.49\textwidth,clip]{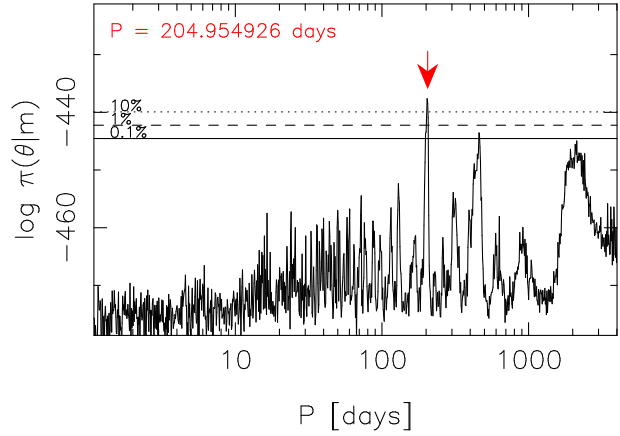}
\caption{As in Fig. \ref{fig:GJ588_psearch} for the one-Keplerian model but when assuming that all HARPS data points from all three observing campaigns are equally distributed and not modelled by assuming that they were taken by independent instruments.}\label{fig:GJ588_psearch2}
\end{figure}

When assuming that all the HARPS observations are equally distributed, the 206-day signal was amplified considerably and dominated the period space as a highly significant maximum, together with it's aliases (Fig. \ref{fig:GJ588_psearch2}). This encouraged us to interpret both periodicities as evidence for candidate planets orbiting the star.

\begin{figure}
\center
\includegraphics[angle=270, width=0.49\textwidth,clip]{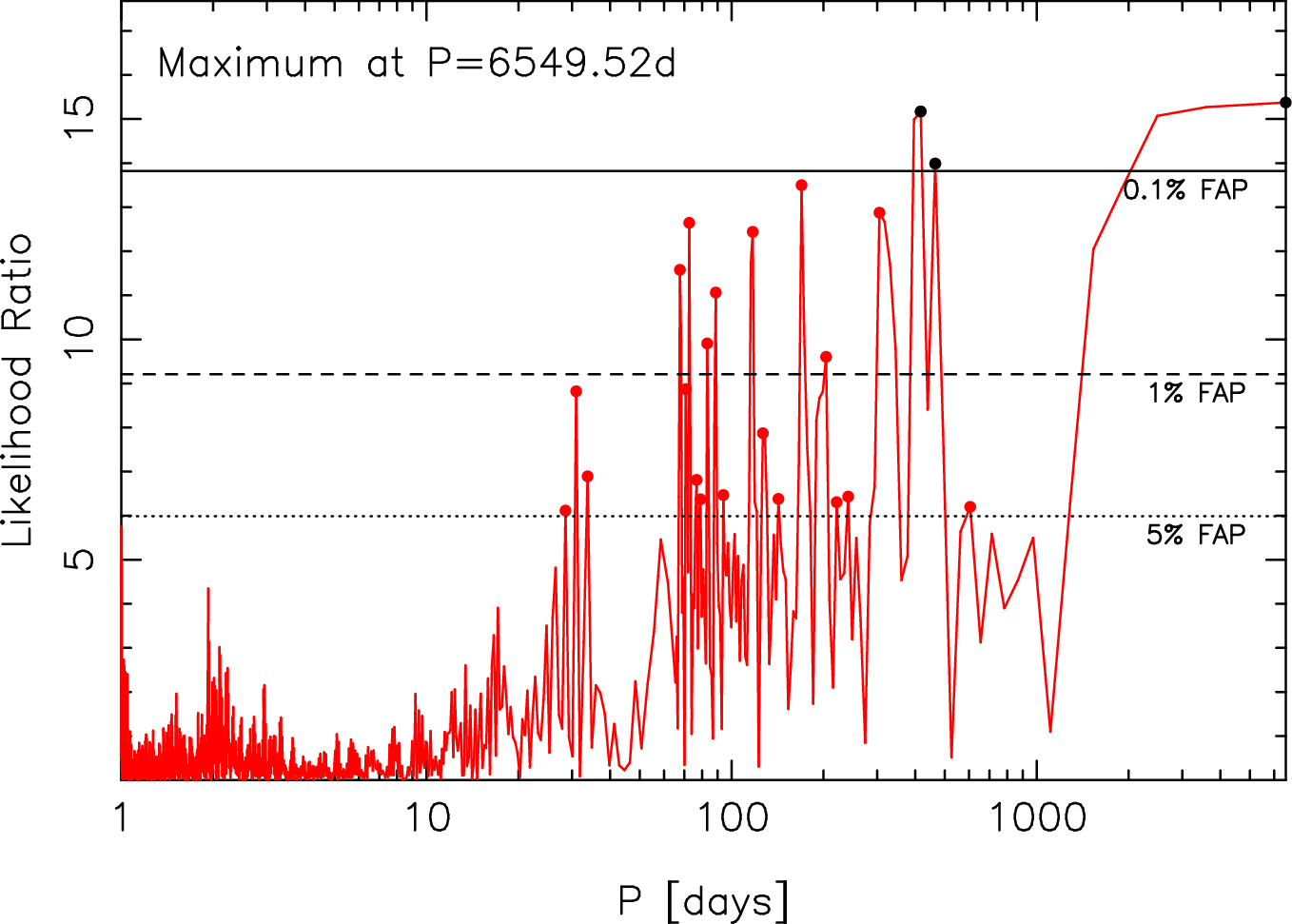}

\includegraphics[angle=270, width=0.49\textwidth,clip]{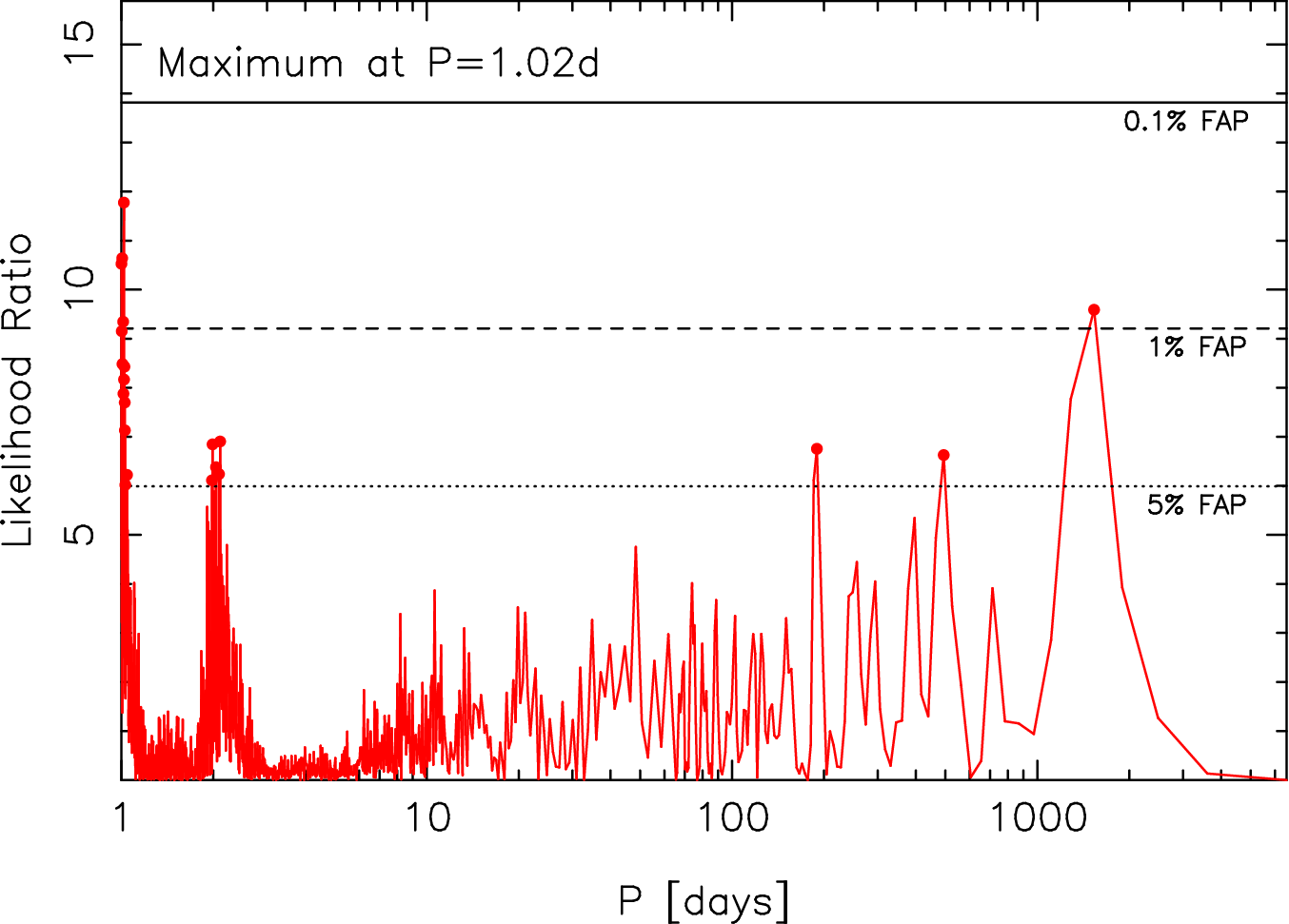}
\caption{Likelihood-ratio periodogram of the HARPS S-indices of GJ 588 (top panel) and the periodogram of residuals after subtracting the long-period signal (bottom panel).}\label{fig:GJ588_S_periodogram}
\end{figure}

We did not observe any periodicities in the HARPS BIS and FWHM values. However, there was a long-period cycle in the corresponding S-indices (Fig. \ref{fig:GJ588_S_periodogram}). Yet, we could not identify any periodicities corresponding to the signals detected in the radial velocities. It is thus our interpretation that the radial velocity signals are caused by planets orbiting the star rather than activity. It is noteworthy that the data sampling and different cadences of the three separate observing campaigns create a ``jungle of aliases'' in the periodogram of the S-index in Fig. \ref{fig:GJ588_S_periodogram} (top panel) that are almost completely removed by subtracting the long-period variability.

The set of 451 ASAS V-band photometry measurements suggested the presence of low-level (amplitudes below 4 mmag) signals at periods of 1.91 and 870 days with significances exceeding 1\% FAP threshold but not 0.1\% FAP threshold. However, there was not even suggestive evidence for periodicities at or near the observed radial velocity signals.

\clearpage

\subsection{GJ 628}\label{sec:GJ628}

GJ 628 (Wolf 1061, HIP 80824) has been reported to be a host to a system of up to three low-mass, potentially rocky, planets based on HARPS radial velocities \citep{wright2016,astudillo2017}. Although \citet{wright2016} could not rule out the longest periodicity of 67.27$\pm$0.12 days as a rotation-induced signal arising from the stellar surface rather than a planet, the other two at periods of 4.8876$\pm$0.0014 and 17.867$\pm$0.011 days were interpreted as signals of \emph{bona fide} planetary companions to the star. Moreover, \citet{astudillo2017} identified a signal at a period of 217 days. We obtained the TERRA reduction of the 162 HARPS spectra, some of which were analysed in \citet{wright2016} and \citet{astudillo2017}, and also a set of 97 HIRES velocities of the star and analysed the combined data in order to verify the results of \citet{wright2016}, to search for additional signals, and to calculate the detection threshold for this target given the available data.

However, the HIRES radial velocities of GJ 628 appeared to suffer a ``shift'' in reference velocity between JDs 2451755 and 2452004. The first four HIRES velocities are on average 26.6 ms$^{-1}$ below the average of the subsequently obtained 93 velocities. We thus removed the first four velocities from the HIRES set.

\begin{figure}
\center
\includegraphics[angle=270, width=0.49\textwidth,clip]{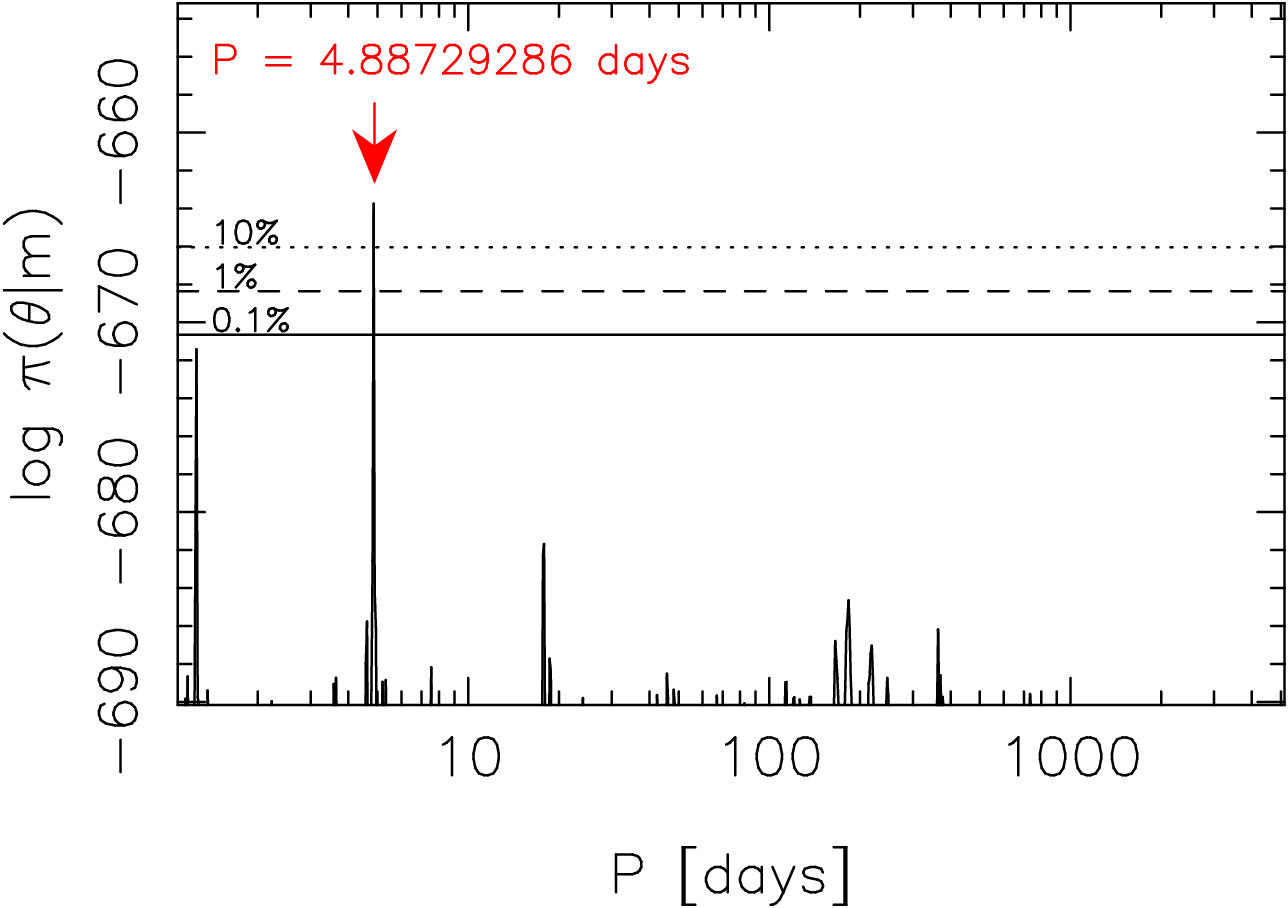}

\includegraphics[angle=270, width=0.49\textwidth,clip]{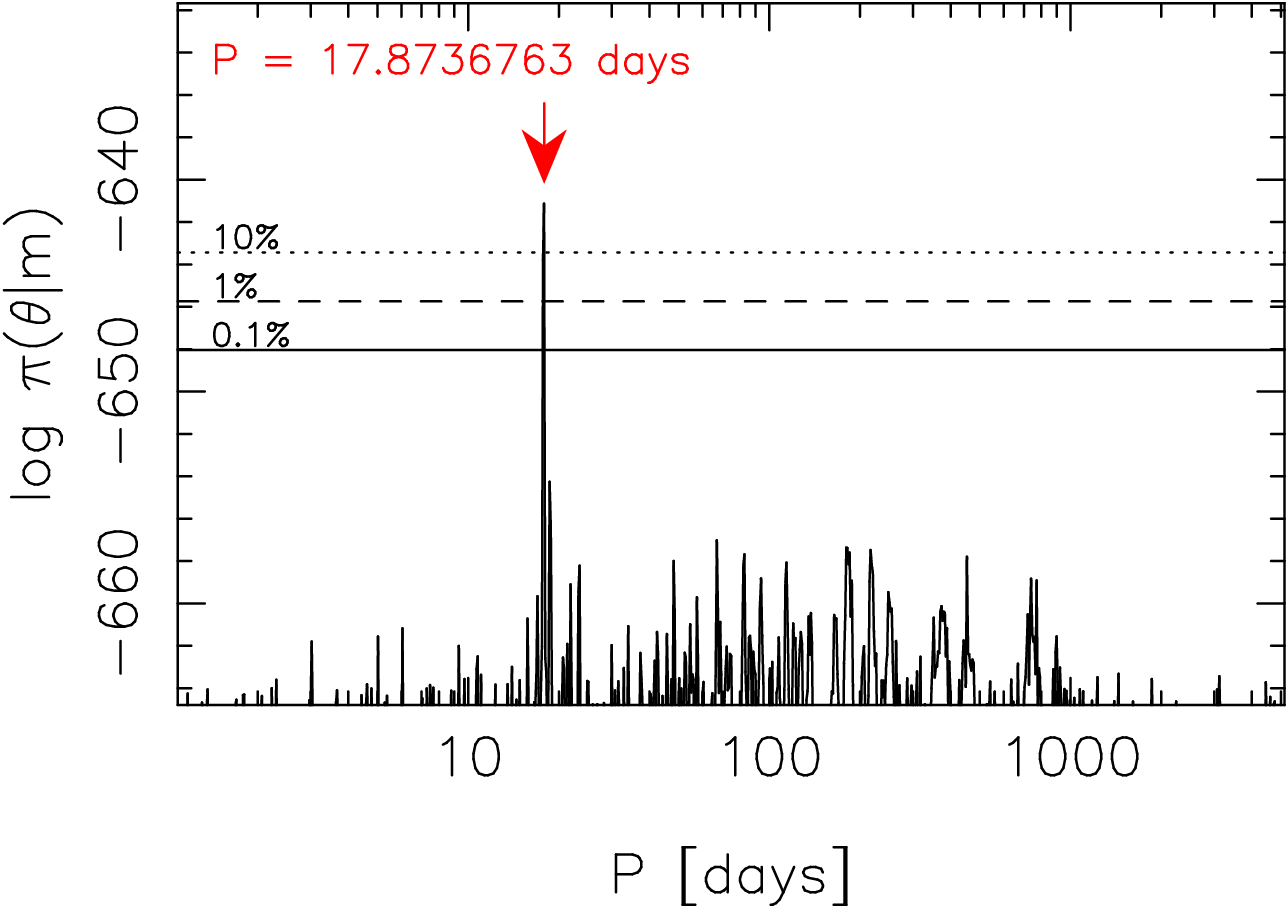}

\includegraphics[angle=270, width=0.49\textwidth,clip]{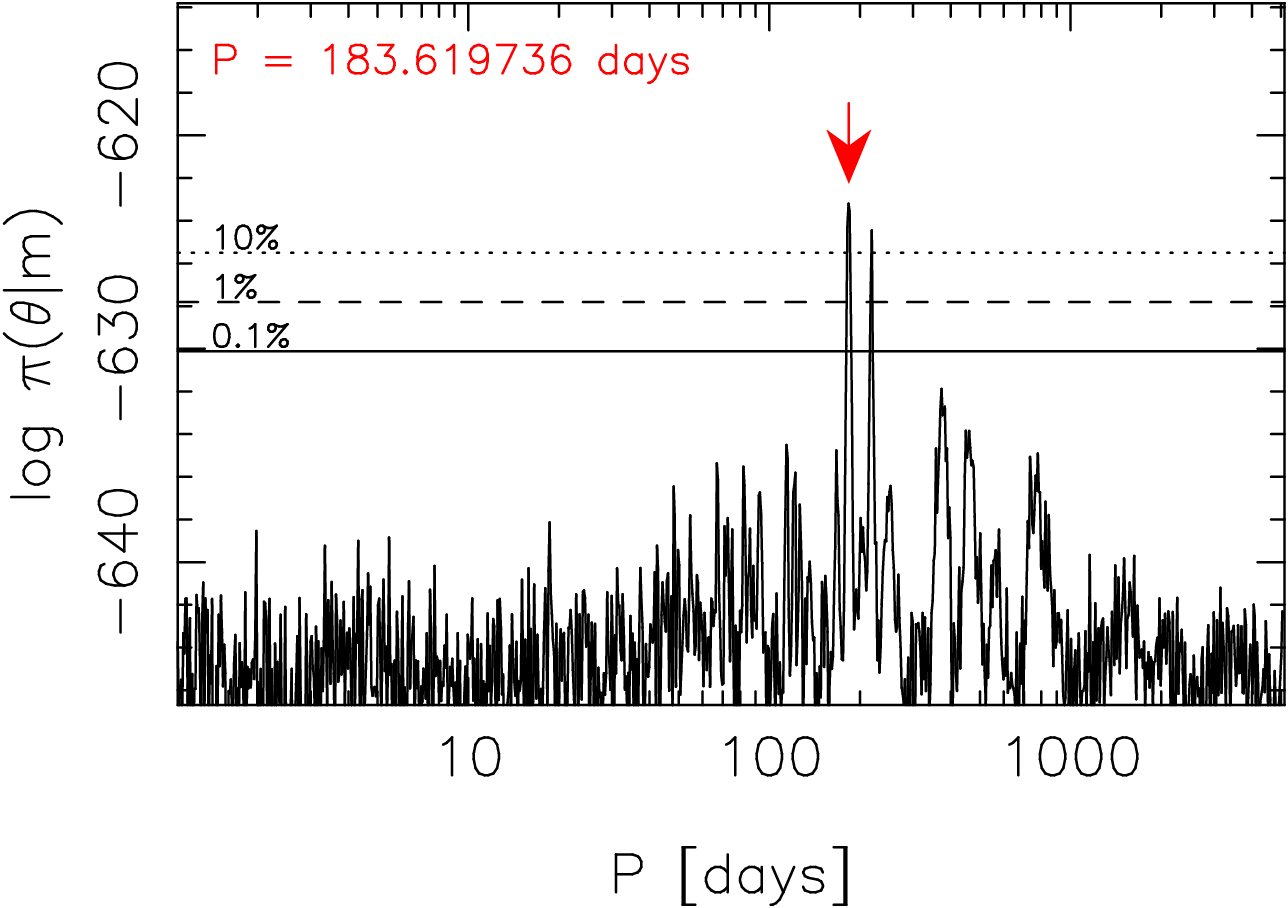}
\caption{Estimated posterior probability densities given GJ628 radial velocities as functions of the period of the $k$th signal of a model with $k$ Keplerian signals. From top to bottom: $k=1, 2, 3$. Red arrows indicate the locations of the global probability maxima in the period space and the horizontal lines denote the 10\% (dotted), 1\% (dashed), and 0.1\% (solid) equiprobability contours with respect to the maxima.}\label{fig:GJ628_psearch}
\end{figure}

\begin{figure}
\center
\includegraphics[angle=270, width=0.49\textwidth]{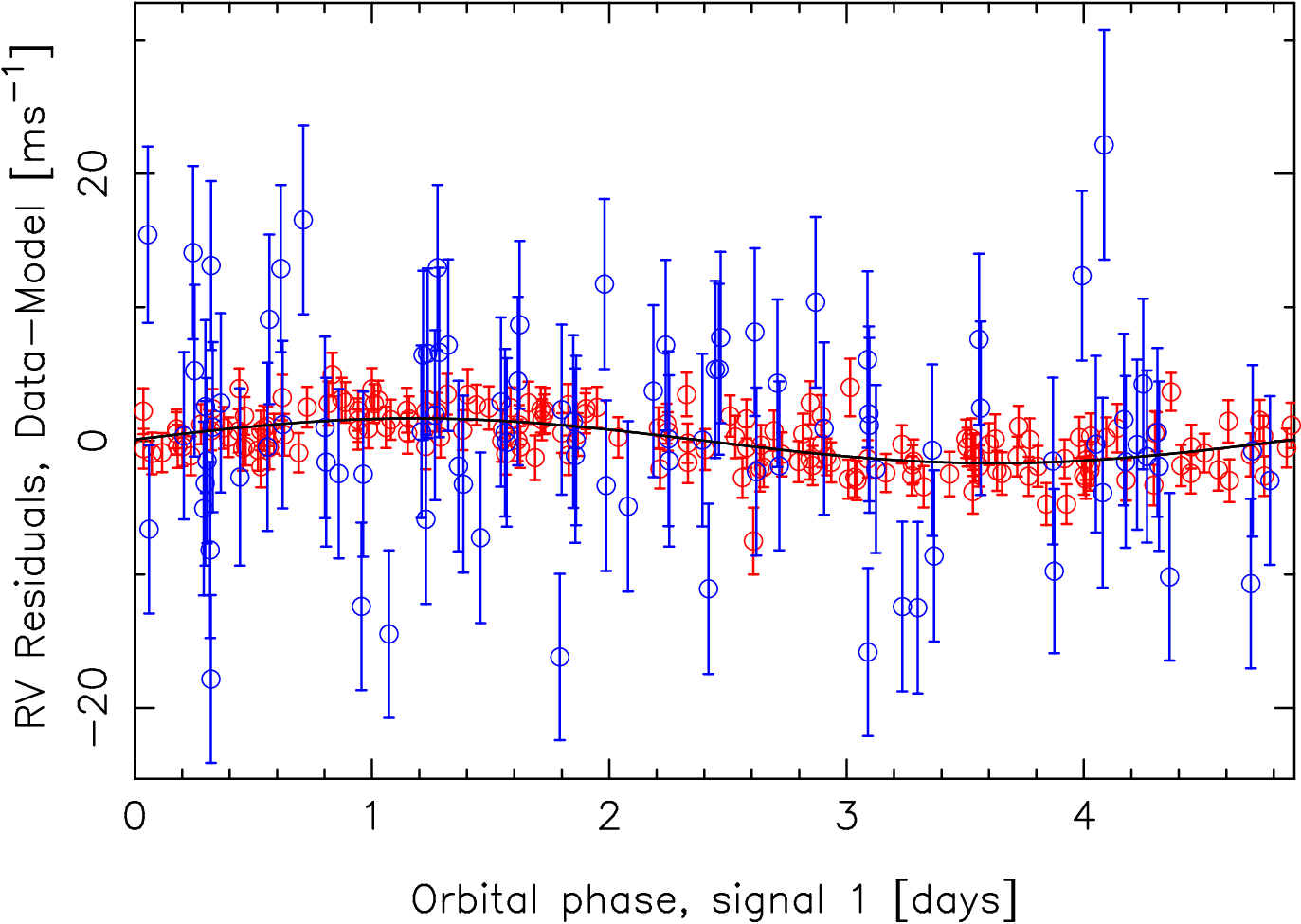}
\includegraphics[angle=270, width=0.49\textwidth]{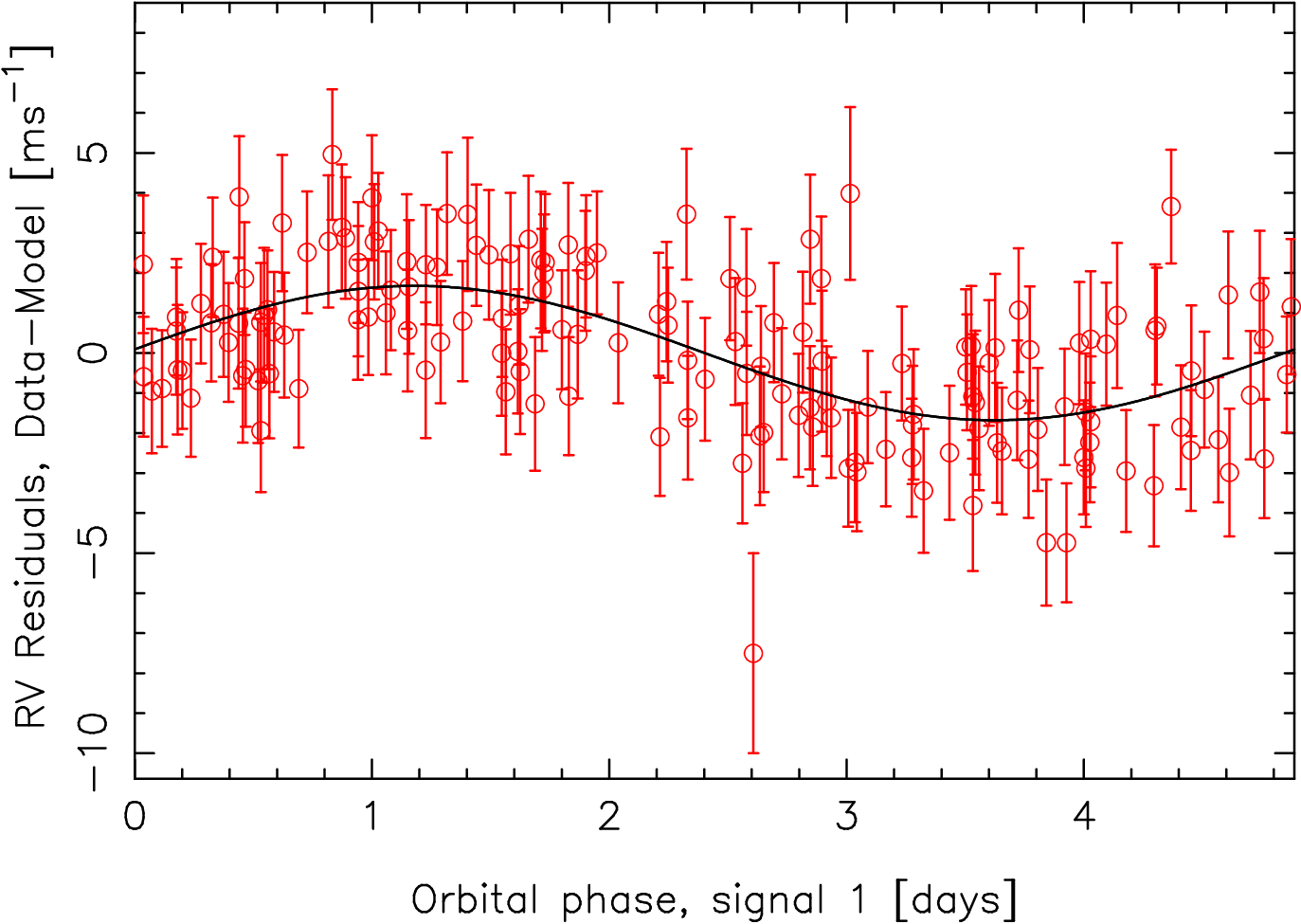}

\includegraphics[angle=270, width=0.49\textwidth]{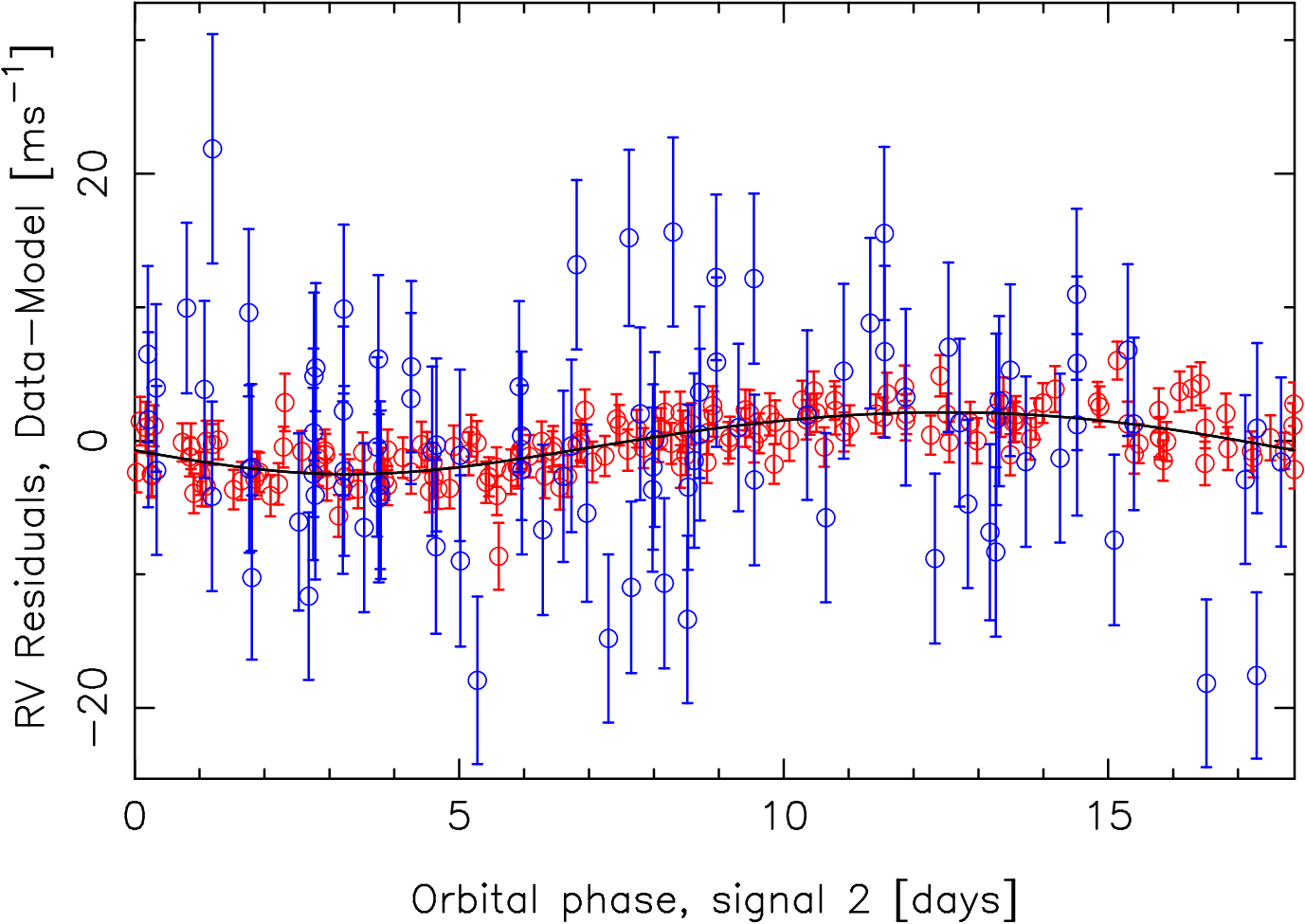}
\includegraphics[angle=270, width=0.49\textwidth]{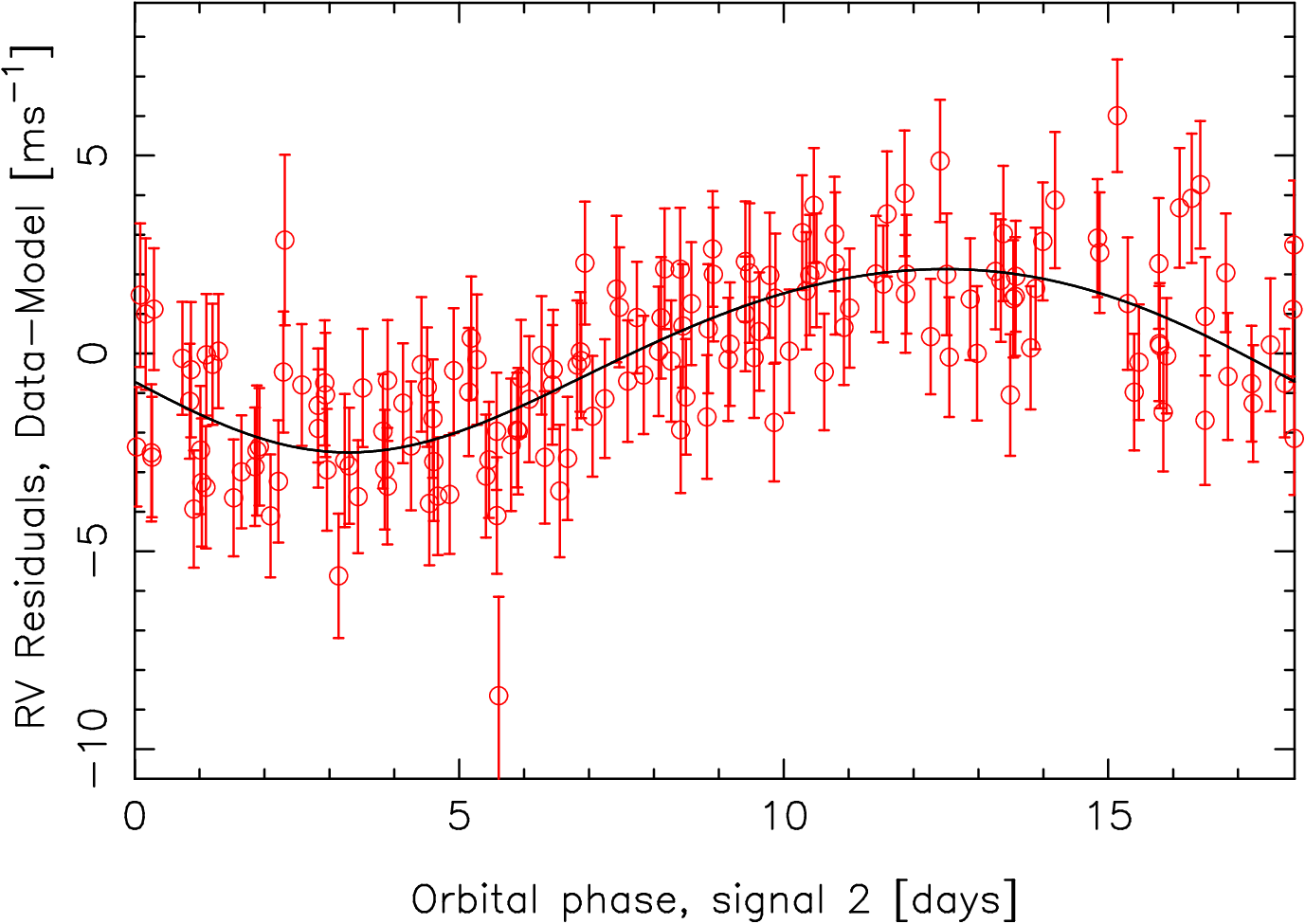}
\
\includegraphics[angle=270, width=0.49\textwidth]{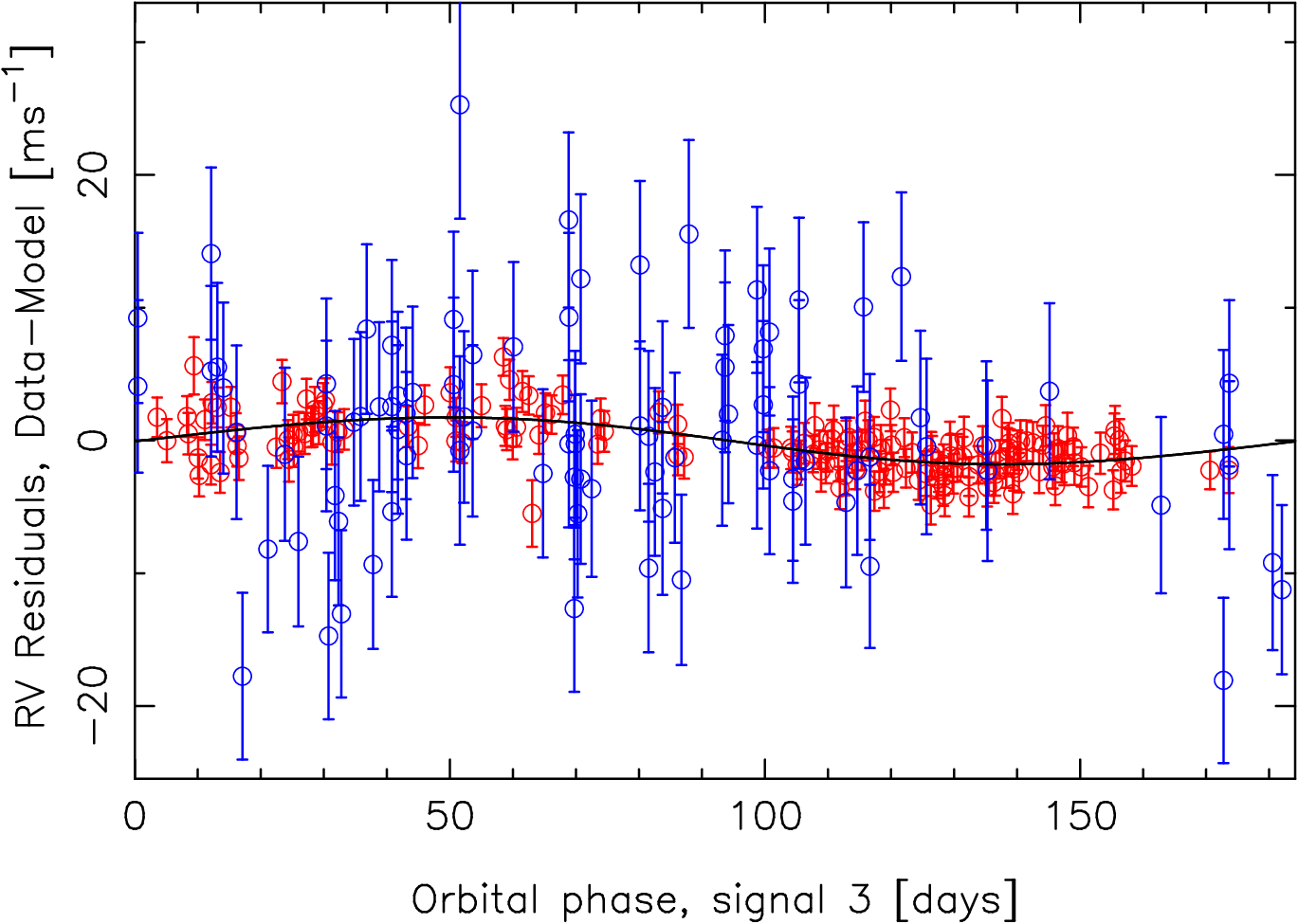}
\includegraphics[angle=270, width=0.49\textwidth]{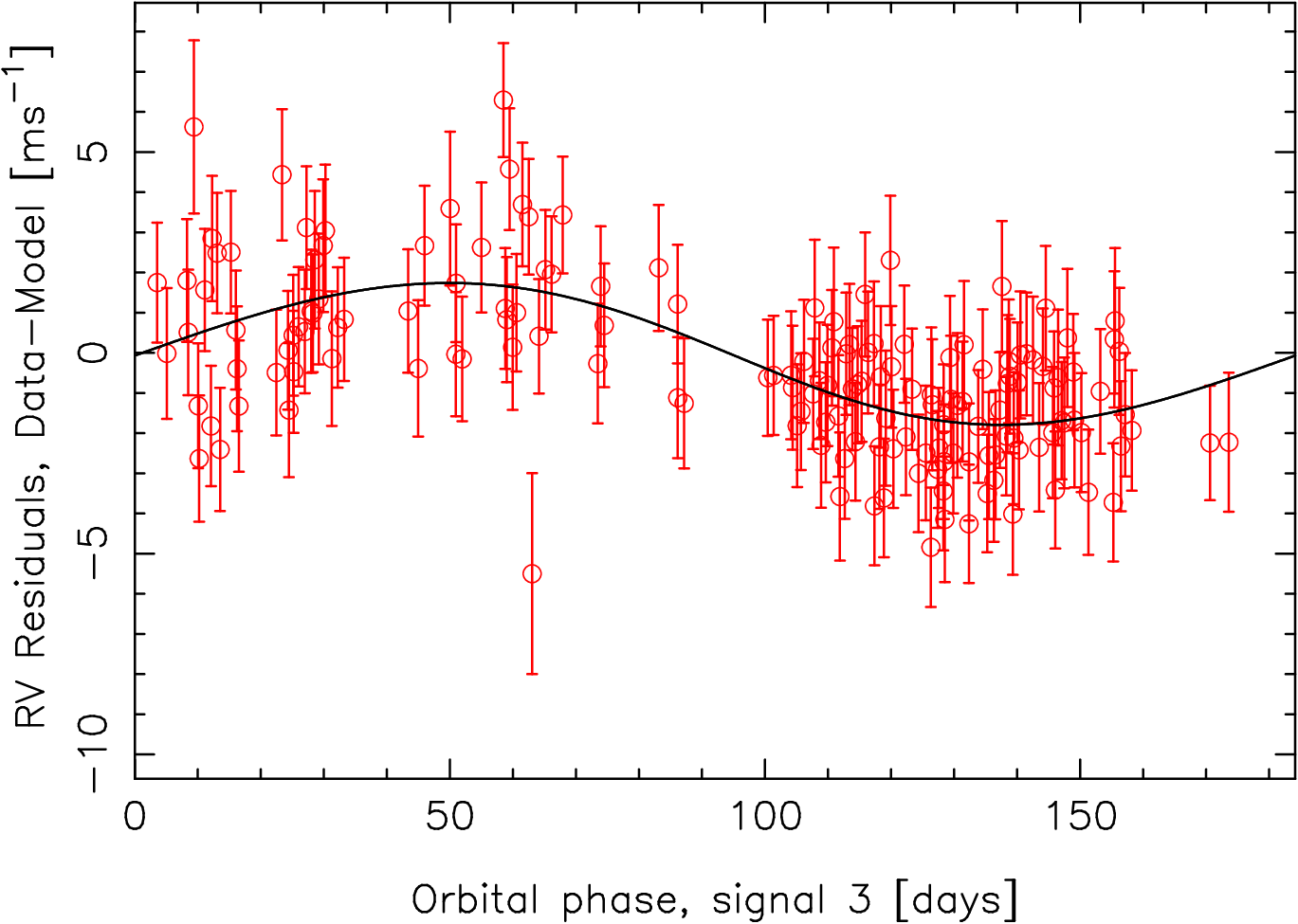}
\caption{Phase-folded signals in the combined HARPS (red) and HIRES (blue) radial velocity data of GJ 628. The solid curves denote the MAP estimated Keplerian curves. The panels on the right hand side show the same phase-folded curves but with the HARPS velocities alone.}\label{fig:GJ628_curves}
\end{figure}

In combination with the 162 high-precision HARPS velocities, the 93 HIRES radial velocities reveal the existence of three significant periodic signals in the radial velocity movement of GJ 628. Two of these signals are unique and detected according to our criteria at periods of 4.88693 [4.88591, 4.88838] and  17.8703 [17.8547, 17.8859], whereas the third was present as a bimodal probability maximum with its highest mode at a period of 183.95 [181.14, 187.14] days. The amplitudes of these signals were estimated to be 1.65 [1.18, 2.17], 2.07 [1.30, 2.84] and 1.70 [0.82, 2.58] ms$^{-1}$, respectively (Figs. \ref{fig:GJ628_curves} and \ref{fig:GJ628_psearch}).

The probability maximum corresponding to the signal at a period of 184 days is not as unique as the others because there is a local maximum in the period space at a period of 220 days also identified by \citet{astudillo2017}. However, this maximum is likely caused by strong sampling frequencies corresponding to periods of 365 and 2000 days and arising from annual gaps in the data as well as a couple of longer gaps in the HARPS data corresponding to observing seasons when no data were taken. Moreover, the local maximum cannot be detected with a model containing four Keplerian signals, which indicates that, together with the global maximum, it is caused by the same underlying periodicity via aliasing. We could not find any evidence in favour of the 67.27-day signal reported by \citep{wright2016}.

\begin{figure}
\center
\includegraphics[angle=270, width=0.49\textwidth,clip]{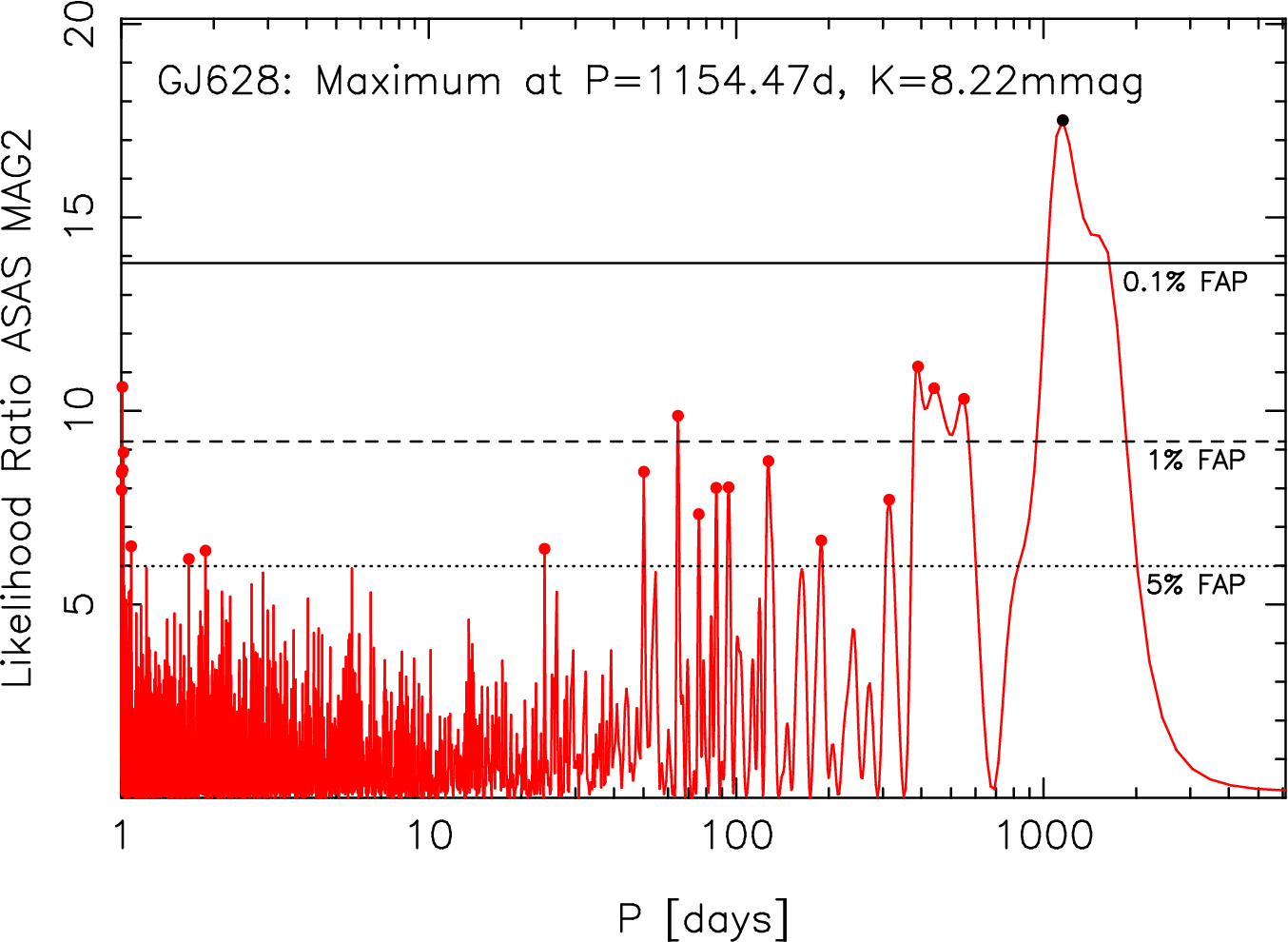}

\includegraphics[angle=270, width=0.49\textwidth,clip]{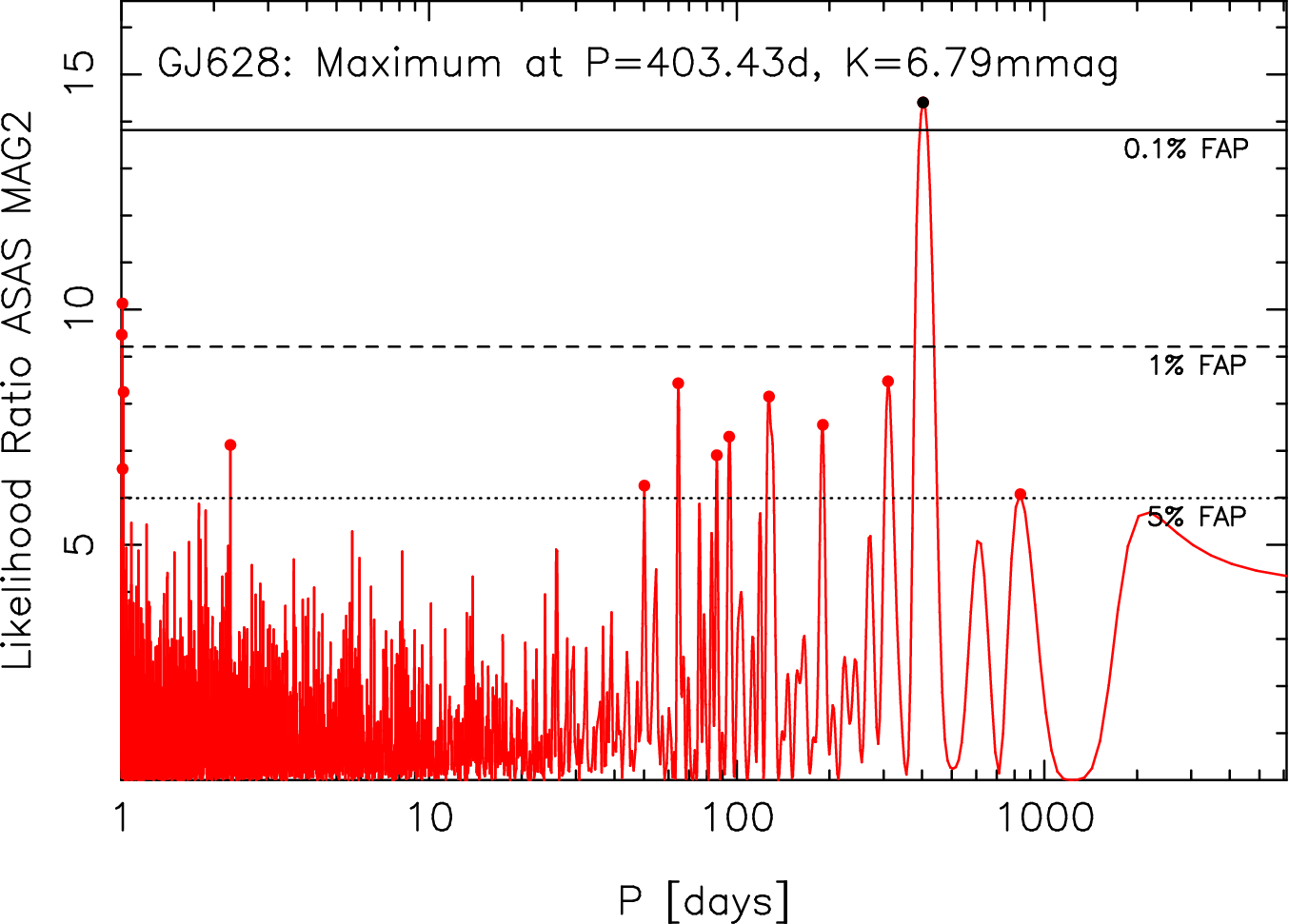}
\caption{Likelihood periodogram of the ASAS V-band photometry measurements of GJ 628 (top panel) and the residual periodogram after subtracting the dominant frequency (bottom panel). The red (black) dots indicate significant likelihood ratios exceeding the 5\% (0.1\%) FAP threshold.}\label{fig:GJ628_asas}
\end{figure}

The ASAS V-band photometry data shows strong periodicities at periods of 1150 and 400 days (Fig. \ref{fig:GJ628_asas}). However, they do not appear to be connected to any of the signals in the radial velocities. Moreover, apart from a suggestive signal at a period of 297 days in the HARPS S-index that only barely exceeded the 1\% FAP, we could not observe any periodic signals in the HARPS and HIRES activity indicators. The radial velocity variations, periodic and otherwise, were also found to be independent of the variations in the activity indicators, which indicates that the three signals in the velocity data are likely caused by candidate planets orbiting the star.

The three candidate planets with minimum masses of 1.9 [1.2, 2.7], 3.6 [2.2, 5.4], and 6.5 [3.1, 10.3] M$_{\oplus}$, respectively, enables us to classify them as a hot super-Earth, a warm super-Earth \citep[in accordance with][]{kane2017}, and a cool Mini-Neptune.

\clearpage

\subsection{GJ 649}

\begin{figure}
\center
\includegraphics[angle=270, width=0.49\textwidth,clip]{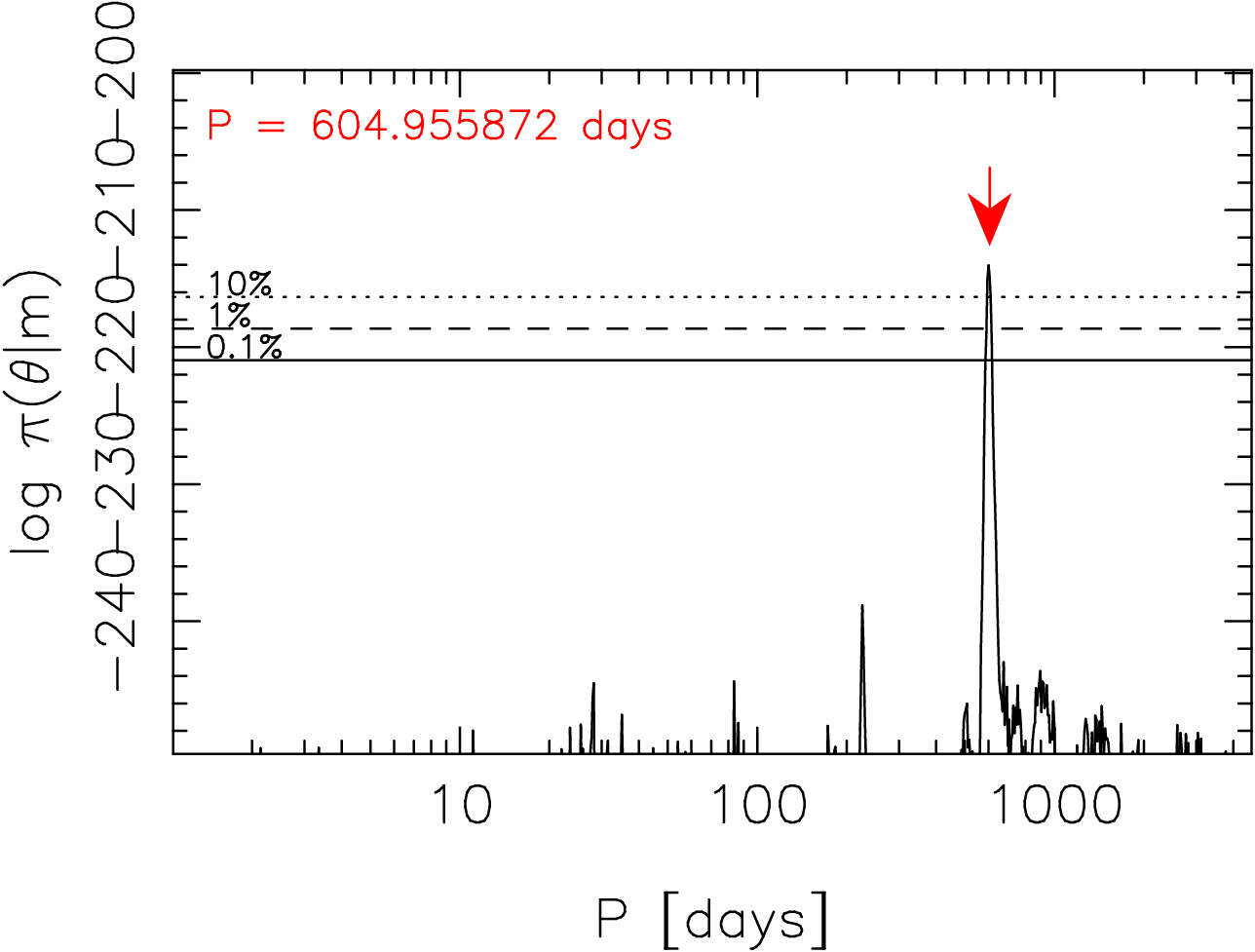}

\includegraphics[angle=270, width=0.49\textwidth,clip]{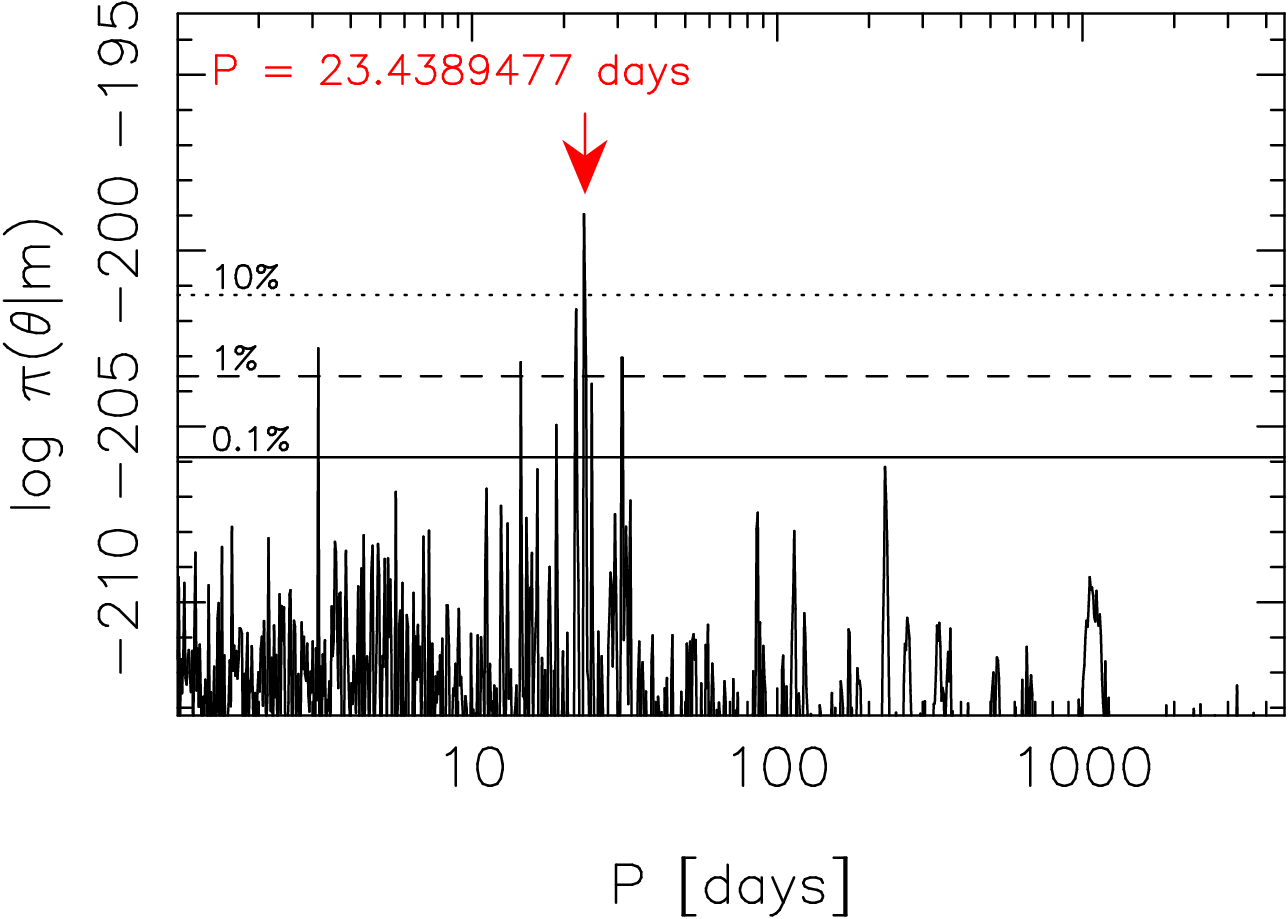}
\caption{Estimated posterior probability densities given GJ 649 dradial velocity data of the one- (top panel) and two-Keplerian (bottom panel) models as functions of the period parameters of the first and second Keplerian signals, respectively. The red arrow denotes the position of the global maximum of the posterior density and the horizontal lines denote the 10\% (dotted), 1\% (dashed) and 0.1\% (solid) equiprobability thresholds with respect to the maximum.}\label{fig:GJ649_psearch}
\end{figure}

According to \citet{johnson2010}, GJ 649 (HIP 83043) is a host to a Saturn-mass planet orbiting the star with an orbital period of 1.638$\pm$0.011 years. Their report was based on 44 HIRES radial velocities taken before JD 2455112. We could verify the existence of this planet candidate based on our updated set of 75 HIRES radial velocities (Fig. \ref{fig:GJ649_psearch}).

\citet{johnson2010} reported a rotation period of approximately 24.8$\pm$1.0 days based on their photometric monitoring of the star. Although we could not verify this result based on ASAS V-band photometry that did not show evidence in favour of periodicities, it appears that this is indeed the stellar rotation period as there is a clear counterpart in the HIRES radial velocities as seen in Fig. \ref{fig:GJ649_psearch} (bottom panel). It is noteworthy that we observed a probability maximum at a period of 23.4 days in the posterior density of the second period parameter of the two-Keplerian model but it is not a unique one as a Keplerian signal should. We did not observe any periodicities in the HIRES S-indices.

\clearpage

\subsection{GJ 674}

A planet with a minimum mass of 11.09 M$_{\oplus}$ has been reported orbiting GJ 674 (HIP 85523) with an orbital period of 4.6938$\pm$0.007 days \citep{bonfils2007}. The discoverers also noted that they detected another periodic signal in the HARPS velocities at a period of $\sim$ 35 days that corresponded to a quasiperiodic variability around the same period in the photometric data. \citet{bonfils2007} thus interpreted this signal as being caused by stellar activity. \citet{kiraga2007} also reported a periodicity of 33.29 days in the ASAS data, which indicates that the star's rotation period is indeed at or near this period.

\begin{figure}
\center
\includegraphics[angle=270, width=0.49\textwidth,clip]{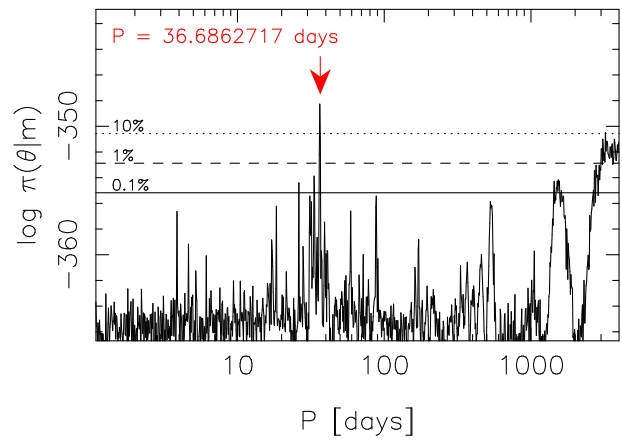}
\caption{Estimated posterior density of the period parameter of the second signal in a model with two Keplerian signals given the HARPS radial velocities of GJ 674. The red arrow indicates the maximum and the horizontal lines denote the 10\% (dotted), 1\% (dashed), and 0.1\% (solid) probability thresholds with respect to this maximum.}\label{fig:GJ674_period_search}
\end{figure}

\begin{figure}
\center
\includegraphics[angle=270, width=0.49\textwidth,clip]{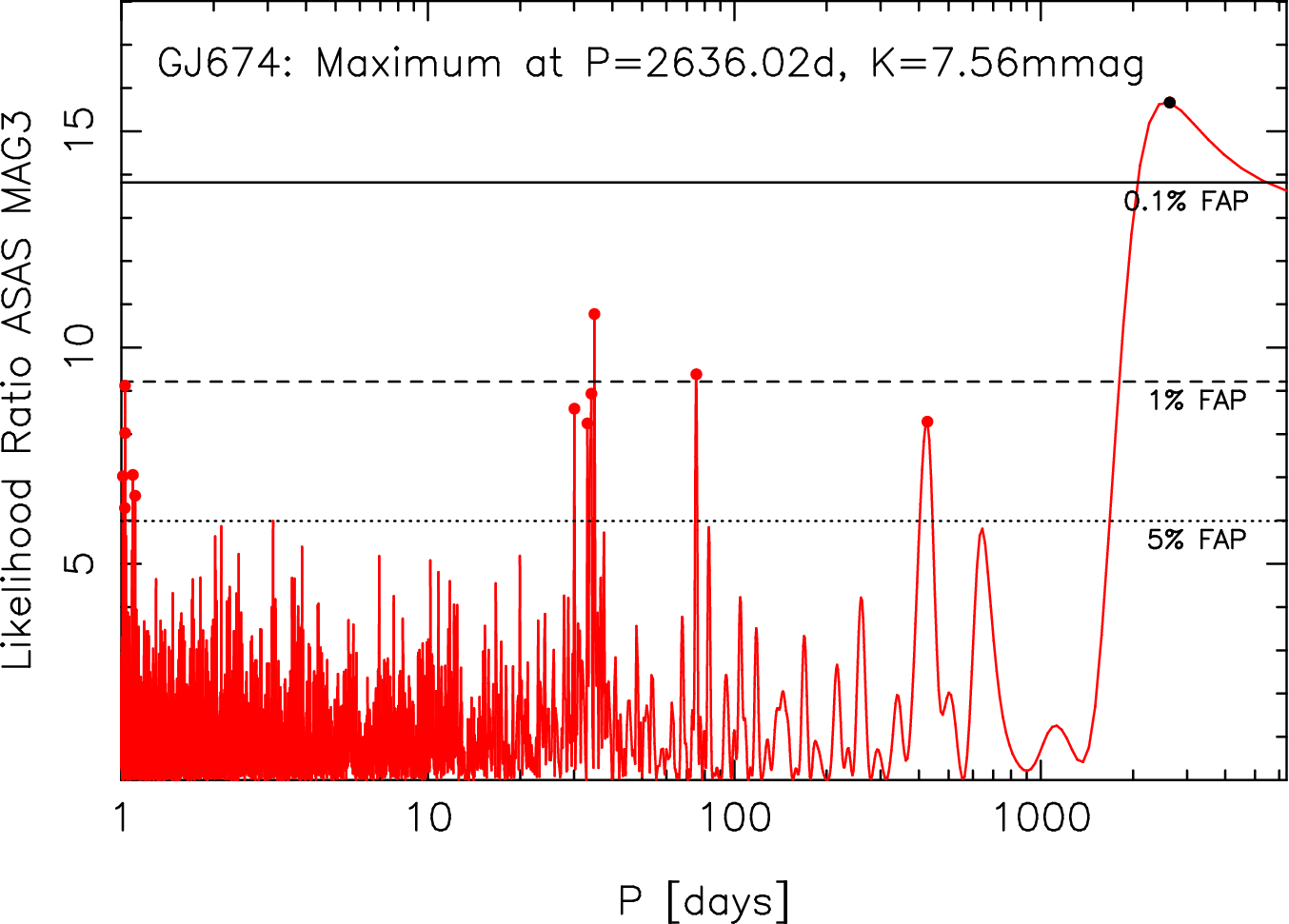}

\includegraphics[angle=270, width=0.49\textwidth,clip]{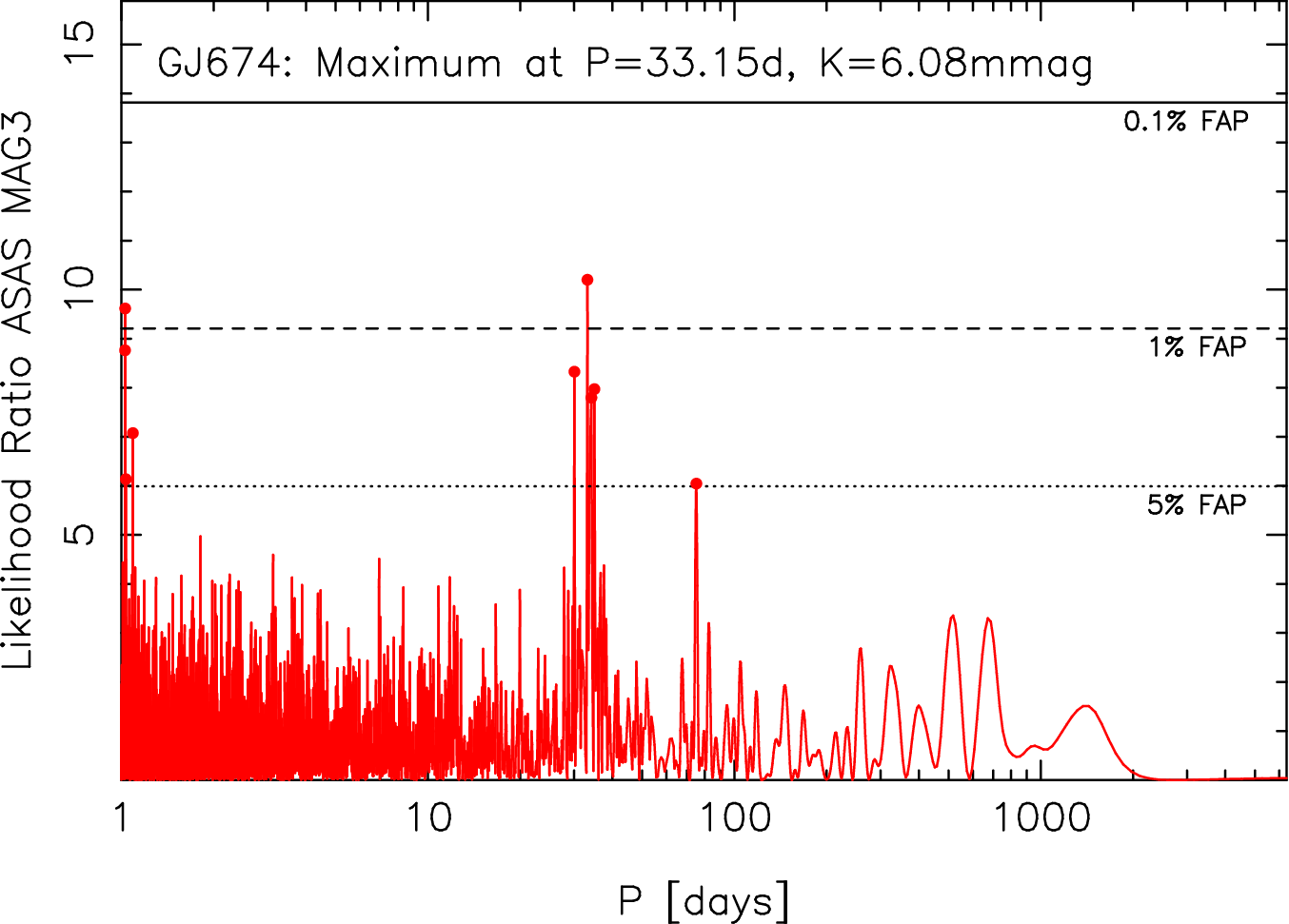}
\caption{Likelihood periodograms of ASAS photometry data of GJ 674 (top panel) and the residual periodogram after subtracting the most significant frequency (bottom panel). The red filled circles denote maxima exceeding the 5\% FAP threshold.}\label{fig:GJ674_asas}
\end{figure}

We also observed a signal at a period of 36.653 [36.581, 36.718] days in the HARPS radial velocities (Fig. \ref{fig:GJ674_period_search}) as well as a somwhat inconclusive photometric counterpart at a very nearby period (Fig. \ref{fig:GJ674_asas}) making it appear likely that the two are indeed connected. We thus interpret the second signal as a signature of stellar rotation coupled with activity. We did not observe any significant correlations between the HARPS velocities and activity indicators.

It is worth noting that we did not see any probability maxima in the vicinity of a period of 25 days, where \citet{bonfils2013} reported a power excess when looking at their periodograms of the HARPS data. However, the candidate planet GJ 674 b with an orbital period of 4.69502 [4.69488, 4.69519] days has an eccentricity of 0.226 [0.181, 0.271] even though its orbit should have been circularized due to tidal forces of the primary. It is thus likely that there is another planetary companion at a nearby orbit disturbing GJ 674 b gravitationally giving rise to non-zero eccentricity. We did not obtain any evidence for such a companion.

\clearpage

\subsection{GJ 676A}\label{sec:GJ676A}

\begin{figure}
\center
\includegraphics[angle=270, width=0.49\textwidth,clip]{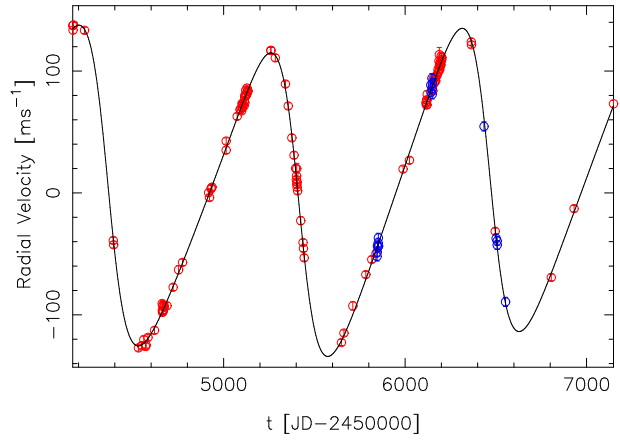}
\caption{Dominant variations in the GJ 676A combined HARPS (red) and PFS (blue) radial velocities corresponding to a superposition of two Keplerian signals interpreted by \citet{forveille2009} and \citet{anglada2012c} as giant planets orbiting the star.}\label{fig:GJ676A_long_signals}
\end{figure}

GJ 676A (HIP 85647) is reportedly a host to two long-period giant planets and two inner super-Earths orbiting the star \citep{forveille2011,anglada2012c}. With our updated HARPS and PFS data sets and the signal detection techniques based on DRAM samplings, we could detect the corresponding signals in a straightforward manner. The variations corresponding to the two long-period planets are shown for reference in Fig. \ref{fig:GJ676A_long_signals}. We note that there is also a strong linear trend in the data of 20.02 [19.09, 21.03] ms$^{-1}$year$^{-1}$ likely corresponding to the gravitational pull of a stellar companion.

\begin{figure}
\center
\includegraphics[angle=270, width=0.49\textwidth,clip]{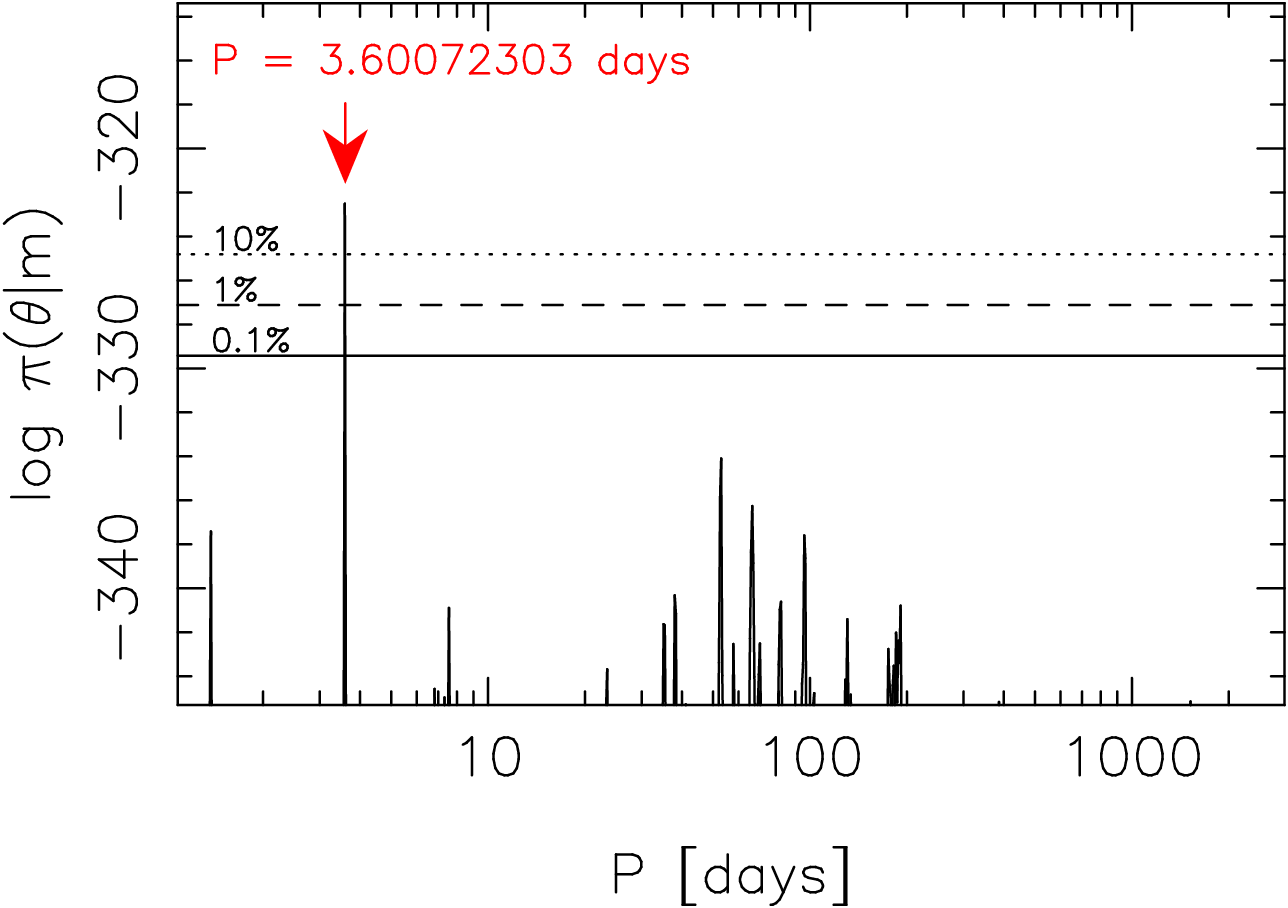}

\includegraphics[angle=270, width=0.49\textwidth,clip]{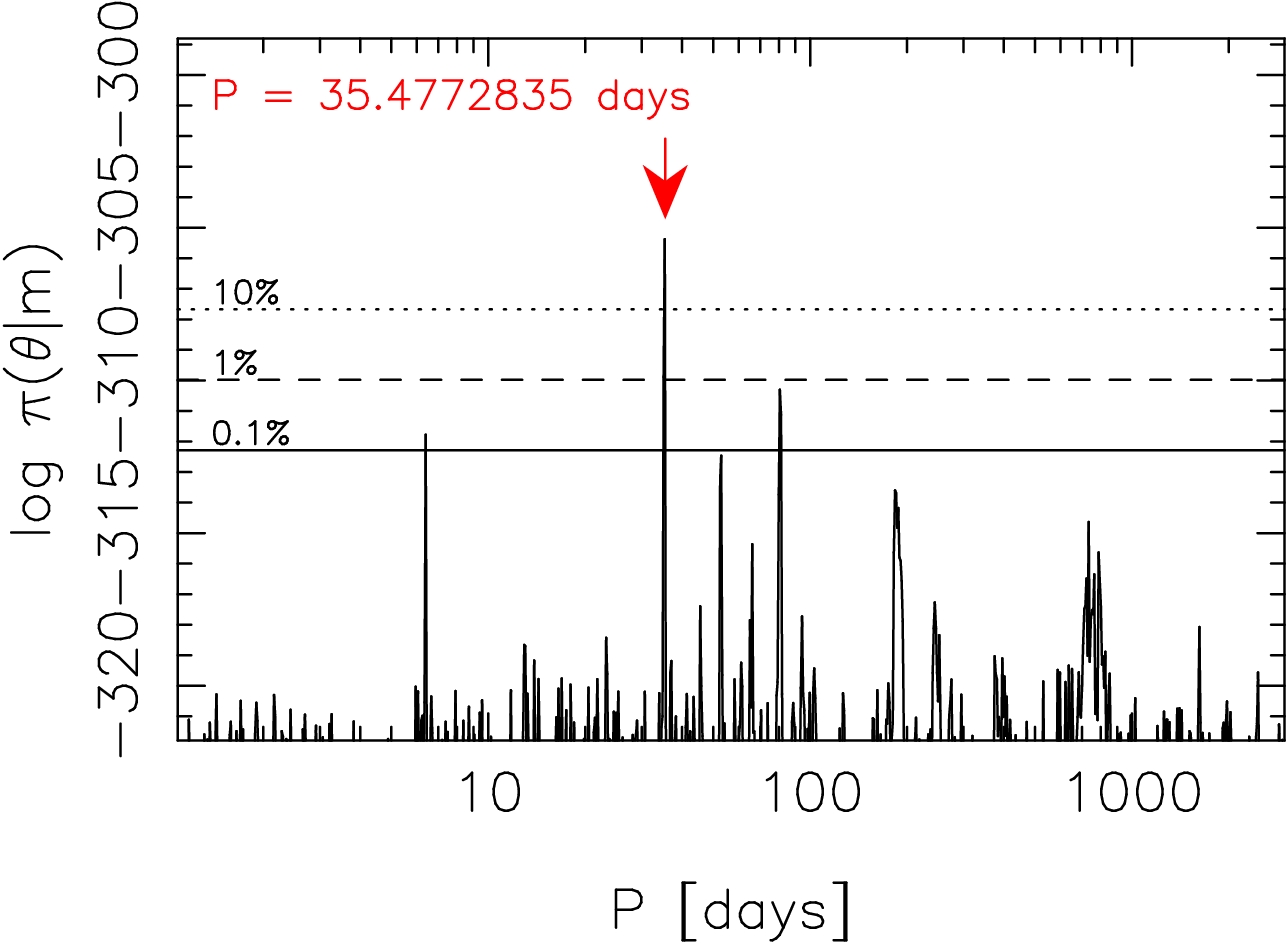}
\caption{Estimated posterior probability densities of the models with $k$ Keplerian signals as functions of the period of the $k$th signal given GJ 676A radia lvelocities. From top to bottom: $k = 3,4$. The red arrows indicate the locations of the global maxima in the period space and the horizontal lines denote the 10\% (dotted), 1\% (dashed), and 0.1\% (solid) equiprobability thresholds with respect to the maxima.}\label{fig:GJ676A_psearch}
\end{figure}

\begin{figure}
\center
\includegraphics[angle=270, width=0.49\textwidth,clip]{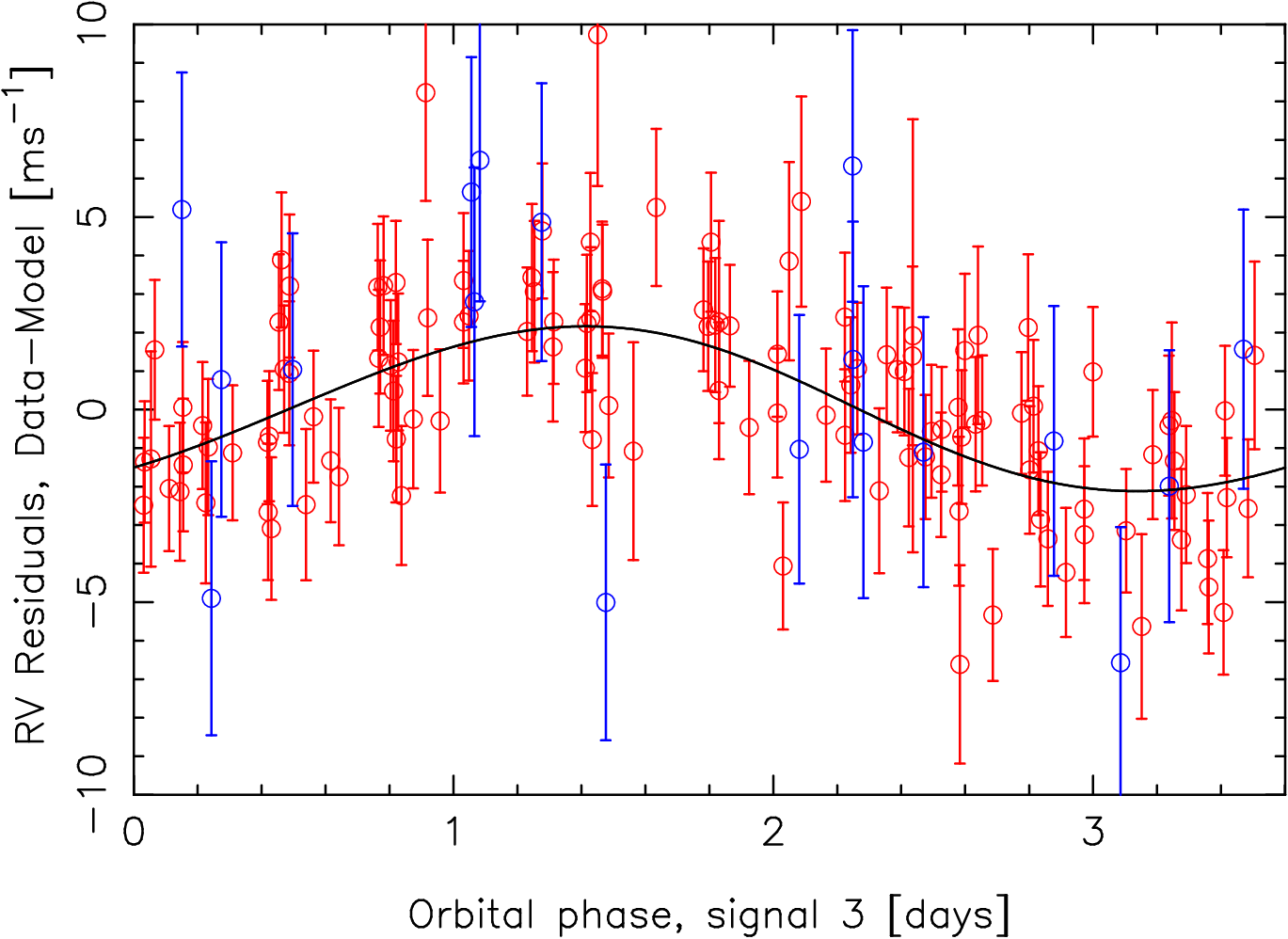}
\includegraphics[angle=270, width=0.49\textwidth,clip]{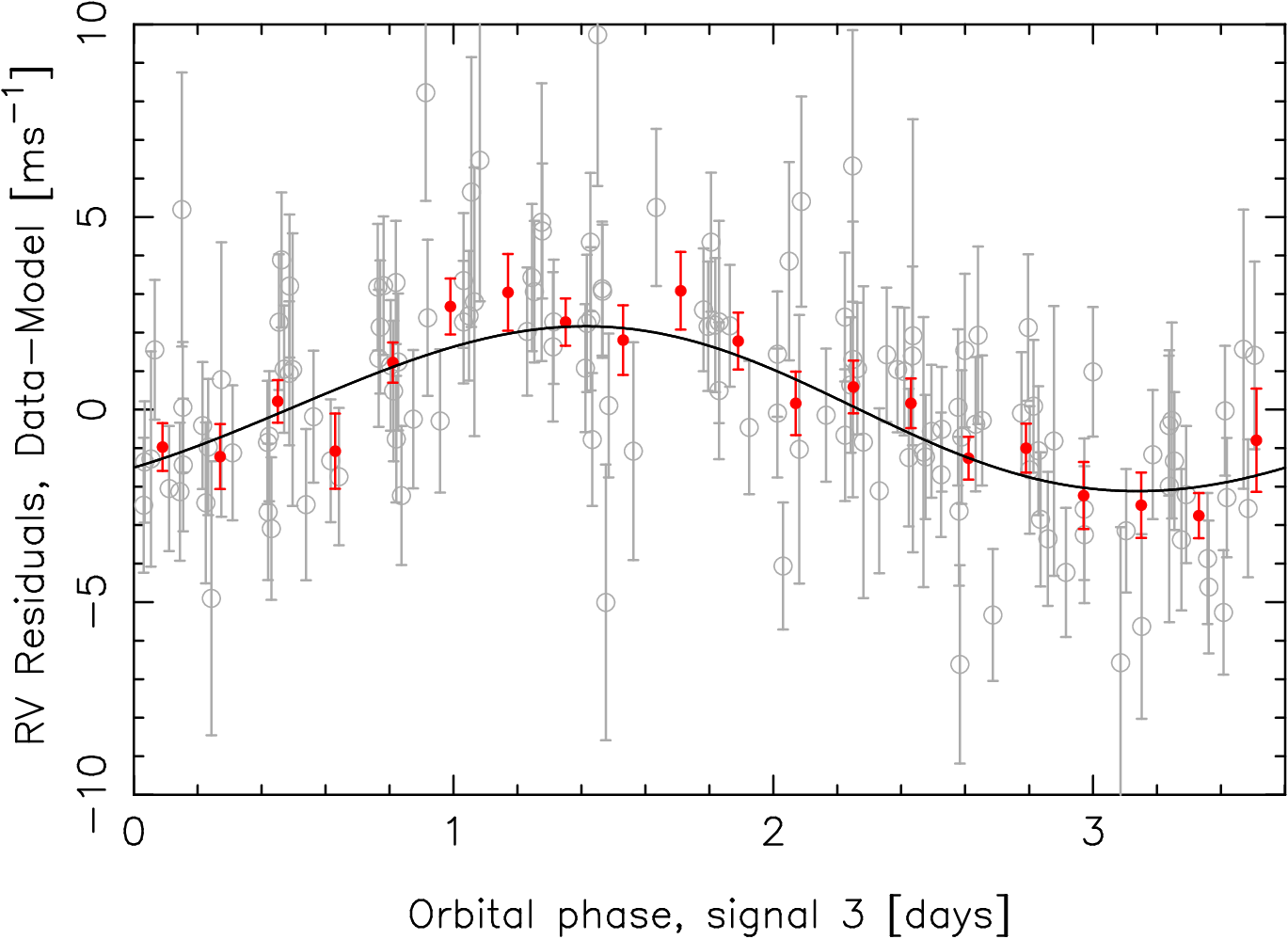}

\includegraphics[angle=270, width=0.49\textwidth,clip]{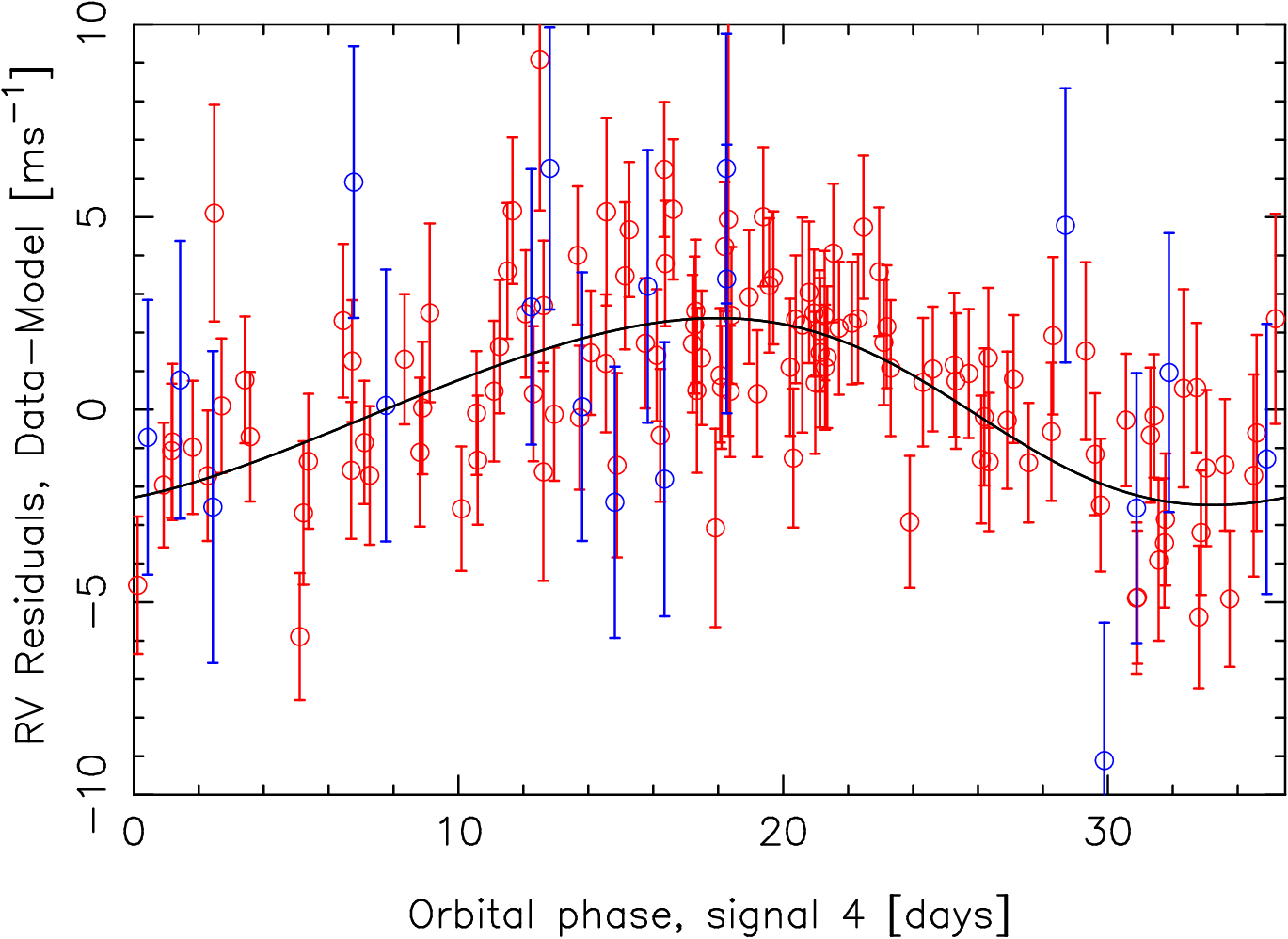}
\includegraphics[angle=270, width=0.49\textwidth,clip]{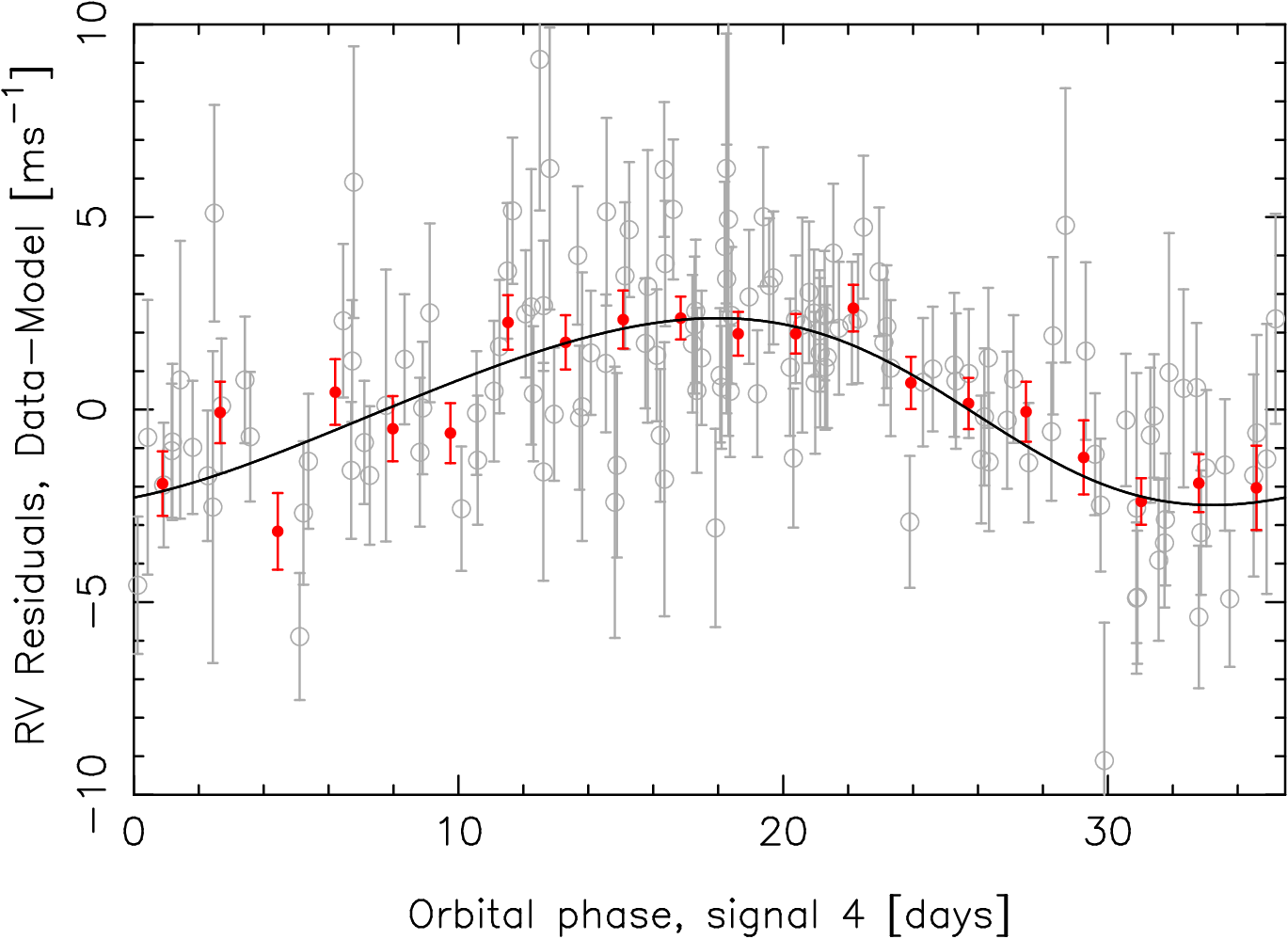}
\caption{HARPS (red) and PFS (blue) radial velocities folded on the phases of the two short-period signals of GJ 676A with all the other signals subtracted from each panel. Left panels show the same phase-folded velocities binned into 20 bins in the phase-space (red filled circles).}\label{fig:GJ676A_signals}
\end{figure}

With our statistical model that is more general than the one applied by \citet{anglada2012c} in terms of accounting for intrinsic correlations and dependence of the radial velocities on activity indices, we could replicate their results and obtained evidence for the two short-period signals corresponding to super-Earths orbiting the star. We demonstrate the presence of these signals in the data in Figs. \ref{fig:GJ676A_psearch} and \ref{fig:GJ676A_signals}.

\begin{figure}
\center
\includegraphics[angle=270, width=0.49\textwidth,clip]{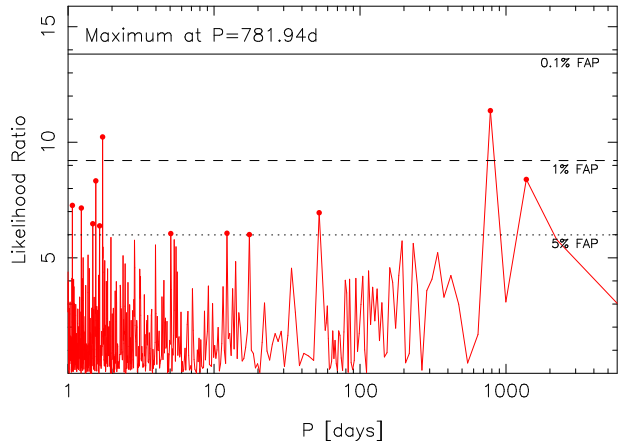}
\caption{Likelihood-ratio periodogram of the HARPS S-indices of GJ 676A.}\label{fig:GJ676A_S}
\end{figure}

\begin{figure}
\center
\includegraphics[angle=270, width=0.49\textwidth,clip]{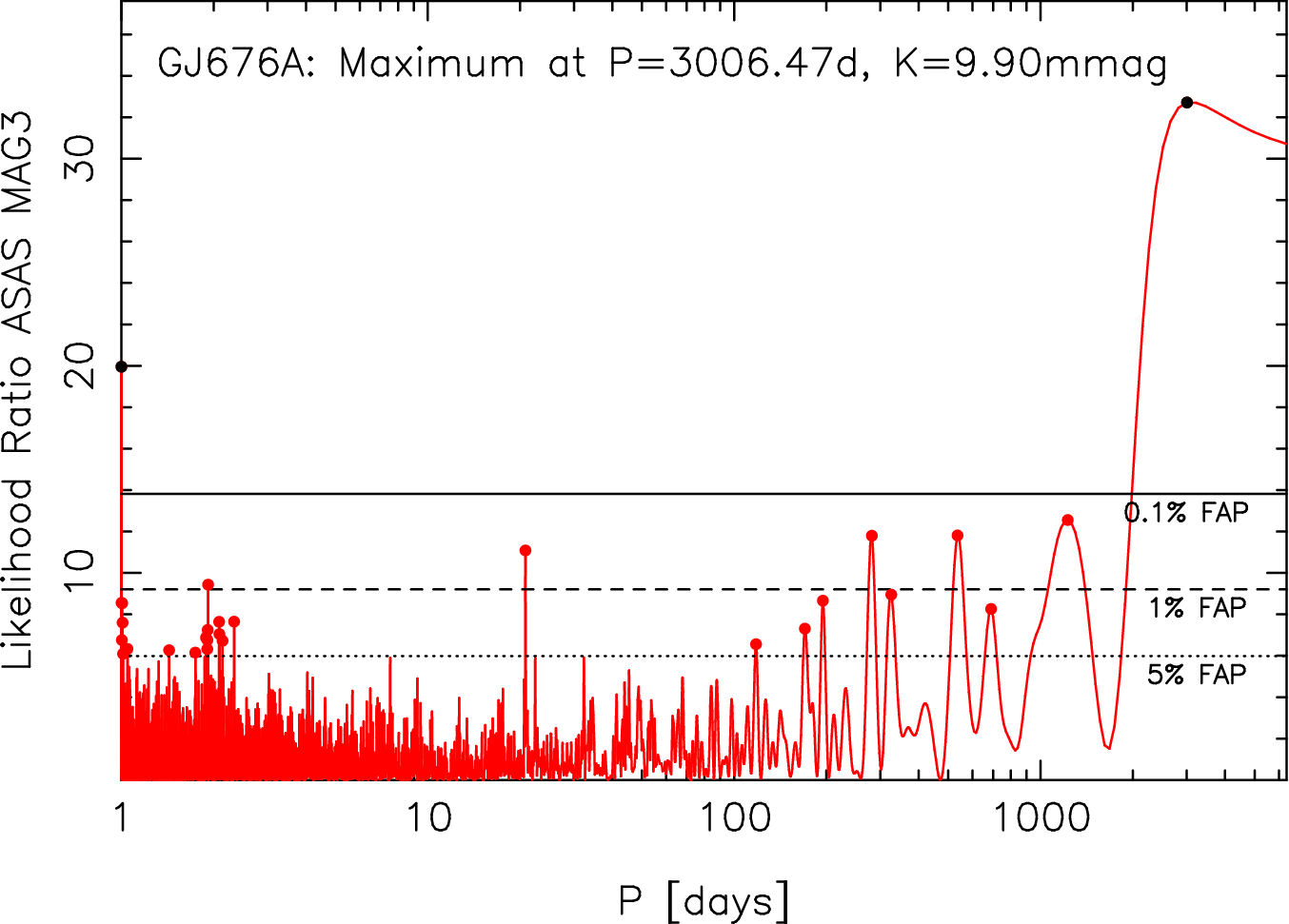}

\includegraphics[angle=270, width=0.49\textwidth,clip]{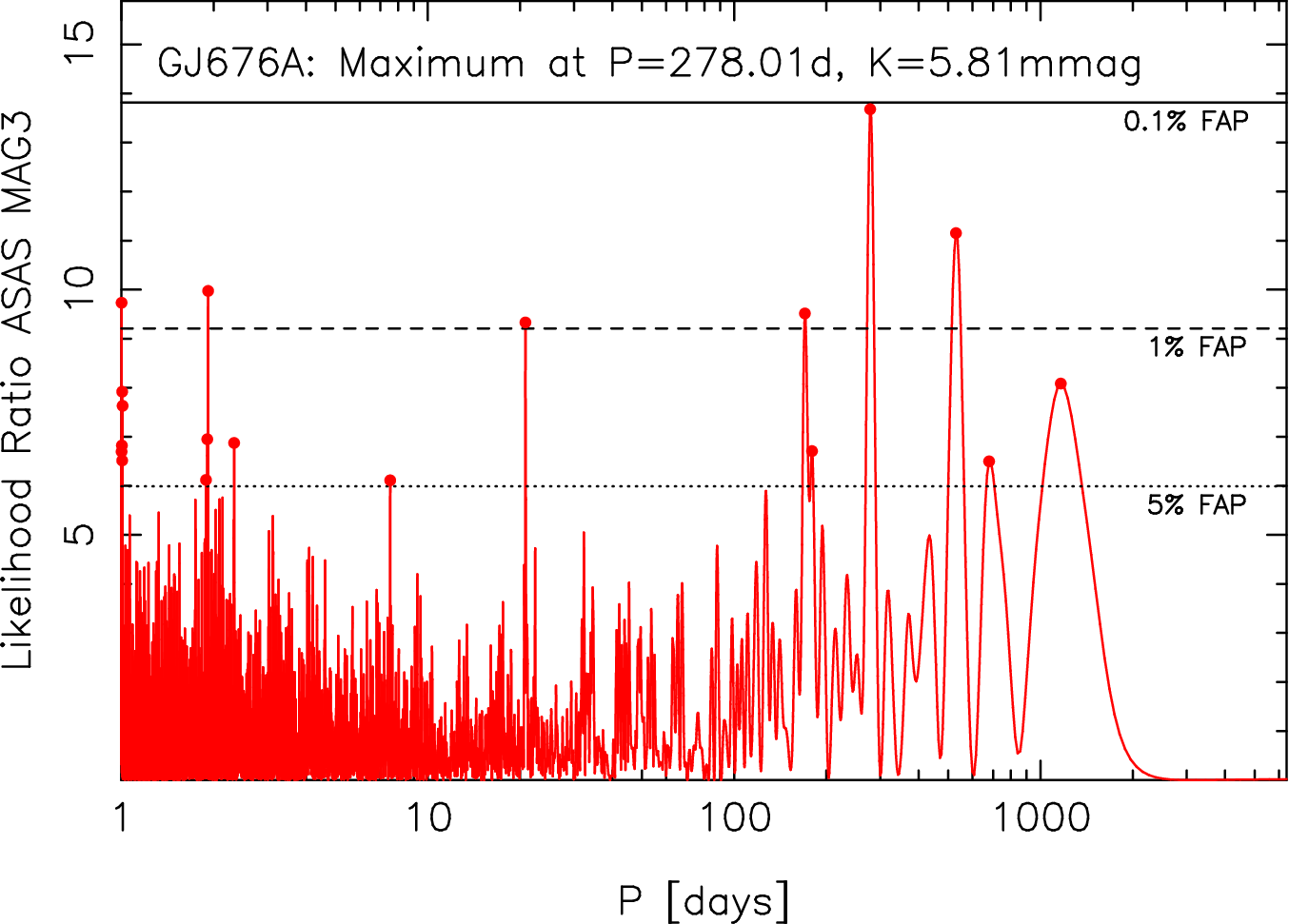}
\caption{Likelihood-ratio periodogram of the ASAS V-band photometry of GJ 676A (top) and the periodogram of residuals after subtracting the dominant periodicity (bottom).}\label{fig:GJ676A_ASAS}
\end{figure}

We observed two periodicities in the HARPS S-indices that exceeded the 1\% FAP in our likelihood-ratio periodogram analysis (Fig. \ref{fig:GJ676A_S}) However, neither of them coincided with the periodic signals in the radial velocities, which suggests that these signals do not arise from stellar periodicities connected to magnetic activity and/or rotation. Moreover, although there was a clear periodic signal in the ASAS V-band photometry (Fig. \ref{fig:GJ676A_ASAS}), as well as several weaker periodicities based on analysis of the residuals, none of them correspond to the signals in the radial velocities. We thus interpret the radial velocity signals as candidate planets orbiting the star.

On top of the two longer periodicities \citep{forveille2011,anglada2012c}, there are thus two low-mass candidate planets orbiting GJ 676A as already postulated by \citet{anglada2012c}. These two signals correspond to a hot super-Earth and a hot mini-Neptune with orbital periods of 3.60 and 35.3 days, respectively.

\clearpage

\subsection{GJ 682}

GJ 682 (HIP 86214) is an interesting target because \citet{tuomi2014} reported, based on radial velocity data from HARPS and UVES, that there is evidence for two candidate planets orbiting it on orbits with periods of 17.478 [17.438, 17.540] and 57.32 [56.84, 57.77] days, respectively. However, with an improved statistical model and new HARPS radial velocities, we could not reach the same conclusion.

\begin{figure}
\center
\includegraphics[angle=270, width=0.49\textwidth,clip]{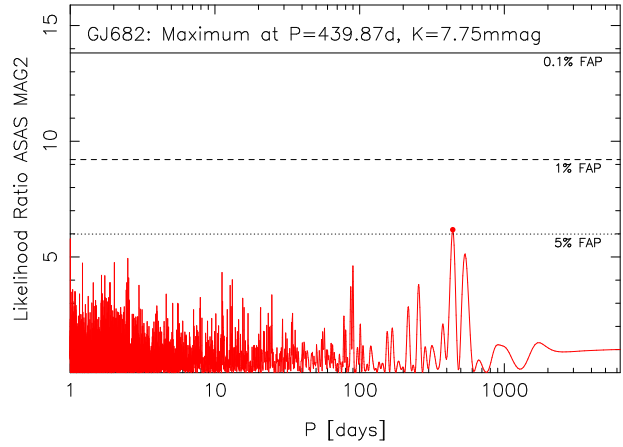}
\caption{Likelihood-ratio periodogram of the ASAS V-band photometry of GJ 682.}\label{fig:GJ682_asas}
\end{figure}

First of all, the ASAS V-band photometry of the star appears to lack photometric variations (Fig. \ref{fig:GJ682_asas}) apart from a barely noteworthy (5\% FAP) periodicity at 440 days. However, we did not find any evidence for variations in the combined HARPS and UVES radial velocities at or around the same period. There was also no evidence for connections between the HARPS radial velocities and the corresponding activity indicators.

\begin{figure}
\center
\includegraphics[angle=270, width=0.49\textwidth,clip]{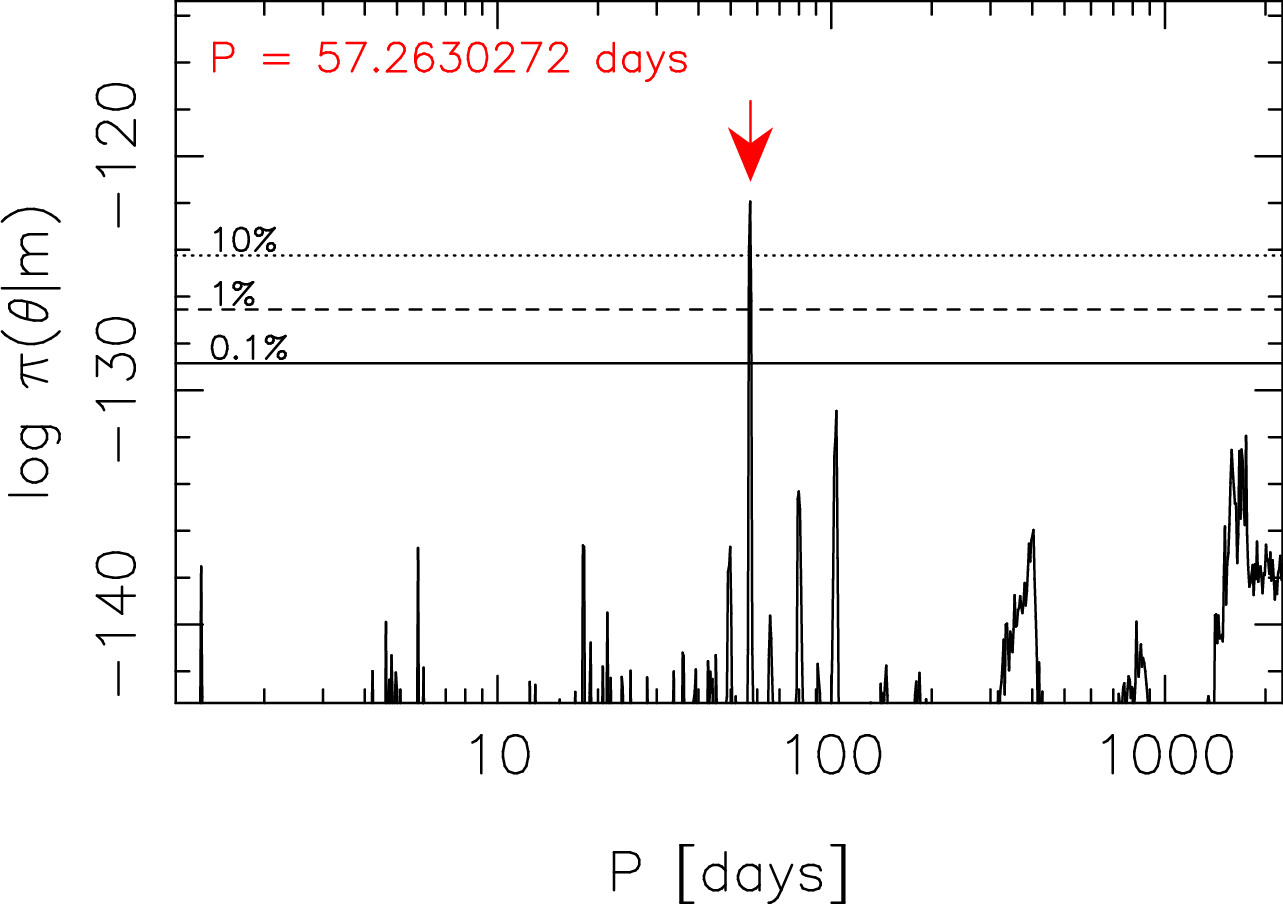}

\includegraphics[angle=270, width=0.49\textwidth,clip]{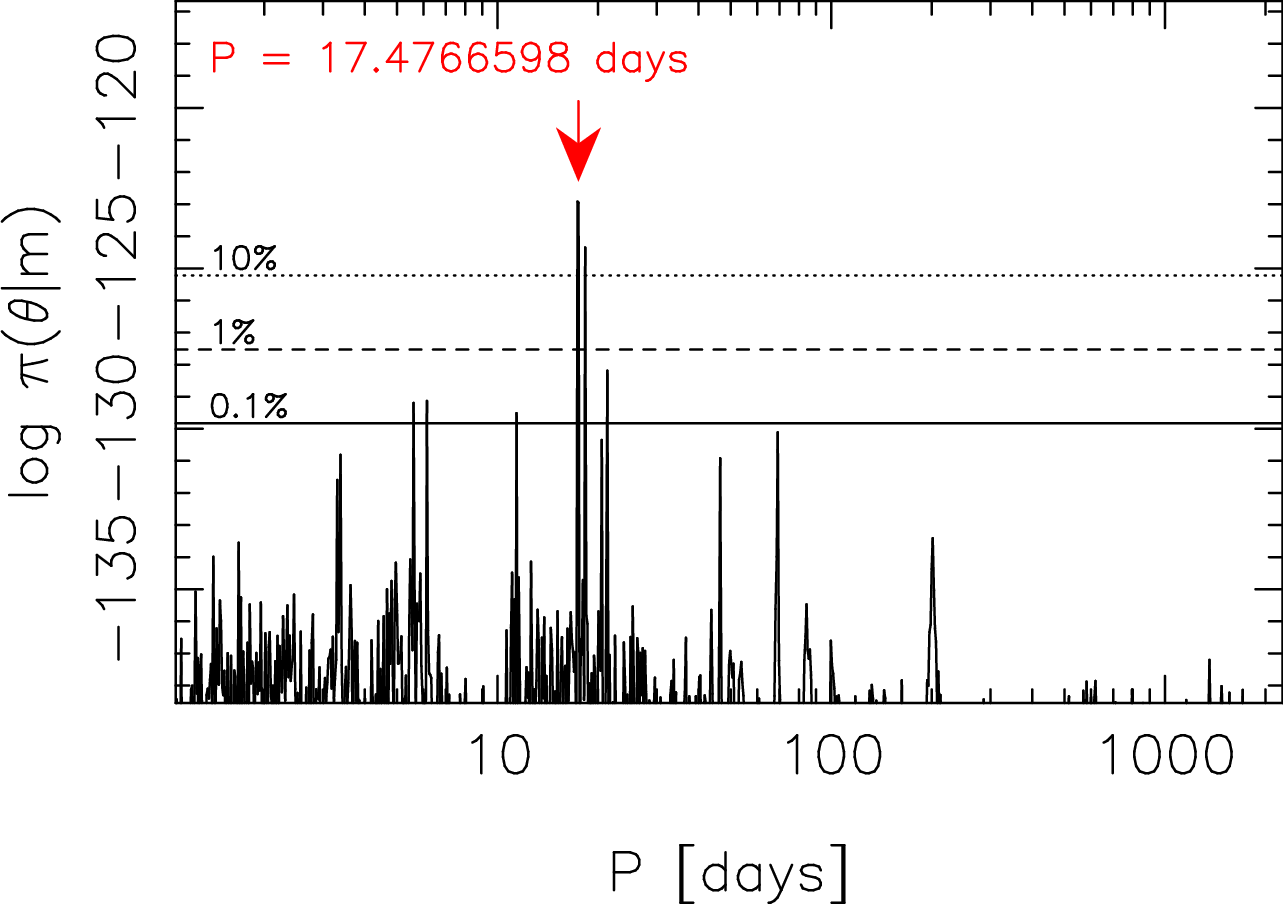}
\caption{Estimated posterior density of the period parameters of the two signals in a model given the HARPS and UVES radial velocities of GJ 682. The red arrow indicates the maximum and the horizontal lines denote the 10\%, 1\%, and 0.1\% probability thresholds with respect to this maximum.}\label{fig:GJ682_period_search}
\end{figure}

Re-analysing the data available to \citet{tuomi2014}, we could detect the same two signals in the combined HARPS and UVES radial velocities (Fig. \ref{fig:GJ682_period_search}). However, as discussed in \citet{tuomi2014}, this could only be done by first estimating the positions of the highest maxima by searching for two signals simultaneously, and then fixing one of the signals to each such maximum at the time in order to search for a second signal given the first one. The rationale behind this approach is to avoid biased results arising from the fact that a one-Keplerian model cannot explain the data much better than a model without signals whereas a model with two Keplerian signals is superior to both due to the fact that the superposition of two Keplerian signals cannot be modelled accurately without them both.

We could not detect the probability maximum corresponding to the aforementioned two signals when analysing the combined UVES and updated HARPS data, with 20 HARPS radial velocities. In fact, we could not identify any signals in the combined data, which suggests that the signals reported by \citet{tuomi2014} were not caused by planets orbiting the star. We postulate four potential explanations for this apparent discrepancy. First, it is possible that additional signals of similar amplitude in the data prevent the detections unless the number of signals in the model matches the number of signals in the data. Second, the signals at periods of 17 and 57 days might have been caused by activity-induced quasiperiodic variability and additional data might then contradict them. Finally, the signals detected by \citet{tuomi2014} could have been spurious ones caused by pure noise. This possibility is rather unlikely as the logarithm of the maximum-likelihood values increase from -22.5 and -124.8 to -8.8 and -112.3 for HARPS and UVES, respectively. Finally, it is possible that some HARPS radial velocities are biased.

\begin{figure}
\center
\includegraphics[angle=270, width=0.49\textwidth,clip]{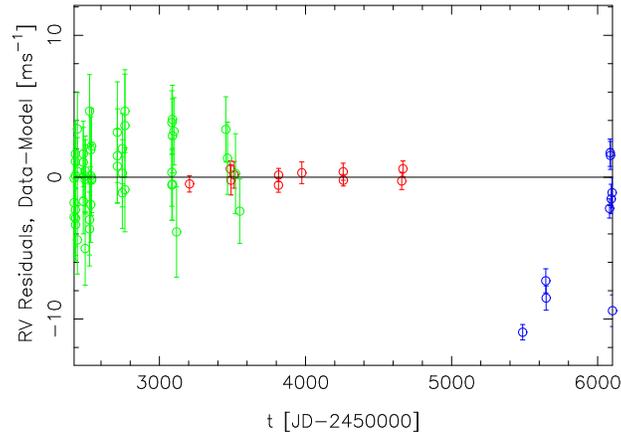}
\caption{Residuals after subtracting the two signals of \citet{tuomi2014} from the UVES (green), old HARPS (red), and new HARPS data (blue) of GJ 682.}\label{fig:GJ682_residuals}
\end{figure}

We further tested whether we could identify some differences between the old HARPS velocities and the new ones that have been made publicly available in the ESO archive after \citet{tuomi2014} published their work. We subtracted the two signals, such as they were reported in \citet{tuomi2014}, from the HARPS and UVES data sets and plotted the remaining residuals in Fig. \ref{fig:GJ682_residuals}. Based on this plot, it is evident that out of the nine new velocities, four are considerably off of the predicted mean and are thus the ones in contradictiong with the \citet{tuomi2014} solution. However, we could not identify any problems with these four HARPS measurements and therefore do not have a reliable explanation for the fact that they deviate from the model considerably more than would be expected based on the rest of the data.

To conclude, it is apparent that unless there is something wrong with the four HARPS velocities that are not distributed according to the rest of them (Fig. \ref{fig:GJ682_residuals}), the interpretation of \citet{tuomi2014} that there are two low-mass planets orbiting the star is invalid. We cannot resolve the corresponding discrepancies here and thus assume conservatively that these planets do not exist because their signals cannot be identified in the full data set. However, should it turn out that some of the HARPS data are indeed biased, this interpretation would require reassessment.

\clearpage

\subsection{GJ 686}\label{sec:GJ686}

\begin{figure}
\center
\includegraphics[angle=270, width=0.49\textwidth,clip]{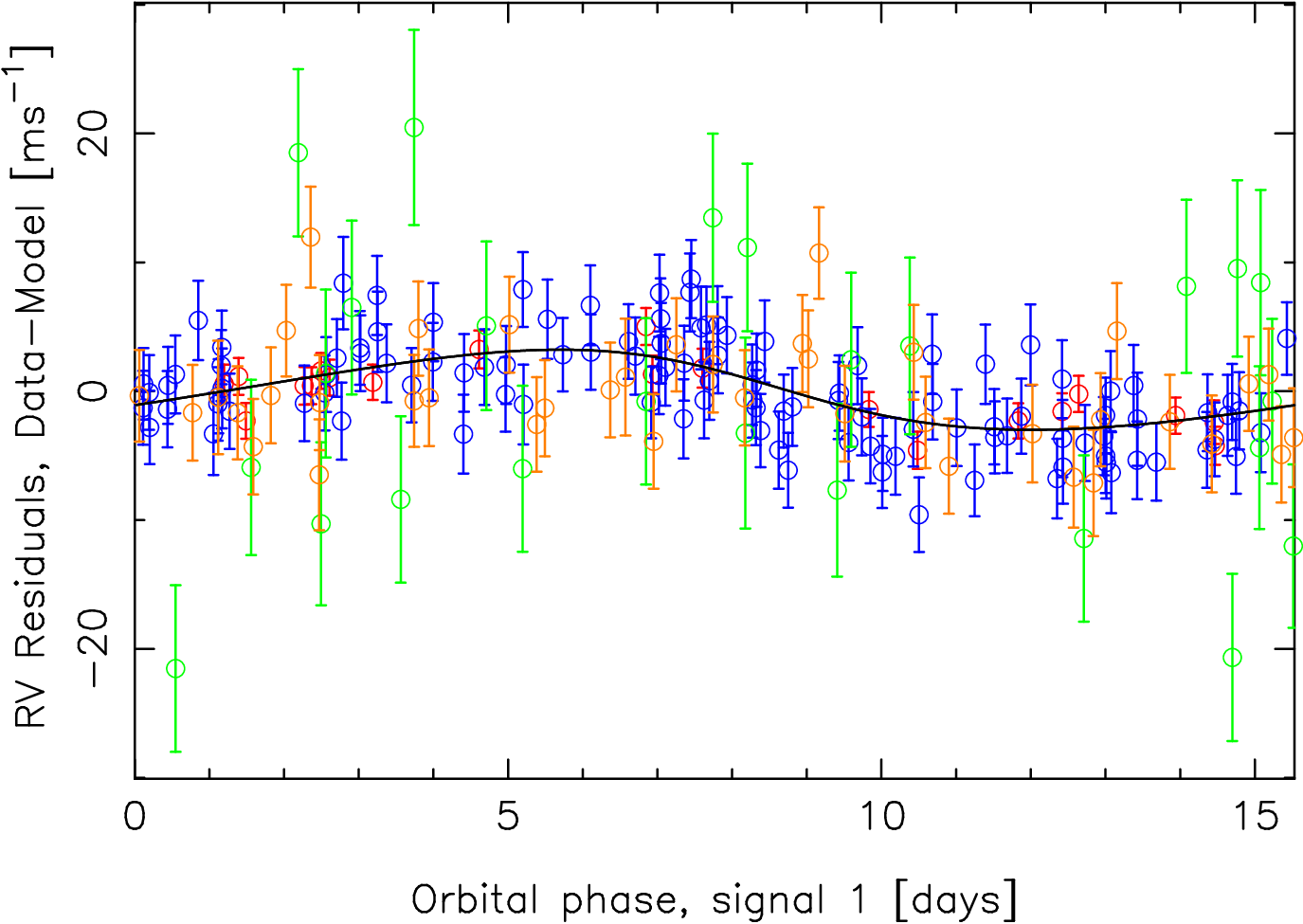}

\includegraphics[angle=270, width=0.49\textwidth,clip]{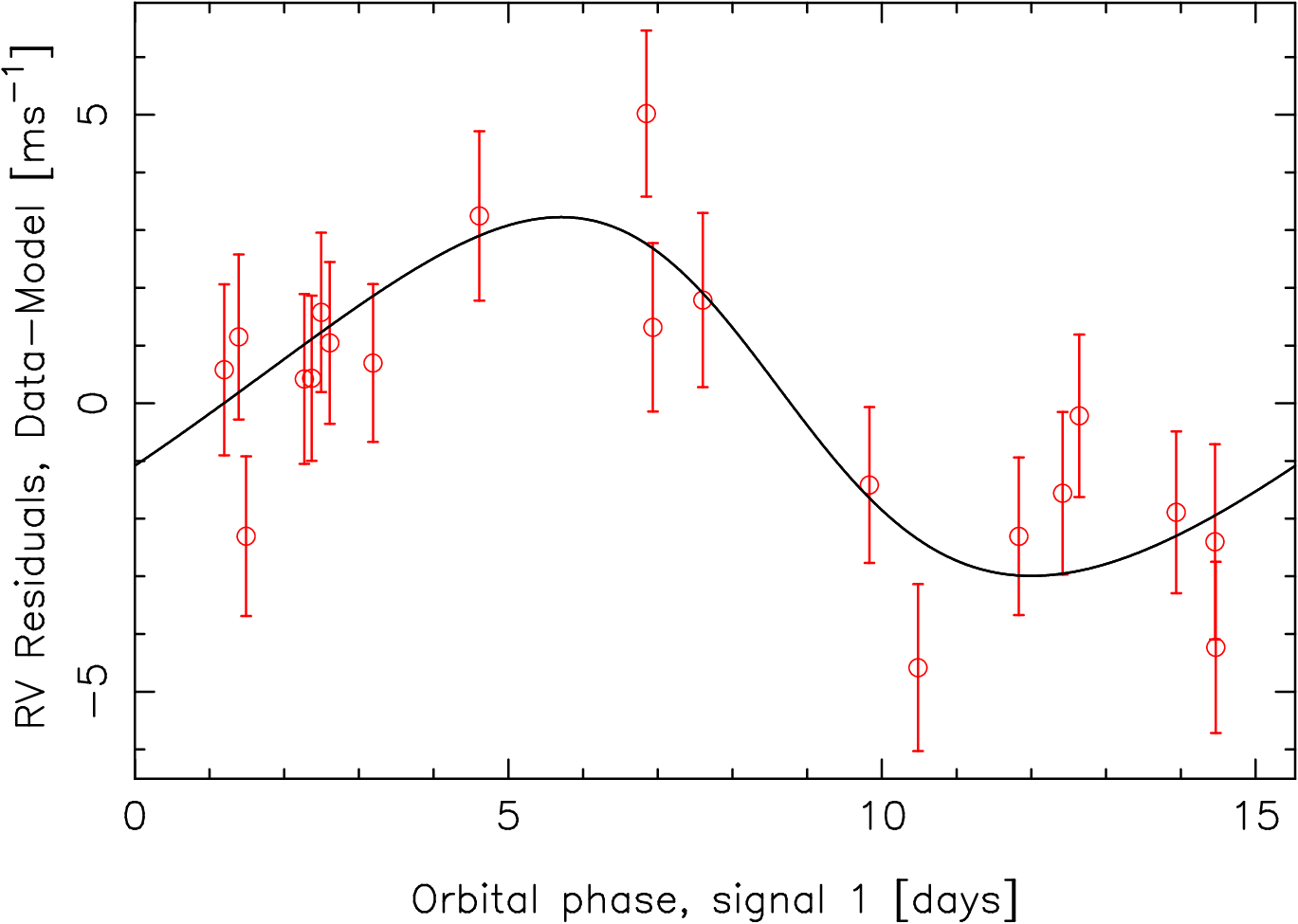}

\includegraphics[angle=270, width=0.49\textwidth,clip]{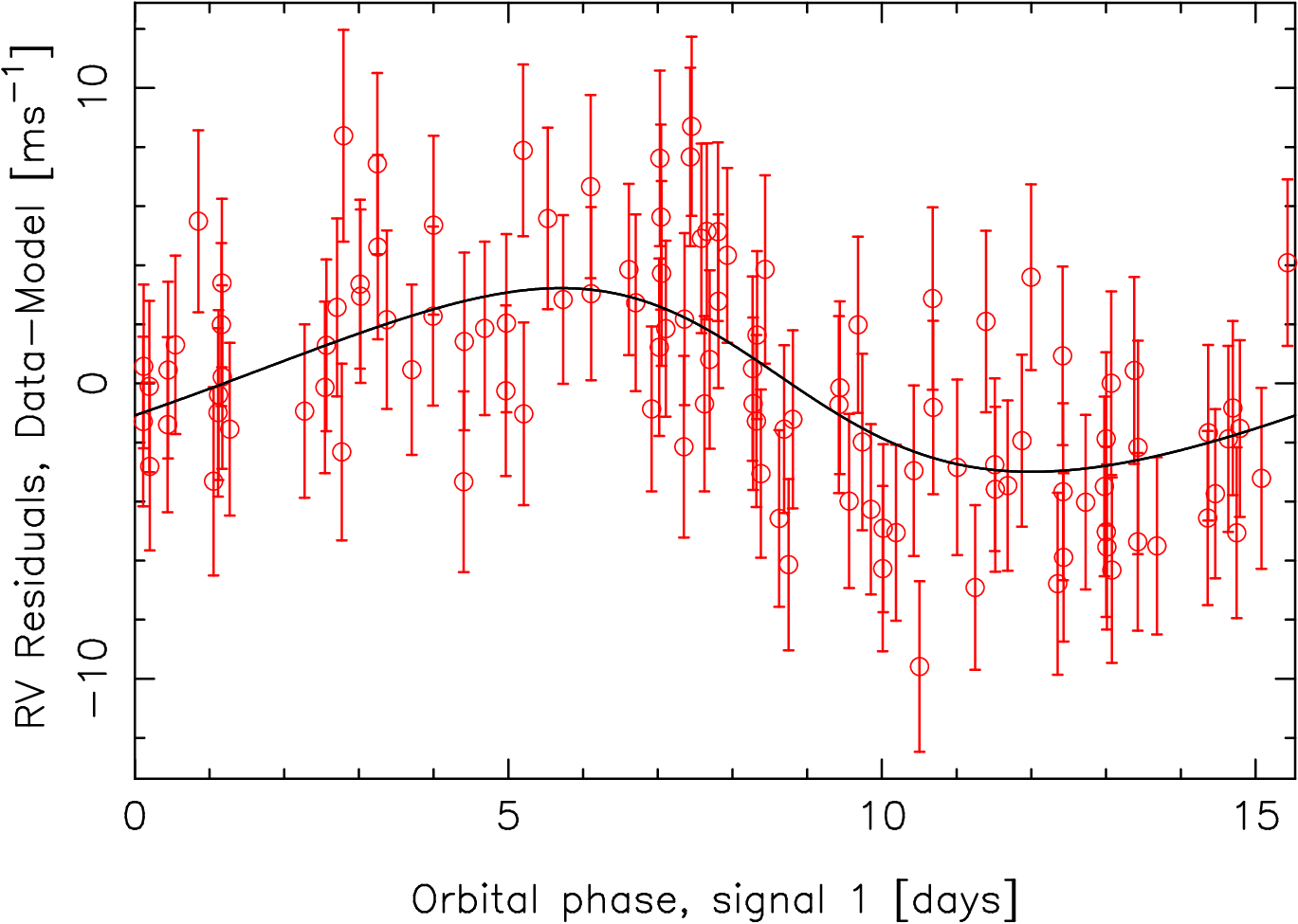}
\caption{HARPS (red), APF (orange), HIRES (blue), and SOPHIE (green) radial velocities of GJ 686 folded on the phase of the signal (top panel). The middle (bottom) panel shows the same given HARPS (HIRES) data alone.}\label{fig:GJ686_curve}
\end{figure}

\begin{figure}
\center
\includegraphics[angle=270, width=0.49\textwidth,clip]{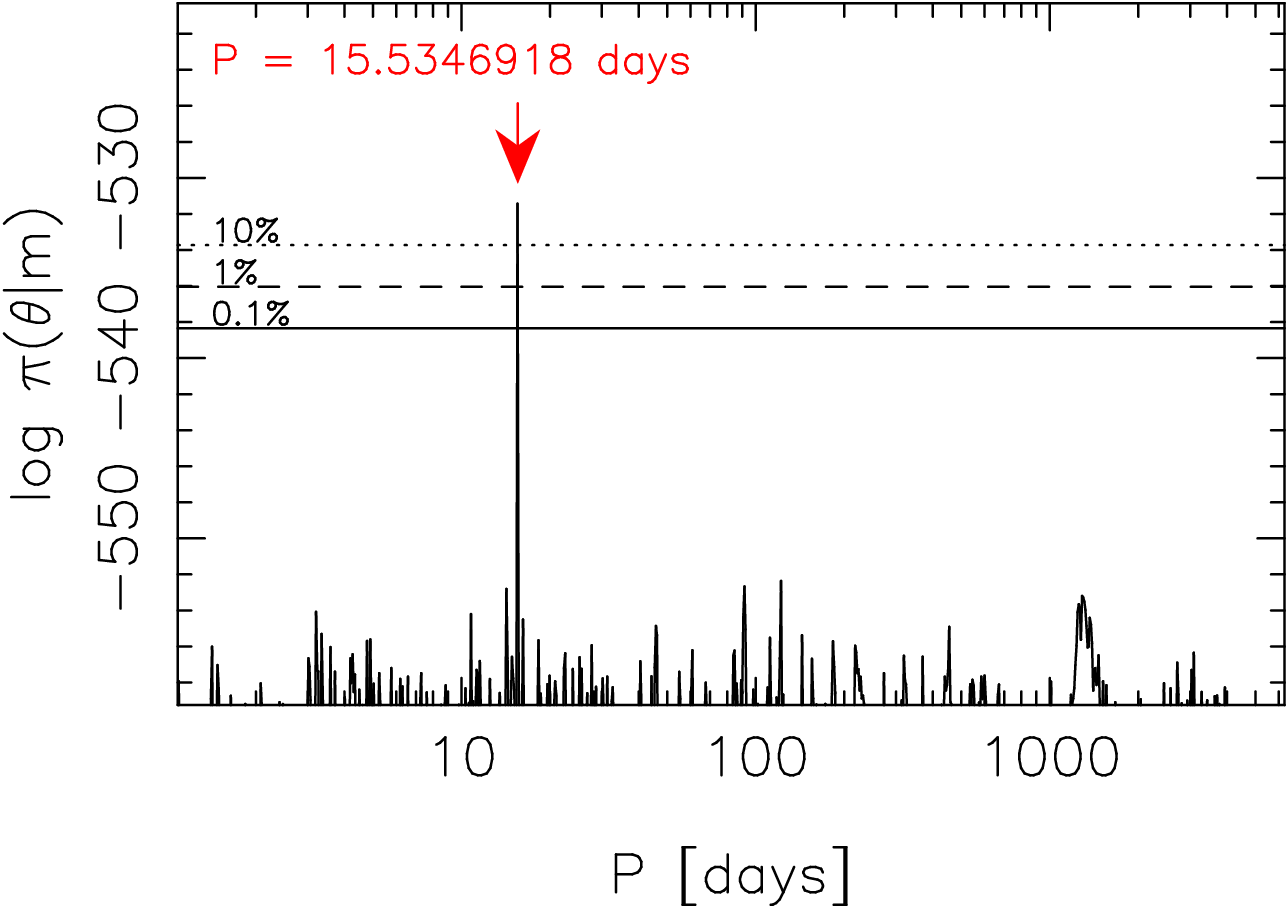}
\caption{Estimated posterior probability density as a function of the period parameter of the Keplerian signal in the combined GJ 686 radial velocities.}\label{fig:GJ686_psearch}
\end{figure}

We obtained strong evidence for a hot super-Earth orbiting GJ 686 (HIP 86287) with an orbital period of 15.5313 [15.5225, 15.5401] days. This signal was already detected by \citet{butler2016} but is essentially confirmed here based on a combined set of radial velocities from HARPS, HIRES, APF, and SOPHIE, although the APF and SOPHIE data sets did not help constraining the signal much due to their low numbers of observations and lower precision than that of HARPS (Fig. \ref{fig:GJ686_curve}). The signal is shown as a unique probability maximum in the period space in Fig. \ref{fig:GJ686_psearch}.

The HIRES S-indices show suggestive evidence for a periodicity at 143 days \citep[see also][]{butler2016} but there are not even emerging signals at or near the period of the radial velocity signal. The HARPS and APF (S-index) activity indicators show no evidence for periodicities. Moreover, the ASAS V-band photometry measurements contain no evidence in favour of photometric periods. It is thus our interpretation that the radial velocity signal indeed corresponds to a hot super-Earth with a minimum mass of 7.1 [3.9, 10.4] M$_{\oplus}$ orbiting the star.

\clearpage

\subsection{GJ 687}\label{sec:GJ687}

\begin{figure}
\center
\includegraphics[angle=270, width=0.49\textwidth,clip]{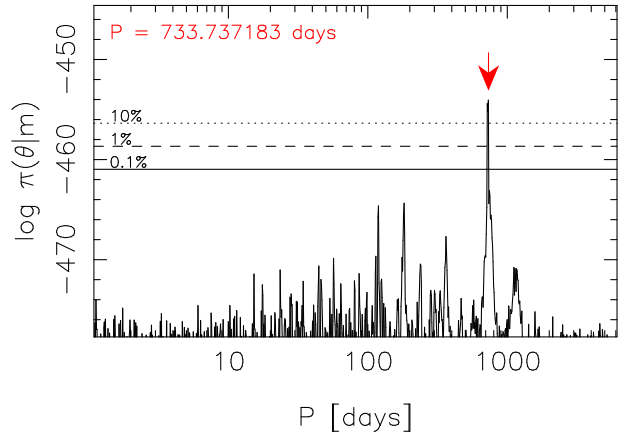}
\caption{Estimated posterior probability density as a function of the period parameter of the second Keplerian signal of a two-Keplerian model given the combined GJ 687 radial velocities from APF and HIRES.}\label{fig:GJ687_psearch}
\end{figure}

\begin{figure}
\center
\includegraphics[angle=270, width=0.49\textwidth,clip]{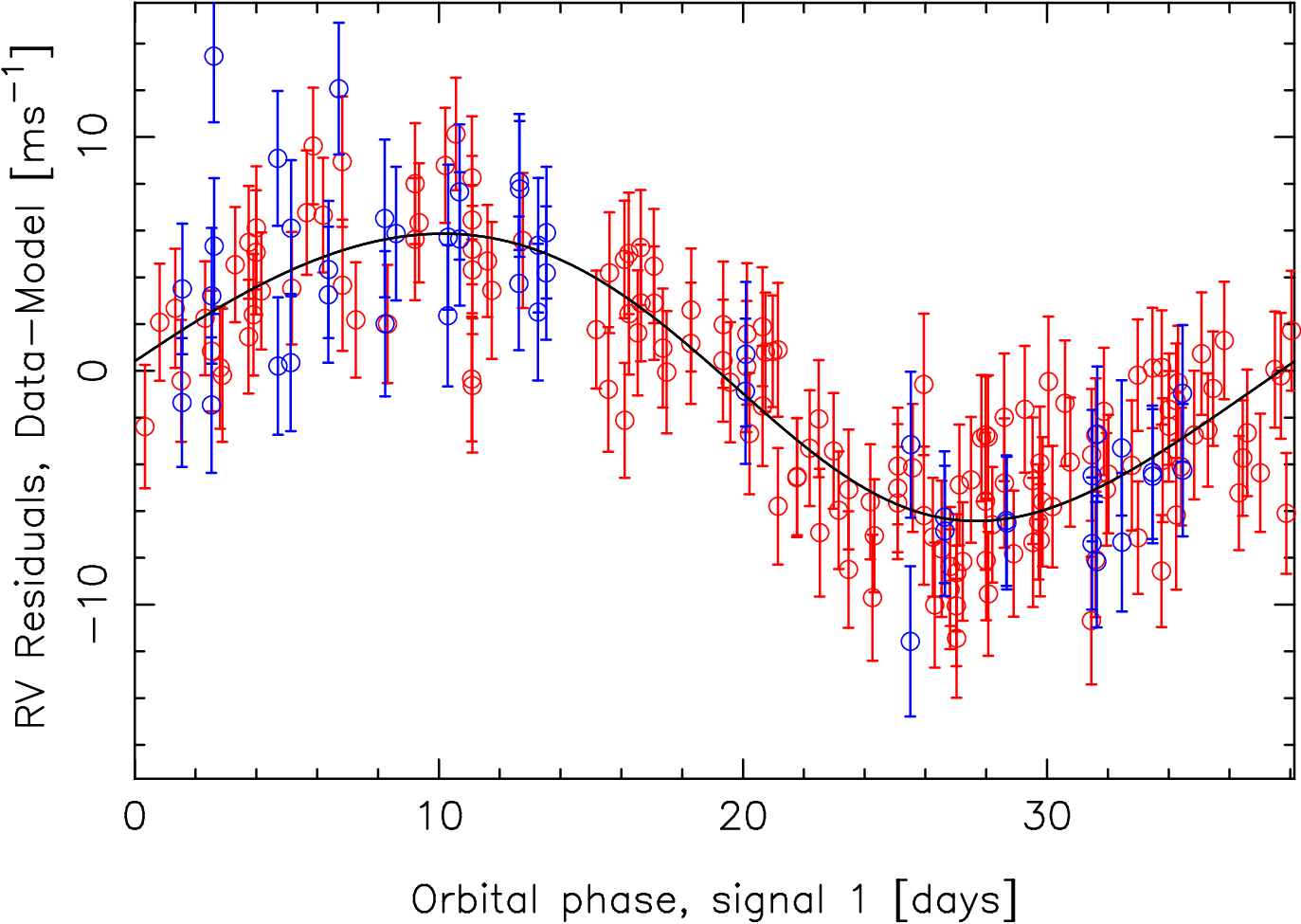}

\includegraphics[angle=270, width=0.49\textwidth,clip]{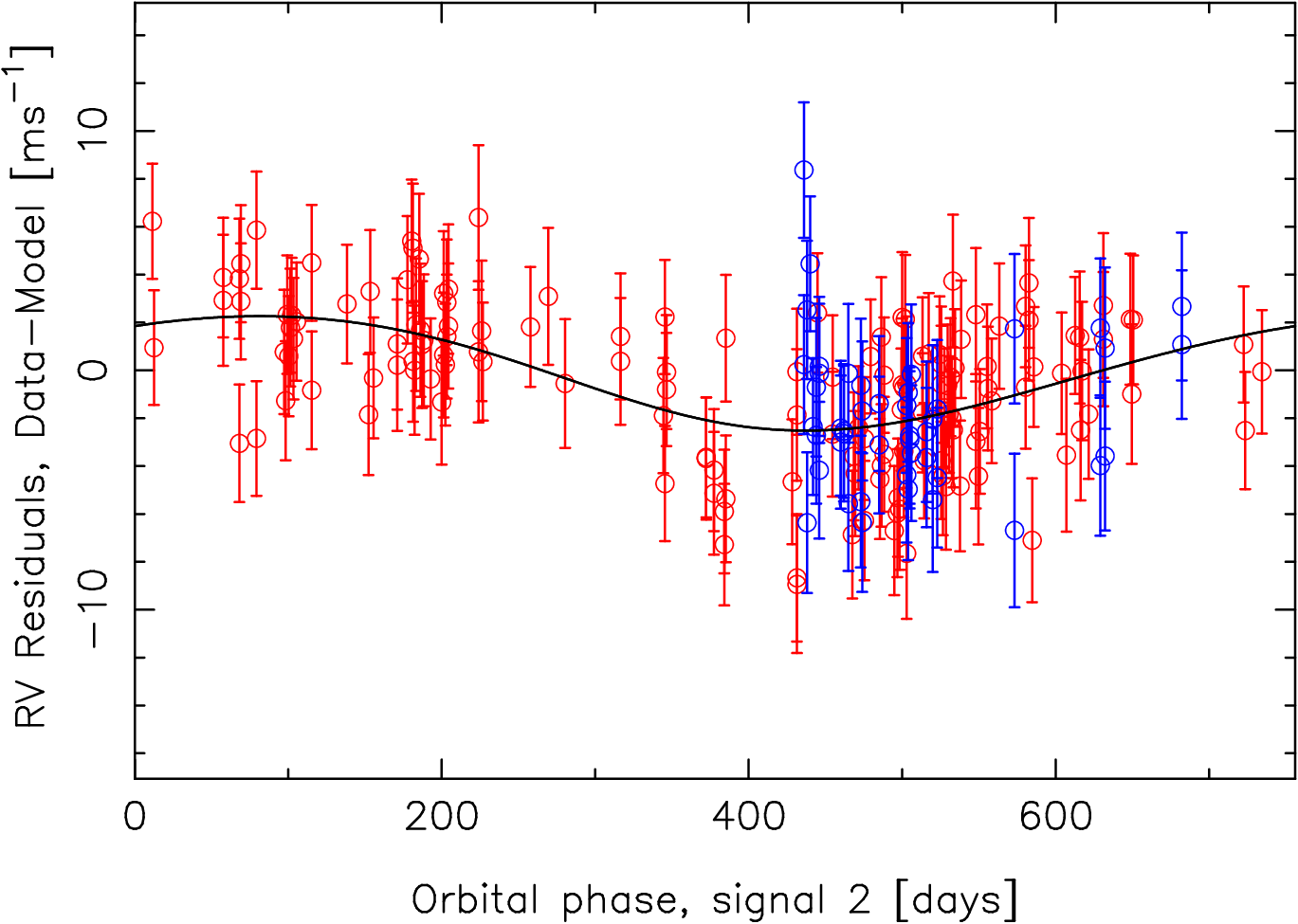}
\caption{HIRES (red) and APF (blue) radial velocities of GJ 687 folded on the phases of the two signals detected in the data.}\label{fig:GJ687_phased}
\end{figure}

\citet{burt2014} reported that there is a minimum mass 18.4$\pm$2.2 M$_{\oplus}$ planet orbiting GJ 687 (HIP 86162) with a period of 38.140$\pm$0.015 days based on combined HIRES and APF radial velocities. With new observations from both instruments, we could easily find the signal corresponding to GJ 687 b with our posterior samplings. However, we also obtained evidence for another signal in the combined data at a period of 758 [692, 796] days and an amplitude of 2.00 [0.81, 3.32] ms$^{-1}$ (Fig. \ref{fig:GJ687_psearch}). We have plotted the radial velocities folded on the phases onf the two signals in Fig. \ref{fig:GJ687_phased}.

\begin{figure}
\center
\includegraphics[angle=270, width=0.49\textwidth,clip]{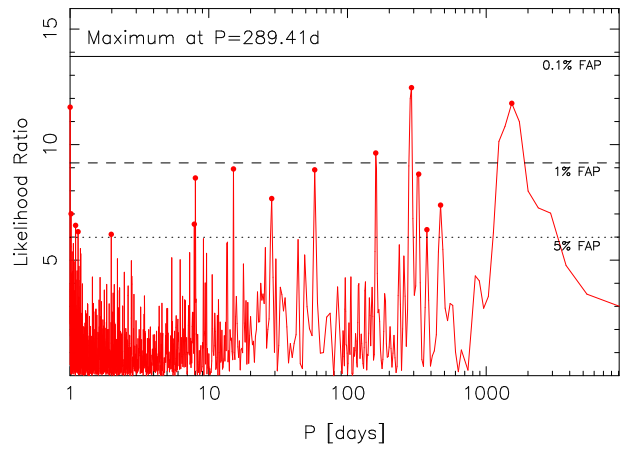}
\caption{Likelihood-ratio periodogram of the HIRES S-indices of GJ 687.}\label{fig:GJ687_S}
\end{figure}

There was weak evidence for periodicities in the HIRES S-indices of GJ 687 at period of 290 and 1500 days (Fig. \ref{fig:GJ687_S}) exceeding the 1\% FAP but not the 0.1\% FAP threshold. However, there is a visible lack of high likelihood ratios at or near the 760-day signal enabling us to interpret this signal as evidence for a planet orbiting the star. It is thus our interpretation that there is a warm Neptune and a cool Neptune orbiting GJ 687.

No ASAS data was available for GJ 687 and we could thus not confirm the photometric rotation period of 61.8$\pm$1.0 days reported by \citet{burt2014} with an independent light curve.

\clearpage

\subsection{GJ 699}\label{sec:GJ699}

Barnard's star (GJ 699, HIP 87937) is famous for the early claims of being a planet host by \citet{vandekamp1963} and \citet{vandekamp1969}. Although these claims, based on astrometry, were later proven false positive detections \citep{gatewood1973}, Barnard's star has been targeted by several Doppler spectroscopy surveys \citep[e.g.][]{choi2013} and is known to have a planet orbiting it \citep{ribas2018}.

\begin{figure}
\center
\includegraphics[angle=270, width=0.49\textwidth,clip]{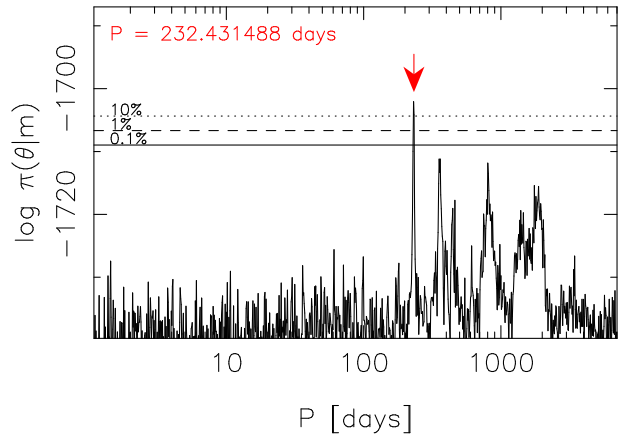}
\caption{Estimated posterior probability density as a function of the period parameter of the Keplerian signal given the combined GJ 699 radial velocities.}\label{fig:GJ699_psearch}
\end{figure}

\begin{figure}
\center
\includegraphics[angle=270, width=0.49\textwidth,clip]{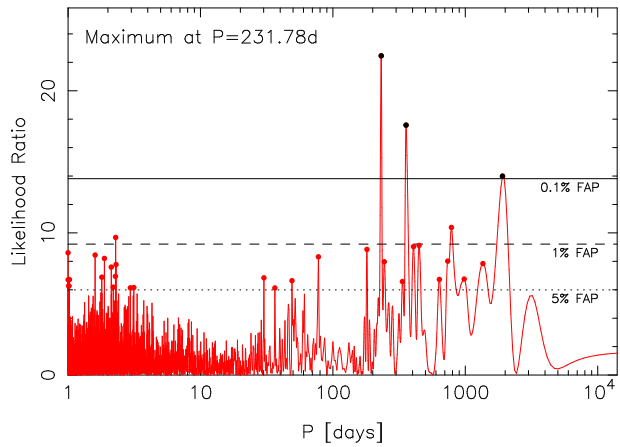}
\caption{Likelihood-ratio periodogram of the combined residuals of the model without Keplerian signals.}\label{fig:GJ699_resid_per}
\end{figure}

\begin{figure}
\center
\includegraphics[angle=270, width=0.49\textwidth,clip]{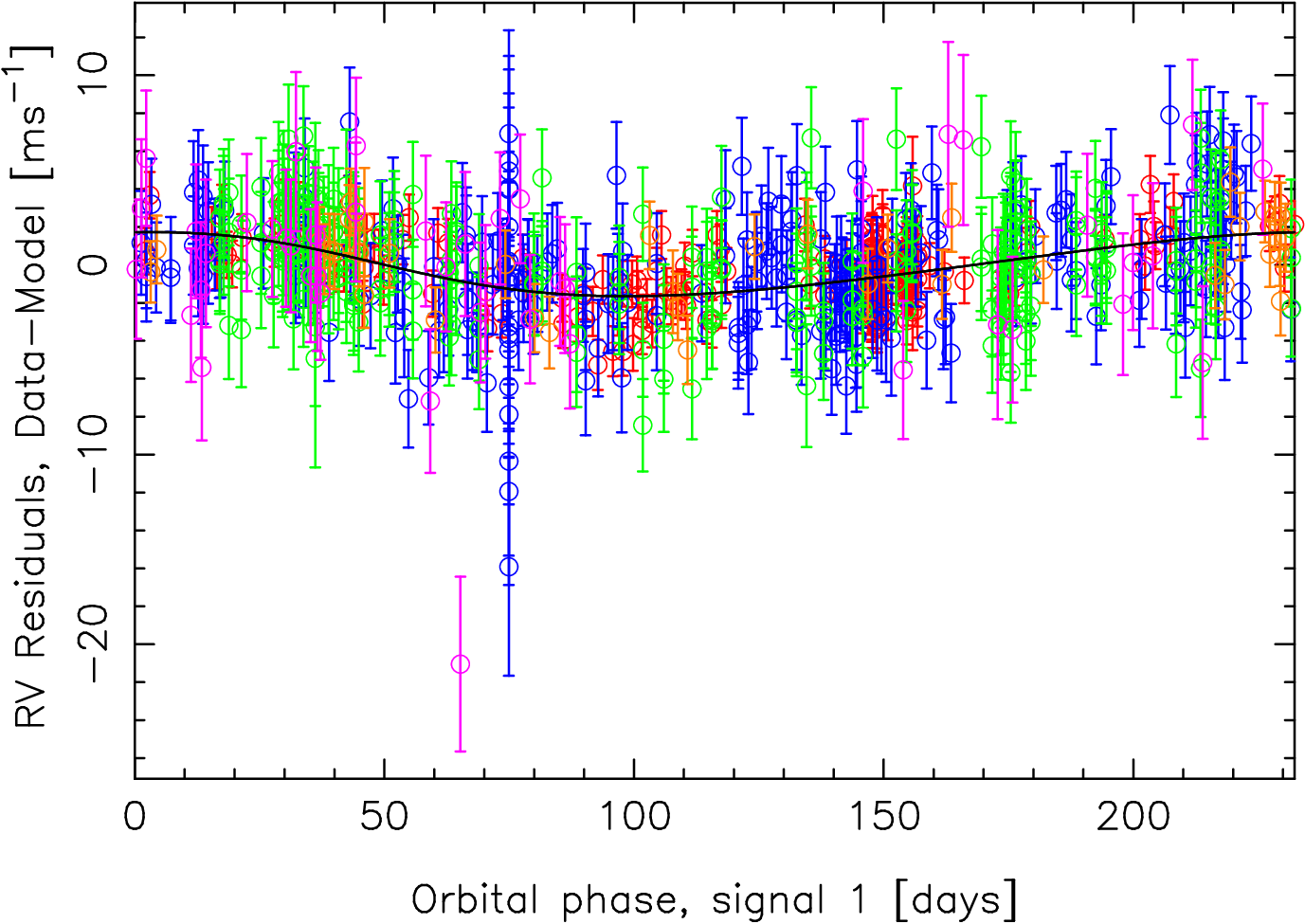}

\includegraphics[angle=270, width=0.49\textwidth,clip]{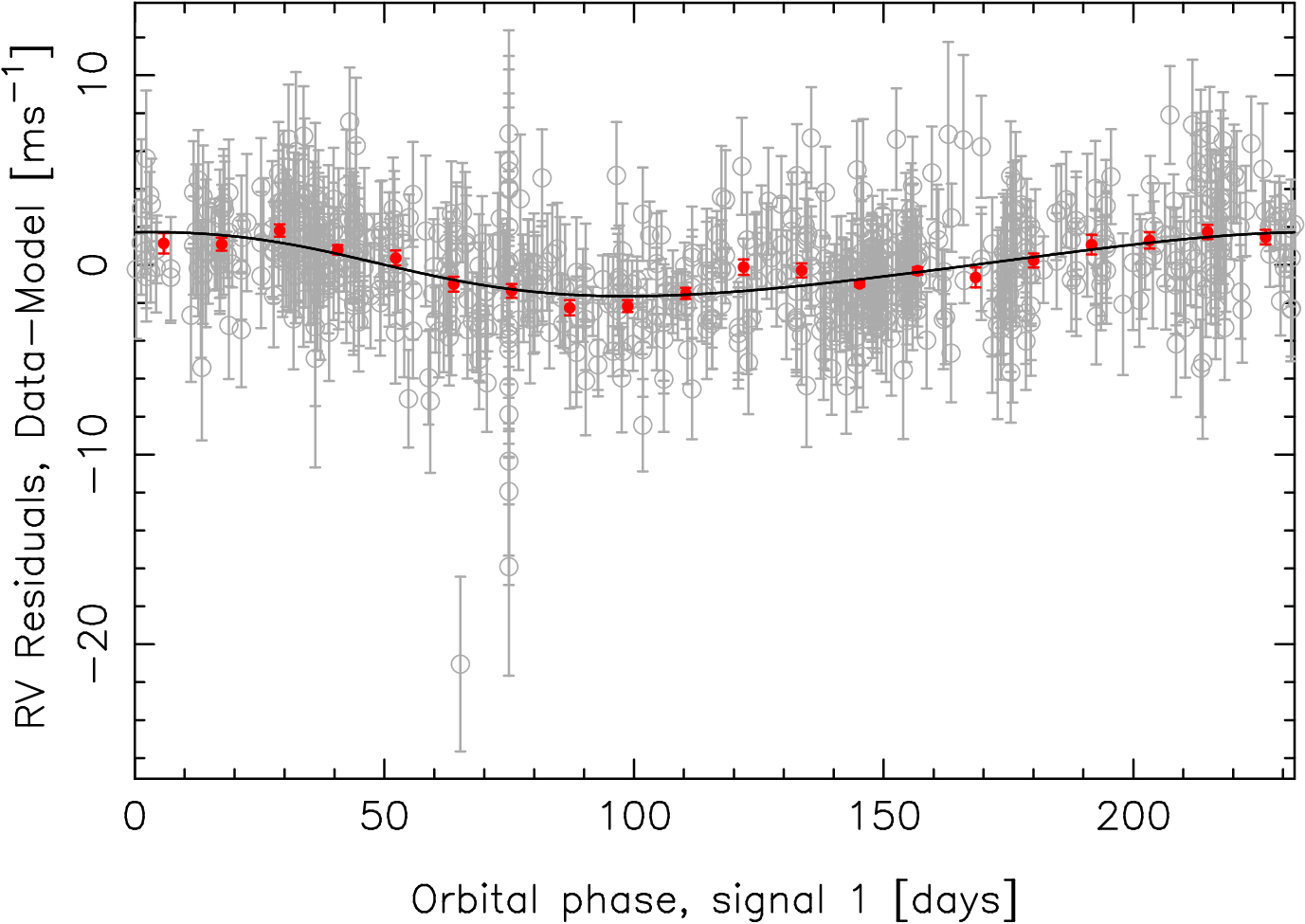}
\caption{HARPS (red), HIRES (blue), UVES (green), PFS (orange), and APF (purple) radial velocities of GJ 699 folded on the period of the signal (top panel). The bottom panel shows the same velocities binned into 20 bins (red filled circles).}\label{fig:GJ699_phased}
\end{figure}

We analysed the combined HARPS, HIRES, PFS, APF, UVES, and LICK data of GJ 699 and obtained evidence for a periodic signal \citet[see also][]{ribas2018}. This signal, at a period of 232.10 [229, 19, 234.00] days and with an amplitude of 1.30 [0.75, 1.80] ms$^{-1}$ was uniquely detected in the combined data in accordance with our signal detection criteria (Fig. \ref{fig:GJ699_psearch}). This signal was so strong in the combined data that it was also detected with the common residual periodogram of the model without Keplerian signals (Fig. \ref{fig:GJ699_resid_per}). The phase-folded radial velocities (excluding LICK data due to its higher uncertainties) of GJ 699 are shown in Fig. \ref{fig:GJ699_phased} to visually illustrate the existence of the signal in the data. We note that the existence of the signal is strongly supported by HARPS ($N = 235$) and HIRES ($N = 231$) data sets whereas APF, PFS and UVES data sets provide only slightly additional support due to either lower precision or number of data points -- yet, none of these datasets disagree with the interpretation that there are periodic variations with a period of 232 days in the radia lvelocities of GJ 699.

\begin{figure}
\center
\includegraphics[angle=270, width=0.49\textwidth,clip]{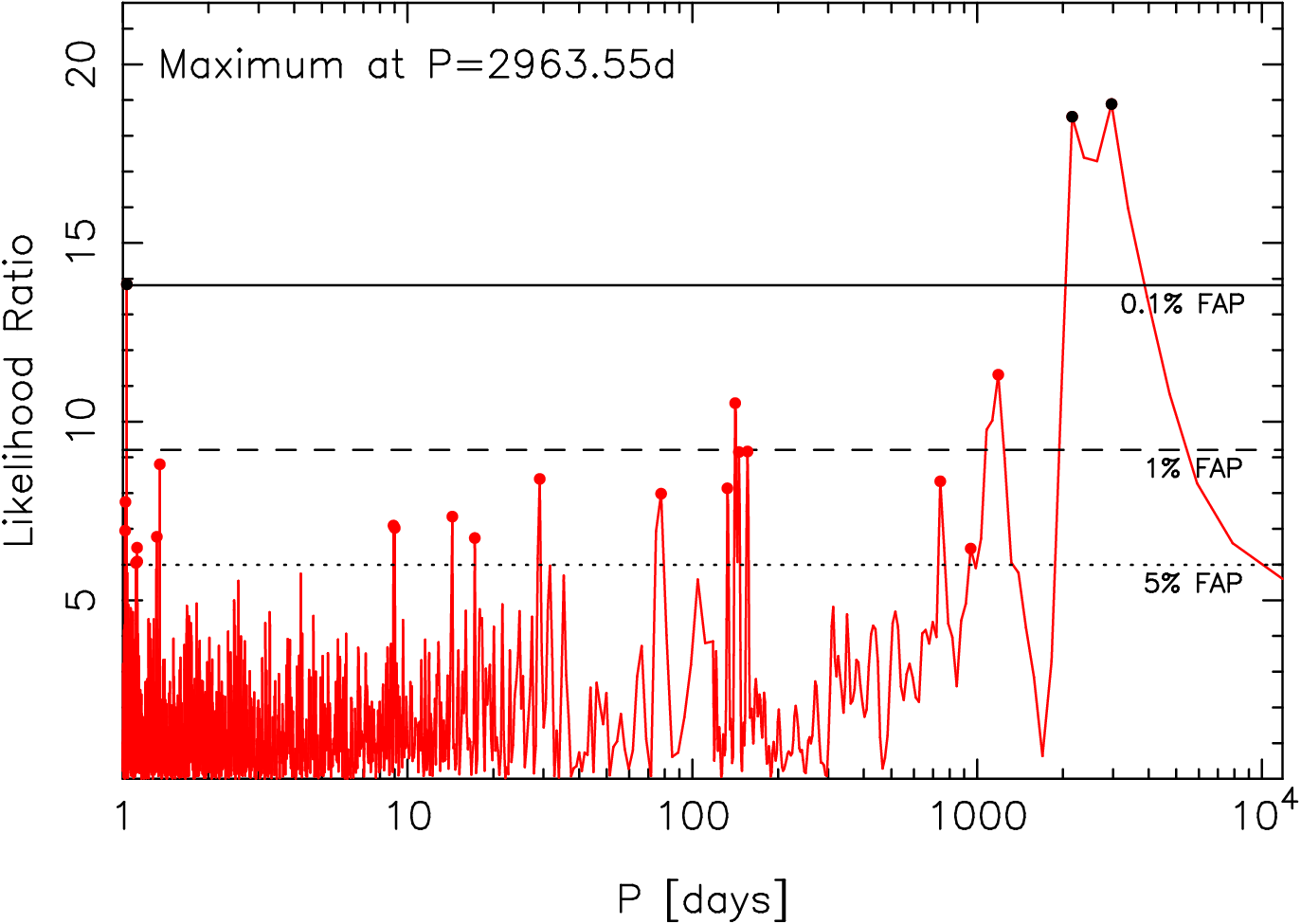}

\includegraphics[angle=270, width=0.49\textwidth,clip]{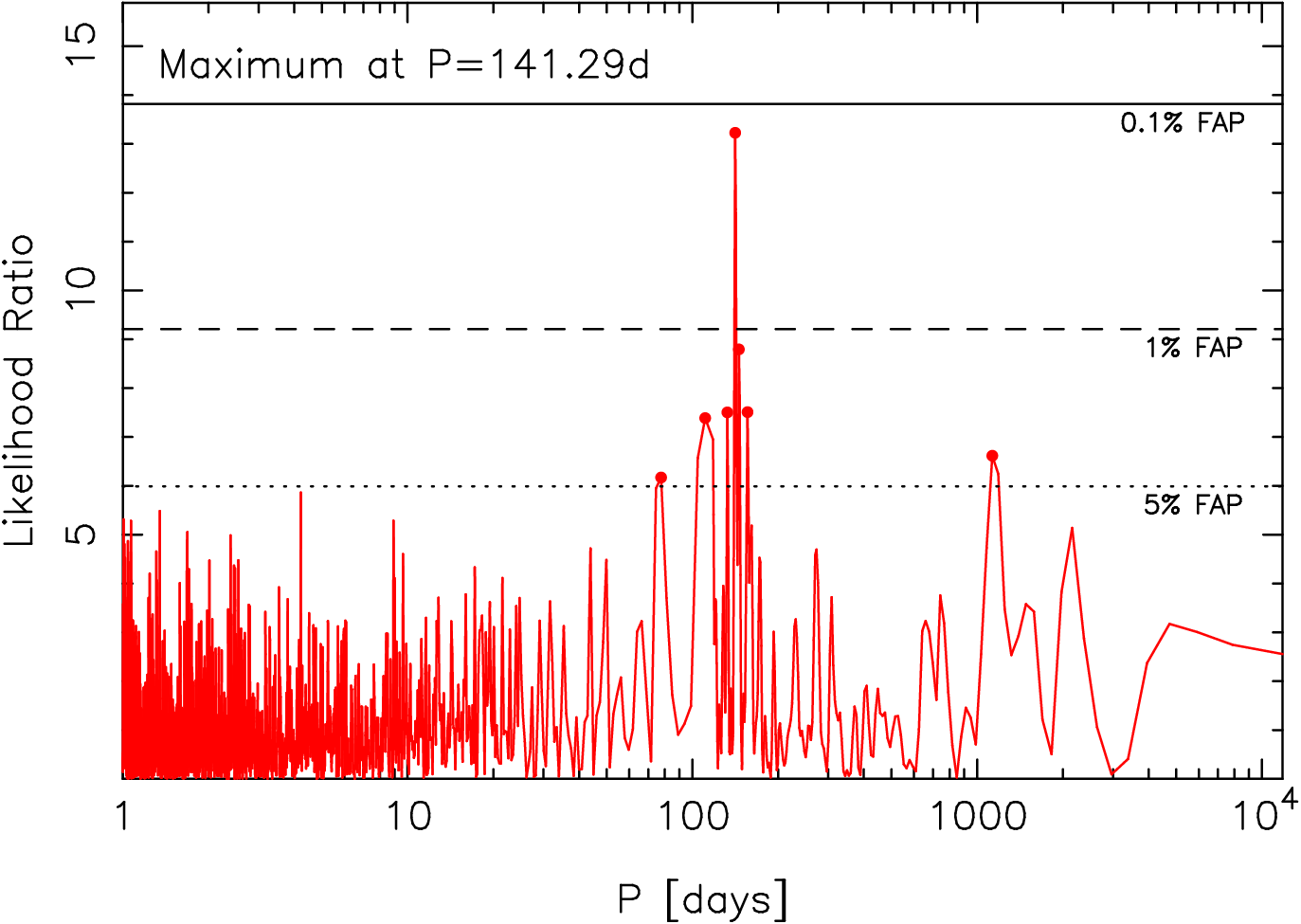}
\caption{Likelihood ratio periodogram of the HIRES S-indices of GJ 699 (top panel) and the periodogram of residuals after subtracting the long-period signal (bottom panel).}\label{fig:GJ699_HIRES_S}
\end{figure}

We obtained strong evidence for periodicity of $\sim$ 3000 days and weaker evidence for another one of 141.29 days in the HIRES S-indices (Fig. \ref{fig:GJ699_HIRES_S}). Moreover, as \citet{mascareno2015} reported a periodicity of 148.6$\pm$0.1 days in their analysis of HARPS CaII H\&K data, we conclude that the rotation period of the star is likely between 140 and 150 days whereas the magnetic activity cycle of the star is likely close to the period of 3000 days detected in the HIRES S-indices. Our analysis of the HARPS S-indices revealed a long-period cycle of roughly 4000 days but we could not detect additional periodicities in HARPS activity data. We could not verify the rotation period of the star based on photometry because ASAS photometry data was not available.

It seems evident that there is a periodic signal in the radial velocities of GJ 699 without counterparts in spectroscopic activity data. This signal, at a period of 232 days, is also significantly different and unlikely to be connected to the observed periodic variability between 140 and 150 days likely caused by rotation of the star. We thus conclude that there is evidence for a cool super-Earth orbiting GJ699 with a minimum mass of 4.15 [2.29, 6.20] M$_{\oplus}$ \citep[see also][]{ribas2018}. This detection is consistent with the results of \citet{choi2013} who demonstrated that HIRES data could rule out planets with minimum masses above roughly 5-8 M$_{\oplus}$ at and near the period of 230 days.

\clearpage

\subsection{GJ 725B}

\begin{figure}
\center
\includegraphics[angle=270, width=0.49\textwidth,clip]{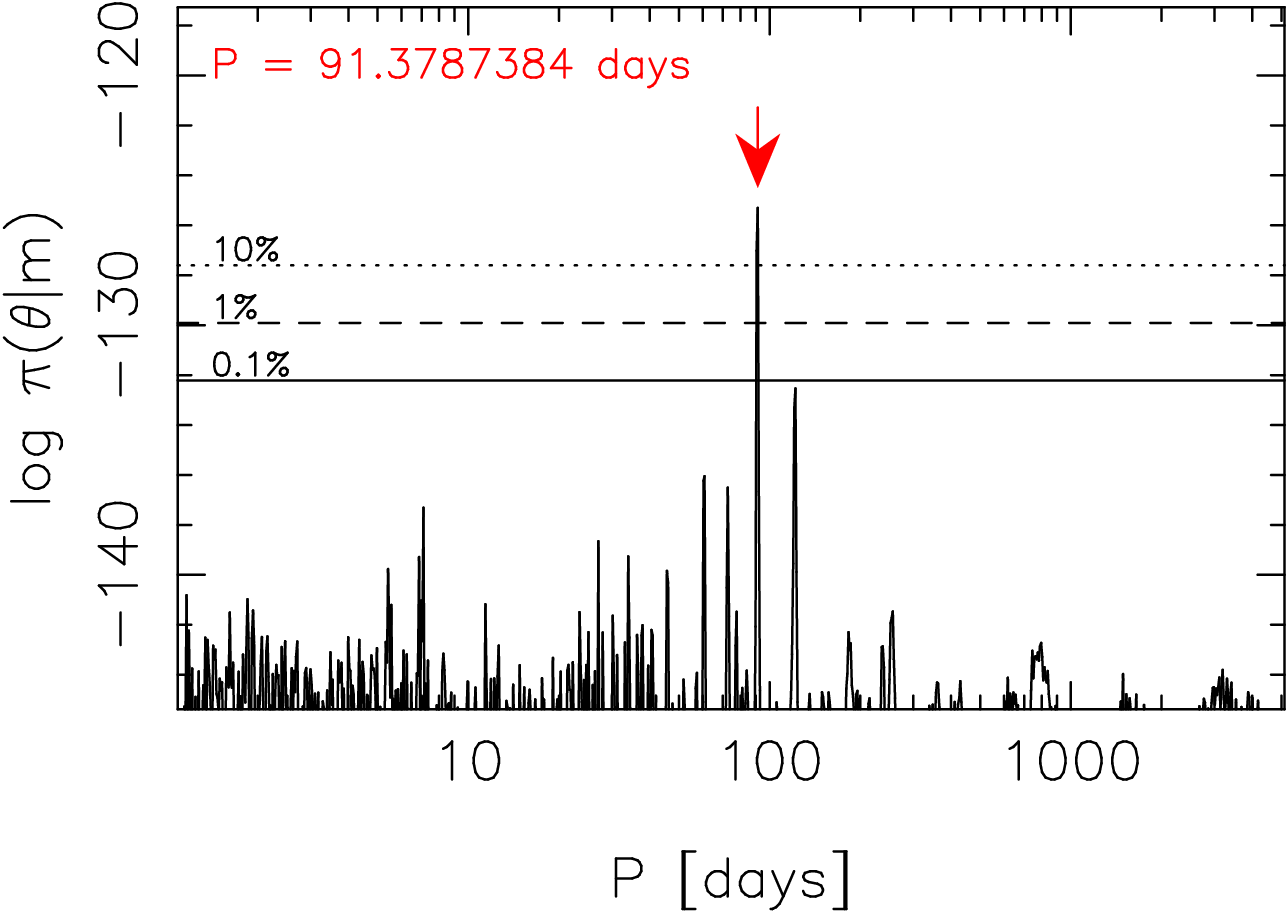}

\includegraphics[angle=270, width=0.49\textwidth,clip]{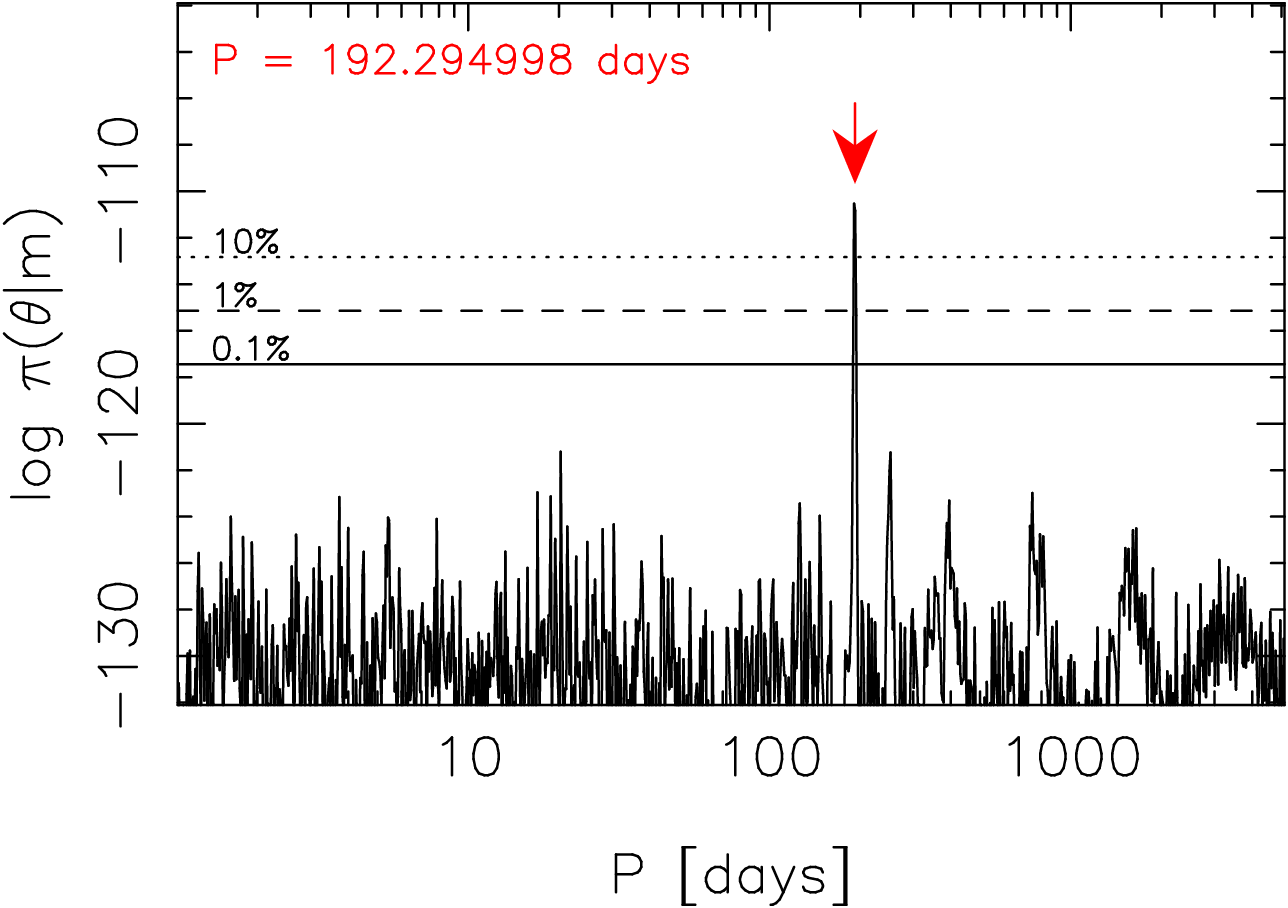}
\caption{Estimated posterior probability density as a function of the period parameter of the Keplerian signal in a one-Keplerian model given the GJ 725B radial velocities (top panel) and the probability density of the two-Keplerian model as a function of the period parameter of the second signal (bottom panel). The red arrows indicate the positions of the global maxima in the period space and the horizontal lines denote the 10\% (dotted), 1\% (dashed), and 0.1\% (solid) equiprobability thresholds of the maxima.}\label{fig:GJ725B_psearch}
\end{figure}

\begin{figure}
\center
\includegraphics[angle=270, width=0.49\textwidth,clip]{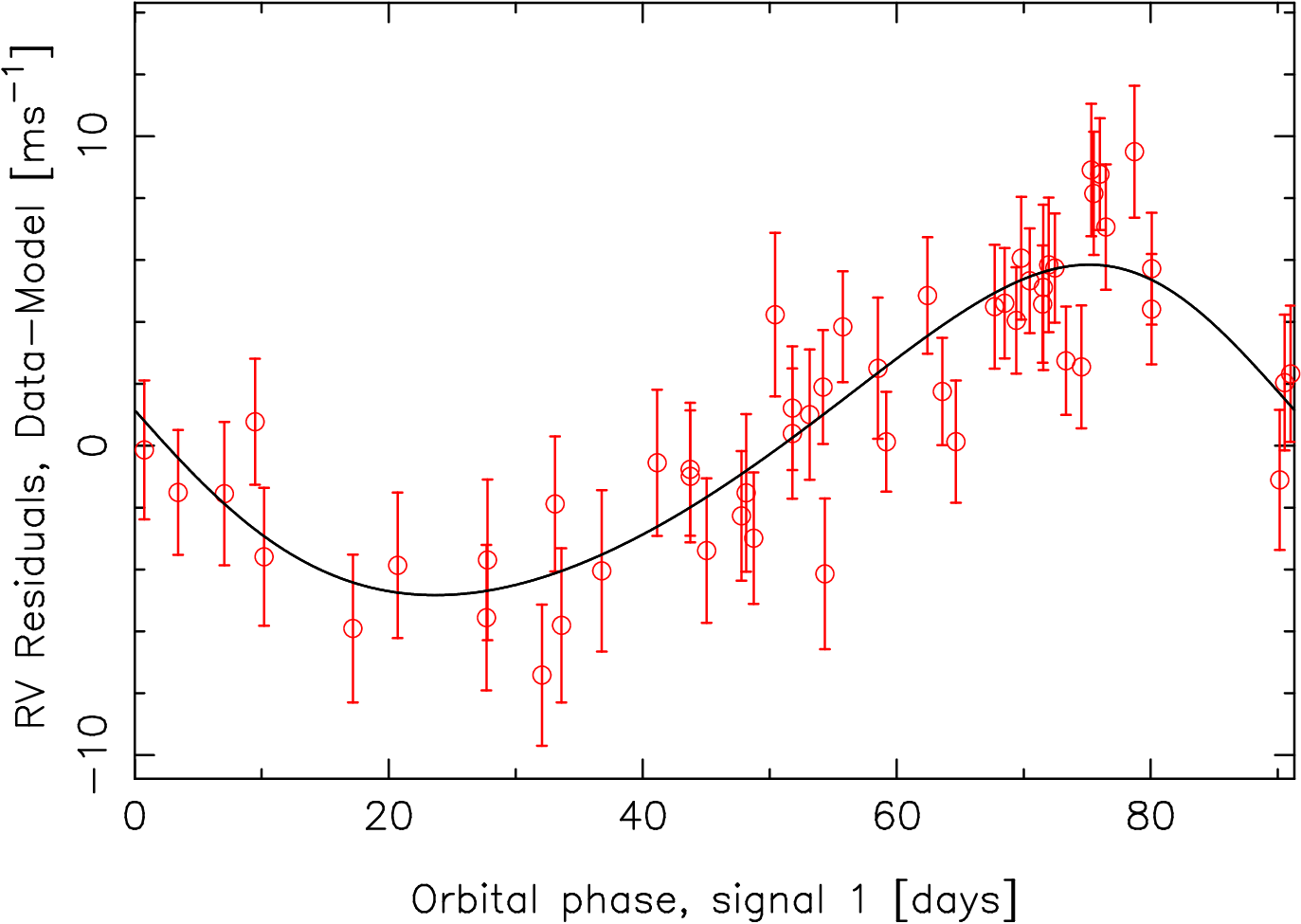}

\includegraphics[angle=270, width=0.49\textwidth,clip]{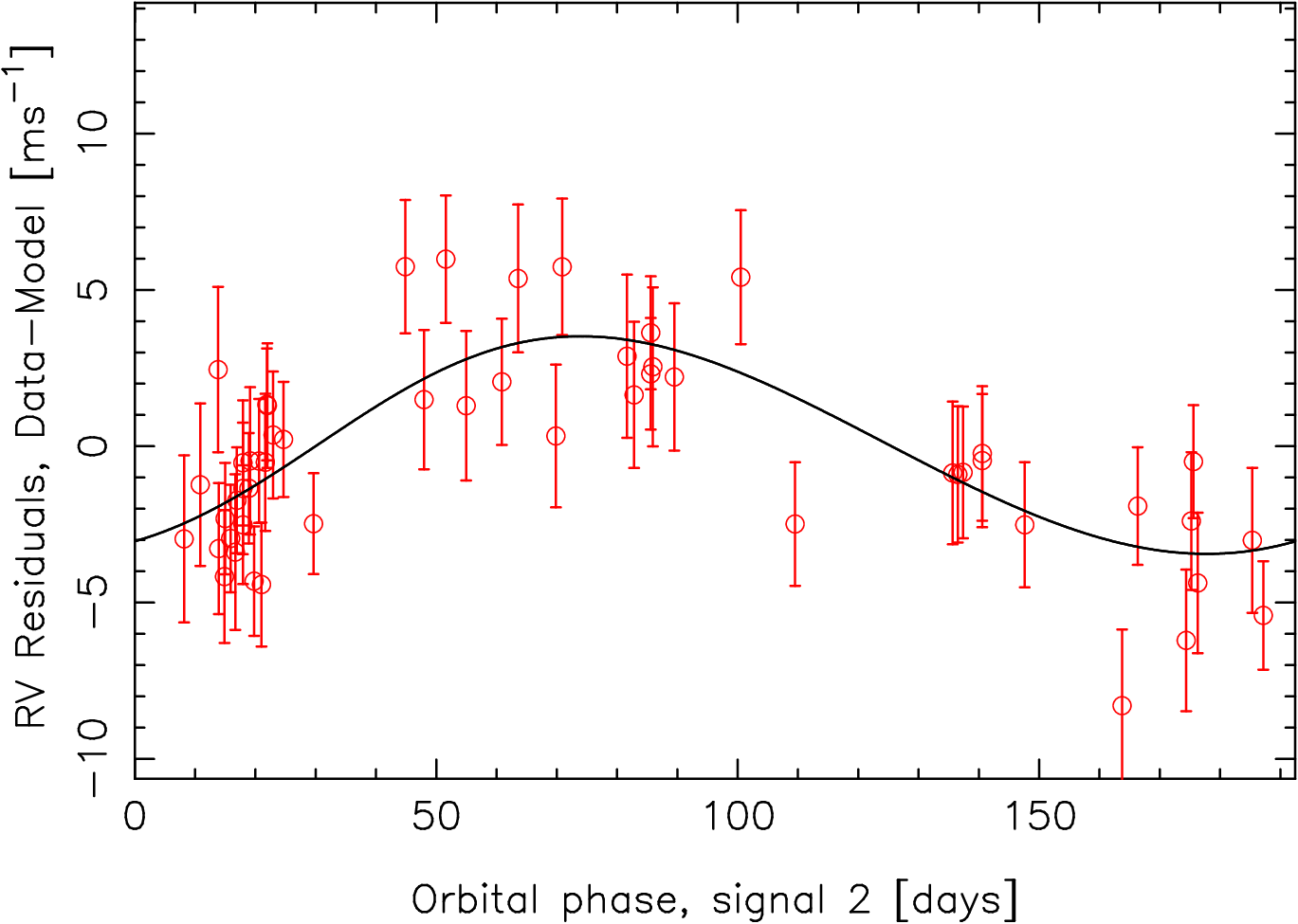}
\caption{HIRES radial velocities of GJ 725B folded of the phases of the two signals. The other signal and linear acceleration have been subtracted from both panels.}\label{fig:GJ725B_curve}
\end{figure}

On top of deceleration of -7.79$\pm$0.12 ms$^{-1}$year$^{-1}$, most likely due to the stellar companion GJ 725A, the 53 HIRES radial velocities of GJ 725B (HD 173740, HIP 91768) contain signals at periods of 91.289 [91.050, 91.601] and 192.37 [190.47, 194.53] days (Figs. \ref{fig:GJ725B_psearch} and \ref{fig:GJ725B_curve}). These signals are rather weak but still detected according to our criteria and present in the data as unique probability maxima as illustrated in Fig. \ref{fig:GJ725B_psearch}. We have no reason not to conclude that these signals are connected to stellar activity as there are no corresponding periodicities in the HIRES S-indices. Moreover, ASAS photometry of this target was not available. We thus interpret the two signals as candidate planets orbiting the star. These candidates, with minimum masses of 15.7 [10.0, 21.5] and 13.1 [6.6, 21.3] M$_{\oplus}$, are classified as cool Neptunes orbiting the star.

We note that \citep{butler2016} detected the probability maximum in the period space corresponding to the 91-day signal but it was not significant enough to satisfy their conservative detection criteria.

\clearpage

\subsection{GJ 739}

\begin{figure}
\center
\includegraphics[angle=270, width=0.49\textwidth,clip]{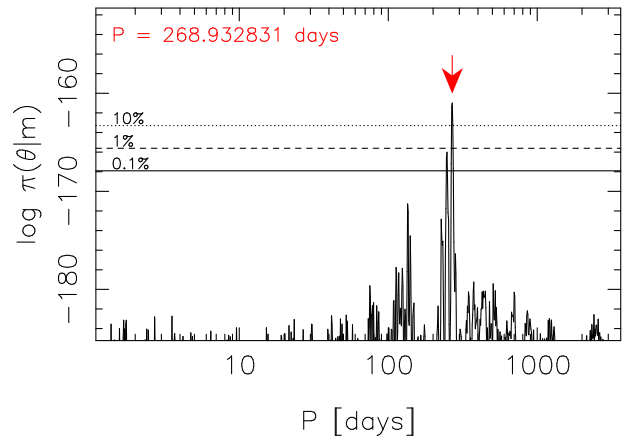}
\caption{Estimated posterior probability density as a function of the period of the Keplerian signal given the HARPS and UVES radial velocities of GJ 739. The red arrow denotes the global probability maximum and the horizontal lines illustrate the 10\% (dotted), 1\% (dashed), and 0.1\% (solid) equiprobability thresholds with respect to the maximum.}\label{fig:GJ739_psearch}
\end{figure}

\begin{figure}
\center
\includegraphics[angle=270, width=0.49\textwidth,clip]{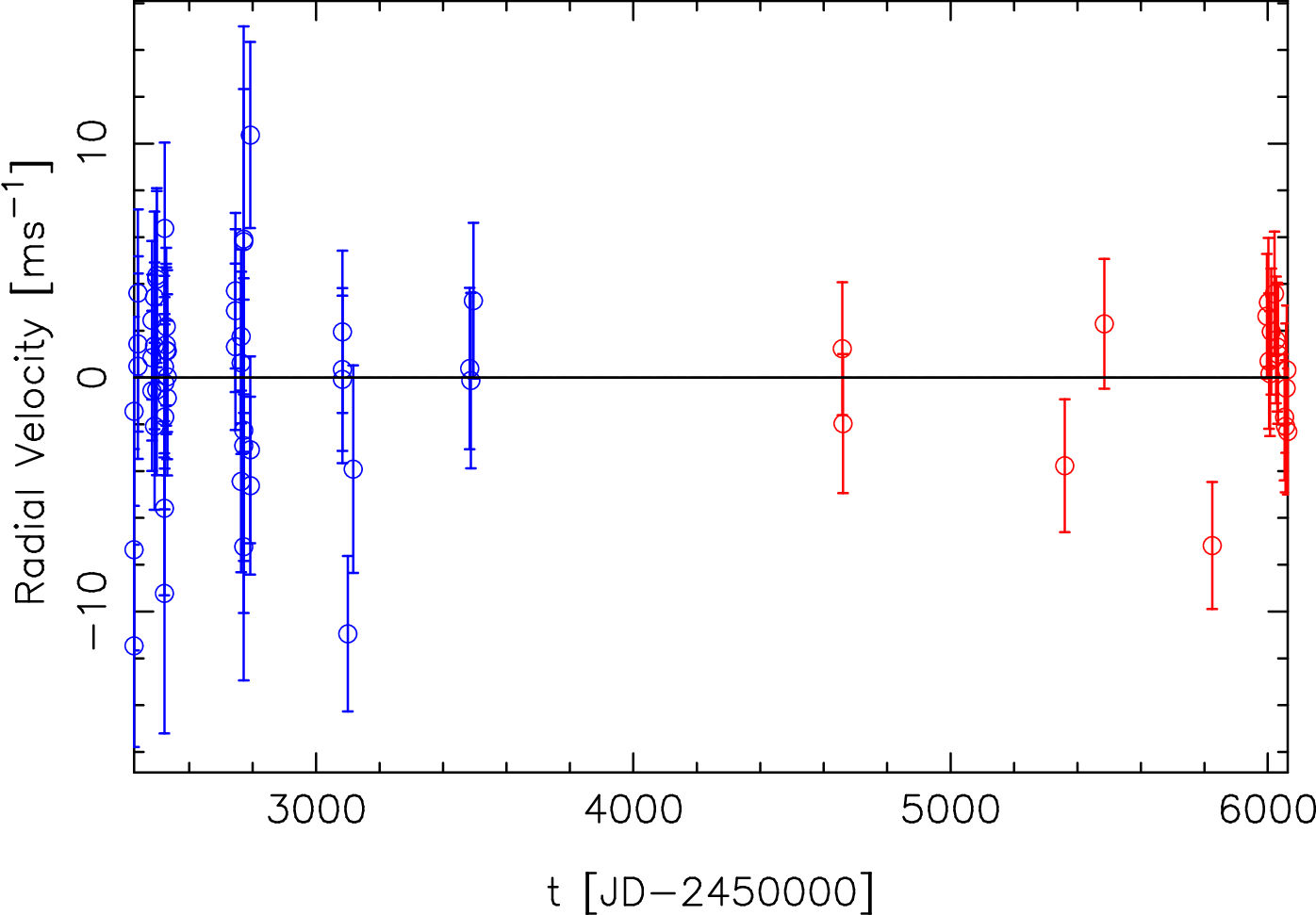}

\includegraphics[angle=270, width=0.49\textwidth,clip]{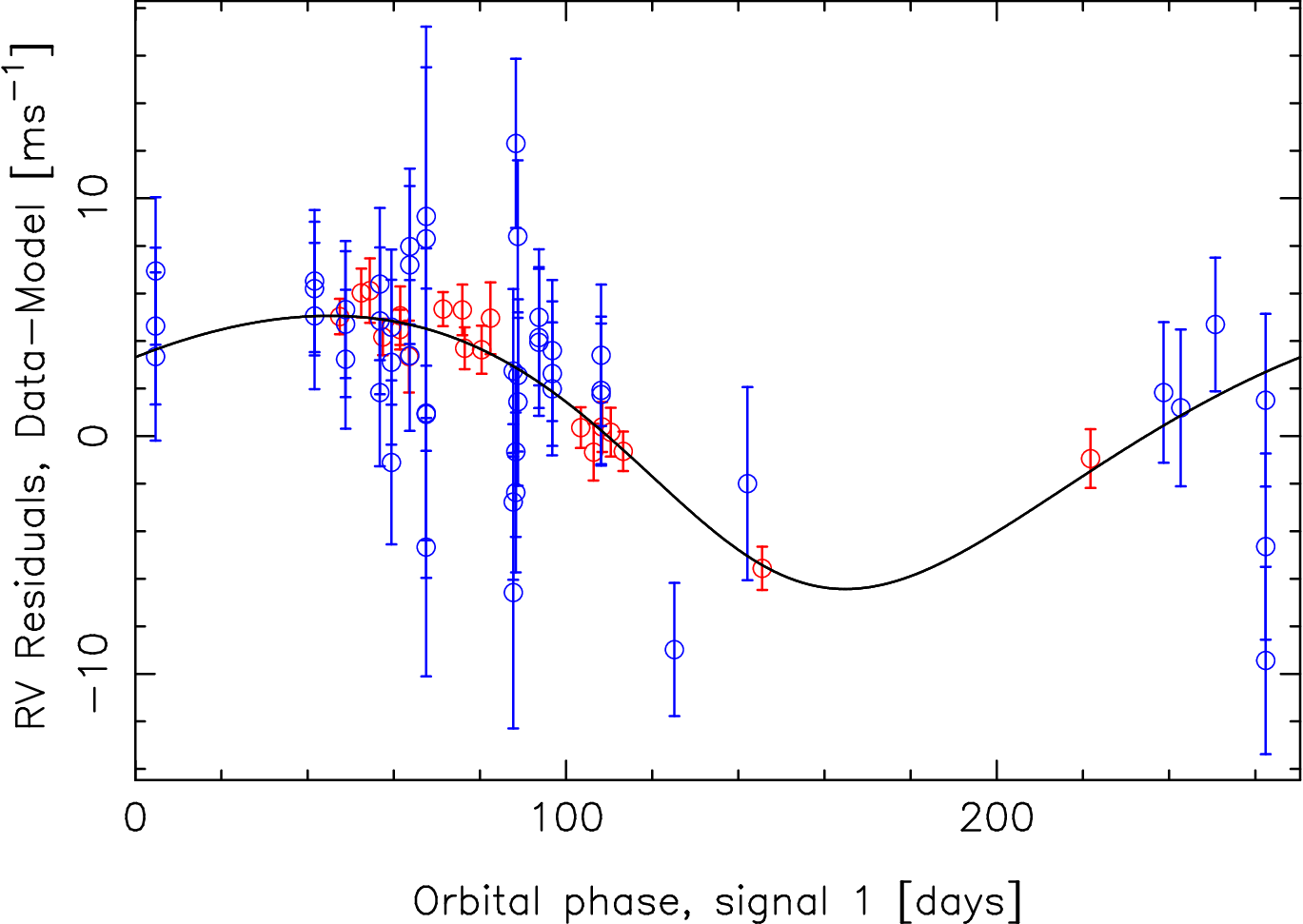}
\caption{Combined HARPS (red) and UVES (blue) radial velocities with a linear trend and reference velocities subtracted (top panel) and the phase-folded radial velocities demonstrating the presence of the Keplerian signal corresponding to GJ 739 b in the combined data.}\label{fig:GJ739_curve}
\end{figure}

GJ 739 (HIP 93206) was already included in the sample of \citet{zechmeister2009}, and consequently re-visited by \citet{tuomi2014}, but no signals have been reported in the radial velocities of the star. In addition to the two HARPS radial velocities that were available to \citet{tuomi2014}, we have obtained 17 HARPS velocities enabling the revision of the results. Consequently, we have obtained evidence in favour of a periodic signal in the combined UVES and HARPS velocities at a period of 269.59 [266.53, 273.48] days (Fig. \ref{fig:GJ739_psearch}). Although there are two prominent local maxima in the period space at periods of 142 and 250 days, these are clear aliases caused by the gap in the data between the HARPS and UVES data sets that have no overlap (see Fig. \ref{fig:GJ739_curve}, top panel) and annual gaps and the consequent aliasing frequencies, respectively. As can be seen in Fig. \ref{fig:GJ739_curve} (bottom panel), the phase coverage of the signal is not optimal. Yet, the probability maximum in the period space corresponding to the signal is strong and unique enough to qualify as a signal that satisfies our detection criteria.

We did not obtain evidence for photometric periodicities in the ASAS V-band data of GJ 739. There are thus no photometric counterparts of the radial velocity signal enabling us to interpret it as a candidate planet. Similarly, HARPS activity indicators were not found to contain evidence in favour of periodicities leaving us with a Keplerian signal in the radial velocities with no evidence pointing towards it originating from stellar activity. We thus conclude that there is evidence for a cool mini-Neptune orbiting the star.

\clearpage

\subsection{GJ 752A}\label{sec:GJ752A}

\begin{figure}
\center
\includegraphics[angle=270, width=0.49\textwidth,clip]{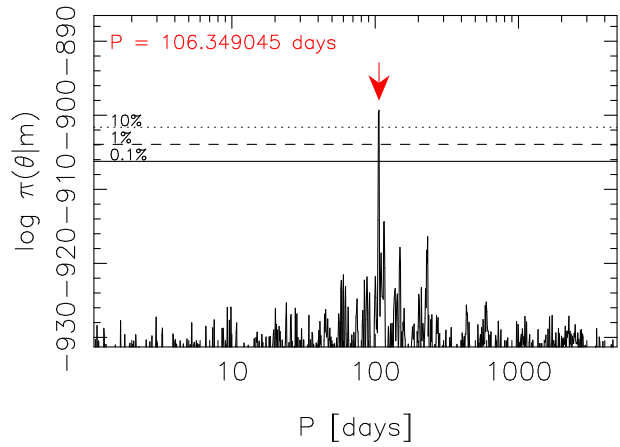}
\caption{Estimated posterior probability density of a one-Keplerian model as a function of the period parameter given the combined HARPS, HIRES, and APF data of GJ 752A.}\label{fig:GJ752A_psearch}
\end{figure}

\begin{figure}
\center
\includegraphics[angle=270, width=0.49\textwidth,clip]{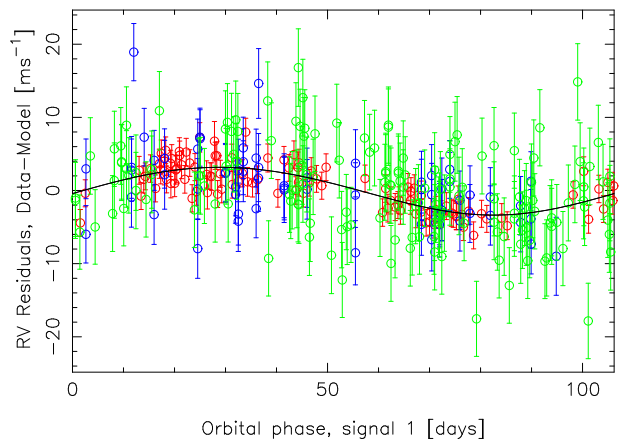}
\caption{Radial velocity curve overplotted on the phase folded HARPS (red), APF (blue) and HIRES (green) radial velocities.}\label{fig:GJ752A_curve}
\end{figure}

The posterior probability density given the combined HARPS ($N = 118$), HIRES ($N = 157$) and APF ($N = 65$) radial velocities of GJ 752A (HD 180617, HIP 94761) was found to have a clear maximum in the period space corresponding to a significant Keplerian signal (Fig. \ref{fig:GJ752A_psearch}). This signal, at a period of 106.18 [105.70, 106.66] has an amplitude of 3.20 [2.02, 4.26] ms$^{-1}$ and is detected according to our detection criteria. The phase-folded radial velocities are plotted in Fig. \ref{fig:GJ752A_curve} for visual inspection.

\begin{figure}
\center
\includegraphics[angle=270, width=0.49\textwidth,clip]{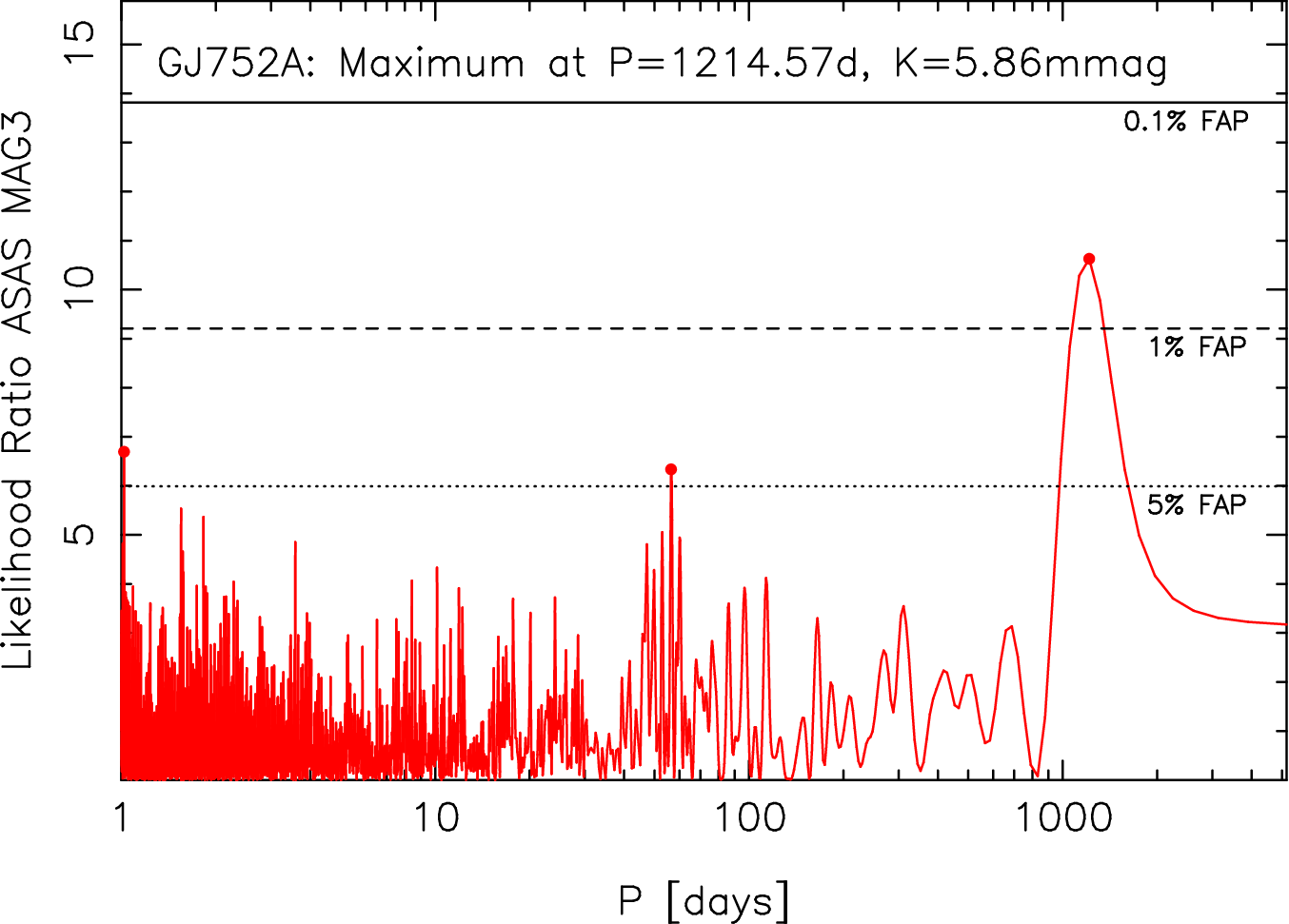}

\includegraphics[angle=270, width=0.49\textwidth,clip]{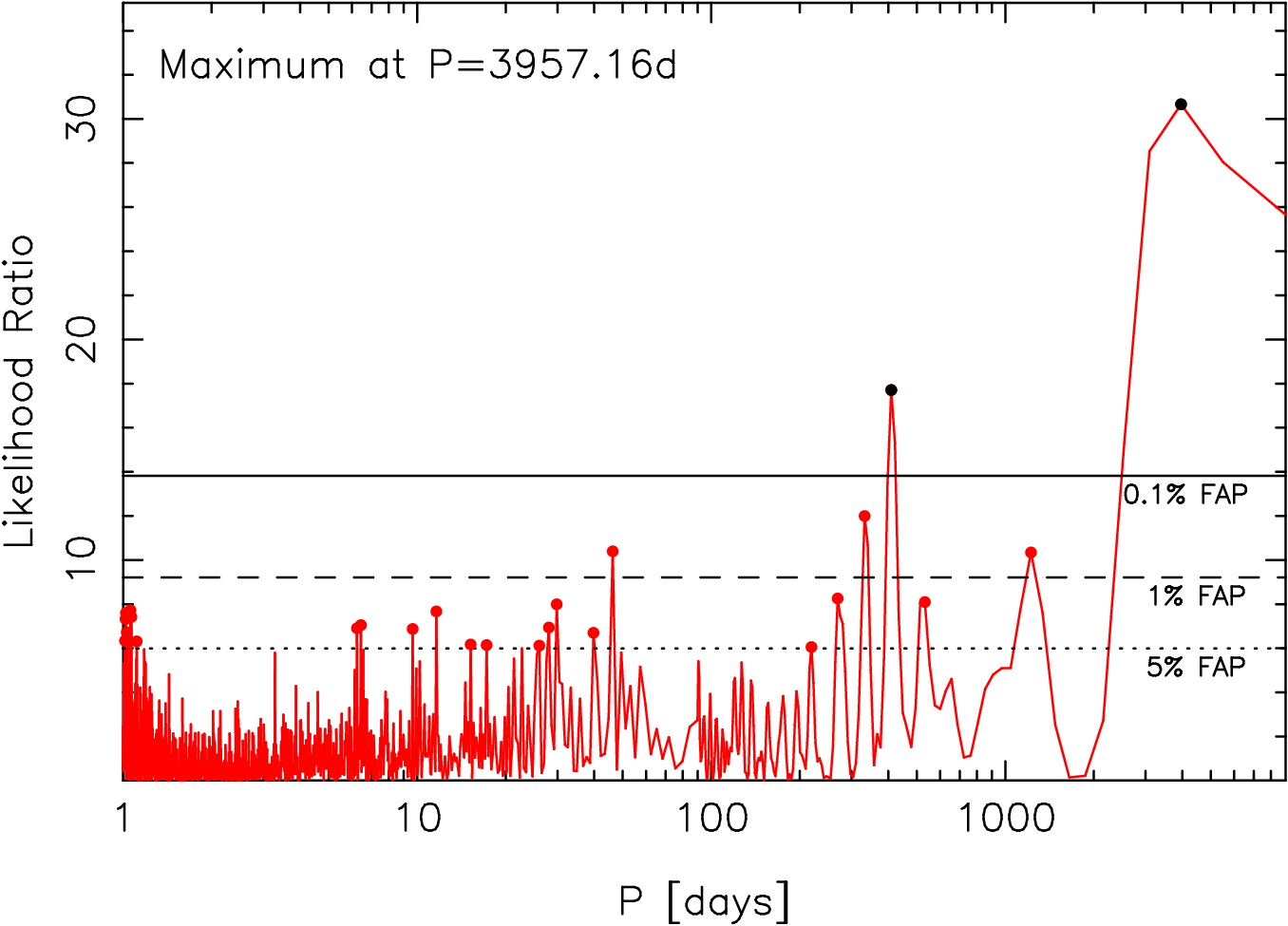}
\caption{Likelihood-ratio periodogram of the ASAS V-band photometry of GJ 752A (top) and the likelihood-ratio periodogram of the HIRES S-indices of the same star.}\label{fig:GJ752A_asas}
\end{figure}

We did not find any counterparts of the radial velocity signal in the activity indicators or photometry. However, we identified suggestive evidence for a photometric activity cycle with a period of roughly 1200 days (Fig. \ref{fig:GJ752A_asas}, top panel) that might have a counterpart in the HIRES S-indices. However, this correspondence is not entirely obvious as the most significant cycle in the S-indices is almost 4000 days long (Fig. \ref{fig:GJ752A_asas} top panel) as also reported by \citet{butler2016}.

We interpret the signal as evidence for a candidate planet orbiting the star. This planet has a minimum mass of 13.6 [8.3, 19.6] M$_{\oplus}$ and is classified as a cool Neptune.

\clearpage

\subsection{GJ 754}

We obtained a set of 138 HARPS data products from the ESO archive and calculated the corresponding radial velocities for GJ 754  based on the algorithms of \citet{anglada2012b}. However, visual inspection of the data suggested that the last epochs from JD 2457185 to 2457192 were systematically lower than the data mean. The first 130 velocities had a mean of 0.24 ms$^{-1}$ and a standard deviation of 3.53 ms$^{-1}$ whereas the last 8 velocities had a mean 7.68 ms$^{-1}$ lower than that. This was accompanied by a significant offset in the BIS and FWHM values -- the FWHM values were systematically 31.9 ms$^{-1}$ higher for the last 8 observations when first 130 FWHM measurements has a standard deviation of only 8.6 ms$^{-1}$. Because of this offset, we base our results on the first 130 radial velocities.

\begin{figure}
\center
\includegraphics[angle=270, width=0.49\textwidth,clip]{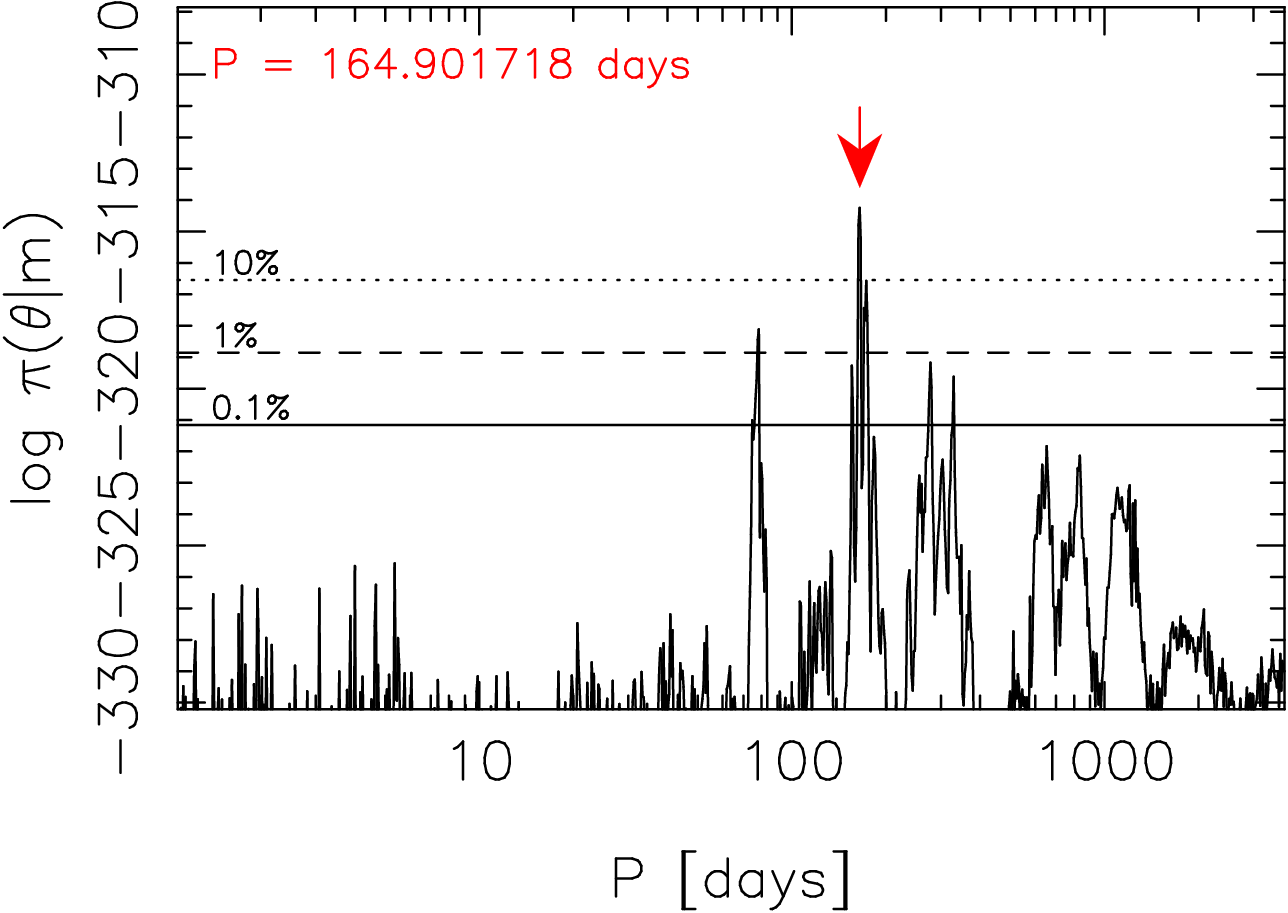}

\includegraphics[angle=270, width=0.49\textwidth,clip]{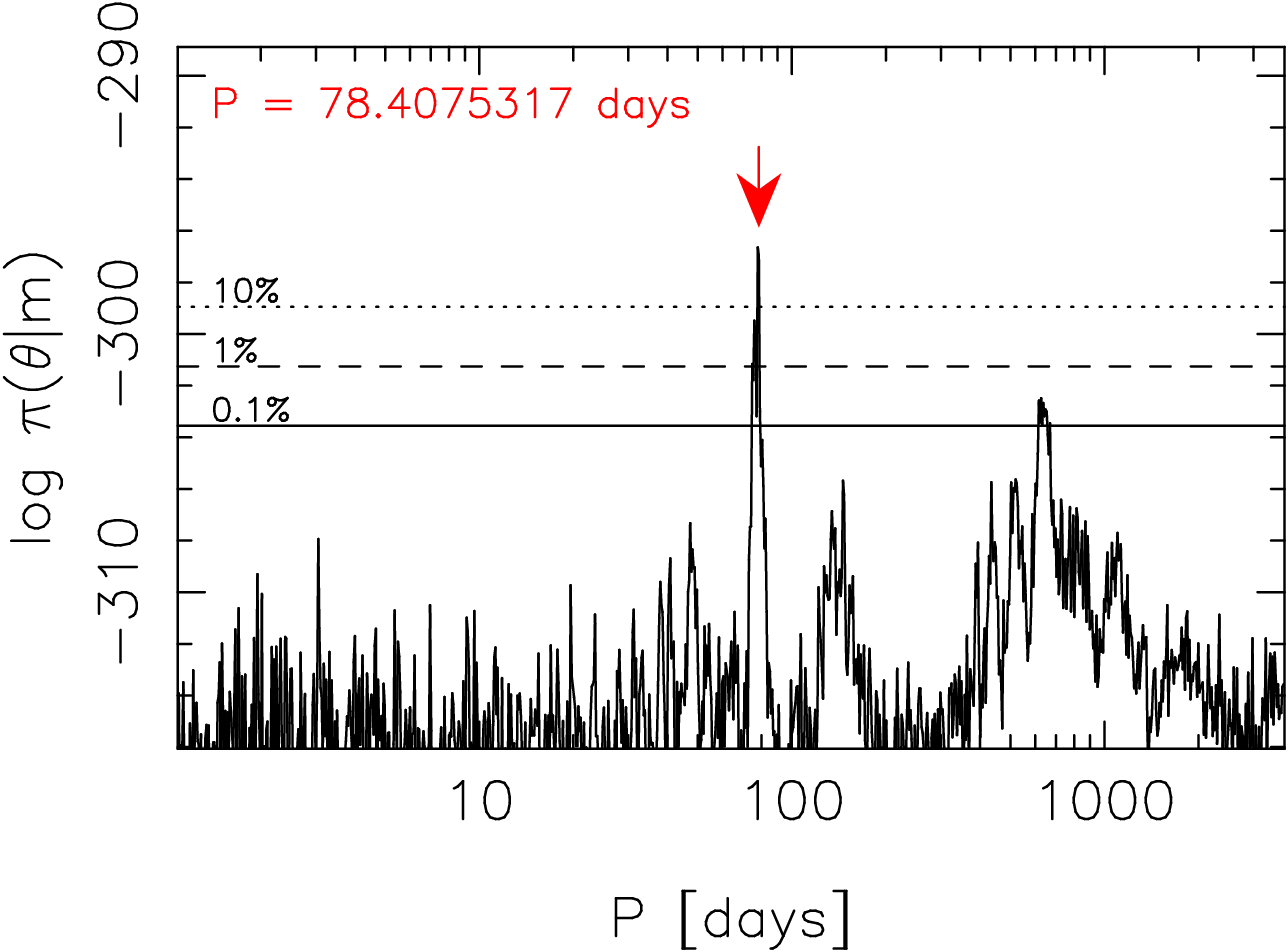}
\caption{Estimated posterior probability density given the GJ 754 radial velocities as a function of the period parameter of the signal in the one-Keplerian model (top panel) and the second signal in the two-Keplerian model (bottom panel). The red arrows indicate the positions of the global maxima and the horizontal lines denote the 10\% (dotted), 1\% (dashed), and 0.1\% (solid) equiprobability contours with respect to the maxima.}\label{fig:GJ754_psearch}
\end{figure}

The HARPS radial velocities showed evidence in favour of two signals at a periods of 165.28 [162.83, 167.28] and 78.37 [77.90, 78.93] days that satisfied our signal detection criteria (Fig. \ref{fig:GJ754_psearch}). However, there were also local maxima in the posterior density of the one-Keplerian model at 280 and 320 days almost reaching the 1\% probability threshold (Fig. \ref{fig:GJ754_psearch}, top panel).

\begin{figure}
\center
\includegraphics[angle=270, width=0.49\textwidth,clip]{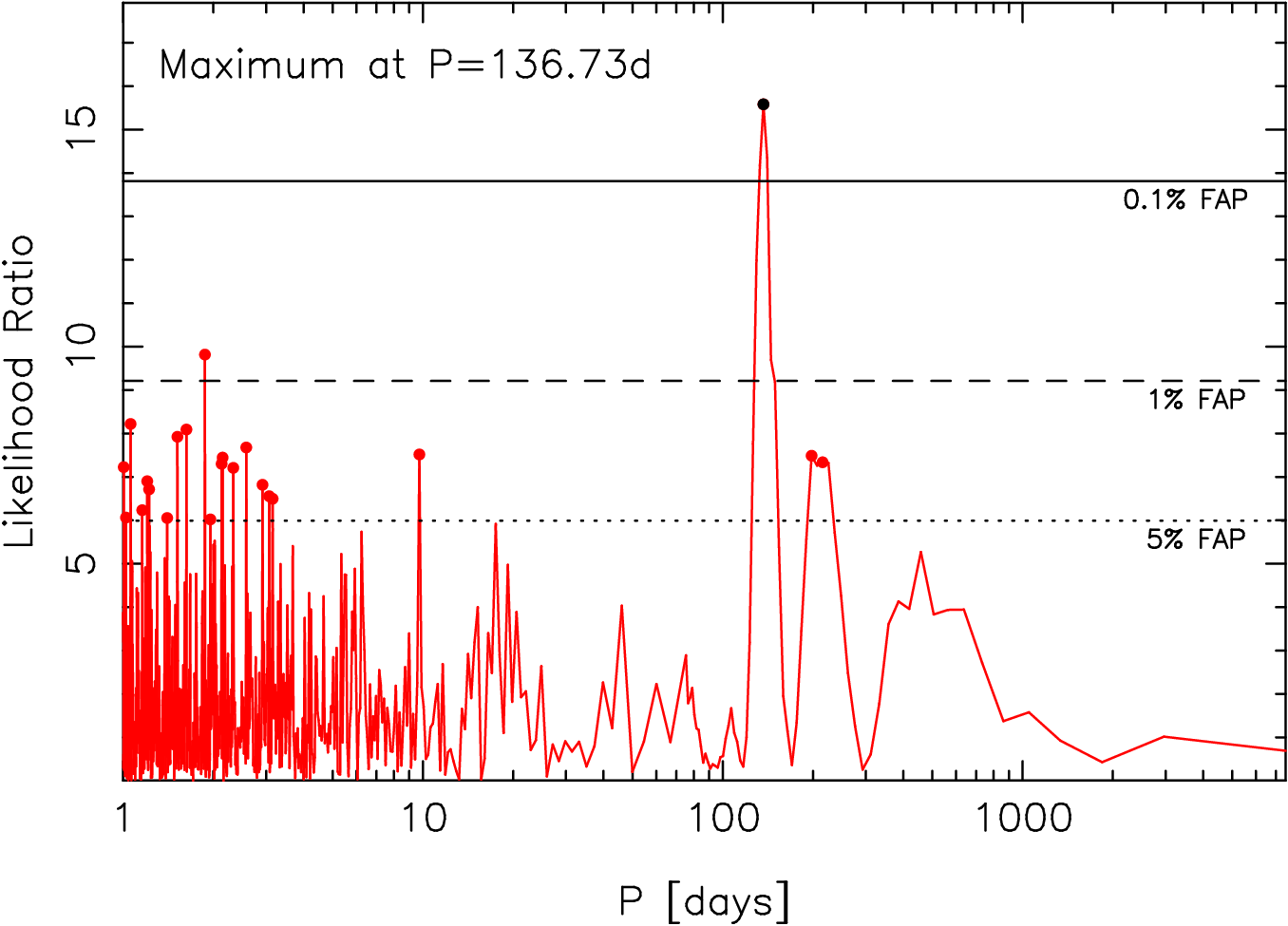}

\includegraphics[angle=270, width=0.49\textwidth,clip]{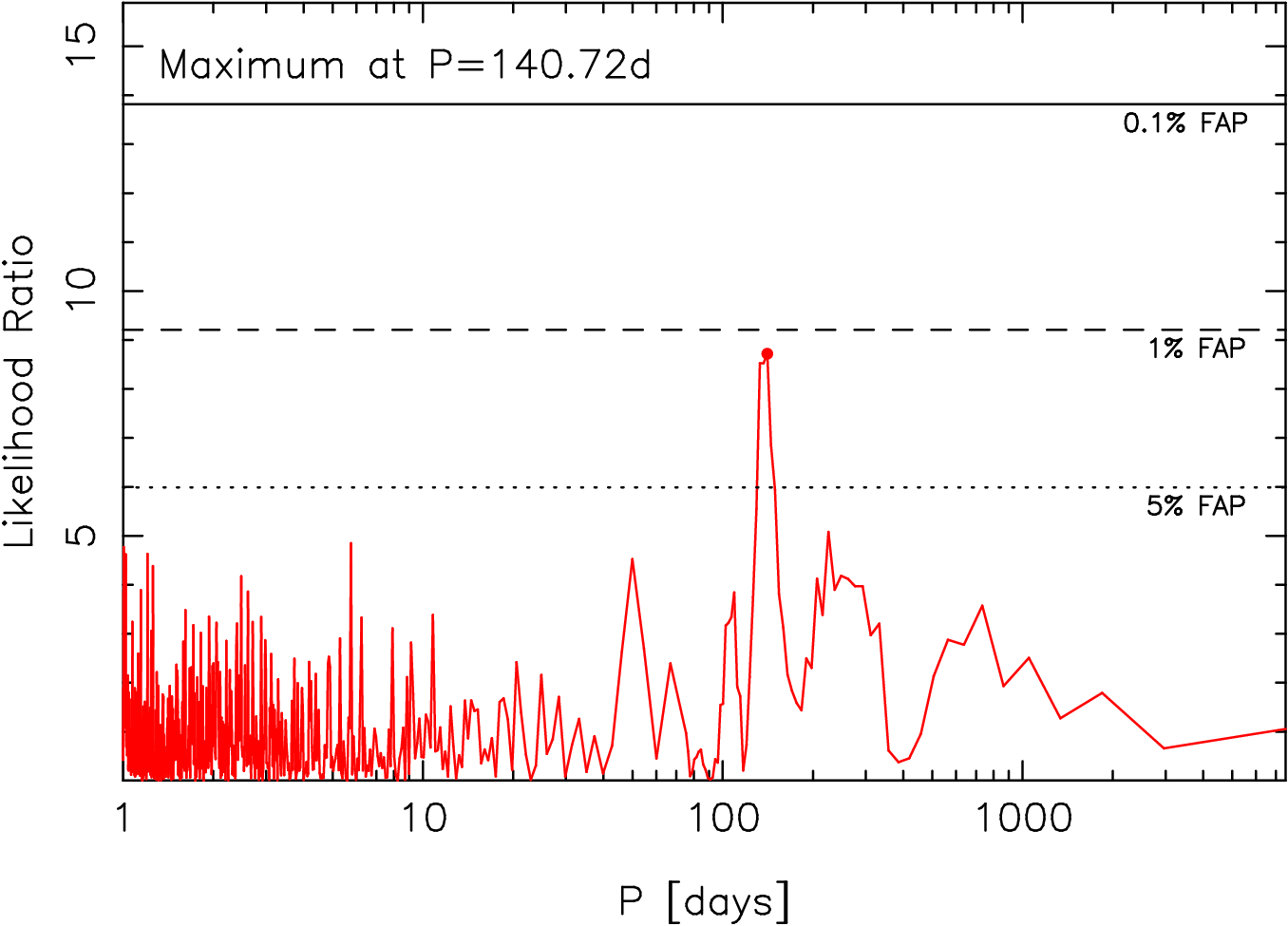}
\caption{Likelihood-ratio periodograms of the HARPS S-indices (top) and FWHM (bottom) of GJ 754.}\label{fig:GJ754_activity}
\end{figure}

When analysing the HARPS activity indices, we observed a significant signal in the S-indices at a period of 137 days (Fig. \ref{fig:GJ754_activity}, top panel). Although this signal appears to have a shorter period than the first signal detected in the radial velocities at a period of 165 days, it is rather broad and coincides with the ``sidelobe'' of the global maximum seen in Fig. \ref{fig:GJ754_psearch} (top panel). We interpret this as an indication that the 165-day radial velocity signal is caused by stellar activity rather than a planet orbiting the star. There is also tentative evidence for a similar periodicity in the HARPS FWHM values (Fig. \ref{fig:GJ754_activity}, bottom panel)

\begin{figure}
\center
\includegraphics[angle=270, width=0.49\textwidth,clip]{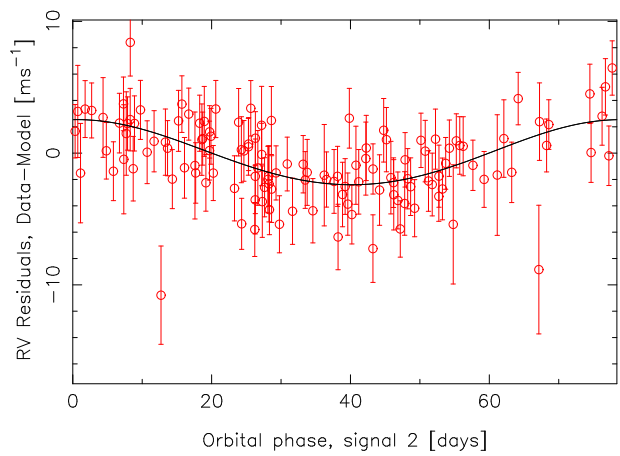}
\caption{Phase-folded HARPS radial velocities of GJ 754 with the Keplerian curve overplotted.}\label{fig:GJ754_curve}
\end{figure}

Apart from a suggestive periodicity of 200 days, the available 244 ASAS V-band photometry measurements of GJ 754 were not found contain any evidence in favour of periodic signals at or near the radial velocity signals. We therefore interpret the radial velocity signal at a period of 78 days (Fig. \ref{fig:GJ754_curve}) as evidence in favour of a planet candidate, classified as a cool mini-Neptune with a minimum mass of 9.8 [4.2, 14.5] M$_{\oplus}$, orbiting the star.

\clearpage

\subsection{GJ 784}

\begin{figure}
\center
\includegraphics[angle=270, width=0.49\textwidth,clip]{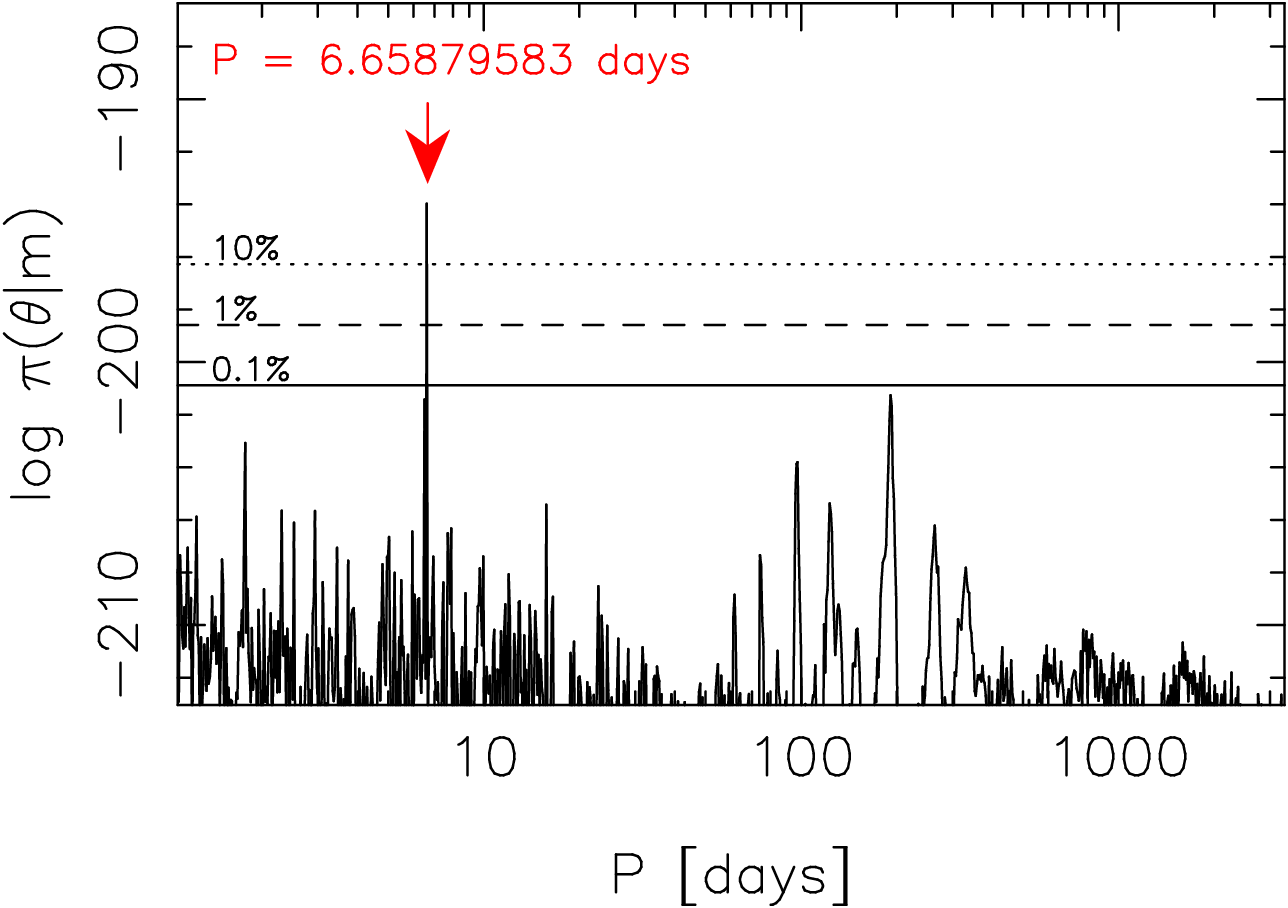}
\caption{Estimated posterior probability density as a function of the period parameter of the Keplerian signal given the combinbed HARPS and PFS radial velocities of GJ 784.}\label{fig:GJ784_psearch}
\end{figure}

\begin{figure}
\center
\includegraphics[angle=270, width=0.49\textwidth,clip]{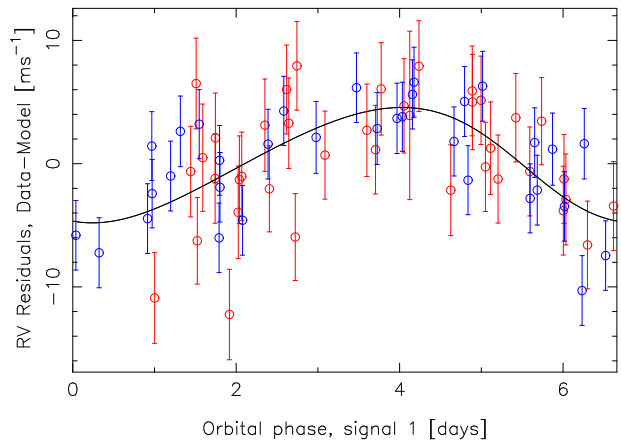}
\caption{HARPS (red) and PFS (blue) radial velocities of GJ 784 folded on the phase of the signal.}\label{fig:GJ784_curve}
\end{figure}

GJ 784 (HD 191849, HIP 99701) has been observed with PFS on 33 nights and we were able to combine the resulting radial velocities with another 39 HARPS velocity measurements. Together, these data sets contain complementary evidence in favour of a Keplerian signal at a period of 6.6591 [6.6554, 6.6625] days with an amplitude of 4.41 [2.56, 6.48] ms$^{-1}$ corresponding to a previously unknown hot mini-Neptune, with a minimum mass of 9.4 [5.2, 14.0] M$_{\oplus}$, orbiting the star. We show this signal as a unique probability maximum in Fig. \ref{fig:GJ784_psearch} and demonstrate that the signal is indeed traced by both HARPS and PFS velocities in Fig. \ref{fig:GJ784_curve}. There was no evidence for additional signals in the combined data set.

\begin{figure}
\center
\includegraphics[angle=270, width=0.49\textwidth,clip]{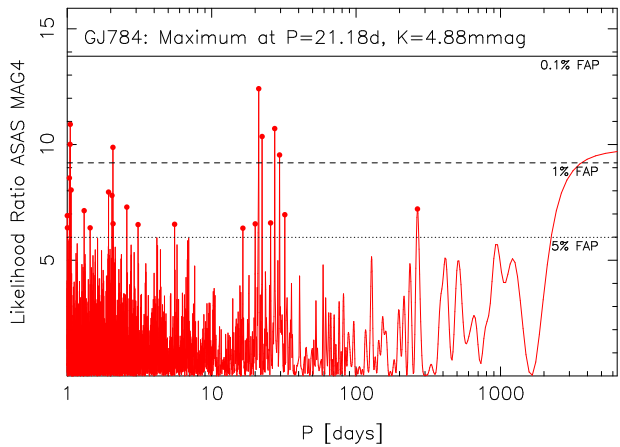}
\caption{Likelihood-ratio periodogram of the ASAS V-band photometry of GJ 784.}\label{fig:GJ784_asas}
\end{figure}

We did not find any periodicities at or near the period of the radial velocity signal in the ASAS photometry data. However, there was a prominent likelihood maximum at a period of 21.18 days in the V-band, suggestive of stellar photometric rotation at that period (Fig. \ref{fig:GJ784_asas}). This signal, together with likelihood maxima at nearby periods, were significant at a 1\% FAP level but not at 0.1\% and we thus did not interpret the perioogram analyses as evidence for a photometric rotation period of the star. But given that the prominent periodogram maxima  in the ASAS photometry data are sufficiently far from the observed period of the radial velocity signal, it is our interpretation that the radial velocity signal is indeed very likely caused by a planet orbiting the star rather than activity. Moreover, HARPS activity indicators did not show any periodic variability making the periodic variations in the radial velocities appear independent of the stellar activity thus reinforcing the planetary interpretation.

\clearpage

\subsection{GJ 821}

\begin{figure}
\center
\includegraphics[angle=270, width=0.49\textwidth,clip]{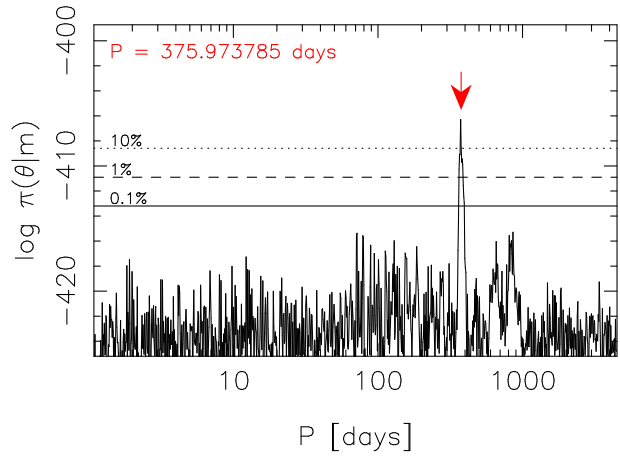}
\caption{Estimated posterior density as a function of the period parameter of the Keplerian signal given HARPS, HIRES, and UVES velocities of GJ 821.}\label{fig:GJ821_period_search}
\end{figure}

The combined HARPS ($N = 11$), HIRES ($N = 31$), and UVES ($N = 106$) radial velocities of GJ 821 (Wolf 918, HIP 104432) provided evidence in favour of a significant signal at a period of 371.8 [357.6, 393.6] days and with an amplitude of 2.79 [1.15, 4.88] ms$^{-1}$. Although this signal was found uniquely in the combined data (Fig. \ref{fig:GJ821_period_search}), it also appears to be consistent with one year given its 99\% credibility interval. This in turn makes it possible that the signal could be connected to the potential annual variations in the data rather than a genuine planet orbiting the star.

\begin{figure}
\center
\includegraphics[angle=270, width=0.49\textwidth,clip]{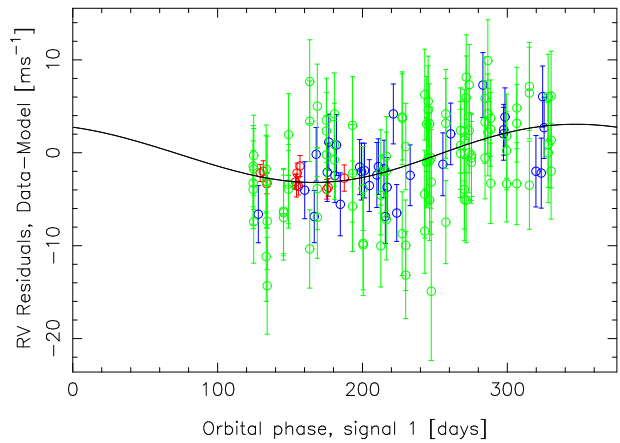}
\caption{HARPS (red), HIRES (blue), and UVES (green) radial velocities of GJ 821 folded on the phase of the signal.}\label{fig:GJ821_phased}
\end{figure}

Due to the fact that the star is not visible throughout the year, the data suffers severe annual gaps resulting in poor phase-coverage (Fig. \ref{fig:GJ821_phased}). In fact, given such a pathological phase-coverage, it is remarkable that the signal has such a unique probability maximum in the period space (Fig. \ref{fig:GJ821_period_search}). Without any evidence for periodic signals in the HARPS and HIRES activity indices we thus interpret the signal as a candidate planet orbiting the star. This is especially likely because none of the activity indicators are connected to the corresponding radial velocities and the signal thus appears to be independent of possible activity-induced variations. We thus interpret these results such that there is evidence for a cool Neptune, with a minimum mass of 18.4 [6.1, 32.1] M$_{\oplus}$ orbiting GJ 821.

The ASAS V-band photometry did not provide any additional insights into the star as we could not detect any significant periodicities in the photometric data.

\clearpage

\subsection{GJ 846}\label{sec:GJ846}

\begin{figure}
\center
\includegraphics[angle=270, width=0.49\textwidth,clip]{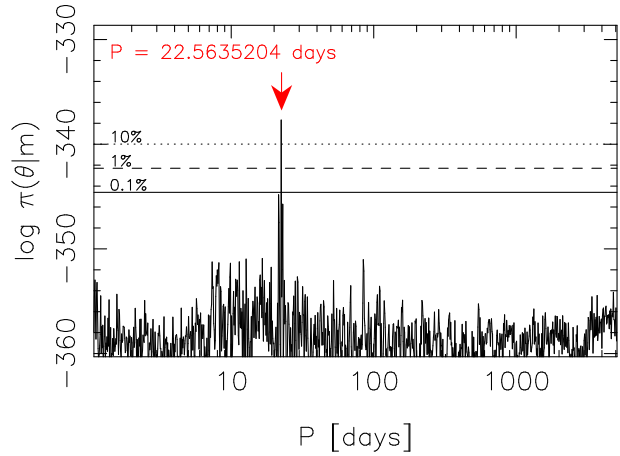}
\caption{Estimated posterior density of the period parameter of the signal in the combined HARPS, HIRES, and APF radial velocities of GJ846. The red arrow indicates the maximum and the horizontal lines denote the 10\%, 1\%, and 0.1\% probability thresholds with respect to this maximum.}\label{fig:GJ846_period_search}
\end{figure}

\begin{figure}
\center
\includegraphics[angle=270, width=0.49\textwidth,clip]{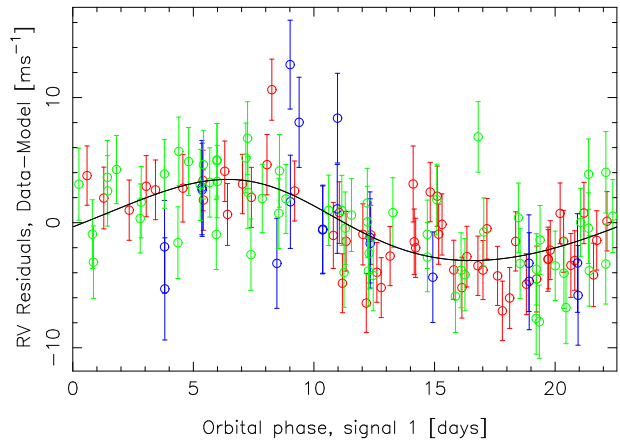}
\caption{Phase-folded radial velocities of GJ846. The red, blue, and green circles denote that HARPS, APF, and HIRES radial velocities, respectively.}\label{fig:GJ846_phased}
\end{figure}

GJ 846 (HD 209290, HIP 108782) appears to be a host to a previously unknown candidate planet that can be classified as a hot mini-Neptune. This candidate is orbiting the star with a period of 22.566 [22.538, 22.595] days. This result is evident because the combined set of radial velocities from HARPS ($N = 55$), HIRES ($N = 62$), and APF ($N = 19$) contains a clear periodic signal without counterparts in neither activity indices of HARPS or HIRES nor ASAS photometry. The existence of this signal is demonstrated in Figs. \ref{fig:GJ846_period_search} and \ref{fig:GJ846_phased} and supported by data from all three instruments.

\citet{bonfils2013} reported that they found significant periodogram powers at periods of 7.4, 7.9, and 10.6 days. However, although we also observed promising maxima in the posterior density of a two-Keplerian model at periods of 7.4 and 15.9 days, they did not correspond to unique and significant solutions and did not satisfy our signal detection criteria. \citet{bonfils2013} also claimed that the HARPS BIS values were correlated with the corresponding radial velocities. This indeed appears to be the case as the parameter $c_{\rm BIS}$ of HARPS data was found to have an estimate of -0.178 [-0.448, 0.063], which indicates that although the correlation is not significantly different from zero with a 99\% credibility, it is significant with 95\% credibility.

\clearpage

\subsection{GJ 880}

\begin{figure}
\center
\includegraphics[angle=270, width=0.49\textwidth,clip]{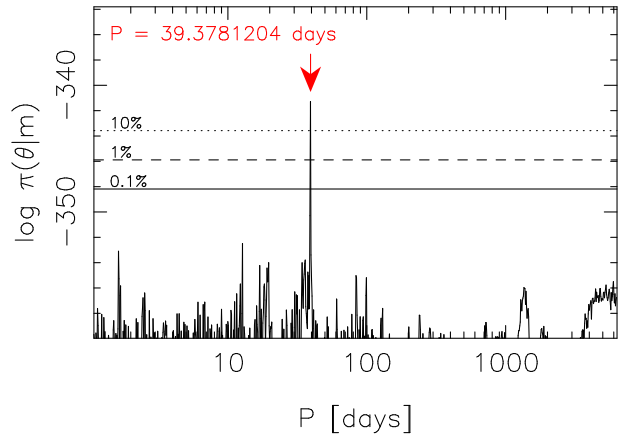}
\caption{Estimated posterior probability density as a function of the period parameter of a Keplerian signal given the GJ 880 radial velocities. The red arrow denotes the position of the global maximum and the horizontal lines denote the 10\% (dotted), 1\% (dashed), and 0.1\% (solid) equiprobability thresholds with respect to the maximum.}\label{fig:GJ880_psearch}
\end{figure}

\begin{figure}
\center
\includegraphics[angle=270, width=0.49\textwidth,clip]{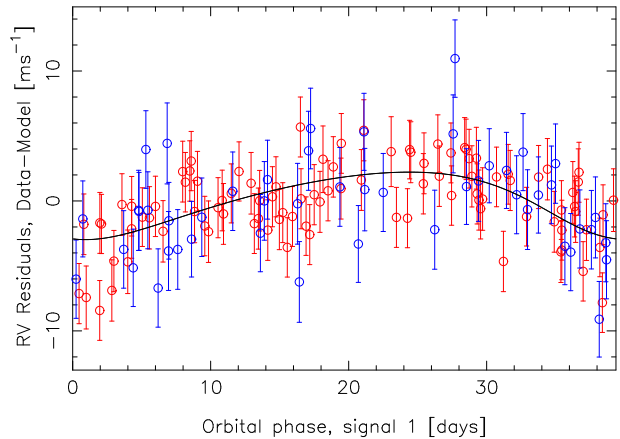}
\caption{HARPS (red) and HIRES (blue) radial velocities of GJ 880 folded on the phase of the signal.}\label{fig:GJ880_signal}
\end{figure}

We observed a clear periodic signal in the combined HARPS ($N = 105$) and HIRES ($N = 49$) radial velocity data of GJ 880 (HD 216899, HIP 113296). This signal, at a period of 39.372 [39.312, 39.423] days and an amplitude of 2.44 [1.04, 3.85] was detected according to our criteria and it was identified as a unique probability maximum in the period space (Fig. \ref{fig:GJ880_psearch}). This signal was supported by data from both instruments and the corresponding radial velocities folded on the phase of the signal are shown in Fig. \ref{fig:GJ880_signal}.

\begin{figure}
\center
\includegraphics[angle=270, width=0.49\textwidth,clip]{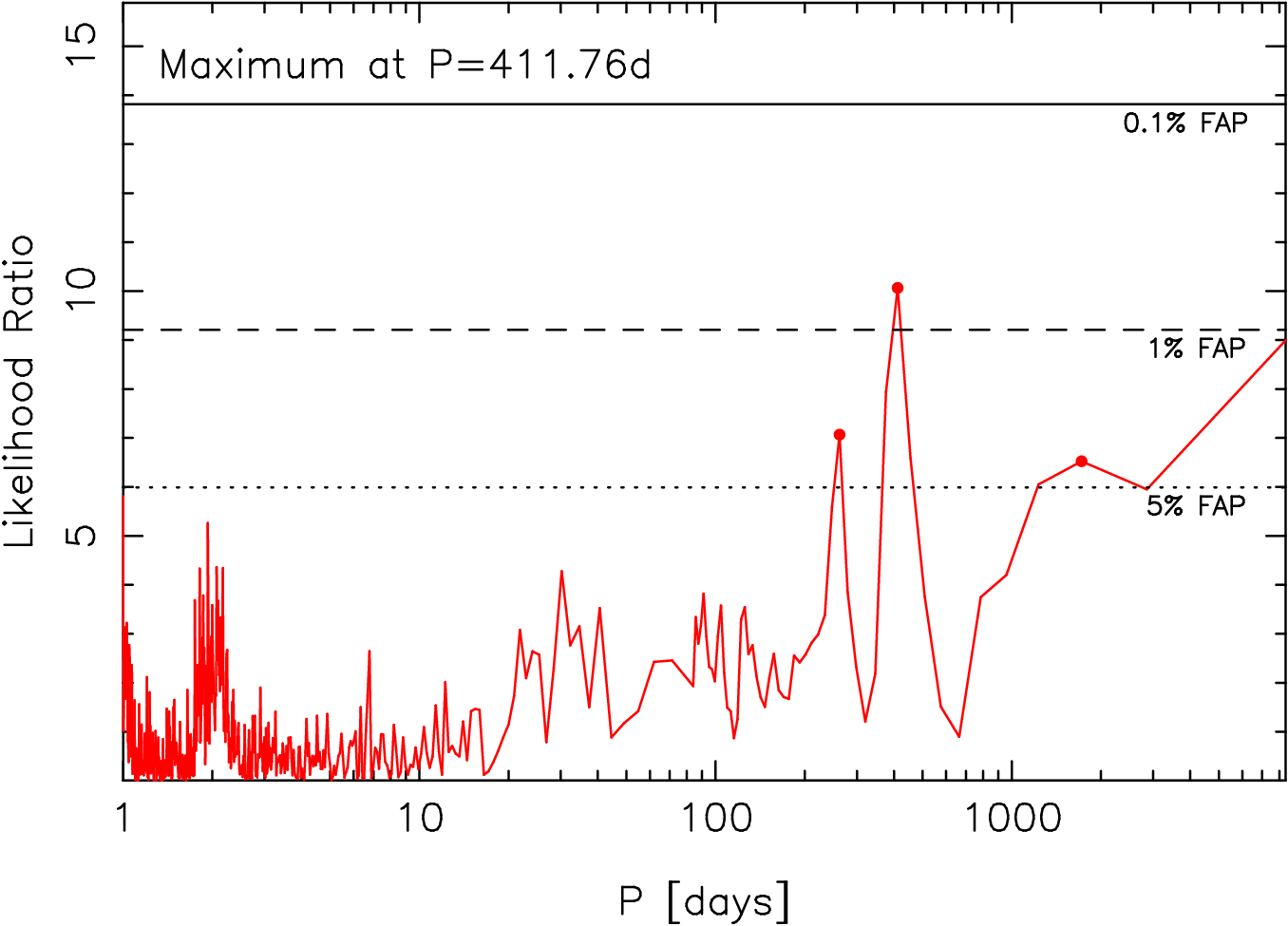}
\includegraphics[angle=270, width=0.49\textwidth,clip]{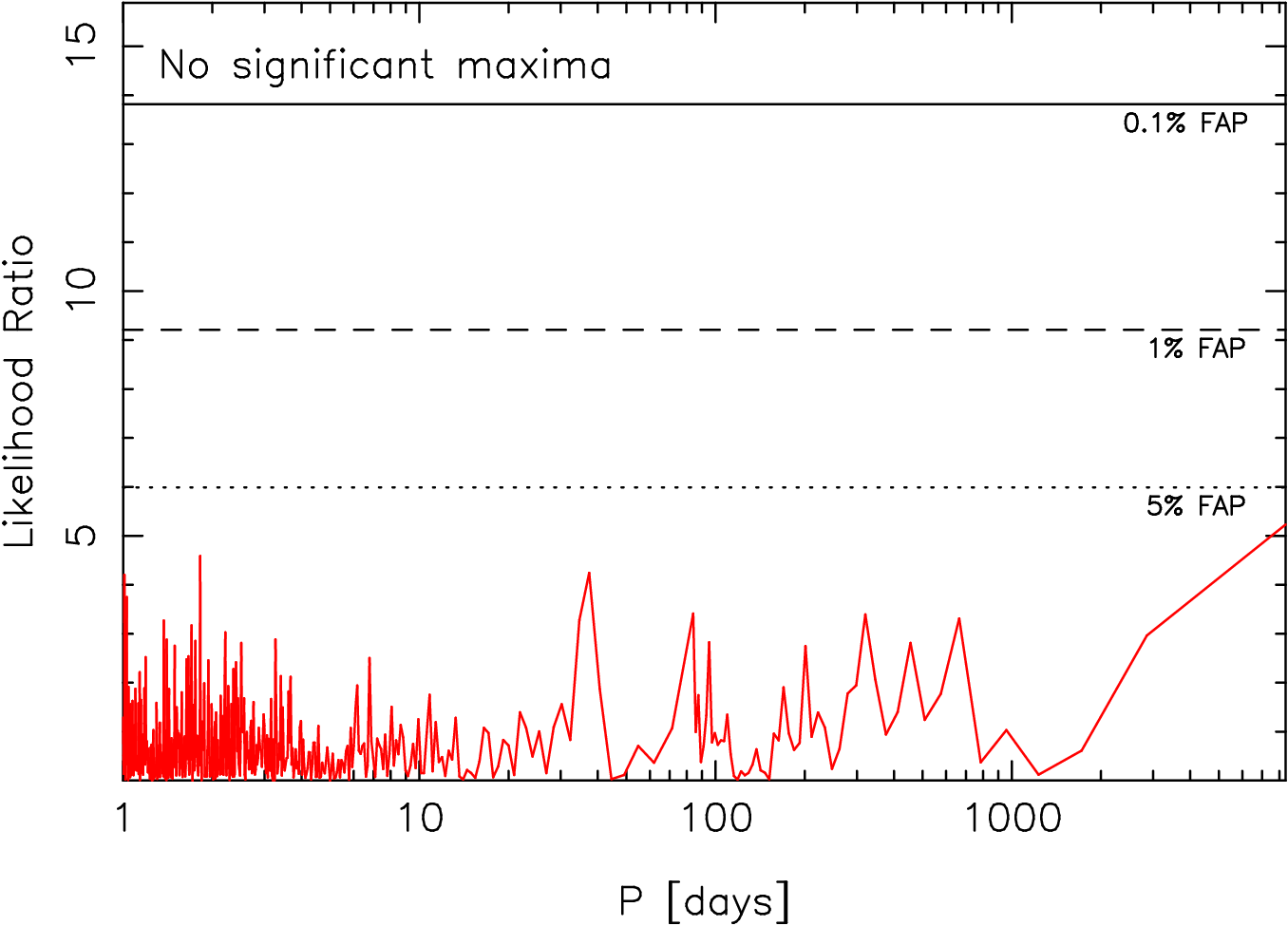}

\includegraphics[angle=270, width=0.49\textwidth,clip]{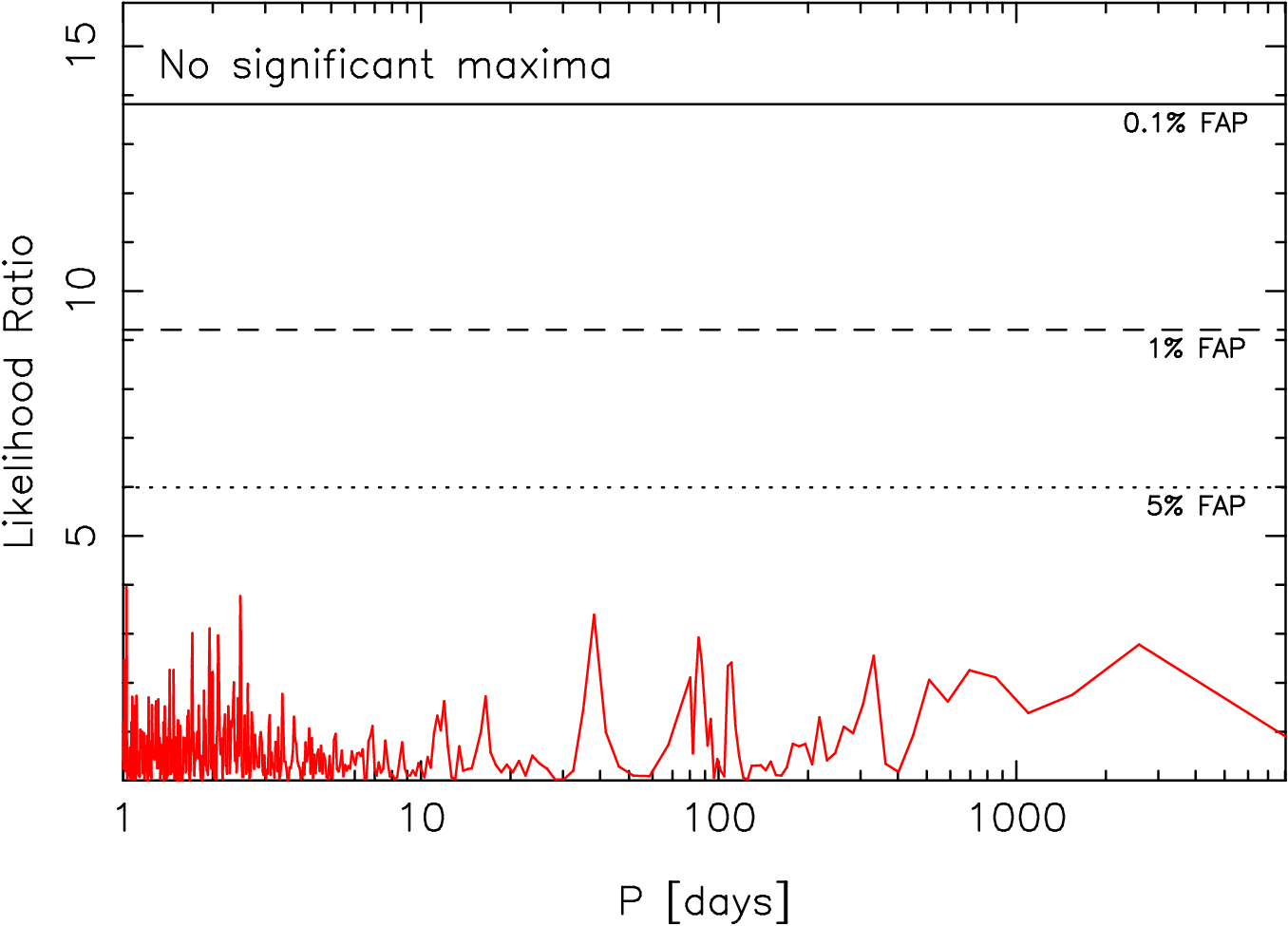}
\includegraphics[angle=270, width=0.49\textwidth,clip]{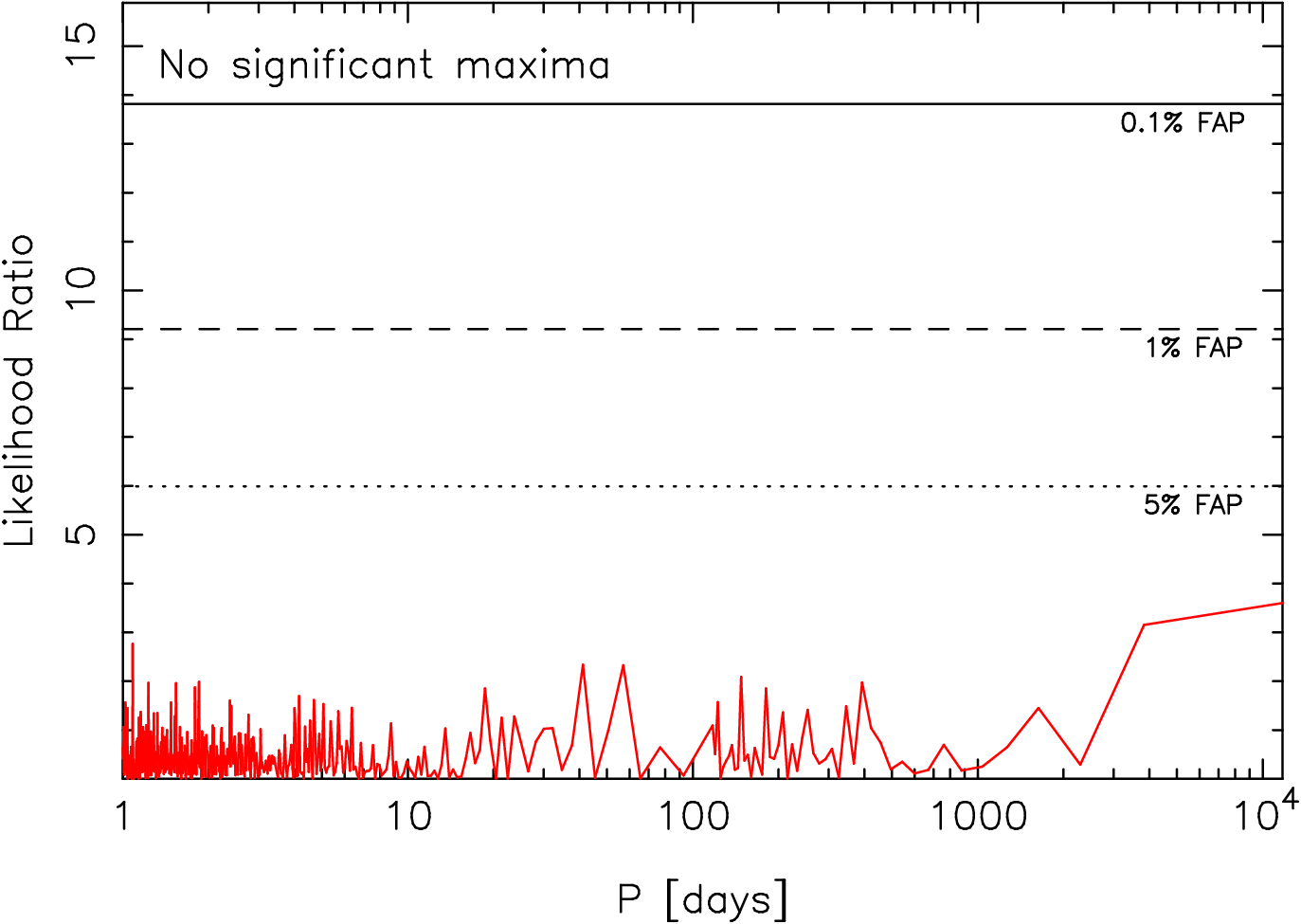}
\caption{Likelihood-ratio periodograms of the HARPS S-index (top left panel), BIS (top right), and FWHM (bottom left), as well as the HIRES S-index (bottom right) for GJ 880.}\label{fig:GJ880_activity}
\end{figure}

However, \citet{mascareno2015} reported a periodicity in the HARPS CaII H\&K emissions, as measured by their $R_{\rm HK}$ index, at a period of 37.5$\pm$0.1 days that they detected with 3$\sigma$ credibility. This signal has a period that is reasonably close to the radial velocity signal making the interpretation of the latter as a signature of a planet orbiting the star unlikely. With the corresponding S-index and the other activity proxies available, we attempted to replicate the result of \citet{mascareno2015} but failed to find counterparts for the radial velocity signal. We have plotted the periodograms of HARPS S-index, BIS, and FWHM as well as the HIRES S-index in Fig. \ref{fig:GJ880_activity}. The only significant maximum was found in the periodogram of the HARPS BIS values at a period of 410 days. This maximum was accompanied with a local maximum at a period of 230 days. We note that these are likely connected to stellar activity as we also spotted a photometric signal in the ASAS V-band data at a period of 270 days. However, we did not obtain any evidence for signals at or near the radial velocity signal and thus interpret theis signal as evidence for a candidate planet orbiting the star. This candiate planet, with a minimum mass of 8.5 [3.8, 14.3] M$)_{\oplus}$, is classified as a warm mini-Neptune.

The HARPS FWHM values are correlated with the corresponding radial velocities with a 95\% credibility but not with 99\% credibility. The parameter $c_{\rm FWHM}$ quantifying the dependence of velocities on FWHM has a value of 0.097 [-0.005, 0.211]. However, the radial velocity signal is independent of this correlation. Moreover, there is no evidence for correlations with other indices. This further suggests that the radial velocity signal is not connected to the stellar activity.

However, Due to the fact that \citet{mascareno2015} reported a periodicity in the CaII H\&K emission close to that observed in the radial velocities makes our interpretation that there is evidence for a candidate planet orbiting GJ 880 questionable. Yet, we have proceeded according to our detection criteria and cannot ignore the rather strong radial velocity signal, supported by data from two independent instruments, or conclusively interpret it as an activity-induced one in the current work.

\clearpage

\subsection{GJ 891}

\begin{figure}
\center
\includegraphics[angle=270, width=0.49\textwidth,clip]{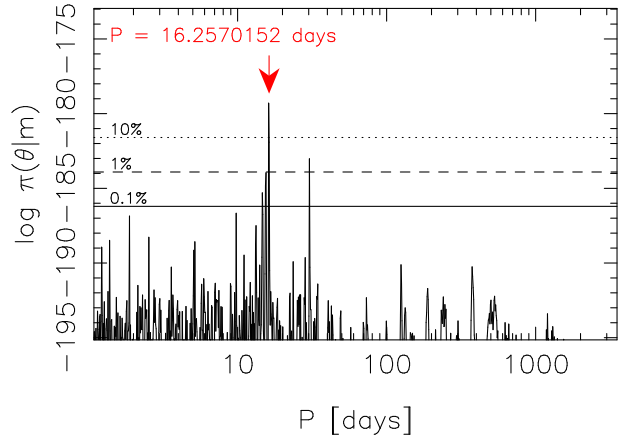}
\caption{As in Fig. \ref{fig:GJ846_period_search} but for the combined HARPS, HIRES, and UVES data of GJ 891.}\label{fig:GJ891_period_search}
\end{figure}

\begin{figure}
\center
\includegraphics[angle=270, width=0.49\textwidth,clip]{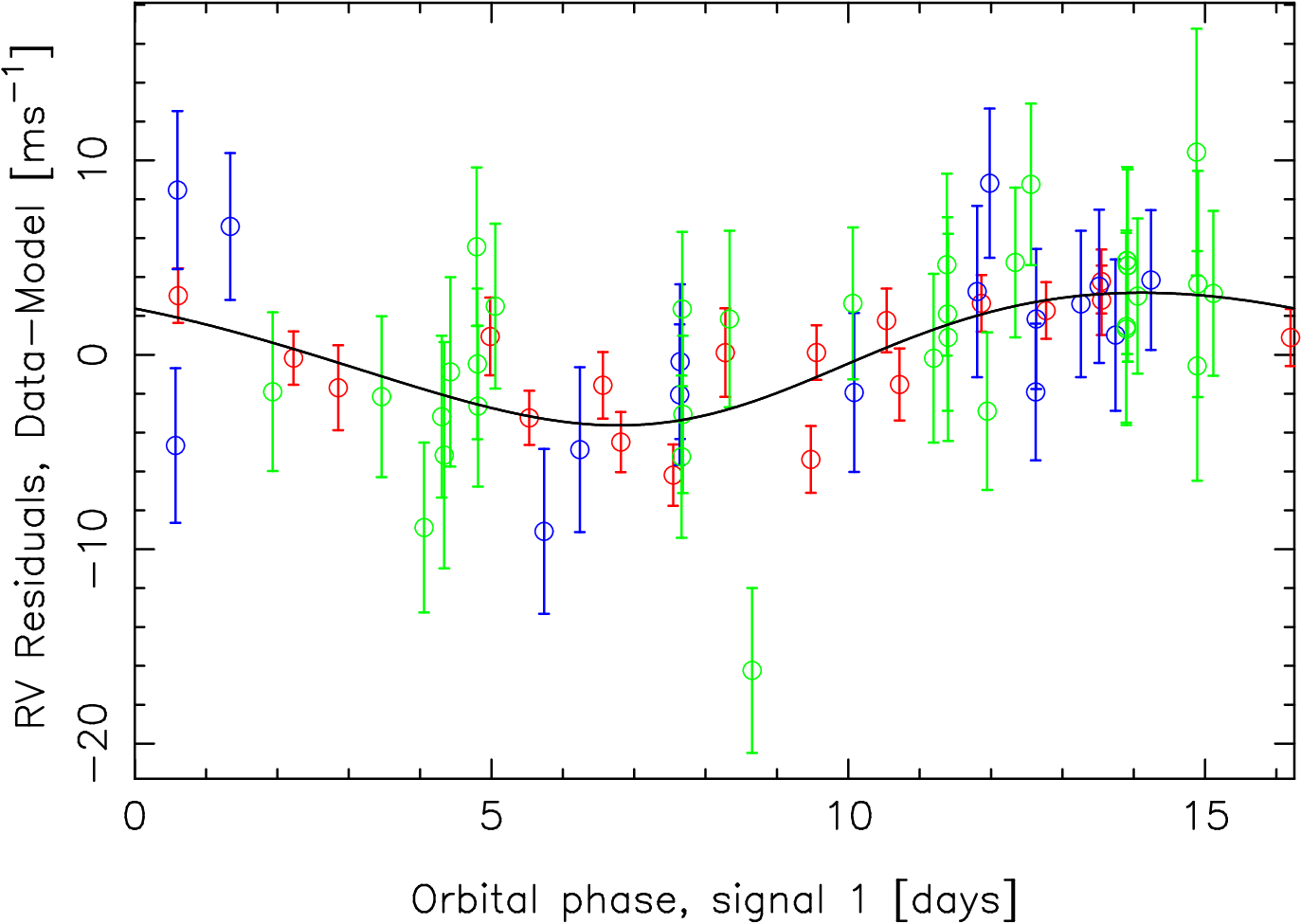}

\includegraphics[angle=270, width=0.49\textwidth,clip]{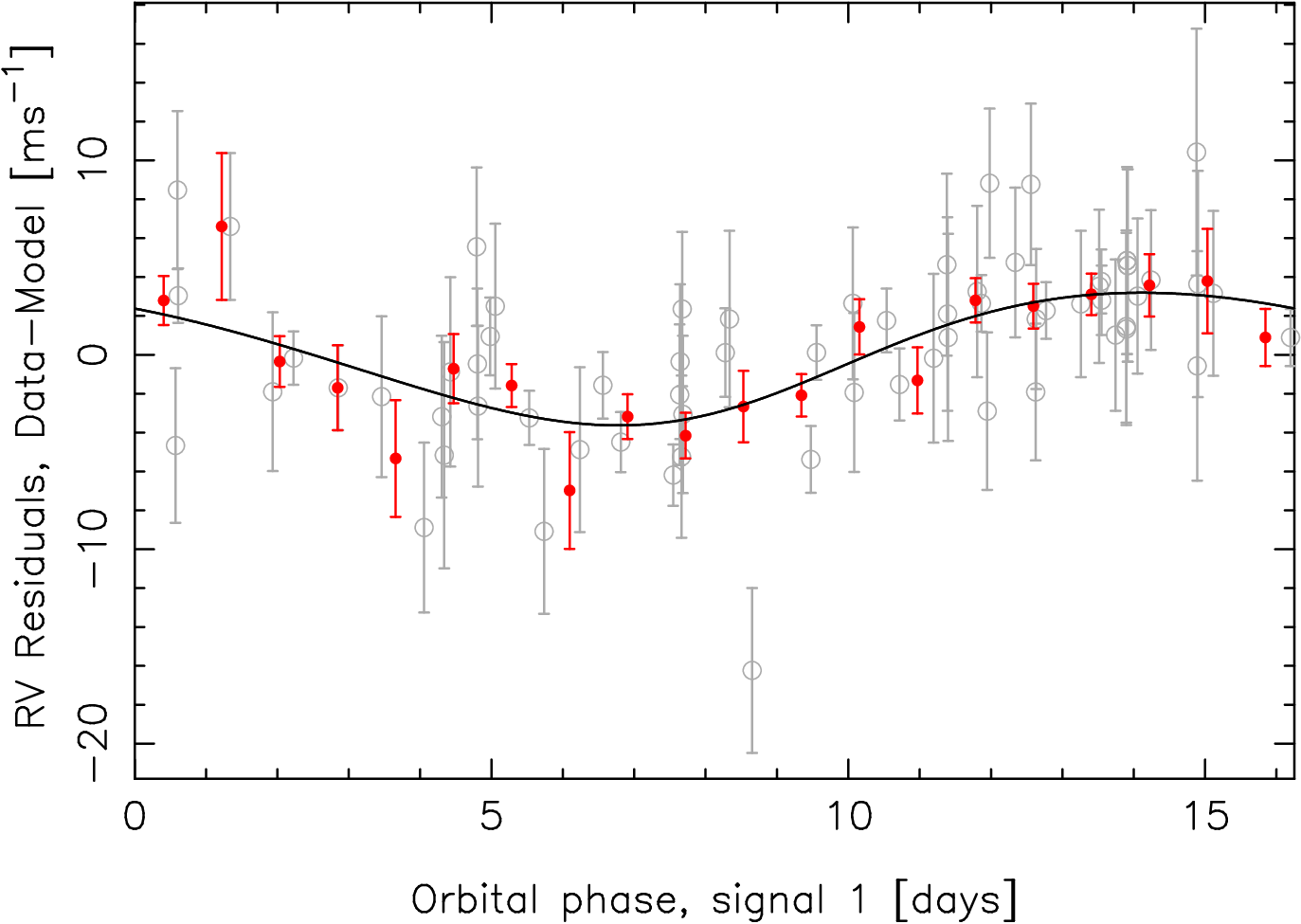}
\caption{Top panel: combined HARPS (red), HIRES (blue), and UVES (green) radial velocities of GJ 891 folded on the phase of the signal. Bottom panel shows the same for binned velocities (red filled circles). 12 UVES velocities have been ignored in this figure because they had uncertainties in excess of 10 ms$^{-1}$.}\label{fig:GJ891_phased}
\end{figure}

The combined HARPS, HIRES, and UVES radial velocities of GJ 891 (HIP 114411) support the existence of a periodic signal at a period of 16.2479 [16.2345, 16.2649] days with an amplitude of 4.85 [3.14, 6.76] ms$^{-1}$ (Fig. \ref{fig:GJ891_period_search}). Although not entirely unique with a local probability maximum at a period 30.6 days, this signal is supported independently by all three data sets with the maximum log-likelihood values of HARPS, HIRES and UVES increasing from -43.9, -47.7 and -159.1 to -35.2, -43.8, and -154.0, respectively. The period is different from the one discussed by \citet{tuomi2014} who reported that they found a signal at a period of 30.55 days in the combined HARPS and UVES data. However, the solution of \citet{tuomi2014} is present in the data as the strongest local probability maximum (Fig. \ref{fig:GJ891_period_search}). Moreover, as this local maximum is not independent from the global one because it is not possible to adjust two Keplerian signals to the data in a meaningful way and because our DRAM searches for a second signal failed to identify any significant solutions, it appears likely that the two maxima are representative of the same signal. We have plotted the radial velocities folded on the signal phase in Fig. \ref{fig:GJ891_phased}.

We could not identify any periodic signals in the HARPS or HIRES activity indicators of GJ 891. Moreover, the radial velocities appeared to be independent of the radial velocities and thus of the signal. We thus interpret this signal as evidence for a hot mini-Neptune with a minimum mass of 11.5 [7.1, 16.5] M$_{\oplus}$ orbiting the star.

\begin{figure}
\center
\includegraphics[angle=270, width=0.49\textwidth,clip]{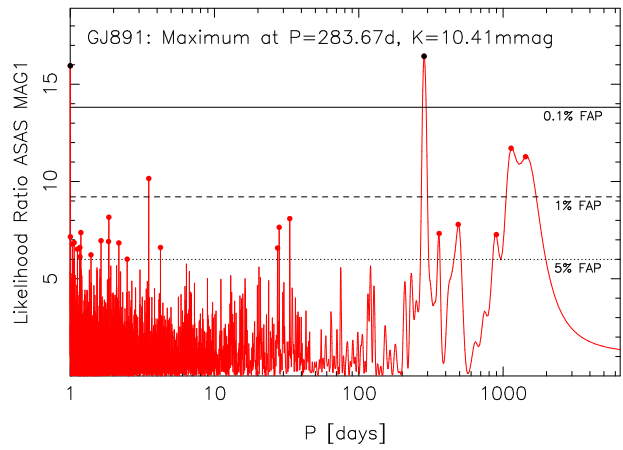}
\caption{Likelihood periodograms of ASAS photometry data of GJ 891.}\label{fig:GJ891_asas}
\end{figure}

The 596 ASAS V-band photometry observations suggest that there is a photometric cycle at or around a period of 284 days (Fig. \ref{fig:GJ891_asas}). However, there were no photometric periodicities at or near the period of the radial velocity signal.

\clearpage

\subsection{GJ 1044}

\begin{figure}
\center
\includegraphics[angle=270, width=0.49\textwidth,clip]{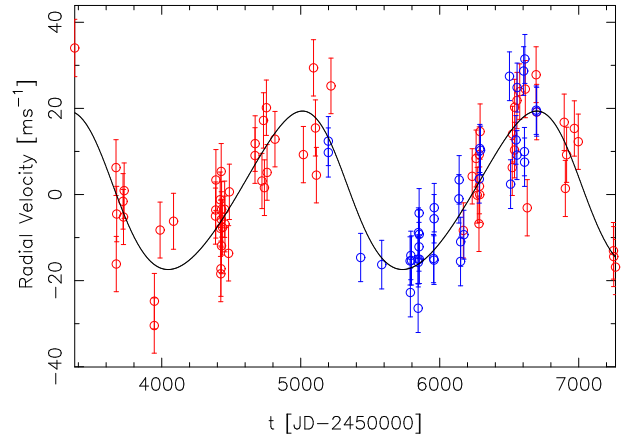}
\caption{Signal in the combined HARPS (red) and PFS (blue) velocities of GJ 1044 corresponding to a giant planet orbiting the star.}\label{fig:GJ1044_signal}
\end{figure}

The dominant feature in the combined HARPS and PFS radial velocities of GJ 1044 (HIP 10337) is strong linear acceleration of 7.58$\pm$0.60 ms$^{-1}$year$^{-1}$. However, in addition to this trend there is also clear evidence for a long-period cycle corresponding to the signal of a giant planet with a minimum mass of 252.4 [179.6, 332.9] M$_{\oplus}$ and an orbital period of 1680 [1580, 1810] days. Given the large amplitude of this signal of 16.91 [12.35, 21.03] ms$^{-1}$ it is remarkable that this signal has not been reported previously \citep[e.g.][]{gaidos2013}. However, this is likely due to the fact that neither the HARPS data available for \citet{gaidos2013} nor PFS data covers the full orbital cycle of the planet and the current work is the first one to combine these two sources of information in such a manner that there are no significant gaps left in the phase coverage of the signal. We show the radial velocity signal of GJ 1044 b in Fig. \ref{fig:GJ1044_signal} demonstrating its existence and the phase-coverage of bothe HARPS and PFS data sets.

\begin{figure}
\center
\includegraphics[angle=270, width=0.49\textwidth,clip]{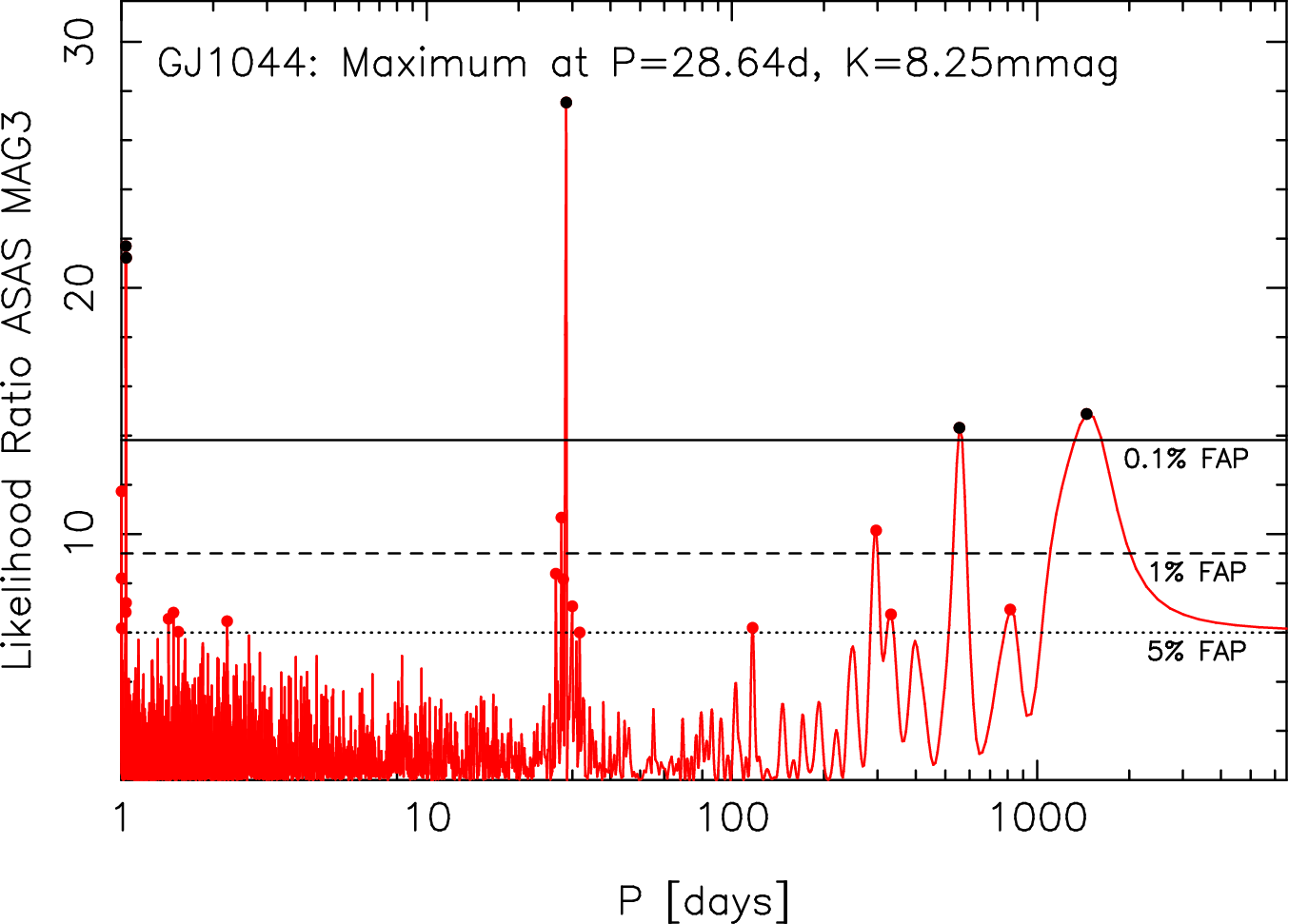}
\caption{Likelihood-ratio periodogram of ASAS V-band photometry of GJ 1044. The red (black) filled circles denote maxima that exceed the 5\% (0.1\%) FAP hreshold.}\label{fig:GJ1044_asas}
\end{figure}

The set of 429 ASAS V-band photometry measurements contain strong evidence for a photometric rotation periodicity at a period of 28.64 days (Fig. \ref{fig:GJ1044_asas}) as well as a probable activity cycle of 1500 days. We interpret this signal as evidence for the rotation period of the star.

\clearpage

\subsection{GJ 1148}

\begin{figure}
\center
\includegraphics[angle=270, width=0.49\textwidth,clip]{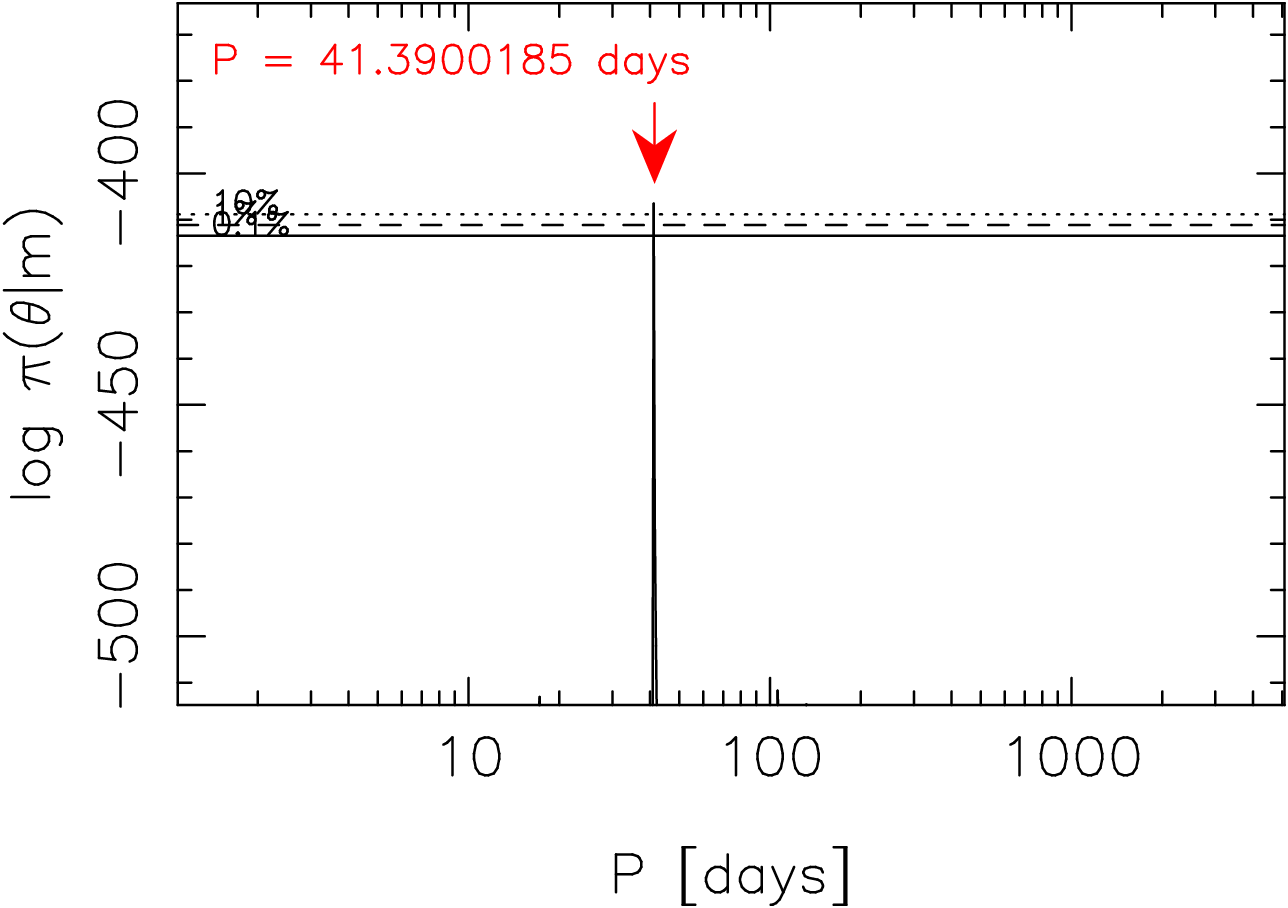}

\includegraphics[angle=270, width=0.49\textwidth,clip]{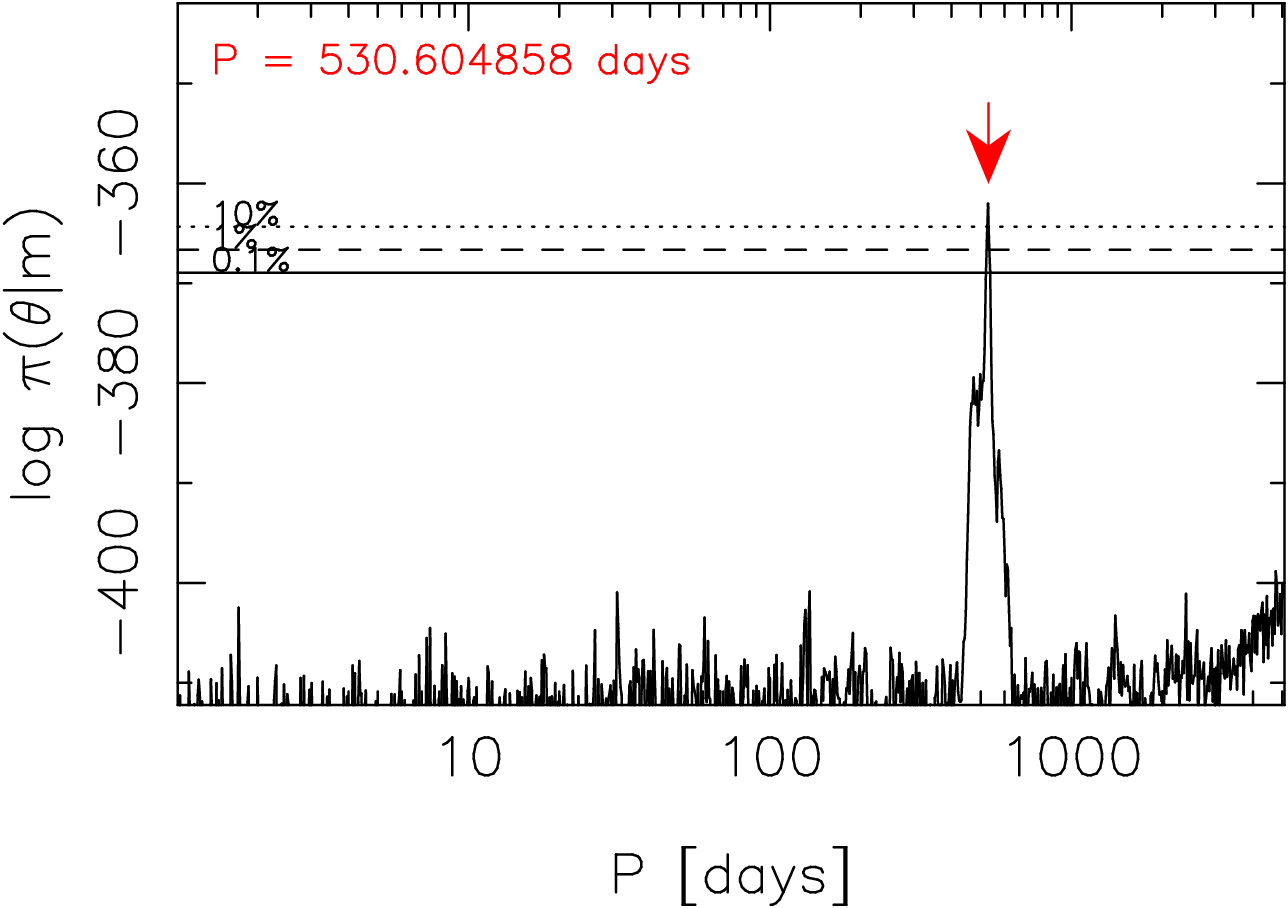}
\caption{Posterior densities as functions of the period of the $k$th signal in a $k$-Keplerian model. From top to bottom: $k=1$ and $k=2$. Red arrows denote the locations of the global maxima and the horizontal lines denote the equiprobability thresholds corresponding to 10\% (dotted line), 1\% (dashed line), and 0.1\% (solid line) of the maxima.}\label{fig:GJ1148_period_search}
\end{figure}

\citet{haghighipour2010} reported the discovery of a planet orbiting GJ 1148 (Ross 1003, HIP 57050) with a minimum mass of 0.3 M$_{\rm J}$ and an orbital period of 41.1 days bsed on HIRES radial velocities ($N = 37$) of the star. We encountered no difficulties in identifying the corresponding signal in an updated set of HIRES data with $N = 123$ (Fig. \ref{fig:GJ1148_period_search}, top panel). However, as also observed by \citet{haghighipour2010}, the candidate planet has an orbital eccentricity that is statistically significantly different from zero -- our estimate for this eccentricity is 0.36 [0.30, 0.42]. This, in turn, suggests that another companion might be orbiting the star elevating the eccentricity of GJ 1148 b due to gravitational planet-planet interactions. We therefore searched for additional signals in the data.

There is indeed strong evidence for an additional signal in the HIRES radial velocities of GJ 1148 at a period of 532.3 [520.8, 540.2] days. This signal was also reported in \citep{butler2016}. We interpret this signal as a planet candidate orbiting the star corresponding to cool planet with a minimum mass of 58.8 [40.7, 78.8] M$_{\oplus}$. We did not obtain evidence for additional candidate planets orbiting the star.

The HIRES S-indices showed weak evidence for periodicities at periods of 16.44 and 255 days but they did not exceed the 0.1\% FAP. Moreover, the S-indices were not connected to the radial velocities in s statistically significant manner. It is thus apparent that the two signals are indeed caused by planets orbiting the star rather than stellar activity cycles and/or rotation. This interpretation is supported by the fact that \citet{haghighipour2010} identified a photometric rotation period of the star to be 98 days, which neither coincides nor is very close to either of the two radial velocity signals. We could not verify this result based on ASAS data because none was available for GJ 1148.

\begin{figure}
\center
\includegraphics[angle=270, width=0.49\textwidth,clip]{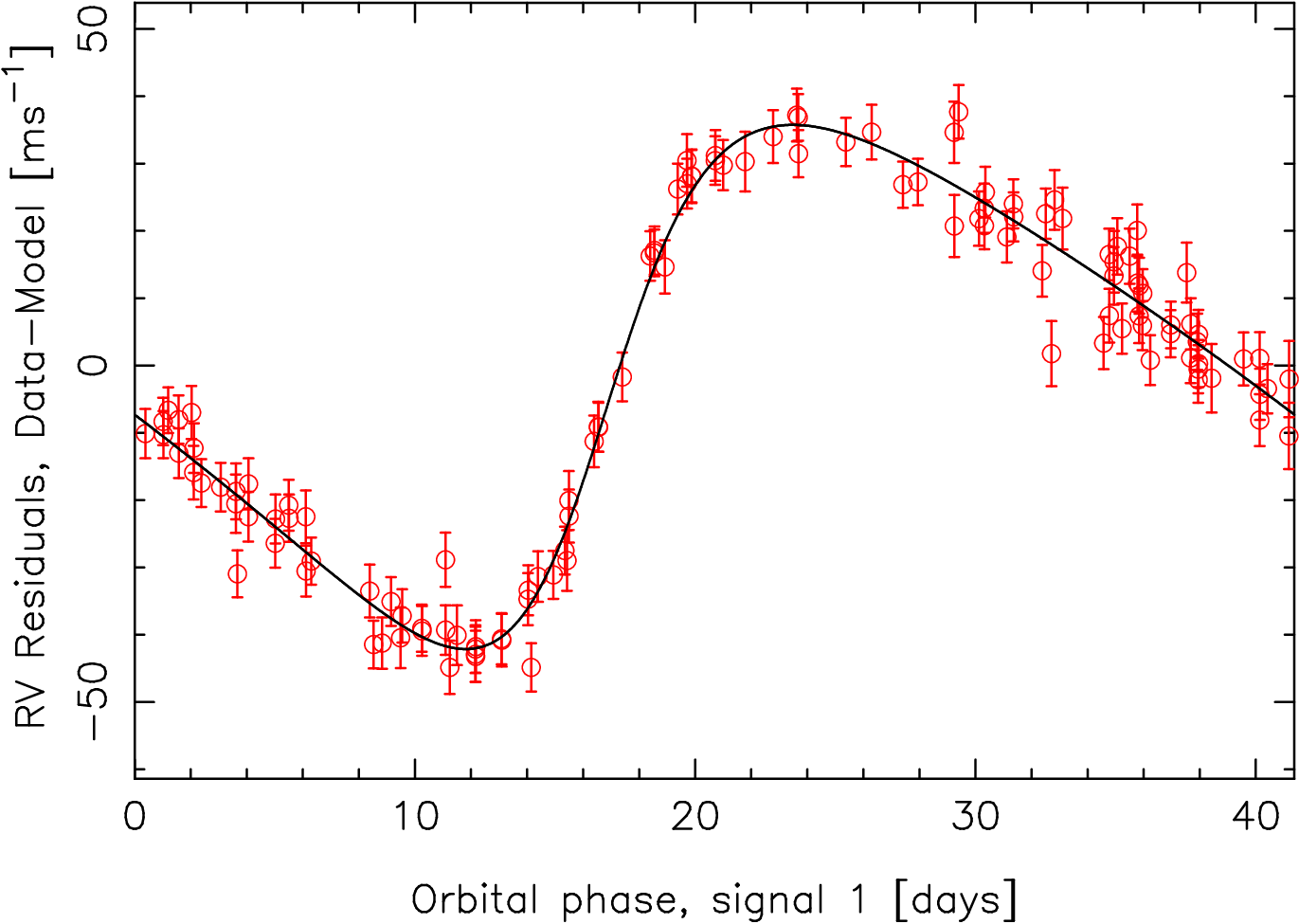}

\includegraphics[angle=270, width=0.49\textwidth,clip]{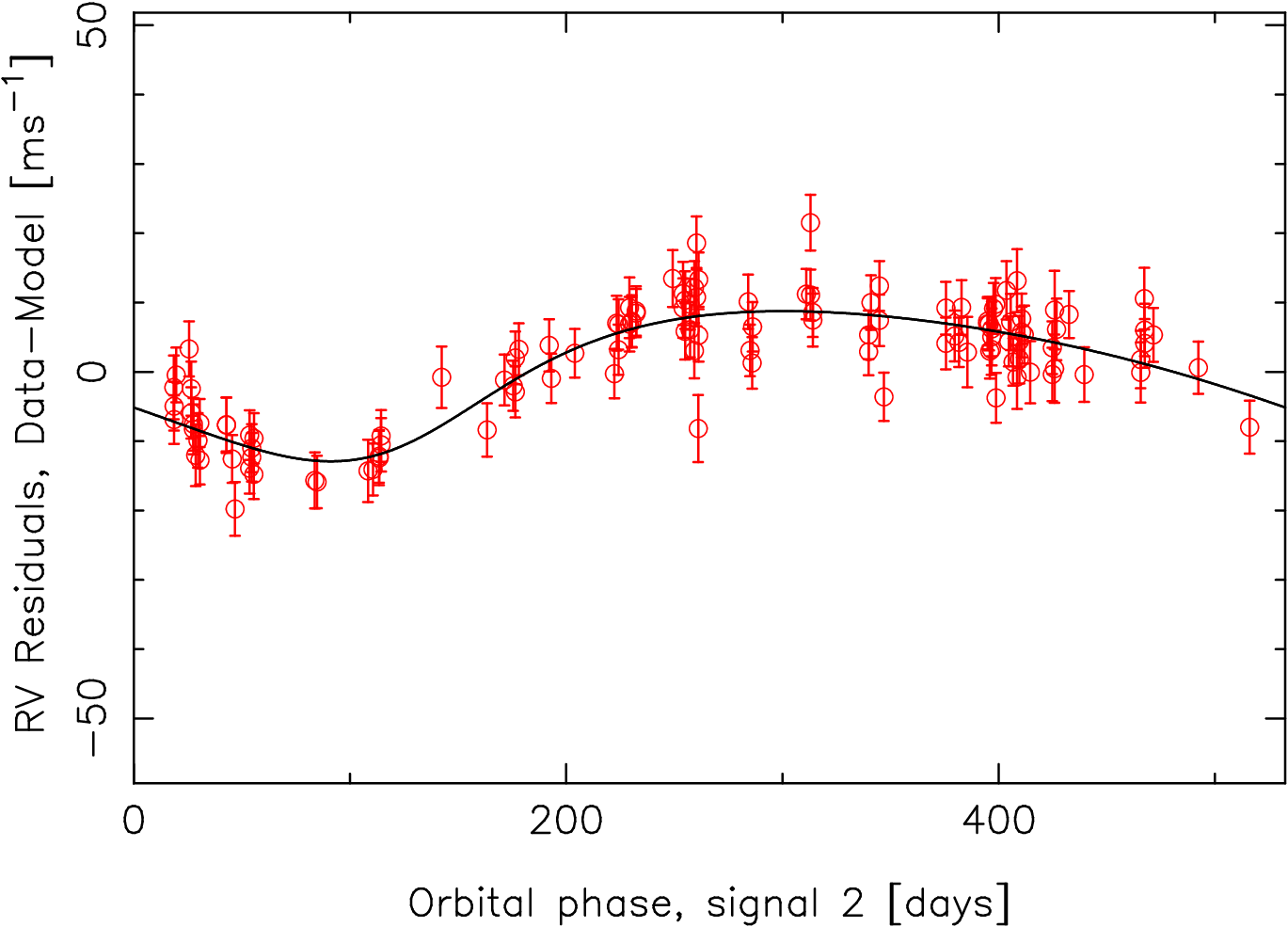}
\caption{Phase-folded HIRES radial velocities and Keplerian curves of GJ 1148 b (top panel) and c (bottom panel).}\label{fig:GJ1148_phased}
\end{figure}

We have plotted the phase-folded radial velocities together with MAP Keplerian curves in Fig. \ref{fig:GJ1148_phased}.

\clearpage

\subsection{GJ 1177B}

\begin{figure}
\center
\includegraphics[angle=270, width=0.49\textwidth,clip]{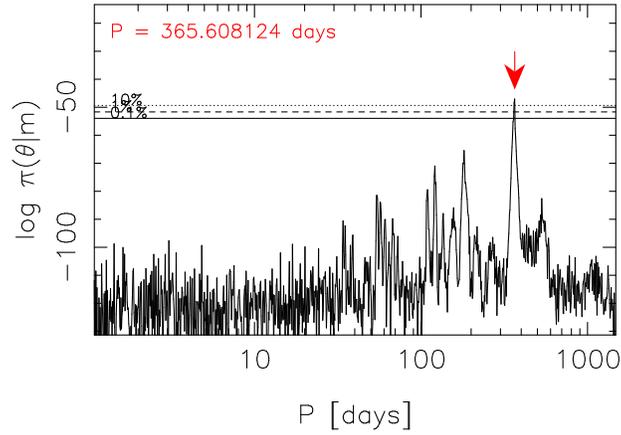}
\caption{Estimated posterior probability density as a function of the period parameter of a Keplerian signal given the GJ 1177B radial velocities. The red arrow denotes the location of the global maximum and the horizontal lines denote the equiprobability thresholds corresponding to 10\% (dotted line), 1\% (dashed line), and 0.1\% (solid line) of the maximum.}\label{fig:GJ1177B_period_search}
\end{figure}

\begin{figure}
\center
\includegraphics[angle=270, width=0.49\textwidth,clip]{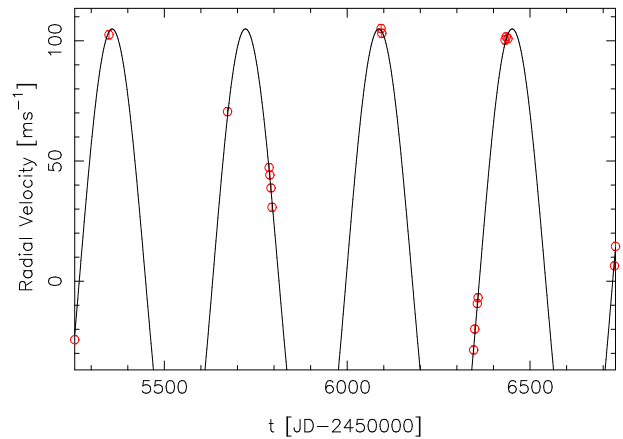}
\caption{Signal in the PFS radial velocities corresponding to a giant planet orbiting GJ 1177B.}\label{fig:GJ1177B_signal}
\end{figure}

GJ 1177B (HD 120036B) has been targeted by PFS and we have obtained a set of 18 velocities covering a baseline of 1479 days. Yet, even with such a low number of radial velocities we were able to identify a strong signature of a massive giant planet orbiting the star. With an amplitude of 109.42 [102.76, 126.40] ms$^{-1}$ and a period of 364.75 [362.73, 367.07] days this signal satisfies all our criteria for a candidate planet (Figs. \ref{fig:GJ1177B_period_search} and \ref{fig:GJ1177B_signal}). 

Although the period of the signal in the PFS velocities of GJ 1177B coincides with one year, given its 99\% credibility interval, it is certainly a genuine signal in the data rather than a signal induced by the fact that the star cannot be observed throughout the year. This is evident because the logarithms of the maximum likelihood values increase from -145.1 to -32.9 when including the Keplerian signal in the model.This increase is considerably more significant that would be caused by spurious signals amplified by poor sampling of the data. The signal is also very unique in the parameter space (Fig. \ref{fig:GJ1177B_period_search}) indicating that it is the only reasonable explanation of the variations in the PFS velocities.

We observed a noteworthy periodicity in the ASAS V-band photometry data of GJ 1177B at a period of 284 days. However, this periodicity cannot be interpreted as a counterpart of the radial velocity signal. It is thus our interpretation that there is a cool giant planet, with minimum mass of 2.78 [2.16, 3.47] M$_{\rm Jup}$, orbiting GJ 1177B.

\begin{figure}
\center
\includegraphics[angle=270, width=0.49\textwidth,clip]{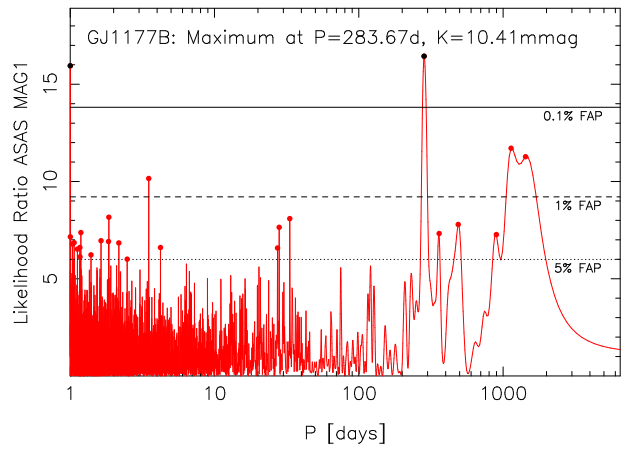}
\caption{Likelihood periodogram of the ASAS V-band photometry measurements of GJ 1177B.}\label{fig:GJ1177B_asas}
\end{figure}

\clearpage

\subsection{GJ 2049}

\begin{figure}
\center
\includegraphics[angle=270, width=0.49\textwidth,clip]{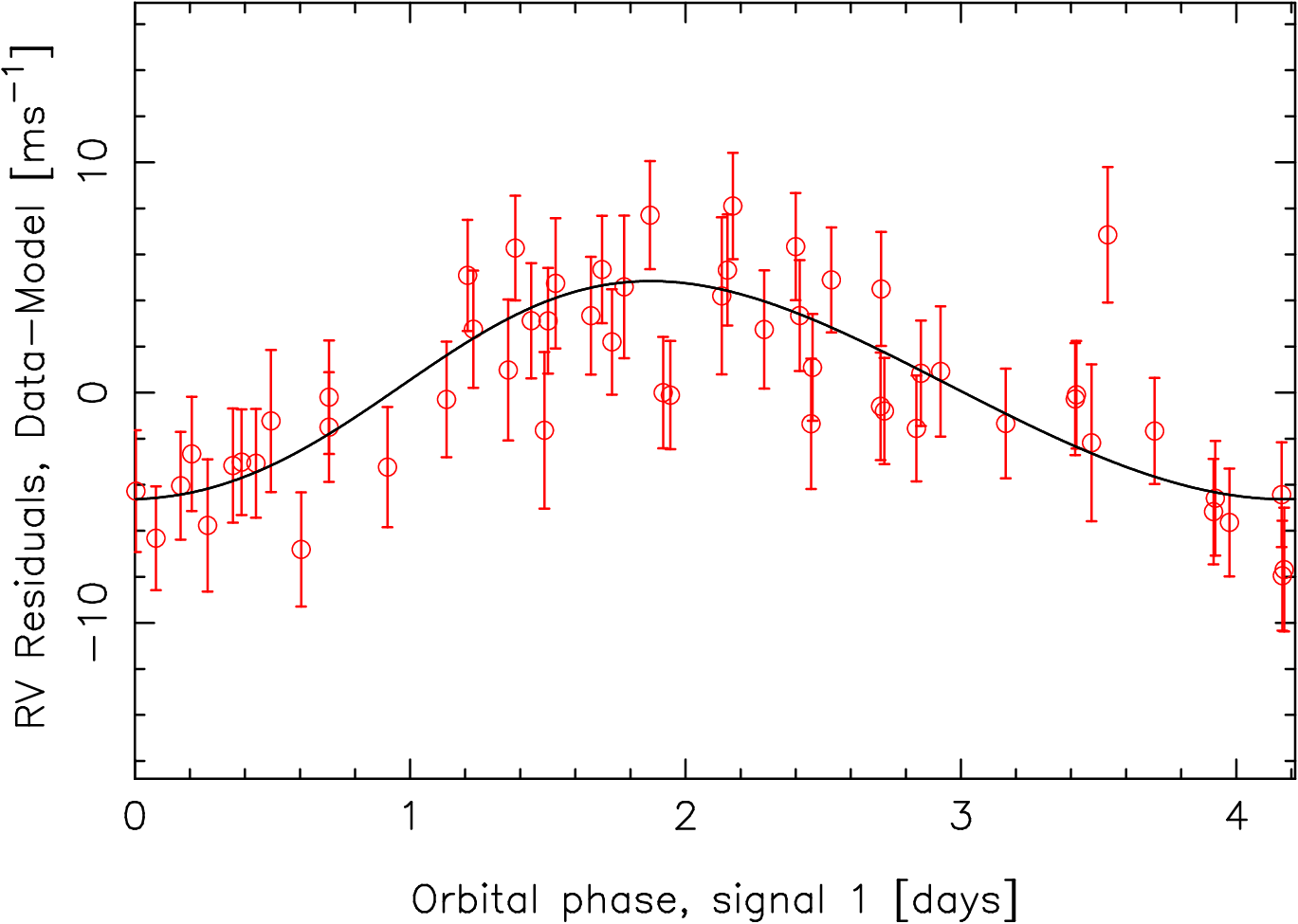}

\includegraphics[angle=270, width=0.49\textwidth,clip]{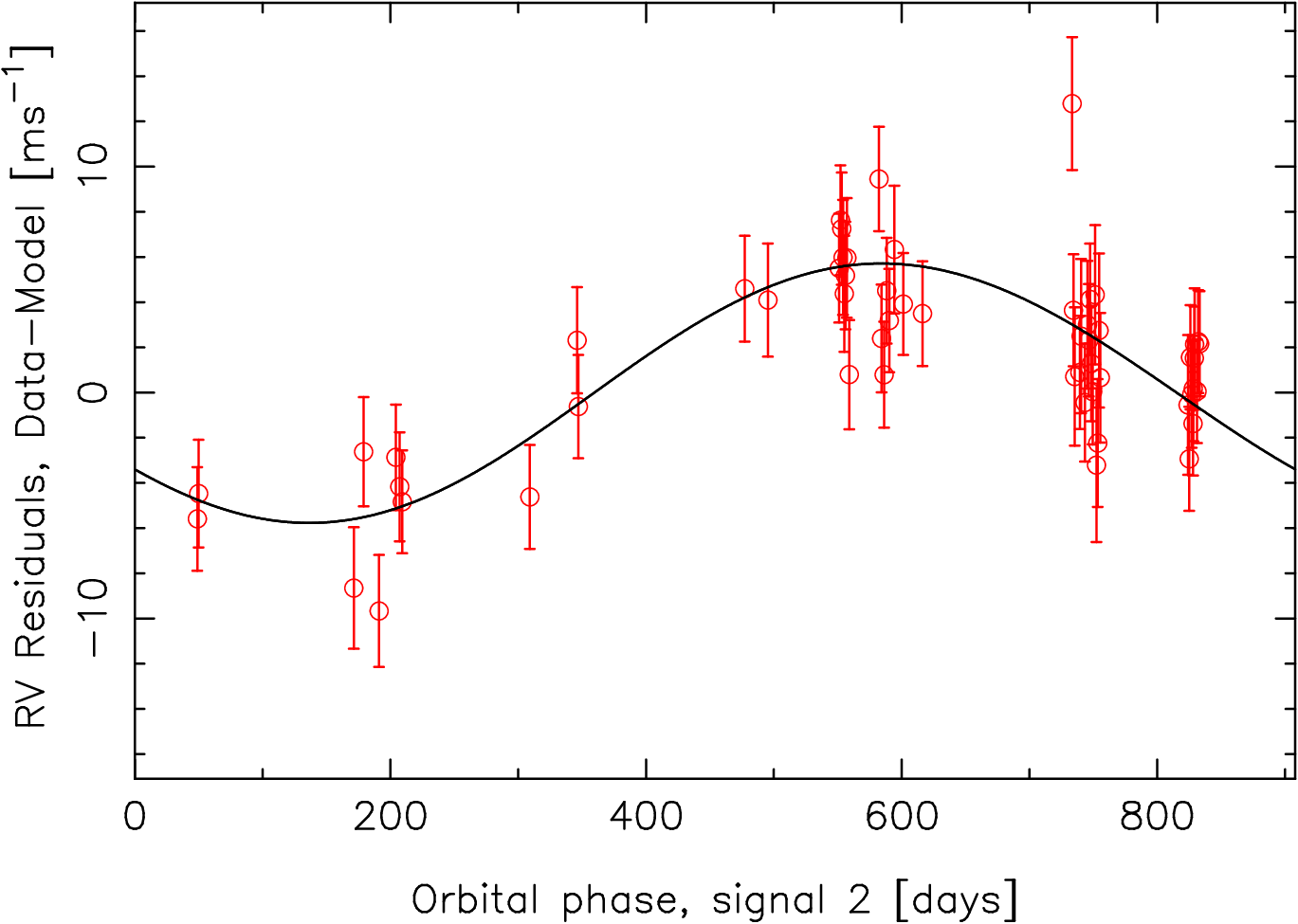}
\caption{HARPS radial velocities folded on the phases of the two signals with the other signal subtracted from each panel.}\label{fig:GJ2049_curve}
\end{figure}

\begin{figure}
\center
\includegraphics[angle=270, width=0.49\textwidth,clip]{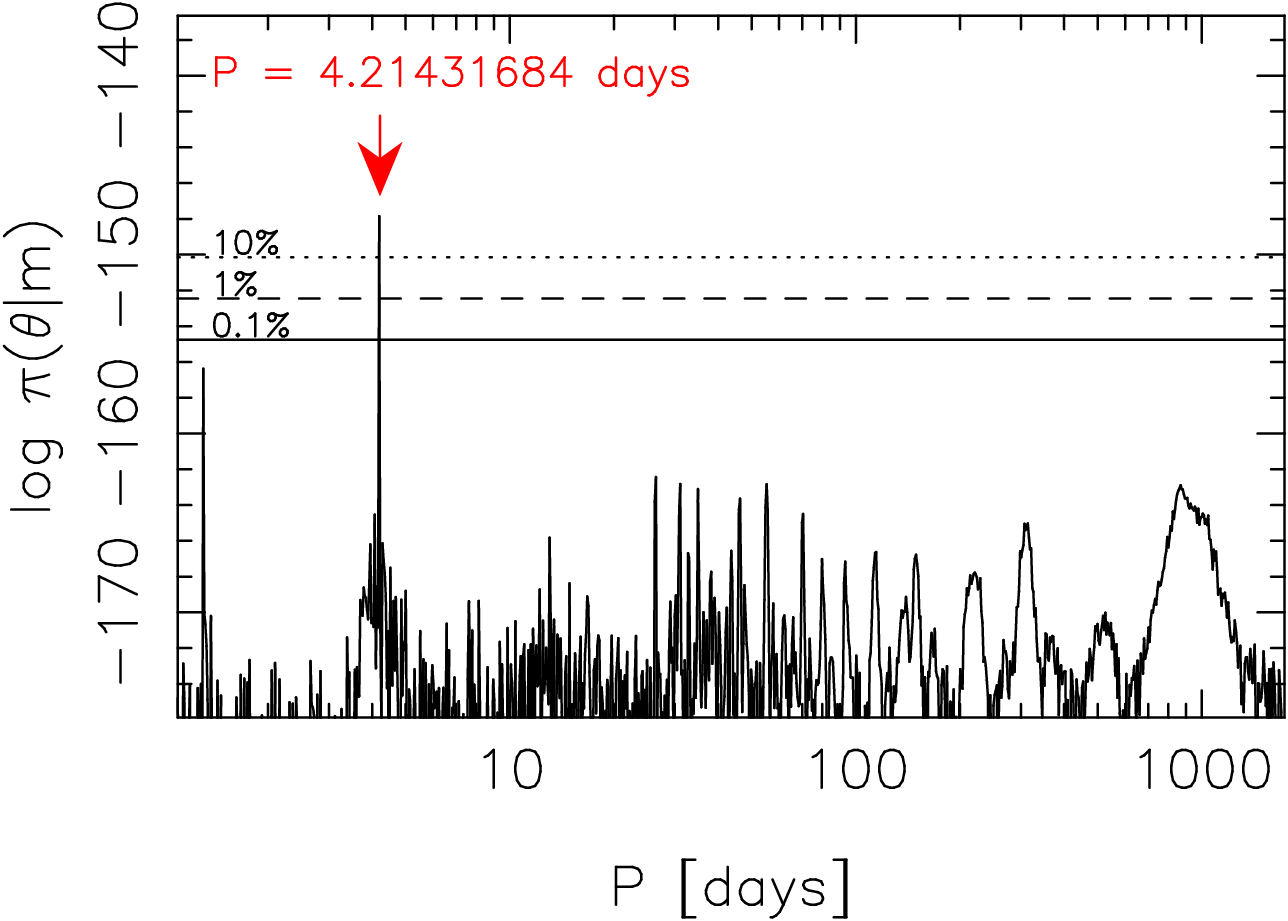}

\includegraphics[angle=270, width=0.49\textwidth,clip]{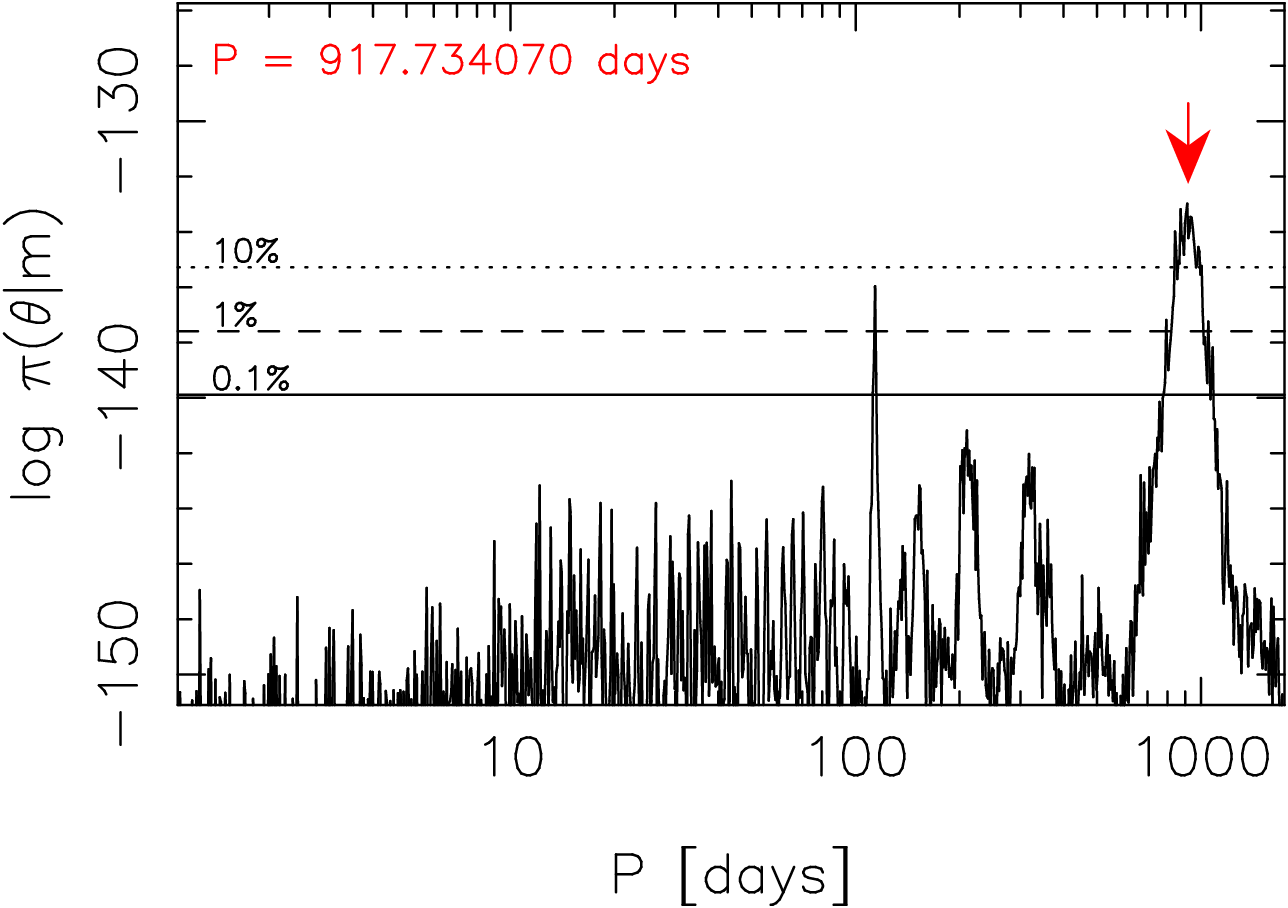}
\caption{Estimated posterior probability densities as functions of the period parameters of the first signal in the one-Keplerian model (top panel) and the second signal in the two-Keplerian model (bottom panel). The red arrows indicate the positions of the global maxima and the horizontal lines denote the 10\% (dotted), 1\% (dashed), and 0.1\% (solid) probability thresholds with respect to the maxima.}\label{fig:GJ2049_psearch}
\end{figure}

The HARPS target, GJ 2049 (HIP 30256), was found to have a clear signal in the set of 56 radial velocities at a period of 4.21346 [4.21214, 4.21492] days that, with an amplitude of 4.33 [3.27, 5.39] ms$^{-1}$, implies the presence of a hot mini-Neptune orbiting the star with a minimum mass of 8.3 [5.7, 10.8] M$_{\oplus}$ (Figs. \ref{fig:GJ2049_curve} and \ref{fig:GJ2049_psearch}). We also found evidence for a long-period signal at a period of 919 [819, 1030] days that we interpret as a signal caused by another candidate planet with a minimum mass of 63.4 [35.6, 94.4] M$_{\oplus}$ orbiting the star (Figs. \ref{fig:GJ2049_curve} and \ref{fig:GJ2049_psearch}). Although there is a local maximum in the period space of the second Keplerian signal at a period of 110 days (Fig. \ref{fig:GJ2049_psearch}, bottom panel), we consider the longer periodicity unique enough to justify interpreting it as a signal caused by a candidate planet.

\begin{figure}
\center
\includegraphics[angle=270, width=0.49\textwidth,clip]{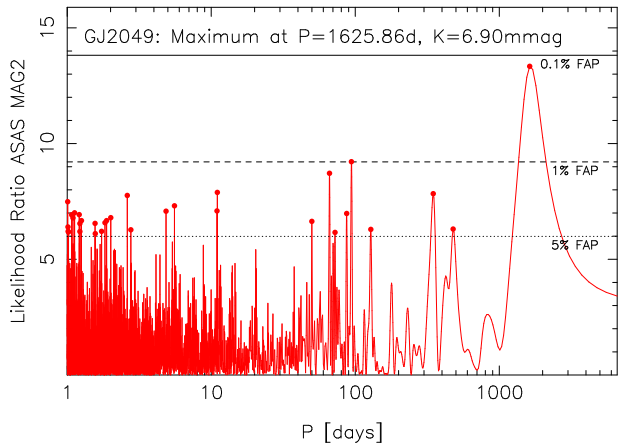}
\caption{Likelihood-ratio periodogram of the ASAS V-band photometry data of GJ 2049.}\label{fig:GJ2049_asas}
\end{figure}

We could not identify any counterparts for the two signals in the HARPS activity indices. Moreover, the only prominent periodicity in the ASAS V-band photometry data was identified at a period of 1630 days, suggestive of the presence of an activity cycle at that period (Fig. \ref{fig:GJ2049_asas}). However, there was a clear minimum at the period of the longer radial velocity signal of 920 days, which indicates that there is no evidence suggesting activity-induced origin for the radial velocity signals.

\clearpage

\subsection{GJ 3293}\label{sec:GJ3293}

\begin{figure}
\center
\includegraphics[angle=270, width=0.49\textwidth,clip]{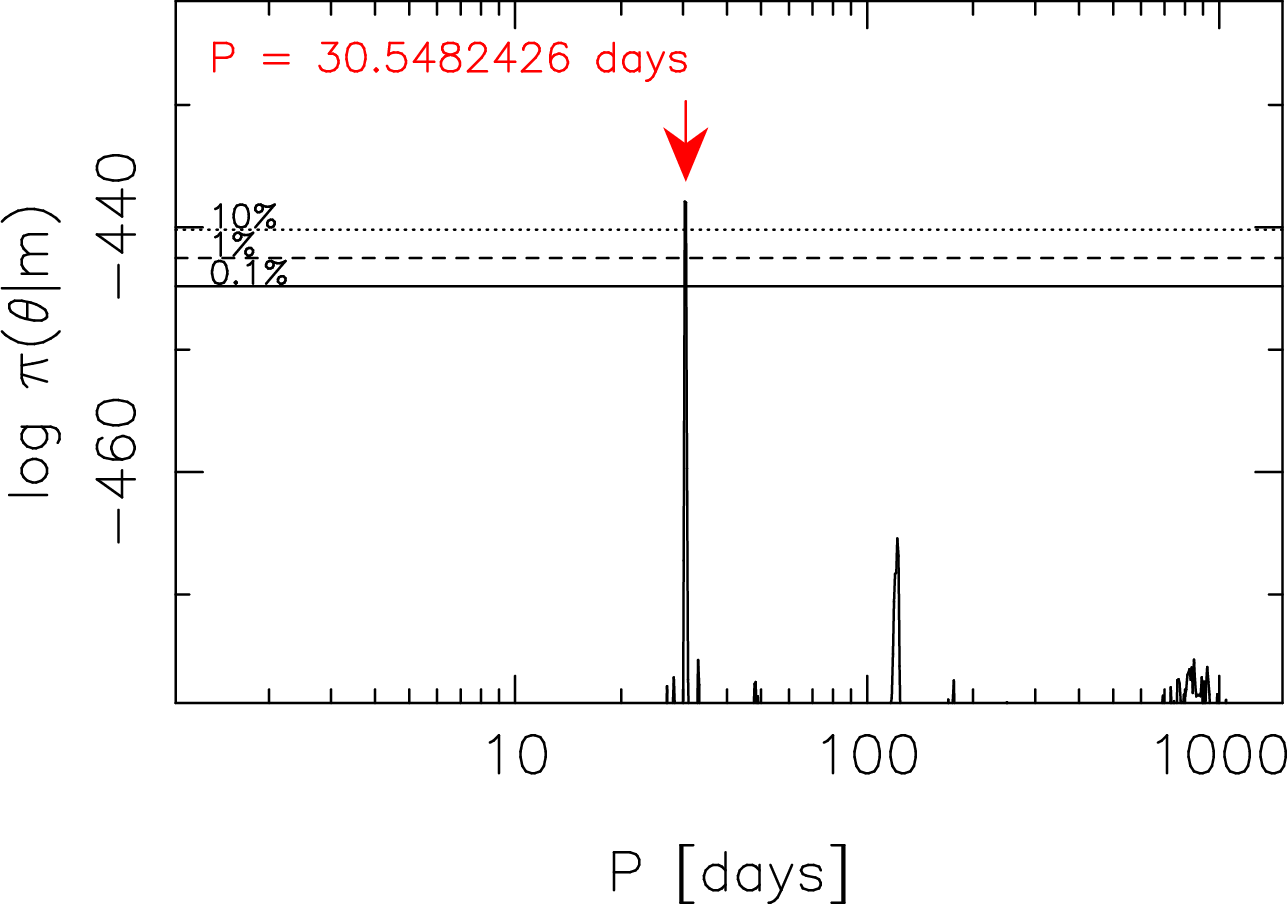}

\includegraphics[angle=270, width=0.49\textwidth,clip]{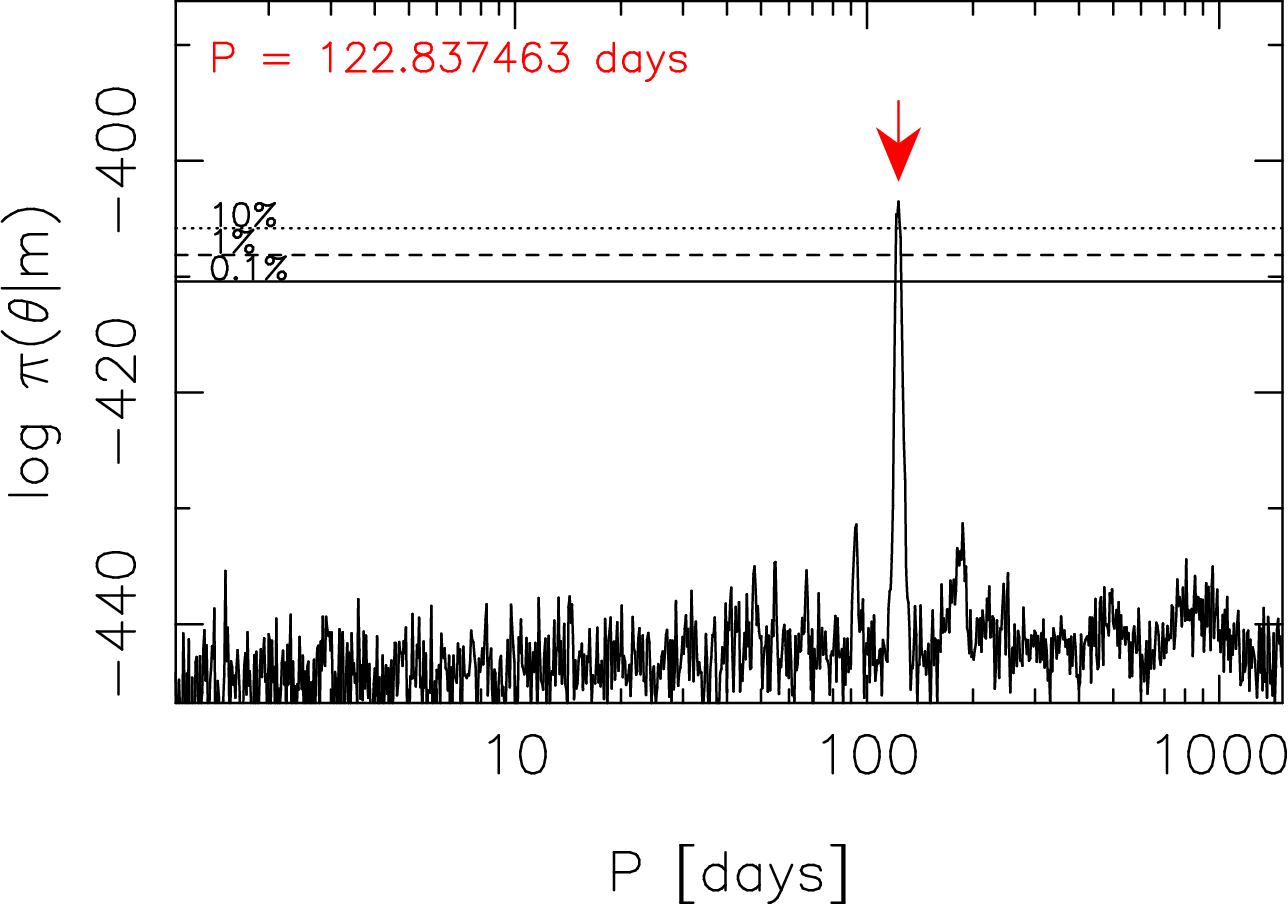}

\includegraphics[angle=270, width=0.49\textwidth,clip]{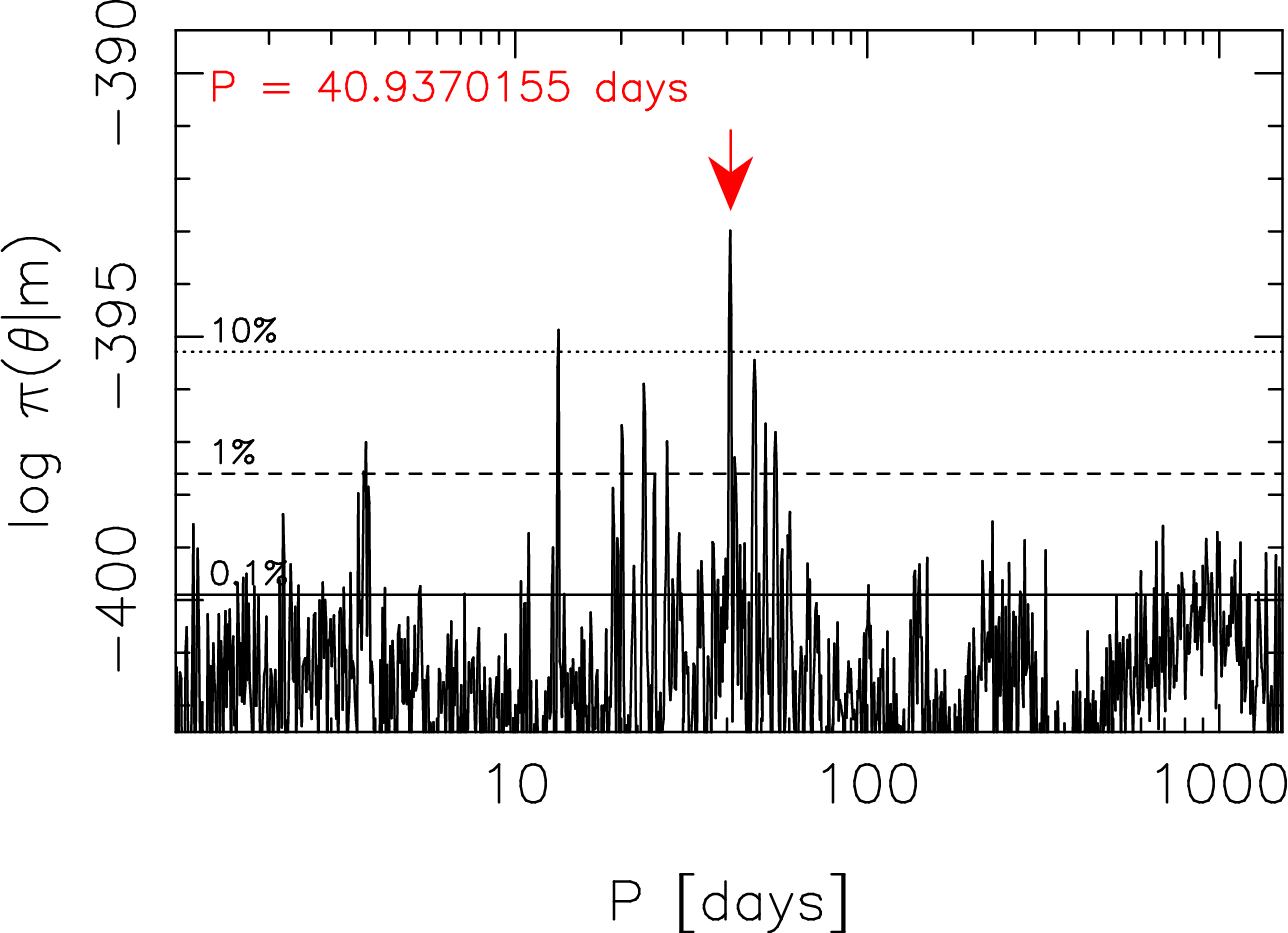}
\caption{Posterior probability densities as functions of the period parameters of the first (top), second (middle) and thhird (bottom panel) Keplerian signals. The red arrod indicates the position of the global maximum in the period space and the horizontal lines represent the 10\% (dotted), 1\% (dashed) and 0.1\% (solid) probability thresholds of this maximum.}\label{fig:GJ3293_psearch}
\end{figure}

As reported by \citet{astudillo2015}, there is evidence for two, possibly three, candidate planets orbiting GJ 3293 based on HARPS radial velocity data. We could confirm the presence of the strongest two signals in the HARPS data but could not identify the third signal reported by \citet{astudillo2015} at a period of 48.14$\pm$0.12 days. We have plotted the coresponding estimated posterior densities based on our DRAM samplings in Fig. \ref{fig:GJ3293_psearch} indicating that while the first two signals, at periods of 30.5902 [30.5400, 30.6495] and 123.27 [121.54, 125.00] days, respectively, are certainly significantly present in the data, the third one cannot be identified as a probability maximum of the model with three Keplerian signals (Fig. \ref{fig:GJ3293_psearch}, bottom panel). This implies that it is an artifact caused by correlated noise and/or activity-induced variations that our statistical model is capable of distinguishing from a Keplerian signal.

\begin{figure}
\center
\includegraphics[angle=270, width=0.49\textwidth,clip]{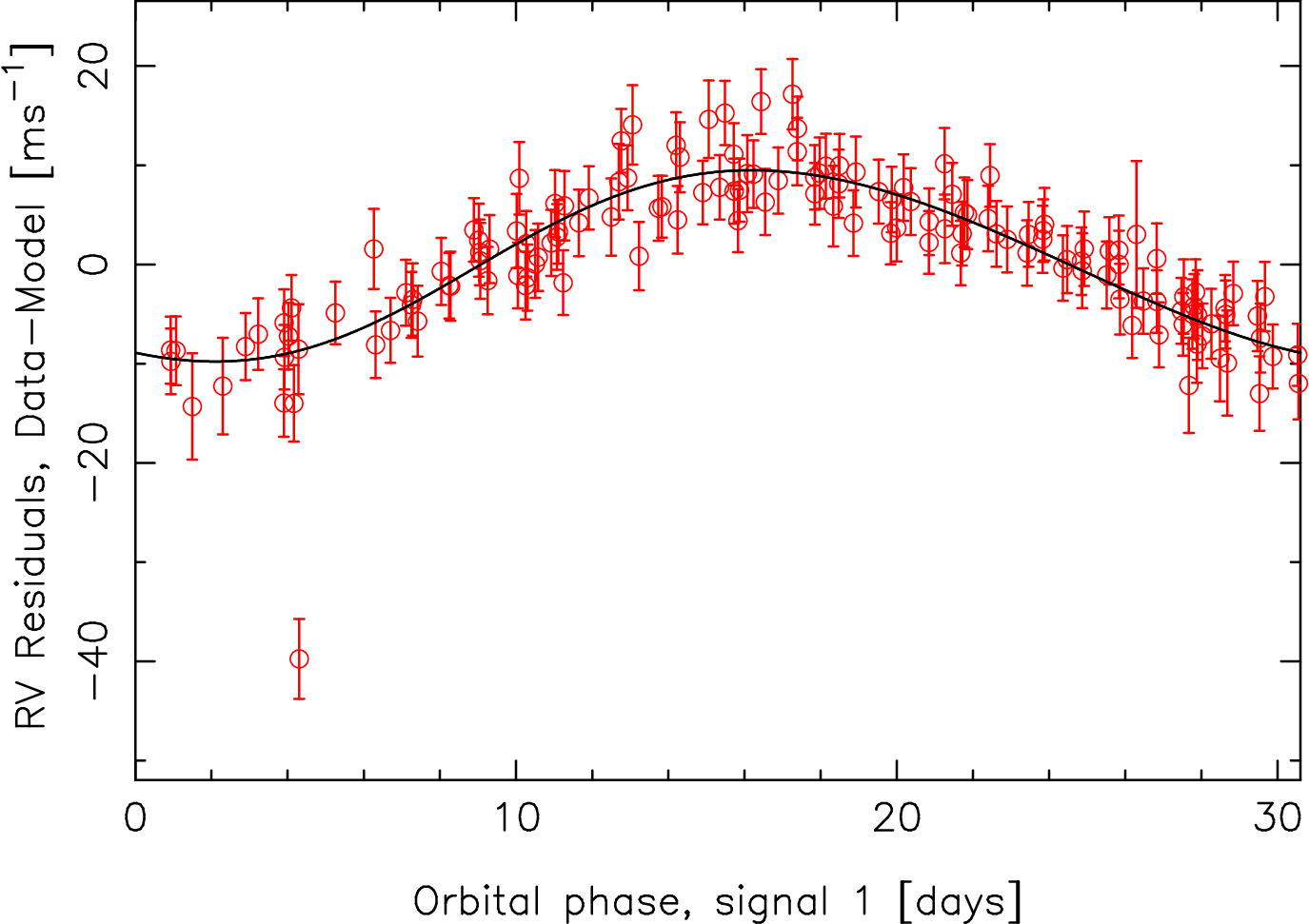}

\includegraphics[angle=270, width=0.49\textwidth,clip]{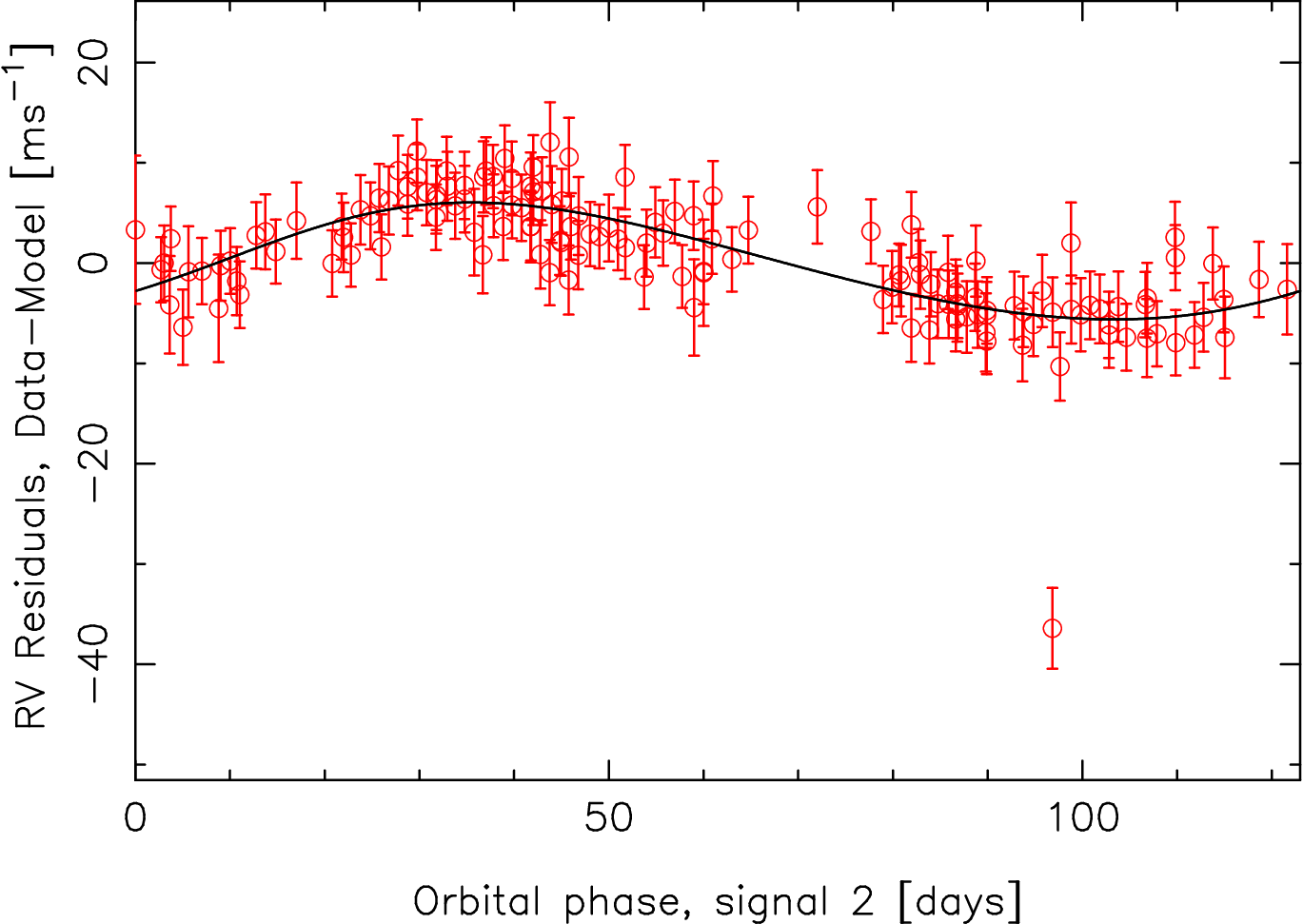}
\caption{Phase-folded radial velocities of GJ 3293 given the signals corresponding to candidate planets b (top) and c (bottom).}\label{fig:GJ3293_curve}
\end{figure}

\citet{astudillo2015} reported that they discovered a signal at a period of 41 days in the H-$\alpha$ indices of the star, likely indicative of stellar rotation period. When searching for a third periodic signal in the HARPS velocities, we observe probability maxima at and near that period (Fig. \ref{fig:GJ3293_psearch}, bottom panel) supporting the fact that the stellar rotation period might be around 41 days. However, we could not find periodicities in the ASAS V-band photometry and could thus not confirm this observation.

\begin{figure}
\center
\includegraphics[angle=270, width=0.49\textwidth,clip]{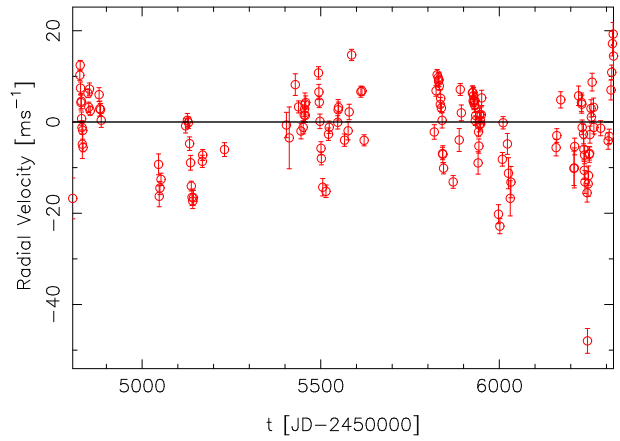}
\caption{HARPS radial velocities of GJ 3293 with respect to the data mean.}\label{fig:GJ3293_velocities}
\end{figure}

We note that \citet{astudillo2015} reported a ``formally significant'' quadratic drift in the GJ 3293 radial velocities. We cannot verify this observation and it is not apparent in the HARPS radial velocities despite some variations on the long time-scale (Fig. \ref{fig:GJ3293_velocities}). When adding such a polynomial term in the radial velocity model, we could obtain an estimate for the corresponding quadratic parameter of -0.13 [-0.63, 0.88] ms$^{-1}$year$^{-2}$, which indicates that the quadratic term is not statistically significantly different from zero.

\clearpage

\subsection{GJ 3325}

\begin{figure}
\center
\includegraphics[angle=270, width=0.49\textwidth,clip]{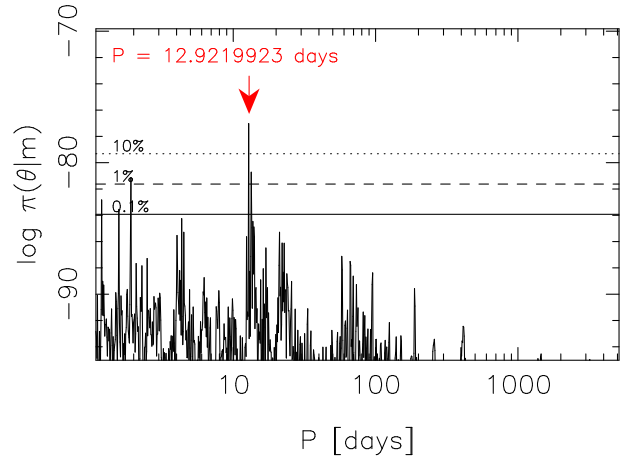}
\caption{As in Fig. \ref{fig:GJ1177B_period_search} but fot the combined HARPS and HIRES velocities of GJ 3325.}\label{fig:GJ3325_period_search}
\end{figure}

\begin{figure}
\center
\includegraphics[angle=270, width=0.49\textwidth,clip]{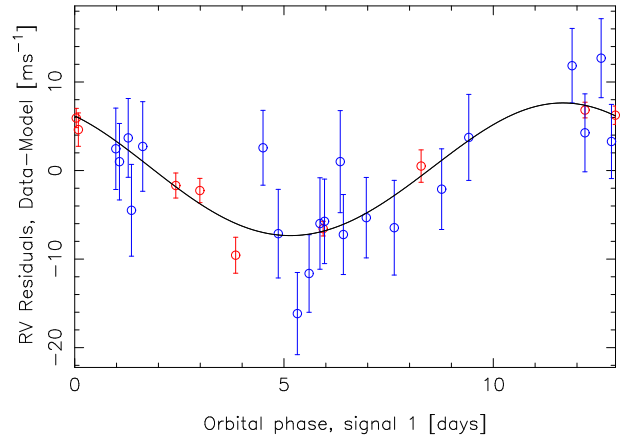}
\caption{Phase-folded radial velocity signal in the combined HARPS (red) and HIRES (blue) data of GJ 3325.}\label{fig:GJ3325_phased}
\end{figure}

\begin{figure}
\center
\includegraphics[angle=270, width=0.49\textwidth,clip]{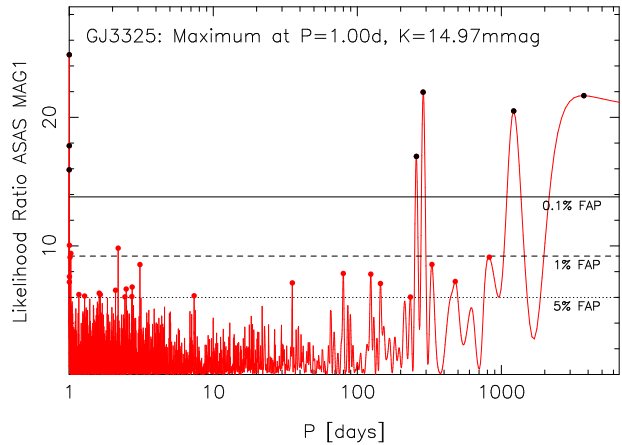}
\caption{Likelihood-ratio periodogram of ASAS V-band photometry of GJ 3325. The red (black) filled circles denote maxima that exceed the 5\% (0.1\%) FAP threshold.}\label{fig:GJ3325_asas}
\end{figure}

With only 9 HARPS and 21 HIRES velocities the prospects of finding significant signals in the combined velocities of GJ 3325 (HIP 32512) appeared slim. However, it was possible to identify a signal at a period of 12.9221 [12.9110, 12.9307] days with an amplitude of 7.24 [4.18, 10.30] ms$^{-1}$ in the combined velocities in accordance with our signal detection criteria (Figs. \ref{fig:GJ3325_period_search} and \ref{fig:GJ3325_phased}). Although there was a local maximum in the period space next to the global maximum and close to two days, we consider these local maxima to be representative of the sampling issues in the combined data set. Moreover, as the ASAS V-band photometry measurements indicated the presence of likely magnetic activity cycles of 286 and 3000 days and did not contain periodicities at or near the period of the radial velocity signal (Fig. \ref{fig:GJ3325_asas}), we interpret this signal as a candidate planet orbiting the star.

The radial velocity signal could not be identified in the data unless the correlations between HARPS velocities and the corresponding activity indices were accounted for. We observed a strong (99\% credibility) positive correlation between HARPS velocities and FWHM values and a moderate (95\% credibility) one with S-indices based on only nine points. HIRES velocities appeared to be independent of the corresponding S-indices. This example indicates that activity-induced variations can mask planetary signals in small data sets and should thus always be accounted for by the statistical model.

Because the radial velocity signal is supported by both instruments and is detected according to all our criteria, we interpret it as a hot Neptune-mass planet candidate orbiting the star.

\clearpage

\subsection{GJ 3341}

\begin{figure}
\center
\includegraphics[angle=270, width=0.49\textwidth,clip]{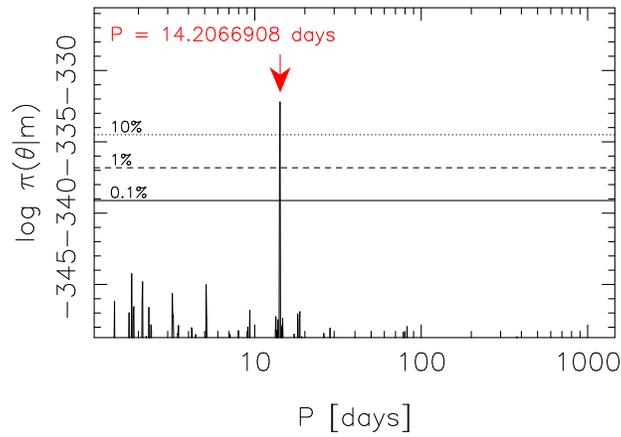}
\caption{Estimated posterior probability density as a function of the period of the Keplerian model.}\label{fig:GJ3341_psearch}
\end{figure}

\begin{figure}
\center
\includegraphics[angle=270, width=0.49\textwidth,clip]{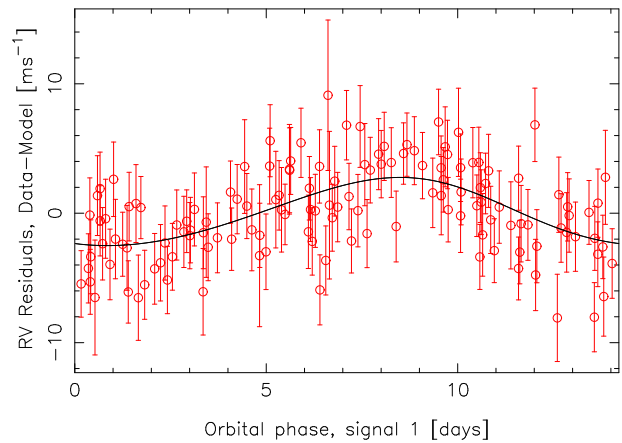}
\caption{Phase-folded Keplerian signal corresponding to GJ 3341 b overplotted on top of the corresponding HARPS radial velocities.}\label{fig:GJ3341_curve}
\end{figure}

We observed a signal at a period of 14.210 [14.178, 14.236] days in the HARPS radial velocities of GJ 3341 (Figs. \ref{fig:GJ3341_psearch} and \ref{fig:GJ3341_curve}). This signal has already been reported by \citet{astudillo2015}, who interpreted it as evidence in favour of a candidate planet orbiting the star. We agree with this interpretation as we could not find counterparts for the radial velocity signal in the HARPS activity indicators or the ASAS photometry. It thus seems apparent that there is a candidate planet with a minimum mass of 6.0 [3.0, 9.7] M$_{\oplus}$ orbiting the star. We classify this candidate as a hot super-Earth.

The photometric variability of this star, based on ASAS V-band photometry, was higher than expected with a standard deviation of 171 mmag. This value is higher than expected for ASAS based on the typical M dwarf in the current work and indicates that either the data is contaminated or the star is photometrically variable.

\clearpage

\subsection{GJ 3543}\label{sec:GJ3543}

The periodic variations in the HARPS radial velocity data of GJ 3543 were reported by \citet{astudillo2015}. They discovered two significant periodogram powers at periods of 1.1 and 9.2 days and concluded that, since these two are the one-day aliases of one another, and because the latter is roughly one half of a prominent periodicity in the activity indicators, that they are thus caused by this periodic activity-related phenomenon. This was based on the discovery of a periodogram power in the HARPS S-indices at a period of 22 days with one-$\sigma$ and a detection of a signal in the H-$\alpha$ index at 19 with a similarly low credibility.

\begin{figure}
\center
\includegraphics[angle=270, width=0.49\textwidth,clip]{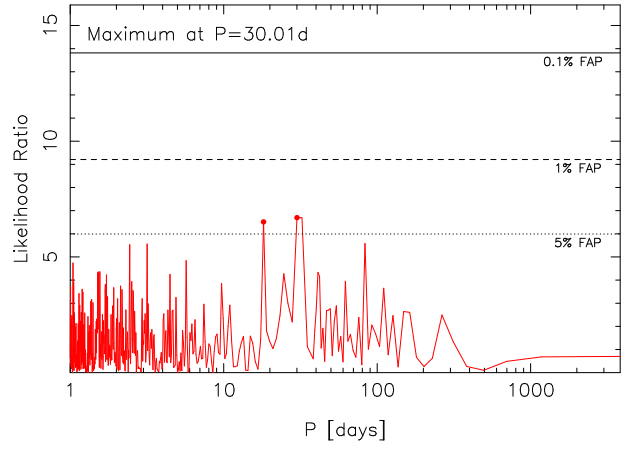}
\caption{Likelihood-ratio periodogram of the HARPS S-indices of GJ 3543.}\label{fig:GJ3543_S}
\end{figure}

We did not observe any convincing signals in the HARPS activity indicators apart from two suggestive likelihood maxima at periods of approximately 18 and 30 days (Fig. \ref{fig:GJ3543_S}). However, as these maxima barely exceeded the 5\% FAP likelihood ratio and are not much higher than the background likelihood ratio, we do not consider them as reliable evidence for periodicities in the S-indices.

\begin{figure}
\center
\includegraphics[angle=270, width=0.49\textwidth,clip]{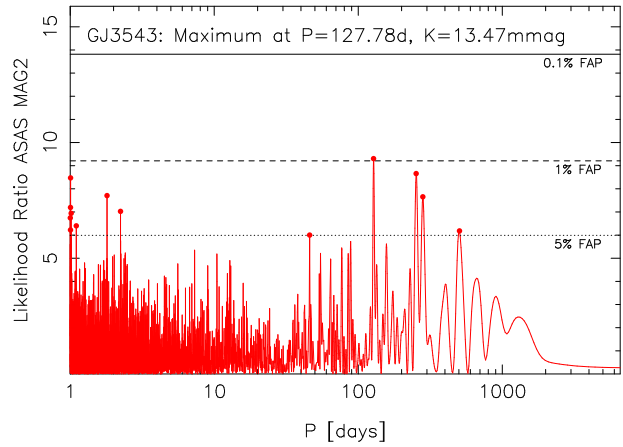}
\caption{Likelihood-ratio periodogram of the ASAS V-band photometry data of GJ 3543.}\label{fig:GJ3543_asas}
\end{figure}

We also analysed a set of 604 ASAS V-band photometry measurements. This data set suggested that there is a photometric periodicity of 128 years but did not provide conclusive evidence for photometric rotation or magnetic activity cycles (Fig. \ref{fig:GJ3543_asas}).

\begin{figure}
\center
\includegraphics[angle=270, width=0.49\textwidth,clip]{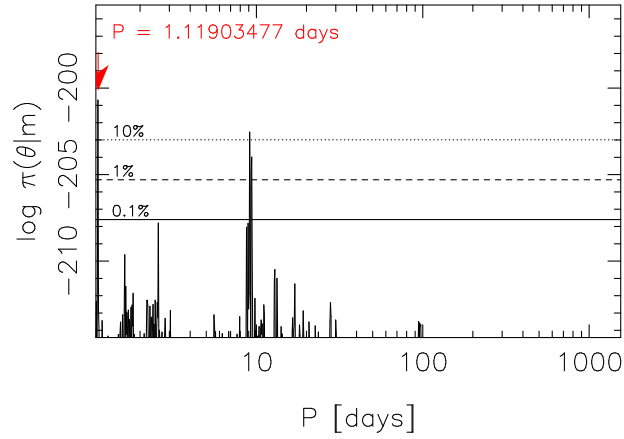}
\caption{Estimated posterior probability density as a function of the period of the Keplerian function given the HARPS radial velocity data of GJ 3543. The red arrow indicates the position of the global maximum in the period space whereas the horizontal lines denote the 10\% (dotted), 1\% (dashed), and 0.1\% (solid) probability thresholds with respect to this maximum.}\label{fig:GJ3543_psearch}
\end{figure}

\begin{figure}
\center
\includegraphics[angle=270, width=0.49\textwidth,clip]{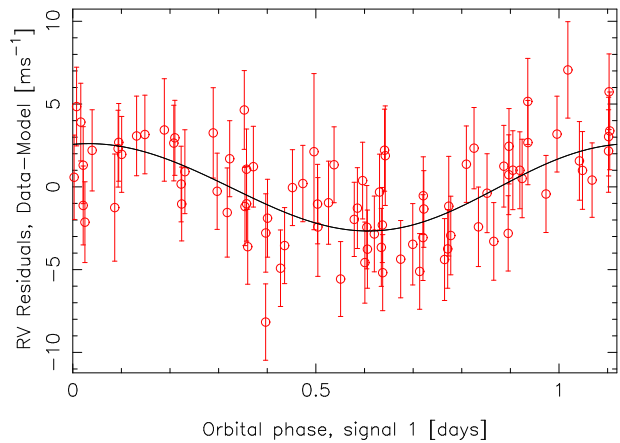}
\caption{Phase-folded radial velocities of GJ 3543.}\label{fig:GJ3543_curve}
\end{figure}

Yet, regardless of the lack of signals in the spectroscopic activity or photometric data, there are two strong periodic signals in the HARPS radial velocities (Fig. \ref{fig:GJ3543_psearch}). The global maximum is at a period of 1.12 days in the period space, as also observed by \citet{astudillo2015} with periodogram analysis, and the daily alias of this signal at 9.2 days shows as the most probable local maximum. Because we have no evidence supporting the view that these signals would be connected to stellar activity, we interpret them as being induced by a candidate planet orbiting the star with orbital period of 11.11909 [1.11877, 1.11941] days and a minimum mass of 2.2 [0.9, 3.6] M$_{\oplus}$ enabling us to classify it as a hot Earth due to its minimum mass estimate being consistent with one Earth mass given its 99\% credibility interval. We have plotted the phase-folded radial velocities in Fig. \ref{fig:GJ3543_curve}.

\clearpage

\subsection{GJ 3822}

\begin{figure}
\center
\includegraphics[angle=270, width=0.49\textwidth,clip]{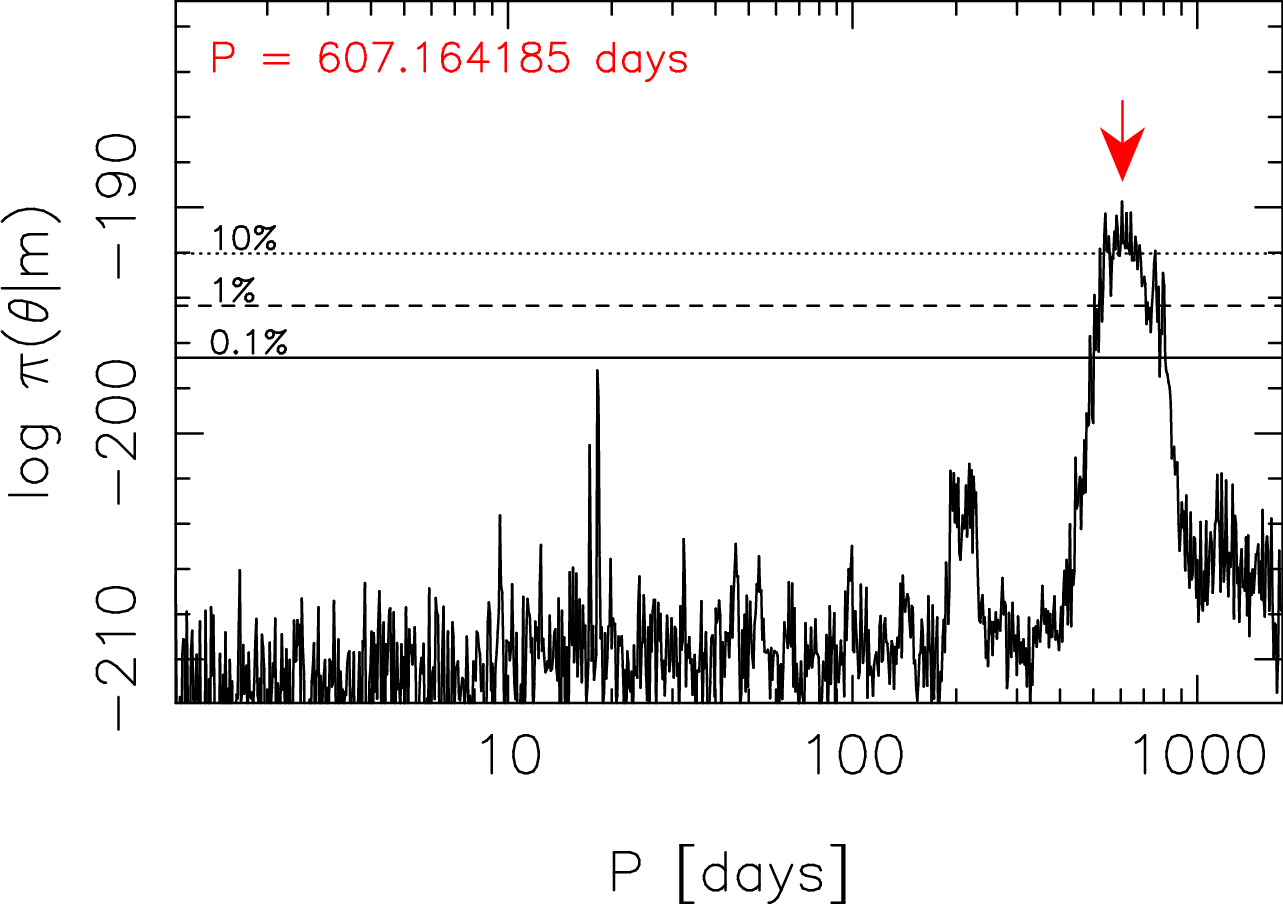}

\includegraphics[angle=270, width=0.49\textwidth,clip]{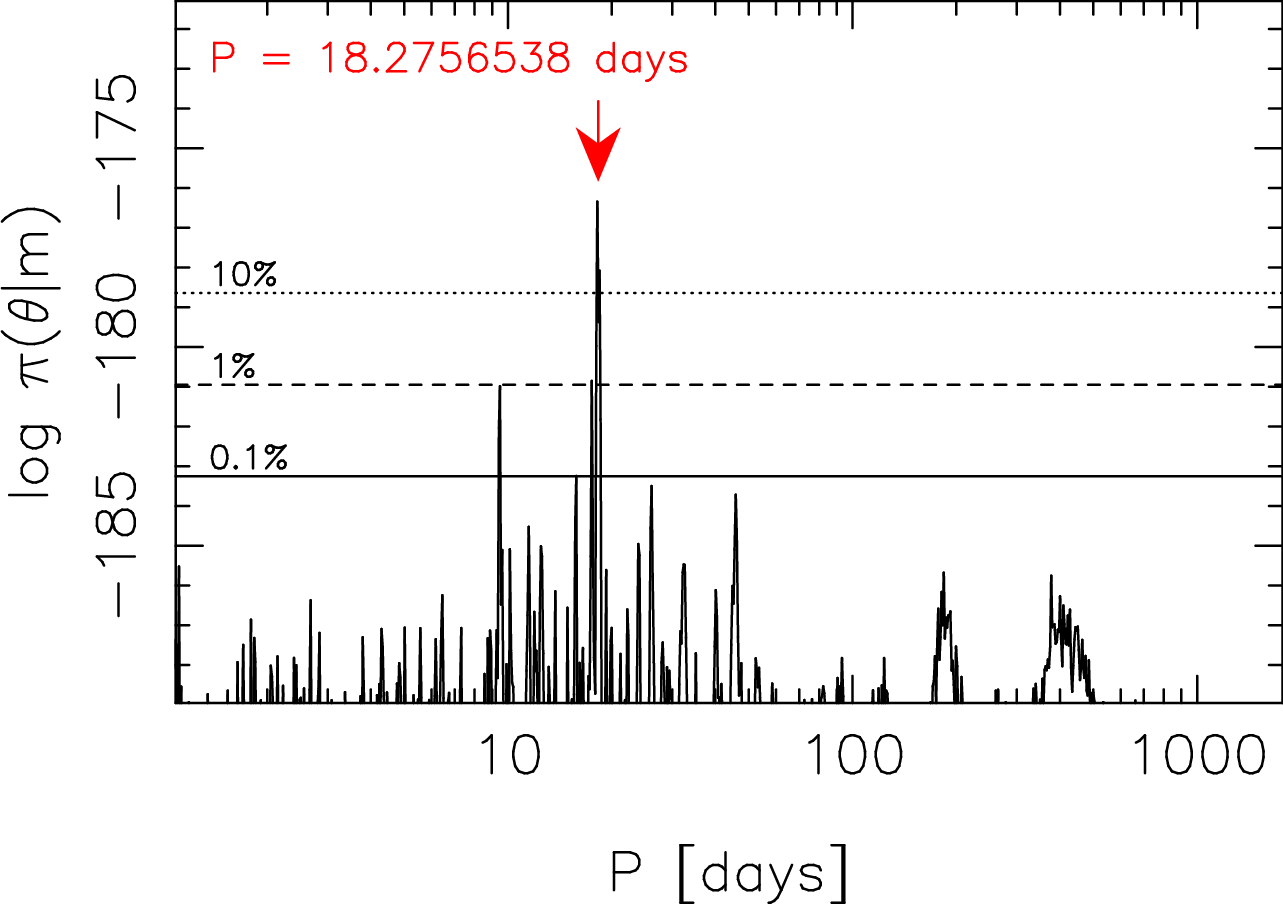}
\caption{Estimated posterior probability densities as functions of the period parameter of the $k$th Keplerian signal for k$k=1$ (top panel) and $k=2$ (bottom panel). The red arrows and horizontal lines denote the positions of the global maxima and selected equiprobability thresholds with respect to the global maxima, respectively.}\label{fig:GJ3822_psearch}
\end{figure}

\begin{figure}
\center
\includegraphics[angle=270, width=0.49\textwidth,clip]{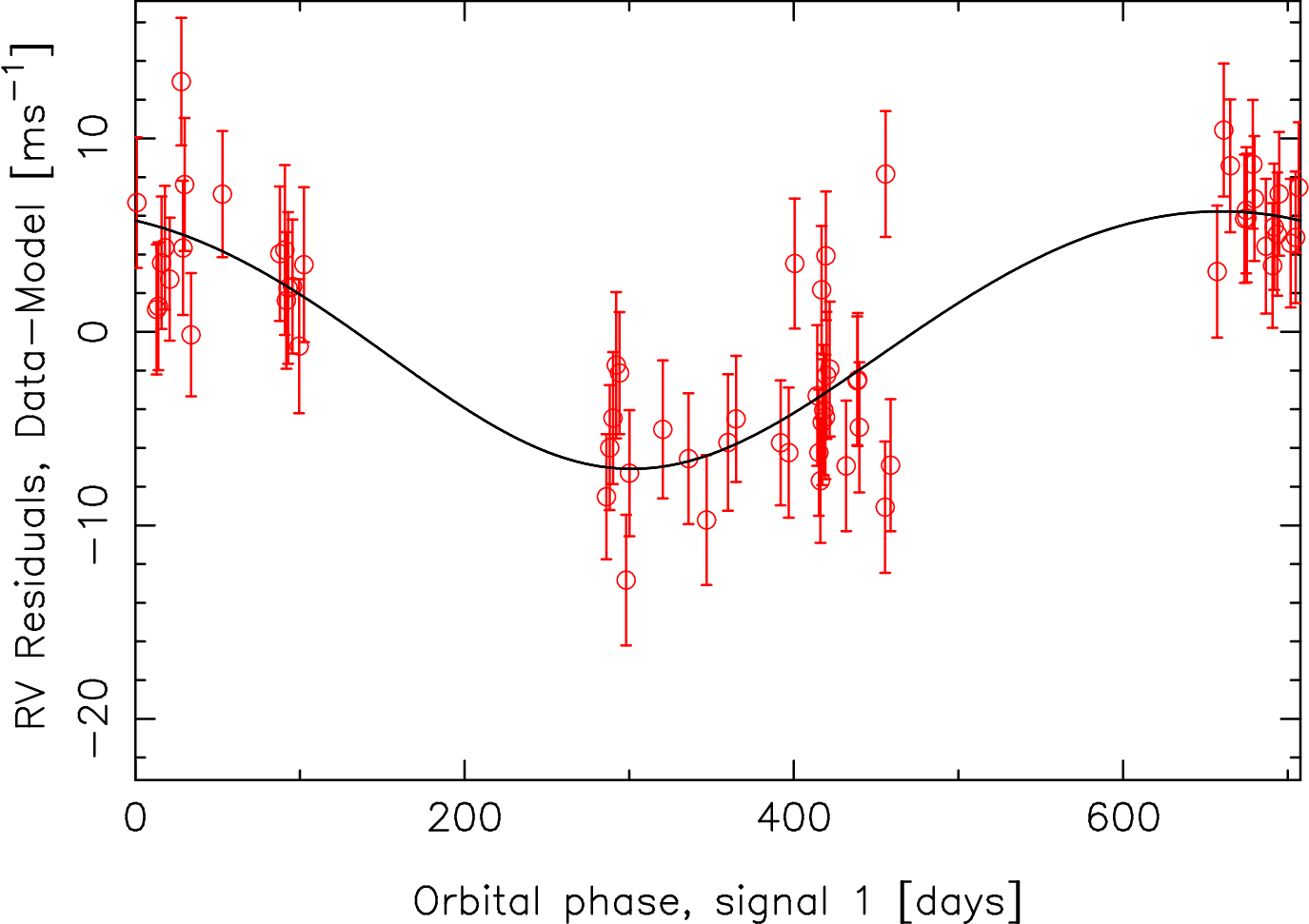}

\includegraphics[angle=270, width=0.49\textwidth,clip]{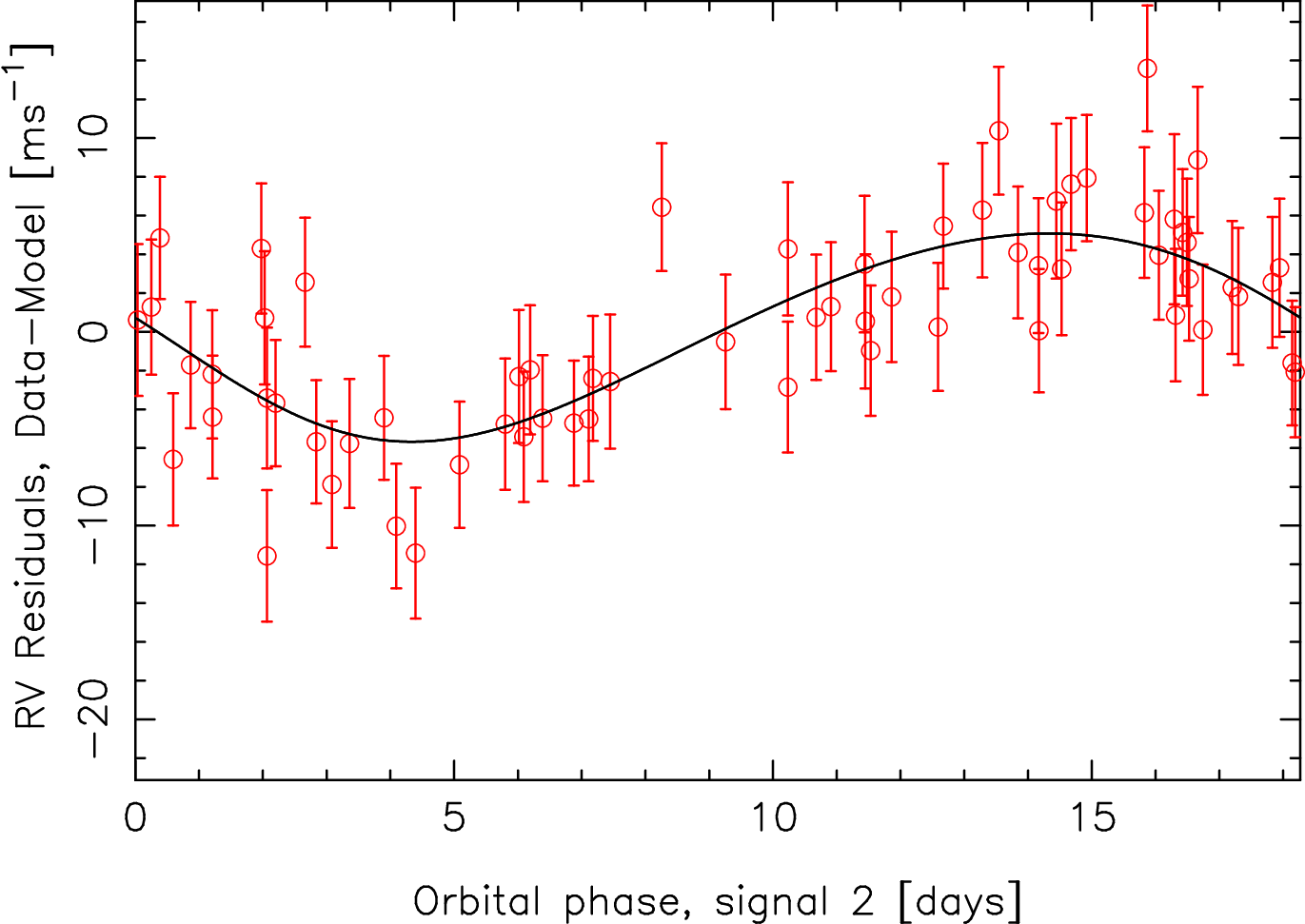}
\caption{Phase-folded radial velocities corresponding to the signals of GJ 3822 b (top) and c (bottom).}\label{fig:GJ3822_curve}
\end{figure}

As a HARPS target GJ 3822 (HIP 68570) has been observed 66 times according to the available data in the ESO archive. We observed two clear signals in these radial velocities and interpret them as candidate planets orbiting the star with orbital periods of 661 [514, 778] and 18.264 [18.185, 18.344] days, respectively (Figs. \ref{fig:GJ3822_psearch} and \ref{fig:GJ3822_curve}). It is noteworthy that the period of the former signal is highly uncertain due to a poor phase-coverage but the signal remains securely and uniquely detected according to our criteria.

We did not identify any significant periodicities in the HARPS activity indicators. Moreover, we did not find periodic signals in the ASAS V-band photometry data either. This means the radial velocity signals have counterparts in neither and that we thus interpret them as candidate planets orbiting the star. These candidate planets with minimum masses of 41.1 [10.8, 66.9] and 9.9 [4.1, 16.5] M$_{\oplus}$ are classified as cool super-Neptune and a hot mini-Neptune, respectively.

\clearpage

\subsection{GJ 4079}

\begin{figure}
\center
\includegraphics[angle=270, width=0.49\textwidth,clip]{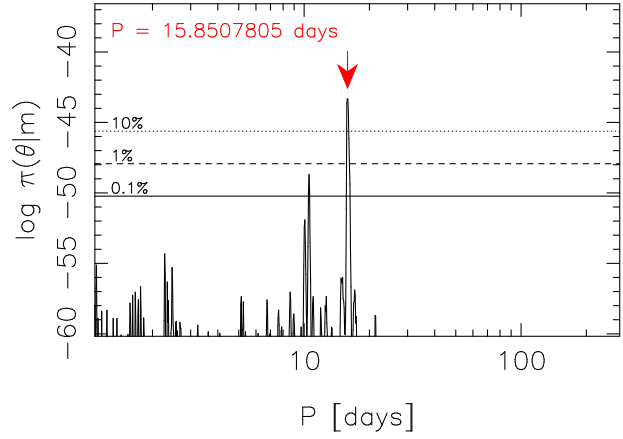}
\caption{As in Fig. \ref{fig:GJ3543_psearch} but for the signal given the HARPS radial velocities of GJ 4079.}\label{fig:GJ4079_psearch}
\end{figure}

\begin{figure}
\center
\includegraphics[angle=270, width=0.49\textwidth,clip]{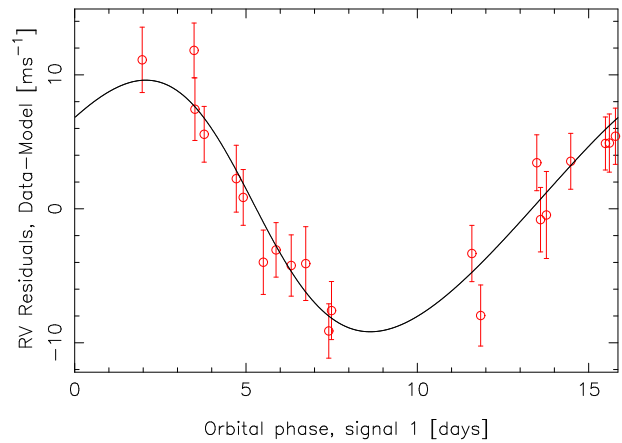}
\caption{Phase-folded radial velocities of GJ 4079.}\label{fig:GJ4079_curve}
\end{figure}

The relatively small data set of GJ 4079 (HIP 92451) HARPS radial velocities with $N = 21$ was found to be a rather interesting one because we discovered a reasonably strong signal at a period of 15.842 [15.772, 16.208] days (Figs. \ref{fig:GJ4079_psearch} and \ref{fig:GJ4079_curve}).

\begin{figure}
\center
\includegraphics[angle=270, width=0.49\textwidth,clip]{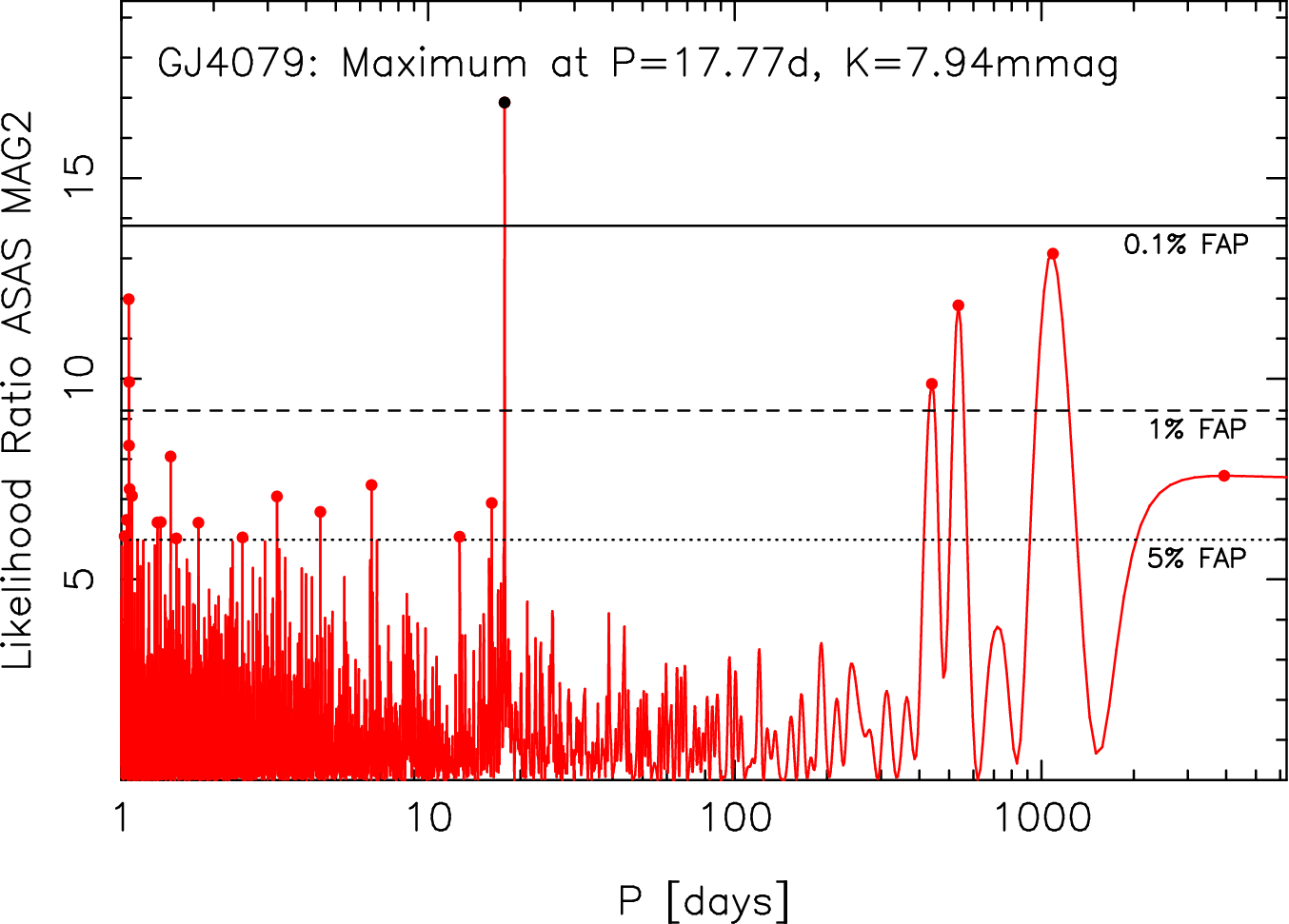}
\caption{Likelihood-ratio periodogram of the ASAS V-band photometry of GJ 4079.}\label{fig:GJ4079_asas}
\end{figure}

However, there was also a strong signal in the ASAS V-band photometry data of GJ 4079 (Fig. \ref{fig:GJ4079_asas})-- at a period of 17.77 days, which is very close to the radial velocity signal in period space. We consider the photometric signal to be likely to be connected to the radial velocity signal and thus assume that they are both caused by a single phenomenon, i.e. the co-rotation of active and inactive regions on the stellar surface. 

\clearpage

\subsection{GJ 4303}

\begin{figure}
\center
\includegraphics[angle=270, width=0.49\textwidth,clip]{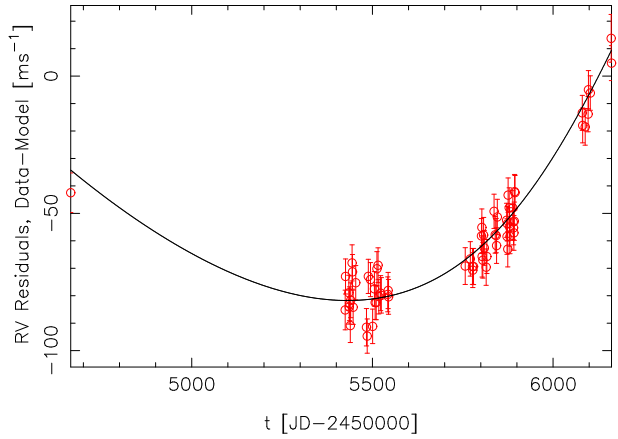}
\caption{Long-period variations in the HARPS radial velocities of GJ 4303 as modelled with a Keplerian function (solid curve).}\label{fig:GJ4303_long}
\end{figure}

The most noticeable feature in the HARPS radial velocity data of GJ 4303 (HIP 113201) is an apparent acceleration pattern resembling one caused by a substellar companion on a long-period orbit (Fig. \ref{fig:GJ4303_long}). However, because it is impossible to constrain the upper limit of the corresponding period parameter, we do not interpret this variation as a candidate companion.

\begin{figure}
\center
\includegraphics[angle=270, width=0.49\textwidth,clip]{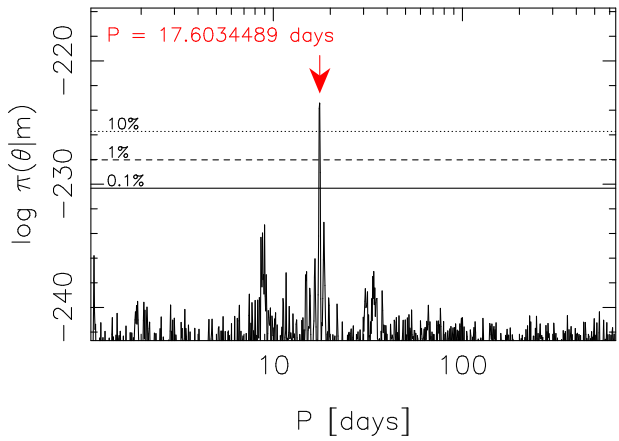}
\caption{As in Fig. \ref{fig:GJ3543_psearch} but for the second signal given the HARPS radial velocities of GJ 4303.}\label{fig:GJ4303_psearch}
\end{figure}

\begin{figure}
\center
\includegraphics[angle=270, width=0.49\textwidth,clip]{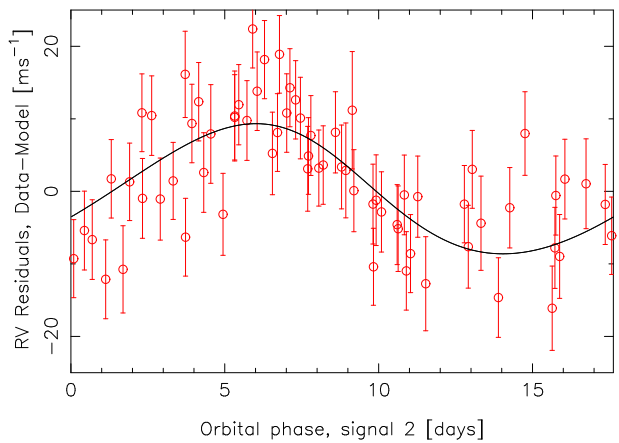}
\caption{Radial velocities of GJ 4303 shown as a function of the phase of the signal corresponding to GJ 4303 b.}\label{fig:GJ4303_curve}
\end{figure}

However, there appears to be evidence in favour of a periodicity in the HARPS data as well -- we observed a signal at a period of 17.615 [17.538, 17.703] days with an amplitude of 9.02 [5.08, 12.55] ms$^{-1}$. As this signal satisfied our detection criteria, and corresponds to a very unique maximum in the period space (Fig. \ref{fig:GJ4303_psearch}), we interpret it as a candidate hot Neptune planet with a minimum mass of 16.7 [9.3, 24.2] M$_{\oplus}$ (see also Fig. \ref{fig:GJ4303_curve}). This interpretation is possible because we did not find any significant periodicities in the HARPS activity indicators or ASAS photometry data of the star. There is thus no evidence that the radial velocity signal is connected to similar variations caused by stellar activity and/or rotation cycles.

\clearpage
\newpage

\section{Sample and data properties}\label{sec:sample_app}

\begin{deluxetable}{lccccc}
\tablewidth{0pt}
\tablecaption{General properties of the ASAS V-band photometry data sets: selected aperture (Ap.), number of grade A data ($N$), mean ($\mu_{\rm V}$) and standard deviation ($\sigma_{\rm V}$) of the V-band magnitude, and data set baseline ($T_{\rm obs}$).\label{tab:asas}}
\tablehead{
Star & Ap. & $N$ & $\mu_{\rm V}$ & $\sigma_{\rm V}$ & $T_{\rm obs}$ \\
 & & & (mmag) & (mmag) & (days)
}
\startdata
 CD-44836A &  0 &  518 &  12353.9 &     79.1 &  3243.09 \\
       GJ1 &  4 &   42 &   8555.6 &     10.3 &   351.99 \\
    GJ1002 &  0 &  162 &  13821.2 &    135.6 &  2870.15 \\
    GJ1008 &  3 &  340 &   9886.1 &     18.6 &  3231.17 \\
    GJ1009 &  1 &  452 &  11130.2 &     29.8 &  3248.23 \\
    GJ1036 &  1 &  491 &  11281.3 &     30.4 &  3243.07 \\
    GJ1044 &  3 &  429 &   9827.2 &     16.7 &  3279.16 \\
    GJ1046 &  1 &  604 &  11588.1 &     37.3 &  3242.11 \\
\enddata
\end{deluxetable}

\begin{deluxetable}{lcclcccc}
\tablewidth{0pt}
\tablecaption{List of target stars, their Gliese, HD and Hipparcos catalog numbers, spectral classifications (SC), parallaxes and $V$ and $J$ magnitudes. \label{tab:targets}}
\tablehead{
Target & HD & HIP & SC & $\pi$ & $\sigma_{\pi}$ & $V$ & $J$ \\
& & & & (mas) & (mas)}
\startdata
CD-44836A  &            &            &         M5 & 113.90 &  38.70 & 12.200 & 8.141 \\
GJ1        & 225213     & 439        &      M1.5V & 230.42 &   0.90 & 8.562  & 5.328 \\
GJ1002     &            &            &      M5.5V & 213.00 &   3.60 & 13.770 & 8.323 \\
GJ1008     &            & 1532       &        M0V & 447.43 &   1.65 & 9.897  & 7.376 \\
GJ1009     &            & 1734       &      M1.5V & 55.62  &   2.32 & 11.130 & 7.674 \\
GJ1036     &            & 6005       &      M2.5V & 61.68  &   3.43 & 11.284 & 7.845 \\
GJ1044     &            & 10337      &         K5 & 44.87  &   1.61 & 9.831  & 7.296 \\
GJ1046     &            & 10812      &      M2.5V & 71.06  &   3.23 & 11.595 & 7.924 \\
\enddata
\end{deluxetable}

\begin{deluxetable}{lcccccccc}
\tablewidth{0pt}
\tablecaption{General properties of the radial velocity data. Inclusion in the \citet{zechmeister2009} (Z) and \citep{bonfils2013} (B) samples; $N_{\rm X}$ for UVES, HARPS, PFS, and HIRES; total number ($N$) of data and data baseline ($T$). Additional data is from UCLES, HET, LICK, CORALIE, ELODIE, APF, HARPN, and/or SOPHIE. We have not reported data baselines for sets that were not analysed because of very low number of data or because the data were not considered trustworthy due to spectral contamination.\label{tab:data_velocities}}
\tablehead{
Target & Sample & $N_{\rm UVES}$ & $N_{\rm HARPS}$ & $N_{\rm PFS}$ & $N_{\rm HIRES}$ & $N$ & $T$ (days) }
\startdata
CD-44836A  &   B &      & 8    &      &      &    8 &  1774.04 \\
GJ1        & Z,B & 37   & 44   & 13   & 65   &  233 &  5574.44 \\
GJ1002     &   B &      & 7    &      &      &    7 &  2535.02 \\
GJ1008     &     &      & 6    & 23   &      &   29 &  3667.99 \\
GJ1009     &   Z & 34   & 12   &      & 8    &   54 &  3686.29 \\
GJ1036     &     &      & 1    &      &      &    1 &          \\
GJ1044     &     &      & 59   & 39   &      &   98 &  3323.89 \\
GJ1046     &   Z & 14   & 7    &      &      &   21 &  2644.75 \\
\enddata
\tablenotetext{a}{Highly variable radial velocities and activity indicators suggestive of spectral contamination.}
\end{deluxetable}

\begin{deluxetable}{lccccccccccccc}
\tablewidth{0pt}
\tablecaption{Estimated physical properties of the targets and inner ($a_{\rm min}$) and outer ($a_{\rm max}$) limits of the liquid-water habitable-zone based on the ''moist greenhouse`` and ''maximum greenhouse`` limits of \citet{kopparapu2013}.\label{tab:targets_physics}}
\tablehead{
\colhead{Target} & $M_{\star}^{1}$ & $T_{\rm eff}^{2}$ & $L_{\star}^{3}$ & [Fe/H]$^{4}$ & [Fe/H]$^{5}$ & $a_{\rm min}$ & $a_{\rm max}$ \\
 & (M$_{\odot}$) & (K) & (L$_{\odot}$) & & & (AU) & (AU)}
\startdata
CD-44836A  & 0.24 & 3137 & 0.0041 & -0.08 &       & 0.06  & 0.10  \\
GJ1        & 0.45 & 3522 & 0.0243 & -0.45 &       & 0.15  & 0.26  \\
GJ1002     & 0.12 & 2735 & 0.0002 & -0.19 &       & 0.01  & 0.02  \\
GJ1008     & 0.20 & 4048 & 0.0961 &       &       & 0.33  & 0.60  \\
GJ1009     & 0.51 & 3401 & 0.0153 & 0.11  & 0.17  & 0.12  & 0.20  \\
GJ1036     & 0.46 & 3410 & 0.0159 &       &       & 0.12  & 0.21  \\
GJ1044     &      & 4035 & 0.0940 &       &       & 0.32  & 0.59  \\
GJ1046     & 0.39 & 3298 & 0.0096 &       &       & 0.09  & 0.16  \\
\enddata
\tablerefs{(1) \citet{delfosse2000}; (2) \citet{casagrande2008}; (3) \citet{boyajian2012}; (4) \citet{neves2012}; (5) \citet{gaidos2014}}
\end{deluxetable}

\begin{deluxetable}{lccccccccccccc}
\tablewidth{0pt}
\tablecaption{Linear accelerations ($\dot{\gamma}$), stellar jitters ($\sigma_{J} = \min \big\{ \sigma_{1}, \sigma_{2}, ... \big\}$), and correlations with activity indicators of the HARPS ($c_{\rm BIS}$, $c_{\rm FWHM}$, and $c_{\rm S}$: S-index) and HIRES ($c_{\rm S}$) data. MAP and 99\% uncertainty estimates corresponding to 99\% credibility intervals.\label{tab:targets_estimates}}
\tablehead{
\colhead{Target} & $\dot{\gamma}$ & $\sigma_{J}$ & $c_{\rm BIS}$ & $c_{\rm FWHM}$ & $c_{\rm S}$ (HARPS) & $c_{\rm S}$ (HIRES) \\
 & (ms$^{-1}$year$^{-1}$) & (ms$^{-1}$) & (--) & (--) & (ms$^{-1}$) & (ms$^{-1}$)
}
\startdata
CD-44836A  & 1.07$\pm$0.96        & 1.25$_{-1.25}^{+2.42}$     &                      &                      &                      &                      \\
GJ1        & 0.05$\pm$0.38        & 1.79$_{-0.51}^{+0.71}$     & -0.266$\pm$0.511     & 0.093$\pm$0.139      & 5.4$\pm$13.2         & 2.3$\pm$10.2         \\
GJ1002     & 0.52$\pm$1.36        & 2.48$_{-1.63}^{+2.40}$     &                      &                      &                      &                      \\
GJ1008     & -0.25$\pm$0.96       & 1.89$_{-1.27}^{+2.44}$     &                      &                      &                      &                      \\
GJ1009     & -0.90$\pm$1.24       & 1.69$_{-1.24}^{+2.01}$     &                      &                      &                      &                      \\
GJ1036     &                      &                            &                      &                      &                      &                      \\
GJ1044     & 7.33$\pm$1.41        & 4.23$_{-1.13}^{+1.53}$     & -0.285$\pm$0.268     & -0.008$\pm$0.021     & 13.9$\pm$27.6        &                      \\
GJ1046     & -2.38$\pm$6.15       & 1.48$_{-1.48}^{+2.86}$     & 0.882$\pm$0.761      & 4.711$\pm$5.106      & -10.1$\pm$49.3       &                      \\
\enddata
\end{deluxetable}



\begin{thebibliography}{100}
\bibitem[\protect\astroncite{Alibert \& Benz}{2017}]{alibert2016} Alibert, Y. \& Benz, W. 2017, A\&A, 598, L5
\bibitem[\protect\astroncite{Anglada-Escud\'e et al.}{2012a}]{anglada2012} Anglada-Escud\'e, G., Arriagada, P., Vogt, S. S., et al. 2012a, ApJ, 751, L16
\bibitem[\protect\astroncite{Anglada-Escud\'e et al.}{2012b}]{anglada2012d} Anglada-Escud\'e, G., Boss, A. P., Weinberger, A. J., et al. 2012b, ApJ, 746, 37
\bibitem[\protect\astroncite{Anglada-Escud\'e et al.}{2013}]{anglada2013} Anglada-Escud\'e, G., Tuomi, M., Gerlach, E., et al. 2013, A\&A, 556, A126
\bibitem[\protect\astroncite{Anglada-Escud\'e et al.}{2014}]{anglada2014} Anglada-Escud\'e, G., Arriagada, P., Tuomi, M., et al. 2014, MNRAS, 443, L89
\bibitem[\protect\astroncite{Anglada-Escud\'e et al.}{2016}]{anglada2016} Anglada-Escud\'e, G., Amado, P. J., Barnes, J., et al. 2016a, Nature, 536, 437
\bibitem[\protect\astroncite{Anglada-Escud\'e et al.}{2016}]{anglada2016b} Anglada-Escud\'e, G., Tuomi, M., Arriagada, P., et al. 2016, ApJ, 830, 74
\bibitem[\protect\astroncite{Anglada-Escud\'e \& Butler}{2012}]{anglada2012b} Anglada-Escud\'e, G. \& Butler, R. P. 2012, ApJS, 200, 15
\bibitem[\protect\astroncite{Anglada-Escud\'e \& Tuomi}{2012}]{anglada2012c} Anglada-Escud\'e, G. \& Tuomi, M. 2012, A\&A, 548, A58
\bibitem[\protect\astroncite{Arriagada et al.}{2013}]{arriagada2013} Arriagada, P., Anglada-Escud\'e, G., Butler, R. P., et al. 2013, ApJ, 771, 42
\bibitem[\protect\astroncite{Astudillo-Defru et al.}{2015}]{astudillo2015} Astudillo-Defru, N., Bonfils, X., Delfosse, X., et al. 2015, A\&A, 575, A119
\bibitem[\protect\astroncite{Astudillo-Defru et al.}{2017}]{astudillo2017} Astudillo-Defru, N., Forveille, T., Bonfils, X., et al. 2017, A\&A, 602, A88
\bibitem[\protect\astroncite{Bailey et al.}{2009}]{bailey2009} Bailey, J., Butler, R. P., Tinney, C. G., et al. 2009, ApJ, 690, 743
\bibitem[\protect\astroncite{Ballard \& Johnson}{2016}]{ballard2016} Ballard, S. \& Johnson, J. A. 2016, ApJ, 816, 66
\bibitem[\protect\astroncite{Baluev}{2009}]{baluev2009} Baluev, R. V., 2009, MNRAS, 393, 969
\bibitem[\protect\astroncite{Baluev}{2013}]{baluev2013} Baluev, R. V., 2013, MNRAS, 429, 2052
\bibitem[\protect\astroncite{Batalha et al.}{2013}]{batalha2013} Batalha, N. M., Rowe, J. F., Bryson, S. T., et al. 2013, ApJS, 204, 24
\bibitem[\protect\astroncite{Biddle et al.}{2014}]{biddle2014} Biddle, L. I., Pearson, K. A., Crossfield, I. J. M., et al. 2014, MNRAS, 443, 1810
\bibitem[\protect\astroncite{Boisse et al.}{2011}]{boisse2011} Boisse, I., Bouchy, F., H\'ebrard, G., et al. 2011, A\&A, 528, A4
\bibitem[\protect\astroncite{Bonfils et al.}{2005}]{bonfils2005} Bonfils, X., Forveille, T., Delfosse, X., et al. 2005, A\&A, 443, L15
\bibitem[\protect\astroncite{Bonfils et al.}{2007}]{bonfils2007} Bonfils, X., Mayor, M., Delfosse, X., et al. 2007, A\&A, 474, 293
\bibitem[\protect\astroncite{Bonfils et al.}{2011}]{bonfils2011} Bonfils, X., Gillon, M., Forveille, T., et al. 2011, A\&A, 528, A111
\bibitem[\protect\astroncite{Bonfils et al.}{2012}]{bonfils2012} Bonfils, X., Gillon, M., Udry, S., et al. 2012, A\&A, 546, A27
\bibitem[\protect\astroncite{Bonfils et al.}{2013}]{bonfils2013} Bonfils, X., Delfosse, X., Udry, S., et al. 2013, A\&A, 549, A109
\bibitem[\protect\astroncite{Bonfils et al.}{2013b}]{bonfils2013b} Bonfils, X., Lo Curto, G., Correia, A. C. M., et al. 2013b, A\&A, 556, A110
\bibitem[\protect\astroncite{Bonomo et al.}{2014}]{bonomo2014} Bonomo, A. S., Sozzetti, A., Lovis, C., et al. 2014, A\&A, 572, A2
\bibitem[\protect\astroncite{Boyajian et al.}{2012}]{boyajian2012} Boyajian, T. S., von Braun, K., vn Belle, G., et al. 2012, ApJ, 757, 112
\bibitem[\protect\astroncite{Burt et al.}{2014}]{burt2014} Burt, J., Vogt, S. S., Butler, R. P., et al. 2014, ApJ, 789, 114
\bibitem[\protect\astroncite{Butler et al.}{2001}]{butler2001} Butler, R. P., Tinney, C. G., Marcy, G. W., et al. 2001, ApJ, 555, 410
\bibitem[\protect\astroncite{Butler et al.}{2004}]{butler2004} Butler, R. P., Vogt, S. S., Marcy, G. W., et al. 2004, ApJ, 617, 580
\bibitem[\protect\astroncite{Butler et al.}{2006}]{butler2006} Butler, R. P., Johnson, J. A., Marcy, G. W. et al. 2006, PASP, 118, 1685
\bibitem[\protect\astroncite{Butler et al.}{2009}]{butler2009} Butler, R. P., Howard, A. W., Vogt, S. S., \& Wright, J. T. 2009, ApJ, 691, 1738
\bibitem[\protect\astroncite{Butler et al.}{2017}]{butler2016} Butler, R. P., Vogt, S. S., Laughlin, G., et al. 2017, AJ, 153, 208
\bibitem[\protect\astroncite{Casagrande et al.}{2008}]{casagrande2008} Casagrande, L., Flynn, C., \& Bessel, M. 2008, MNRAS, 389, 585
\bibitem[\protect\astroncite{Chabrier \& Baraffe}{2000}]{chabrier2000} Chabrier, G. \& Baraffe, I. 2000, Ann. Rev. Astron. Astrophys., 38, 337
\bibitem[\protect\astroncite{Charbonneau et al.}{2009}]{charbonneau2009} Charbonneau, D., Berta, Z. K., Irwin, J., et al. 2009, Nature, 462, 891
\bibitem[\protect\astroncite{Chen \& Johns-Krull}{2013}]{chen2013} Chen, W. \& Johns-Krull, C. M. 2013, ApJ, 776 ,113
\bibitem[\protect\astroncite{Choi et al.}{2013}]{choi2013} Choi, J., McCarthy, C., Marcy, G. W. et al. 2013, ApJ, 764, 131
\bibitem[\protect\astroncite{Clanton \& Gaudi}{2015}]{clanton2015} Clanton, C. \& Gaudi, B. S. 2015, ApJ, 791, 91
\bibitem[\protect\astroncite{Clanton \& Gaudi}{2016}]{clanton2016} Clanton, C. \& Gaudi, B. S. 2015, ApJ, 819, 125
\bibitem[\protect\astroncite{Collins et al.}{2017}]{collins2017} Collins, J. M., Jones, H. R. A., \& Barnes, J. R. 2017, A\&A, 602, A48
\bibitem[\protect\astroncite{Cosentino et al.}{2012}]{cosentino2012} Cosentino, R., Lovis, C., Pepe, F., et al. 2012, Proc. SPIE, 8446, 84461V
\bibitem[\protect\astroncite{Crane et al.}{2010}]{crane2010} Crane, J. D., Shectman, S. A., Butler, R. P. et al. 2010, Proc. SPIE, 7735, 773553
\bibitem[\protect\astroncite{Cumming}{2004}]{cumming2004} Cumming, A. 2004, MNRAS, 354, 1165
\bibitem[\protect\astroncite{Cumming et al.}{2008}]{cumming2008} Cumming, A. Butler, R. P., Marcy, G. W., et al. 2008, PASP, 120, 531
\bibitem[\protect\astroncite{Dekker et al.}{2000}]{dekker2000} Dekker H., D'Odorico ,S., Kaufer, A., et al. 2000, Proc. SPIE, 4008, 534
\bibitem[\protect\astroncite{Delfosse et al.}{1998}]{delfosse1998} Delfosse, X., Forveille, T., Mayor, M., et al. 1998, A\&A, 338, L67
\bibitem[\protect\astroncite{Delfosse et al.}{2000}]{delfosse2000} Delfosse, X., Forveille, T., S\'egransan, D., et al. 2000, A\&A, 364, 217
\bibitem[\protect\astroncite{Delfosse et al.}{2013}]{delfosse2013} Delfosse, X., Bonfils, X., Forveille, T., et al. 2013, A\&A, 553, A8
\bibitem[\protect\astroncite{Diego et al.}{1990}]{diego1990} Diego, F., Charalambous, A., Fish, A. C., \& Walker, D. D. 1990, Proc. SPIE, 1235, 562
\bibitem[\protect\astroncite{Dressing \& Charbonneau}{2013}]{dressing2013} Dressing, C. D. \& Charbonneau, D. 2013, ApJ, 767, 95
\bibitem[\protect\astroncite{Dressing \& Charbonneau}{2015}]{dressing2015} Dressing, C. D. \& Charbonneau, D. 2015, ApJ, 807, 45
\bibitem[\protect\astroncite{Dressing et al.}{2015}]{dressing2015b} Dressing, C. D., Charbonneau, D., Dumusque, X., et al. 2015, ApJ, 800, 135
\bibitem[\protect\astroncite{Dumusque et al.}{2011}]{dumusque2011} Dumusque, X., Lovis, C., S\'egransan, D., et al. 2011, A\&A, 535, A55
\bibitem[\protect\astroncite{Dumusque et al.}{2012}]{dumusque2012} Dumusque, X., Pepe, F., Lovis, C., et al. 2012, Nature, 491, 207
\bibitem[\protect\astroncite{Dumusque et al.}{2014}]{dumusque2014} Dumusque, X., Bonomo, A. S., Haywood, R. D., et al. 2014, ApJ, 789, 154
\bibitem[\protect\astroncite{Dumusque et al.}{2017}]{dumusque2016} Dumusque, X., Borsa, F., Damasso, M., et al. 2017, A\&A, 598, 133
\bibitem[\protect\astroncite{Endl et al.}{2003}]{endl2003} Endl, M., Cochran, W. D., Tull, R. G., \& MacQueen, P. J. 2003, AJ, 126, 3099
\bibitem[\protect\astroncite{Endl et al.}{2006}]{endl2006} Endl, M., Cochran, W. D., K\"{u}rster, M. et al. 2006, ApJ, 649, 436
\bibitem[\protect\astroncite{Endl et al.}{2008}]{endl2008} Endl, M., Cochran, W. D., Wittenmyer, R. A., \& Boss, A. P. 2008, ApJ, 673, 1165
\bibitem[\protect\astroncite{Engle et al.}{2009}]{engle2009} Engle, S. G., Guinan, E. F., \& Mizusawa, T. 2009, in ''Future directions in ultraviolet spectroscopy``, AIP Conference Proceedings, Vol 1135, pp. 221-224
\bibitem[\protect\astroncite{Espinoza et al.}{2016}]{espinoza2016} Espinoza, N., Brahm, R., Jord\'an, A., et al. 2016, ApJ, 830, 43
\bibitem[\protect\astroncite{Feng et al.}{2016}]{feng2016} Feng, F., Tuomi, M., Jones, H. R. A., et al. 2016, MNRAS, 461, 2440
\bibitem[\protect\astroncite{Feroz et al.}{2011}]{feroz2011} Feroz, F., Balan, S. T., \& Hobson, M. P. 2011, MNRAS, 415, 3462
\bibitem[\protect\astroncite{Feroz \& Hobson}{2014}]{feroz2014} Feroz, F. \& Hobson, M. P. 2013, MNRAS, 437, 3540
\bibitem[\protect\astroncite{Forveille et al.}{2009}]{forveille2009} Forveille, T., Bonfils, X., Delfosse, X., et al. 2009, A\&A, 493, 645
\bibitem[\protect\astroncite{Forveille et al.}{2011}]{forveille2011} Forveille, T., Bonfils, X., Lo Curto, G., et al. 2011, A\&A, 526, A141
\bibitem[\protect\astroncite{Fressing et al.}{2013}]{fressin2013} Fressin, F., Torres, G., Charbonneau, D., et al. 2013, ApJ, 766, 81
\bibitem[\protect\astroncite{Gaidos et al.}{2013}]{gaidos2013} Gaidos, E., Fischer, D. A., Mann, A. W., \& howard, A. W. 2014, ApJ, 771, 18
\bibitem[\protect\astroncite{Gaidos et al.}{2014}]{gaidos2014} Gaidos, E., Mann, A. W., L\'epine, S., et al. 2014, MNRAS, 443, 2561
\bibitem[\protect\astroncite{Gatewood \& Eichhorn}{1973}]{gatewood1973} Gatewood, G. \& Eichhorn, H. 1973, AJ, 78, 769
\bibitem[\protect\astroncite{Gelman et al.}{1996}]{gelman1996} Gelman, A. G., Roberts, G. O., \& Gilks, W. R. 1996, Efficient Metropolis jumping rules. In Bernardo, J. M., Berger, J. O., David, A. F., \& Smith, A. F. M. (eds.), Bayesian Statistics V, pp. 599
\bibitem[\protect\astroncite{Gillon et al.}{2017}]{gillon2017} Gillon, M., Demory, B.-O., Van Grootel, V., et al. 2017, Nature Astronomy, 1, 0056
\bibitem[\protect\astroncite{Gomez da Silva et al.}{2011}]{gomez2011} Gomez da Silva, J., Santos, N. C., Bonfils, X., et al. 2011, A\&A, 534, A30
\bibitem[\protect\astroncite{Gomez da Silva et al.}{2012}]{gomez2012} Gomez da Silva, J., Santos, N. C., Bonfils, X., et al. 2012, A\&A, 541, A9
\bibitem[\protect\astroncite{Gregory}{2011}]{gregory2011} Gregory, P. C. 2011, MNRAS, 415, 2523
\bibitem[\protect\astroncite{Haario et al.}{2001}]{haario2001} Haario, H., Saksman, E., \& Tamminen, J. 2001, Bernoulli, 7, 223
\bibitem[\protect\astroncite{Haario et al.}{2006}]{haario2006} Haario, H., Laine, M., Mira, A., \& Saksman, E. 2006, Stat. Comp., 16, 339 
\bibitem[\protect\astroncite{Haghighipour et al.}{2010}]{haghighipour2010} Haghighipour, N., Vogt, S. S., Butler, R. P., et al. 2010, ApJ, 715, 271
\bibitem[\protect\astroncite{Hartman et al.}{2013}]{hartman2013} Hartman, J. D., Bakos, G. \'A., Kov\'acs, G., \& Noyes, R. W. 2013, MNRAS, 408, 475
\bibitem[\protect\astroncite{Hastings}{1970}]{hastings1970} Hastings, W. 1970, Biometrika 57, 97
\bibitem[\protect\astroncite{H\'ebrard et al.}{2013}]{hebrard2013} H\'evbrard, G., Almenara, J.-M., Santerne, A., et al. A\&A, 554, A114
\bibitem[\protect\astroncite{H\'ebrard et al.}{2014}]{hebrard2014} H\'evbrard, G., Santerne, A., Montagnier, G., et al. 2014, A\&A, 572, A93
\bibitem[\protect\astroncite{Henry et al.}{1994}]{henry1994} Henry, T. J., Kirkpatrick, J. D., \& Simons, D. A. 1994, AJ, 108, 1437
\bibitem[\protect\astroncite{H\o{}g et al.}{2000}]{hog2000} H\o{}g, E., Fabricius, C., Makarov, V. V., et al. 2000, A\&A, 355, L27
\bibitem[\protect\astroncite{Howard et al.}{2010}]{howard2010} Howard, A. W., Johnson, J. A., Marcy, G. W., et al. 2010, ApJ, 721, 1467
\bibitem[\protect\astroncite{Howard et al.}{2012}]{howard2012} Howard, A. W., Marcy, G. W., Bryson, S. T., et al. 2012, ApJS, 201, 15
\bibitem[\protect\astroncite{Howard et al.}{2014}]{howard2014} Howard, A. W., Marcy, G. W., Fischer, D. A., et al. 2014, ApJ, 794, 51
\bibitem[\protect\astroncite{Hunt-Walker et al.}{2012}]{hunt2012} Hunt-Walker, N. M., Hilton, E. J., Kowalski, A. F. et al. 2012, PASP, 124, 545
\bibitem[\protect\astroncite{Jenkins \& Tuomi}{2014}]{jenkins2014} Jenkins, J. S. \& Tuomi, M. 2014, ApJ, 794, 110
\bibitem[\protect\astroncite{Jenkins et al.}{2013}]{jenkins2013} Jenkins, J. S., Tuomi, M., Brasser, R., et al. 2013, ApJ, 771, 41
\bibitem[\protect\astroncite{Jenkins et al.}{2014}]{jenkins2014b} Jenkins, J. S., Yoma, N. B., Rojo, P., et al. 2014, MNRAS, 441, 2253
\bibitem[\protect\astroncite{Jenkins et al.}{2016}]{jenkins2016} Jenkins, J. S., Jones, H. R. A., Tuomi, M., et al. 2016, MNRAS, 466, 443
\bibitem[\protect\astroncite{Johnson et al.}{2007}]{johnson2007} Johnson, J. A., Butler, R. P., Marcy, G. W., et al. 2007, ApJ, 670, 833
\bibitem[\protect\astroncite{Johnson et al.}{2010a}]{johnson2010} Johnson, J. A., Howard, A. W., Marcy, G. W., et al. 2010a, PASP, 122, 149
\bibitem[\protect\astroncite{Johnson et al.}{2010b}]{johnson2010b} Johnson, J. A., Aller, K. M., Howard, A. W., \& Crepp, J. R. 2010b, PSAP, 122, 905
\bibitem[\protect\astroncite{Kane et al.}{2016}]{kane2016} Kane, S. R., Hill, M. L., Kasting, J. F., et al. 2016, ApJ, 830, 1
\bibitem[\protect\astroncite{Kane et al.}{2017}]{kane2017} Kane, S. R., von Braun, K., Henry, G. W. et al. 2017, ApJ, 835, 200
\bibitem[\protect\astroncite{Kass \& Raftery}{1995}]{kass1995} Kass, R. E. \& Raftery, A. E. 1995, J. Am. Stat. Ass., 430, 773
\bibitem[\protect\astroncite{Kasting et al.}{1993}]{kasting1993} Kasting, J. F., Whitmire, D. P., \& Reynolds, R. T. 1993, Icarus, 101, 108
\bibitem[\protect\astroncite{Kiraga \& Stepien}{2007}]{kiraga2007} Kiraga, M. \& Stepien, K. 2007, Acta Astronomica, 57, 149
\bibitem[\protect\astroncite{Kiraga}{2012}]{kiraga2012} Kiraga, M. 2012, Acta Astronomica, 62, 67
\bibitem[\protect\astroncite{Koen et al.}{2010}]{koen2010} Koen, C., Kilkenny, D., van Wyk, F., \& Marang, F. 2010, MNRAS, 403, 1949
\bibitem[\protect\astroncite{Kopparapu et al.}{2013}]{kopparapu2013} Kopparapu, R. K., Ramirez, R., Kasting, J. F., et al. 2013, ApJ, 765, 131
\bibitem[\protect\astroncite{Kopparapu}{2013}]{kopparapu2013b} Kopparapu, R. K. 2013, ApJ, 767, L8
\bibitem[\protect\astroncite{K\"urster et al.}{2008}]{kurster2008} K\"urster, M., Endl, M., \& Reffert, S. 2008, A\&A, 483, 869
\bibitem[\protect\astroncite{Liddle}{2007}]{liddle2007} Liddle, A. R. 2007, MNRAS, 377, L74
\bibitem[\protect\astroncite{Lo Curto et al.}{2013}]{locurto2013} Lo Curto, G., Mayor, M., Benz, W., et al. 2013, A\&A, 551, A59
\bibitem[\protect\astroncite{Lomb}{1976}]{lomb1976} Lomb, N. R. 1976, Astrophys. Space Sci., 39, 447
\bibitem[\protect\astroncite{Lovis et al.}{2011}]{lovis2011} Lovis, C., S\'egransan, D., Mayor, M., et al. 2011, A\&A, 528, A112
\bibitem[\protect\astroncite{Luger et al.}{2015}]{luger2015} Luger, R., Barnes, R., Lopez, E. et al. 2016, Astrobiology, 15, 57
\bibitem[\protect\astroncite{Mann et al.}{2015}]{mann2015} Mann, A. W., Feiden, G. A., Gaidos, E., et al. 2015, ApJ, 804, 64
\bibitem[\protect\astroncite{Marcy et al.}{1998}]{marcy1998} Marcy, G. W., Butler, R. P., Vogt, S. S., et al. 1998, ApJ, 505, L147
\bibitem[\protect\astroncite{Marcy et al.}{2001}]{marcy2001} Marcy, G. W., Butler, R. P., Fischer, D., et al. 2001, ApJ, 556, 296
\bibitem[\protect\astroncite{Mayor et al.}{2003}]{mayor2003} Mayor, M., Pepe, F., Queloz, D., et al. 2003, ESO Messenger, 114, 20
\bibitem[\protect\astroncite{Mayor et al.}{2009}]{mayor2009} Mayor, M., Bonfils, X., Forveille, T., et al. 2009, A\&A, 507, 487
\bibitem[\protect\astroncite{McQuillan et al.}{2014a}]{mcquillan2014} McQuillan, A., Mazeh, T., \& Aigrain, S. 2014a, ApJS, 211, 24
\bibitem[\protect\astroncite{McQuillan et al.}{2014b}]{mcquillan2014b} McQuillan, A., Aigrain, S. \& Mazeh, T. 2014b, MNRAS, 432, 1203
\bibitem[\protect\astroncite{Messina \& Guinan}{2003}]{messina2003} Messina, S. \& Guinan, E. F. 2003, A\&A, 409, 1017
\bibitem[\protect\astroncite{Metropolis et al.}{1953}]{metropolis1953} Metropolis, N., Rosenbluth, A. W., Rosenbluth, M. N., et al. 1953, J. Chem. Phys., 21, 1087
\bibitem[\protect\astroncite{Montet et al.}{2014}]{montet2014} Montet, B. T., Crepp, J. R., Johnson, J. A., et al. 2014, ApJ, 781, 28
\bibitem[\protect\astroncite{Morin et al.}{2008}]{morin2008} Morin, J., Donati, J.-F., Petit, P., et al. 2008, MNRAS, 390, 567
\bibitem[\protect\astroncite{Morton \& Swift}{2014}]{morton2014} Morton, T. D. \& Swift, J. 2014, ApJ, 791, 10
\bibitem[\protect\astroncite{Morton et al.}{2016}]{morton2016} Morton, T. D. Bryson, S. T., Coughlin, J. L., et al. 2016, ApJ, 822, 86
\bibitem[\protect\astroncite{Moutou et al.}{2011}]{moutou2011} Moutou, C., Mayor, M., Lo Curto, G., et al. 2011, A\&A, 527, A63
\bibitem[\protect\astroncite{Nakajima et al.}{1995}]{nakajima1995} Nakajima, T., Oppenheimer, B. R., Kulkarni, S. R., et al. 1995, Nature, 378, 463
\bibitem[\protect\astroncite{Nelson et al.}{2016}]{nelson2016} Nelson, B. E., Robertson, P. M., Payne, M. J., et al. 2016, MNRAS, 455, 2484
\bibitem[\protect\astroncite{Neves et al.}{2012}]{neves2012} Neves, V., Bonfils, X., Santos, N. C. et al. 2012, A\&A, 538, A25
\bibitem[\protect\astroncite{Neves et al.}{2013}]{neves2013} Neves, V., Bonfils, X., Santos, N. C. et al. 2013, A\&A, 551, A36
\bibitem[\protect\astroncite{Newton \& Raftery}{1994}]{newton1994} Newton, M. A. \& Raftery, A. E. 1994, J. R. Stat. Soc. B, 56, 3
\bibitem[\protect\astroncite{Newton et al.}{2016}]{newton2015} Newton, E. R., Irwin, J., Charbonneau, D., et al. 2016a, ApJ, 821, 93
\bibitem[\protect\astroncite{Newton et al.}{2016}]{newton2016} Newton, E. R., Irwin, J., Charbonneau, D., et al. 2016b, ApJ, 821, L19
\bibitem[\protect\astroncite{Perruchot et al.}{2008}]{perruchot2008} Perruchot, S., Kohler, D., Bouchy, F., et al. 2008, Proc. SPIE, 7014, 70140
\bibitem[\protect\astroncite{Petigura et al.}{2013}]{petigura2013} Petigura, E. A., Howard, A. W., \& Marcy, G. W. 2013, PNAS, 110, 19273
\bibitem[\protect\astroncite{Pojma\'nski}{1997}]{pojmanski1997} Pojma\'nski, G. 1997, Acta Astronomica, 47, 467
\bibitem[\protect\astroncite{Pojma\'nski}{2002}]{pojmanski2002} Pojma\'nski, G. 2002, Acta Astronomica, 52, 397
\bibitem[\protect\astroncite{Quintana et al.}{2015}]{quintana2015} Quintana, E. V., Barclay, T., Raymond, S. N., et al. 2015, Science, 6181, 277
\bibitem[\protect\astroncite{Radovan et al.}{2014}]{radovan2014} Radovan, M. V., Lanclos, K., Holden, B. P., et al. 2014, Proc. SPIE, 9145, 91452B
\bibitem[\protect\astroncite{Reiners et al.}{2013}]{reiners2013} Reiners, A., Shulyak, D., Anglada-Escud\'e, G., et al. 2013, A\&A, 552, A103
\bibitem[\protect\astroncite{Reinhold \& Gizon}{2015}]{reinhold2015} Reinhold, T. \& Gizon, L. 2015, A\&A, 583, A65
\bibitem[\protect\astroncite{Ribas et al.}{2018}]{ribas2018} Ribas, I., Tuomi, M., Reiners, A., et al. 2018, Nature, 563, 365
\bibitem[\protect\astroncite{Rivera et al.}{2005}]{rivera2005} Rivera, E. J., Lissauer, J. J., Butler, R. P., et al. 2005, ApJ, 634, 625
\bibitem[\protect\astroncite{Rivera et al.}{2010}]{rivera2010} Rivera, E. J., Laughlin, G., Butler, R. P., et al. 2010, ApJ, 719, 890
\bibitem[\protect\astroncite{Robertson et al.}{2014}]{robertson2014} Robertson, P., Mahadevan, S., Endl, M., \& Roy, A. 2014, Science, 345, 440
\bibitem[\protect\astroncite{Robertson et al.}{2015a}]{robertson2015} Robertson, P., Endl, M., Henry, G. W., et al. 2015a, ApJ, 801, 79
\bibitem[\protect\astroncite{Robertson et al.}{2015b}]{robertson2015b} Robertson, P., Roy, A. \& Mahadevan, S. 2015b, ApJ, 805, L22
\bibitem[\protect\astroncite{Santos et al.}{2010}]{santos2010} Santos, N. C., Gomes da Silva, J., Lovis, C., \& Melo, C. 2010, A\&A, 511, A54
\bibitem[\protect\astroncite{Santos et al.}{2014}]{santos2014} Santos, N. C., Mortier, R., Faria, J. P., et al. 2014, A\&A, 566, A35
\bibitem[\protect\astroncite{Scargle}{1982}]{scargle1982} Scargle, J. D. 1982, ApJ, 263, 835
\bibitem[\protect\astroncite{Selsis et al.}{2007}]{selsis2007} Selsis, F., Kasting, J. F., Levrard, B., et al. 2007, A\&A, 476, 1373
\bibitem[\protect\astroncite{Silburt et al.}{2015}]{silburt2015} Silburt, A., Gaidos, E., \& Wu, Y. 2015, ApJ, 799, 180
\bibitem[\protect\astroncite{Spiesman \& Hawley}{1986}]{spiesman1986} Spiesman, W. J. \& Hawley, S. L. 1986, AJ, 92, 664
\bibitem[\protect\astroncite{Su\'arez Mascare\~{n}o et al.}{2015}]{mascareno2015} Su\'arez Mascare\~{n}o, A. S., Rebolo, R., Gon\'alez Hern\'andez, J. I., \& Esposito, M. 2015, MNRAS, 452, 2745
\bibitem[\protect\astroncite{Su\'arez Mascare\~{n}o et al.}{2016}]{suarezmascareno2016} Su\'arez Mascare\~{n}o, A. S., Gon\'alez Hern\'andez, J. I., Rebolo, R., et al. 2016, A\&A, accepted (arXiv:1611:12022)
\bibitem[\protect\astroncite{Suzuki et al.}{2016}]{suzuki2016} Suzuki, D., Bennett, D. P., Sumi, T., et al. 2016, ApJ, 833, 145
\bibitem[\protect\astroncite{Tian \& Ida}{2015}]{tian2015} Tian, F. \& Ida, S. 2015, Nature Geoscience, 8, 177
\bibitem[\protect\astroncite{Tikhonov \& Arsenin}{1977}]{tikhonov1977} Tikhonov, A. N. \& Arsenin, V. Y. 1977, Solutions of ill-posed problems, Scripta series in mathematics (Vh Winston)
\bibitem[\protect\astroncite{Tinney et al.}{2001}]{tinney2001} Tinney, C. G., Butler, R. P., Marcy, G. W., et al. 2001, ApJ, 551, 507
\bibitem[\protect\astroncite{Tinney et al.}{2011}]{tinney2011} Tinney, C. G., Wittenmyer, R. A., Butler, R. P., et al. 2011, ApJ, 732, 31
\bibitem[\protect\astroncite{Torres et al.}{2006}]{torres2006} Torres, C. A. O., Quast, G. R., Da Silva, L., et al. 2006, A\&A, 460, 695
\bibitem[\protect\astroncite{Tuomi}{2011}]{tuomi2011} Tuomi, M. 2011, A\&A, 528, L5
\bibitem[\protect\astroncite{Tuomi}{2012}]{tuomi2012} Tuomi, M. 2012, A\&A, 543, A52
\bibitem[\protect\astroncite{Tuomi}{2014}]{tuomi2014b} Tuomi, M. 2013, MNRAS, 440, L1
\bibitem[\protect\astroncite{Tuomi \& Anglada-Escud\'e}{2013}]{tuomi2013c} Tuomi, M. \& Anglada-Escud\'e 2013, A\&A, 556, A111
\bibitem[\protect\astroncite{Tuomi et al.}{2011}]{tuomi2011b} Tuomi, M., Pinfield, D., \& Jones, H. R. A. 2011, A\&A, 532, A116
\bibitem[\protect\astroncite{Tuomi et al.}{2013a}]{tuomi2013a} Tuomi, M., Anglada-Escud\'e, G., Gerlach, E., et al. 2013a A\&A, 549, A48
\bibitem[\protect\astroncite{Tuomi et al.}{2013b}]{tuomi2013b} Tuomi, M., Jones, H. R. A., Jenkins, J. S., et al. 2013b, A\&A, 551, A79
\bibitem[\protect\astroncite{Tuomi et al.}{2014}]{tuomi2014} Tuomi, M., Jones, H. R. A., Barnes, J. R., et al. 2014, MNRAS, 441, 1545
\bibitem[\protect\astroncite{Tuomi et al.}{2018}]{tuomi2018} Tuomi, M., Jones, H. R. A., Barnes, J. R., et al. 2018, AJ, 155, 192
\bibitem[\protect\astroncite{Udry et al.}{2007}]{udry2007} Udry, S., Bonfils, X., Delfosse, X., et al. 2007, A\&A, 469, L43
\bibitem[\protect\astroncite{Youdin}{2011}]{youdin2011} Youdin, A. N. 2011, ApJ, 742, 38
\bibitem[\protect\astroncite{van de Kamp}{1963}]{vandekamp1963} van de Kamp, P. 1963, AJ, 68, 515
\bibitem[\protect\astroncite{van de Kamp}{1969}]{vandekamp1969} van de Kamp, P. 1969, AJ, 74, 757
\bibitem[\protect\astroncite{van Leeuwen}{2007}]{vanleeuwen2007} van Leeuwen, F., 2007, A\&A, 474, 653
\bibitem[\protect\astroncite{Vogt}{1987}]{vogt1987} Vogt, S. S. 1987, PASP, 99, 1214
\bibitem[\protect\astroncite{Vogt et al.}{1994}]{vogt1994} Vogt, S. S., Allen, S. L., Bigelow, B. C., et al. 1994, Proc. SPIE, 2198, 362
\bibitem[\protect\astroncite{Vogt et al.}{2010}]{vogt2010} Vogt, S., Butler, P., Rivera, E., et al. 2010. ApJ, 723, 954
\bibitem[\protect\astroncite{Vogt et al.}{2014}]{vogt2014} Vogt, S. S., Radovan, M., Kibrick, R., et al. 2014, PASP, 126, 359 
\bibitem[\protect\astroncite{Walkowicz \& Basri}{2013}]{walkowicz2013} Walkowicz, L. M. \& Hawley, S. L. 2009, MNRAS, 436, 1883
\bibitem[\protect\astroncite{Weiss \& Marcy}{2014}]{weiss2014} Weiss, L. M., \& Marcy G. W. 2014, ApJ, 783, L6
\bibitem[\protect\astroncite{Winters et al.}{2015}]{winters2015} Winters, J. G., Henry, T., J., Lurie, J. C., et al. 2015, AJ, 149, 5
\bibitem[\protect\astroncite{Wittenmyer et al.}{2012}]{wittenmyer2012} Wittenmyer, R. A., Horner, J., Tuomi, M., et al. 2012, ApJ, 753, 169
\bibitem[\protect\astroncite{Wittenmyer et al.}{2014}]{wittenmyer2014} Wittenmyer, R. A., Tuomi, M., Butler, R. P., et al. 2014, ApJ, 719, 114
\bibitem[\protect\astroncite{Wolfgang et al.}{2016}]{wolfgang2016} Wolfgang, A., Rogers, L. A., \& Ford, E. B. 2016, ApJ, 825, 19
\bibitem[\protect\astroncite{Wright}{2005}]{wright2005} Wright, J. T. 2005, PASP, 117, 657
\bibitem[\protect\astroncite{Wright et al.}{2016}]{wright2016} Wright, D. J., Wittenmyer, R. A., Tinney, C. G., et al. 2016, ApJ, 817, L20
\bibitem[\protect\astroncite{Zakamska et al.}{2011}]{zakamska2011} Zakamska, N. L., Pan, M., \& Ford, E. B. 2011, MNRAS, 410, 1895
\bibitem[\protect\astroncite{Zechmeister et al.}{2009}]{zechmeister2009} Zechmeister, M., K\"urster, M., \& Endl, M. 2009, A\&A, 505, 859
\bibitem[\protect\astroncite{Zeng \& Jacobsen}{2016}]{zeng2016} Zeng, L., Sasselov, D. D., \& Jacobsen S. B. 2016, ApJ, 819, 127
\end{thebibliography}
\end{document}